\documentclass{article}
 \usepackage{JOEconfig}

\usepackage[T1]{fontenc}

\usepackage{graphicx}

\usepackage{hyperref}
\hypersetup{
    linktocpage,  
    colorlinks,
    citecolor=black,
    filecolor=blue,
    linkcolor=blue,
    urlcolor=blue
}

\usepackage{accents} 
\usepackage{bm} 

\usepackage{amsmath, amsthm} 
\usepackage{array}
\usepackage{longtable,tabularx} 
\usepackage{enumitem} 
\setlist[itemize,1]{left=3ex,labelsep=0.75ex,itemsep=1pt,topsep=1pt,label={\scriptsize{\textbullet}}}
\setlist[enumerate,1]{left=3ex,labelsep=0.75ex,itemsep=1pt,topsep=1pt}

\usepackage{stackengine}
\usepackage{xfrac}
\usepackage{tensor}
\usepackage{mathtools} 
\usepackage{bigints}
\usepackage{xcolor} 
\usepackage{cancel} 

 \usepackage[trueslanted]{newtxtext}

\usepackage[slantedGreek,libertine,vvarbb]{newtx}

\DeclareRobustCommand{\textsb}[1]{{\small\textbf{#1}}} 

\usepackage[frak=euler,scr=cm]{mathalpha} 

\usepackage{upgreek} 

\usepackage[artemisia]{textgreek}

\title{Linear and Regular Kepler-Manev Dynamics via Projective Transformations: A Geometric Perspective} 
\author{Joseph T.A. Peterson\footnote{Graduate Research Assistant, Aerospace Engineering, Texas A\&M University, College Station, Texas.}, 
Manoranjan Majji\footnote{Edward "Pete" Aldridge Endowed Professor of Aerospace Engineering, Director of LASR Laboratory, Texas A\&M University, College Station, Texas.}, and
John L.~Junkins\footnote{Distinguished Professor of Aerospace Engineering, Director of The Hagler Institute for Advanced Study, Texas A\&M University, College Station, Texas.}
}

\begin{document}
\date{}
\maketitle



\newcommand{\pgrf}[1]{\noindent\textbf{#1}}
\newcommand\blfootnote[1]{%
  \begingroup
  \renewcommand\thefootnote{}\footnote{#1}%
  \addtocounter{footnote}{-1}%
  \endgroup
}

\newcommand{\cf}[0]{f}%
\newcommand{\cff}[0]{{\sl{f}}}

\newcommand{\tx}[1]{{\text{#1}}}
\newcommand{\txi}[1]{{\textnormal{\textit{#1}}}}
\newcommand{\tbf}[1]{{\textbf{\textup{\textrm{#1}}}}}
\newcommand{\trm}[1]{{\textup{\textrm{#1}}}}
\newcommand{\txnonb}[1]{{\textmd{#1}}} 
\newcommand{\txup}[1]{{\textup{#1}}} 
\newcommand{\bfi}[1]{{\textbf{\textit{#1}}}}
\newcommand{\Bfi}[1]{{ \scalebox{1.18}{\textbf{\textit{#1}}}  }}

\newcommand{\rmsb}[1]{{\textup{\textsb{#1}}}}%
\newcommand{\itsb}[1]{{\textit{\textsb{#1}}}}

\renewcommand{\sl}[1]{{\textnormal{\textrm{\textsl{#1}}}}}
\newcommand{\slb}[1]{{\textbf{\textsl{#1}}}}

\newcommand{\txsfb}[1]{\textbf{\textit{\textsf{#1}}}}
\newcommand{\sfi}[1]{{\textit{\textsf{#1}}}}%
\newcommand{\sfup}[1]{{\textup{\textsf{#1}}}}
\newcommand{\sfbup}[1]{ {\textbf{\textup{\textsf{#1}}}} }%

\renewcommand{\sc}[1]{{\textsc{#1}}}
\newcommand*{\rmsc}[1]{{\textrm{\textup{\textsc{#1}}}}}
\newcommand*{\bsc}[1]{{\textbf{\textsc{#1}}}}
\newcommand*{\rmbsc}[1]{{\textbf{\textrm{\textsc{#1}}}}}
\newcommand*{\itsc}[1]{{\textit{\textrm{\textsc{#1}}}}}
\newcommand*{\bisc}[1]{{\textit{\textbf{\textsc{#1}}}}}
\newcommand*{\sfsc}[1]{\textsf{\textit{\textsc{#1}}}}
\newcommand*{\sfbsc}[1]{\textbf{\textsf{\textit{\textsc{#1}}}}}
\newcommand{\scg}[0]{{\trm{\textsc{g}}}}
\newcommand{\sck}[0]{{\trm{\textsc{k}}}}
\newcommand{\scm}[0]{{\trm{\textsc{m}}}}
\newcommand{\scb}[0]{{\trm{\textsc{b}}}}

\newcommand{\isc}[1]{{\scriptstyle{\textit{\textrm{\textsc{#1}}}}}} 
\newcommand{\ssc}[1]{{\scalebox{0.6}{$#1$}}} 

\newcommand{\eq}[1]{\text{$#1$}}

\newcommand{\bemph}[1]{\itsb{#1}} 
 \newcommand{\sbemph}[1]{\rmsb{#1}} 


\newcommand{\smsize}[1]{ {\text{\small{#1}}} }
\newcommand*{\fnsize}[1]{ {\text{\footnotesize{#1}}} }%
\newcommand{\scrsize}[1]{ {\text{\scriptsize{#1}}} }
\newcommand{\nssize}[1]{ {\text{\normalsize{#1}}} }

\newcommand*{\nmsz}[1]{{\displaystyle{#1}}}
\newcommand*{\scrsz}[1]{{\scriptstyle{#1}}}
\newcommand*{\ssz}[1]{{\scriptstyle{#1}}}
\newcommand*{\ii}[1]{{\scriptscriptstyle{#1}}} 
\renewcommand*{\ss}[1]{ {\scalebox{0.7}{$#1$}} }

\newcommand*{\smsz}[1]{\text{\small{$#1$}} }
\newcommand*{\fnsz}[1]{\text{\footnotesize{$#1$}} }
\newcommand*{\ns}[1]{{\scalebox{1.15}{$#1$}} }
\renewcommand*{\lg}[1]{{\scalebox{1.25}{$#1$}} }

\newcommand{\mscale}[2][.7]{ {\text{\scalebox{#1}{$#2$}}} } 

\newcommand{\ttfrac}[2]{{\scriptstyle{\frac{#1}{#2}}}}


\newcommand{\til}[1]{\tilde{#1}}
\newcommand{\wt}[1]{ \widetilde{#1}  }
\newcommand{\wh}[1]{\widehat{#1}}
\newcommand{\ol}[1]{\overline{#1}}

\newcommand{\uln}[1]{{\underline{\smash{#1}\mkern-1mu}{\mkern1mu}}}
\newcommand{\ubar}[1]{ \underbar{$#1\hspace{-0.15mm}$}\hspace{0.15mm} }
\newcommand{\ub}[1]{{\underaccent{\text{—}}{\smash{#1}}}} 

\newcommand{\dt}[1]{\accentset{\mbox{\scriptsize\textbf{.}}}{#1}}
\newcommand*{\ddt}[1]{ \accentset{\mbox{\scriptsize\bfseries..}}{#1} }
\newcommand*{\Dt}[1]{\accentset{\mbox{\normalsize\bfseries .}}{#1}}
\newcommand*{\DDt}[1]{ \accentset{\mbox{\bfseries .\hspace{-0.03ex}.}}{#1} }

\newcommand{\rng}[1]{ \mathring{#1} }
\newcommand{\rrng}[1]{{\rlap{$\,\mathring{\vphantom{#1}}\mkern4.5mu\mathring{\vphantom{#1}}$}}#1}
\newcommand{\rring}[1]{ \accentset{\circ\circ}{#1} }
\newcommand{\ringp}[1]{\accentset{\circ'}{#1} }
\newcommand{\rringp}[1]{\accentset{\circ\circ'}{#1}}

\newcommand{\pdt}[1]{ \acute{#1} }
\newcommand{\pddt}[1]{ \rlap{$\;\,\acute{\vphantom{#1}}\!\acute{\vphantom{#1}}$}#1 }
\newcommand{\pdot}[1]{{\'{#1}}}
\newcommand{\pddot}[1]{{\H{#1}}}

\newcommand{\tridot}[1]{ \accentset{\triangle}{#1} }
\newcommand{\triddot}[1]{ \accentset{\triangle\triangle}{#1} }
\newcommand{\tridt}[1]{ \accentset{\blacktriangle}{#1} }
\newcommand{\triddt}[1]{ \accentset{\blacktriangle\!\blacktriangle}{#1} }
\newcommand{\deldot}[1]{ \accentset{\triangledown}{#1} }
\newcommand{\delddot}[1]{ \accentset{\triangledown\triangledown}{#1} }

\newcommand{\bxdot}[1]{ \accentset{\square}{#1} }
\newcommand{\bxddot}[1]{ \accentset{\square\square}{#1} }
\newcommand{\sqdt}[1]{ \accentset{\blacksquare}{#1} }
\newcommand{\sqddt}[1]{ \accentset{\blacksquare\blacksquare}{#1} }

\newcommand{\hdot}[1]{ \accentset{\mbox{\hspace{0.1mm}.}}{#1} }
\newcommand{\hddot}[2][.2ex]{\ddot{\raisebox{0pt}[\dimexpr\height+#1][\depth]{$#2$}}}
\newcommand{\ggvec}[2][-2pt]{\vec{\raisebox{0pt}[\dimexpr\height+#1][\depth]{$#2$}}}

\newcommand{\harp}[1]{ {\accentset{\rightharpoonup}{#1}} }
\newcommand{\varvec}[1]{ {\accentset{\to}{#1}} }


\let\oldcdot\cdot
\renewcommand*{\cdot}{{\mkern1.5mu\oldcdot\mkern1.5mu}}
\newcommand{\slot}[0]{{\oldcdot}} 
\newcommand{\cdt}[0]{ \pmb{\cdot} }
\newcommand{\bdt}[0]{ \bs{\cdot} }

\newcommand{\txblt}[0]{{\text{\small{\textbullet\hspace{.7ex}}}}}
\newcommand*{\sblt}[0]{ \raisebox{0.25ex}{$\scriptscriptstyle{\bullet}$}  }
\newcommand*{\bltt}[0]{ \text{\large{${\,\,\bullet\,\,}$}} }
\newcommand*{\nmblt}[0]{ \raisebox{0.15ex}{$\scriptstyle{\bullet}$}  }

\newcommand{\cddt}[0]{{\mkern2mu\pmb{\cdot}\mkern2mu}}
\newcommand{\cddot}[0]{ {\mkern2mu\uln{\pmb{\cdot}}\mkern2mu} }

\newcommand*{\otms}[0]{ {\raisebox{0.15ex}{$\scriptstyle{\,\otimes\,}$}}  }
\newcommand*{\opls}[0]{ \raisebox{0.15ex}{$\scriptstyle{\,\oplus\,}$}  }
\newcommand*{\wdg}[0]{ \raisebox{0.15ex}{$\scriptstyle{\,\wedge\,}$}  }
\newcommand{\botimes}[0]{ \scalebox{1.3}{$\otimes$} }
\newcommand{\boplus}[0]{ \scalebox{1.3}{$\oplus$} }
\newcommand{\bwedge}[1]{ {\scalebox{1}[1.3]{$\wedge$}}^{\!#1} }

\newcommand*{\ssqr}[0]{ \raisebox{0.15ex}{$\scriptstyle{\blacksquare}$}  }
\newcommand*{\iisqr}[0]{ \raisebox{0.25ex}{$\,\scriptscriptstyle{\blacksquare}\,$}  }
\newcommand*{\sbx}[0]{ \raisebox{0.15ex}{$\scriptstyle{\square}$}  }
\newcommand*{\iibx}[0]{ \raisebox{0.25ex}{$\,\scriptscriptstyle{\square}\,$}  }

\newcommand{\dimd}[0]{\diamond}%
\newcommand*{\tri}[0]{\triangle}%
\newcommand*{\trid}[0]{\triangledown} %
\newcommand*{\rtri}[0]{\triangleright} %
\newcommand*{\ltri}[0]{\triangleleft} %
\newcommand*{\stri}[0]{ \raisebox{0.25ex}{$\scriptstyle{\,\triangle\,}$} }
\newcommand*{\strid}[0]{\raisebox{0.25ex}{$\scriptstyle{\,\triangledown\,}$}}%
\newcommand*{\srtri}[0]{ \raisebox{0.25ex}{$\scriptstyle{\,\triangleright\,}$} }
\newcommand*{\slri}[0]{\raisebox{0.25ex}{$\scriptstyle{\,\triangleleft\,}$}}%

\newcommand*{\rnlarrow}[0]{ \overset{\nLeftarrow\;}{\Rightarrow}}
\newcommand*{\Rnlarrow}[0]{ \overset{\Leftarrow \!/ \!=\,}{\Longrightarrow}}
\newcommand*{\lnrarrow}[0]{ \overset{\,\nRightarrow}{\Leftarrow}}
\newcommand*{\Lnrarrow}[0]{ \overset{=\! /\!\Rightarrow\,}{\Longleftarrow}}
\newcommand{\nDownarrow}[0]{\rotatebox[origin=c]{-90}{$\nRightarrow$}}
\newcommand{\nUparrow}[0]{\rotatebox[origin=c]{90}{$\nRightarrow$}}


\newcommand*{\shrp}[0]{{\scalebox{0.55}{$\sharp$}}}
\newcommand*{\flt}[0]{{\scalebox{0.6}{$\flat$}}}

\newcommand{\iio}[0]{{\scalebox{0.45}{$\bs{\circ}$}} }
\newcommand{\iix}[0]{{\scriptscriptstyle{\times}} }
\newcommand{\iistr}[0]{{\scriptscriptstyle{\star}} }
\newcommand{\str}[0]{{\scriptscriptstyle{\star}} }
\newcommand{\drg}[0]{{\ss{\dagger}}} 

\newcommand{\hdge}[1]{ #1^{\scriptscriptstyle{\star}}  }
\newcommand{\hodge}[0]{\star}
\newcommand{\varhodge}[0]{ {\,^{\scriptscriptstyle{\bar{\star}}}} }
\newcommand{\ax}[1]{ {#1^{\scriptscriptstyle{\times}}} }
\newcommand{\axx}[1]{ {#1^{\scriptstyle{\times}}} }

\newcommand{\inv}[1]{ #1^{\scriptscriptstyle{-\!1}} }
\newcommand{\negg}[0]{{\textnormal{-}}}

 \newcommand{\trn}[1]{ #1^{\ii{\textsf{\textup{T}}}}  }
\newcommand{\invtrn}[1]{ #1^{\ii{\textsf{\textup{--T}}}} }
 
\newcommand*{\zer}[0]{ {\textrm{\tiny{\textit{0}}}} } %
\newcommand*{\zr}[0]{ {\ss{0}} }%
\newcommand{\nozer}[0]{ {\scriptscriptstyle{\emptyset}} }

\newcommand*{\sodot}[0]{ \raisebox{0.12ex}{$\scriptstyle{\odot}$}  }


\newcommand*{\mrm}[1]{ {\mathrm{#1}} } 
\newcommand*{\iimrm}[1]{ {\ii{\mathrm{#1}}} }
\newcommand*{\iirm}[1]{ {\ii{\textrm{#1}}} }

\newcommand*{\mbf}[1]{{\mathbf{#1}}} 

\newcommand{\rmb}[1]{\tbf{#1}}


\newcommand{\bs}[1]{{\boldsymbol{#1}}}
\newcommand{\bsi}[1]{\boldsymbol{\mit{#1}}} 
\newcommand{\accalign}[2]{ \mkern2mu #1{\mkern-2mu #2} }
\newcommand{\tbs}[1]{ {\accalign{\tilde}{\bs{#1}}} }%
\newcommand{\wtbs}[1]{ {\accalign{\widetilde}{\bs{#1}}} }%
\newcommand{\hbs}[1]{ {\accalign{\hat}{\bs{#1}}} }%
\newcommand{\whbs}[1]{ {\accalign{\widehat}{\bs{#1}}} }%
\newcommand{\barbs}[1]{ {\accalign{\bar}{\bs{#1}}} }%
\newcommand{\vecbs}[1]{ {\accalign{\vec}{\bs{#1}}} }%
\newcommand*{\dtbs}[1]{ {\accalign{\dt}{\bs{#1}}} }%
\newcommand*{\ddtbs}[1]{ {\accalign{\ddt}{\bs{#1}}} }%

\newcommand{\tbm}[1]{\tilde{\bm{#1}}}
\newcommand{\wtbm}[1]{\widetilde{\bm{#1}}}
\newcommand{\hbm}[1]{\hat{\bm{#1}}}
\newcommand{\whbm}[1]{\widehat{\bm{#1}}}
\newcommand{\barbm}[1]{\bar{\bm{#1}}}

\newcommand*{\nbs}[1]{ \scalebox{1.15}{$\boldsymbol{#1}$} }
\newcommand*{\lbs}[1]{ \scalebox{1.4}{$\boldsymbol{#1}$} }
\newcommand*{\nbm}[1]{ \scalebox{1.2}{$\bm{#1}$} }
\newcommand*{\lbm}[1]{ \scalebox{1.4}{$\bm{#1}$} }


\newcommand{\msfb}[1]{{\mathsfb{#1}}} 

\newcommand{\sfb}[1]{{\txsfb{#1}}} 

\renewcommand{\accalign}[2]{ \mkern1.5mu #1{\mkern-1.5mu #2} }
\newcommand{\tsfb}[1]{ {\accalign{\tilde}{\sfb{#1}}} }%
\newcommand{\wtsfb}[1]{ {\accalign{\widetilde}{\sfb{#1}}} }%
\newcommand{\hsfb}[1]{ {\accalign{\hat}{\sfb{#1}}} }%
\newcommand{\whsfb}[1]{ {\accalign{\widehat}{\sfb{#1}}} }%
\newcommand{\barsfb}[1]{ {\accalign{\bar}{\sfb{#1}}} }%
\newcommand{\bsfb}[1]{ {\accalign{\bar}{\sfb{#1}}} }%
\newcommand{\vecsfb}[1]{ {\accalign{\vec}{\sfb{#1}}} }%
\newcommand*{\dtsfb}[1]{ {\accalign{\dt}{\sfb{#1}}} }%
\newcommand*{\ddtsfb}[1]{ {\accalign{\ddt}{\sfb{#1}}} }%

\newcommand*{\sfg}[0]{{\bs{g}}}


\newcommand{\tn}[1]{{\sfb{#1}}} 

\newcommand*{\mbb}[1]{ {\mathbb{#1}} } 
\newcommand*{\tmbb}[1]{ \tilde{\mbb{#1}} }
\newcommand*{\wtmbb}[1]{ \widetilde{\mbb{#1}} }
\newcommand*{\hmbb}[1]{ \hat{\mbb{#1}} }
\newcommand*{\whmbb}[1]{ \widehat{\mbb{#1}} }
\newcommand*{\bbk}[0]{\Bbbk} 

\newcommand*{\mfrak}[1]{ {\mathfrak{#1}} }
\newcommand*{\bfrak}[1]{ {\bs{\mathfrak{#1}}} }
\newcommand*{\tmfrak}[1]{ \tilde{\mfrak{#1}} }
\newcommand*{\wtmfrak}[1]{ \widetilde{\mfrak{#1}} }
\newcommand*{\hmfrak}[1]{ \hat{\mfrak{#1}} }
\newcommand*{\whmfrak}[1]{ \widehat{\mfrak{#1}} }


\newcommand*{\mscr}[1]{{\mathscr{#1}}}%
\newcommand*{\bscr}[1]{ \bs{\mathscr{#1}} }
\newcommand*{\tscr}[1]{\tilde{\mathscr{#1}}}
\newcommand*{\wtscr}[1]{\widetilde{\mathscr{#1}}}
\newcommand*{\hscr}[1]{ \hat{\mathscr{#1}} }
\newcommand*{\whscr}[1]{ \widehat{\mathscr{#1}} }
\newcommand*{\barscr}[1]{ \bar{\mathscr{#1}} }
\newcommand*{\olscr}[1]{ \overline{\mathscr{#1}} }

\newcommand*{\mcal}[1]{ {\mathcal{#1}} }
\newcommand*{\bcal}[1]{ \bs{\mathcal{#1}} }
\newcommand*{\tcal}[1]{\tilde{\mathcal{#1}}}
\newcommand*{\wtcal}[1]{\widetilde{\mathcal{#1}}}
\newcommand*{\hcal}[1]{ \hat{\mathcal{#1}} }
\newcommand*{\whcal}[1]{ \widehat{\mathcal{#1}} }
\newcommand*{\barcal}[1]{ \bar{\mathcal{#1}} }
\newcommand*{\olcal}[1]{ \overline{\mathcal{#1}} }

\newcommand{\iical}[1]{{\scriptscriptstyle{\mcal{#1}}}}
\newcommand{\sscal}[1]{{{}^{{}_{\mathcal{#1}}}}} 
\newcommand{\iiscr}[1]{ {\scriptscriptstyle{\mscr{#1}}} }
\newcommand*{\sscr}[1]{ {\scalebox{0.7}{$\mscr{#1}$}} } 



\newcommand{\en}[0]{{\itsc{n}}}
\newcommand{\envec}[0]{ \hbe_\en }
\newcommand{\enform}[0]{\,\hat{\!\bs{\epsilon}}^\en}
\newcommand{\tilp}[0]{{\mkern2mu \til{\mkern-2mu\smash{p}}}}

\newcommand*{\tp}[1]{{\vphantom{#1}\uln{#1}}} 
\newcommand*{\btp}[1]{ {\bar{\tp{#1}}} }
\newcommand*{\bartp}[1]{ {\bar{\tp{#1}}} }
\newcommand*{\tiltp}[1]{ {\til{\tp{#1}}} }
\newcommand*{\ttp}[1]{ {\til{\tp{#1}}} }
\newcommand*{\wttp}[1]{ {\wt{\tp{#1}}} }
\newcommand*{\htp}[1]{ {\hat{\tp{#1}}} }
\newcommand*{\whtp}[1]{ {\wh{\tp{#1}}} }
\newcommand{\dottp}[1]{{\dot{\tp{#1}}} }

\newcommand*{\tup}[1]{{\vphantom{#1}\uln{#1}}} 
 \newcommand*{\bartup}[1]{ {\bar{\tup{#1}}} } 
\newcommand*{\tiltup}[1]{ {\til{\tup{#1}}} }
\newcommand*{\ttup}[1]{ {\til{\tup{#1}}} }
\newcommand*{\wttup}[1]{ {\wt{\tup{#1}}} }
\newcommand*{\htup}[1]{ {\hat{\tup{#1}}} }
\newcommand*{\whtup}[1]{ {\wh{\tup{#1}}} }
\newcommand{\dottup}[1]{{\dot{\tup{#1}}} }
\newcommand{\nrmtup}[1]{ {#1} } %


\newcommand*{\pt}[1]{{#1}}
\newcommand{\tilpt}[1]{ {\til{\pt{#1}}} }
\newcommand*{\barpt}[1]{ {\bar{\pt{#1}}} }
\newcommand{\hatpt}[1]{ {\hat{\pt{#1}}} }
\newcommand{\hpt}[1]{ {\hat{\pt{#1}}} }
\newcommand*{\ptvec}[1]{ {\vec{\pt{#1}}} }
\newcommand*{\dtpt}[1]{\dt{\pt{#1}}}


\renewcommand{\d}[0]{\text{d}}
\newcommand*{\rmd}[0]{{\text{\rmsb{d}}}}
\newcommand*{\dif}[0]{ {\textbf{\textsf{\textsl{d}}}} }
\newcommand*{\Dif}[0]{{ \scriptstyle{\textbf{\textsf{\textsl{D}}}} }}

\newcommand{\difbar}[0]{{ \txsfb{\kern0.6ex\ooalign{\hidewidth\raisebox{0.63ex}{--}\hidewidth\cr{\kern-0.6ex\textsf{d}}\cr}} \vphantom{d} }}
\newcommand{\itdifbar}{%
   {\mkern4.5mu \txsfb{\ooalign{\hidewidth\raisebox{0.65ex}{--}\hidewidth\cr$\mkern-4.5mu \bs{d}$\cr}} \vphantom{d} }%
}

\newcommand*{\hdif}[0]{  \hspace{1mm}\hat{\hspace{-1mm}\dif} }
\newcommand*{\tdif}[0]{  \hspace{1mm}\tilde{\hspace{-1mm}\dif} }
\newcommand*{\deldif}[0]{\bs{\delta}}

\newcommand{\tinywedge}[0]{ \scalebox{0.45}{$\pmb{\wedge}$}}
\newcommand{\exd}[0]{\dif_{\hspace{-0.25mm}{\tinywedge}}}
\newcommand{\exdbar}[0]{\difbar_{\hspace{-0.25mm}{\tinywedge}}}

\newcommand{\codif}[0]{ {\exd^{\scriptscriptstyle{\star}}} }

\newcommand*{\fibd}[0]{{\sfb{F}}}%

\newcommand{\pderiv}[2]{ \tfrac{\uppartial #1}{\uppartial #2} }
\newcommand{\ppderiv}[3]{\tfrac{\uppartial^2 #1}{\uppartial #2 \uppartial #3}}

\newcommand{\pder}[2]{ \tfrac{ \displaystyle{\uppartial #1}}{\displaystyle{\uppartial #2}} }  
\newcommand{\pderr}[2]{ \tfrac{ \textstyle{\uppartial #1}}{\textstyle{\uppartial #2}} }  
\newcommand{\ppder}[3]{ \tfrac{\displaystyle{\uppartial^2 #1}}{\displaystyle{\uppartial #2 \uppartial #3}} }

\newcommand{\Pderiv}[2]{\frac{\uppartial #1}{\uppartial #2}}
\newcommand{\PPderiv}[3]{\frac{\uppartial^2 #1}{\uppartial #2 \uppartial #3}}

 \newcommand*{\diff}[2]{ \tfrac{\mathrm{d} #1}{\mathrm{d} #2} }
\newcommand{\ddiff}[2]{ \tfrac{\mathrm{d}^2{#1}}{\mathrm{d} {#2}^2} }
\newcommand{\Diff}[2]{ \frac{\mathrm{d}#1}{\mathrm{d}\,#2} }
\newcommand{\DDiff}[2]{ \frac{\mathrm{d}^2{#1}}{\mathrm{d}\,{#2}^2} }

\newcommand{\fdiff}[1]{ \,\tfrac{ {\!\!\!\!}^{ \text{\footnotesize{$#1$}} }\d }{\;\d\, t} }
\newcommand{\fDiff}[3]{ \,\frac{ \!\!\!\!^{ \text{\footnotesize{$#3$}} }\d #1 }{\;\d\, #2} }
\newcommand{\fddiff}[1]{ \,\tfrac{ \!\!\!\!^{ \text{\scriptsize{$#1$}} }\d^2 }{\;\d\, t^2} }
\newcommand{\fDDiff}[3]{ \,\frac{ \!\!\!\!^{ \text{\footnotesize{$#3$}} }\d^2 #1 }{\;\d\, #2^2} }

\newcommand{\mdiff}[2]{ \tfrac{ \mathrm{D} #1}{\mathrm{d} #2}   }
\newcommand{\nabdiff}[2]{ \tfrac{\nabla #1}{\mathrm{d} #2}   }
\newcommand{\nabpderiv}[2]{ \tfrac{\nabla #1}{\uppartial #2}   }

%
\newcommand{\nab}[0]{ {\nabla} }
\newcommand{\del}[1]{ \nabla_{\hspace{-0.6mm} {#1}}  }


\newcommand*{\kd}[0]{1}
\newcommand*{\lc}[0]{{\itsc{e}}}
\newcommand{\ibase}[0]{\hat{\oldstylenums{1}}}
\newcommand*{\imat}[0]{{\text{1}}\hspace{-.2mm}}
\newcommand*{\jmat}[0]{{\textsl{J}}}

\newcommand{\Id}[0]{{\mathrm{Id}}}%
\newcommand{\iden}[0]{\sfb{1}}
\newcommand*{\gvol}[0]{\epsilon}
\newcommand*{\spvol}[0]{\sigma} 
\newcommand*{\lagvol}[0]{\varsigma}
\newcommand*{\spform}[0]{\theta}  
\newcommand*{\spformup}[0]{\thetaup}
\newcommand*{\lagform}[0]{\vartheta} 
\newcommand*{\lagformup}[0]{\varthetaup}

\newcommand{\vf}[0]{v} 
\newcommand{\pf}[0]{p} 
\newcommand{\vlin}[0]{{\scalebox{0.7}{$\mathcal{V}$}}}
\newcommand{\plin}[0]{{\pi}} 
\newcommand{\uplin}[0]{{\piup}} 
\newcommand{\tilplin}[0]{{\mkern2mu \til{\mkern-2mu\smash{\pi}}}} 


\newcommand{\rmat}[1]{ \mathrm{M}^{#1}_\ii{\mbb{R}} }
\newcommand{\mats}[1]{ \rmat{#1} }
\newcommand{\mat}[2]{ \mathrm{M}^{#1}_\ii{#2} }
\newcommand{\lgrp}[3]{\mathrm{#1}^{#2}_\ii{\mbb{#3}}}
\newcommand{\lalg}[3]{\mathfrak{#1}^{#2}_\ii{\mbb{#3}}}
\newcommand{\Afmat}[1]{ \mathrm{Af}^{#1}_\ii{\mbb{R}} }
\newcommand{\afmat}[1]{ \mathfrak{af}^{#1}_\ii{\mbb{R}} }
\newcommand{\Glmat}[1]{ \mathrm{Gl}^{#1}_\ii{\mbb{R}} }
\newcommand{\Slmat}[1]{ \mathrm{Sl}^{#1}_\ii{\mbb{R}} }
\newcommand{\glmat}[1]{ \mathfrak{gl}^{#1}_\ii{\mbb{R}} }
\newcommand{\slmat}[1]{ \mathfrak{sl}^{#1}_\ii{\mbb{R}} }
\newcommand{\Spmat}[1]{ \mathrm{Sp}^{#1}_\ii{\mbb{R}} }
\newcommand{\spmat}[1]{ \mathfrak{sp}^{#1}_\ii{\mbb{R}} }
\newcommand{\Omat}[1]{ \mathrm{O}^{#1}_\ii{\mbb{R}} }
\newcommand{\Somat}[1]{ \mathrm{SO}^{#1}_\ii{\mbb{R}} }
\newcommand{\omat}[1]{ \mathfrak{o}^{#1}_\ii{\mbb{R}} }
\newcommand{\somat}[1]{ \mathfrak{so}^{#1}_\ii{\mbb{R}} }
\newcommand{\Umat}[1]{ \mathrm{U}^{#1}_\ii{\mbb{C}} }
\newcommand{\Sumat}[1]{ \mathrm{SU}^{#1}_\ii{\mbb{C}} }
\newcommand{\umat}[1]{ \mathfrak{u}^{#1}_\ii{\mbb{C}} }
\newcommand{\sumat}[1]{ \mathfrak{su}^{#1}_\ii{\mbb{C}} }
\newcommand{\Emat}[1]{ \mathrm{E}^{#1}_\ii{\mbb{R}} }
\newcommand{\Semat}[1]{ \mathrm{SE}^{#1}_\ii{\mbb{R}} }
\newcommand{\emat}[1]{ \mathfrak{e}^{#1}_\ii{\mbb{R}} }
\newcommand{\semat}[1]{ \mathfrak{se}^{#1}_\ii{\mbb{R}} }

\newcommand{\Aften}[0]{ \mathrm{Af} }
\newcommand{\aften}[0]{ \mathfrak{af} }
\newcommand{\Glten}[0]{ \mathrm{Gl} }
\newcommand{\Slten}[0]{ \mathrm{Sl} }
\newcommand{\glten}[0]{ \mathfrak{gl} }
\newcommand{\slten}[0]{ \mathfrak{sl} }
\newcommand{\Spten}[0]{ \mathrm{Sp}}
\newcommand{\spten}[0]{ \mathfrak{sp} }
\newcommand{\Oten}[0]{ \mathrm{O} }
\newcommand{\Soten}[0]{ \mathrm{SO} }
\newcommand{\oten}[0]{ \mathfrak{o} }
\newcommand{\soten}[0]{ \mathfrak{so} }
\newcommand{\Eten}[0]{ \mathrm{E} }
\newcommand{\Seten}[0]{ \mathrm{SE} }
\newcommand{\eten}[0]{ \mathfrak{e} }
\newcommand{\seten}[0]{ \mathfrak{se} }

\newcommand{\Spism}[0]{\mathscr{S}\mkern-1mu p}
\newcommand{\Dfism}[0]{\mathscr{D}\mkern-2mu f}
\newcommand{\Hmism}[0]{\mathscr{H}m }
\newcommand{\Ctism}[0]{\mathscr{C}t}
\newcommand{\Isom}[0]{\mathcal{I}\mkern-3mu s}


\newcommand{\man}[1]{ \mathcal{#1} }  
\newcommand{\tman}[1]{ \til{\man{#1}} } 
\newcommand{\barman}[1]{ \bar{\man{#1}} }
\newcommand{\bman}[1]{ \bar{\man{#1}} }
\newcommand{\hman}[1]{ \hat{\man{#1}} }
\newcommand{\whman}[1]{ \widehat{\man{#1}} }
\newcommand{\sman}[1]{\sl{#1}} 
\newcommand{\tint}[0]{\mathcal{I}} 

\newcommand{\chart}[2]{ \man{#1}_{\hspace{-0.2mm}\textup{\tiny(} {\scriptscriptstyle{#2}} \textup{\tiny)}} }
\newcommand{\rchart}[2]{ \mathbb{#1}_{\hspace{-0.2mm}\textup{\tiny(} {\scriptscriptstyle{#2}} \textup{\tiny)}} }
\newcommand{\chrt}[2]{ {#1}_{\hspace{-0.2mm}\textup{\tiny(} {\scriptscriptstyle{#2}} \textup{\tiny)}} }

\newcommand{\vsp}[1]{ \mathbb{#1} } 
\newcommand{\aff}[1]{\man{#1}} 
\newcommand{\tvsp}[1]{ \tilde{\vsp{#1}} } 
\newcommand{\bvsp}[1]{ \bar{\vsp{#1}} } 
\newcommand{\taff}[1]{ \tilde{\aff{#1}} } 
\newcommand{\baff}[1]{ \bar{\aff{#1}} } 

\newcommand{\affE}[0]{\man{E}}
\newcommand{\vecE}[0]{\vsp{E}}
\newcommand{\Eaf}[0]{\man{E}}
\newcommand{\Evec}[0]{\vsp{E}}
\newcommand{\emet}[0]{I} 
\newcommand{\bemet}[0]{\sfb{I}}
\newcommand{\rfun}[0]{{r}} 
\newcommand{\lang}[0]{{\ell}} 
\newcommand{\Lang}[0]{{L}}
\newcommand{\langb}[0]{\bm{\ell}} 
\newcommand{\Langb}[0]{\sfb{L}} 
\newcommand{\slang}[0]{{\ldash}} 
\newcommand{\Slang}[0]{{\Ldash}} 
\newcommand{\slangb}[0]{\ldashb}
\newcommand{\Slangb}[0]{{\sfb{\L}}}

\newcommand{\blang}[0]{\langb} 
\newcommand{\bLang}[0]{\Langb} 
\newcommand{\bslang}[0]{\slangb} 
\newcommand{\bSlang}[0]{\Slangb}

\newcommand*{\fun}[0]{{\scalebox{0.85}[1]{$\mathcal{F}\mkern-2mu$}}}  
\newcommand*{\tens}[0]{ \mathscr{T} } 
\newcommand*{\forms}[0]{ \Lambdaup }
\newcommand{\formsex}[0]{ \Lambdaup_\ii{\mrm{ex}\!}}
\newcommand{\formscl}[0]{ \Lambdaup_\ii{\mrm{cl}\!}}

\newcommand*{\veckl}[0]{ \mathfrak{X}_{\scriptscriptstyle{\!\mathfrak{i\!s}}\!} }
\newcommand*{\vect}[0]{ \mathfrak{X} } 
\newcommand*{\vecsp}[0]{ \mathfrak{X}_{\scriptscriptstyle{\!\mathfrak{s\!p}}\!\!} } 
\newcommand*{\vechm}[0]{ \mathfrak{X}_{\scriptscriptstyle{\!\mathfrak{h\!\!m}}\!\!} } 
\newcommand{\ham}[1]{{\mathscr{#1}}}
\newcommand{\tham}[1]{\widetilde{\mathscr{#1}}}

 \newcommand{\Tspacefont}[1]{ {\mathrm{#1}} } 
\newcommand{\Tan}[0]{ \Tspacefont{T}   }
\newcommand{\Tanh}[0]{  \Tspacefont{H}  }
\newcommand{\Tanv}[0]{  \Tspacefont{V}  }
\newcommand{\tsp}[1][\mkern2mu]{ \Tspacefont{T}_{\mkern-2mu #1} }
\newcommand{\tspv}[1][\mkern2mu]{\Tspacefont{V}_{\!#1}}
\newcommand{\tsph}[1][]{\Tspacefont{H}_{#1}}
\newcommand{\tspn}[1][\hspace{.5mm}]{\Tspacefont{N}_{\hspace{-.2mm}#1}}
\newcommand{\cotsp}[1][\hspace{1mm}]{ {\Tspacefont{T}^* {\hspace{-1ex}}_{\hspace{-.4mm}#1\hspace{0.5mm}}} } 
\newcommand{\cotspv}[1][\hspace{1mm}]{ \Tspacefont{V}^* {\hspace{-1ex}}_{\!#1\hspace{0.5mm}} }
\newcommand{\cotsph}[1][\hspace{1mm}]{ \Tspacefont{H}^* {\hspace{-1ex}}_{#1\hspace{0.1mm}} }
\newcommand{\cotspn}[1][\hspace{1mm}]{ \Tspacefont{N}^* {\hspace{-1ex}}_{#1\hspace{0.1mm}} }

\newcommand{\prj}[0]{ \scalebox{0.7}{$\Pi$} }  
 \newcommand{\tpr}[0]{ \scalebox{0.5}[0.68]{$\bs{\mcal{T}}$} }
 \newcommand{\copr}[0]{{ \mkern1mu\hat{\mkern-1mu\smash{\tpr}} }}

\newcommand{\vfun}[1][]{{V^{#1}}}
\newcommand{\pfun}[1][]{{P^{#1}}}
\newcommand{\varpfun}[1][]{{l^{#1}}}

\newcommand*{\tlift}[0]{ T\hspace{-.3mm} }
\newcommand*{\colift}[0]{ \hat{T}\hspace{-.3mm} }
\newcommand*{\uptlift}[0]{ \mrm{T}\hspace{-.3mm} }
\newcommand*{\upcolift}[0]{ \hat{\mrm{T}}\hspace{-.3mm} }

\newcommand{\formsh}[0]{\Lambdaup_\ii{\mrm{h}\!}} 
\newcommand{\formsv}[0]{\Lambdaup_\ii{\mrm{v}\!}} 
\newcommand{\vectv}[0]{\mathfrak{X}_\ii{\mrm{v}\!}} 
\newcommand{\vecth}[0]{\mathfrak{X}_\ii{\mrm{h}\!}} 
\newcommand{\formsbh}[0]{\Lambdaup_\ii{\mrm{bh}\!}} 
\newcommand{\vectbv}[0]{\mathfrak{X}_\ii{\mrm{bv}\!}}

\newcommand{\lft}[1]{ #1^{\hspace{-0.2mm}\scriptscriptstyle\upharpoonright} } 
\newcommand{\lift}[1]{ #1^{\hspace{-0.2mm}\scriptscriptstyle\uparrow} } 
\newcommand{\cotlft}[1]{  #1^{\hspace{-0.2mm}\hat{\scriptscriptstyle{\uparrow}}} }
\newcommand{\invlift}[1]{ #1^{\hspace{-0.2mm}\scriptscriptstyle\downarrow} }

\newcommand{\vlift}[1]{ #1^{\hspace{-0.2mm}\scriptscriptstyle\uparrow} }
\newcommand{\Vlift}[1]{  #1^{\scriptscriptstyle\Uparrow} }
\newcommand{\coVlift}[1]{  #1^{\hat{\scriptscriptstyle{\Uparrow}}} }%

\newcommand{\covlift}[1]{ #1^{\hat{\textup{\tiny{\textsf{V}}}}}}
\newcommand{\verlift}[1]{ #1^{\textup{\tiny{\textsf{V}}}} }%
\newcommand{\coverlift}[1]{ #1^{\hat{\textup{\tiny{\textsf{V}}}}}}

\newcommand{\hlift}[1]{ #1^{\textup{\tiny{\textsf{H}}}} }%
\newcommand{\horlift}[1]{ #1^{\textup{\tiny{\textsf{H}}}} }%
\newcommand{\cohlift}[1]{ #1^{\hat{\textup{\tiny{\textsf{H}}}}} }%

 \newcommand{\dubuparrow}[0]{\rotatebox[origin=c]{90}{$\ii{\twoheadrightarrow}$}}
\newcommand{\tailuparrow}[0]{\rotatebox[origin=c]{90}{$\ii{\rightarrowtail}$}}
\newcommand{\lftt}[1]{#1^{\dubuparrow}}
\newcommand{\llft}[1]{#1^{\tailuparrow}}

\newcommand{\rhk}[0]{\hookrightarrow}
\newcommand{\lhk}[0]{\hooklefttarrow}
\newcommand{\subman}[0]{ \,\accentset{\scriptstyle{\subset}}{\scriptstyle{\hookrightarrow}}\, }

\newcommand{\pbrak}[2]{\{#1  ,  #2 \}}
\newcommand{\Pbrak}[2]{\big\{#1\;,\;#2\big\}}
\newcommand{\ppbrak}[2]{\{ \hspace{-0.8mm} \{#1 ,  #2 \} \hspace{-0.8mm} \}}

\newcommand{\lbrak}[2]{ [ #1 , #2 ] }
\newcommand{\Lbrak}[2]{ \big[ #1\;,\;#2 \big] }
\newcommand{\llbrak}[2]{ [\hspace{-0.8mm}[#1 ,  #2 ]\hspace{-0.8mm}] }

\newcommand{\lderiv}[1]{ {\textit{\textrm{\pounds}}}_{\scalebox{0.72}{$#1$}} } 

\newcommand{\altbrak}[2]{ {\lfloor #1 ,  #2 \rfloor} }
\newcommand{\aaltbrak}[2]{ {\lfloor \hspace{-0.8mm} \lfloor #1 ,  #2 \rfloor \hspace{-0.8mm} \rfloor} }
\newcommand{\lagbrak}[2]{ {\lfloor #1 ,  #2 \rfloor}  }
\newcommand{\llagbrak}[2]{ {\lfloor \hspace{-0.8mm} \lfloor #1 ,  #2 \rfloor \hspace{-0.8mm} \rfloor} }

\newcommand{\inner}[2]{\langle#1  , #2\rangle}
\newcommand{\iinner}[2]{\langle\!\langle#1\,,\,#2\rangle\!\rangle}

\newcommand{\ang}[1]{\langle #1 \rangle}
\newcommand{\Ang}[1]{\big\langle #1 \big\rangle }


\newcommand{\abs}[1]{ \big| #1 \big| }
\newcommand{\Abs}[1]{ \left| #1 \right| }
\renewcommand{\mag}[1]{ |\!| #1 |\!| }
\newcommand{\nrm}[1]{{\lvert #1 \rvert} }
\renewcommand{\det}[0]{ {\textrm{\textup{\footnotesize{det}}}\hspace{0.5mm}} }
\renewcommand{\dim}[0]{ {\textrm{\textup{\footnotesize{dim}}}\hspace{0.5mm}} }
\newcommand{\codim}[0]{ {\textrm{\textup{\footnotesize{codim}}}\hspace{0.5mm}} } %
\newcommand{\rnk}[0]{ {\textrm{\textup{\footnotesize{rnk}}}\hspace{0.5mm}} }
\newcommand{\sgn}[0]{ {\textrm{\textup{\footnotesize{sgn}}}\hspace{0.5mm}} }
\newcommand{\tr}[0]{ {\textrm{\textup{\footnotesize{tr}}}\hspace{0.5mm}} }
\renewcommand{\ker}[0]{ {\textrm{\textup{\footnotesize{ker}}}\hspace{0.5mm}} }
\newcommand{\img}[0]{ {\textrm{\textup{\footnotesize{im}}}\hspace{0.5mm}} }
\newcommand{\spn}[0]{ {\textrm{\textup{\footnotesize{span}}}\hspace{0.5mm}} }
\renewcommand{\div}[1][\,]{ {{\textrm{\textup{\footnotesize{div}}}}_{\!#1}}}

\newcommand{\crd}[2]{ \tensor*[^{\ss{#2}}]{[#1]}{} }
\newcommand{\crdl}[2]{ \tensor*[^{\ss{#2}}]{#1}{} }
\newcommand{\cordl}[2]{  #1_{\scriptscriptstyle{\!#2}}  } 
\newcommand{\cord}[2]{  #1_{\scriptscriptstyle{#2}}  }

\newcommand{\eval}[2]{\left.{#1}\right|_{#2}}

\newcommand{\pdup}[0]{\uppartial} %
\newcommand{\upd}[0]{{\rotatebox[origin=t]{10}{$\partial$} \mkern-2mu}}
\newcommand{\pd}[0]{{\rotatebox[origin=t]{10}{$\partial$} \mkern-2mu}}


\newcommand{\mypd}[0]{\rotatebox[origin=t]{10}{$\partial$}} 

\newcommand*{\ffsize}[1]{{\scalebox{0.7}{$#1$}}}

\newcommand*{\bpart}[1]{\bs{\mypd}_{\hspace{-.3mm}#1}}
\newcommand*{\bpartup}[1]{\bs{\mypd}^{#1}}
\newcommand*{\tbpart}[1]{\tilde{\bs{\mypd}}_{\hspace{-.3mm} #1}}
\newcommand*{\tbpartup}[1]{{\tilde{\bs{\mypd}}\vphantom{l}^{#1}}}
\newcommand*{\hbpart}[1]{\hbs{\pd}_{\mkern-0.5mu #1}}
\newcommand*{\hbpartup}[1]{{\hbs{\pd}\vphantom{l}^{#1}}}

\newcommand*{\pdii}[1]{ \bs{\mypd}_{\ffsize{\!#1}}}
\newcommand*{\pdupii}[1]{ \bs{\mypd}^{\ffsize{#1}} }
\newcommand*{\pdiiup}[1]{ \bs{\mypd}^{\ffsize{#1}} }
\newcommand*{\tpdii}[1]{ \tilde{\bs{\mypd}}_{\hspace{-.5mm}\ffsize{#1}} }
\newcommand*{\tpdupii}[1]{ \tilde{\bs{\mypd}}{}^{\ffsize{#1}} }
\newcommand*{\hpdii}[1]{ \hbs{\pd}_{\ffsize{\mkern-1mu #1}} }
\newcommand*{\hpdiiup}[1]{ {\hbs{\pd}}^{\ffsize{#1}} }
\newcommand*{\barpdii}[1]{ \bar{\bs{\mypd}}_{\hspace{-.6mm}\ffsize{#1}} }
\newcommand*{\barpdiiup}[1]{ {\bs{\bar{\pd}}}^{\ffsize{#1}} }

\newcommand{\bpd}[1][]{ {\bs{\mypd}_{\ffsize{\!#1}}} }%
\newcommand{\bpdup}[1][]{ {\bs{\mypd}^{\ffsize{#1}}} }
\newcommand{\hbpd}[1][]{ {\hbs{\pd}_{\ffsize{\!#1}}} }%
\newcommand{\hbpdup}[1][]{ {\hbs{\pd}^{\ffsize{#1}}} }
\newcommand{\tbpd}[1][]{ {\tbs{\pd}_{\ffsize{\!#1}}} }%
\newcommand{\tbpdup}[1][]{ {\tbs{\pd}^{\ffsize{#1}}} }
\newcommand{\barbpd}[1][]{ {\barbs{\pd}_{\ffsize{\!#1}}} }%
\newcommand{\barbpdup}[1][]{ {\barbs{\pd}^{\ffsize{#1}}} }

\newcommand*{\bpdh}[1]{ {\bs{\mypd}_{\mkern-1mu\hat{#1}}} }
\newcommand*{\bpdhup}[1]{ {\bs{\mypd}^{\hat{#1}}} }
\newcommand*{\bpdt}[1]{ {\bs{\mypd}_{\mkern-1mu\til{#1}}} }
\newcommand*{\bpdtup}[1]{ {\bs{\mypd}^{\til{#1}}} }
\newcommand*{\bpdbar}[1]{ {\bs{\mypd}_{\mkern-1mu\bar{#1}}} }
\newcommand*{\bpdbarup}[1]{ {\bs{\mypd}^{\bar{#1}}} }
\newcommand*{\bdelh}[1]{ \bs{\delta}^{\hat{#1}}\vphantom{l} }
\newcommand{\bdelhdn}[1][]{ {\bs{\delta}_{\mkern-1mu\hat{#1}}\vphantom{l}} }
\newcommand*{\bdelt}[1]{ \bs{\delta}^{\til{#1}}\vphantom{l} }
\newcommand{\bdeltdn}[1][]{ {\bs{\delta}_{\mkern-1mu\til{#1}}\vphantom{l}} }
\newcommand*{\bdelbar}[1]{ \bs{\delta}^{\bar{#1}}\vphantom{l} }
\newcommand{\bdelbardn}[1][]{ {\bs{\delta}_{\mkern-1mu\bar{#1}}\vphantom{l}} }



\newcommand{\bdel}[1][]{ {\bs{\delta}^{\ffsize{#1}}} }%
\newcommand{\bdeldn}[1][]{ {\bs{\delta}_{\ffsize{\!#1}}} }
\newcommand{\hbdel}[1][]{ \hbs{\delta}^{\ffsize{#1}}\vphantom{l} }%
\newcommand{\hbdeldn}[1][]{ {\hbs{\delta}_\ffsize{\!#1}\vphantom{l}} } 
\newcommand{\tbdel}[1][]{ \tbs{\delta}^{\ffsize{#1}}\vphantom{l} }%
\newcommand{\tbdeldn}[1][]{ {\tbs{\delta}_\ffsize{\!#1}\vphantom{l}} }
\newcommand{\barbdel}[1][]{ \barbs{\delta}^{\ffsize{#1}}\vphantom{l} }%
\newcommand{\barbdeldn}[1][]{ {\barbs{\delta}_\ffsize{\!#1}\vphantom{l}} }

\newcommand{\bD}[1][]{{ \fnsz{\sfb{D}}_{\ffsize{\!#1}}} }
\newcommand{\bDel}[1][]{{ \fnsz{\bs{\Delta}}^{\mkern-1mu\ffsize{#1}}}}%
\newcommand{\bDeldn}[1][]{{ \fnsz{\bs{\Delta}}_{\ffsize{#1}}}}%

\newcommand{\ff}[2]{\sfb{#1}_{\ffsize{\!#2}}} 
\newcommand{\ffup}[2]{\sfb{#1}^{\ffsize{#2}}} 
\newcommand{\coff}[2]{\bs{#1}^{\ffsize{#2}}}
\newcommand{\coffdn}[2]{\bs{#1}_{\ffsize{\!#2}}}

\newcommand{\bi}[1][]{{\sfb{i}_{\ffsize{\!#1}}}}
\newcommand{\bio}[1][]{{\bs{\iota}^{\ffsize{#1}}} \vphantom{\iota}}
\newcommand{\hbi}[1][]{{\hsfb{\i}_{\ffsize{\!#1}}}} 
\newcommand{\hbio}[1][]{{\hbs{\iota}^{\ffsize{#1}}} \vphantom{\iota}}

\newcommand{\be}[1][]{{\sfb{e}_{\ffsize{\!#1}}}}
\newcommand{\beup}[1][]{{\sfb{e}^{\ffsize{#1}}}} 
\newcommand{\tbe}[1][]{ {\tsfb{e}_{\ffsize{\!#1}}} }
\newcommand{\hbe}[1][]{ {\hsfb{e}_{\ffsize{\!#1}}} }
\newcommand{\bep}[1][\!]{{\bs{\epsilon}^{\ffsize{#1}}}}
\newcommand{\bepdn}[1][]{{\bs{\epsilon}_{\ffsize{\!#1}}}}
\newcommand{\tbep}[1][\!]{\tbs{\epsilon}^{\ffsize{#1}} \vphantom{\epsilon}}
\newcommand{\hbep}[1][\!]{{\hbs{\epsilon}^{\ffsize{#1}}} \vphantom{\epsilon}}

\newcommand{\bt}[1][]{ \sfb{t}_{\ffsize{\!#1}} \vphantom{t}}
\newcommand{\tbt}[1][]{\tsfb{t}_{\ffsize{\!#1}} \vphantom{t}}
\newcommand{\hbt}[1][]{\hsfb{t}_{\ffsize{\!#1}} \vphantom{t}}
\newcommand{\btup}[1][\,]{ \sfb{t}^{\ffsize{#1}\!} \vphantom{t}}
\newcommand{\tbtup}[1][\,]{\tsfb{t}^{\ffsize{#1}\!} \vphantom{t}}
\newcommand{\hbtup}[1][\,]{\hsfb{t}^{\ffsize{#1}\!} \vphantom{t}}

\newcommand{\btau}[1][\!]{{\bm{\tau}^{\ffsize{#1}}}}
\newcommand{\tbtau}[1][\!]{{\tbm{\tau}^{\ffsize{#1}}}}
\newcommand{\hbtau}[1][\!]{{\hbm{\tau}^{\ffsize{#1}}}}
\newcommand{\btaudn}[1][]{{\bm{\tau}_{\ffsize{\!#1}}}}
\newcommand{\tbtaudn}[1][]{{\tbm{\tau}_{\ffsize{\!#1}}}}
\newcommand{\hbtaudn}[1][]{{\hbm{\tau}_{\ffsize{\!#1}}}}


\newcommand{\Zvar}{{ \text{\ooalign{\hidewidth\raisebox{0.25ex}{--}\hidewidth\cr$Z$\cr}}\vphantom{Z}} }

\newcommand{\zvar}{%
  \text{\ooalign{\hidewidth -\kern-.3em-\hidewidth\cr$z$\cr}} \vphantom{z}%
}

\renewcommand{\Zbar}{{\textnormal{\ooalign{\hidewidth\raisebox{0.3ex}{\footnotesize{--}}\hidewidth\cr$Z$\cr}} \vphantom{Z}} }

\newcommand{\zbar}{ \mkern0.5mu
  {\text{\ooalign{\hidewidth\raisebox{0.1ex}{\scriptsize{--}}\hidewidth\cr$\mkern-0.5muz$\cr}} \vphantom{z}}%
} 

\newcommand{\Jbar}{ \mkern2mu {\textnormal{\ooalign{\hidewidth\raisebox{0.4ex}{\scriptsize{--}}\hidewidth\cr$\mkern-2mu J$\cr}} \vphantom{J}} }

\newcommand{\Ibar}{{\textnormal{\ooalign{\hidewidth\raisebox{0.4ex}{\scriptsize{--}}\hidewidth\cr$I$\cr}} \vphantom{I}} }

\newcommand{\Sbar}{{\mkern0mu \textnormal{\ooalign{\hidewidth\raisebox{0.4ex}{\scriptsize{\textbf{---}}}\hidewidth\cr$\mkern-0mu S$\cr}} \vphantom{S}} }

\newcommand{\sbar}{ \mkern-0.5mu
  {\text{\ooalign{\hidewidth\raisebox{0.1ex}{\scriptsize{--}}\hidewidth\cr$\mkern0.5mu s$\cr}} \vphantom{s}}%
}

\newcommand{\Cbar}{{\mkern-7mu \textnormal{\ooalign{\hidewidth\raisebox{0.35ex}{\footnotesize{--}}\hidewidth\cr$\mkern7muC$\cr}} \vphantom{C}} }

\newcommand{\cbar}{%
  {\mkern-3mu\text{\ooalign{\hidewidth\raisebox{0.1ex}{\scriptsize{--}}\hidewidth\cr$\mkern3muc$\cr}} \vphantom{c}}%
}

\newcommand{\ldash}[0]{{\mkern-1mu \textit{\l}\mkern0.5mu}} 
\newcommand{\Ldash}[0]{{\textit{\L}}} 
\newcommand{\ldashb}[0]{{\itsb{\l}}}
\newcommand{\Ldashb}[0]{{\txsfb{\L}}} 
\newcommand{\Lbar}{ {\mkern-2mu \textnormal{\ooalign{\hidewidth\raisebox{0.35ex}{\scriptsize{--}}\hidewidth\cr$\mkern2mu L$\cr}} \vphantom{L}}  }

\newcommand{\lbar}{%
    {\textnormal{\ooalign{\hidewidth\raisebox{0.3ex}{-}\hidewidth\cr$l$\cr}} \vphantom{l}}%
} 

\newcommand{\blbar}{%
    {\textbf{\ooalign{\hidewidth\raisebox{0.4ex}{\scriptsize{--}}\hidewidth\cr$\bs{l}$\cr}} \vphantom{l}}%
} 

\newcommand{\kbar}{%
   {\mkern-1mu \text{\ooalign{\hidewidth\raisebox{0.8ex}{\scriptsize{--}}\hidewidth\cr$\mkern1mu k$\cr}} \vphantom{k} }%
}
\newcommand{\txkbar}{%
   {\mkern-4mu \textnormal{\ooalign{\hidewidth\raisebox{0.8ex}{\scriptsize{--}}\hidewidth\cr$\mkern4mu \mrm{k}$\cr}} \vphantom{k} }%
}

\newcommand{\hvar}{%
   {\mkern-1mu \text{\ooalign{\hidewidth\raisebox{0.8ex}{\scriptsize{--}}\hidewidth\cr$\mkern1mu h$\cr}} \vphantom{h} }%
}
\newcommand{\txhbar}{%
   {\mkern-4mu \textnormal{\ooalign{\hidewidth\raisebox{0.8ex}{\scriptsize{--}}\hidewidth\cr$\mkern4mu \mrm{h}$\cr}} \vphantom{k} }%
}

\newcommand{\bbar}{%
   {\mkern-1mu \text{\ooalign{\hidewidth\raisebox{0.8ex}{\scriptsize{--}}\hidewidth\cr$\mkern1mu b$\cr}} \vphantom{b} }%
}
\newcommand{\txbbar}{%
   {\mkern-4mu \textnormal{\ooalign{\hidewidth\raisebox{0.8ex}{\scriptsize{--}}\hidewidth\cr$\mkern4mu \mrm{b}$\cr}} \vphantom{b} }%
} 

\newcommand{\dbar}{%
   {\mkern4.5mu \text{\ooalign{\hidewidth\raisebox{0.8ex}{\scriptsize{--}}\hidewidth\cr$\mkern-4.5mu d$\cr}} \vphantom{d} }%
}
\newcommand{\txdbar}{%
   {\mkern3mu \textnormal{\ooalign{\hidewidth\raisebox{0.8ex}{\scriptsize{--}}\hidewidth\cr$\mkern-3mu \mrm{d}$\cr}} \vphantom{d} }%
} 

\newcommand{\Gambar}{%
   {\mkern-4mu \text{\ooalign{\hidewidth\raisebox{0.4ex}{\scriptsize{--}}\hidewidth\cr$\mkern4mu \Gamma$\cr}} \vphantom{h} }%
}


\renewcommand{\v}[0]{{\nu}} 
\newcommand{\txv}[0]{{\txi{v}}}
\newcommand{\altv}[0]{{\scriptstyle{\mathcal{V}}}}
\newcommand{\newv}[0]{{\scriptstyle{\mathscr{V}}}}

\newcommand{\one}[0]{{\ss{1}}}
\newcommand{\two}[0]{{\ss{2}}}
\newcommand{\three}[0]{{\ss{3}}}
\newcommand{\four}[0]{{\ss{4}}}
\newcommand{\six}[0]{{\ss{6}}}
\newcommand{\eight}[0]{{\ss{8}}}


\renewcommand{\a}[0]{{\alpha}}
\renewcommand{\b}[0]{{\ii{\beta}}}
\newcommand{\g}[0]{{\ss{\gamma}}}
\newcommand{\gam}[0]{{\gamma}}
\newcommand{\Gam}[0]{{\Gamma}}
\newcommand{\y}[0]{{\ss{\lambda}}}
\newcommand{\lam}[0]{{\lambda}}
\newcommand{\Lam}[0]{{\Lambda}}
\newcommand{\sig}[0]{{\sigma}}
\newcommand{\Sig}[0]{{\Sigma}}
\newcommand{\varsig}[0]{{\varsigma}}
\newcommand{\ep}[0]{{\epsilon}}
\newcommand*{\varep}[0]{{\varepsilon}}
\newcommand{\omg}[0]{{\omega}}
\newcommand{\Omg}[0]{{\Omega}}
\newcommand{\kap}[0]{{\kappa}}
\newcommand{\varkap}[0]{{\varkappa}}
\newcommand{\ro}[0]{{\varrho}}
\renewcommand{\th}[0]{{\theta}}
\newcommand{\Th}[0]{{\Theta}}
\newcommand{\varth}[0]{{\vartheta}}

\DeclareRobustCommand{\rchi}{{\mathpalette\irchi\relax}}
\newcommand{\irchi}[2]{\raisebox{\depth}{$#1\chi$}} 
\DeclareRobustCommand{\pup}{{\mathpalette\newp\relax}}
\newcommand{\newp}[2]{\raisebox{0.8\depth}{$#1p$}}
\DeclareRobustCommand{\jup}{{\mathpalette\newj\relax}}
\newcommand{\newj}[2]{\raisebox{0.8\depth}{$#1\jmath$}}

\newcommand{\upa}[0]{{\upalpha}}
\newcommand{\upb}[0]{{\upbeta}}
\newcommand{\upgam}[0]{{\upgamma}}
\newcommand{\upGam}[0]{{\upGamma}}
\newcommand{\Upgam}[0]{{\Upgamma}}
\newcommand{\uplam}[0]{{\uplambda}}
\newcommand{\upLam}[0]{{\upLambda}}
\newcommand{\Uplam}[0]{{\Uplambda}}
\newcommand{\updel}[0]{{\updelta}}
\newcommand{\upsig}[0]{{\upsigma}}
\newcommand{\upSig}[0]{{\upSigma}} 
\newcommand{\Upsig}[0]{{\Upsigma}}
\newcommand{\upomg}[0]{{\upomega}}
\newcommand{\upOmg}[0]{{\upOmega}}
\newcommand{\Upomg}[0]{{\Upomega}}
\newcommand{\upep}[0]{\upepsilon}
\newcommand{\upth}[0]{\uptheta}
\newcommand{\upvarth}[0]{\upvartheta}
\newcommand{\upTh}[0]{\upTheta}
\newcommand{\Upth}[0]{\Uptheta}
\newcommand{\upkap}[0]{\upkappa}
\newcommand{\upvark}[0]{\upvarkappa}
\newcommand{\upn}[0]{\upeta}
\newcommand{\upi}[0]{\upiota}

\newcommand{\aup}[0]{{\alphaup}}
\newcommand{\bup}[0]{{\betaup}}
\newcommand{\gamup}[0]{{\gammaup}}
\newcommand{\Gamup}[0]{{\Gammaup}}
\newcommand{\lamup}[0]{{\lambdaup}}
\newcommand{\Lamup}[0]{{\Lambdaup}}
\newcommand{\delup}[0]{{\deltaup}}
\newcommand{\sigup}[0]{{\sigmaup}}
\newcommand{\Sigup}[0]{{\Sigmaup}}
\newcommand{\wup}[0]{{\omegaup}} 
\newcommand{\omgup}[0]{{\omegaup}}
\newcommand{\Omgup}[0]{{\Omegaup}}
\newcommand{\epup}[0]{\epsilonup}
\newcommand{\thup}[0]{\thetaup}
\newcommand{\varthup}[0]{\varthetaup}
\newcommand{\kup}[0]{\kappaup}
\newcommand{\kapup}[0]{\kappaup}
\newcommand{\varkup}[0]{\varkappaup}
\newcommand{\nup}[0]{\etaup}
\newcommand{\iup}[0]{\iotaup}

\newcommand{\txa}[0]{{\text{\textalpha}}}
\newcommand{\txb}[0]{{\text{\textbeta}}}
\newcommand{\txbet}[0]{{\text{\textbeta}}}
\newcommand{\txg}[0]{{\text{\textgamma}}}
\newcommand{\txgam}[0]{{\text{\textgamma}}}
\newcommand{\txGam}[0]{{\text{\textGamma}}}
\newcommand{\txy}[0]{{\text{\textlambda}}}
\newcommand{\txlam}[0]{{\text{\textlambda}}}
\newcommand{\txr}[0]{{\text{\textrho}}}
\newcommand{\txrho}[0]{{\text{\textrho}}}
\newcommand{\txsig}[0]{{\text{\textsigma}}}
\newcommand{\txSig}[0]{{\text{\textSigma}}}
\newcommand{\txw}[0]{{\text{\textomega}}}
\newcommand{\txomg}[0]{{\text{\textomega}}}
\newcommand{\txep}[0]{{\text{\textepsilon}}}
\newcommand{\txth}[0]{{\text{\texttheta}}}
\newcommand{\txvarth}[0]{{\text{\textvartheta}}}
\newcommand{\txphi}[0]{{\text{\textphi}}}
\newcommand{\txvarphi}[0]{{\text{\textvarphi}}}
\newcommand{\txpsi}[0]{{\text{\textpsi}}}
\newcommand{\txd}[0]{{\text{\textdelta}}}
\newcommand{\txdel}[0]{{\text{\textdelta}}}
\newcommand{\txiota}[0]{{\text{\textiota}}}
\newcommand{\txn}[0]{{\text{\texteta}}}
\newcommand{\txeta}[0]{{\text{\texteta}}}
\newcommand{\txxi}[0]{{\text{\textxi}}}
\newcommand{\txchi}[0]{{\text{\textchi}}}
\newcommand{\txpi}[0]{{\text{\textpi}}}
\newcommand{\txtau}[0]{{\text{\texttau}}}
\newcommand{\txz}[0]{{\text{\textzeta}}}
\newcommand{\txzeta}[0]{{\text{\textzeta}}}
\newcommand{\txnu}[0]{{\text{\textnu}}}
\newcommand{\txmu}[0]{{\text{\textmugreek}}} 
\newcommand{\txups}[0]{{\text{\textupsilon}}} 
\newcommand{\txk}[0]{{\text{\textkappa}}}
\newcommand{\txkap}[0]{{\text{\textkappa}}}

\newcommand{\tximu}[0]{ {\textit{\textmugreek}} } %
\newcommand{\txiu}[0]{ {\textit{\textupsilon}} } 
\newcommand{\txitau}[0]{ {\textit{\texttau}} } %
\newcommand{\txipi}[0]{ {\textit{\textpi}} }%
\newcommand{\txirho}[0]{ {\textit{\textrho}} } 
\newcommand{\txilam}[0]{ {\textit{\textlambda}} }

\newcommand{\wtxt}[0]{{\textomega}}
\newcommand{\ytxt}[0]{{\textlambda}}
\newcommand{\thtxt}[0]{{\texttheta}}
\newcommand{\btxt}[0]{{\textbeta}}
\newcommand{\atxt}[0]{{\textalpha}}
\newcommand{\gtxt}[0]{{\textgamma}}
\newcommand{\tautxt}[0]{{\texttau}}

\newcommand{\gvgr}[1]{\vec{\tbf{#1}}}
\newcommand{\uvgr}[1]{\hat{\tbf{#1}}}
\newcommand{\tvgr}[1]{\widetilde{\tbf{#1}}}
\newcommand{\bvgr}[1]{\bar{\tbf{#1}}}
\newcommand{\gvbb}[1]{\vec{\pmb{#1}}}
\newcommand{\uvbb}[1]{\hat{\pmb{#1}}}
\newcommand{\tvbb}[1]{\widetilde{\pmb{#1}}}
\newcommand{\bvbb}[1]{\bar{\pmb{#1}}}

\newcommand{\sigvec}[0]{\vec{\pmb{\upsigma}} }
 \newcommand{\sigtvec}[0]{\widetilde{\pmb{\upsigma}} }

\newcommand{\gvec}[1]{\vec{\bs{\mathrm{#1}}}}
\newcommand{\gv}[1]{\vec{\bm{#1}}}
\newcommand{\gvb}[1]{\vec{\bs{#1}}}
\newcommand*{\bsvec}[1]{\vec{\bs{#1}}}
\newcommand*{\sfvec}[1]{\,\vec{\!\sfb{#1}}}

\newcommand{\uvec}[1]{\hat{\bs{#1}}}
\newcommand{\uv}[1]{\hat{\bs{#1}}}
\newcommand{\ihat}[0]{\hat{\bs{\iota}}}
\newcommand{\ehat}[0]{\hat{\bs{e}}}
\newcommand{\bhat}[0]{\hat{\bs{b}}}
\newcommand{\shat}[0]{\hat{\bs{s}}}
\newcommand{\nhat}[0]{\uvec{n}}
\newcommand{\evec}[0]{\vec{\bs{e}}}
\newcommand{\epvec}[0]{\vec{\bs{\epsilon}}}

\newcommand{\uvecb}[1]{ {\hbs{#1}} }
\newcommand{\uvb}[1]{ {\hbs{#1}} }
\newcommand{\ihatb}[0]{\hat{\pmb{\upiota}}} 
\newcommand{\ehatb}[0]{\uvecb{e}} 
\newcommand{\bhatb}[0]{\!\!\uvecb{\;b}}
\newcommand{\hhatb}[0]{\!\!\uvecb{\;h}}
\newcommand{\shatb}[0]{\uvecb{s}}
\newcommand{\nhatb}[0]{\uvecb{n}}
\newcommand{\ohatb}[0]{\uvecb{o}}
\newcommand{\ahatb}[0]{\uvecb{a}}
\newcommand{\epvecb}[0]{\vec{\bs{\epsilon}}}
\newcommand{\evecb}[0]{ \hspace{0.3mm}\vec{\mbf{e}}\hspace{0.1mm} }

\newcommand{\tvb}[1]{\widetilde{\mbf{#1}}}
\newcommand{\tvecb}[1]{\bar{\bs{#1}}}
\newcommand{\bv}[1]{\bar{\bs{#1}}}
\newcommand{\bvb}[1]{\bar{\bs{\mathrm{#1}}}}


\newcommand{\rnote}[1]{\noindent{\footnotesize\textit{{\color{red}${\circ}$}  #1}}}
\newcommand{\red}[1]{{\color{red}#1}}

\newcommand{\bnote}[1]{\noindent{\footnotesize\textit{{\color{blue}${\circ}$} #1}}}

\newcommand{\gnote}[1]{\noindent{\footnotesize\textit{$\circ$  #1}}}

 \newcommand{\note}[1]{\noindent{\footnotesize\textit{{\color{darkgray} #1}}}}

\newcommand{\nsp}[0]{\!\!\!\!}    
\newcommand{\nquad}[0]{\hspace{-1em}} 
\newcommand{\nqquad}[0]{\hspace{-2em}} 


\theoremstyle{plain} 
\newtheorem{thrm}{Theorem}[section]
\newtheorem{defn}{Def.}[section]

\theoremstyle{plain} 
\newtheorem{remit}{Remark}[section]

\theoremstyle{definition} 
\newtheorem{remark}{Remark}[section]


\newtheoremstyle{remsans}
{8pt} 
{12pt} 
{\rmfamily\slshape\small}
{}
{\sffamily\bfseries\small}
{.}
{.5em}
{}

\theoremstyle{remsans}
\newtheorem{remsf}{$\iisqr\,$Remark}[section]

\newtheoremstyle{remrmsmall}
{8pt} 
{12pt} 
{\rmfamily\small\slshape}
{}
{\small\bfseries}
{.}
{.3em}
{}

\theoremstyle{remrmsmall}
\newtheorem{remrm}{$\iisqr\,$\rmsb{Remark}}[section]
\newtheorem{noether}[remark]{$\iisqr\,$\rmsb{Noether}}

\newtheoremstyle{remslant}
{8pt} 
{12pt} 
{\rmfamily\slshape\small}
{}
{\rmfamily\slshape\bfseries\small}
{.}
{.3em}
{}

\theoremstyle{remslant} 
\newtheorem{remsl}{Remark}[section]

\newtheoremstyle{remnopunc}
{8pt} 
{12pt} 
{\rmfamily\small}
{}
{\bfseries\small}
{}
{.2em}
{}
\theoremstyle{remnopunc} 
\newtheorem*{noteblt}{$\nmblt$}
\newtheorem*{notesq}{$\iisqr$}
\newtheorem*{notestr}{\raisebox{0.1ex}{$\star$}}
\newtheorem*{noteast}{$\bs{*}$}

\newtheoremstyle{remslantnopunc}
{8pt} 
{12pt} 
{\slshape\rmfamily\small}
{}
{\slshape\bfseries\small}
{}
{.2em}
{}
\theoremstyle{remslantnopunc} 
\newtheorem*{notesl}{$\iisqr$}

\newtheoremstyle{remrmnonbold}
{8pt} 
{12pt} 
{\rmfamily\footnotesize}
{}
{\rmfamily\itshape\footnotesize}
{.}
{.2em}
{}
\theoremstyle{remrmnonbold}
\newtheorem*{notation}{Notation}







\begin{abstract}
  This work presents a geometric formulation for transforming nonconservative mechanical Hamiltonian systems and introduces a new method for regularizing and linearizing central force dynamics — in particular, Kepler and Manev dynamics — through a projective transformation. 
  The transformation is formulated as a configuration space diffeomorphism (rather than a submersion) that is lifted to a cotangent bundle (phase space) symplectomorphism and used to pullback the original mechanical Hamiltonian system, Riemannian kinetic energy metric, and other key geometric objects. Full linearization of both Kepler and Manev dynamics (in any finite dimension) is achieved by a subsequent conformal scaling of the projectively-transformed Hamiltonian vector field. Two such conformal scalings are given, both achieving linearization.
   Arbitrary conservative and non-conservative perturbations are included, with closed-form solutions readily obtained in the unperturbed Kepler or Manev cases.   
\end{abstract}


\renewcommand{\contentsname}{}
\tableofcontents  


\phantomsection
\addcontentsline{toc}{section}{INTRODUCTION}
\section*{INTRODUCTION}

\paragraph{Background.} The study of regularization in celestial mechanics has traditionally been carried out within a classic coordinate-based, analytical  framework.\footnote{For brief historical tour of regularization and linearization in celestial mechanics, see the introduction of Deprit at al. \cite{deprit1994linearization}, and references therein.}
In recent decades, however, aspects of the subject have been re-examined and re-formulated using the tensor-based, coordinate-agnostic, perspective of differential geometry.  For instance, Cariñena et al.~have explored 
evolution parameter transformations\footnote{What we refer to as a transformation of the evolution parameter, many others refer to as a ``time'' reparameterization.}
and applications to regularized Kepler dynamics in the tangent bundle/Lagrangian setting \cite{carinena2022infinitesimal,carinena2023sundman,carinena2024sundman2,carinena2017tangent}; Marmo et al.~have investigated the relationship between the Kepler problem and the harmonic oscillator via the Kustaanheimo–Stiefel transformation in the tangent bundle/Lagrangian setting \cite{marmo2005reduction}, and Cordani has written extensively about the geometry of the Kepler problem, including several regularizing transformations \cite{cordani2003kepler}.

\vspace{1ex}

 \noindent \sloppy The most well-known example of Hamiltonian-based regularization and linearization of Kepler dynamics is likely the \sbemph{Kustaanheimo-Stiefel (KS) transformation} \cite{KS1kustaanheimo1965perturbation,stiefel1973linear,stiefel1971linear,kurcheeva1977kustaanheimo,ferrer2017alternative,breiter2017KS}, which employs quaternionic coordinates to transform the 3-dim Kepler problem to a 4-dim harmonic oscillator (as does the present work, via a different route). While the physical or geometric interpretation of the KS transformation\footnote{Originally \cite{KS1kustaanheimo1965perturbation}, the KS transformation was presented in relation to Spinors.}
is often obscured by its typical matrix-based presentation, several sources have clarified the close connection to quaternions \cite{deprit1994linearization,breiter2017KS,saha2009interpreting}.\footnote{Chelnokov is notably prolific on the subject (e.g., two separate series of papers with initial entries \cite{chelnokov2013quatTraj1} and \cite{chelnokov2017quat3BP1}).}

\vspace{1ex}

\noindent Rather than using quaternionic transformations, this work instead focuses on 
\textit{projective}\footnote{\textit{\textbf{Note on terminology.}} 
    Although the term "projective" appears frequently in this work, it should not be interpreted too formally. 
    In the mathematical context of projective spaces and projective geometry, terms like "projective transformation" and "projective coordinates" have definitions which do not always align with our, more relaxed, use of such terms. 
    This clash of terminology arises because, in the celestial mechanics literature,  point transformations
    which decompose a displacement vector, \eq{\ptvec{r}}, into a unit vector and a scalar are commonly referred to as a ``projective decomposition'' or a ``projective transformation'' (we use either interchangeably). That is, a projective transformation/decomposition often refers to something of the form \eq{\vec{r} = z^b \hat{r}} for unit vector \eq{\hat{r}}, some coordinate \eq{z}, and some number \eq{0\neq b \in\mbb{R}} (often, \eq{b=\pm 1}). When interpreted as a coordinate transformation, \eq{\tup{r}=z^b \tup{y}} (with \eq{\tup{y}=\htup{r}}), then the set \eq{(\tup{y},z)} are often referred to as projective coordinates in the celestial mechanics literature. While there is indeed a connection to drawn with the mathematical use of such terms, the language is not always in one-to-one correspondence.}
transformations for regularizing and linearizing central force dynamics.
To the authors' knowledge, the only prior work investigating projective regularizations in a \textit{Hamiltonian}\footnote{A number of others have investigated some form of projective transformation for regularized Kepler dynamics in a \textit{non}-Hamiltonian framework (cf.~Burdet, Sperling, Vitins and Bond \cite{burdet1969mouvement,Burdet+1969+71+84,vitins1978keplerian,bond1985transformation}). Much of this historically significant work was detailed and expanded upon by Schumacher \cite{schumacher1987results}. }
framework is that of Ferrándiz et al.~\cite{ferrandiz1987general,ferrandiz1988extended,ferrandiz1992increased}, Deprit et al.~\cite{deprit1994linearization}, and the present authors \cite{peterson2025prjCoord,peterson2022nonminimal,peterson2023regularized}. 
Among these, Ferrándiz et al.~were the first to realize a canonical/symplectic extension of Burdet's projective regularization of the Kepler problem and re-formulate it in a classical Hamiltonian framework \cite{ferrandiz1987general}.  We refer to this as the \sbemph{Burdet-Ferrándiz (BF) transformation}, which follows the naming convention of Deprit et al.~\cite{deprit1994linearization}, who re-visited Ferrándiz's BF transformation, offering comparisons to the KS transformation as well as their own formulation, 
which they dub the \rmsb{DEF transformation}\footnote{DEF stands for Deprit-Elipe-Ferrer.}.
In the authors' previous work \cite{peterson2025prjCoord,peterson2022nonminimal}, we presented a family of generalized ``BF-like'' projective transformations — ultimately preferring a version differing from the usual BF (and DEF) transformation — and derived a related set of non-singular\footnote{In all cases other than rectilinear motion.}
orbit elements for perturbed Kepler dynamics.
While the present paper is certainly inspired by these earlier works in Hamiltonian-based projective regularization, the developments presented here are self contained and differ considerably in mathematical presentation as well as generality of application (more on this further down in the \textit{outline \& contributions} paragraph).

\vspace{1ex}
\noindent The quaternion-based KS transformation has received notably more interaction and use compared to the projective-based BF transformation and its subsequent modifications and generalizations. On this note, Deprit et al.~make the the following remark: 

 \begin{small}
 \begin{quote}
          \textit{We claim [the projective BF/DEF transformation] achieves equally well all the objectives of the KS transformation — linearization, regularization and canonicity [i.e., Hamiltonian/symplic structure] — although, we are inclined to believe, in a simpler and more intuitive way.} 
          $\quad$ — Deprit et al. \cite{deprit1994linearization}
 \end{quote}
 \end{small}

\noindent The authors are inclined to agree with Deprit's comment and are of the opinion that their own twist on the BF projective transformation (solidified here in geometric terms) is perhaps even more simple and intuitive, though this is subjective.

\paragraph{Kepler and Manev Dynamics.}
The inclusion of Manev dynamics in this work is notable and warrants some comments. The Manev potential is an augmentation of the Kepler potential that includes an additional inverse square term:
\begin{small}
\begin{align}
    \begin{array}{cc}
         \fnsize{Kepler}  \\[-1pt]
          \fnsize{potential}
    \end{array}
    \!\!:\quad 
    V^0 = -\tfrac{\sck}{r}
    \qquad\quad,\qquad\quad
    \begin{array}{cc}
         \fnsize{Manev}  \\[-1pt]
          \fnsize{potential}
    \end{array}
    \!\!:\quad 
     V^0 = -\tfrac{\sck_1}{r} - \tfrac{1}{2} \tfrac{\sck_2}{r^2}
\end{align}
\end{small}
(the negative signs and factor of \eq{\sfrac{1}{2}} are included for later convenience) for scalars \eq{\sck_1,\sck_2\in\mbb{R}} and where \eq{r} is the scalar-valued radial distance/magnitude function on some finite-dimensional inner product space of interest (classically, \eq{\Evec^3}, Euclidean 3-space\footnote{or, some finite-dimensional Euclidean \eq{n}-space, \eq{\Evec^n}.}). 
Originally introduced as a classical approximation to certain relativistic corrections, the Manev potential  maintains key features of the Kepler potential such as conic orbits and integrability, while simultaneously introducing key relativistic features like perihelion precession 
(a hallmark prediction of general relativity).\footnote{The Manev potential is not a true model of relativistic effects near the classical limit. It is not derived from general relativity. Rather, the additional term \textit{qualitatively} reproduces the leading-order relativistic correction to the Kepler problem predicted by the relativistic Schwarzschild solution, while also still maintaining the integrability of the classic Kepler problem.} 
A noteworthy feature of the regularizing transformation presented in this work is that it is equally effective at linearizing both Kepler-type and and Manev-type dynamics (the former being a special case of the later). 
However, other than incorporating the above mathematical form of the Manev potential field, we provide little discussion of the physical implications and practical applications in regards to relativity-motivated corrections in orbital mechanics. We refer the reader to Cordani \cite{cordani2003kepler}, Marmo et al.~\cite{marmo2006manev1,marmo2007manev2symmetries,marmo2008manev3kepler}, and Diacu et al.~\cite{diacu1995manev1,diacu1996manev2,diacu2000manevPhase3,diacu2004manev4}, who have discussed the subject more thoroughly. 



\paragraph{Outline \& Contributions.}
We (re)develop and (re)interpret a modified BF-like projective regularization for central force dynamics — originally developed in \cite{peterson2025prjCoord,peterson2022nonminimal}, with key parts summarized in Appx.~\ref{sec:prj_sum2} of this work — but, now, within the context of symplectic and Riemannian geometry.
By framing mechanics in the coordinate-agnostic tensorial language of differential geometry — this is an extensive field in and of itself, which we loosely refer to as \textit{geometric mechanics} (cf. \cite{abraham2008foundations,marsden2013introduction,abraham2012manifolds,fecko2006differential,frankel2011geometry,libermann2012symplectic,carinena2015geoFromDyn,munoz2024geometry,deLeon2011methods,arnold2013mathematical,arnold2006mathematical}) —
we are able to more clearly and rigorously define transformations of general nonconservative Hamiltonian systems and their dynamical properties. Ultimately, this framework offers greater clarity and coherence in what has been described as the “unwieldy and cumbersome” treatment of regularized and linearized point-mass dynamics.\footnote{Deprit et al.~attribute the description of “unwieldy and cumbersome” to Stiefel and Scheifele.} 
For instance, we are able to realize the ``canonical projective transformation'' of earlier works as a remarkably ``simple'' transformation characterized by two geometric objects:~a Riemannian metric and a potential function. 

\vspace{1ex}
\noindent While the present paper shares themes with the works mentioned earlier\footnote{In particular, work from Ferrándiz et al.~\cite{ferrandiz1987general,ferrandiz1988extended,ferrandiz1992increased}, Deprit et al.~\cite{deprit1994linearization}, and the preset authors \cite{peterson2025prjCoord,peterson2022nonminimal,peterson2023regularized}.}
— namely, projective transformations, regularized and linearized central force dynamics, and Hamiltonian formulations — it differs in several significant ways:
\begin{small}
\begin{itemize}
    \item \textit{Generality of dimensions and forces.} Previous work has been specific the Kepler-type dynamics (forces from a potential \eq{V^0=-\sck/r}) in 2 or 3 dimensional coordinate space (\eq{\mbb{R}^2} or \eq{\mbb{R}^3}).\footnote{An exception to this is the authors' previous work \cite{peterson2025prjCoord}, where Manev dynamics were also linearized, though this was given little attention and the mathematical presentation was again limited to 2 or 3 dimensions.}
    This is true of both the BF and the KS transformations (the latter is, further, mathematically limited to 2 or 3 dimensions). In contrast, the regularizing transformation presented here is defined on any finite-dimensional real inner product space and, furthermore, it fully linearizes not only Kepler-type dynamics, but more generally Manev-type dynamics (forces from a potential \eq{V^0=-\sck_1/r -\tfrac{1}{2}\sck_2/r^2}).
  \item \textit{Mathematical perspective/presentation.} Previous work has been formulated in coordinate-based language with little to no connection to more general geometric concepts.\footnote{Appx.~\ref{sec:prj_sum2} is an example of such a ``coordinate-based'' presentation.}
   Though this work does not shun coordinates, we adopt a geometric perspective and frame the regularization process as the transformation of of key geometric, coordinate-agnostic objects (e.g., metrics, symplectic forms, curves, submanifolds, symmetries, etc.).\footnote{We do not mean to say that the inclusion or exclusion of differential geometry is inherently good or bad; only that it provides different perspectives and tools. At times, we feel the geometric framework offers greater clarity and coherence. While, at other times, the abundance of notation and terminology endemic to differential geometry can bog down the presentation of simple ideas.}
    \item \textit{Properties of the transformation.} Previous work has framed the the initial projective point transformation as a submersion, making its extension/lift to an invertible symplectic transformation on phase space a non-trivial problem involving redundant coordinates, Lagrange multipliers, and other complications. In this work, we avoid the issue by formulating the projective point transformation as a diffeomorphism such that the cotangent-lift (which is \textit{automatically} a phase space symplectomorphism) is readily obtained. Everything else needed to transform the Hamiltonian dynamics then follows in a clear, prescribed, manner. The issue of redundancy/over-parameterization then resolved by restricting attention to certain invariant submanifolds with clear meaning.  
\end{itemize}
\end{small}

\vspace{1ex}
\noindent The outline of this paper is adequately reflected in the table of contents, with the developments of each main section summarized as follows (the main contributions are in sections \ref{sec:prj_geomech} and \ref{sec:prj_regular}, with the general theory in section \ref{sec:Hmech_xform_gen} also of interest):
\begin{small}
\begin{enumerate}
    \item[] \textit{Section \ref{sec:Hmech_xform_gen}.} Before addressing anything related to central force dynamics or regularization, we first outline the theory for transforming general mechanical Hamiltonian systems on cotangent bundles (phase space) using a point transformation on the base manifold (configuration space). Arbitrary conservative and nonconservative forces are included and we connect these developments back to Riemannian dynamics on the configuration space. 
    While this material may not be mathematically novel, it has not previously been presented in this consolidated form for the specific purpose of transforming nonconservative mechanical Hamiltonian systems.
    \item[]  \textit{Section \ref{sec:NOM}.} Here, we "set the stage" for the main part of this paper in the subsequent sections. 
    In section \ref{sec:Hnom_prj}, we define the original Hamiltonian system in which we are interested: a simple, general, single-particle system which eventually becomes the perturbed Manev problem (this includes the Kepler problem as a special case). However, we embed the simple original system in a higher-dimensional space — this allows us to later define our projective transformation as a diffeomorphism. The price we pay for this luxury is that we must spend some time (section \ref{sec:prj_prelim}) establishing terms and notation for the decomposition of configuration and phase space into hyperplanes and a normal direction. Much of this can be skimmed and used for reference as needed. 
    \item[] \textit{Section \ref{sec:prj_geomech}.} Here, we start the main construction of this paper. We define our ``projective transformation'' as a diffeomorphism on configuration space, along with its cotangent lift (which defines the momentum transformation) as a symplectomorphism on phase space. This is used to transform the original mechanical Hamiltonian system to its ``projectively-transformed'' counterpart. The new system is simply another mechanical Hamiltonian system for a non-Euclidean kinetic energy metric and transformed potential. We examine the relation to Riemannian dynamics, the transformation of forces and integrals of motion, and identify important invariant submanifolds. We also briefly illustrate the analogous ``passive'' view as a coordinate transformation (paralleling the authors' previous work, as summarized in section \ref{sec:prj_sum2}).
    \item[]  \textit{Section \ref{sec:prj_regular}.} 
     Lastly, we show that when the original system is Manev-type (with Kepler as a special case), then the projectively transformed system can be fully linearized via a conformal scaling of the Hamiltonian vector field (i.e., a transformation of  the evolution parameter). 
     The linearized dynamics readily admit closed-form solutions when restricted to a family of invariant submanifolds — restrictions that do not correspond to any physically meaningful limitations on the original system. The transformed solutions, and  their mapping back to solutions of the original Manev system,  are particularly simple in the Kepler case.
    \item[]  \textit{Appendices.} There are several appendices (listed in the table of contents). Many of these give the geometric treatment, in a general context, for material used in specific contexts within the body of the paper. For instance, we make frequent use of cotangent bundle geometry, the cotangent-lift of various objects, nonconservative forces, conformal scalings, etc. So as not to burden the main presentation with numerous mathematical detours, we relegate the relevant material to appendices and make free use that material during the main developments of the paper (sometimes without explicit mention). An exception to this is Appx.~\ref{sec:prj_sum2} which is less mathematical and summarizes some previous, relevant work by the authors. 
\end{enumerate}
\end{small}

\section{GENERAL THEORY:~POINT TRANSFORMATIONS OF MECHANICAL HAMILTONIAN SYSTEMS} \label{sec:Hmech_xform_gen}

 Many developments in subsequent section of this work boil down to using some specified diffeomorphism \eq{\psi:\man{Q}\to\man{Q}} to transform some original mechanical Hamiltonian system on \eq{\cotsp\man{Q}} to some ``new'' Hamiltonian system on \eq{\cotsp\man{Q}} such that the new system is, in some sense, ``better'' than the original system. As such, before delving into the specific case in which we are interested (central force dynamics in Euclidean space), we will first develop the general theory in the context of mechanical systems on arbitrary smooth manifolds (but still real and finite-dimensional). This begins in section \ref{sec:Hmech_xform_active}. First, a short review of some basic terms and notation for Hamiltonian dynamics in the language of symplectic geometry:

\subsubsection*{Review:~Hamiltonian Dynamics on Symplectic Manifolds}

Recall that a \rmsb{Hamiltonian system} is a triple \eq{(\man{P},\nbs{\omg},\mscr{H})} where \eq{(\man{P},\nbs{\omg})} is a symplectic manifold and \eq{\mscr{H}\in\fun(\man{P})} is \textit{the} Hamiltonian function which, through the Hamiltonian vector field \eq{\sfb{X}^\sscr{H}:=\inv{\nbs{\omg}}(\dif \mscr{H},\slot)\in\vechm(\man{P},\nbs{\omg})},  dictates the dynamics of the system of interest. 
 A \rmsb{non-conservative Hamiltonian system} (detailed further in section \ref{sec:noncon}) is a quadruple \eq{(\man{P},\nbs{\omg},\mscr{H},\bs{\alpha})} where \eq{\bs{\alpha}\in\forms(\man{P})} is a  \textit{non-exact} (i.e., \eq{\exd\bs{\alpha}\neq \dif f}) 1-form — which is also \textit{horizontal} in the case \eq{\man{P}} is a cotangent bundle —   that corresponds to any non-conservative forces that do work on the system. The dynamics are given by the non-symplectic vector field \eq{\sfb{X}^\ss{\mscr{H},\bs{\alpha}}= \sfb{X}^\sscr{H} + \inv{\nbs{\omg}}\cdt \bs{\alpha}\in\vect(\man{P})}.

 In the following, we are primarily concerned with the particular case the the symplectic manifold is the (canonically) exact symplectic cotangent bundle, \eq{(\cotsp\man{Q},\nbs{\omg}=-\exd\bs{\spform})}, of some \eq{n}-dim configuration manifold \eq{\man{Q}}. In this context, we define a \rmsb{mechanical Hamiltonian system} as a Hamiltonian system \eq{(\cotsp\man{Q},\nbs{\omg},\mscr{H})} for which the Hamiltonian function can be written as \eq{\mscr{H} = T_{\bs{\sfg}} + U} for some \textit{basic} function \eq{U\equiv\copr^* U \in\fun(\cotsp\man{Q})} (i.e., the potential function), and a
 kinetic energy function\footnote{For a metric tensor \eq{\sfg\in\tens^0_2(\man{Q})}, the associated cotangent bundle kinetic energy function, \eq{T_{\bs{\sfg}}\in\fun(\cotsp\man{Q})}, is defined by \eq{T_{\bs{\sfg}}(\mu_{\pt{q}}):=\inv{\sfg}_{\pt{q}}(\bs{\mu},\bs{\mu})} for any \eq{\mu_{\pt{q}}=(\pt{q},\bs{\mu})\in\cotsp\man{Q}}.} 
 \eq{T_{\bs{\sfg}}\in\fun(\cotsp\man{Q})} for a (pseudo)Riemannian metric tensor field \eq{\sfg\in\tens^0_2(\man{Q})} (though it is actually \eq{\inv{\sfg}} which is used explicitly in \eq{\mscr{H}}). In other words, we could say that a mechanical Hamiltonian system on phase space is described by \eq{(\cotsp\man{Q},\nbs{\omg},\sfg,U)} where \eq{\sfg} and \eq{U} (tensor fields on \eq{\man{Q}}) define a Hamiltonian function \eq{\mscr{H} = T_{\bs{\sfg}} + U}, which determines the Hamiltonian vector field as usual, \eq{\sfb{X}^\sscr{H}=\inv{\nbs{\omg}}(\dif \mscr{H},\slot)}.

\vspace{1ex}
Let us illustrate how a general Hamiltonian vector field is related to \sbemph{Hamilton's canonical equations of motion} from classic analytical dynamics. 
Consider any Hamiltonian system  
\eq{(\man{P},\nbs{\omg},\mscr{H})} on an arbitrary \eq{2n}-dim symplectic manifold (\eq{\man{P}} could be a cotangent bundle, but it need not be). The dynamics are governed by the Hamiltonian vector field \eq{\sfb{X}^\sscr{H}= -\inv{\nbs{\omg}}(\dif \mscr{H}) \in \vechm(\man{P},\nbs{\omg})} with symplectic flow \eq{\phi_t\in\Spism(\man{P},\nbs{\omg})}.  
Let \eq{\rho_t  =\phi_t(\rho_{\zr})} be the integral curve starting from point \eq{\rho_{\zr}=\rho(0)}. Then the ``velocity'' vector \eq{\dt{\rho}_t := \diff{}{t} \rho_t\in\tsp[\rho_t]\man{P}} satisfies 
\begin{small}
\begin{align} \label{Hw_eom}
    \dt{\rho}_t
    \;=\; \sfb{X}^\sscr{H}_{\rho_t} \;=\; \inv{\nbs{\omg}}_{\rho_t}(\dif \mscr{H},\slot)
\end{align}
\end{small}
The above is the coordinate-agnostic version of Hamilton's canonical equations of motion, expressed in tensor form on a symplectic manifold.  
Let \eq{(\chart{P}{\zeta},\tp{\zeta})} be any local coordinate chart with basis \eq{\bpart{\ssc{I}}} (not necessarily symplectic) for which \eq{\inv{\nbs{\omg}}} has components \eq{\omega^{\ssc{IJ}}} (in general, \eq{\pderiv{}{\zeta^\ssc{K}}\omega^{\ssc{IJ}}\neq 0}). 
Let  \eq{(\chart{P}{z},\tp{z})} be any local symplectic coordinate chart with basis \eq{\hbpart{\ssc{I}}} satisfying \eq{\hat{\omega}^{\ssc{JI}} = \hat{\omega}_{\ssc{IJ}} = J^{\ssc{JI}} = J_{\ssc{IJ}}}. Letting \eq{\zeta^\ssc{I}(t) :=\zeta^\ssc{I}(\rho_t)} and  \eq{z^\ssc{I}(t) :=z^\ssc{I}(\rho_t)} and xpressing the above in these coordinate bases leads to 
\begin{small}
\begin{align} \label{ndot_0_gen}
   &\dt{\rho}_t \;=\;
     \dot{\zeta}^\ssc{I}(t) \bpart{\ssc{I}} 
     \;=\; \omega^{\ssc{JI}}\pderiv{\mscr{H}}{\zeta^\ssc{J}} \bpart{\ssc{I}} 
 && \Rightarrow &&
    \dot{\zeta}^\ssc{I}(t) \,=\, -\omega^{\ssc{IJ}}  \pderiv{\mscr{H}}{\zeta^\ssc{J}} \big|_{\tp{\zeta}_t}
    \,=\;
    \pbrak{\zeta^\ssc{I}}{\zeta^\ssc{J}}\pderiv{\mscr{H}}{\zeta^\ssc{J}} \big|_{\tp{\zeta}_t}
&&
     \begin{array}{cc}   \scrsize{arbitrary} \\[-3pt]  \scrsize{coordinates}  \end{array}  
\\[4pt] \label{ndot_0}
    &\dt{\rho}_t  \;=\;
    \dot{z}^\ssc{I}(t) \hbpart{\ssc{I}} \,=\,  J^{\ssc{JI}}  \pderiv{\mscr{H}}{z^\ssc{J}}  \hbpart{\ssc{I}} 
&& \Rightarrow &&
    \dot{z}^\ssc{I}(t) \,=\, -J^{\ssc{IJ}} \pderiv{\mscr{H}}{z^\ssc{J}} \big|_{\tp{z}_t}
&&
    \begin{array}{cc}     \scrsize{symplectic} \\[-3pt]      \scrsize{coordinates}  \end{array}
\end{align}
\end{small}
where  \eq{\dot{\zeta}^\ssc{I}(t)} is taken to mean \eq{\diff{}{t}(\zeta^\ssc{I} \circ \rho)_t}, and likewise for \eq{\dot{z}^\ssc{I}(t)}. It is implied that the above holds with all terms evaluated along the curve \eq{\rho_t} (and likewise for any other integral curve). 
If we split the symplectic coordinates into conjugate pairs as \eq{\tp{z}=(\tp{q},\tp{\pf})}, then the second of the above expression for \eq{\sfb{X}^\sscr{H}} leads to the classic form of Hamilton's equations of motion:
\begin{small}
\begin{align} \label{qpdot_0}
    \dt{\rho}_t \;=\;
     \dot{q}^i(t) \hbpart{i} \,+\, \dot{p}_i(t) \hbpartup{i} \;=\;
     \partial^{i} \mscr{H} \, \hbpart{i}
     \;-\;
     \pd_{i} \mscr{H} \, \hbpartup{i}  
 && \Rightarrow &&
    \begin{array}{llll}
        \dot{q}^i (t) \;=\; \pderiv{\mscr{H}}{\pf_i}\big|_{\tp{z}_t}
  \\[6pt]
        \dot{p}_i (t) \;=\; -\pderiv{\mscr{H}}{q^i}\big|_{\tp{z}_t}
    \end{array}
&&
 \begin{array}{cc}     \scrsize{symplectic} \\[-3pt]      \scrsize{coordinates}  \end{array} 
\end{align}
\end{small} 
That is, any \emph{symplectic} coordinate representation of any integral curve of \eq{\sfb{X}^\sscr{H}} obeys Hamilton's canonical equations of motion. The same dynamics in arbitrary coordinates are given by Eq.\ref{ndot_0_gen} where the Poisson brackets \eq{\omega^\ssc{IJ}=-\pbrak{\zeta^\ssc{I}}{\zeta^\ssc{J}}\in\fun(\man{P})} will be some functions of the coordinates.
This bracket has a close relationship with Hamiltonian dynamics worth mentioning. 
Recall  that the Poisson bracket, \eq{ \pbrak{f}{g}=-\pbrak{g}{f}}, of any \eq{f,g\in\fun(\man{P})} is given by any of the following:
\begin{small}
\begin{align}
      \pbrak{f}{g}  :=\,   -\inv{\nbs{\omg}}(\dif f,\dif g)  
      \,=\, \dif f \cdt  \sfb{Z}^g  
      \,=\, \lderiv{\sfb{Z}^g} f 
    \,=\   \omega^\ssc{JI} \pd_\ssc{I} f  \pd_\ssc{J} g
    \,=\, \pd_i f \partial^i g \,-\, \pd_i g \partial^i f 
\end{align}
\end{small}
where the last equality holds locally only for symplectic coordinates. 
The above shows that for a
``dynamical variable'' \eq{f\in\fun(\man{P})} of a Hamiltonian system \eq{\sfb{X}^\sscr{H}}, 
then the bracket \eq{\pbrak{f}{\mscr{H}}} gives the derivative of \eq{f} along the flow of \eq{\sfb{X}^\sscr{H}}. This is obvious from the relation \eq{\pbrak{f}{\mscr{H}}=\lderiv{\sfb{X}^{\mscr{H}}} f} but can also be verified explicitly as follows: 
if \eq{\phi_t} is the flow of \eq{\sfb{X}^\sscr{H}} and \eq{\rho_t=\phi_t(\rho_{\zr})} is any integral curve such that \eq{\dt{\rho}_t=\sfb{X}^\sscr{H}_{\rho_t}}, then
\eq{\dot{f}(t) := \diff{}{t}f(\rho_t)} is given by
\begin{small}
\begin{align} \label{EOM_pbrak}
  \dt{f}(t) :=  \diff{}{t}(f\circ \rho)_t 
  \,=\, \dif f_{\rho_t} \cdt  \diff{}{t}\rho_t \,=\,  (\dif f \cdt  \sfb{X}^\sscr{H})_{\rho_t} \,=\, \lderiv{\sfb{X}^{\mscr{H}}} f_{\rho_t}
    \,=\,
     - \inv{\nbs{\omg}}_{\rho_t}( \dif f, \dif \mscr{H})
    \,=\, \pbrak{f}{\mscr{H}}_{\rho_t}
     && \fnsize{i.e., } \;\; \dt{f} = \pbrak{f}{\mscr{H}}
\end{align}
\end{small}
where the above integral curve is arbitrary and one might simply write \eq{\dot{f} = \pbrak{f}{\mscr{H}}} when it is understood that we are only concerned with the dynamics of \eq{\sfb{X}^\sscr{H}}.
If \eq{\phi_t\in\Spism(\man{P},\nbs{\omg})} is the flow of \eq{\sfb{X}^\sscr{H}}, 
then:\footnote{The relation \eq{\pbrak{f}{\mscr{H}} \circ \phi_t = \pbrak{f\circ \phi_t}{\mscr{H}\circ \phi_t}}  follows from \eq{\phi_t} being a symplectomorphism.}
\begin{small}
\begin{align}
    \diff{}{t}(f\circ \phi_t) = \phi_t^* \lderiv{\sfb{X}^{\mscr{H}}} f = \phi_t^*\pbrak{f}{\mscr{H}} = \pbrak{f}{\mscr{H}} \circ \phi_t \,=\, \pbrak{f\circ \phi_t}{\mscr{H}\circ \phi_t}
\end{align}
\end{small}
\begin{small}
\begin{itemize}[topsep=1pt,itemsep=1pt]
    \item \textit{Integrals of Motion.}\footnote{We note that some sources use the term \textit{constant of motion} for the more general case of a possibly time-dependent \eq{f\in\fun(\man{P}\times\mbb{R})} which is likewise constant along the flow of some \eq{\sfb{X}\in\vect(\man{P})}. I.e., a constant of motion of \eq{\sfb{X}^\sscr{H}} satisfies \eq{\pbrak{f}{\mscr{H}} + \pd_t f =0}.}
    In general, an integral of motion of some \eq{\sfb{X}\in\vect(\man{P})} is a function \eq{f\in\fun(\man{P})} that is constant along the flow \eq{\phi_t} of \eq{\sfb{X}}. That is, \eq{f=\phi_t^* f=f\circ\phi_t} or, equivalently,  \eq{\lderiv{\sfb{X}} f =0}. 
    In the case of a Hamiltonian vector field, \eq{\sfb{X}^\sscr{H}\in\vechm(\man{P},\nbs{\omg})}, we further have:
    \begin{small}
    \begin{itemize}[nosep]
        \item  \eq{f} is an integral of motion of \eq{\sfb{X}^\sscr{H}} iff \eq{\pbrak{f}{\mscr{H}}=0} (assuming \eq{\pd_t f =0}).
        \item if \eq{f} is an integral of motion of \eq{\sfb{X}^\sscr{H}}, then \eq{\mscr{H}} is an integral of motion of \eq{\sfb{Z}^f}. 
        \item if \eq{f} and \eq{g} are both integrals of motion of \eq{\sfb{X}^\sscr{H}}, then  so is \eq{\pbrak{f}{g}} (this follows from the Jacobi identity). 
    \end{itemize}
    \end{small}
    Note the the function \eq{\mscr{H}\in\fun(\man{P})} itself is an integral of motion of \eq{\sfb{X}^\sscr{H}} (assuming \eq{\pd_t\mscr{H}=0}). If \eq{\phi_t} is the flow of \eq{\sfb{X}^\sscr{H}}, and \eq{\rho_t} an arbitrary integral curve,
    then:\footnote{In the context of mechanics, \eq{\mscr{H}} is often, but not always, the total energy of the system such that  \eq{\pbrak{\mscr{H}}{\mscr{H}} =0} corresponds to conservation of energy. Regardless of \eq{\mscr{H}}'s  physical interpretation, it is always an integral of motion for any \textit{autonomous} Hamiltonian system. 
    However, if \eq{\mscr{H}\in\fun(\man{P}\times\mbb{R})} is time-dependent, then \eq{\dot{\mscr{H}}=\pbrak{\mscr{H}}{\mscr{H}} +\pd_t \mscr{H} =\pd_t \mscr{H}} is generally non-zero and \eq{\mscr{H}} is \textit{not} an integral of motion of \eq{\sfb{X}^\sscr{H}} for such systems. }
    \begin{small}
    \begin{align} \label{h_conservation}
          \dot{\mscr{H}}(t) \,=\, \pbrak{\mscr{H}}{\mscr{H}}_{\rho_t} \,=\, 0 \qquad \Rightarrow \qquad 
           \phi_t^* \mscr{H} \,=\, \mscr{H}\circ \phi_t \,=\, \mscr{H}
           \qquad \qquad 
          \fnsize{i.e.,} \;\;  \mscr{H}(\rho_t) = \mscr{H}(\rho_{\zr}) 
    \end{align}
    \end{small}
\end{itemize}
\end{small}

\subsection{Transformation of Hamiltonian Systems on \txi{T*Q}} \label{sec:Hmech_xform_active}

 \begin{small}
 \begin{notation}
     In this section, \eq{\cotsp\man{Q}} is the cotangent bundle (phase space) of a smooth \eq{n}-dim manifold \eq{\man{Q}} (configuration space), with the canonical bundle projection denoted \eq{\copr:\cotsp\man{Q}\to\man{Q}}.
     Let \eq{(\tup{r},\tup{\plin}) =\colift\tup{r}: \cotsp\chart{Q}{\tup{r}}\to\mbb{R}^{2n}} be any cotangent-lifted coordinates with corresponding  basis vector fields denoted \eq{\hbpart{i}\equiv\hpdii{r^i},\hbpartup{i}\equiv\hpdiiup{\plin_i}\in \vect(\cotsp\man{Q})}, with dual basis 1-forms \eq{\hbdel^i\equiv \dif r^i, \hbdeldn_i\equiv \dif \plin_i \in\forms(\cotsp\man{Q})}. The \eq{r^i} frame fields on the base configuration manifold, \eq{\man{Q}}, are denoted \eq{\be_i\equiv\be[r^i]\in\vect(\man{Q})} and \eq{\bep^i\equiv\bep[r^i]\in\forms(\man{Q})}.
     Elsewhere in this work, \eq{(\tup{r},\tup{\plin})} will denote linear coordinates on the cotangent bundle of a (Euclidean) vector space but, here, they are arbitrary cotangent bundle coordinates on an arbitrary \eq{\cotsp\man{Q}}. 
     Unlike some other parts of this work, here we will also use \eq{\pt{x},\pt{q}\in\man{Q}} for points (or curves) in configuration space, rather than generalized coordinate functions on configuration space.  
 \end{notation}
 \end{small}

\noindent Here, we detail the general procedure for using a diffeomorphism on configuration space (a point transformation) to transform a Hamiltonian system on the cotangent bundle (phase space).  We will first take the ``active'' view of everything and the ``passive'' view as a coordinate transformation then follows as a re-interpretation. Though we are primarily concerned with \textit{mechanical} Hamiltonian systems on \eq{\cotsp\man{Q}} (those equivalent to a Newton-Riemann system on \eq{\man{Q})}), much of the following also holds for more general Hamiltonian systems on a symplectic manifold.



\paragraph{The Original System.} Suppose we have some \eq{2n}-dim original Hamiltonian system \eq{(\cotsp\man{Q},\nbs{\omg},\mscr{K})} with \eq{\nbs{\omg}=-\exd\bs{\theta}\in\formsex^2(\cotsp\man{Q})} the canonical symplectic form and with \eq{\mscr{K}\in\fun(\cotsp\man{Q})} a mechanical Hamiltonian for a Riemannian metric tensor (kinetic energy metric), \eq{\sfb{m}\in\tens^0_2(\man{Q})}, and potential function \eq{V\in\fun(\man{Q})}. That is, the Hamiltonian system \eq{(\cotsp\man{Q},\nbs{\omg},\mscr{K})} is simply the symplectic formulation of a Riemannian Newtonian system \eq{(\man{Q},\sfb{m},V)}.   
Then, for any  \eq{\kap_\pt{x}=(\pt{x},\bs{\kap})\in\cotsp\man{Q}},  the function  \eq{\mscr{K}} is given as follows, along with \eq{\sfb{X}^\sscr{K}\in\vechm(\cotsp\man{Q},\nbs{\omg})}:
 \begin{small}
 \begin{align} \label{Ksystem_gen}
 \begin{array}{cccc}
         \mscr{K}(\kap_\pt{x}) \,=\, \tfrac{1}{2}\inv{\sfb{m}}_\pt{x}(\bs{\kap},\bs{\kap}) \,+\, V(\pt{x})
         \qquad\qquad \fnsize{i.e., } \;\; \mscr{K} = \tfrac{1}{2}m^{ij}\plin_i \plin_j \,+\, V 
 \\[2pt] 
    \sfb{X}^\sscr{K} :=\, \inv{\nbs{\omg}}(\dif \mscr{K},\cdot)  
       \,=\, \upd^i\mscr{K}\hbpart{i} \,-\,  \pd_i\mscr{K}\hbpartup{i} 
    \;=\; 
    m^{ij}\plin_j \hbpart{i} \,+\, (  m^{sl}\Omega^j_{is} \plin_j \plin_l \,-\, \pd_i V) \hbpartup{i}
 \end{array}
 \end{align}
 \end{small}
where \eq{m^{ij}:=\inv{\sfb{m}}(\bep^i,\bep^j)} are the inverse metric components in the \eq{r^i} basis and \eq{\Omega^i_{jk}=\Omega^i_{kj}} are \eq{\sfb{m}}'s Levi-Civita connection coefficients (Christoffel symbols) for the \eq{r^i} frame fields on \eq{\man{Q}}. In the above, \eq{m^{ij},V\in\fun(\man{Q})} are regarded as basic functions, \eq{m^{ij}\equiv \copr^* m^{ij},V\equiv \copr^* V\in\fun(\cotsp\man{Q})}. Note that if \eq{\kap_t=(\pt{x}_t,\bs{\kap}_t)} is an integral curve of \eq{\sfb{X}^\sscr{K}} — that is, \eq{\dt{\kap}_t=\sfb{X}^\sscr{K}_{\kap_t}} with \eq{\copr(\kap_t)=\pt{x}_t} — then \eq{\bs{\kap}_t=\sfb{m}_{\pt{x}}(\dtsfb{x}_t)\in\tsp[\pt{x}_t]^*\man{Q}} is the kinematic
(and conjugate\footnote{As shown previously, the kinematic and conjugate momentum coincide for mechanical Lagrangians/Hamiltonians.})
momentum covector along \eq{\pt{x}_t\in\man{Q}}:
\begin{small}
\begin{align}
 \dt{\kap}_t = \sfb{X}^\sscr{K}_{\kap_t} 
   \quad \Rightarrow \quad 
   \bs{\kap}_t = \sfb{m}_{\pt{x}}(\dtsfb{x}_t)
\end{align}
\end{small}

 \paragraph{The \eq{\colift\psi}-Transformed System.}
 In subsequent sections, we will wish to use some given  diffeomorphism (perhaps local) on configuration space, \eq{\psi\in\Dfism(\man{Q})}, to transform some original Hamiltonian system to a new Hamiltonian system. 
In the geometric formulation, this is rather straightforward; we simply use the 
\textit{cotangent lift}\footnote{For any \eq{\psi\in\Dfism(\man{Q})}, recall the tangent lift, \eq{\tlift\psi}, and the cotangent lift, \eq{\colift\psi}, are given as follows for any \eq{\pt{u}_\pt{q}=(\pt{q},\sfb{u})\in\tsp\man{Q}} and \eq{\mu_\pt{q}=(\pt{q},\bs{\mu})\in\cotsp\man{Q}}:
\begin{align} \label{colifts_again}
\begin{array}{llllll}
     \tlift\psi \,=\, (\psi\circ\tpr, \dif \psi(\slot,\sblt) ) \in\Dfism(\tsp\man{Q}) 
     &,\qquad 
     \tlift\psi (\pt{q},\sfb{u}) \,=\, ( \psi(\pt{q}), \dif \psi_\ss{\pt{q}}\cdot\sfb{u}) \,=\, ( \psi(\pt{q}), \psi_*\sfb{u})
      &,\qquad 
      \tpr \circ \tlift\psi= \psi\circ\tpr
\\[2pt] 
      \colift\psi \,=\, (\psi\circ\copr, \inv{\dif \psi_\ii{\psi}} (\sblt,\slot)) \in \Spism(\cotsp\man{Q},\nbs{\omg})
      &,\qquad 
      \colift\psi (\pt{q},\bs{\mu}) 
      \,=\, 
       ( \psi(\pt{q}), \bs{\mu}\cdot \inv{ (\dif \psi_\ss{\pt{q}})} )
      \,=\, ( \psi(\pt{q}), \psi_*\bs{\mu})
       &,\qquad 
      \copr \circ \colift\psi= \psi\circ\copr
\end{array}
\end{align}
 where we have used \eq{\dif \inv{\psi}_\ii{\psi(\pt{q})} = \inv{(\dif \psi_\ss{\pt{q}})}}. 
},
\eq{\colift\psi\in\Spism(\cotsp\man{Q},\nbs{\omg})}, 
to pullback the original Hamiltonian system, \eq{(\cotsp\man{Q},\nbs{\omg},\mscr{K})}, to a new Hamiltonian system \eq{(\cotsp\man{Q},\colift\psi^*\nbs{\omg},\colift\psi^*\mscr{K})}. 
However, it automatically holds for any \eq{\psi\in\Dfism(\man{Q})} that 
\eq{\colift \psi^* \bs{\theta} = \bs{\theta}} and thus \eq{\colift \psi^* \nbs{\omg}=\nbs{\omg}} — this property is rather important and is derived in Appx.~\ref{sec:phase space}. 
That is, 
\eq{\colift\psi} is a cotangent bundle symplectomorphism (with the additional property that it preserves the canonical 
1-form):\footnote{This is one advantage of formulating the dynamics on \eq{\cotsp\man{Q}} rather than \eq{\tsp\man{Q}}. Cotangent lifts are always symplectic with respect to the canonical symplectic form, \eq{\nbs{\omg}=-\exd\bs{\theta}\in\formsex^2(\cotsp\man{Q})}, but the same is not true of tangent lifts which may not be symplectic with respect to the \eq{\mscr{L}}-adapted symplectic form, \eq{\nbs{\varpi}^\sscr{L}:=-\exd \bs{\varth}^\sscr{L} = -\lft{\iden}(\dif \mscr{L},\cdot)\in\formsex^2(\tsp\man{Q})}.}
\begin{small}
\begin{align} \label{colift_sp_gen}
\begin{array}{llllll}
\forall \, \psi\in\Dfism(\man{Q}):
\qquad\qquad 
   \bs{\theta} = \colift \psi^* \bs{\theta}  
 \qquad,\qquad 
      \nbs{\omg} = \colift \psi^* \nbs{\omg} 
\qquad,\qquad 
 \fnsize{i.e., } \;\; \colift\psi\in\Spism(\cotsp\man{Q},\nbs{\omg})
\end{array}
\end{align}
\end{small}
Thus,  \eq{\colift\psi} transforms the original Hamiltonian system, \eq{(\cotsp\man{Q},\nbs{\omg},\mscr{K})}, into a new Hamiltonian system, \eq{(\cotsp\man{Q},\nbs{\omg},\mscr{H})}, where the transformation is fully described by the transformed Hamiltonian function \eq{\mscr{H}:=\colift\psi^*\mscr{K}}: 
\begin{small}
\begin{align}
\begin{array}{llllll}
     \mscr{H}:= \colift{\psi}^* \mscr{K} \in\fun(\cotsp\man{Q})
\qquad,\qquad  
    \sfb{X}^\sscr{H} := \inv{\nbs{\omg}}(\dif \mscr{H},\cdot) \,=\, \sfb{X}^\ss{(\colift\psi^*\mscr{K})} = \colift\psi^* \sfb{X}^\sscr{K} \in \vechm(\cotsp\man{Q},\nbs{\omg})
\end{array}
\end{align}
\end{small}
where the relation \eq{\sfb{X}^\sscr{H}=\colift\psi^* \sfb{X}^\sscr{K}} means that the integral curves, flows, and integrals of motion of \eq{\sfb{X}^\sscr{H}} are related to those of \eq{\sfb{X}^\sscr{K}} as below:

\begin{small}
\begin{remrm} \label{rem:HK_related}
   The Hamiltonian vector fields \eq{\sfb{X}^\sscr{K}} and \eq{\sfb{X}^\sscr{H}=\colift\psi^*\sfb{X}^\sscr{K}} are \eq{\colift \psi}-related and, therefore: 
    \begin{small}
    \begin{itemize}[nosep]
        \item If \eq{\kap_t\in\cotsp\man{Q}} is an integral curve of \eq{\sfb{X}^\sscr{K}}, then \eq{\mu_t = \inv{\colift \psi}(\kap_t)\in\cotsp\man{Q}} is an integral curve of \eq{\sfb{X}^\sscr{H}}. That is,
        \begin{small}
        \begin{align} \label{intcurv_xform_gen}
          \fnsize{if: } \;\; \dt{\kap}_t = \sfb{X}^\sscr{K}_{\kap_t} 
            \;\;\, \fnsize{\&} \;\;\, 
            \dt{\mu}_t = \sfb{X}^\sscr{H}_{\mu_t}
            &&\;&&
            \fnsize{then: } \;\;
            \kap_t = \colift\psi(\mu_t)  \;\;\, \leftrightarrow \;\;\, \mu_t = \inv{\colift\psi}(\kap_t)
            &&,&&
             \mscr{K}(\kap_t) = \mscr{H}(\mu_t)
        \end{align}
        \end{small}
        \item If \eq{\varphi_t\in\Spism(\cotsp\man{Q},\nbs{\omg})} is the flow of \eq{\sfb{X}^\sscr{K}}, then the flow of \eq{\sfb{X}^\sscr{H}} is \eq{\phi_t =  \inv{\colift \psi} \circ \varphi_t \circ \colift \psi\in\Spism(\cotsp\man{Q},\nbs{\omg}) }. That is,
        \begin{small}
        \begin{align} \label{flows_xform_gen}
           \varphi^\sscr{K}_t \,=\, \colift\psi \circ \phi^\sscr{H}_t \circ \inv{\colift\psi} 
            \quad \leftrightarrow \quad 
            \phi^\sscr{H}_t =  \inv{\colift \psi} \circ \varphi^\sscr{K}_t \circ \colift \psi
        \end{align}
        \end{small} 
        \item If  \eq{f\in\fun(\cotsp\man{Q})} is an integral of motion of \eq{\sfb{X}^\sscr{K}}, then \eq{\colift \psi^*f} is an integral of motion of \eq{\sfb{X}^\sscr{H}} 
        (proof in footnote\footnote{From \eq{\sfb{X}^\sscr{H}= \colift \psi^* \sfb{X}^\sscr{K}}, it follows that \eq{\lderiv{\sfb{X}^{\mscr{H}}}(\colift \psi^*f) = \lderiv{\colift \psi^* \sfb{X}^\sscr{K}}(\colift \psi^*f) = \colift \psi^* \lderiv{\sfb{X}^{\mscr{K}}} f } (the last equality is a known property of Lie derivatives). Therefore, if \eq{\lderiv{\sfb{X}^{\mscr{K}}} f=0} then \eq{\lderiv{\sfb{X}^{\mscr{H}}}(\colift \psi^*f) = \colift \psi^*\lderiv{\sfb{X}^{\mscr{K}}} f=0} such that \eq{\colift \psi^*f} is an integral of motion of  \eq{\sfb{X}^\sscr{H}}. Equivalently, \eq{\colift\psi} is automatically a symplectomorphism such that \eq{\pbrak{\colift \psi^*g}{\colift \psi^*h} = \colift \psi^* \pbrak{g}{h}} for any \eq{g,h\in\fun(\cotsp\man{Q})}. Thus, if \eq{\lderiv{\sfb{X}^{\mscr{K}}} f \equiv \pbrak{f}{\mscr{K}}=0}, then \eq{\pbrak{\colift \psi^*f}{\mscr{H}} = \pbrak{\colift \psi^*f}{\colift \psi^*\mscr{K}} = \colift \psi^* \pbrak{f}{\mscr{K}} =0 }. }).
         That is,
         \begin{small}
         \begin{align} \label{iom_xform_gen}
              \lderiv{\sfb{X}^{\mscr{K}}} f = \pbrak{f}{\mscr{K}} = 0 
              \quad \Leftrightarrow \quad 
               \lderiv{\sfb{X}^{\mscr{H}}}( \colift\psi^* f) = \pbrak{\colift\psi^* f}{\mscr{H}} = 0 
                &&\big|&&
            \varphi_t^* f = f 
             \quad \Leftrightarrow \quad 
             \phi_t^*(\colift\psi^* f) = \colift\psi^* f
         \end{align}
         \end{small}
         \item[{}] $\;\;$
    \end{itemize}
    \end{small}
\end{remrm}
\end{small}


\noindent Recall now that \eq{\mscr{K}} is a mechanical Hamiltonian given by Eq.\eqref{Ksystem_gen}. 
\eq{\mscr{H}=\colift\psi^*\mscr{K}} is then given as follows  for any \eq{\mu_\pt{q}=(\pt{q},\bs{\mu})\in\cotsp\man{Q}}:
\begin{small}
\begin{align} \label{Hmech_gen}
    \mscr{H} (\mu_\pt{q}) \,=\, \mscr{K}\circ \colift\psi (\mu_\pt{q}) \;=\;  \tfrac{1}{2}\inv{\sfb{m}}_\ii{\psi(\pt{q})}(\psi_*\bs{\mu},\psi_*\bs{\mu}) \,+\, V(\psi(\pt{q}))
    &&=\;
    \tfrac{1}{2}\inv{\sfb{m}}_{\pt{x}}(\bs{\kap},\bs{\kap}) \,+\, V(\pt{x}) \,=\, \mscr{K}(\kap_\pt{x})
\end{align}
\end{small}
where the relations on the right hold for \eq{\kap_\pt{x}=\colift\psi(\mu_\pt{q})}.  
We see \eq{\mscr{H}} is also a mechanical Hamiltonian, but for a new  potential \eq{U:=\psi^* V \in\fun(\man{Q})} (treated as \eq{U\equiv \copr^* U\in\fun(\cotsp\man{Q})}), and a new kinetic energy metric \eq{\sfg:=\psi^*\sfb{m}\in\tens^0_2(\man{Q})}. For the latter, recall that the pullback of a \eq{(0,2)}-tensor field \eq{\sfb{m}} by \eq{\psi} is defined as follows for any \eq{\sfb{u},\sfb{v}\in\tsp[\pt{q}]\man{Q}} and \eq{\bs{\mu},\bs{\eta}\in\tsp[\pt{q}]^*\man{Q}}:
\begin{small}
\begin{align} \label{metric_pback_gen}
\begin{array}{rlllll}
      \sfg := &\!\!\!\! \psi^* \sfb{m} \,=\, \trn{\dif \psi} \cdot \sfb{m}_\ii{\psi} \cdot\dif \psi  
    &\quad,\qquad 
    \sfg_\ss{\!\pt{q}}(\sfb{u},\sfb{v}) \,=\, \sfb{m}_\ii{\psi(\pt{q})}(\psi_*\sfb{u},\psi_*\sfb{v}) 
\\[2pt] 
     \inv{\sfg} 
      = &\!\!\!\! \inv{\psi}_* \inv{\sfb{m}} \,=\, \dif \inv{\psi}_\ii{\psi} \cdot \inv{\sfb{m}}_\ii{\psi}\cdot \dif \invtrn{\psi}_\ii{\psi}  
     &\quad,\qquad 
     \inv{\sfg}_\ss{\!\pt{q}}(\bs{\mu},\bs{\eta}) \,=\, \inv{\sfb{m}}_\ii{\psi(\pt{q})}( \psi_* \bs{\mu},\psi_*\bs{\eta})  
\end{array}
\end{align}
\end{small}
where \eq{(\slot)_\ii{\psi} \equiv (\slot)\circ\psi}. 
As such, Eq.\eqref{Hmech_gen} leads to 
\begin{small}
\begin{align} \label{Hsystem_gen}
 \left. \begin{array}{lllll}
    \mscr{H}:= \colift\psi^* \mscr{K} = \mscr{K}\circ \colift\psi
\\[2pt] 
      \sfg := \psi^* \sfb{m} 
  \\[2pt] 
      \inv{\sfg} = \psi^* \inv{\sfb{m}} 
\\[2pt] 
      U:= \psi^* V = V\circ \psi
 \end{array}
 \right\} \quad \Rightarrow 
 &&
 \begin{array}{cccc}
         \mscr{H}(\mu_\pt{q}) \,=\, \tfrac{1}{2}\inv{\sfg}_\pt{q}(\bs{\mu},\bs{\mu}) \,+\, U(\pt{q})
           \qquad\qquad \fnsize{i.e., } \;\; \mscr{H} = \tfrac{1}{2}g^{ij}\plin_i \plin_j \,+\, U 
 \\[2pt] 
       \sfb{X}^\sscr{H} 
       \,=\, \upd^i\mscr{H}\hbpart{i} \,-\,  \pd_i\mscr{H}\hbpartup{i} 
    \;=\; 
    g^{ij} \plin_j \hbpart{i} \,+\, (  g^{sl}\Gamma^j_{is} \plin_j \plin_l \,-\, \pd_i U) \hbpartup{i}
 \end{array}
 \end{align}
 \end{small}
 where \eq{(\tp{r},\tp{\plin})} are any local cotangent-lifted coordinates (perhaps the same as in Eq.~\ref{Ksystem_gen}), where \eq{g^{ij}=\inv{\sfg}(\bep^i,\bep^j)} are the components in the \eq{r^i} basis,  
 and  where \eq{\Gamma^i_{jk}=\Gamma^i_{kj}} are \eq{\sfg}'s Levi-Civita connection coefficients for the \eq{r^i} frame fields on \eq{\man{Q}}.
We see from the above that an integral curve \eq{\mu_t=(\pt{q}_t,\bs{\mu}_t)} of \eq{\sfb{X}^\sscr{H}} has the property that  \eq{\bs{\mu}_t=\sfg_\ss{\pt{q}}(\dtsfb{q}_t)\in\cotsp[\pt{q}_t]\man{Q}} is the kinematic momentum —  as defined by the \eq{\psi}-induced metric, \eq{\sfg} — along the base curve \eq{\pt{q}_t=\copr(\mu_t)\in\man{Q}}:
\begin{small}
\begin{align}
    \dt{\mu}_t = \sfb{X}^\sscr{H}_{\mu_t} 
    \quad \Rightarrow \quad 
    \bs{\mu}_t = \sfg_\ss{\!\pt{q}}(\dtsfb{q}_t)
\end{align}
\end{small}
We also know from  Eq.\eqref{intcurv_xform_gen} that for any such \eq{\mu_t \in \cotsp\man{Q}}, then 
\eq{\colift\psi(\mu_t)=:\kap_t} is an integral curve of \eq{\sfb{X}^\sscr{K}} with the property that \eq{\bs{\kap}_t=\sfb{m}_{\pt{x}}(\dtsfb{x}_t)\in\cotsp[\pt{x}_t]\man{Q}} is the kinematic momentum — defined by the original metric, \eq{\sfb{m}} — along the base curve \eq{\pt{x}_t=\copr(\kap_t)\in\man{Q}}. 
Therefore, any integral curve \eq{\mu_t=(\pt{q}_t,\bs{\mu}_t)} of \eq{\sfb{X}^\sscr{H}} is \eq{\colift\psi}-related to some integral curve  \eq{\kap_t=(\pt{x}_t,\bs{\kap}_t)} of \eq{\sfb{X}^\sscr{K}}, such that the following relations hold along these curves (suppressing the argument \eq{t}): 
\begin{small}
\begin{align} \label{Hp_rels_gen}
\!\!\!\!
\begin{array}{rlllll}
       \fnsize{for:} \!& \dt{\kap}_t = \sfb{X}^\sscr{K}_{\kap_t} 
       \;\; \fnsize{\&} \;\;
            \dt{\mu}_t = \sfb{X}^\sscr{H}_{\mu_t}
\\[2pt] 
     \fnsize{with:} \!&   (\pt{x}_t,\bs{\kap}_t) \,=\, \colift\psi  (\pt{q}_t,\bs{\mu}_t) 
\end{array}
\;\;
\left\{\qquad  \begin{array}{lllll}
        \pt{x} \,=\, \psi(\pt{q}) 
\\[2pt] 
         \bs{\kap} = \sfb{m}_{\pt{x}}(\dtsfb{x}_t) = \psi_*(\bs{\mu})
 \\[2pt] 
          \bs{\mu}_t = \sfg_\ss{\!\pt{q}}(\dtsfb{q}) = \psi^*(\bs{\kap})
\end{array} \right.
&&,&&
\begin{array}{lllll}
    V(\pt{x}) \,=\, U(\pt{q})
\\[2pt] 
        \sfb{m}_{\pt{x}}(\dtsfb{x}, \dtsfb{x}) =\inv{\sfb{m}}_{\pt{x}}(\bs{\kap},\bs{\kap})  = \inv{\sfg}_\ss{\!\pt{q}}(\bs{\mu},\bs{\mu}) = \sfg_\ss{\!\pt{q}}(\dtsfb{q}, \dtsfb{q})
\\[2pt] 
     \mscr{H}(\pt{q},\bs{\mu}) \,=\,  \mscr{K}(\pt{x},\bs{\kap}) 
\end{array}
\end{align}
\end{small}

\begin{small}
\begin{notesq}
   \noindent \rmsb{The ``direction'' of transformation.}  In the above developments, we started with some original Hamiltonian system on \eq{(\cotsp\man{Q},\nbs{\omg})} characterized by a Hamiltonian function \eq{\mscr{K}\in\fun(\cotsp\man{Q})}. Then, for a given point transformation  \eq{\psi\in\Dfism(\man{Q})},  we chose to define a new Hamiltonian system on \eq{(\cotsp\man{Q},\nbs{\omg})} by defining \eq{\mscr{H}:=\colift{\psi}^*\mscr{K}}. The corresponding transformations of the Hamiltonian vector field, metric tensor, potential,  integral curves, etc.~all then follow. However, we could just as easily go the ``reverse direction''  and define some other Hamiltonian \eq{\til{\mscr{H}}:=\colift\psi_*\mscr{K}}. Everything 
   for the new Hamiltonian system \eq{(\cotsp\man{Q},\nbs{\omg},\mscr{\til{H}})} would then be defined from the original system in a similar manner as above, but under the exchange \eq{\psi \leftrightharpoons \inv{\psi} } (or \eq{\psi^* \leftrightharpoons \psi_*} where \eq{\inv{\psi}_* = \psi^*} for diffeomorphisms), 
   and likewise for \eq{\colift\psi}.\footnote{That is, \eq{\til{\mscr{H}}:= \colift\psi_* \mscr{K} =\mscr{K}\circ \inv{\colift\psi}} 
   would be a mechanical Hamiltonian for an induced kinetic energy metric \eq{\til{\sfg}:=\psi_*\sfb{m}\in\tens^0_2(\man{Q})} and potential function \eq{\til{U}:=\psi_*V=V\circ\inv{\psi}\in\fun(\man{Q})}. The corresponding Hamiltonian vector field \eq{\sfb{X}^\sscr{\wt{H}} = \colift\psi_*\sfb{X}^\sscr{K}\in\vechm(\cotsp\man{Q})} would have integral curves \eq{(\tilpt{q}_t,\tbs{\mu}_t)=\colift\psi(\pt{x}_t,\bs{\kap}_t)}, where \eq{(\pt{x}_t,\bs{\kap}_t)} is an integral curve of the original dynamics \eq{\sfb{X}^\sscr{K}}. }  
\end{notesq}
\end{small}

\subsection{Transformation of Conservative \& Nonconservative Forces} \label{sec:Hxform_noncon}

\paragraph{Conservative Forces.}
Including additional conservative forces in the above is easy if the corresponding potential is known; one simply adds the additional potential function (i.e., basic function) to the Hamiltonian.  
That is, for the Hamiltonian system \eq{(\cotsp\man{Q},\nbs{\omg},\mscr{K})}, all conservative forces can be built into the potential \eq{V\in\fun(\man{Q})} appearing in \eq{\mscr{K} = \tfrac{1}{2}m^{ij} \plin_i \plin_j + V\in\fun(\cotsp\man{Q})}. 
Here, for the dynamics formulated on \eq{\cotsp\man{Q}}, the potential is treated as a basic function which we will temporarily denote by \eq{\hat{V}:=\copr^*V\in\fun(\cotsp\man{Q})} (we usually make no distinction). The conservative forces are given by the exact 1-form \eq{\dif V\in\formsex(\man{Q})} which is, again, lifted to a horizontal exact  1-form, \eq{\copr^*\dif V=\dif \hat{V}\in \formsh\cap\formsex(\cotsp\man{Q})}. For the potential \eq{U=\psi^* V\in\fun(\man{Q})} in the \eq{\colift\psi}-related Hamiltonian system, \eq{(\cotsp\man{Q},\nbs{\omg},\mscr{H})}, the corresponding transformations 
are:\footnote{We use the relations \eq{\copr\circ \colift\psi = \psi\circ \copr} and \eq{(\varphi_1\circ\varphi_2)^* = \varphi_2^* \circ \varphi_1^*}. }
\begin{small}
\begin{align} \label{Fcon_gen}
\begin{array}{rlllll}
    \fnsize{on } \man{Q}: &\quad 
    U := \psi^* V 
    &,\qquad  \dif U \,=\, \psi^*\dif V = \dif V_\ii{\psi} \cdot \dif \psi 
\\[2pt] 
    \fnsize{on } \cotsp\man{Q}: &\quad 
    \hat{U}:=\copr^* U = \copr^*(\psi^* V) = \colift\psi^* \hat{V}
    &,\qquad 
   \dif \hat{U} = \colift \psi^* \dif \hat{V}  \,=\,  \copr^* (\psi^* \dif V) \,=\, \copr^*  \dif U
\end{array}
\end{align}
\end{small}
That is, the only non-trivial transformation of the conservative forces is on  the base, \eq{\dif U = \psi^*\dif V}. The horizontal lift to the cotangent bundle is then trivial (\eq{\dif \hat{U} = \copr^*\dif U}). 
Note the conservative forces  appear in the original Hamiltonian vector field as a \textit{vertical} Hamiltonian vector field, \eq{\sfb{X}^\ss{\hat{V}}= -\inv{\nbs{\omg}}(\dif \hat{V})= -\pd_i \hat{V}\hbpartup{i}\in\vectv\cap \vechm}, which transforms to \eq{\colift\psi^*\sfb{X}^\ss{\hat{V}} = \sfb{X}^\ss{\hat{U}} = -\inv{\nbs{\omg}}(\dif \hat{U}) = -\pd_i \hat{U}\hbpartup{i}\in\vectv\cap \vechm}. All of this is detailed in section \ref{sec:noncon}. 

\begin{small}
\begin{notesq}
    Everything just said about transforming potential functions and conservative forces is automatically built into the relation \eq{\mscr{H}=\colift\psi^*\mscr{K}}. This guarantees that \eq{\dif \mscr{H}} and \eq{\sfb{X}^\sscr{H}} already include the appropriately-transformed conservative force 1-form and vector field, respectively. 
\end{notesq}
\end{small}



\paragraph{Nonconservative Forces.} 
The following uses the symplectic geometric treatment of nonconservative forces that was detailed in section \ref{sec:noncon}. 
Suppose that the original Hamiltonian system is nonconservative, that is, \eq{(\cotsp\man{Q},\nbs{\omg},\mscr{K},\bfi{f})} where \eq{\bfi{f}\in \formsh(\cotsp\man{Q})} is  a horizontal (i.e., semi-basic), but non-exact,
1-form accounting for nonconservative forces 
(recall that non-exact$\,\Leftrightarrow\,$nonconservative\footnote{A horizontal 1-form that is also exact corresponds to a conservative force and its potential function can simply be added to the Hamiltonian.}).
The total dynamics are then given by the non-Hamiltonian vector field \eq{\sfb{X}^\sscr{K} + \inv{\nbs{\omg}}(\bfi{f})\in\vect(\cotsp\man{Q})} where \eq{\inv{\nbs{\omg}}(\bfi{f})=f_i\hbpartup{i}\in\vectv(\cotsp\man{Q})} is vertical. In local cotangent-lifted coordinate frame fields:
\begin{small}
\begin{align} \label{Knoncon_gen}
   \sfb{X}^\ss{\mscr{K},\,\bfi{f}} \,:=\,   \inv{\nbs{\omg}}(-\dif \mscr{K} + \bfi{f}) \,=\, \sfb{X}^\sscr{K} + \sfb{F}^\ss{\,\bfi{f}} \,=\,    \upd^i\mscr{K} \hbpart{i} \,-\,  ( \pd_i \mscr{K} - f_i) \hbpartup{i}
\end{align}
\end{small}
such that the time derivative of some \eq{h\in\fun(\cotsp\man{Q})} along the flow is no longer given by the Poisson bracket with \eq{\mscr{K}} but rather:
\begin{small}
\begin{align}
  \lderiv{\sfb{X}^{\mscr{K},\bfi{f}}} h \,=\, \pbrak{h}{\mscr{K}} + \lderiv{\sfb{F}^{\,\bfi{f}}}h \,=\, \pbrak{h}{\mscr{K}} \,+\, f_i \upd^i h
\end{align}
\end{small}
The transformed \eq{\colift\psi}-related Hamiltonian system is then \eq{(\cotsp\man{Q},\nbs{\omg},\mscr{H},\bs{\alpha})} with \eq{\mscr{H}:=\colift\psi^*\mscr{K}} the same as detailed in Eqs.~\ref{Hmech_gen} – \ref{Hsystem_gen} and with \eq{\bs{\alpha}:=\colift\psi^*\bfi{f}} the transformed  \eq{\tsph^*\!\pmb{.}(\cotsp\man{Q})}-valued nonconservative forces such that the total dynamics of the transformed system are then given by the non-Hamiltonian vector field:
\begin{small}
 \begin{align} \label{Hnoncon_gen}
    \colift\psi^* \sfb{X}^\ss{\mscr{K},\,\bfi{f}} =:\;  \sfb{X}^\ss{\mscr{H},\bs{\alpha}} \,=\, 
   \inv{\nbs{\omg}}(-\dif \mscr{H} + \bs{\alpha}) \,=\, \sfb{X}^\sscr{H} + \sfb{F}^\ss{\bs{\alpha}} \,=\,    \upd^i\mscr{H} \hbpart{i} \,-\,  ( \pd_i \mscr{H} - \alpha_i) \hbpartup{i}
&&
   \begin{array}{llll}
        \mscr{H}:=\colift\psi^* \mscr{K}
    \\[2pt] 
        \bs{\alpha} := \colift\psi^* \bfi{f}
   \end{array}
 \end{align}
\end{small}
The relation for the nonconservative force transformation can actually be simplified to a transformation of covectors on \eq{\man{Q}}. 
Let us limit consideration to two cotangent bundle points related by \eq{\kap_\pt{x} = \colift\psi(\mu_\pt{q})}. 
We then have: 
\begin{small}
\begin{align}
  \fnsize{for } \, \mu_\pt{q}  = \inv{\colift\psi}(\kap_\pt{x}): 
    \qquad \quad 
    \bs{\alpha}_{\mu_\pt{q}} := \colift\psi^*(\bfi{f}_{\!\kap_\pt{x}}) \,=\, \bfi{f}_{\!\kap_\pt{x}} \cdot  \dif (\colift\psi)_{\mu_\pt{q}} \in \tsph[\mu_\pt{q}]^*(\cotsp\man{Q})
\end{align}
\end{small}
But, at a point, the covector \eq{\bfi{f}_{\!\kap_\pt{x}}\in \tsph[\kap_\pt{x}]^*(\cotsp\man{Q})} can always be written as the horizontal lift, \eq{\bfi{f}_{\!\kap_\pt{x}} = \copr^* \sfb{f}_{\!\pt{x}}},  
of some \eq{\sfb{f}_{\!\pt{x}}\in\tsp[\pt{x}]^*\man{Q}}.\footnote{For instance, let \eq{\bfi{f}_{\!\kap_\pt{x}}\in \tsph[\kap_\pt{x}]^*(\cotsp\man{Q})} have local expression \eq{\bfi{f}_{\!\kap_\pt{x}} = f_i\hbdel[r^i]_{\kap_\pt{x}}}, then \eq{\bfi{f}_{\!\kap_\pt{x}} = \copr^* \sfb{f}_{\!\pt{x}}} where \eq{\sfb{f}_{\!\pt{x}}\in\tsp[\pt{x}]^*\man{Q}} has local expression \eq{\sfb{f}_{\!\pt{x}} = f_i\bep[r^i]_{\pt{x}}}. Here, \eq{f_i\in\mbb{R}} are the same components, \eq{\bep[r^i]=\dif r^i\in\forms(\man{Q})} are the \eq{r^i} basis 1-forms on the base, and \eq{\bdel[r^i] = \copr^* \bep[r^i] = \dif (\copr^* r^i)\in\formsh(\cotsp\man{Q})} are their basic cotangent-lift. }
The corresponding vertical vector is then \eq{ \sfb{F}^\ss{\,\bfi{f}}_{\!\kap_\pt{x}}  :=\inv{\nbs{\omg}}_{\kap_\pt{x}}(\bfi{f}) = \vlift{\sfb{f}_{\!\pt{x}}}\in\tspv[\kap_\pt{x}](\cotsp\man{Q})}.
The above relation \eq{\bs{\alpha}_{\mu_\pt{q}} = \colift\psi^*(\bfi{f}_{\!\kap_\pt{x}})} is then equivalent to the more simple \eq{\sfb{a}_\pt{q}:=\psi^*(\sfb{f}_{\!\pt{x}})\in\tsp[\pt{q}]^*\man{Q}}, followed by a trivial horizontal lift to get \eq{\bs{\alpha}_{\mu_\pt{q}} = \copr^* \sfb{a}_\pt{q} \in \tsph[\mu_\pt{q}]^*(\cotsp\man{Q})}. 
That is:\footnote{Derivation of Eq.\eqref{Fnoncon_gen}: We can always write the horizontal covector  \eq{\bfi{f}_{\!\kap_\pt{x}}\in \tsph[\kap_\pt{x}]^*(\cotsp\man{Q})} as \eq{\bfi{f}_{\!\kap_\pt{x}} = \copr^* \sfb{f}_{\!\pt{x}}} for some \eq{\sfb{f}_{\!\pt{x}}\in\tsp[\pt{x}]^*\man{Q}}. If we define \eq{\sfb{a}_\pt{q}:=\psi^*(\sfb{f}_{\!\pt{x}})\in\tsp[\pt{q}]^*\man{Q}} then we  see that \eq{ \bs{\alpha}_{\mu_\pt{q}} :=  \colift\psi^*(\bfi{f}_{\!\kap_\pt{x}})} is equivalent to:\\
\eq{\qquad\qquad \bs{\alpha}_{\mu_\pt{q}} :=\,   \colift\psi^*(\bfi{f}_{\!\kap_\pt{x}}) \,=\, 
\colift\psi^*(\copr^* \sfb{f}_{\!\pt{x}}) \,=\,
(\colift\psi^*\circ\copr^*)\sfb{f}_{\!\pt{x}} \,=\, (\copr \circ \colift\psi)^*\sfb{f}_{\!\pt{x}} \,=\, (\psi \circ\copr)^* \sfb{f}_{\!\pt{x}} \,=\, (\copr^*\circ \psi^*)\sfb{f}_{\!\pt{x}}
\,=\, \copr^*(\psi^* \sfb{f}_{\!\pt{x}}) \,=\, \copr^* \sfb{a}_\pt{q}   }. \\
where we have used the relations \eq{\copr\circ \colift\psi = \psi\circ \copr} and \eq{(\varphi_1\circ\varphi_2)^* = \varphi_2^* \circ \varphi_1^*}.  } 
\begin{small}
\begin{align} \label{Fnoncon_gen}
\begin{array}{rlllll}
    \fnsize{on } \man{Q}: &\quad 
    \sfb{a}_\pt{q} :=\psi^*(\sfb{f}_{\!\pt{x}}) \,=\, \sfb{f}_{\!\pt{x}} \cdot \dif \psi_\pt{q} \in\tsp[\pt{q}]^*\man{Q}
\\[2pt] 
    \fnsize{on } \cotsp\man{Q}: &\quad 
    \bs{\alpha}_{\mu_\pt{q}} :=\,   \colift\psi^*(\bfi{f}_{\!\kap_\pt{x}}) \,=\, \copr^* ( \psi^*\sfb{f}_{\!\pt{x}}) \,=\, \copr^* \sfb{a}_\pt{q}  \in \tsph[\mu_\pt{q}]^*(\cotsp\man{Q})
    &\Rightarrow \qquad 
    \sfb{F}^\ss{\bs{\alpha}}_{\mu_\pt{q}} \,=\, \vlift{\sfb{a}_\pt{q}} \in \tspv[\mu_\pt{q}](\cotsp\man{Q})
\end{array}
\end{align}
\end{small}
Where \eq{\sfb{f}_{\!\pt{x}}} and \eq{\sfb{a}_\pt{q}} are \textit{not} necessarily 1-forms on \eq{\man{Q}} evaluated at points; they are simply covectors at the indicated points. 
The above transformation applies to the covectors at the indicated points. That is, the components of \eq{\bs{\alpha}_{\mu_\pt{q}}}, which are elements of \eq{\mbb{R}}, transform the same as those of \eq{\sfb{a}_\pt{q}}. 
This is the same as the transformation of the conservative force covector \eq{\dif \hat{U}_{\mu_\pt{q}}} given by  Eq.\eqref{Fcon_gen}. Note however, that Eq.\eqref{Fcon_gen} is valid for the covector \textit{field}, \eq{\dif \hat{U}\in\formsh\cap\formsex(\cotsp\man{Q})}, whose components are functions. 
We have given the above nonconservative force transformation in terms of covectors at a point (rather than 1-forms, i.e., covector fields) for the following reason:

\begin{small}
\begin{notesq}
    Above, we said that nonconservative forces are modeled as a horizontal non-exact 1-form, \eq{\bfi{f}\in\formsh(\cotsp\man{Q})}, which gives a non-Hamiltonian vertical vector field, \eq{\sfb{F}^\ss{\,\bfi{f}}:=\inv{\nbs{\omg}}(\bfi{f})\in\vectv(\cotsp\man{Q})}. However, in general, the domain of a nonconservative force 1-form need not be \eq{\cotsp\man{Q}} (nor \eq{\tsp\man{Q}}) such that, technically, \eq{\bfi{f}} may not be a true 1-form on \eq{\cotsp\man{Q}} (and \eq{ \sfb{F}^\ss{\,\bfi{f}}} not a true vector field on \eq{\cotsp\man{Q}}). Yet, even in such general cases, \eq{\bfi{f}} \textit{does} take values in \eq{\tsph^*\!\pmb{.}(\cotsp\man{Q})}. That is, nonconservative forces are more generally modeled as a \eq{\tsph^*\!\pmb{.}(\cotsp\man{Q})}-valued covector, \eq{\bfi{f}}, 
     whose domain may or may not be \eq{\cotsp\man{Q}}.\footnote{Dissipative forces or velocity-dependent potential forces can indeed be modeled as horizontal 1-forms on \eq{\cotsp\man{Q}} (or on \eq{\tsp\man{Q}}). However, more general nonconservative forces are simply \eq{\tsph^*\!\pmb{.}(\cotsp\man{Q})}-valued covector fields with ``some domain''. E.g., a non-feedback control force or some nonautonomous nonconservative force may have domain differing from \eq{\cotsp\man{Q}} (or \eq{\tsp\man{Q}}). For the present context, it only matters that such forces take values in \eq{\tsph^*\!\pmb{.}(\cotsp\man{Q})} (or in \eq{\tsph\pmb{.}(\cotsp\man{Q})}).  }
    The corresponding \eq{ \sfb{F}^\ss{\,\bfi{f}}} is then still a \eq{\tspv\pmb{.}(\cotsp\man{Q})}-valued vector whose domain may or may not be  \eq{\cotsp\man{Q}}. 
    These generalizations change nothing about the above expressions for \eq{\sfb{X}^\ss{\mscr{K},\bfi{f}}} and \eq{\sfb{X}^\ss{\mscr{H},\bs{\alpha}}} nor the above transformation of nonconservative forces; it was the codomain/target space, not the domain, of the nonconservative forces which was important for Eq.\eqref{Knoncon_gen}-Eq.\eqref{Fnoncon_gen}.   
    Note, however, that the notation \eq{\bfi{f}_{\!\kap_\pt{x}}} or \eq{\bs{\alpha}_{\mu_\pt{q}}} should generally \textit{not} be taken to mean \eq{1}-forms on \eq{\cotsp\man{Q}} evaluated at the indicated points; they are simply covectors at the indicated points.  
\end{notesq}
\end{small}

 The main point of the above developments is that conservative or nonconservative forces take values in \eq{\tsph\pmb{.}\!^*(\cotsp\man{Q})} and thus thus transform the same as elements of \eq{\tsp[\cdt]^*\man{Q}}. That is, if \eq{\pt{x}=\psi(\pt{q})} then some original force \eq{f_i\dif r^i|_{\pt{x}}} is pulled back by \eq{\psi} to a new force \eq{\alpha_i \dif r^i|_\ss{\pt{q}} } given by 
\begin{small}
\begin{align}
    \alpha_i(\pt{q}) = f_j(\pt{x}) \pd_i\psi^j(\pt{q})
\end{align}
\end{small}
the above is same whether we view \eq{f_i\dif r^i|_{\pt{x}}} and \eq{\alpha_i \dif r^i|_\ss{\pt{q}}} as covectors on \eq{\man{Q}} or as horizontal covectors on \eq{\cotsp\man{Q}}.

\begin{small}
\begin{remrm} \label{rem:HK_related_noncon}
Remark \ref{rem:HK_related} mostly still holds in the presence of nonconservative forces. The only difference is that the vector fields and flows are no longer symplectic:
 \begin{small}
    \begin{itemize}[nosep]
        \item If \eq{\kap_t\in\cotsp\man{Q}} is an integral curve of \eq{\sfb{X}^\ss{\mscr{K},\bfi{f}}}, then \eq{\mu_t = \inv{\colift \psi}(\kap_t)\in\cotsp\man{Q}} is an integral curve of \eq{\sfb{X}^\ss{\mscr{H},\bs{\alpha}}}. 
        \item If \eq{\varphi_t\in\Dfism(\cotsp\man{Q})} is the flow of \eq{\sfb{X}^\ss{\mscr{K},\bfi{f}}}, then the flow of \eq{\sfb{X}^\ss{\mscr{H},\bs{\alpha}}} is \eq{\phi_t =  \inv{\colift \psi} \circ \varphi_t \circ \colift \psi\in\Dfism(\cotsp\man{Q}) }. 
        \item If  \eq{h\in\fun(\cotsp\man{Q})} is an integral of motion of \eq{\sfb{X}^\ss{\mscr{K},\bfi{f}}}, then \eq{\colift \psi^*h} is an integral of motion of \eq{\sfb{X}^\ss{\mscr{H},\bs{\alpha}}}
        (proof in footnote\footnote{From \eq{\sfb{X}^\ss{\mscr{H},\bs{\alpha}}= \colift \psi^* \sfb{X}^\ss{\mscr{K},\,\bfi{f}}}, it follows that \eq{\lderiv{\sfb{X}^{\mscr{H},\bs{\alpha}}}(\colift \psi^*h) = \lderiv{\colift \psi^* \sfb{X}^\ss{\mscr{K},\,\bfi{f}}}(\colift \psi^*h) = \colift \psi^* \lderiv{\sfb{X}^{\mscr{K},\bfi{f}}} h } (the last equality is a known property of Lie derivatives). Therefore, if \eq{\lderiv{\sfb{X}^{\mscr{K},\bfi{f}}} h=0} then \eq{\lderiv{\sfb{X}^{\mscr{H},\bs{\alpha}}}(\colift \psi^*h) = \colift \psi^*\lderiv{\sfb{X}^{\mscr{K},\bfi{f}}} h=0} such that \eq{\colift \psi^*h} is an integral of motion of  \eq{\sfb{X}^\ss{\mscr{H},\bs{\alpha}}}. }).
         That is,
         \begin{small}
         \begin{align} \label{iom_xform_noncon_gen}
              \lderiv{\sfb{X}^{\mscr{K},\bfi{f}}} (h) \,=\, \pbrak{h}{\mscr{K}} \,+\,  \lderiv{\sfb{F}^{\,\bfi{f}}}h \,=\, 0 
              \qquad \Leftrightarrow \qquad 
        \lderiv{\sfb{X}^{\mscr{H},\bs{\alpha}}}( \colift\psi^* h) \,=\, \pbrak{\colift\psi^* h}{\mscr{H}} \,+\, \lderiv{\sfb{F}^{\bs{\alpha}}} (\colift\psi^*h) \,=\, 0 
         \end{align}
         \end{small}
         In other words, \eq{\varphi_t^* h = h 
             \;\; \Leftrightarrow \;\;
             \phi_t^*(\colift\psi^* h) = \colift\psi^* h}. 
    \end{itemize}
    \end{small}
\end{remrm}
\end{small}

\subsection{Transformation of Corresponding Newton-Riemann Dynamics on \txi{Q}}

Let \eq{\sfb{X}^\ss{\mscr{K},\bfi{f}} = \sfb{X}^\sscr{K}+\sfb{F}^\ss{\,\bfi{f}}\in\vect(\cotsp\man{Q})} be the vector field for some given nonconservative mechanical Hamiltonian system, \eq{(\cotsp\man{Q},\nbs{\omg},\mscr{K},\bfi{f})}, and
consider the \eq{\colift\psi}-related system \eq{(\cotsp\man{Q},\nbs{\omg},\mscr{H},\bs{\alpha})} with vector field \eq{\sfb{X}^\ss{\mscr{H},\bs{\alpha}}:= \colift\psi^* \sfb{X}^\ss{\mscr{K},\,\bfi{f}} = \sfb{X}^\sscr{H}+\sfb{F}^\ss{\bs{\alpha}}\in\vect(\cotsp\man{Q})}, as detailed previously.  
Let
\eq{\kap_t=(\pt{x}_t,\bs{\kap}_t)\in\cotsp\man{Q}} be an integral curve of \eq{\sfb{X}^\ss{\mscr{K},\bfi{f}}} such that \eq{\mu_t=(\pt{q}_t,\bs{\mu}_t):=\inv{\colift\psi}(\kap_t)\in\cotsp\man{Q}} is an integral curve of \eq{\sfb{X}^\ss{\mscr{H},\bs{\alpha}}= \colift\psi^* \sfb{X}^\ss{\mscr{K},\,\bfi{f}}}. This is equivalent to the following \eq{\psi}-related second-order dynamics on \eq{\man{Q}}:
\begin{small}
\begin{align} \label{Qsystem_gen}
\begin{array}{rlllll}
     &\del{\dtsfb{x}}^\ss{\sfb{m}} \dtsfb{x}_t \,=\, \inv{\sfb{m}}_\pt{x}(-\dif V + \sfb{f})
\\[4pt] 
   \fnsize{or:}  &\del{\dtsfb{x}}^\ss{\sfb{m}} \bs{\kap}_t = -\dif V_{\pt{x}} \,+\, \sfb{f}_{\!\pt{x}}
\\[4pt] 
   \fnsize{with:} &\bs{\kap}_t = \sfb{m}_{\pt{x}}(\dtsfb{x}_t)
\end{array}
&& \xrightarrow{\;\;\psi\;\;} &&
\begin{array}{lllll}
     \del{\dtsfb{q}}^\ss{\sfg} \dtsfb{q}_t \,=\, \inv{\sfg}_\ss{\!\pt{q}}(-\dif U + \sfb{a})
\\[4pt] 
   \del{\dtsfb{q}}^\ss{\sfg} \bs{\mu}_t = -\dif U_\ss{\pt{q}} \,+\, \sfb{a}_\ss{\pt{q}}
\\[4pt] 
    \bs{\mu}_t = \sfg_\ss{\!\pt{q}}(\dtsfb{q}_t)
\end{array}
\qquad 
\left\{ \;\;\begin{array}{llll}
     \pt{q}_t = \inv{\psi}(\pt{x}_t) 
 \\[2pt] 
     \bs{\mu}_{t} = \psi^*( \bs{\kap}_{t})
 \\[2pt] 
     \sfg = \psi^* \sfb{m}
 \\[2pt] 
     U = \psi^* V
 \\[2pt] 
    \sfb{a}_\ss{\pt{q}} = \psi^*(\sfb{f}_{\!\pt{x}})
\end{array} \right.
\end{align}
\end{small}
where \eq{\bs{\kap}_t \in\tsp[\pt{x}_t]^*\man{Q}} and \eq{\bs{\mu}_t\in\tsp[\pt{q}_t]^*\man{Q}} are the kinematic momentum covectors along the base curves.\footnote{For mechanical systems, the kinematic and conjugate momentum are the same.}
Note they are defined not only from different base curves but also from different metrics. Likewise, the above \eq{\nab^\ss{\sfb{m}}} and \eq{\nab^\ss{\sfg}} are the Levi-Civita connections for two different metrics.  
To clarify, the momentum curves, \eq{\kap_t=(\pt{x}_t,\bs{\kap}_t)} and  \eq{\mu_t=(\pt{q}_t,\bs{\mu}_t)=\inv{\colift\psi}(\kap_t)} then satisfy
\begin{small}
\begin{align} \label{T*Qsystem_gen}
\begin{array}{rllllll}
     &\dt{\kap}_t \,=\, \sfb{X}^\ss{\mscr{K},\,\bfi{f}}_{\kap_t} \,=\, \inv{\nbs{\omg}}_{\kap_t}( -\dif \mscr{K} + \bfi{f}) 
\\[4pt] 
    \fnsize{with} &\mscr{K}(\kap_t) = \tfrac{1}{2}\inv{\sfb{m}}_{\pt{x}}(\bs{\kap},\bs{\kap}) + V(\pt{x})
\end{array}
&& \xrightarrow{\;\;\colift\psi\;\;} &&
\begin{array}{llllll}
     &\dt{\mu}_t \,=\, \sfb{X}^\ss{\mscr{H},\bs{\alpha}}_{\mu_t} \,=\, \inv{\nbs{\omg}}_{\mu_t}( -\dif \mscr{H} + \bs{\alpha}) 
\\[4pt] 
     &\mscr{H}(\mu_t) = \tfrac{1}{2}\inv{\sfg}_\ss{\!\pt{q}}(\bs{\mu},\bs{\mu}) + U(\pt{q})
\end{array}
\qquad 
\left\{ \;\;\begin{array}{llll}
     \mu_t = \inv{\colift\psi}(\kap_t) 
 \\[2pt] 
    \mscr{H} = \colift\psi^*\mscr{K}
 \\[2pt] 
    \bs{\alpha}_{\mu} = \colift\psi^* \bfi{f}_{\!\kap} 
\end{array} \right.
\end{align}
\end{small}
where \eq{\bfi{f}_{\!\kap} = \copr^* \sfb{f}_{\!\pt{x}} \in \tsph[\kap_{\pt{x}}]^*(\cotsp\man{Q})} and  \eq{ \bs{\alpha}_{\mu} = \copr^* \sfb{a}_\ss{\pt{q}} \in \tsph[\mu_\ss{\pt{q}}]^*(\cotsp\man{Q}) } are the horizontal lifts and where all the relations in Eq.\eqref{Qsystem_gen} still apply in the above.

\subsection{Passive Interpretation as a Coordinate Transformation} \label{sec:Hmech_xform_passive} 

\begin{small}
\begin{notation}
    Above, \eq{\pt{q}\in\man{Q}} denoted a point (or curve) on the configuration manifold. In the following, we will instead return to our usual convention of using \eq{\tp{q}=(q^1,\dots, q^n):\chart{Q}{\tp{q}}\to\mbb{R}^n} to denote coordinate functions with associated frame fields on the base denoted \eq{\bt_i\equiv \bt[q^i]\in\vect(\man{Q})} and  \eq{\btau^i\equiv \btau[q^i]\in\forms(\man{Q})}. 
     We will also re-use notation such as \eq{g_{ij}} and \eq{\Gamma^i_{jk}} but with at a different meaning than as above.  
\end{notation}
\end{small}

So far, we have used some given \eq{\psi\in\Dfism(\man{Q})}, along with its cotangent lift, to transform a original mechanical Hamiltonian system, \eq{(\cotsp\man{Q},\nbs{\omg},\mscr{K})}, to a different mechanical Hamiltonian system, \eq{(\cotsp\man{Q},\nbs{\omg},\mscr{H})}.  The symplectic form was preserved but most other relevant objects (e.g., the Hamiltonian, metric, potential, Hamiltonian vector field, forces, integral curves, flow, integrals of motions, etc.) were all transformed to entirely new objects. Although coordinate were not actually necessary to characterize the resultant transformed objects, when we used local coordinates (for clarity), we used the \textit{same} cotangent-lifted coordinates, \eq{(\tp{r},\tp{\plin})}, for both the original \eq{\mscr{K}}-system  and transformed \eq{\mscr{H}}-system. 

Now, we will instead consider the same original system \eq{(\cotsp\man{Q},\nbs{\omg},\mscr{K})} expressed in some original  cotangent-lifted coordinates, \eq{(\tp{r},\tp{\plin})}, and then use a configuration coordinate transformation \eq{\uppsi\in\Dfism(\rchart{R}{\tp{q}}^n ; \rchart{R}{\tp{r}}^n)}, along with its cotangent lift, to express the \textit{same} system in \textit{different} cotangent-lifted coordinates, \eq{(\tp{q},\tp{\pf})}. Unlike the previous sections, nothing about the original system \eq{(\cotsp\man{Q},\nbs{\omg},\mscr{K})} ever changes. Only our coordinate description of it changes. Coordinate transformations do not ``do'' anything to a system; the Hamiltonian, metric, potential, vector field, integral curves, flow,  etc.~all stay exactly the same. We simply express them using different coordinates and their associated frame fields.
This can always be done using any arbitrary coordinate transformation. However, we wish to define a specific coordinate transformation \eq{(\tp{q},\tp{\pf})\mapsto (\tp{r},\tp{\plin})} such that the resulting \eq{(\tp{q},\tp{\pf})}-representation of the \textit{original} \eq{\mscr{K}}-system ``looks the same'' as the \eq{(\tp{r},\tp{\plin})}-representation of the transformed \eq{\mscr{H}}-system given in the preceding sections.  
In particular, the ODEs for the \eq{(\tp{q},\tp{\pf})}-representation of the original vector field, \eq{\sfb{X}^\sscr{K}}, will be the same as the ODEs for the \eq{(\tp{r},\tp{\plin})}-representation of the transformed vector field \eq{\sfb{X}^\sscr{H}=\colift\psi^*\sfb{X}^\sscr{K}} of the previous sections.


\paragraph{Interpretation of a Diffeomorphism as a Coordinate Transformation.}
We now re-interpret the maps \eq{\psi\in\Dfism(\man{Q})} and \eq{\colift\psi\in\Spism(\cotsp\man{Q},\nbs{\omg})} as passive coordinate transformations. 
For any given \eq{\psi\in\Dfism(\man{Q})} and any given coordinates \eq{\tp{r}=(r^1,\dots,r^n):\chart{Q}{\tp{r}}\to\mbb{R}^n}, one may define new coordinates by \eq{\tiltup{q}:=\tp{r}\circ \psi} and also by \eq{\tp{q}:= \tp{r}\circ \inv{\psi}}. 
The developments of the previous, and subsequent,  sections will correspond to the latter; new  coordinates \eq{\tp{q}=(q^1,\dots,q^n)} are defined from given coordinates \eq{\tp{r}} by 
\begin{small}
\begin{align} \label{pass1}
    \tp{q} := \psi_* \tp{r} = \tp{r}\circ \inv{\psi} :\chart{Q}{\tp{q}} \to \rchart{R}{\tp{q}}^n 
    && \leftrightarrow &&
    \tp{r} = \psi^* \tp{q} = \tp{q}\circ \psi : \chart{Q}{\tp{r}} \to \rchart{R}{\tp{r}}^n 
\end{align}
\end{small}
Now consider the cotangent lift, \eq{\Psi:=\colift\psi\in\Spism(\cotsp\man{Q},\nbs{\omg})}. Let \eq{\tp{\xi}=(\tp{r},\tp{\plin}):=\colift\tp{r}:\cotsp\chart{Q}{\tp{r}}\to\mbb{R}^{2n}} be the original cotangent-lifted coordinates defined from \eq{\tp{r}}. If we define new cotangent bundle coordinates by \eq{\tp{z} := \tp{\xi}\circ \inv{\Psi}}, then this is equivalent to the cotangent-lifted coordinates \eq{\tp{z}=\colift\tp{q}} for the above \eq{\tp{q}=\tp{r}\circ\inv{\psi}}. That is, 
\begin{small}
\begin{align}
\begin{array}{lllll}
     \colift\psi \in\Spism(\cotsp\man{Q},\nbs{\omg})
\end{array}
&&
\begin{array}{ccccc}
     \tp{q} := \tp{r} \circ \inv{\psi}
     &\leftrightarrow & 
     \tp{r} = \tp{q} \circ \tp{\psi}
\\[2pt] 
     \tp{z} :=\, \tp{\xi}\circ \inv{\colift\psi} \,=\, \colift\tp{q} 
 & \leftrightarrow & 
 \tp{\xi} := \colift\tp{r} \,=\, \tp{z}\circ \colift\psi  
\end{array}
\end{align}
\end{small}

\begin{small}
\begin{remrm} \label{rem:act_v_pass}
    With \eq{\tp{q}\leftrightarrow \tp{r}} as above, 
    the \eq{\tp{q}}-representation of any \eq{\sfb{Q}\in\tens^r_s(\man{Q})} looks the same as the \eq{\tp{r}}-representation of \eq{\psi^*\sfb{Q}} and, likewise, the \eq{\tp{z}=\colift\tp{q}}-representation of any \eq{\sfb{T}\in\tens^r_s(\cotsp\man{Q})} looks the same as the  \eq{\tp{\xi}=\colift\tp{r}}-representation of 
    \eq{\,\colift \psi^*\sfb{T}}:\footnote{Eq.\eqref{act_v_pass} is quick to verify:  using \eq{\tp{q}=\tp{r}\circ\inv{\psi}} and \eq{\tp{z}=\tp{\xi}\circ\inv{\colift\psi}}: \\
        \eq{\qquad\qquad \tp{q}_*\sfb{Q} = (\tp{r} \circ \inv{\psi})_*\sfb{Q} = \tp{r}_*(\inv{\psi}_*\sfb{Q}) = \tp{r}_*( \psi^* \sfb{Q})
        \qquad,\qquad 
        \tp{z}_*\sfb{T} = (\tp{\xi} \circ \inv{\colift\psi})_*\sfb{T} =  \tp{\xi}_*(\inv{\colift\psi}_*\sfb{T})  =
        \tp{\xi}_*( \colift\psi^*\sfb{T}) } .
    } 
    \begin{small}
    \begin{align}\label{act_v_pass}
    \begin{array}{rllll}
        \tp{q}:=\tp{r}\circ\inv{\psi}: 
         &\qquad 
        \tp{q}_*\sfb{Q} 
        \,=\, \tp{r}_*( \psi^*\sfb{Q}) 
        &\qquad \qquad 
        \fnsize{i.e.,} \quad \crd{\sfb{Q}}{q} \,=\, \crd{\psi^*\sfb{Q}}{r} 
    \\[2pt] 
        \tp{z} := \tp{\xi}\circ \inv{\colift \psi}:
         &\qquad 
        \tp{z}_*\sfb{T} 
        \,=\, \tp{\xi}_*( \colift\psi^*\sfb{T}) 
        &\qquad \qquad 
        \fnsize{i.e.,} \quad \crd{\sfb{T}}{z} \,=\, \crd{\colift\psi^*\sfb{T}}{\xi}  
    \\[2pt] 
        \;\;
    \end{array}
    \end{align}
    \end{small}
\end{remrm}
\end{small}

Now, for a coordinate transformation \eq{\tp{q}\leftrightarrow\tp{r}}, there necessarily exists a transition function \eq{\ns{c}^\ss{r}_\ss{q}=\tp{r}\circ\inv{\tp{q}}:\rchart{R}{\tp{q}}^n\to\rchart{R}{\tp{r}}^n} such that \eq{\tp{r}=\ns{c}^\ss{r}_\ss{q}\circ\tp{q}} which, by slight abuse of notation, we write as \eq{\tp{r}=\ns{c}^\ss{r}_\ss{q}(\tp{q})}. 
With \eq{\tp{q}:=\tp{r}\circ\inv{\psi}} 
defined as above, one finds that \eq{\ns{c}^\ss{r}_\ss{q}=\cord{\uppsi}{r}=\cord{\uppsi}{q}} is equivalent to both the  \eq{\tp{r}}-representation and the \eq{\tp{q}}-representation  of  \eq{\psi\in\Dfism(\man{Q})} 
(shown in footnote\footnote{The relations in Eq.\eqref{Qdif_passive} are easily verified using the defined relation \eq{\tp{q} :=  \tp{r}\circ \inv{\psi} \,\leftrightarrow \, \tp{r} = \tp{q}\circ \psi }, leading to:
\begin{align}\nonumber 
  \begin{array}{rllllll}
     \uppsi &\!\!\!\! :=\;  \cord{\uppsi}{q} \,=\, \tp{q} \circ \psi \circ \inv{\tp{q}} \,=\, \tp{r}\circ \inv{\tp{q}} \,=\, \ns{c}^\ss{r}_\ss{q} 
      &\equiv &
        \cord{\uppsi}{r} \,=\, \tp{r} \circ \psi \circ \inv{\tp{r}} \,=\, \tp{r} \circ \inv{(\tp{r}\circ\inv{\psi})} \,=\, \tp{r} \circ \inv{\tp{q}}  \,=\, \ns{c}^\ss{r}_\ss{q} : \rchart{R}{\tp{q}}^n \to \rchart{R}{\tp{r}}^n
\\[2pt] 
     \inv{\uppsi}  &\!\!\!\! :=\;  \cord{\inv{\uppsi}}{r}  \,=\, \tp{r}\circ \inv{\psi}\circ \inv{\tp{r}} \,=\, \tp{q}\circ \inv{\tp{r}} \,=\, \ns{c}^\ss{q}_\ss{r}  
        &\equiv &
       \cord{\inv{\uppsi}}{q} \,=\, \tp{q}\circ \inv{\psi}\circ \inv{\tp{q}} \,=\, \tp{q}\circ \inv{(\tp{q}\circ \psi)} \,=\, \tp{q}\circ \inv{\tp{r}}  \,=\, \ns{c}^\ss{q}_\ss{r}  : \rchart{R}{\tp{r}}^n \to \rchart{R}{\tp{q}}^n
\end{array}  
\end{align} }):
\begin{small}
\begin{align} \label{Qdif_passive}
\begin{array}{rllllllllll}
     \uppsi &\!\!\!\! :=\;  
     \tp{r} \circ \psi \circ \inv{\tp{r}} 
     &\!\!\!\!=\, \tp{q} \circ \psi \circ \inv{\tp{q}} 
       &\!\!\!\!=\, \tp{r} \circ \inv{\tp{q}}  \,=\, \ns{c}^\ss{r}_\ss{q} : \rchart{R}{\tp{q}}^n \to \rchart{R}{\tp{r}}^n
       &,\qquad\qquad 
       \tp{r}= \tp{q}\circ \psi=\uppsi \circ\tp{q} \equiv \uppsi (\tp{q})
\\[2pt] 
     \inv{\uppsi}  &\!\!\!\! :=\;  
     \tp{r}\circ \inv{\psi}\circ \inv{\tp{r}} 
     &\!\!\!\!=\,
      \tp{q}\circ \inv{\psi}\circ \inv{\tp{q}}   &\!\!\!\!=\, \tp{q}\circ \inv{\tp{r}}  \,=\, \ns{c}^\ss{q}_\ss{r}  : \rchart{R}{\tp{r}}^n \to \rchart{R}{\tp{q}}^n
       &,\qquad\qquad
       \tp{q} := \tp{r}\circ\inv{\psi} = \inv{\uppsi}\circ\tp{r} \equiv \inv{\uppsi}(\tp{r})
\end{array}
\end{align}
\end{small}
Where we have denoted the coordinate transformation by \eq{\uppsi=\ns{c}^\ss{r}_\ss{q}\in\Dfism(\rchart{R}{\tp{q}}^n ; \rchart{R}{\tp{r}}^n)} since it can be thought of as a reinterpretation of the actual map \eq{\psi\in\Dfism(\man{Q})}. 
Similarly, for any cotangent lifted coordinate transformation  \eq{\tp{\xi}=\colift\tp{r} \leftrightarrow \tp{z}=\colift\tp{q} } there exists some transition function \eq{C^\ss{\xi}_\ss{z}=\tp{\xi}\circ \inv{\tp{z}}} such that  \eq{\tp{\xi}= C^\ss{\xi}_\ss{z}\circ \tp{z} \equiv C^\ss{\xi}_\ss{z}(\tp{z})} and it then also holds that  \eq{C^\ss{\xi}_\ss{z}=\colift \ns{c}^\ss{r}_\ss{q}}. 
With \eq{\tp{q}:=\tp{r}\circ\inv{\psi}} and  \eq{\tp{z}:=\tp{\xi}\circ\inv{\colift\psi}}  defined as above (which implies \eq{\tp{z}=\colift\tp{q}}), we again find that \eq{C^\ss{\xi}_\ss{z} = \cord{\upPsi}{\xi} = \cord{\upPsi}{z}}  is  equivalent to both the \eq{\tp{\xi}}-representation and the \eq{\tp{z}}-representation  of \eq{\Psi=\colift{\psi} \in\Spism(\cotsp\man{Q})} 
(shown in the footnote\footnote{For \eq{\Psi:=\colift\psi} and \eq{\tp{\xi}:=\colift\tp{r}} and \eq{\tp{z}:=\tp{\xi}\circ\inv{\Psi}=\colift\tp{q}}, then we have:
\begin{align} \nonumber 
\begin{array}{lllllllll}
  \upPsi &\!\!\!\! :=\;  \cord{\upPsi}{z} \,=\, \tp{z} \circ \Psi \circ \inv{\tp{z}} \
  ,=\, \tp{\xi}\circ \inv{\tp{z}} \,=\, C^\ss{\xi}_\ss{z} 
      &\equiv &
        \cord{\upPsi}{\xi} \,=\, \tp{\xi} \circ \Psi \circ \inv{\tp{\xi}} \,=\, \tp{\xi} \circ \inv{(\tp{\xi}\circ\inv{\Psi})} \,=\, \tp{\xi} \circ \inv{\tp{z}}  \,=\, C^\ss{\xi}_\ss{z} : \rchart{R}{\tp{z}}^{2n} \to \rchart{R}{\tp{\xi}}^{2n}
\\[2pt] 
    \inv{\upPsi}  &\!\!\!\! :=\;  \cord{\inv{\upPsi}}{\xi}  \,=\, \tp{\xi}\circ \inv{\Psi}\circ \inv{\tp{\xi}} \,=\, \tp{z}\circ \inv{\tp{\xi}} \,=\, C^\ss{z}_\ss{\xi}  
        &\equiv &
       \cord{\inv{\upPsi}}{z} \,=\, \tp{z}\circ \inv{\Psi}\circ \inv{\tp{z}} \,=\, \tp{z}\circ \inv{(\tp{z}\circ \Psi)} \,=\, \tp{z}\circ \inv{\tp{\xi}}  \,=\, C^\ss{z}_\ss{\xi}  : \rchart{R}{\tp{\xi}}^{2n} \to \rchart{R}{\tp{z}}^{2n}
\end{array}
\end{align} }): 
\begin{small}
\begin{align} \label{T*Qdif_passive}
\begin{array}{rllllllllll}
     \upPsi &\!\!\!\! :=\;  
     \tp{\xi} \circ \colift\psi\circ \inv{\tp{\xi}} 
     &\!\!\!\!=\, \tp{z} \circ \colift\psi \circ \inv{\tp{z}} 
       &\!\!\!\!=\, \tp{\xi} \circ \inv{\tp{z}}  \,=\, C^\ss{\xi}_\ss{z} : \cotsp\rchart{R}{\tp{q}}^{n} \to \cotsp\rchart{R}{\tp{r}}^{n}
       &,\quad\;\;
       \tp{\xi}= \tp{z}\circ\colift\psi =\upPsi \circ\tp{z} \equiv\upPsi(\tp{z})
\\[2pt] 
     \inv{\upPsi}  &\!\!\!\! :=\; 
     \tp{\xi}\circ \inv{\colift\psi}\circ \inv{\tp{\xi}} 
     &\!\!\!\!=\, \tp{z}\circ \inv{\colift\psi}\circ \inv{\tp{z}}   
    &\!\!\!\!=\, \tp{z}\circ \inv{\tp{\xi}}  \,=\, C^\ss{z}_\ss{\xi}  : \cotsp\rchart{R}{\tp{r}}^{n} \to \cotsp\rchart{R}{\tp{q}}^{n}
       &,\quad\;\;
       \tp{z} := \tp{\xi}\circ\inv{\colift\psi} = \inv{\upPsi}\circ\tp{\xi} \equiv \inv{\upPsi}(\tp{\xi})
\end{array}
\end{align}
\end{small}
where, again, \eq{\upPsi= C^\ss{\xi}_\ss{z} \in \Spism(\cotsp\rchart{R}{\tp{q}}^n ; \cotsp\rchart{R}{\tp{r}}^n)} can be thought of as a reinterpretation of \eq{\Psi=\colift\psi\in\Spism(\cotsp\man{Q})}. 
In addition to the above, we also know that, for cotangent-lifted coordinates, the transition function \eq{C^\ss{\xi}_\ss{z}}  is simply the cotangent lift of the base  transition function. That is, \eq{C^\ss{\xi}_\ss{z} =\colift \ns{c}^\ss{r}_\ss{q}} or, equivalently, \eq{\upPsi = \upcolift\uppsi}.
Note when we write \eq{\tp{r}=\uppsi(\tp{q})} or \eq{\tp{\xi}=\upPsi(\tp{z})}, we are treating the transition functions not as maps between literal points in coordinate space (which is what they technically are) but instead as maps between coordinate maps (which how we actually  use them): 
\begin{small}
\begin{align}
 \uppsi 
 = \ns{c}^\ss{r}_\ss{q}  :\fun^n(\chart{Q}{\tp{q}}) \to \fun^n(\chart{Q}{\tp{r}})
 &&,&&
  \upPsi \,=\, \upcolift\uppsi 
       \,=\,  C^\ss{\xi}_\ss{z} \,=\, \colift \ns{c}^\ss{r}_\ss{q}   :\fun^{2n}(\cotsp\chart{Q}{\tp{q}}) \to \fun^{2n}(\cotsp\chart{Q}{\tp{r}})
\end{align}
\end{small}
Now, let the cotangent-lifted fiber coordinates (i.e., momenta coordinates) be denoted \eq{\tp{\plin}} and \eq{\tp{\pf}}, that is,  \eq{\tp{\xi}=(\tp{r},\tp{\plin})=\colift\tp{r}} and \eq{\tp{z}=(\tp{q},\tp{\pf})=\colift\tp{q}}, where \eq{\tp{r}=\uppsi(\tp{q})} and \eq{\tp{\xi} =\upPsi(\tp{z}) = \upcolift\uppsi(\tp{z}) }. Then, as previously shown, the fiber coordinates transform as below
\begin{small}
\begin{align} \label{crd_colift_gen}
\tp{\xi} = (\tp{r},\tp{\plin}) = \upPsi(\tp{z}) 
\quad \leftrightarrow \quad 
\tp{z} = (\tp{q},\tp{\pf})  = \inv{\upPsi}(\tp{\xi}) 
\quad \left\{\quad
\begin{array}{lllll}
        \tp{r} = \uppsi(\tp{q}) 
\\[2pt]  
 \tp{\plin} = \tp{\pf} \cdot \dif \inv{\uppsi} \equiv \tp{\pf} \cdot \pderiv{\tp{q}}{\tp{r}}
 \end{array} \right.
 \quad \leftrightarrow \quad 
  \begin{array}{llllll}
        \tp{q} = \inv{\uppsi}(\tp{r})
 \\[2pt] 
      \tp{\pf} = \tp{\plin} \cdot \dif \uppsi \equiv  \tp{\plin} \cdot \pderiv{\tp{r}}{\tp{q}}
 \end{array}  
\end{align}
\end{small}

\begin{small}
\begin{notesq}
\rmsb{Generating Function.} Any cotangent-lifted coordinate transformation (such as the above) is a canonical transformations with type-2 generating function \eq{G(\tp{q},\tp{\plin})\in\fun(\cotsp\man{Q})} satisfying the following relations (in agreement with the above):
\begin{small}
\begin{align}
    G(\tp{q},\tp{\plin}) \,=\, \uppsi(\tp{q}) \cdot \tp{\plin} \,\equiv\, r^j(\tp{q})\cdot \plin_j
    \qquad,\qquad
    \pf_i = \pderiv{G}{q^i} \,=\,  \pderiv{r^j}{q^i}\plin_j 
     \qquad,\qquad
     r^i \,=\, \pderiv{G}{\plin_i} \,=\, \uppsi^i(\tp{q})
\end{align}
\end{small} 
\end{notesq}
\end{small}

\paragraph{Coordinate Transformation of Hamilton-Symplectic Dynamics.}
As before, suppose we have some original Hamiltonian system \eq{(\cotsp\man{Q},\nbs{\omg},\mscr{K})} with \eq{\mscr{K}} a mechanical Hamiltonian for metric \eq{\sfb{m}\in\tens^0_2(\man{Q})} and potential \eq{V\in\fun(\man{Q})}.  For some original cotangent-lifted coordinates, \eq{(\tp{r},\tp{\plin})=\colift\tp{r}}, then \eq{\mscr{K}} and \eq{\sfb{X}^\sscr{K}:=\inv{\nbs{\omg}}(\dif \mscr{K},\cdot)\in\vechm(\cotsp\man{Q},\nbs{\omg})} have local expressions:
\begin{small}
\begin{align} \label{raj1}
    \mscr{K}_\ii{r,\plin} \,=\, \tfrac{1}{2}m^{ij} \plin_i \plin_j \,+\, V_\ss{r}
    \,=\, \tfrac{1}{2}\tp{\plin}\cdot \cord{\inv{m}}{r} \cdot\tp{\plin} \,+\, V_\ss{r}
    &&,&&
    \sfb{X}^\sscr{K} 
       \,=\, \pderiv{\mscr{K}}{\plin_i}\hpdii{r^i} \,-\,  \pderiv{\mscr{K}}{r^i}\hpdiiup{\plin_i} 
    \;=\; 
    m^{ij}\plin_j \hpdii{r^i} \,+\, (  m^{sl}\Omega^j_{is} \plin_j \plin_l \,-\, \pderiv{V}{r^i}) \hpdiiup{\plin_i} 
\end{align}
\end{small}
where \eq{m^{ij}:= \inv{\sfb{m}}(\bep[r^i],\bep[r^j])} are collected into the symmetric matrix \eq{\cord{\inv{m}}{r} = [m^{ij}] = \tp{r}_*\inv{\sfb{m}}\in\Glmat{n}} (really a \eq{\Glmat{n}}-valued matrix of functions).
Note we are now making the coordinates \eq{(\tp{r},\tp{\plin})} more explicit in the above.  We now wish to obtain the local expressions for the above using the new coordinates \eq{(\tp{q},\tp{\pf})=\inv{\upPsi}(\tp{r},\tp{\plin})}. There is nothing special about this; one simply uses the well-known formulas for transformations of coordinate frame fields and tensor components. We will quickly go through the details. 
The Hamiltonian \eq{\mscr{K}} is expressed in the new coordinates simply by direct substitution of Eq.\eqref{crd_colift_gen}:
\begin{small}
\begin{align}
    \mscr{K}_\ii{q,\pf} \,=\,  \mscr{K}_\ii{r,\plin} \circ \upPsi
    \;=\; \tfrac{1}{2} (m^{kl}\circ\uppsi) \pderiv{q^i}{r^k} \pderiv{q^j}{r^l} \pf_i  \pf_j \,+\, V_\ss{r}\circ\uppsi 
\end{align}
\end{small} 
Note the above is not a ``new Hamiltonian'' but, rather, \eq{\mscr{K}_\ii{q,\pf}} and \eq{\mscr{K}_\ii{r,\plin}} are both coordinate expressions for the exact same Hamiltonian \eq{\mscr{K}\in\fun(\cotsp\man{Q})}. The first term in the above is simply the formula for the components of \eq{\inv{\sfb{m}}\in\tens^2_0(\man{Q})} in the \eq{q^i}-basis. The last expression, \eq{V_\ss{r}\circ\uppsi }, is simply the potential (not a new potential) expressed in the new coordinates. 
Eq.\eqref{raj1} is then expressed using \eq{(\tp{q},\tp{\pf})} as follows:
\begin{small}
\begin{align} \label{Hsystem_passive_gen}
\left.\begin{array}{llllll}
    \mscr{K}_\ii{q,\pf} \,=\,  \upPsi^*\mscr{K}_\ii{r,\plin}  
\\[2pt] 
     g^{ij} := \inv{\sfb{m}}(\btau[q^i],\btau[q^j]) =  \pderiv{q^i}{r^k} \pderiv{q^j}{r^l}  \uppsi^* m^{kl}
\\[2pt] 
    V_\ii{q} = \uppsi^* V_\ss{r}
\end{array} \right\}
\quad \Rightarrow 
&&
\begin{array}{llllll}
       \mscr{K}_\ii{q,\pf} \,=\,  \tfrac{1}{2}g^{ij}\pf_i \pf_j \,+\, V_\ii{q} \,=\, \tfrac{1}{2}\tp{\pf}\cdot \cord{\inv{m}}{q} \cdot\tp{\pf} \,+\, V_\ii{q}
\\[2pt] 
        \sfb{X}^\sscr{K} 
       \,=\, \pderiv{\mscr{K}}{\pf_i}\hpdii{q^i} \,-\,  \pderiv{\mscr{K}}{q^i} \hpdiiup{\pf_i}
    \;=\; 
    g^{ij} \pf_j \hpdii{q^i} \,+\, (  g^{sl}\Gamma^j_{is} \pf_j \pf_l \,-\, \pderiv{V}{q^i}) \hpdiiup{\pf_i}
\end{array}
\end{align}
\end{small}
where \eq{g^{ij} := \inv{\sfb{m}}(\btau[q^i],\btau[q^j])} are collected into the matrix \eq{\cord{\inv{m}}{q} = [g^{ij}] = \tp{q}_*\inv{\sfb{m}}}.
 Often,  we would not bother writing any of the compositions with \eq{\upPsi=C^\ss{\xi}_\ss{z}} or \eq{\uppsi=\ns{c}^\ss{r}_\ss{q}}; any function can clearly be expressed in any coordinates we like.
 Compare the above to the ``active'' transformation in Eq.\eqref{Hsystem_gen}. In contrast to the active version, we now have the exact same system just expressed in two different coordinate charts: 
 \begin{small}
 \begin{align}
 \begin{array}{rlllll}
        \mscr{K} &\!\!\! =\,  \tfrac{1}{2}m^{ij} \plin_i \plin_j \,+\, V_\ss{r} 
\\[2pt] 
       &\!\!\! =\,  \tfrac{1}{2}g^{ij}\pf_i \pf_j \,+\, V_\ii{q}
\end{array}
&&
\begin{array}{rlllll}
        \sfb{X}^\sscr{K} =\inv{\nbs{\omg}}(\dif \mscr{K},\cdot)  
          &\!\!\! =\, \pderiv{\mscr{K}}{\plin_i}\hpdii{r^i} \,-\,  \pderiv{\mscr{K}}{r^i}\hpdiiup{\plin_i} 
     \;=\; 
        m^{ij}\plin_j \hpdii{r^i} \,+\, (  m^{sl}\Omega^j_{is} \plin_j \plin_l \,-\, \pderiv{V}{r^i}) \hpdiiup{\plin_i} 
\\[2pt] 
       &\!\!\!  =\, \pderiv{\mscr{K}}{\pf_i}\hpdii{q^i} \,-\,  \pderiv{\mscr{K}}{q^i} \hpdiiup{\pf_i}
    \;=\; 
    g^{ij} \pf_j \hpdii{q^i} \,+\, (  g^{sl}\Gamma^j_{is} \pf_j \pf_l \,-\, \pderiv{V}{q^i}) \hpdiiup{\pf_i}
\end{array}
 \end{align}
 \end{small}
 where the above \eq{m_{ij}} and \eq{g_{ij}} are components of the same metric, \eq{\sfb{m}}, in different coordinate bases (\eq{r^i} and \eq{q^i}, respectively). Similarly, \eq{\Omega^i_{jk}} and \eq{\Gamma^i_{jk}} are \eq{\sfb{m}}'s Levi-Civita connection coefficients for different coordinate bases. These are thus related in the usual way, as follows (with all indices ranging from \eq{1} to \eq{n=\dim\man{Q}} and suppressing compositions with \eq{\uppsi}): 
 \begin{small}
 \begin{align} \label{raj_end}
 \begin{array}{lllll}
        g_{ij} \,=\, \pderiv{r^a}{q^i}\pderiv{r^b}{q^j} m_{ab} 
 &,\qquad 
      g^{ij} 
      =  \pderiv{q^i}{r^a} \pderiv{q^j}{r^b}   m^{ab}
 &,\qquad 
      \Gamma^i_{jk} \,=\, 
      \pderiv{q^i}{r^a}  (\ppderiv{r^a}{q^j}{q^k} + \Omega^a_{bc} \pderiv{r^b}{q^j} \pderiv{r^c}{q^k}) 
 \\[2pt] 
        m_{ab} \,=\, \pderiv{q^i}{r^a}\pderiv{q^j}{r^b} g_{ij}  
    &,\qquad 
        m^{ab}  =  \pderiv{r^a}{q^i} \pderiv{r^b}{q^j}   g^{ij}
    &,\qquad 
      \Omega^a_{bc} \,=\, \pderiv{r^a}{q^i}  (\ppderiv{q^i}{r^a}{r^b} +  \Gamma^i_{jk} \pderiv{q^j}{r^b} \pderiv{q^k}{r^c}) 
\end{array}
\end{align}
\end{small}
Now, Eqs.~\ref{raj1} – \ref{raj_end} above apply for any arbitrary cotangent-lifted coordinate transformation, \eq{(\tp{r},\tp{\plin})=\upPsi(\tp{q},\tp{\pf})}, where \eq{\upPsi=\upcolift\uppsi}. But, when \eq{\uppsi} and  \eq{\upPsi} are defined from  diffeomorphisms \eq{\psi\in\Dfism(\man{Q})} and \eq{\colift\psi\in\Spism(\cotsp\man{Q})} as in Eq.~\ref{pass1} – \ref{crd_colift_gen}, then we have the correspondence given in Eq.\eqref{act_v_pass}. In particular, we see that:

\begin{small}
\begin{remrm}
    The \eq{(\tp{q},\tp{\pf})}-representation of \eq{\sfb{X}^\sscr{K}} ``looks the same'' as the \eq{(\tp{r},\tp{\plin})}-representation of the transformed dynamics \eq{\sfb{X}^\sscr{H}} derived previously in section \ref{sec:Hmech_xform_active}. That is, the original dynamics \eq{\sfb{X}^\sscr{K}} expressed in the transformed coordinates, \eq{(\tp{q},\tp{\pf}):=(\tp{r},\tp{\plin})\circ \inv{\colift\psi}}, are of exactly the same form as the transformed dynamics, \eq{\sfb{X}^\sscr{H}:=\colift\psi^* \sfb{X}^\sscr{K}}, expressed in the original coordinates \eq{(\tp{r},\tp{\plin})}.
\end{remrm}
\end{small}


\subsection{Example:~Linear Transformations}

 We consider the simple example of linear transformations of a mechanical  Hamiltonian system on an affine space (using its associated vector space). We will be a bit more meticulous than needed for the case at hand as our goal is not to simply execute the linear transformation but, rather,  to illustrate the more general methods detailed in the preceding sections.

\begin{small}
\begin{notation}
   \sloppy In the following, \eq{\affE} is an \eq{n}-dim real affine space with associated vector space \eq{\vsp{E}} equipped with a Euclidean metric/inner product, \eq{\sfb{\emet}\equiv\inner{\cdot}{\cdot}\in\botimes^0_2 \vsp{E}}. 
    Let \eq{\be_i\in\vsp{E}} be any fixed inertial (homogeneous and autonomous) basis, but not necessarily orthogonal,  with dual basis \eq{\bep^i\in\vsp{E}^*}. We denote by \eq{\tp{r}=(r^1,\dots,r^n):(\chrt{\affE}{r} =\affE)\to\mbb{R}^n} the global linear (but not necessarily cartesian) coordinates defined for a  fixed frame \eq{(\pt{o},\be_i)}. Using \eq{\tsp[\cdt]\affE\cong \vsp{E}}, we will recycle the notation \eq{\be_i} and \eq{\bep^i} to denote the \eq{r^i}  homogeneous frame fields again as \eq{\be_i\equiv \be[r^i]\in\vect(\affE)} and \eq{\bep^i\equiv \bep[r^i]=\dif r^i\in\forms(\affE)}. We denote by \eq{(\tp{r},\tp{\plin})=\colift\tp{r}:\cotsp\affE\to\mbb{R}^{2n}} the corresponding cotangent-lifted linear coordinates with homogeneous frame fields denoted in the usual way; \eq{\hbpart{i}=\hpdii{r^i}, \hbpartup{i}=\hpdiiup{\plin_i}\in\vect(\cotsp\man{Q})} and \eq{\hbdel^i = \hbdel[r^i], \hbdeldn_i = \hbdeldn[\plin_i]\in\forms(\cotsp\man{Q})}.    
\end{notation}
\end{small}

\subsubsection*{As an Active Diffeomorphism}

\paragraph{The Original System.} Consider the case of a particle of mass \eq{m} moving in \eq{\affE}. For some chosen origin \eq{\pt{o}\in\affE}, we identify any point in the affine space, \eq{\pt{x}\in\affE}, with a displacement vector \eq{\ptvec{x}=\pt{x}-\pt{o}\in\vsp{E}}. Consider some original mechanical Hamiltonian system \eq{(\cotsp\vsp{E},\nbs{\omg},\mscr{K})} with \eq{\mscr{K}} the Hamiltonian for the kinetic energy metric \eq{\sfb{m}:=m\sfb{\emet}} and some potential function \eq{V\in\fun(\vsp{E})}. 
The original Hamiltonian, and its corresponding vector field \eq{\sfb{X}^\sscr{K}\in\vechm(\cotsp\vsp{E},\nbs{\omg})}, are:
\begin{small}
 \begin{align} \label{Ksystem_rot}
 \begin{array}{cccc}
         \mscr{K}(\kap_\pt{x}) \,=\, \tfrac{1}{2}\inv{\sfb{m}}_\pt{x}(\bs{\kap},\bs{\kap}) \,+\, V(\ptvec{x})
          \qquad,\qquad \sfb{m}:=m \sfb{\emet}\in\tens^0_2(\affE)
 \\[2pt] 
   \fnsize{i.e., } \;\; \mscr{K} = \tfrac{1}{2}m^{ij}\plin_i \plin_j \,+\, V 
   \qquad \Rightarrow \qquad 
       \sfb{X}^\sscr{K} :=\, \inv{\nbs{\omg}}(\dif \mscr{K},\cdot)  
       \,=\, \upd^i\mscr{K}\hbpart{i} \,-\,  \pd_i\mscr{K}\hbpartup{i} 
    \;=\; 
    m^{ij}\plin_j \hbpart{i} \,+\, (  m^{sl}\Omega^j_{is} \plin_j \plin_l \,-\, \pd_i V) \hbpartup{i}
 \end{array}
 \end{align}
 \end{small}
 where \eq{\sfb{m}} may be treated as a homogeneous tensor field and where \eq{\kap_\pt{x}=(\ptvec{x},\bs{\kap})\in\cotsp\vsp{E}} is any phase space point. The second line above holds for any arbitrary cotangent-lifted coordinates \eq{(\tp{r},\tp{\plin})=\colift\tp{r}}.
 Since we are working on a vector space, we have the luxury of using linear coordinates.  So, let \eq{(\tp{r},\tp{\plin})} be cotangent-lifted linear coordinates as descried above.  
 The \eq{r^i}-basis components of the metric, \eq{m_{ij}= \sfb{m}(\be_i,\be_j) = m\inner{\be_i}{\be_j}}, are constant such that the  Christoffel symbols for the \eq{r^i} basis, \eq{\Omega^i_{jk}}, vanish and the above simplifies to
 \begin{small}
 \begin{align} \label{Knom_lin}
 \begin{array}{cc}
      \fnsize{linear}  \\[-3pt]
       \fnsize{coordiantes}
 \end{array} \!\!: 
 && 
      \mscr{K} = \tfrac{1}{2}m^{ij}\plin_i \plin_j \,+\, V 
       &&,&& 
      \pd_k m^{ij} = 0 
   &&,&&
       \sfb{X}^\sscr{K} 
       \,=\, \upd^i\mscr{K}\hbpart{i} \,-\,  \pd_i\mscr{K}\hbpartup{i} 
    \;=\; 
    m^{ij}\plin_j \hbpart{i} \,-\, \pd_i V\hbpartup{i}
 \end{align}
 \end{small}
 If \eq{\be_i} is an \eq{\sfb{\emet}}-orthonormal basis, then \eq{m_{ij}=m\kd_{ij}} and \eq{m^{ij}=(1/m)\kd^{ij}}.

\paragraph{A Linearly Transformed System.}
Now, let our configuration space diffeomorphism (point transformation), \eq{\psi\in\Dfism(\vsp{E})}, be any general \textit{linear} transformation: 
\begin{small}
\begin{align}
     \psi \equiv \sfb{R}(\slot,\sblt):\vsp{E}\to \vsp{E} 
     \qquad,\qquad 
     \ptvec{q} \;\mapsto \; \psi(\ptvec{q}) = \sfb{R}\cdot\ptvec{q} 
     \qquad\qquad 
     \sfb{R}\in \Glten(\vsp{E}) \subset \botimes^1_1\vsp{E}
\end{align}
\end{small}
where linearity means that \eq{\psi} can be identified with a  \eq{(1,1)}-tensor, \eq{\sfb{R}}, and the requirement that \eq{\psi} be bijective corresponds to non-degeneracy of \eq{\sfb{R}} (i.e., the inverse exists) such that it is an element of \eq{\vsp{E}}'s general linear group, \eq{\Glten(\vsp{E})}. That is, for any \eq{\ptvec{q}\in\vsp{E}}, we may define a one-to-one linear relation 
\begin{small}
\begin{align}
    \qquad\qquad 
     \ptvec{x} \,=\, \psi(\ptvec{q}) = \sfb{R}\cdot\ptvec{q} \qquad \leftrightarrow \qquad \ptvec{q} \,=\, \inv{\psi}(\ptvec{x}) = \sfb{Q}\cdot \ptvec{x} 
      \qquad\quad ,  \qquad\quad 
     \sfb{Q}=\inv{\sfb{R}}
\end{align}
\end{small}

\begin{small}
\begin{notesq}
    For linear transformations as above, we would often do away with \eq{\psi} and just work  with the tensor \eq{\sfb{R}}. However, for the present purposes,  we will continue to use \eq{\psi} in order to clearly demonstrate the relation between the present, linear, example and the general case outlined previously. 
\end{notesq}
\end{small}

\noindent Transforming the original Hamiltonian system by \eq{\psi} requires the cotangent lift, \eq{\colift\psi\in\Spism(\cotsp\vsp{E},\nbs{\omg})}, which involves the differential. 
For linear maps on vector spaces, the identifications \eq{\tsp[\cdt]\vsp{E}\cong\vsp{E}} and \eq{\tsp[\cdt]^*\vsp{E}\cong\Evec^*} lead to the following 
identifications:\footnote{It is obvious that \eq{\psi\cong\sfb{R}(\slot,\sblt)} but perhaps less obvious that \eq{ \dif \psi \cong \sfb{R}}.   To see the latter, note we can express \eq{\psi} in the \eq{r^i} basis, \eq{\be_i}, as a sort of ``vector field'', \eq{\psi\cong\sfb{R}(\slot,\sblt) = R^i_j r^j\be_i}. Then, using \eq{r^i\circ\psi = R^i_j r^j} and  \eq{\be[r^i]\equiv\be_i} are homogeneous, the usual formula for the differential of smooth maps then gives:
\begin{align}
     \psi\cong\sfb{R}(\slot,\sblt) = R^i_j r^j\be_i =: \psi^i \be_i
 &&\Rightarrow &&
    \dif \psi \,=\, \pderiv{r^i\circ\psi}{r^j}(\be[r^i])_\ii{\psi} \otms \bep[r^j] \,=\, \pderiv{\psi^i}{r^j}(\be[r^i])_\ii{\psi} \otms \bep[r^j]  \,=\, R^i_j (\be[r^i])_\ii{\psi} \otms \bep[r^j] \,\equiv\, R^i_j \be_i \otms \bep^j \,=\, \sfb{R}
\end{align}
}
\begin{small}
\begin{align} \label{dmap_lin}
    \psi \cong \sfb{R} 
   &&,&&
    \inv{\psi} \cong \inv{\sfb{R}} = \sfb{Q} 
     &&,&&
    \dif \psi \cong \sfb{R} 
    &&,&&
    \dif \inv{\psi} \cong \inv{\sfb{R}} = \sfb{Q}
      &&,&&
     \tlift\psi \cong ( \sfb{R}, \sfb{R} ) 
      &&,&&
     \colift\psi \cong ( \sfb{R}, \trn{\sfb{Q}} ) 
\end{align}
\end{small}
Where the maps correspond to the right-hand ``slots'' of the tensors. 
To clarify, for any \eq{\colift\psi}-related points \eq{\mu_\pt{q}=(\ptvec{q},\bs{\mu})\in\cotsp\vsp{E}} and \eq{\kap_\pt{x}=(\ptvec{x},\bs{\kap})=\colift\psi(\mu_\pt{q})}, we have the one-to-one transformations:
\begin{small}
\begin{align} \label{colift_lin}
\begin{array}{rrclll}
  \cotsp\vsp{E} \ni&    \kap_\pt{x} \,=\, \colift \psi (\mu_\pt{q}) \,=\, (\sfb{R}\cdot\ptvec{q}, \,\trn{\sfb{Q}}\!\cdot\bs{\mu})
    &\qquad \leftrightarrow &\qquad
    \mu_\pt{q} \,=\, \colift \inv{\psi} (\kap_\pt{x}) \,=\, (\sfb{Q}\cdot\ptvec{x}, \, \trn{\sfb{R}}\!\cdot\bs{\kap}) 
    &\in  \cotsp\vsp{E} 
\\[2pt] 
     \vsp{E} \ni &  \ptvec{x} = \psi(\ptvec{q}) = \sfb{R}\cdot\ptvec{q}
       &\qquad \leftrightarrow &\qquad \ptvec{q} = \inv{\psi}(\ptvec{x}) = \sfb{Q}\cdot\ptvec{x}
       & \in \vsp{E}
\\[2pt] 
    \tsp[\pt{x}]^*\vsp{E} \ni& 
    \bs{\kap} \,=\,  \psi_* \bs{\mu} \,=\, \bs{\mu} \cdot \inv{(\dif \psi_\ss{\!\pt{q}})} \,=\, \bs{\mu}\cdot \sfb{Q}
 &\qquad \leftrightarrow &\qquad
     \bs{\mu} \,=\, \psi^*  \bs{\kap} \,=\, \bs{\kap} \cdot \dif \psi_\ss{\!\pt{q}}  \,=\, \bs{\kap} \cdot\sfb{R}
     & \in   \tsp[\pt{q}]^*\vsp{E} 
\end{array}
\end{align}
\end{small}
We now  transform the original mechanical Hamiltonian system \eq{(\cotsp\vsp{E},\nbs{\omg},\mscr{K})} using \eq{\psi} and \eq{\colift\psi\in\Spism(\cotsp\vsp{E},\nbs{\omg})} as outlined in section \ref{sec:Hmech_xform_active}. This leads to a new mechanical Hamiltonian system \eq{(\cotsp\vsp{E},\nbs{\omg},\mscr{H})}  with \eq{\mscr{H}} and \eq{\sfb{X}^\sscr{H}} given as in  Eq.\eqref{Hsystem_gen}: 
\begin{small}
\begin{align} \label{Hsystem_rot}
 \left. \begin{array}{lllll}
    \mscr{H}:= \colift\psi^* \mscr{K} 
\\[2pt] 
      \sfg := \psi^* \sfb{m} 
  \\[2pt] 
      \inv{\sfg} = \psi^* \inv{\sfb{m}} 
\\[2pt] 
      U:= \psi^* V 
 \end{array}
 \right\} \quad \Rightarrow 
\qquad 
 \begin{array}{cccc}
         \mscr{H}(\mu_\pt{q}) \,=\, \tfrac{1}{2}\inv{\sfg}_\pt{q}(\bs{\mu},\bs{\mu}) \,+\, U(\pt{q}) \;\;=\, \mscr{K}(\kap_\pt{x})
 \\[2pt] 
          \fnsize{i.e., } \;\; \mscr{H} = \tfrac{1}{2}g^{ij}\plin_i \plin_j \,+\, U 
 \\[2pt] 
       \sfb{X}^\sscr{H} 
       \,=\, \upd^i\mscr{H}\hbpart{i} \,-\,  \pd_i\mscr{H}\hbpartup{i} 
    \;=\; 
    g^{ij} \plin_j \hbpart{i} \,-\,  \pd_i U \hbpartup{i}
 \end{array}
 \end{align}
 \end{small}
 where \eq{ \mscr{H}(\mu_\pt{q}) = \mscr{K}(\kap_\pt{x})} holds for \eq{\colift\psi}-related points \eq{\kap_\pt{x}=\colift\psi(\mu_\pt{q})\in\cotsp\vsp{E}} as in Eq.\eqref{colift_lin}. 
 In the above, \eq{(\tp{r},\tp{\plin})} are the same linear coordinates as before and \eq{g^{ij}=\inv{\sfg}(\bep^i,\bep^j)} are the components of the induced (inverse) metric in the \eq{r^i} basis. Since \eq{\psi} is a linear transformation, and \eq{r^i} are linear coordinates, these components are again constant (\eq{\pd_k g_{ij} = 0 =\pd_k g^{ij}}). As such, \eq{\sfg}'s Levi-Civita connection coefficients (Christoffel symbols) for the \eq{r^i}-basis vanish. 
Further,  using the identifications in Eq.\eqref{dmap_lin}, the above \eq{\psi}-induced metric is homogeneous and equivalent to the following: 
 \begin{small}
 \begin{align}
\begin{array}{lllllll}
 \sfg := \psi^* \sfb{m} 
 &\!\!\!\! =\,  \trn{\dif \psi}\cdot\sfb{m}_\ii{\psi} \cdot \dif \psi 
 &\!\!\!\!=\, \trn{\sfb{R}}\cdot\sfb{m}\cdot\sfb{R}
 \\[2pt] 
    \inv{\sfg} = \psi^* \inv{\sfb{m}} 
    &\!\!\!\! =\, \inv{\dif \psi}_\ii{\psi} \cdot \inv{\sfb{m}}_\ii{\psi} \cdot \invtrn{\dif \psi}_\ii{\psi} 
    &\!\!\!\! =\,  \sfb{Q} \cdot \inv{\sfb{m}} \cdot \trn{\sfb{Q}}
\end{array}
\end{align}
\end{small}
The above is actually implicit in the definition \eq{ \mscr{H}:= \colift\psi^* \mscr{K}} (as is \eq{U:=\psi^* V});  for any   \eq{\mu_\pt{q}=(\ptvec{q},\bs{\mu})\in\cotsp\vsp{E}} we have,
\begin{small}
\begin{align} \nonumber 
      \mscr{H}(\ptvec{q},\bs{\mu}) \,=\, \mscr{K}\circ\colift\psi(\ptvec{q},\bs{\mu})  \,=\, \mscr{K}(\sfb{R}\cdot\ptvec{q}, \trn{\sfb{Q}}\cdot\bs{\mu}) 
      &\,=\, \tfrac{1}{2}\inv{\sfb{m}}(\trn{\sfb{Q}}\cdot\bs{\mu},\trn{\sfb{Q}}\cdot\bs{\mu}) \,+\, V(\sfb{R}\cdot\ptvec{q})
      \;=\;
      \tfrac{1}{2} \bs{\mu}\cdot ( \sfb{Q}\cdot\inv{\sfb{m}} \cdot \trn{\sfb{Q}})\cdot\bs{\mu} \,+\, V(\sfb{R}\cdot\ptvec{q})
\\[2pt] \nonumber 
     &\,=:\, 
      \tfrac{1}{2} \bs{\mu}\cdot\inv{\sfg}\cdot\bs{\mu} \,+\, U(\ptvec{q}) \,=\, \tfrac{1}{2} \inv{\sfg}(\bs{\mu},\bs{\mu}) \,+\, U(\ptvec{q})
\end{align}
\end{small}
Compare the original system in Eq.\eqref{Knom_lin} to the new system in Eq.\eqref{Hsystem_rot}. Using the \textit{same} linear coordinates, \eq{(\tp{r},\tp{\plin})}:
\begin{small}
 \begin{align} \label{Hlin_act}
 \begin{array}{rrlllll}
     \fnsize{original:}&   \mscr{K} &\!\!\! =\,  \tfrac{1}{2}m^{ij} \plin_i \plin_j \,+\, V
\\[2pt] 
    \fnsize{transformed:}&   \mscr{H} &\!\!\! =\,  \tfrac{1}{2}g^{ij}\plin_i \plin_j \,+\, U
\end{array}
&&
\begin{array}{rlllll}
        \sfb{X}^\sscr{K} =\inv{\nbs{\omg}}(\dif \mscr{K},\cdot)  
          &\!\!\! =\,  \upd^i\mscr{K}\hbpart{i} \,-\,  \pd_i\mscr{K}\hbpartup{i}  \;=\; 
    m^{ij}\plin_j \hbpart{i} \,-\, \pd_i V\hbpartup{i}
\\[2pt] 
      \sfb{X}^\sscr{H} =\inv{\nbs{\omg}}(\dif \mscr{H},\cdot)   &\!\!\!  =\, \upd^i\mscr{H}\hbpart{i} \,-\,  \pd_i\mscr{H}\hbpartup{i}  \;=\; 
    g^{ij}\plin_j \hbpart{i} \,-\, \pd_i U\hbpartup{i}
\end{array}
 \end{align}
 \end{small}
where \eq{m^{ij}} and \eq{g^{ij}} are components of two different metrics (both in the same \eq{r^i} coordinate basis) and \eq{V} and \eq{U} are two different functions (both expressed in terms of the same \eq{r^i}). 
Solving the following coordinate ODEs
\begin{small}
\begin{align}
 \begin{array}{rrlllll}
     \fnsize{original:}&   \quad  \dot{r}^i = m^{ij}\plin_j \quad,\quad \dot{\plin}_i = -\pd_\ss{r^i} V
     &\qquad \quad \fnsize{or,} & \ddot{r}^i = - m^{ij}\pd_\ss{r^j} V
\\[2pt] 
    \fnsize{transformed:}&  \quad 
    \dot{r}^i = g^{ij}\plin_j \quad,\quad \dot{\plin}_i = -\pd_\ss{r^i} U
     &\qquad \quad \fnsize{or,} & \ddot{r}^i = - g^{ij}\pd_\ss{r^j} U
\end{array}
\end{align}
\end{small}
leads to the coordinate representations, in the same chart,  of two \textit{different} integral curves; one for \eq{\sfb{X}^\sscr{K}} (original) and one for \eq{\sfb{X}^\sscr{H}} (transformed). To make this abundantly clear, let \eq{\kap_t=(\ptvec{x}_t,\bs{\kap}_t)\in\cotsp\vsp{E}} be an integral curve of \eq{\sfb{X}^\sscr{K}} and let \eq{\mu_t=(\ptvec{q}_t,\bs{\mu}_t)=\inv{\colift\psi}(\kap_t)} be an integral curve of \eq{\sfb{X}^\sscr{H}}. Let \eq{x^i(t):= r^i(\ptvec{x}_t)=\bep^i\cdot\ptvec{x}_t} and \eq{\kap_i(t):=\plin_i(\kap_t)= \be_i\cdot\bs{\kap}_t} be the linear coordinates (components in the \eq{\be_i \equiv \be[r^i]} basis) of \eq{\kap_t} and let   \eq{q^i(t):= r^i(\ptvec{q}_t)=\bep^i\cdot\ptvec{q}_t} and \eq{\mu_i(t):=\plin_i(\mu_t)= \be_i\cdot\bs{\mu}_t} be the linear coordinates \eq{\mu_t} (in the same basis). 
We then have the following:
\begin{small}
\begin{align} \label{odes_lin_act}
\begin{array}{lllllll}
     \dt{\kap}_t \,=\, \sfb{X}^\sscr{K}_{\kap_t}
     &\Rightarrow \qquad 
     \dot{x}^i = m^{ij} \kap_j &,\quad 
     \dot{\kap}_i = -\pd_\ss{r^i}V_{\ptvec{x}}
\\[2pt] 
        \dt{\mu}_t \,=\, \sfb{X}^\sscr{H}_{\mu_t}
     &\Rightarrow \qquad 
     \dot{q}^i = g^{ij} \mu_j &,\quad 
     \dot{\mu}_i = -\pd_\ss{r^i}U_{\ptvec{q}}
\end{array}
\end{align}
\end{small}
where \eq{\pd_\ss{r^i}V_{\ptvec{x}}\equiv \pderiv{V}{r^i}|_{\ptvec{x}}} and  where 
\eq{ \kap_t=\colift\psi(\mu_t)\leftrightarrow \mu_t=\inv{\colift\psi}(\kap_t)} are related as in Eq.\eqref{colift_lin}.

 \begin{small}
 \begin{notesq}
 \rmsb{Orthogonal Case.}
    Consider the more specific case that \eq{\psi=\sfb{R}(\slot,\sblt)} corresponds to a special orthogonal transformation — that is, \eq{\sfb{R}\in\Soten(\vsp{E},\sfb{m})\subset\Glten(\vsp{E})} such that \eq{\sfb{Q}=\inv{\sfb{R}}=\sfb{R}^{\dagger}=\inv{\sfb{m}}\cdot\trn{\sfb{R}}\cdot\sfb{m}}. In this case, \eq{\psi^*\sfb{m}=\sfb{m}} such that the original \eq{\mscr{K}=\tfrac{1}{2}m^{ij} \plin_i \plin_j +V} is transformed to \eq{\mscr{H}=\colift\psi^*\mscr{K}=\tfrac{1}{2}m^{ij} \plin_i \plin_j +U}. That is, the metric is preserved or, in other words, \eq{\psi} is a linear isometry:
    \begin{small}
    \begin{align}
         \psi\in\Isom(\vsp{E},\sfb{m}) \qquad \Leftrightarrow \qquad \sfb{R}\in\Soten(\vsp{E},\sfb{m}) 
         \qquad \Leftrightarrow \qquad 
         \begin{array}{rlllll}
               \psi^*\sfb{m} &\!\!\!\! =  \trn{\sfb{R}}\cdot\sfb{m}\cdot\sfb{R} = \sfb{m} 
          \\[2pt] 
           \psi^* \inv{\sfb{m}} &\!\!\!\! = 
           \sfb{Q} \cdot \inv{\sfb{m}} \cdot \trn{\sfb{Q}} = \inv{\sfb{m}} 
         \end{array}
    \end{align}
    \end{small}
\end{notesq}
\end{small}

\subsubsection*{As a Passive Coordinate Transformation}

\begin{small}
\begin{notation}
    Again, let \eq{\tp{r}:\affE\to\mbb{R}^n} be linear coordinates on the base and \eq{(\tp{r},\tp{\plin})=\colift\tp{r}} be the same  cotangent-lifted linear coordinates appearing above. Unlike the above, the following will use \eq{\tp{q}=(q^1,\dots,q^n):\affE\to\mbb{R}^n} to denote new coordinates (rather than a point \eq{\ptvec{q}\in\vsp{E}}). 
\end{notation}
\end{small}

\noindent We now illustrate the ``passive'' interpretation of the (linear) map \eq{\psi:\vsp{E}\to\vsp{E}} as a (linear) coordinate transformation, \eq{\uppsi:\rchart{R}{\tp{q}}^n\to \rchart{R}{\tp{r}}^n}, \eq{\tp{q}\mapsto\tp{r}}. The general case was detailed in section \ref{sec:Hmech_xform_passive}. 
First, note that \eq{\psi \cong \sfb{R}(\slot,\sblt):\vsp{E}\to\vsp{E}} can be expressed as a ``vector field''  in the \eq{r^i} basis, \eq{\be_i}: 
 \begin{small}
 \begin{align} \label{xform_lin_vec}
    \phantom{XXX} &&
     \psi \,=\, R^i_j r^j \be_i
     \qquad,\qquad 
     \inv{\psi} \,=\, Q^i_j r^j \be_i
     &&
      R^i_j Q^j_{\,k} \,=\, \kd^i_k
 \end{align}
 \end{small}
 where \eq{R^i_j=\sfb{R}(\bep^i,\be_j)} and likewise for the components of \eq{\sfb{Q}=\inv{\sfb{R}}}. We now define new configuration coordinates \eq{q^i:= r^i\circ \inv{\psi}}, as per section \ref{sec:Hmech_xform_passive}. We see from the above that this is simply:
\begin{small}
\begin{align} \label{qx_xform_lin}
\begin{array}{rcll}
      q^i :=\, r^i\circ \inv{\psi} \,=\, Q^i_j r^j 
     &\quad \leftrightarrow &\quad 
     r^i \,=\, q^i \circ \psi \,=\, R^i_j q^j
 \\[2pt] 
     \fnsize{i.e., } \qquad 
     \tp{q} \,=\, \inv{\uppsi}(\tp{r}) \,=\, Q\cdot\tp{r}
      &\quad \leftrightarrow &\quad 
     \tp{r} \,=\, \uppsi(\tp{q}) \,=\, R\cdot\tp{q}
\end{array}
\end{align}
\end{small}
The second line is just the matrix version of the first using the \eq{\tp{r}}-representations  \eq{R=\tp{r}_*\sfb{R}\in \Glmat{n}}, and \eq{\inv{R}=Q=\tp{r}_*\sfb{Q}\in \Glmat{n}}, and \eq{\uppsi =\cord{\uppsi}{r} \in\Dfism(\rchart{R}{\tp{q}}^n; \rchart{R}{\tp{r}}^n)}. As shown previously
(and again\footnote{Direct substitution shows \eq{ \uppsi:=\;  \cord{\uppsi}{r} \,=\, \tp{r} \circ \psi \circ \inv{\tp{r}} \,=\, \tp{r} \circ \inv{(\tp{r}\circ\inv{\psi})} \,=\, \tp{r} \circ \inv{\tp{q}} \;\equiv \; 
\cord{\uppsi}{q} \,=\, \tp{q} \circ \psi \circ \inv{\tp{q}} \,=\, \tp{r}\circ \inv{\tp{q}} \,=\, \ns{c}^\ss{r}_\ss{q} : \rchart{R}{\tp{q}}^n \to \rchart{R}{\tp{r}}^n} . }),
the \eq{\tp{r}}-representation, \eq{\cord{\uppsi}{r}},  of \eq{\psi} is equivalent to the \eq{\tp{q}}-representation, \eq{\cord{\uppsi}{q}};  both are simply the transition function, \eq{\uppsi=\cord{\uppsi}{r}=\cord{\uppsi}{q}=\ns{c}^\ss{r}_\ss{q}=\tp{r}\circ\inv{\tp{q}}}. Equivalently, \eq{R=\tp{r}_*\sfb{R}=\tp{q}_*\sfb{R}} 
(and likewise for \eq{\sfb{Q}}).\footnote{This is equivalent to the well-known fact that, for a general basis transformation \eq{\tbe_i:=\sfb{T}\cdot \be_i}, the components of \eq{\sfb{T}} are the same in both bases.} 
Note the above is the usual relation for a linear transformation of vector components in different bases. The linear coordinate functions \eq{r^i,q^i\in\fun(\affE)} can indeed  be seen as components of the displacement ``vector field'', \eq{\bs{\rho}:\affE\to\vsp{E}}, in different bases: 
\begin{small}
\begin{align}
     \bs{\rho} \,=\, r^i\be_i \,=\, q^i \bt_i : \affE\to\vsp{E}
\end{align}
\end{small}
where \eq{(\pt{o},\bt_i)} is an inertial frame corresponding to the \eq{q^i} coordinates in the same way that \eq{r^i} corresponds to the  inertial frame \eq{(\pt{o},\be_i)}. That is, the basis transformation corresponding the coordinate transformation in Eq.\eqref{qx_xform_lin} is the  
standard one:\footnote{We also have the usual relations for \eq{\sfb{R},\sfb{Q}\in\Glten(\vsp{E})}:
 \begin{align}
   \begin{array}{lllll}
         \sfb{R} \,=\, \bt_{i} \otms \bep^i  \,=\,  R^i_j \be_{i}\otms\bep^j \,=\,  R^i_j \bt_{i}\otms\btau^j 
 \qquad\quad,\qquad\quad 
         \sfb{Q} :=\, \inv{\sfb{R}} =\, \be_{i}\otms\btau^i \,=\,  Q^i_j \be_{i}\otms\bep^j \,=\, Q^i_j \bt_{i}\otms\btau^j  
 \end{array}
\end{align}
 }
\begin{small}
\begin{align} \label{basis_xform_lin}
 \begin{array}{rlll}
    \bt[q^i] \equiv  &  \bt_{i} \,=\, \sfb{R}\cdot\be_i \,=\,    R^{j}_{\,i} \be_{j}
\\[2pt] 
    \btau[q^i] \equiv&   \btau^i \,=\, \bep^i\cdot \sfb{Q}  \,=\,   Q^i_j \bep^j 
\end{array}
\quad \leftrightarrow \quad
\begin{array}{rlll}
    \be[r^i] \equiv&   \be_{i}  \,=\, \sfb{Q}\cdot\bt_i  \,=\,  Q^{j}_{\,i} \bt_{j}
\\[2pt] 
    \bep[r^i]  \equiv &   \bep^i  \,=\,  \btau^i\cdot\sfb{R} \,=\,  R^i_j \btau^j 
\end{array} 
\end{align}
\end{small}
Note that \eq{\bt_i\in\vsp{E}} can be identified with the \eq{q^i} frame fields, \eq{\bt_i\equiv\bt[q^i]\in\vect(\affE)}, for all the same reasons that \eq{\be_i\equiv \be[r^i]\in\vect(\affE)}.
Let the metric components in the \eq{r^i} basis be denoted \eq{m_{ij}:=\sfb{m}(\be_i,\be_j)=m\inner{\be_i}{\be_j}}. Then the components of the same metric in the \eq{q^i} basis (which we will denote \eq{g_{ij}} in analogy to the previous ``active'' transformation) are:
\begin{small}
\begin{align}
    g_{ij} := \sfb{m}(\bt_i,\bt_j) = R^k_i R^l_j \sfb{m}(\be_k,\be_l) = R^k_i R^l_j m_{kl}
    \qquad,\qquad 
    g^{ij} := \inv{\sfb{m}}(\btau^i,\btau^j) \,=\, Q^i_k Q^j_l \inv{\sfb{m}}(\bep^k,\bep^l) \,=\, Q^i_k Q^j_l m^{kl}
\end{align}
\end{small}
We want to express the mechanical Hamiltonian system on \eq{\cotsp\affE} in new coordinates. 
For the original cotangent-lifted coordinates, \eq{\tp{\xi}=(\tp{r},\tp{\plin})=\colift\tp{r}}, recall the new cotangent bundle coordinates defined by \eq{\tp{z}:=\tp{\xi}\circ\colift\psi = \upcolift\uppsi(\tp{\xi})} are simply the cotangent-lifted coordinates \eq{\tp{z}=(\tp{q},\tp{\pf})=\colift\tp{q}}. In a similar manner to the identifications in Eq.\eqref{dmap_lin}, we now  have the following identifications for the coordinate representations of \eq{\psi} and \eq{\colift\psi} (they are the same in both sets of of coordinates):
\begin{small}
\begin{align}
    \uppsi \cong R &&,&&
    \inv{\uppsi} \cong Q 
     &&,&&
     \dif \uppsi = \pderiv{\tp{r}}{\tp{q}} = R
      &&,&&
       \inv{\dif \uppsi} = \pderiv{\tp{q}}{\tp{r}} =  Q 
       &&,&&
       \uptlift\uppsi \cong (R,R)
        &&,&&
         \upcolift\uppsi \cong (R,\trn{Q})
\end{align}
\end{small}
That is, the cotangent-lifted linear coordinate transformation between \eq{\tp{\xi}=(\tp{r},\tp{\plin})} and \eq{\tp{z}=(\tp{q},\tp{\pf})} is simply:
\begin{small}
\begin{align} \label{colift_crd_lin}
\begin{array}{rclll}
  \tp{\xi} = (\tp{r},\tp{\plin}) \,=\, \upcolift\uppsi (\tp{z}) \,=\, (R\cdot\tp{q}, \,\trn{Q}\!\cdot\tp{\pf})
    &\qquad \leftrightarrow &\qquad
   \tp{z}=(\tp{q},\tp{\pf}) \,=\, \inv{\upcolift\uppsi} (\tp{z}) \,=\, (Q\cdot\tp{r}, \, \trn{R}\!\cdot\tp{\plin}) 
\\[2pt] 
     \tp{r} = \uppsi(\tp{q}) = R\cdot\tp{q}
       &\qquad \leftrightarrow &\qquad \tp{q} = \inv{\uppsi}(\tp{r}) = Q\cdot\tp{r} 
\\[2pt] 
    \tp{\plin} \,=\, \tp{\pf} \cdot \pderiv{\tp{q}}{\tp{r}} \,=\, \tp{\pf}\cdot Q \,=\, \trn{Q}\!\cdot\tp{\pf}
 &\qquad \leftrightarrow &\qquad
    \tp{\pf}   \,=\, \tp{\plin} \cdot \pderiv{\tp{r}}{\tp{q}}   \,=\, \tp{\plin} \cdot R \,=\, \trn{R}\!\cdot \tp{\plin}
\end{array}
\end{align}
\end{small}
which corresponds to the following linear transformation of the co-tangent-lifted frame fields:
\begin{small}
\begin{align} \label{ff_colift_lin}
\begin{array}{lllllll}
     \hpdii{q^i} \,=\,  R^j_i \hpdii{r^j}  
&,\quad
     \hpdiiup{\pf_i} \,=\, Q^i_j  \hpdiiup{\plin_j}
 \\[2pt] 
    \hbdel[q^i] \,=\, Q^i_j \hbdel[r^j] 
&,\quad
    \hbdeldn[\pf_i] \,=\, R^j_i \hbdeldn[\plin_j]
\end{array}
\qquad \leftrightarrow\qquad 
\begin{array}{lllllll}
     \hpdii{r^i} \,=\,  Q^j_i \hpdii{q^j}  
&,\quad
     \hpdiiup{\plin_i} \,=\, R^i_j  \hpdiiup{\pf_j}
 \\[2pt] 
    \hbdel[r^i] \,=\, R^i_j \hbdel[q^j] 
&,\quad
    \hbdeldn[\plin_i] \,=\, Q^j_i \hbdeldn[\pf_j]
\end{array}
\end{align}
\end{small}
The Hamiltonian \eq{\mscr{K}\in\fun(\cotsp\affE)} given in Eq.\eqref{Ksystem_rot} is then expressed in the two sets of cotangent-lifted linear coordinates as
\begin{small}
\begin{align}
    \mscr{K} \,=\, \tfrac{1}{2}m^{ij}\plin_i \plin_j + V_\ss{r} \;=\; \tfrac{1}{2} m^{ij} Q^k_i Q^l_j \pf_k \pf_l + V_\ii{q} \,=\, \tfrac{1}{2} g^{kl}\pf_k \pf_l  + V_\ii{q}
\end{align}
\end{small}
The above is still a function on \eq{\cotsp\affE}, not a function on coordinate space. The notation \eq{V_\ss{r}} and \eq{V_\ii{q}} just implies that \eq{V} should be expressed in terms of the indicated 
coordinates.\footnote{If we instead defined it to mean \eq{V_\ss{r}:=V\circ\inv{\tp{r}}\in\fun(\rchart{R}{\tp{r}}^n)} then we would have \eq{V_\ss{r}(\tp{r})=V_\ss{r}(\uppsi(\tp{q})) = V_\ss{r}(R\cdot\tp{q}) = V_\ii{q}(\tp{q})} with \eq{ V_\ii{q}=V_\ss{r}\circ\uppsi}. But, above, \eq{\mscr{K}} and \eq{V} are coordinate-agnostic functions (i.e., \eq{(0,0)}-tensor fields).}
In the last expression, note that \eq{g^{ij}} are the components of the \textit{same} inverse metric, \eq{\inv{\sfb{m}}=\tfrac{1}{m}\inv{\sfb{\emet}}\in\tens^2_0(\affE)}, in the \eq{q^i} basis, \eq{\bt[q^i]\equiv \bt_i}, as shown above. 
Since \eq{r^i} and \eq{q^i=Q^i_j r^j} are both linear coordinates corresponding homogeneous and autonomous bases, the components \eq{m_{ij}} and \eq{g_{ij}} are constant and the Hamiltonian vector field \eq{\sfb{X}^\sscr{K}:=\inv{\nbs{\omg}}(\dif \mscr{K},\cdot)\in\vechm(\cotsp\affE,\nbs{\omg})} is expressed in the two sets of cotangent-lifted coordinates as:\footnote{Eq.\eqref{Hlin_pass} also follows directly from the frame field transformation in Eq.\eqref{ff_colift_lin}. }
\begin{small}
\begin{align} \label{Hlin_pass}
\begin{array}{rlllll}
     \mscr{K} &\!\!\! =\,  \tfrac{1}{2}m^{ij}\plin_i \plin_j \,+\, V
\\[2pt] 
     &\!\!\! =\,  \tfrac{1}{2}g^{ij}\pf_i \pf_j \,+\, V
\end{array}
&&
\begin{array}{rlllll}
        \sfb{X}^\sscr{K} =\inv{\nbs{\omg}}(\dif \mscr{K},\cdot)  
          &\!\!\! =\, \pderiv{\mscr{K}}{\plin_i}\hpdii{r^i} \,-\,  \pderiv{\mscr{K}}{r^i}\hpdiiup{\plin_i} 
     \;=\; 
        m^{ij}\plin_j \hpdii{r^i}  \,-\, \pderiv{V}{r^i} \hpdiiup{\plin_i} 
\\[2pt] 
       &\!\!\!  =\, \pderiv{\mscr{K}}{\pf_i}\hpdii{q^i} \,-\,  \pderiv{\mscr{K}}{q^i} \hpdiiup{\pf_i}
    \;=\; 
    g^{ij} \pf_j \hpdii{q^i}  \,-\, \pderiv{V}{q^i} \hpdiiup{\pf_i}
\end{array}
\end{align}
\end{small}
(Compare the above to the active transformation in Eq.\eqref{Hlin_act}). Solving either of the following coordinate ODEs
\begin{small}
\begin{align}
\begin{array}{rllllll}
     \fnsize{original coordinates:}& 
      \dot{r}^i = m^{ij}\plin_j \quad,\quad \dot{\plin}_i = -\pd_\ss{r^i} V
       &\qquad\qquad \fnsize{or,} &
       \ddot{r}^i = - m^{ij}\pd_\ss{r^j} V
\\[2pt] 
      \fnsize{new coordinates:}&  
       \dot{q}^i = g^{ij}\pf_j \quad,\quad \dot{\pf}_i = -\pd_\ii{q^i} V
       &\qquad\qquad \fnsize{or,} & \ddot{q}^i = -g^{ij} \pd_\ii{q^j} V
\end{array}
\end{align}
\end{small}
leads to the coordinate representation, in two different charts,  of the \textit{same} integral curve of \eq{\sfb{X}^\sscr{K}}. To make this clear, let  \eq{\kap_t=(\ptvec{x}_t,\bs{\kap}_t)\in\cotsp\vsp{E}} be an integral curve of \eq{\sfb{X}^\sscr{K}} and let \eq{x^i(t):= r^i(\ptvec{x}_t)=\bep^i\cdot\ptvec{x}_t} and \eq{\kap_i(t):=\plin_i(\kap_t)= \be_i\cdot\bs{\kap}_t} be its \eq{(\tp{r},\tp{\plin})}-representation (for the present case,  this is simply the (co)vector components in the \eq{\be_i \equiv \be[r^i]} basis). Let us denote the coordinate representation of the same curve,  \eq{\kap_t=(\ptvec{x}_t,\bs{\kap}_t)}, in the new coordinates \eq{(\tp{q},\tp{\pf})=\inv{\upcolift\uppsi}(\tp{r},\tp{\plin})}  by 
 \eq{\til{x}^i(t):=q^i(\ptvec{x}_t) = \btau^i\cdot \ptvec{x}_t} and \eq{\til{\kap}_i(t):=\pf_i(\kap_t) = \bt_i\cdot\bs{\kap}_t} (that is, the  (co)vector components in the \eq{\bt_i\equiv\bt[q^i]} basis).  
We then have the following:
\begin{small}
\begin{align} \label{odes_lin_pass}
 \dt{\kap}_t \,=\, \sfb{X}^\sscr{K}_{\kap_t}
     \quad \Rightarrow \quad 
\left\{\begin{array}{rlll}
    \fnsize{original coordinates:}& 
      \dot{x}^i = m^{ij}\kap_j \quad,\quad \dot{\kap}_i = -\pd_\ss{r^i} V_{\ptvec{x}}
\\[2pt] 
      \fnsize{new coordinates:}&  
       \dot{\til{x}}^i = g^{ij}\til{\kap}_j \quad,\quad \dot{\til{\kap}}_i = -\pd_\ii{q^i} V_{\ptvec{x}}
\end{array} \right.
\end{align}
\end{small}

\begin{small}
\begin{notesq}
    Compare the effects of the ``active'' transformation in Eqs.~\ref{Hlin_act} – \ref{odes_lin_act} to those of the ``passive'' transformation in Eqs.~\ref{Hlin_pass} – \ref{odes_lin_pass}.
\end{notesq}
\end{small}

\section{DEFINING THE ORIGINAL HAMILTONIAN SYSTEM} \label{sec:NOM}



The system in which we are ultimately interested is the standard case of a particle of mass \eq{m} moving in Euclidean \eq{n}-space \eq{\affE^n} (with inner product space \eq{(\vsp{E}^n,\sfb{\emet})}), that is subject to both conservative and non-conservative forces. In particular, we will consider the case that the conservative forces contain a dominant central force term (which will eventually become the Manev potential, including the Kepler potential as a special case). We will exclude non-conservative forces at first since we can, and will, add them in later (section \ref{sec:prj_noncon}). The original system, defined thoroughly in section \ref{sec:Hnom_prj}, is briefly described as follows:

\begin{small}
\begin{notesl}
  \sloppy Although subsequent developments will involve several transformations, one should keep in mind that the Hamiltonian system we are interested in at the end of the day is described by a mechanical Hamiltonian \eq{\mscr{K}\in\fun(\cotsp\affE^n)} (classically, \eq{n=3}), which may be expressed in cotangent-lifted coordinates, \eq{(\tup{r},\tup{\plin})=(r^1,\dots,r^n,\plin_1,\dots,\plin_n):\cotsp\affE^n \to \cotsp\mbb{R}^n \cong \mbb{R}^{2n}},  as:\footnote{In the case that \eq{(\tup{r},\tup{\plin})} are  \textit{cartesian} coordinates, then we have simply  \eq{\emet^{ij} = \emet_{ij} = \kd^i_j}.}
    \begin{small}
    \begin{align} \label{Hnom_actual}
         \mscr{K} = \tfrac{1}{2}m^{ij}\plin_i \plin_j \,+\, V^\zr + V^\ss{1}
         \qquad\qquad,\qquad\qquad
         m_{ij} = m \emet_{ij} \qquad m^{ij} = \tfrac{1}{m} \emet^{ij} 
    \end{align}
    \end{small}
    where \eq{V^\zr} accounts for all conservative central forces\footnote{For any \eq{\ptvec{x}\in\vsp{E}^n}, then \eq{V^\zr(\ptvec{x}) \equiv V^\zr(\mag{\ptvec{x}})} is a ``function of'' only the magnitude \eq{\mag{\ptvec{x}}}}
    (later, it will be the Manev potential \eq{V^0=-\sck_1/r - \ttfrac{1}{2}\sck_2/r^2}),
    and \eq{V^\ss{1}} for all other arbitrary conservative forces (non-conservative forces are also included). 
\end{notesl}
\end{small}

\noindent Though we are interested in dynamics on \eq{\affE^n}, we are going to instead start with \eq{\affE^{n+1}=\affE^{\en}} as our configuration manifold where  \eq{\en:=n+1} (classically, \eq{n=3}). 
This will allow us to treat the projective transformation as a diffeomorphism (rather than a submersion). With one assumption on the external forces, we will easily recover the ``actual'' system in \eq{(\en-1)}-dim space (rather, \eq{(2\en-2)}-dim phase space) as an affine subspace of  \eq{\cotsp\affE^{\en}} which is invariant under the dynamics on \eq{\cotsp\affE^{\en}}. 
 Conceptually, this is no different than considering motion in the cotangent bundle of a plane, \eq{\cotsp\Sig^2\subset\cotsp\affE^3}, or any hyperplane, \eq{\cotsp\Sig^{\en-\ii{1}}\subset\cotsp\affE^{\en}}. The only difference is the dimension which is of little consequence for the following developments.\footnote{So long as the dimension is finite.} 
 
 \begin{notation}
    We let \eq{\affE \equiv \affE^{\en}} denote Euclidean \eq{\en}-space (with vector space \eq{\Evec\equiv\Evec^\en}), where the configuration of the system we ultimately care about evolves in an \eq{(\en-1)}-dim hyperplane, \eq{\Sig \subset \affE}.  The case that \eq{\en-1=3} corresponds to classical dynamics in 3-space (viewed as a hyperplane in \eq{4}-space).
 \end{notation}

 Our reason for introducing the extra dimension to begin with is to facilitate use of the projective transformation. Specifically, it allows us to define the transformation as a diffeomorphism (rather than a submersion). 
 The specifics of the the original Hamiltonian system on the \eq{2 \en}-dim phase space \eq{\cotsp\affE} are given in section \ref{sec:Hnom_prj}. First, in section \ref{sec:prj_prelim}, we introduce some preliminary ideas and establish notation. Then, in section \ref{sec:prj_geomech}, we introduce the geometric view of the projective transformation and use it to transform the original system into some ``other'' system which, ultimately, will have more desirable properties for central force motion (namely, the Manev and Kepler problems). Yet, the benefits will not be obvious until section \ref{sec:prj_regular} when the projective transformation is combined with a Sundman-like transformation of the evolution parameter.

\subsection{Setting the Stage:~Configuration \& Phase Space, Angular Momentum, etc.} \label{sec:prj_prelim}

For the entirety of this section, our configuration manifold is Euclidean affine \eq{\en}-space, \eq{\affE \equiv \affE^{\en}} (we suppress the superscript), with associated \eq{\en}-dim Euclidean inner product space, \eq{(\Evec,\bsfb{\emet}_\ii{\!\Evec}=\inner{\slot}{\slot})}.\footnote{The configuration manifold will actually be taken as an \eq{\en}-dim region of \eq{\vsp{E}} excluding the origin as well as the case \eq{r^\en<0}.}
However, we will impose upon \eq{\affE} the structure of a product manifold,  \eq{\affE =\Sig \times\aff{N}}, where \eq{\Sig \cong \affE^{\en-\ii{1}}} and \eq{\aff{N}\cong\affE^1\cong\mbb{R}} are Euclidean spaces of of dimensions \eq{\en-1} and {1}, respectively.
 The associated inner product space then has the analogous form \eq{(\vsp{E} =\bs{\Sigup}\oplus\vsp{N},\bsfb{\emet}_\ii{\!\Evec}=\sfb{\emet}_\ii{\!\Sigup}\oplus \sfb{\emet}_\ii{\vsp{N}})}. 
Much of what follows is somewhat intuitive and serves primarily to establish notation. Those lacking time or interest may skip this section and come back to it as needed for reference. 
 The key point is the decomposition of the  (co)tangent spaces of \eq{\affE =\Sig\times \aff{N}} into orthogonal \eq{(\en-1)}-dim  and \eq{1}-dim vector (sub)spaces seen in Eq.\eqref{Eaf_split}–\ref{Eaf_split2};  any (co)vector on \eq{\affE} splits into a part that is tangent to \eq{\Sig} (and normal to \eq{\aff{N}}) and  a part that is normal to \eq{\Sig} (and tangent to \eq{\aff{N}}). The same is also true of any displacement vector in \eq{\vsp{E} =\bs{\Sigup}\oplus\vsp{N}}. Essentially, the space \eq{\affE =\Sig\times \aff{N}} and its (co)tangent spaces are treated similarly to how one might treat \eq{\mbb{R}^\en} as \eq{\mbb{R}^{\en-\ii{1}}\times \mbb{R}}  or \eq{\mbb{R}^{\en-\ii{1}}\oplus \mbb{R}}.

Any point \eq{\barpt{x}\in \affE =\Sig \times\aff{N}}  may be viewed as \eq{\barpt{x}=(\pt{x},x^\en)\in \Sig \times\aff{N}}. We then define projections onto each subspace:
\begin{small}
\begin{align} \label{E4_x4prj}
\affE=\Sig\times \aff{N}
&&
\begin{array}{lllllll}
     \prj_\ii{\aff{N}}:\affE \to \aff{N}\cong\mbb{R}
    &\qquad 
     \barpt{x}=(\pt{x},x^\en) \;\mapsto \;
      \prj_\ii{\aff{N}}(\barpt{x}) = x^\en
      \;\; \equiv\, r^\en(\barpt{x})
\\[2pt]
      \prj_\ii{\Sig}: \affE \to \Sig
     & \qquad 
     \barpt{x}=(\pt{x},x^\en) \;\mapsto \;
      \prj_\ii{\Sig}(\barpt{x}) = \pt{x}
\end{array}
\end{align}
\end{small}
where, using the identification \eq{\aff{N}\cong\mbb{R}},  we may view \eq{\prj_\ii{\aff{N}}} as a (coordinate) function which we re-name \eq{r^\en \in\fun(\affE)}. This function, in turn, defines a homogeneous unit vector field, \eq{\envec}, as follows:\footnote{Typically, one would define the unit 1-form \eq{\enform := \dif r^\en / \mag{\dif r^\en}} but, in this case, it already holds that \eq{\mag{\dif r^\en}:= \sqrt{\bsfb{\emet}_\ii{\!\Evec}(\dif r^\en,\dif r^\en)}=1}.}
\begin{small}
\begin{align}
    \prj_\ii{\aff{N}}\cong r^\en \in\fun(\affE)
&&,&&
     \enform :=  
     \dif r^\en  
     \in \forms(\affE)
 &&,&&
    \envec := \inv{\bsfb{\emet}_\ii{\!\Evec}}(\enform) 
    \in\vect(\affE)
 &&,&&
    \enform(\envec ) = \inner{\envec}{\envec}=\inner{\enform}{\enform} =1
\end{align}
\end{small}
such that \eq{{\envec}_{|\barpt{x}}} spans each 1-dim line \eq{\tsp[x^\en]\aff{N}\subset \tsp[\barpt{x}]\affE} and is normal to each \eq{(\en-1)}-dim hyperplane \eq{\tsp[\pt{x}]\Sig\subset \tsp[\barpt{x}]\affE} (and similarly for \eq{\enform}). 
That is, the tangent spaces of \eq{\affE=\Sig\times \aff{N}} split accordingly as
\eq{\tsp[\barpt{x}]\affE =  \tsp[\barpt{x}](\Sig \times \aff{N}) \cong \tsp[\pt{x}]\Sig \oplus \tsp[x^\en]\aff{N}} (and similarly for cotangent spaces). 
Thus, any \eq{\barpt{x}\in\affE}, any \eq{\bsfb{v}\in\tsp[\barpt{x}]\affE}, and any \eq{\barbs{\eta}\in\cotsp[\barpt{x}]\affE} can be decomposed as
\begin{small}
\begin{align} \label{Eaf_split}
 \begin{array}{rlll}
     \barpt{x}=(\pt{x},x^\en) &\!\!\! \in\affE = \Sig  \times \aff{N}
     &\quad,\qquad r^\en(\pt{x})=0
 \\[2pt] 
     \bsfb{v}=\sfb{v}+  \sfb{v}^\ii{\perp} &\!\!\! \in \tsp[\cdt]\affE \,\cong\,  \tsp[\cdt]\Sig \oplus \tsp[\cdt]\aff{N}
     &\quad,\qquad \enform(\sfb{v})=0
     &\quad,\qquad  \sfb{v}^\ii{\perp} \,=\, \v^\en\envec
\\[2pt]
     \barbs{\eta}=\bs{\eta}+ \bs{\eta}^\ii{\perp}  &\!\!\! \in \cotsp[\cdt]\affE \,\cong\,  \cotsp[\cdt]\Sig \oplus \cotsp[\cdt]\aff{N}
     &\quad,\qquad \envec (\bs{\eta})=0
      &\quad,\qquad  \bs{\eta}^\ii{\perp} \,=\,  \eta_\en\enform
 \end{array}
\end{align}
\end{small}
where  \eq{\v^\en := \enform(\bsfb{v})} and \eq{\eta_\en :=\envec(\barbs{\eta})}. Note that \eq{\bsfb{v}=\sfb{v}+\sfb{v}^\ii{\perp}} could also be viewed as \eq{\sfb{v}\oplus \sfb{v}^\ii{\perp}} or \eq{(\sfb{v},\v^\en)} (and similarly for \eq{\barbs{\eta}}).
Above, we should technically write \eq{{\envec}_{|\barpt{x}}} but, because \eq{\envec} is  homogeneous
(the same everywhere, \eq{\nab \envec =0}\footnote{That is, for any points \eq{\barpt{x}\neq\barpt{s}\in\affE}, then \eq{{\envec}_{|\barpt{x}}\cong {\envec}_{|\barpt{s}}} are, for all practical purposes, the same vector (in the sense they are are parallel and have the same magnitude).}),
we will seldom bother to indicate the evaluation of \eq{\envec} or \eq{\enform} at a point.
Similarly, we did not bother to specify the base points of various (co)tangent spaces in Eq.\eqref{Eaf_split}; 
since \eq{\affE} is an affine space, all (co)tangent spaces are isomorphic to one another and all can further be identified with \eq{\affE}'s associated  
vector space \eq{\vsp{E}} (or its dual \eq{\Evec^*}).\footnote{E.g., \eq{\tsp[\barpt{x}]\affE \cong \tsp[\barpt{q}]\affE} for any \eq{\barpt{x}\neq \barpt{q}\in\vsp{E}} and, furthermore, \eq{\tsp[\barpt{x}]\affE \cong \tsp[\barpt{q}]\affE \cong \vsp{E}} where \eq{\vsp{E}} is the vector space associated with \eq{\affE}. Likewise, \eq{\cotsp[\barpt{x}]\affE \cong \cotsp[\barpt{q}]\affE \cong \Evec^*}. }
The same is true of the affine (sub)spaces \eq{\Sig} and \eq{\aff{N}}. 
That is, letting \eq{\vsp{E}}, \eq{\bs{\Sigup}}, and \eq{\vsp{N}\cong\mbb{R}} denote the vector spaces associated to the affine spaces in \eq{\affE =\Sig\times\aff{N}}, we have the following identifications: 
\begin{small}
\begin{align} \label{Eaf_split2}
\begin{array}{lllllll}
      \tsp[\cdt]\affE \,=\, \tsp[\cdt](\Sig\times\aff{N}) \,\cong\,  \tsp[\cdt]\Sig \oplus \tsp[\cdt]\aff{N} \,\cong\, \vsp{E} = \bs{\Sigup} \oplus \vsp{N}
    &&,\qquad 
        \bs{\Sigup} = \vsp{N}^\ii{\perp} = \ker \enform
        &,\quad 
        \vsp{N}  = \bs{\Sigup}^\ii{\perp} = \spn  { \envec }
\\[2pt]
      \cotsp[\cdt]\affE \,=\, \cotsp[\cdt](\Sig\times\aff{N})  \,\cong\,  \cotsp[\cdt]\Sig \oplus \cotsp[\cdt]\aff{N} \,\cong\, \Evec^* = \bs{\Sigup}^*\oplus \vsp{N}^* 
     &&,\qquad 
        \bs{\Sigup}^* = {(\vsp{N}^*)}^\ii{\perp} = \ker \envec 
        &,\quad 
        \vsp{N}^* = {(\bs{\Sigup}^*)}^\ii{\perp} = \spn {\enform}
\end{array}
\end{align}
\end{small}
where we are using \eq{\tsp[\cdt]\affE \cong \vsp{E}}  to regard the homogeneous vector field \eq{\envec} as a fixed vector in \eq{\vsp{E}}.   
We then adopt the following notation for points in the (co)tangent bundle of \eq{\affE=\Sig \times \aff{N}}:
\begin{small}
\begin{align} \label{TEaff_prj}
\begin{array}{rlllll}
   \barpt{\v}_{\barpt{x}} &\!\!\!\! := ( \barpt{x}, \bsfb{v}) \,=\, ( (\pt{x},x^\en),\, \sfb{v} + \v^\en \envec) \in \tsp\affE = \tsp(\Sig \times \aff{N}) &,
   \qquad 
   \pt{\v}_{\pt{x}} :=(\pt{x},\sfb{v}) \in \tsp\Sig
\\[2pt]
    \bar{\eta}_{\barpt{x}} &\!\!\!\! :=  ( \barpt{x}, \barbs{\eta}) \,=\, ( (\pt{x},x^\en), \,\bs{\eta}+ \eta_\en \enform) \in \cotsp\affE = \cotsp(\Sig \times \aff{N})&,
    \qquad 
    \eta_{\pt{x}} := (\pt{x},\bs{\eta}) \in \cotsp\Sig
\end{array}
\end{align}
\end{small}

\noindent  To further clarify the developments so far, consider the tangent and normal projection tensors fields \eq{ \bs{\Pi} ,\bs{\Pi}^\ii{\perp} \in \tens^\ss{1}_\ss{1}(\affE)}, which we form in the usual way 
for a hypersurface (in this case, a hyperplane) with normal vector field \eq{\envec}:
\begin{small}
\begin{align} \nonumber 
\bar{\iden}_\ii{\!\Evec} =  \bs{\Pi}  + \bs{\Pi}^\ii{\perp}
 &&,&&
    \bs{\Pi}^\ii{\perp} := \envec \otms\enform  =\bar{\iden}_\ii{\!\Evec} -  \bs{\Pi}  
   &&,&&
     \bs{\Pi}  := \bar{\iden}_\ii{\!\Evec} - \bs{\Pi}^\ii{\perp} 
  &&,&&
          \bs{\Pi} \cdot \bs{\Pi}^\ii{\perp} = 0
 &&,&&   
       \bs{\Pi} \cdot  \bs{\Pi}  =  \bs{\Pi} 
 &&,&& 
      \bs{\Pi}^\ii{\perp}\cdot \bs{\Pi}^\ii{\perp} = \bs{\Pi}^\ii{\perp}
\end{align}
\end{small}
For the case at hand, these projection tensor fields are homogeneous and we will typically not bother to indicate their evaluation at a point. 
Above,  \eq{\bar{\iden}_\ii{\!\Evec}\in\tens^\ss{1}_\ss{1}(\affE)} is the \eq{\en}-space identity and the relation \eq{\bar{\iden}_\ii{\!\Evec} =  \bs{\Pi}  + \bs{\Pi}^\ii{\perp}} 
 is simply a re-statement of Eq.\eqref{Eaf_split}; any \eq{\bsfb{v}\in\tsp[\barpt{x}]\affE} or \eq{\barbs{\eta}\in\cotsp[\barpt{x}]\affE} can be decomposed into tangent and normal parts:
\begin{small}
\begin{align}
\begin{array}{lllll}
    \bsfb{v} \,=\, \bar{\iden}_\ii{\!\Evec}(\bsfb{v}) \,=\, ( \bs{\Pi}  + \bs{\Pi}^\ii{\perp})\cdot \bsfb{v} \,=\, \sfb{v} + \sfb{v}^\ii{\perp}  
    &,\qquad 
    \sfb{v} =  \bs{\Pi}(\bsfb{v}) \in \tsp[\cdt]\Sig 
     &,\qquad 
     \sfb{v}^\ii{\perp}  = \bs{\Pi}^\ii{\perp} (\bsfb{v}) = \v^\en \envec 
     \in \tsp[\cdt]\aff{N}
\\[2pt]
      \barbs{\eta} \,=\,  \trn{\bar{\iden}_\ii{\!\Evec}}(\barbs{\eta}) \,=\, \barbs{\eta}\cdot ( \bs{\Pi}  + \bs{\Pi}^\ii{\perp}) \,=\, \bs{\eta} + \bs{\eta}^\ii{\perp}  
     &,\qquad 
       \bs{\eta} =  \bs{\Pi} (\barbs{\eta},\cdot) \in  \cotsp[\cdt]\Sig 
     &,\qquad 
    \bs{\eta}^\ii{\perp}  = \bs{\Pi}^\ii{\perp}(\barbs{\eta},\cdot) =  \eta_\en \enform  
     \in \cotsp[\cdt]\aff{N}
\end{array}
\end{align}
\end{small}
Note that \eq{\bs{\Pi} \equiv \iden_\ii{\!\Sigup} } is just the identity tensor on \eq{\Sig}, viewed extrinsically on \eq{\affE}.\footnote{That is, if \eq{\iden_\ii{\!\Sigup}\in\tens^\ss{1}_\ss{1}(\Sig)} is the intrinsic identity on \eq{\Sig} (as its own independent manifold), then the extrinsic view of \eq{\iden_\ii{\!\Sigup}} as a tensor on \eq{\affE = \Sig\times\aff{N}} is simply \eq{\imath_*\iden_\ii{\!\Sigup} = \prj_\ii{\Sig}^* \iden_\ii{\!\Sigup} = \iden_\ii{\!\Sigup} \oplus (\sfb{0}\otms\sfb{0}) = \bs{\Pi}}. We will simply re-use the notation \eq{\iden_\ii{\!\Sigup}}.}
The same view is applied to other ``usual'' objects on the Euclidean space \eq{\Sig}. For instance, along with \eq{\iden_\ii{\!\Sigup}\in\tens^\ss{1}_\ss{1}(\Sig)}, the metric/inner product \eq{\sfb{\emet}_\ii{\!\Sigup}\in \tens^0_2(\Sig) }, displacement ``vector field'' (relative to some fixed origin) \eq{\sfb{r}:\Sig \to \bs{\Sigup}}, and the norm \eq{r:=\sqrt{ \sfb{\emet}_\ii{\!\Sigup}(\sfb{r},\sfb{r}) }\in\fun(\Sig)},  are all viewed extrinsically on \eq{\affE}.
By abuse of notation, we will re-use the same symbols such that \eq{\iden_\ii{\!\Sigup}}, \eq{\sfb{\emet}_\ii{\!\Sigup}}, \eq{\sfb{r}}, and \eq{r} are all regarded as tensor 
fields on \eq{\affE}:\footnote{That is, if \eq{\prj_\ii{\Sig}:\affE\to \Sig} is the projection and \eq{\imath:\Sig \hookrightarrow (\Sig \times \{0\} \subset\affE)} is  the inclusion, then the extrinsic view of \eq{ \iden_\ii{\!\Sigup}}, \eq{ \sfb{\emet}_\ii{\!\Sigup}}, and \eq{\sfb{r}} (which we denote by the same symbols) really means:
\eq{\quad}  \eq{ \iden_\ii{\!\Sigup} \equiv  \prj_\ii{\Sig}^* \iden_\ii{\!\Sigup} \equiv \imath_*  \iden_\ii{\!\Sigup} 
\quad,\quad  
\sfb{\emet}_\ii{\!\Sigup} \equiv  \prj_\ii{\Sig}^* \sfb{\emet}_\ii{\!\Sigup} \equiv \imath^*  \bsfb{\emet}_\ii{\!\Evec} 
\quad,\quad 
 \inv{\sfb{\emet}_\ii{\!\Sigup}} \equiv \imath_* \sfb{\emet}_\ii{\!\Sigup} 
 \quad,\quad 
 \sfb{r} \equiv \imath_* \sfb{r} \quad,\quad r \equiv  \prj_\ii{\Sig}^* r  }. }
\begin{small}
\begin{align} \label{tinga}
 \begin{array}{rlllll}
      \iden_\ii{\!\Sigup} \,= &\!\!\!\! 
      \bar{\iden}_\ii{\!\Evec} - \envec \otms\enform \,=\,  \bs{\Pi}  
      &\qquad,\qquad\qquad 
       \iden_\ii{\!\Sigup} \cdot\envec = 0
\\[2pt]
      \sfb{\emet}_\ii{\!\Sigup} \,= &\!\!\!\!
      \bsfb{\emet}_\ii{\!\Evec} - \enform \otms\enform
      \,=\, \trn{\bs{\Pi}}\cdot\bsfb{\emet}_\ii{\!\Evec} \cdot \bs{\Pi}  
       &\qquad,\qquad\qquad 
       \sfb{\emet}_\ii{\!\Sigup}\cdot\envec = 0
\\[2pt]
      \inv{\sfb{\emet}_\ii{\!\Sigup} } = &\!\!\!\!
      \inv{\bsfb{\emet}_\ii{\!\Evec} } - \envec \otms\envec
       &\qquad,\qquad\qquad 
       \inv{\sfb{\emet}_\ii{\!\Sigup}} \cdot\enform = 0
\\[2pt]
    \sfb{r} \,= &\!\!\!\!
    \bsfb{r} \,-\, r^\en \envec \,=\, \bs{\Pi} (\bsfb{r})
     &\qquad,\qquad\qquad 
     \sfb{r}\cdot\enform = 0
\\[2pt]
        r^2 \,= &\!\!\!\! \bar{r}^2 - r_\en^2
         &\qquad,\qquad\qquad  \lderiv{\envec} r = \envec\cdot \dif r = 0
 \end{array}
\end{align}
\end{small}
Where the above \eq{\iden_\ii{\!\Sigup}\equiv\bs{\Pi}\in\tens^\ss{1}_\ss{1}(\affE)} and \eq{\sfb{\emet}_\ii{\!\Sigup} \equiv \prj_\ii{\Sig}^*\sfb{\emet}_\ii{\!\Sigup}\in \tens^0_2(\affE) } are now
\textit{degenerate}\footnote{E.g., the matrix representations of \eq{\iden_\ii{\!\Sigup}\in\tens^\ss{1}_\ss{1}(\affE)} and \eq{\sfb{\emet}_\ii{\!\Sigup}\in \tens^0_2(\affE) } in a coordinate basis for cartesian coordinates \eq{\bartup{r}:\affE \to \mbb{R}^{\en}} are degenerate matrices:
\begin{align}
  \crd{\iden_\ii{\!\Sigup}}{\bartup{r}} \,=\, 
  \begin{pmatrix}
      \imat_{\en-\ii{1}} & \tup{0} \\ \trn{\tup{0}} & 0
  \end{pmatrix}
  \,=\, \crd{\sfb{\emet}_\ii{\!\Sigup}}{\bartup{r}}   \,=\, \crd{ \inv{\sfb{\emet}_\ii{\!\Sigup}} }{\bartup{r}} \in \mbb{R}^{\en \times \en}
\end{align}
The notation \eq{\inv{\sfb{\emet}_\ii{\!\Sigup}}} for \eq{\inv{\sfb{\emet}_\ii{\!\Sigup}}\in\tens^2_0(\vsp{E})} should then be interpreted with care: \eq{\inv{\sfb{\emet}_\ii{\!\Sigup}} \cdot \sfb{\emet}_\ii{\!\Sigup} \neq \bar{\iden}_\ii{\!\Evec} } but, rather,  \eq{\inv{\sfb{\emet}_\ii{\!\Sigup}} \cdot \sfb{\emet}_\ii{\!\Sigup} =\iden_\ii{\!\Sigup} \equiv \bs{\Pi}}. }.
For any \eq{\barpt{x}=(\pt{x},x^\en)\in\affE=\Sig\times\aff{N}}, the above \eq{\sfb{r}} is such that,
\begin{small}
\begin{gather}
\begin{array}{llllll}
       \sfb{r} = \bs{\Pi}(\bsfb{r}): \affE \to \bs{\Sigup}\subset\Evec
    &,\qquad  
    \sfb{r}_{\barpt{x}} \equiv \sfb{r}_{\pt{x}} =: \ptvec{x} 
\\[2pt]
     r = \sqrt{ \inner{\sfb{r}}{\sfb{r}} }  = \sqrt{ \bsfb{\emet}_\ii{\!\Evec}(\sfb{r},\sfb{r}) } \equiv \prj_\ii{\Sig}^*r \in\fun(\affE) 
      &,\qquad  
      r(\barpt{x}) \equiv r(\pt{x}) = \nrm{\ptvec{x}} =: \nrm{\pt{x}}
\end{array}
\end{gather}
\end{small}
The above do not depend on the ``\eq{\aff{N}}-part'' of any \eq{\barpt{x}=(\pt{x},x^\en)} and, as above, we will often write \eq{\sfb{r}_{\pt{x}}} or \eq{r(\pt{x})=\nrm{\pt{x}}} rather than \eq{\sfb{r}_{\barpt{x}}} or \eq{r(\barpt{x})}. 
Similarly, the standard radial coordinate vector field \eq{\be_{\rfun}\in\vect(\Sig)} may also be regarded on \eq{\affE}, with the usual relations:
\begin{small}
\begin{align} \label{E4_rhat_vec}
\begin{array}{lllllll}
       \; \hsfb{r} = \tfrac{1}{r}\sfb{r} 
      \equiv \bpart{r} = \be_{\rfun} \in\vect(\affE)
    &,\quad
    \nab \hsfb{r} 
    =  \tfrac{1}{r}(\iden_\ii{\!\Sigup}- \hsfb{r}\otms\hsfb{r}^\flt) 
     =  \tfrac{1}{r}(\bar{\iden}_\ii{\!\Evec}- \hsfb{r}\otms\hsfb{r}^\flt - \envec \otms \enform) 
    \in\tens^\ss{1}_\ss{1}(\affE)
\\[2pt]
     \hsfb{r}^\flt := \sfb{\emet}_\ii{\!\Sigup}(\hsfb{r}) %
     \equiv \dif r = \bep^{\rfun} \in\forms(\affE)
     &,\quad
      \nab \hsfb{r}^\flt =  \tfrac{1}{r}(\sfb{\emet}_\ii{\!\Sigup} - \hsfb{r}^\flt\otms\hsfb{r}^\flt) 
       =  \tfrac{1}{r}(\bsfb{\emet}_\ii{\!\Evec}- \hsfb{r}^\flt \otms\hsfb{r}^\flt - \enform \otms \enform)
      \in\tens^0_2 (\affE)
\end{array}
&&
\begin{array}{llll}
      \enform\cdot \hsfb{r} =0 
\\[2pt]
     \envec \cdot \hsfb{r}^\flt =0
 \\[2pt]
     \hsfb{r}^\flt\cdot \hsfb{r} = 1
\end{array}
\end{align}
\end{small}
with \eq{\iden_\ii{\!\Sigup}\equiv \bs{\Pi}} and \eq{\sfb{\emet}_\ii{\!\Sigup}} as in Eq.\eqref{tinga} and \eq{\nab \equiv \nab^\ss{\bsfb{\emet}_\ii{\!\Evec}}} the Euclidean LC connection on \eq{(\affE,\bsfb{\emet}_\ii{\!\Evec})}. 
Again, the above do not depend on the ``\eq{\aff{N}}-part'' of  any \eq{\barpt{x}=(\pt{x},x^\en)\in\affE=\Sig\times\aff{N}} such that we often write, for instance, \eq{\hsfb{r}_{\pt{x}}} rather than \eq{\hsfb{r}_{\barpt{x}}}, which is the usual unit radial vector:
\begin{small}
\begin{align}
      \hsfb{r}_{\pt{x}} \,\equiv\,  \hsfb{r}_{\barpt{x}} \,=\,  \tfrac{1}{\nrm{\pt{x}}} \ptvec{x} \,=:\hsfb{x} \in\tsp[\pt{x}]\Sig \subset \tsp[\barpt{x}]\affE
       \qquad,\qquad 
   \hsfb{r}^\flt_{\pt{x}}  \,\equiv\, \hsfb{r}^\flt_{\barpt{x}} \,=\, \hsfb{x}^\flt \,:=\, \inv{\bsfb{\emet}_\ii{\!\Evec}}(\hsfb{x}) \,=\, \inv{\sfb{\emet}_\ii{\!\Sigup}} (\hsfb{x})
    \in\cotsp[\pt{x}]\Sig \subset \cotsp[\barpt{x}]\affE
\end{align}
\end{small}
From Eq.\eqref{tinga}, note the identity, metric, and displacement vector field on \eq{\affE}  decompose as:
\begin{small}
\begin{align} \label{linga}
    \bar{\iden}_\ii{\!\Evec} =  \bs{\Pi}  + \bs{\Pi}^\ii{\perp} 
       = \iden_\ii{\!\Sigup} + \envec \otms\enform \in\tens^\ss{1}_\ss{1}(\affE)
&&,&&
    \bsfb{\emet}_\ii{\!\Evec} = \sfb{\emet}_\ii{\!\Sigup} + \enform \otms\enform  \in\tens^0_2(\affE)
&&,&&
    \bsfb{r} = \sfb{r} + r^\en \envec :\affE \to \vsp{E}
\end{align}
\end{small}
such that the inner product of any (co)tangent vectors (or vector fields and 1-forms), \eq{\bsfb{u},\bsfb{v}\in\tsp[\cdt]\affE} and \eq{\barbs{\mu},\barbs{\eta}\in\cotsp[\cdt]\affE}, satisfies 
\begin{small}
\begin{align} \label{inners_E4}
\begin{array}{llll}
   \inner{\bsfb{u}}{\bsfb{v}} :=\,  \bsfb{\emet}_\ii{\!\Evec}(\bsfb{u},\bsfb{v})  \,=\,  \inner{\sfb{u}}{\sfb{v}} + u^\en \v^\en
    &,\qquad  \inner{\barbs{\mu}}{\barbs{\eta}} :=\,  \inv{\bsfb{\emet}_\ii{\!\Evec}}(\barbs{\mu},\barbs{\eta})  \,=\,  \inner{\bs{\mu}}{\bs{\eta}} + \mu_\en \eta_\en
\\[2pt]
    \inner{\sfb{u}}{\sfb{v}} :=\, \bsfb{\emet}_\ii{\!\Evec}(\sfb{u},\sfb{v})  \,=\,   \sfb{\emet}_\ii{\!\Sigup}(\sfb{u},\sfb{v})   \,=\,  \sfb{\emet}_\ii{\!\Sigup}(\bsfb{u},\bsfb{v}) 
    &,\qquad  \inner{\bs{\mu}}{\bs{\eta}} :=\,  \inv{\bsfb{\emet}_\ii{\!\Evec}}(\bs{\mu},\bs{\eta}) 
     \,=\, \inv{\sfb{\emet}_\ii{\!\Sigup}}(\bs{\mu},\bs{\eta}) 
  \,=\, \inv{\sfb{\emet}_\ii{\!\Sigup}}(\barbs{\mu},\barbs{\eta})  
\end{array}
\end{align}
\end{small}

\paragraph{Using $\vsp{E}$ as the Configuration Manifold.}
We could also  take the vector space \eq{\vsp{E}} as the base manifold rather than the affine space \eq{\affE}. This requires a \textit{choice} 
of origin, \eq{\barpt{o}\in\affE}, such that we may then identify \eq{\affE} with the vector space \eq{\vsp{E}} by sending any point \eq{\barpt{x}\in\affE} to a displacement vector \eq{\bsfb{r}_{\barpt{x}}:=\barpt{x}-\barpt{o}\in\vsp{E}} which, by abuse of notation, we will denote again by \eq{\barpt{x}}. The affine space \eq{\affE = \Sig \times \aff{N}} can then be replaced by the vector space \eq{\vsp{E} = \bs{\Sigup} \oplus \vsp{N}}. That is, a point \eq{\barpt{x}=(\pt{x},x^\en)\in\affE} is replaced by a displacement vector which decomposes in the same way as a tangent vector:  \eq{\barpt{x}=\ptvec{x}+ \ptvec{x}^\ii{\perp}\in\vsp{E}}, with  \eq{\ptvec{x}\in\bs{\Sigup}} and \eq{ \ptvec{x}^\ii{\perp}= x^\en \envec \in\vsp{N}}:
\begin{small}
\begin{align} \label{Evec_split}
     \barpt{x} \,=\, \ptvec{x} + x^\en\envec \in \vsp{E} \,=\, \bs{\Sigup} \oplus \vsp{N}  
      \qquad,\qquad  r^\en(\ptvec{x}) =  \enform (\ptvec{x})=0
      \qquad,\qquad  r^\en(\barpt{x}) =  \enform (\barpt{x}) = x^\en 
        \qquad,\qquad 
        \ptvec{x}^\ii{\perp} = x^\en \envec
\end{align}
\end{small}
The (co)tangent spaces of an affine space are the same as those of its associated vector space such that  Eq.\eqref{Eaf_split}-Eq.\eqref{Eaf_split2} still holds with \eq{\affE} replaced by \eq{\vsp{E}}.
For points in the (co)tangent \textit{bundles} of \eq{\vsp{E}=\bs{\Sigup}\oplus\vsp{N}}, we employ notation analogous to Eq.\eqref{TEaff_prj}:
\begin{small}
\begin{align} \label{TEvec_prj}
\begin{array}{rlllll}
    \barpt{\v}_{\barpt{x}} &\!\!\!\! := ( \barpt{x}, \bsfb{v}) \,=\, ( \ptvec{x}+x^\en \envec,\, \sfb{v} + \v^\en \envec) \in \tsp\vsp{E} = \tsp(\bs{\Sigup} \oplus \vsp{N}) &,
   \qquad 
   \pt{\v}_{\pt{x}} :=(\ptvec{x},\sfb{v}) \in \tsp\bs{\Sigup}
\\[2pt]
    \bar{\eta}_{\barpt{x}} &\!\!\!\! :=  ( \barpt{x}, \barbs{\eta}) \,=\, ( \ptvec{x}+x^\en \envec, \,\bs{\eta}+ \eta_\en \enform) \in \cotsp\vsp{E} = \cotsp(\bs{\Sigup} \oplus \vsp{N}) &,
    \qquad 
    \eta_{\pt{x}} := (\ptvec{x},\bs{\eta}) \in \cotsp\bs{\Sigup}
\end{array}
\end{align}
\end{small}
Note when \eq{\ptvec{x}} is to be written as a subscript, we write \eq{(\cdot)_{\pt{x}}} rather than \eq{(\cdot)_{\ptvec{x}}} so as to avoid possible confusion with \eq{(\cdot)_{\barpt{x}}}.

\begin{small}
\begin{notesq}
    We will not always be careful about the affine vs.~vector space view (\eq{\affE =\Sig\times \aff{N}} vs.~\eq{\vsp{E} =\bs{\Sigup}\oplus \vsp{N}}). From here forward, \rmsb{we will usually adopt the vector space view}. Note that we may use \eq{\tsp[\cdt]\vsp{E} = \tsp[\cdt]\affE \cong \vsp{E}} to treat elements of \eq{\vsp{E}} (displacement vectors) and tangent vectors on equal footing. Likewise, \eq{\cotsp[\cdt]\vsp{E} = \cotsp[\cdt]\affE \cong \vsp{E}^*} allows us to treat elements of \eq{ \vsp{E}^*} and cotangent vectors and on equal footing. We will often do so without making explicit mention of it.\footnote{For instance, consider some displacement vector \eq{\barpt{x}=\ptvec{x}+x^\en \envec\in\vsp{E}\cong\tsp[\cdt]\vsp{E}} and some 1-form/covector \eq{\barbs{\eta}=\bs{\eta}+\eta_\en \enform\in\cotsp[\cdt]\vsp{E}\cong \vsp{E}^*}. Expressions such as \eq{\barpt{x}\cdot\barbs{\eta}} or \eq{\ptvec{x}\cdot\bs{\eta}} then clearly imply that \eq{\barpt{x}} is being treated as \eq{\eq{\barpt{x}}\in\tsp[\cdt]\vsp{E}} or, alternatively, that \eq{\barbs{\eta}} is being treated as \eq{\barbs{\eta}\in\vsp{E}^*}. }
\end{notesq}
\end{small}

\paragraph{Cartesian Coordinates.} 
Let \eq{\hbe_i \equiv ( \hbe_1,\dots, \hbe_{\en-\ii{1}})} be any basis for \eq{\bs{\Sigup}\subset \vsp{E}}. It follows that the homogeneous unit normal vector, \eq{\envec\in\vect(\vsp{E})_\ii{\tx{hmg.}} \cong \vsp{E}},  completes the set such that \eq{\hbe_\a \equiv  ( \hbe_1,\dots, \hbe_{\en-\ii{1}},\envec) } is a basis for \eq{\vsp{E}=\bs{\Sigup}\oplus \vsp{N}}, with dual basis for \eq{\vsp{E}^*} denoted \eq{\hbep^\a} (defined by \eq{ \hbep^\a(\hbe_\b) = \kd^\a_\b}).\footnote{Recall that the unit normal vector field \eq{\envec\in\vect(\vsp{E})} is homogeneous \eq{(\nab \envec =0}) and autonomous \eq{(\pd_t \envec =0}), and the relation \eq{\tsp[\cdt]\vsp{E}\cong \vsp{E}} then allows us to view \eq{\envec} as a fixed vector, \eq{\envec\in\vsp{E}} (and similarly for \eq{\enform=\dif r^\en}).}
Although not required, we will, unless specified otherwise, always take \eq{\hbe_i} to be homogeneous and autonomous (like \eq{\envec}) such that \eq{\hbe_\a  \equiv (\hbe_i,\envec)} is a
fixed inertial basis\footnote{By ``fixed inertial basis'' we mean that all \eq{\hbe_\a} are homogeneous (\eq{\nab\hbe_\a = 0}) and autonomous (\eq{\pd_t \hbe_\a}), and likewise for the dual 1-forms \eq{\hbep^\a}.} 
for \eq{\vsp{E}}
with \eq{\bsfb{\emet}_\ii{\!\Evec}} expressed as: 
\begin{small}
\begin{align}
\begin{array}{llllll}
     \bsfb{\emet}_\ii{\!\Evec} \,=\, \sfb{\emet}_\ii{\!\Sigup} + \enform \otms\enform  \,=\, \emet_{ij}\hbep^i \otms \hbep^j + \enform \otms\enform \,=\,  \emet_{\a\b}\hbep^\a \otms \hbep^\b
\\[2pt]
     \inv{\bsfb{\emet}_\ii{\!\Evec} } \,=\, \inv{\sfb{\emet}_\ii{\!\Sigup} } + \envec \otms\envec  \,=\, \emet^{ij}\hbe_i \otms \hbe_j + \envec \otms\envec \,=\,  \emet^{\a\b}\hbe_\a \otms \hbe_\b
\end{array}
&& 
\begin{array}{cc}
     i,j = 1,\dots, \en =1
 \\[2pt]
     \a, \beta = 1,\dots, \en
\end{array}
\end{align}
\end{small}
where the \eq{\hbe_\a}-components, \eq{\emet_{\a\b}} and \eq{\emet^{\a\b}}, are all constants. Additionally, unless specified otherwise, we always take \eq{\hbe_\a=(\hbe_i,\envec)} to be a \eq{\bsfb{\emet}_\ii{\!\Evec}}-\textit{orthonormal} basis (i.e., cartesian) such that these components are simply:
\begin{small}
\begin{align} \label{emet_cart}
\begin{array}{lllll}
  \emet_{\a\b} :=\, \bsfb{\emet}_\ii{\!\Evec}(\hbe_\a,\hbe_\b) = \kd_{\a\b}
\\[2pt]
     \emet^{\a\b} :=\, \inv{\bsfb{\emet}_\ii{\!\Evec}}(\hbep^\a,\hbep^\b)  = \kd^{\a\b}
\end{array}
&& \fnsize{with:} \qquad 
\begin{array}{llll}
      \emet_{ij} \,=\, \bsfb{\emet}_\ii{\!\Evec}(\hbe_i,\hbe_j) = \sfb{\emet}_\ii{\!\Sigup}(\hbe_i,\hbe_j) \,=\,  \kd^i_j
      &,\qquad \emet_{\en\en} = \bsfb{\emet}_\ii{\!\Evec}(\envec,\envec) = \kd_{\en\en} = 1
\\[2pt]
     \emet^{ij} \,=\, \inv{\bsfb{\emet}_\ii{\!\Evec}}(\hbep^i,\hbep^j) = \inv{\sfb{\emet}_\ii{\!\Sigup}}(\hbep^i,\hbep^j) \,=\,  \kd^i_j
     &,\qquad \emet^{\en\en} = \inv{\bsfb{\emet}_\ii{\!\Evec}}(\enform,\enform) = \kd^{\en\en} = 1
\end{array}
\end{align}
\end{small}
(where \eq{\kd^\a_\b=\kd_{\a\b}=\kd^{\a\b}} is the Kronecker delta \textit{symbol}\footnote{The index placement on \eq{\kd^\a_\b=\kd_{\a\b}=\kd^{\a\b}} has no significance.}).
Now, the orthonormal basis \eq{\hbe_\a} defines an \eq{\en}-tuple of global
\textit{cartesian}\footnote{In this work, ``cartesian'' implies not only linearity but also orthonormality. That is, to say that \eq{r^\a} are cartesian coordinates is to say that \eq{\bsfb{\emet}_\ii{\!\Evec}(\hbe_\a,\hbe_\b)=\kd^\a_\b} (where \eq{\hbe_\a\equiv \be[r^\a]}). 
All coordinates are assumed time-independent unless specified otherwise.}
 coordinates which we will denote \eq{\bartup{r}=(\tup{r},r^\en)=(r^1,\dots,r^{\en-\ii{1}},r^\en):\vsp{E} \to\mbb{R}^{\en} }.   
That is, for any \eq{\barpt{x}\in\vsp{E}}:
\begin{small}
\begin{align}
       r^\a(\barpt{x}) = \hbep^\a(\barpt{x}) \equiv  \hbep^\a\cdot\barpt{x} \,=: x^\a 
\qquad,\qquad 
    \barpt{x} \,=\,  r^\a(\barpt{x})\hbe_\a =: x^\a \hbe_\a=x^i\hbe_i + x^\en\envec   \,=\, \ptvec{x} + x^\en \envec
\end{align}
\end{small}
The displacement vector fields and the ``\eq{\bs{\Sigup}}-norm'' function (regarded on \eq{\vsp{E}}) may then be expressed globally as:\footnote{We will have little need for the actual \eq{\en}-space norm function on \eq{\vsp{E}} but it would be \eq{\bar{r}^2 = \inner{\bsfb{r}}{\bsfb{r}} = \nrm{\bartup{r}}^2 = r^2 + r_\en^2 = \nrm{\tup{r}}^2 + r_\en^2 \in\fun(\vsp{E}) }. }
\begin{small}
\begin{align} \label{E4_disp_vec}
    \bsfb{r} \,=\, \sfb{r} + r^\en \envec  \,=\, r^\a \hbe_\a 
&&,&&
     \sfb{r} \,=\,  r^i \hbe_i
&&,&&
   r^2 = \inner{\sfb{r}}{\sfb{r}} 
      = \emet_{ij}r^i r^j \,\equiv\, \kd_{ij}r^i r^j = \nrm{\tup{r}}^2 =: \nrmtup{r}^2
\end{align}
\end{small}
Next, we note that the \eq{r^\a} frame fields, \eq{\be[r^\a]  \in \vect(\vsp{E})} and \eq{\bep[r^\a] =\dif r^\a  \in\forms(\vsp{E})} are homogeneous and can be identified with \eq{\hbe_\a \in\vsp{E}} and \eq{\hbep^\a\in\vsp{E}^*}. As such, we will simply re-use the notation \eq{\hbe_\a} and \eq{\hbep^\a} for the \eq{r^\a} frame fields:
\begin{small}
\begin{align} \label{ibase_same}
\begin{array}{llllll}
   \hbe_\a \cong   \be[r^\a]  
   \in \vect(\vsp{E})_\ii{\mrm{hmg.}}
 \qquad,\qquad 
      \hbep^\a \cong  \bep[r^\a] = \dif r^\a
      \in\forms(\vsp{E})_\ii{\mrm{hmg.}}  
\end{array}
\end{align}
\end{small}
The radial coordinate (co)vector fields on \eq{\bs{\Sigup}\subset\vsp{E}} (Eq.\eqref{E4_rhat_vec}),  \eq{\hsfb{r}=\be_{\rfun}\in\vect(\vsp{E})} and \eq{\hsfb{r}^\flt =\bep^{\rfun}=\dif r \in\forms(\vsp{E})}, may be expressed as 
\begin{small}
\begin{align} \label{rhat_cart}
\begin{array}{llll}
    \hsfb{r}^\flt = \dif r = \pd_i r \hbep^i = \hat{r}_i \hbep^\a
\\[2pt]
    \hsfb{r}  = \tfrac{1} {r} \sfb{r} = \hat{r}^i \hbe_i
\end{array}
    &&
\begin{array}{llll}
    \hat{r}^i := r^i/\nrm{\tup{r}} = r^i/\rfun 
\\[2pt]
      r_i := \emet_{ij}r^j \quad,\quad \hat{r}_i = r_i/\rfun
\\[2pt]
     \pd_i r = \tfrac{1} {r} \emet_{ij}r^j = \hat{r}_i
\end{array}
\end{align}
\end{small}
Now, let the corresponding cotangent-lifted cartesian coordinates on \eq{\cotsp\vsp{E}}  be denoted \eq{(\bartup{r},\bartup{\plin}):=\colift\bartup{r}:\cotsp\vsp{E} \to\mbb{R}^{2\en}}.  The associated coordinate frame fields on \eq{\cotsp\vsp{E}\cong \vsp{E}\oplus\vsp{E}^*} are again homogeneous tensor fields which can be identified with a fixed inertial basis for \eq{\vsp{E} \oplus \vsp{E}^*} given in terms of \eq{\hbe_\a} as:
 \begin{small}
 \begin{align} \label{T*V_ibase_proj}
 \begin{array}{lllll}
    \vecth(\cotsp\Evec)_\ii{\mrm{hmg.}} \ni  \hpdii{r^\a} \equiv  \hbpart{\a} \cong   \hbe_\a \oplus \sfb{0} 
\\[2pt] 
     \vectv(\cotsp\Evec)_\ii{\mrm{hmg.}}  \ni  \hpdiiup{\plin_\a} \equiv \hbpartup{\a} \cong   \sfb{0}\oplus \hbep^\a 
 \end{array}
 &&
 \begin{array}{lllll}
     \formsh(\cotsp\Evec)_\ii{\mrm{hmg.}} \ni \dif{r^\a} \equiv \hbdel^\a \cong  \hbep^\a \oplus \sfb{0} 
\\[2pt] 
   \formsv(\cotsp\Evec)_\ii{\mrm{hmg.}} \ni \dif{\plin_\a} \equiv \hbdeldn_\a \cong   \sfb{0}\oplus \hbe_\a 
 \end{array}
 \end{align}
 \end{small}
 We will use the Hamiltonian formulation on \eq{(\cotsp\Evec,\nbs{\omg})} with \eq{\nbs{\omg}=-\exd\bs{\theta}\in\formsex^2(\cotsp\Evec)} the canonical symplectic form  and \eq{\bs{\theta}\in\formsh(\cotsp\Evec)} the canonical 1-form.
Like any cotangent-lifted coordinate frame fields, the above are \eq{\nbs{\omg}}-symplectic, with:
\begin{small}
\begin{align} \label{E4_sp_cords}
     \bs{\theta} = \plin_\a \hbdel^\a  
\qquad,\qquad 
   \nbs{\omg} = -\exd\bs{\theta} \,=\, \hbdel^\a \wedge \hbdeldn_\a
 \qquad,\qquad 
    \inv{\nbs{\omg}} = - \hbpart{\a} \wedge \hbpartup{\a} 
     \qquad,\qquad 
     \bscr{P}:=\inv{\nbs{\omg}}\cdot\bs{\theta} = \plin_\a \hbpartup{\a}
\end{align}
\end{small}
such that \eq{\cord{\ns{\txw}}{(\bartup{r},\bartup{\plin})} = J_{\ii{2}\en} = -\cord{\inv{\ns{\txw}}}{(\bartup{r},\bartup{\plin})} \in \Spmat{\ii{2}\en} } with \eq{J_{\ii{2}\en}} the standard symplectic matrix. This is the symplectic analog of Eq.\eqref{emet_cart} which says that \eq{\cord{\bar{\emet}}{\bartup{r}}=\imat_{\en}}.


\paragraph{Angular Momentum.} 
We quickly discuss some aspects of angular momentum as this will be important later on. We define functions \eq{\lang^{ij}\in\fun(\cotsp\vsp{E})}, expressed in cartesian coordinates \eq{(\tup{r},r^\en,\tup{\plin},\plin_\en)} as:
\begin{small}
\begin{align} \label{angMoment_prj_def}
      \lang^{ij} := (r^i \emet^{jk} - r^j \emet^{ik}) \plin_k = r^i \plin^j - \plin^i r^j  
\qquad,\qquad 
     \lang_{ij} := \emet_{ki}\emet_{sj} \lang^{ks}  
     = r_i \plin_j - \plin_i r_j  
\qquad,\qquad 
      \lang^2 :=  
       \ttfrac{1}{2}\lang^{ij}\lang_{ij} =  \nrmtup{r}^2 \nrmtup{\plin}^2 \,-\, (r^i \plin_i)^2
\end{align}
\end{small}
where \eq{\plin^i:=\emet^{ij} \plin_j }  and \eq{r_i:=\emet_{ij}r^j }. Note \eq{\lang^{ij}} are just the angular momentum functions on \eq{\cotsp\bs{\Sigup}}, but regarded on \eq{\cotsp\vsp{E}} such that the above do not depend on \eq{r^\en} or \eq{\plin_\en}.
As such, for any \eq{\bar{\kap}_{\barpt{x}}=(\barpt{x},\barbs{\kap})\in\cotsp\vsp{E}} we often write \eq{\lang^{ij}(\kap_{\pt{x}}) \equiv \lang^{ij}(\bar{\kap}_{\barpt{x}})} where \eq{\kap_{\pt{x}}=(\ptvec{x},\bs{\kap})\in\cotsp\bs{\Sigup}} (as in Eq.\eqref{TEvec_prj}).  
If we disregard the distinction between \eq{\tsp[\cdt]\vsp{E}} and \eq{\vsp{E}} (more specifically, between \eq{\tsp[\cdt]\bs{\Sigup}} and \eq{\bs{\Sigup}}), then we may view \eq{\lang^{ij}(\kap_{\pt{x}})} as the cartesian \eq{\hbe_i}-components of the following 2-vector, with magnitude \eq{\lang (\kap_{\pt{x}})}:\footnote{If we also ignore the distinction between \eq{\bs{\Sigup}} and \eq{\bs{\Sigup}^*}, then \eq{\lang^2(\kap_{\pt{x}}) \approx \tfrac{1}{2} \tr (\trn{\langb_{\kap_{\pt{x}}}}\cdot\langb_{\kap_{\pt{x}}})}. }
\begin{small}
 \begin{align} \label{angMoment_prj_nom}
          \langb_{\pt{x}} :=\, \ptvec{x}  \wedge \bs{\kap}^{\shrp} \,=\, (x^i\kap^j - \kap^i x^j)\hbe_i\otms\hbe_j
     \qquad,\qquad 
          \lang^2(\kap_{\pt{x}}) \,=\, \iinner{\langb_{\kap_{\pt{x}}}}{\langb_{\kap_{\pt{x}}}} 
          \,=\,  \det \begin{Vmatrix}
        \inner{\ptvec{x} }{\ptvec{x} }  & \inner{\ptvec{x} }{\bs{\kap}^{\shrp}} \\ \inner{\bs{\kap}^{\shrp}}{\ptvec{x} } & \inner{\bs{\kap}^{\shrp}}{\bs{\kap}^{\shrp}}
        \end{Vmatrix}
    \,=\, \nrm{\pt{x}}^2 \nrm{\bs{\kap}}^2 - (\ptvec{x}\cdot\bs{\kap})^2 
 \end{align}
\end{small}
where \eq{\bs{\kap}^{\shrp} = \inv{\bsfb{\emet}_\ii{\!\Evec}}(\bs{\kap}) = \inv{\sfb{\emet}_\ii{\!\Sigup}}(\bs{\kap})} such that \eq{\inner{\ptvec{x} }{\bs{\kap}^{\shrp}} = \ptvec{x}\cdot\bs{\kap}}, and where the above \eq{\iinner{\slot}{\slot}} denotes the inner product on \eq{\bwedge{2} \vsp{E} \cong \bwedge{2} \tsp[\cdt]\vsp{E}} induced by the inner product \eq{\bsfb{\emet}_\ii{\!\Evec}\equiv\inner{\slot}{\slot}} on \eq{\vsp{E}}. Note that if \eq{\barbs{\kap}_t=\sfb{m}(\dt{\bsfb{x}}_t)\in\cotsp[\barpt{x}_t]\vsp{E}} is the Euclidean kinematic covector along a curve \eq{\barpt{x}_t} — and thus \eq{\bs{\kap}_t = \sfb{m}(\dtsfb{x}_t) = m\sfb{\emet}(\dtsfb{x}_t) = m\dtsfb{x}_t^\flt\in\cotsp[\ptvec{x}_t]\bs{\Sigup}} — then the above is equivalent to 
\begin{small}
\begin{align}
        \bs{\kap}_t = \sfb{m}(\dtsfb{x}_t)
\quad \Rightarrow \quad 
      \langb_{\pt{x}} \,=\, \ptvec{x}  \wedge m\dtsfb{x} \,=\, m(x^i\dot{x}^j - \dot{x}^i x^j)\hbe_i\otms\hbe_j
 \quad,\quad 
          \lang^2(\kap_{\pt{x}})  \,=\, m^2 \big( \nrm{\pt{x}}^2 \nrm{\dtsfb{x}}^2 - \inner{\ptvec{x}}{\dtsfb{x}}^2 \big) 
\end{align}
\end{small}
Lastly, we note the following useful relations in cartesian coordinates:
\begin{small}
\begin{gather} \nonumber 
     \;\lang^{ij} r_i \plin_j = \lang_{ij} r^i \plin^j = \ttfrac{1}{2} \lang^{ij} \lang_{ij}  \;=\, \lang^2
 \qquad,\qquad 
      \lang^{ij}r_i r_j = \lang^{ij}\plin_i \plin_j = \lang_{ij} \plin^i \plin^j = \lang_{ij} r^i r^j \;=\, 0
\\[2pt]   \label{angmoment_rels_prj}
\begin{array}{rllll}
    \pd_\ii{r^i}  \ttfrac{1}{2}\lang^2 =&  
\\[2pt] 
      \pd_\ii{\plin_i}  \ttfrac{1}{2}\lang^2 =& 
\\[2pt] 
   \pd_\ii{r_i}  \ttfrac{1}{2}\lang^2 =& 
\\[2pt] 
    \pd_\ii{\plin^i}  \ttfrac{1}{2}\lang^2 =& 
\end{array}
\!\!\!
 \boxed{ \begin{array}{lllll}
    \lang_{ij} \plin^j \,=\, \nrmtup{\plin}^2 r_i  - (\tup{r}\cdot\tup{\plin})\plin_i 
    \,=\, (\nrmtup{\plin}^2 \emet_{ij} - \plin_i \plin_j) r^j
\\[2pt] 
   \lang^{ji}r_j  \,=\, \nrmtup{r}^2 \plin^i  - (\tup{r}\cdot\tup{\plin})r^i 
    \,=\, (\nrmtup{r}^2 \emet^{ij} - r^i r^j) \plin_j
\\[2pt] 
      \lang^{ij} \plin_j \,=\, \nrmtup{\plin}^2 r^i - (\tup{r}\cdot\tup{\plin}) \plin^i \,=\, (\nrmtup{\plin}^2 \kd^i_j - \plin^i \plin_j)r^j
\\[2pt] 
      \lang_{ji}r^j \,=\,  \nrmtup{r}^2 \plin_i - (\tup{r}\cdot\tup{\plin}) r_i \,=\, (\nrmtup{r}^2 \kd^j_i - r^j r_i) \plin_j
\end{array} }
\qquad\qquad 
\begin{array}{llll}
       \lang^{is}\lang_{sj} r^j \,=\, -\lang^2 r^i 
\\[2pt] 
       \lang^{is}\lang_{sj} r_i \,=\, -\lang^2 r_j
\\[2pt] 
       \lang^{is}\lang_{sj} \plin^j \,=\, -\lang^2 \plin^i
\\[2pt] 
       \lang^{is}\lang_{sj} \plin_i \,=\, -\lang^2 \plin_j
\end{array}
\end{gather}
\end{small}

\begin{small}
\begin{notesq}
    In the case that \eq{\en-1=3}, then \eq{\lang} is the usual angular momentum magnitude function on \eq{\cotsp\bs{\Sigup}^3 \subset \cotsp\Evec^4}, with Eq.\eqref{angMoment_prj_nom} equivalent to:
 \begin{small}
 \begin{align}
       \fnsize{for } \, \en-1=3: \quad \lang^2(\kap_{\pt{x}})  \,=\,  \nrm{\pt{x}}^2 \nrm{\bs{\kap}}^2 - (\ptvec{x}\cdot\bs{\kap})^2 
 \,=\, \nrm{\ptvec{x}\times \bs{\kap}^{\shrp} }^2 
 &&,&&
  \begin{array}{lllll}
        \langb_{\pt{x}} = \hdge{(\ptvec{x}\times \bs{\kap}^\shrp)} 
  \\[2pt]
    \hdge{\langb}_{\pt{x}} \,=\, \ptvec{x}\times \bs{\kap}^\shrp \,=\, -\hdge{\ptvec{x}}\cdot \bs{\kap}^\shrp \,=\,   \ax{\ptvec{x}}\cdot \bs{\kap}^\shrp
  \end{array}
 \end{align}
 \end{small}
\end{notesq}
\end{small}

\subsubsection{Some Submanifolds}
Let us define a few subsets of \eq{\affE} which will be important later on (this may be skipped and used for later reference as needed). For simplicity, we will take the vector space view and identify the affine space \eq{\affE=\Sig\times\aff{N}} with the vector space \eq{\vsp{E}=\bs{\Sigup} \oplus \vsp{N}} (by identifying points with their displacement vectors relative to some chosen fixed origin). First, for any \eq{b \in \mbb{R}}, we define a \eq{(\en-1)}-dim  hyperplane \eq{\Sig_\ii{b}\subset \vsp{E}} as the following level set:
\begin{small}
\begin{gather} \label{3surf_0}
\begin{array}{rlllll}
      \Sig_\ii{b} &\!\!\!\! :=\, \inv{(r^\en)}\{b\} 
      &\!\!\!\! =\, \big\{ \barpt{x}=\ptvec{x}+x^\en\envec\in\vsp{E}\;\big|\;   r^\en(\barpt{x}) =  b \big\} 
      \,=\, \bs{\Sigup} \oplus \{b\} 
      &\qquad\fnsize{i.e.,} \; \barpt{x} = \ptvec{x} + b \envec  \;,\; x^\en = b 
\\[2pt] 
     \tsp[\barpt{x}] \Sig_\ii{b} &\!\!\!\! \,=\, \ker \enform_{\barpt{x}} 
     &\!\!\!\! =\, \big\{ \bsfb{v}=\sfb{v}+\v^\en \envec\in\tsp[\barpt{x}]\vsp{E} \;\big|\;  \enform (\bsfb{v}) =\inner{ {\envec} }{\bsfb{v}} =0 \big\}
     \,\cong\, \tsp[\cdt]\bs{\Sigup} \oplus \{0\} \,\cong\, \tsp[\cdt]\bs{\Sigup}
      &\qquad\fnsize{i.e.,} \; \bsfb{v} = \sfb{v} \;, \; \v^\en =0
\\[2pt] 
     \cotsp[\barpt{x}] \Sig_\ii{b} &\!\!\!\! \,=\, \ker {\envec}_{|\barpt{x}} 
     &\!\!\!\! =\, \big\{ \barbs{\kap} = \bs{\kap}+\kap_\en \enform \in\cotsp[\barpt{x}]\vsp{E} \;\big|\;  \envec (\barbs{\kap}) = \inner{\enform}{\barbs{\kap}}  =0 \big\}  \,\cong\, \cotsp[\cdt]\bs{\Sigup} \oplus \{0\} 
     \,\cong\, \cotsp[\cdt]\bs{\Sigup}
      &\qquad\fnsize{i.e.,} \; \barbs{\kap} = \bs{\kap}
      \;,\; \kap_\en = 0 
\end{array}
\end{gather}
\end{small}
(where \eq{\enform = \dif r^\en} is homogeneous). That is, \eq{\Sig_\ii{b} \perp \vsp{N}} is just the flat slice of \eq{\vsp{E}=\bs{\Sigup}\oplus \vsp{N}} that orthogonally intersects \eq{\vsp{N}}  (i.e., the \eq{r^\en}  axis, which can be identified with the \eq{\envec} axis) at \eq{r^\en = b}. Every such \eq{\Sig_\ii{b}\subset \vsp{E}} is just an identical, parallel, copy of \eq{\Sig_\zr \equiv \bs{\Sigup}} that has been translated by
by \eq{b \envec}.\footnote{\textit{As a foliated manifold.} In other words, \eq{\vsp{E}} can be seen as a foliated manifold \eq{(\vsp{E}, \{\Sig_\ii{b} \} )}  where, for the present case, the leaves/slices, \eq{\Sig_\ii{b} = \bs{\Sigup} \oplus \{b\} } are all just parallel, non-intersecting, hyperplanes. }
As such, the (co)tangent spaces of any \eq{\Sig_\ii{b}} are, for all practical purposes, the same as the (co)tangent spaces of \eq{\bs{\Sigup}}; we seldom bother to distinguish them, often writing \eq{\tsp[\cdt]\bs{\Sigup}} 
rather than  \eq{\tsp[\cdt]\Sig_\ii{b}}.\footnote{To clarify, consider that a tangent space to \eq{\Sig_\ii{b}} is given by \eq{ \tsp[\barpt{x}] \Sig_\ii{b}= \ker \enform_{\barpt{x}}}. But, \eq{\enform = \dif r^\en \forms(\vsp{E})} is homogeneous such that \eq{\enform_{\barpt{x}} \cong \enform_{\barpt{s}}} are, essentially, the same for any \eq{\barpt{x},\barpt{s}\in\vsp{E}}. And thus, for any \eq{\barpt{x}\in\Sig_\ii{b}} and \eq{\barpt{s}\in\Sig_{k}}, we have  \eq{ \tsp[\barpt{x}] \Sig_\ii{b}= \ker \enform_{\barpt{x}} \cong \ker \enform_{\barpt{s}} = \tsp[\barpt{s}] \Sig_{k} \cong \tsp[\cdt]\bs{\Sigup}}. }
That is, a tangent space \eq{\tsp[\barpt{x}]\Sig_\ii{b}} does not actually depend — in any manner significant for the purposes of this work —  on the point  \eq{\barpt{x}} or the value \eq{b}. Further, each \eq{\Sig_\ii{b}} is parallelizable with trivial (co)tangent bundles:
\begin{small}
\begin{align}
   \tsp \Sig_\ii{b} \,\cong\, \Sig_\ii{b} \times \tsp[\cdt] \Sig_\ii{b} \,\cong\,  \Sig_\ii{b} \times \tsp[\cdt] \bs{\Sigup} \, \cong\,  \Sig_\ii{b} \times  \bs{\Sigup}
   &&,&&
   \cotsp \Sig_\ii{b} \,\cong\, \Sig_\ii{b} \times \cotsp[\cdt] \Sig_\ii{b} \,\cong\,  \Sig_\ii{b} \times \cotsp[\cdt] \bs{\Sigup} \, \cong\,  \Sig_\ii{b} \times  \bs{\Sigup}^*
\end{align}
\end{small}
Lastly, note that each \eq{\Sig_\ii{b}\subset \vsp{E}} is an (\eq{\en-1})-dim affine subspace, with \eq{\Sig_\zr} being naturally identified with the vector space \eq{\bs{\Sigup}}:
\begin{small}
\begin{align} 
    \imath: \bs{\Sigup} \hookrightarrow (\Sig_\zr \subset \vsp{E})
    \qquad,\qquad 
   \imath(\bs{\Sigup} ) = \Sig_\zr = \bs{\Sigup} \oplus \{0\} \cong \bs{\Sigup}
    \qquad,\qquad 
    \Sig_\ii{b} = \Sig_\zr + b\envec = \bs{\Sigup} \oplus \{b\}
\end{align}
\end{small}
Next, for some positive \eq{b \in \mbb{R}_\ii{+} }, we define a \eq{(\en-1)}-dim hypersurface \eq{\man{Q}_\ii{b}\subset \vsp{E}} as the following level set:
\begin{small}
\begin{align} \label{Qsurf_0}
\begin{array}{rlllll}
    \man{Q}_\ii{b} &\!\!\!\! :=\, \inv{\rfun}\{b\} &\!\!\!\!=\, \big\{  \barpt{q}=\ptvec{q}+q^\en \envec  \in \vsp{E}  \;\big|\; 
       r(\ptvec{q}) = \nrm{\ptvec{q}} = b \big\} 
       \,\cong\, \man{S}^{\en-\ii{2}}_\ii{b} \times \vsp{N}  
\\[2pt] 
      \tsp[\barpt{q}]\man{Q}_\ii{b} &\!\!\!\! =\, \ker \hsfb{r}^\flt_\ss{\!\pt{q}}
      &\!\!\!\!=\, \big\{  \bsfb{u} = \sfb{u} +  u^\en\envec  \in \tsp[\barpt{q}]\vsp{E} \;\big|\; \hsfb{r}^\flt_\ss{\!\pt{q}}(\sfb{u}) = \inner{\hsfb{r}_\ss{\!\pt{q}}}{\sfb{u}} =  0    \big\} 
      \,\cong\, \tsp[\pt{q}]\man{S}^{\en-\ii{2}}_\ii{b} \oplus \tsp[\cdt]\vsp{N}
\\[2pt] 
      \cotsp[\barpt{q}]\man{Q}_\ii{b}  &\!\!\!\! =\, \ker \hsfb{r}_\ss{\!\pt{q}} 
      &\!\!\!\!=\, \big\{  \barbs{\mu} = \bs{\mu}+\mu_\en \enform  \in \cotsp[\barpt{q}] \vsp{E}  \;\big|\;  \hsfb{r}_\ss{\!\pt{q}}(\bs{\mu}) = \inner{\hsfb{r}^\flt_\ss{\!\pt{q}}}{\bs{\mu}} =  0    \big\}
       \,\cong\, \cotsp[\pt{q}]\man{S}^{\en-\ii{2}}_\ii{b} \oplus \cotsp[\cdt]\vsp{N}
\end{array}
\end{align}
\end{small}
where \eq{\hsfb{r}^\flt = \dif r = \hat{r}_i \hbep^i\in\forms(\vsp{E})} and \eq{\hsfb{r}=\hat{r}^i \hbe_i\in\vect(\vsp{E})} such that \eq{\hsfb{r}_\ss{\!\pt{q}} = \hsfb{q}\in\tsp[\cdt]\bs{\Sigup}\subset \tsp[\cdt]\vsp{E}}. 
Above,  \eq{\man{S}^{\en-\ii{2}}_\ii{b} \subset \bs{\Sigup} \subset \vsp{E}} is the (\eq{\en-2})-dim sphere of radius \eq{b}, viewed as a submanifold of the (\eq{\en-1})-dim space \eq{\bs{\Sigup}} which is, in turn, viewed as hyperplane in \eq{\vsp{E}\equiv \vsp{E}^\en} 
(example\footnote{For example, if \eq{\en =3} then \eq{\man{Q}_\ii{b}=\man{S}^1_\ii{b}\times\vsp{N} \subset \Evec^3} is a \eq{2}-dim cylinder of radius \eq{b} and centered along \eq{\vsp{N}} (where we might take \eq{\vsp{N}} to be the ``\eq{z}-axis''). }).
Unlike the hyperplanes \eq{\Sig_\ii{b}}, the hypersurfaces \eq{\man{Q}_\ii{b}} are \textit{not} affine subspaces of \eq{\vsp{E}} and different (co)tangent spaces are \textit{not} trivially isomorphic. The distinction between some \eq{\tsp[\barpt{q}]\man{Q}_\ii{b}} and \eq{\tsp[\barpt{r}]\man{Q}_\ii{b}} is significant. 


\begin{small}
\begin{notesq}
\rmsb{An Important Modification.}
When the above hypersurfaces \eq{\Sig_\ii{b}} and \eq{\man{Q}_\ii{b}}  appear in this work, we often impose one further condition which was omitted in Eq.\eqref{3surf_0} – \ref{Qsurf_0}. For any positive \eq{b\in\mbb{R}_{+}}, we usually take \eq{\Sig_\ii{b}} and \eq{\man{Q}_\ii{b}} to be the following hypersurfaces (still of dimension  \eq{\en-1}):
\begin{small}
\begin{align} \label{3surf_3}
\begin{array}{llll}
     \Sig_\ii{b}  :=\, \inv{(r^\en)}\{b\} \cap  \bvsp{E}   \;=\, \bs{\Sigup}_\nozer \oplus \{b\} \subset \bvsp{E} 
 \\[2pt] 
     \man{Q}_\ii{b} :=\, \inv{\rfun}\{b\} \cap \bvsp{E}  \;=\, \man{S}^{\en-\ii{2}}_\ii{b}  \times \vsp{N}_\ii{+} \subset \bvsp{E}
\end{array}
&&,&&
  \bvsp{E} :=\, \bs{\Sigup}_\nozer \oplus \vsp{N}_\ii{+} 
     \,=\, \big\{ \barpt{x} = \ptvec{x} + x^\en\envec \in \vsp{E} \;\big|\;\,  \nrm{\pt{x}} \neq 0 \;\, , \;\,  x^\en >0  \big\}  \,\subset \Evec_\nozer \subset \vsp{E} 
\end{align}
\end{small}
These \eq{\Sig_\ii{b} } and   \eq{ \man{Q}_\ii{b}} are nearly the same as those defined in Eq.\eqref{3surf_0} and Eq.\eqref{Qsurf_0} except they now include the additional properties of \eq{\bvsp{E}} defined above. The (co)tangent spaces are still the same as in Eq.\eqref{3surf_0} and Eq.\eqref{Qsurf_0}.       
\end{notesq}
\end{small}

\subsubsection{Compendium of Notation} \label{sec:notation_prj}

Unless specified otherwise, the below notation applies for all developments in section \ref{sec:NOM} and subsequent sections: 
\begin{small}
\begin{itemize}[nosep]
     \item  \eq{\affE\equiv \affE^\en} denotes affine Euclidean \eq{\en}-space regarded as \eq{\affE=\Sig\times \aff{N}} (cf.~section \ref{sec:prj_prelim}). We almost always identify \eq{\affE} with its associated vector space, \eq{\vsp{E} =\bs{\Sigup}\oplus\vsp{N}}, with Euclidean metric/inner product \eq{\bsfb{\emet}_\ii{\!\Evec}\equiv\inner{\slot}{\slot}\in\botimes^0_2 \vsp{E}}. 
     Our indexing is as follows: 
     \begin{small}
    \begin{align}
        \vsp{E} = \bs{\Sigup}\oplus \vsp{N}
       \qquad,\qquad
           \a,\beta,\gam = 1,\dots, \en  = \dim \vsp{E}
        \qquad,\qquad
           i,j,k = 1,\dots,\en-1 = \dim \bs{\Sigup}
    \end{align}
     \end{small}
    \item For any \eq{\barpt{x}=\ptvec{x}+x^\en\envec\in\vsp{E}=\bs{\Sigup}\oplus\vsp{N}}, any \eq{\barbs{\kap}=\bs{\kap}+\bs{\kap}^\ii{\perp}=\bs{\kap}+\kap_\en\enform\in\cotsp[\cdt]\vsp{E}}, and any \eq{\barsfb{v}=\sfb{v}+\sfb{v}^\ii{\perp}=\sfb{v}+\v^\en \envec\in\tsp[\cdt]\vsp{E}}, we use the metric ``musical isomorphism'' given by \eq{\bsfb{\emet}_\ii{\!\Evec}} (and \eq{\sfb{\emet}_\ii{\!\Sigup}}) to define the following for \eq{\ptvec{x}\in\bs{\Sigup}},   \eq{\bs{\kap}\in\cotsp[\cdt]\bs{\Sigup}},  and  
   \eq{\sfb{v}\in\tsp[\cdt]\bs{\Sigup}}:\footnote{The first line in Eq.\eqref{metric_music_prj} is only ``allowed'' since the base configuration manifold is a vector space and, in the second line, we are regarding \eq{\bsfb{\emet}_\ii{\!\Evec}} as a (homogeneous) tensor field, \eq{\bsfb{\emet}_\ii{\!\Evec}\in\tens^0_2(\vsp{E})}. }
    \begin{small}
    \begin{align} \label{metric_music_prj}
     \begin{array}{rlllll}
        \nrm{\pt{x}}^2 \equiv \!\!\!\! & \nrm{\ptvec{x}}^2 := \bsfb{\emet}_\ii{\!\Evec}(\ptvec{x},\ptvec{x})
        &,\quad  \hpt{x} := \ptvec{x}/\nrm{\pt{x}}
        &,\quad  \ptvec{x}\,^\flt := \bsfb{\emet}_\ii{\!\Evec}(\ptvec{x}) 
        &,\quad  \hpt{x}\,^\flt :=  \bsfb{\emet}_\ii{\!\Evec}(\hpt{x}) = \ptvec{x}\,^\flt /\nrm{\pt{x}}
      \\[2pt] 
        &\nrm{\bs{\kap}}^2 := \inv{\bsfb{\emet}_\ii{\!\Evec}}(\bs{\kap},\bs{\kap})
         &,\quad  \hbs{\kap} := \bs{\kap}/\nrm{\bs{\kap}}
          &,\quad  \bs{\kap}^\shrp :=\inv{\bsfb{\emet}_\ii{\!\Evec}}(\bs{\kap}) 
           &,\quad  \hbs{\kap}^\shrp :=\inv{\bsfb{\emet}_\ii{\!\Evec}}(\hbs{\kap}) = \bs{\kap}^\shrp/ \nrm{\bs{\kap}}
       \\[2pt] 
         &\nrm{\sfb{v}}^2 := \bsfb{\emet}_\ii{\!\Evec}(\sfb{v},\sfb{v})
         &,\quad  \hsfb{v} := \sfb{v}/\nrm{\sfb{v}}
          &,\quad  \sfb{v}^\flt :=\bsfb{\emet}_\ii{\!\Evec}(\sfb{v}) 
           &,\quad  \hsfb{v}^\flt := \bsfb{\emet}_\ii{\!\Evec}(\hsfb{v}) = \sfb{v}^\flt / \nrm{\sfb{v}}
    \end{array}
    \end{align}
    \end{small}
    (we write \eq{\nrm{\pt{x}}^2} rather than \eq{\nrm{\ptvec{x}}^2} to avoid confusion with \eq{\nrm{\barpt{x}}^2=\bsfb{\emet}_\ii{\!\Evec}(\barpt{x},\barpt{x})}). 
   Everything above is defined in the same manner for the ``full'' (co)vectors \eq{\barpt{x}}, \eq{\barbs{\kap}}, and \eq{\bsfb{v}}. Yet, we will make more frequent use of the above.  
   All of this also extends to (co)vector \textit{fields} in the usual way. 
     \item Let \eq{(\bartup{r},\bartup{\plin})=(\tup{r},r^\en,\tup{\plin},\plin_\en):\cotsp\vsp{E}\to\mbb{R}^{\ii{2}\en}} be cotangent-lifted linear coordinates, where \eq{(\tup{r},\tup{\plin}):\cotsp\bs{\Sigup}\to\mbb{R}^{\ii{2}(\en-\ii{1})}}, and let \eq{\emet_{\a\b}:=\bsfb{\emet}_\ii{\!\Evec}(\be[r^\a],\be[r^\b])} be such that \eq{\emet_{i\en}=\emet^{i\en}=0}. We then define the following 
    (in analogy to Eq.\eqref{metric_music_prj}):\footnote{The first line in Eq.\eqref{cord_norm} only makes sense when \eq{r^\a} are linear coordinates. Furthermore, all relations in Eq.\eqref{cord_norm} assume that \eq{\emet_{i\en}=\emet^{i\en}=0}. For instance, \eq{\plin^i} and \eq{\plin^\en} would generally need to be defined as \eq{\plin^i:=\emet^{i\a} \plin_\a} and \eq{\plin^\en:=\emet^{\en \a} \plin_\a} (similarly for \eq{r_i}). This simplifies to the relations in Eq.\eqref{cord_norm} when \eq{\emet_{i\en}=\emet^{i\en}=0}.}
    \begin{small}
    \begin{align} \label{cord_norm}
    \begin{array}{lllll}
        \nrmtup{r}^2 \equiv \nrm{\tup{r}}^2 := \emet_{ij}r^i r^j
        &,\quad  \hat{r}^i := r^i/\nrmtup{r}
        &,\quad   r_i := \emet_{ij}r^j 
        &,\quad  \hat{r}_i := r_i/\nrmtup{r} = \emet_{ij}\hat{r}^j
      \\[2pt] 
        \nrmtup{\plin}^2 \equiv \nrm{\tup{\plin}}^2 := \emet^{ij}\plin_i \plin_j
         &,\quad  \hat{\plin}_i := \plin_i/\nrmtup{\plin}
          &,\quad  \plin^i := \emet^{ij} \plin_j 
           &,\quad  \hat{\plin}^i := \plin^i/\nrmtup{\plin} = \emet^{ij}\hat{\plin}_j
    \end{array}
    &&
    \begin{array}{lllll}
         r_\en = \emet_{\en \en} r^\en 
      \\[2pt] 
        \plin^\en = \emet^{\en \en} \plin_\en 
    \end{array}
    \end{align}
    \end{small} 
    with \eq{\htup{r}} and \eq{\htup{\plin}} defined analogously.\footnote{In matrix notation: for \eq{(\bartup{r},\bartup{\plin})=(\tup{r},r^\en,\tup{\plin},\plin_\en)} we likewise define the following for the \eq{(\en-1)}-tuples \eq{\tup{r}=(r^1,\dots,r^{\en-\ii{1}})} and  \eq{\tup{\plin}=(\plin_1,\dots,\plin_{\en-\ii{1}})}:
     \begin{align} \label{cord_norm0}
     \begin{array}{lllll}
           \htup{r} := \tup{r} / \nrmtup{r} 
         &,\qquad 
              \tup{r}^\flt := \cord{\bar{\emet}}{\bartup{r}}\cdot \tup{r} = [r_i]  \cong \trn{\tup{r}}
            &,\qquad 
          \tup{r}\otms \tup{r}^\flt = [r^i r_j] 
          \cong \tup{r}\trn{\tup{r}}
      \\[2pt] 
          \htup{\plin} := \tup{\plin} / \nrmtup{\plin} 
         &,\qquad 
              \tup{\plin}^\shrp := \cord{\inv{\bar{\emet}}}{\bartup{r}}\cdot \tup{\plin} = [\plin^i]  
            &,\qquad 
          \tup{\plin}^\shrp\otms \tup{\plin} = [\plin^i \plin_j] 
     \end{array}
      \end{align}
      If \eq{\tup{r}=(r^1,\dots,r^{\en-\ii{1}})} is viewed as a column vector (of functions), then \eq{\tup{r}^\flt=(r_1,\dots,r_{\en-\ii{1}})\cong\trn{\tup{r}}} would be viewed as a row vector (of functions). A similar analogy can be said for \eq{\tup{\plin}} but it depends on if we wish to consider \eq{\tup{\plin}} as row or column vector to begin with. Some sources view \eq{\tup{\plin}} as a \textit{row} vector, in which case \eq{\tup{\plin}^\shrp} would be viewed as a column vector. } 
     We stress that \eq{r^\en \neq \nrm{\tup{r}}^\en} is simply the \eq{\en^{\mrm{th}}} coordinate and \textit{never} means exponentiation. Additionally, we will frequently need to write exponentials of \eq{r^\en}, say \eq{(r^\en)^2}, and we will write this as \eq{r_\en^2} which does \textit{not} mean \eq{(\emet_{\en\en}r^\en)^2}:
    \begin{small}
    \begin{align} \label{xn_down}
        r_\en^2 \,:=\, (r^\en)^2
    \end{align}
    \end{small}
    \item  From now on,  \eq{\bartup{r}=(\tup{r},r^\en)=(r^1,\dots,r^{\en-\ii{1}},r^\en):\vsp{E} \to\mbb{R}^{\en} } will, unless states otherwise, always denote global
    \textit{cartesian}\footnote{In this work, ``cartesian'' implies not only linearity but also orthonormality. That is, to say that \eq{r^\a} are cartesian coordinates is to say that \eq{\bsfb{\emet}_\ii{\!\Evec}(\hbe_\a,\hbe_\b)=\kd_{\a\b}} (where \eq{\hbe_\a\equiv \be[r^\a]}). 
    All coordinates are assumed time-independent unless specified otherwise.}
    coordinates for a fixed \eq{\bsfb{\emet}_\ii{\!\Evec}}-orthonormal basis \eq{\hbe_\a\in\vsp{E}} 
    as previously described. The notation in Eq.\eqref{cord_norm} still applies but now simply with \eq{\emet_{\a\b}=\kd_{\a\b}} such that the distinction between \eq{r^\a} and \eq{r_\a}, and between \eq{\plin_\a} and \eq{\plin^\a}, is inconsequential (for computational purposes):
    \begin{small}
    \begin{align}
        \fnsize{for cartesian } r^\a: 
        \qquad\qquad
        \emet_{\a\b} = \emet^{\a\b} = \kd^\a_\b 
        \qquad,\qquad
        r_\a = r^\a 
        \qquad,\qquad
        \plin^\a = \plin_\a 
         \qquad,\qquad
         r := \nrm{\sfb{r}} = \nrm{\tup{r}} 
    \end{align}
    \end{small}
    \begin{small}
    \begin{itemize}[nosep]
        \item[!]  Thus, for any equations expressed in terms of cartesian coordinates, one may safely choose to ignore any distinction between \eq{\emet_{\a\b}, \; \emet^{\a\b}}, and \eq{\kd^\a_\b} — as well the distinction between \eq{r^\a} and \eq{r_\a}, between \eq{\plin_\a} and \eq{\plin^\a}, or any other upper/lower index placement — at no risk to \textit{computational} accuracy. However, for \textit{conceptual} clarity, we will generally continue to make these distinctions. As such, although we often use cartesian coordinates, the majority of the equations in this section still hold in the more general case that \eq{r^\a} are non-cartesian, but still \textit{linear}, coordinates. 
        \item[!] However, we make an exception in the case of \eq{r^\en} which we will often write as \eq{r_\en} but which generally does \textit{not} mean \eq{\emet_{\en\a}r^\a = \emet_{\en\en}r^\en} (e.g., Eq.\eqref{xn_down}). When \eq{r^\a} are cartesian (almost always), the distinction is anyway \textit{numerically} inconsequential.  
    \end{itemize}
    \end{small}
    \item  The corresponding cotangent-lifted cartesian coordinates will be denoted \eq{(\bartup{r},\bartup{\plin})=\colift\bartup{r}:\cotsp\vsp{E} \to\mbb{R}^{2\en}}, with the base coordinates, \eq{\bartup{r}=(\tup{r},r^\en)}, and the fiber coordinates, \eq{\bartup{\plin}=(\tup{\plin},\plin_\en)}, each \eq{\en}-tuples of functions.  
    \textit{No other coordinates are introduced until section \ref{sec:prj_gen_passive}}. Until then, there is no danger of confusion with other coordinate systems and we will often employ the usual abbreviated notation for partial derivatives:
    \begin{small}
    \begin{align}
      \qquad\qquad\qquad 
        \pd_\a := \pd_\ii{r^\a} \equiv \pderiv{}{r^\a} 
        \qquad,\qquad 
         \upd^\a := \pd_\ii{\plin_\a} \equiv \pderiv{}{\plin_\a} 
         \qquad\qquad\qquad \fnsize{also:} \;\;  \pd_{\rfun} \equiv \pderiv{} {r}
    \end{align}
    \end{small}
    \item The \eq{r^\a} frame fields on \eq{\vsp{E}} are  homogeneous and will again be written as \eq{\hbe_\a \equiv \be[r^\a]\in\vect(\Evec)} and \eq{\hbep^\a \equiv \bep[r^\a]\in\forms(\Evec)}, as in Eq.\eqref{ibase_same}. For the cotangent-lifted cartesian coordinates \eq{(r^\a,\plin_\a)}, the corresponding frame 
    fields are denoted \eq{\hbpart{\a}\equiv \hpdii{r^\a},\hbpartup{\a}\equiv\hpdiiup{\plin_\a}\in\vect(\cotsp\vsp{E})} and \eq{\hbdel^\a\equiv \hbdel[r^\a],\hbdel_\a\equiv\hbdeldn[\plin_\a]\in\forms(\cotsp\vsp{E})}. They are likewise homogeneous and can be identified with an inertial basis for \eq{\vsp{E} \oplus \Evec^{*}}  as in Eq.\eqref{T*V_ibase_proj}.
    \item  We will use the Hamiltonian formulation on \eq{(\cotsp\Evec,\nbs{\omg})} with \eq{\nbs{\omg}=-\exd\bs{\theta}\in\formsex^2(\cotsp\Evec)} the canonical symplectic form and \eq{\bs{\theta}\in\formsh(\cotsp\Evec)} the canonical 1-form.  Like any cotangent-lifted coordinate frame fields, the frame fields for \eq{(\bartup{r},\bartup{\plin})} satisfy  \eq{\bs{\theta}= \plin_\a \hbdel^\a} and are \eq{\nbs{\omg}}-symplectic coordinates (i.e., canonical coordinates) as in Eq.\eqref{E4_sp_cords}. 
\end{itemize}
\end{small}

\begin{small}
\begin{notesq}
    \rmsb{A note on mass and the metric ``musical isomorphism''.} 
    We use the standard Euclidean metric/inner product, \eq{\bsfb{\emet}_\ii{\!\Evec}}, to define a metric musical isomorphism in the usual way (Eq.\eqref{metric_music_prj}-Eq.\eqref{cord_norm}). However, generally, configuration manifolds are not vector spaces such that \eq{\bsfb{\emet}_\ii{\!\Evec}} does not exist and it is instead the \textit{kinetic energy metric} which defines the metric musical isomorphism. For the present case of a 1-particle system of mass \eq{m} in \eq{\vsp{E}}, the  kinetic energy metric is simply \eq{\sfb{m}:=m\bsfb{\emet}_\ii{\!\Evec}}. The ``correct'' metric musical isomorphism, which generalizes to any mechanical system, should then be defined using \eq{\sfb{m}} (e.g., \eq{\bs{\kap}^\shrp:=\inv{\sfb{m}}(\bs{\kap})}  and \eq{\plin^i:=m^{ij} \plin_j} are more ``correct'' than our chosen \eq{\bs{\kap}^\shrp:=\inv{\bsfb{\emet}_\ii{\!\Evec}}(\bs{\kap})} and \eq{\plin^i:=\emet^{ij} \plin_j}). 
    Yet, in the present case, for relations at the configuration level, it is more useful to use \eq{\bsfb{\emet}_\ii{\!\Evec}} (e.g., \eq{\nrm{\pt{x}}^2:=\bsfb{\emet}_\ii{\!\Evec}(\ptvec{x},\ptvec{x})} is more useful than defining  \eq{\nrm{\pt{x}}^2:= \sfb{m}(\ptvec{x},\ptvec{x})}). 
    Rather than deal with the isomorphisms and inner products of two different metrics (which only differ by a constant mass scaling), we have elected to simply stick with \eq{\bsfb{\emet}_\ii{\!\Evec}} for the musical isomorphism and the inner product. As a consequence, we must deal with the additional factor of \eq{m} independently. Alternatively, for a single-particle system, it is common to "scale units such that \eq{m=1}" or to simply factor \eq{m} out of the Hamiltonian to begin with such that all quantities are expressed per unit mass; this would make \eq{\sfb{m}} and \eq{\bsfb{\emet}_\ii{\!\Evec}} equivalent. 
\end{notesq}
\end{small}

\subsection{The Original Hamiltonian System (with Redundant Dimensions)} \label{sec:Hnom_prj}

\paragraph{Hamiltonian Dynamics on $\cotsp\bs{\Sigup}$ Formulated on $\cotsp\vsp{E}$.}  
As described at the start of section \ref{sec:NOM}, the system in which we are interested is simply a Newtonian system of a particle of mass \eq{m} moving in Euclidean \eq{(\en-1)}-space, \eq{(\bs{\Sigup}, \sfb{\emet}_\ii{\!\Sigup})}, subject to arbitrary conservative forces (non-conservative forces are considered later) corresponding to some potential \eq{V\in\fun(\bs{\Sigup})}. However, we will instead view the particle as moving in Euclidean \eq{\en}-space, \eq{(\vsp{E},\bsfb{\emet}_\ii{\!\Evec})} — viewed as \eq{\vsp{E} =\bs{\Sigup}\oplus\vsp{N}} — but still subject to the same forces (i.e., the same potential \eq{V} but regarded as \eq{V\in\fun(\Evec)}). \textit{Our reason for introducing this redundant-dimensional formulation is simply to facilitate our subsequent use of the projective transformation  in later sections}. Here, we merely describe the original Newtonian system on \eq{\vsp{E}}, which we formulate as a Hamiltonian system on the canonically symplectic cotangent bundle, \eq{(\cotsp\Evec,\nbs{\omg}, K )}. Thus, our original system is described by a mechanical Hamiltonian \eq{K \in\fun(\cotsp\Evec)}, given as follows for arbitrary \eq{\bar{\kap}_{\barpt{x}}=(\barpt{x},\barbs{\kap}) \in\cotsp\vsp{E}}:
\begin{small}
\begin{align} \label{Kprj_4dim} 
\forall \,  \bar{\kap}_{\barpt{x}}\in\cotsp\vsp{E}:
\qquad 
    K(\bar{\kap}_{\barpt{x}}) \,=\, \tfrac{1}{2}\inv{\sfb{m}}(\barbs{\kap},\barbs{\kap}) \,+\, V(\ptvec{x})
     \qquad,\qquad 
    \sfb{m} := m \bsfb{\emet}_\ii{\!\Evec}
\end{align}
\end{small}
where \eq{\sfb{m}\in\tens^0_2(\Evec)} is the (homogeneous) Euclidean kinetic energy metric. 
Let  \eq{(\bartup{r},\bartup{\plin})=(\tup{r},r^\en,\tup{\plin},\plin_\en)} be any cotangent-lifted 
cartesian\footnote{In this work, ``cartesian'' implies not only linearity but also orthonormality. All coordinates are assumed  time-independent unless specified otherwise.}
coordinates as previously described such that \eq{\sfb{m}} has components in the \eq{r^\a} frame fields given by \eq{m_{\a\b} = m \emet_{\a\b} = m\kd_{\a\b}} and \eq{m^{\a\b} = (1/m) \emet^{\a\b}=(1/m)\kd^{\a\b}}. The cartesian coordinate expression for the original Hamiltonian function is simply
\begin{small}
\begin{align}
     K \,=\, \tfrac{1}{2} m^{\a\b} \plin_\a \plin_\b \,+\, V
    \;=\;  \tfrac{1}{2m}( \nrmtup{\plin}^2 + \plin_\en^2 )   \,+\, V 
    \qquad,\qquad V = V^\zr(\rfun) + V^\ss{1}(\tup{r})
\end{align}
\end{small}
where the expression  \eq{V = V^\zr(\rfun) + V^\ss{1}(\tup{r})} is an abuse of notation meant only to convey the following important conditions imposed on the potential function, \eq{V}:

\begin{small}
\begin{remrm} \label{rem:U4}
 We impose the following properties on the original potential, \eq{V\in\fun(\vsp{E})}:
 \begin{small}
 \begin{enumerate}[nosep]
     \item  We insist that \eq{\pd_\en V =0}. That is, \eq{V} is really some \eq{V\in\fun(\bs{\Sigup})} which we take to mean \eq{V\equiv \prj_\ii{\Sig}^*V\in\fun(\vsp{E})}. In other words, \eq{V=V(\tup{r})} depends only on the ``\eq{\bs{\Sigup}}-part'' of any \eq{\barpt{x}=\ptvec{x}+x^\en\envec\in\vsp{E}=\bs{\Sigup}\oplus\vsp{N}}. 
     Regardless, the point is simply:\footnote{As previously described, the conservative forces are described by the exact 1-form, \eq{-\dif V=-\pd_i V \hbep^i \in\formsex(\vsp{E})}, with \eq{\hbep^i=\dif r^i}. Yet, as usual, when \eq{V} appears in the Hamiltonian function or corresponding dynamics, it is regarded as a basic function \eq{V\equiv\copr^*V\in\fun(\cotsp\Evec)} such that \eq{\upd^\a V =0} and, similarly, \eq{\dif V} is regarded as a basic 1-form \eq{\dif V \equiv \copr^* \dif V = \pd_i V \hbdel^i} (where \eq{\hbdel^i = \copr^* \hbep^i \cong \hbep^i \oplus \bsfb{0}}). That is, for the dynamics formulated on phase space, conservative forces are described by the exact basic horizontal 1-form, \eq{-\dif V \equiv - \copr^*\dif V\in\formsbh\cap\formsex(\cotsp\vsp{E})} which we insist satisfy \eq{\pd_\en V = \dif V \cdot \hbpart{\en} = 0}. }
     \begin{small}
     \begin{align} \label{U4}
          \pd_\en V  =  \dif V \cdot \envec  = 0
        &&
          \fnsize{i.e.,} \quad \dif V = \pd_i V   \dif r^i = \pd_i V  \hbep^i \in\formsex(\vsp{E})
      \quad,\quad
            \dif V \equiv \copr^* \dif V = \pd_i V \hbdel^i  \in\formsbh\cap\formsex(\cotsp\vsp{E})
     \end{align}
     \end{small}  
      \item  We further assume that \eq{V = V^\zr(\rfun) + V^\ss{1}(\tup{r})} where \eq{\rfun =\nrm{\tup{r}}} is the norm on \eq{\bs{\Sigup}\subset\vsp{E}} such that
      \eq{V^\zr} accounts for all central forces\footnote{Later, \eq{V^\zr} will be the Kepler-Coulomb potential, \eq{V^\zr = -\sck/\rfun}, for positive \eq{\sck\in\mbb{R}_\ii{+}}. }
      in  \eq{\bs{\Sigup}}, and where \eq{V^\ss{1}} is arbitrary (though still subject to the above), accounting for all other conservative forces. Thus, with \eq{\hat{r}_i:=r_i/\rfun = \pd_i r }:\footnote{The expression \eq{\dif V^\zr = \pd_{\rfun} V^\zr \hsfb{r}^\flt} uses \eq{\pd_i r = \emet_{ij}r^j/\rfun = \hat{r}_i} and \eq{\dif r = \hsfb{r}^\flt} to obtain \eq{ \dif V^\zr = \pd_i V^\zr \hbep^i =  \pd_{\rfun} V^\zr \pd_i r  \hbep^i =   \pd_{\rfun} V^\zr   \hat{r}_i \hbep^i =  \pd_{\rfun} V^\zr  \dif r =  \pd_{\rfun} V^\zr \hsfb{r}^\flt}. }
     \begin{small}
     \begin{align} \label{U4_r}
           V = V^\zr(\rfun) + V^\ss{1}(\tup{r})
         &&,&&
             \dif V^\zr \,=\, \pd_i V^\zr \hbep^i \,=\, 
              \pd_{\rfun} V^\zr \pd_i r  \hbep^i
             \,=\,  \pd_{\rfun} V^\zr \hsfb{r}^\flt
        &&,&&
             \pd_i V \,=\, \hat{r}_i  \pd_{\rfun} V^\zr + \pd_i V^\ss{1}
     \end{align}
     \end{small}
    \item Later, when the particular form of  \eq{V^{0}} becomes relevant in section \ref{sec:prj_regular}, then we consider the \textit{Manev potential}, given as follows for \eq{\sck_1,\sck_2\in\mbb{R}} (the Kepler potential is the special case \eq{\sck_1>0} and \eq{\sck_2=0}):\footnote{The negative signs and factor of \eq{\ttfrac{1}{2}} are included simply for convenience.}
        \begin{small}
        \begin{align} \label{Vmanev_intro}
            V^{0}   \,=\, -\tfrac{\sck_1}{\rfun} \,-\, \tfrac{1}{2} \tfrac{\sck_2}{\rfun^2}
            &&,&&
            \dif V^0 \,=\, (\tfrac{\sck_1}{\rfun^2} \,+\,  \tfrac{\sck_2}{\rfun^3}) \hsfb{r}^\flt
            &&,&&
            \begin{array}{cc}
                 \fnsize{Manev}  \\[-1pt]
                 \fnsize{potential}
            \end{array}
        \end{align}
        \end{small}
        
 \end{enumerate}
 \end{small}
\end{remrm}
\end{small}

As mentioned, for dynamics on \eq{\cotsp\vsp{E}}, we will re-use \eq{V} and \eq{\dif V} to mean the basic forms \eq{V\equiv \copr^* V\in\fun(\cotsp\vsp{E})} and  \eq{\dif V \equiv \copr^* \dif V\in\forms(\cotsp\vsp{E})}. 
The Hamiltonian in Eq.\eqref{Kprj_4dim}, with \eq{V} as above, then gives the following Hamiltonian vector field \eq{\sfb{X}^\ss{K} \in \vechm(\cotsp\Evec,\nbs{\omg})}:
\begin{small}
\begin{align}       
    \sfb{X}^\ss{K} :=\, \inv{\nbs{\omg}}(\dif K,\cdot) \,=\,  m^{\a\b} \plin_\b \hbpart{\a} \,-\, \pd_i V \hbpartup{i} \;=\; \tfrac{1}{m} \plin^\a \hbpart{\a} \,-\, \pd_i V \hbpartup{i} 
    && \Rightarrow &&
    \begin{array}{lllll}
         \dot{r}^i = \tfrac{1}{m} \plin^i 
         &,\quad \dot{\plin}_i = -\pd_i V
     \\[2pt] 
         \dot{r}^\en  = \tfrac{1}{m} \plin^\en 
         &,\quad \dot{\plin}_\en = 0
    \end{array}
\end{align}
\end{small}
where \eq{\plin^\a:=\emet^{\a\b} \plin_\b (\equiv \plin_\a)} and
where the ODEs on the right determine the (inertial cartesian coordinate representation of) integral curves of \eq{\sfb{X}^\ss{K}}. One does not need coordinates to realize that if \eq{\bar{\kap}_t=(\barpt{x}_t,\barbs{\kap}_t)\in\cotsp\vsp{E}} is an integral curve (i.e., \eq{\dt{\bar{\kap}}_t=\sfb{X}^\ss{K}_{\bar{\kap}_t}}) then \eq{\barbs{\kap}_t=\sfb{m}(\dt{\bsfb{x}}_t)\in\cotsp[\barpt{x}_t]\vsp{E}} is the (Euclidean) kinematic momentum covector along the base curve \eq{\barpt{x}\in\vsp{E}}:
\begin{small}
\begin{align}
    \fnsize{if} \;\; \dt{\bar{\kap}}_t = \sfb{X}^\ss{K}_{\bar{\kap}_t} \qquad \Rightarrow \qquad \barbs{\kap}_t \,=\, \sfb{m}_{\barpt{x}}(\dt{\bsfb{x}}_t) \,=\, m \bsfb{\emet}_\ii{\!\Evec}(\dt{\bsfb{x}}_t) \,=\, m\dt{\bsfb{x}}^\flt_t
\end{align}
\end{small}

\paragraph{Integrals of Motion \& Invariant Submanifolds.}
Recall the angular momentum functions,
\eq{\lang^{ij}=r^i \plin^j - \plin^i r^j\in\fun(\cotsp\vsp{E})}, with norm \eq{\lang^2=\nrmtup{r}^2 \nrmtup{\plin}^2 = (r^i \plin_i)^2}, as detailed at Eq.\eqref{angMoment_prj_def}.
We then note the following Poisson bracket expressions for the Hamiltonian function \eq{K = \tfrac{1}{2}m^{\a\b} \plin_\a \plin_\b + V^\zr(\rfun) + V^\ss{1}(\tup{r})}: 
\begin{small}
\begin{align} \label{pbraks_nom}
\begin{array}{lllll}
      \pbrak{\plin_\en}{K} = 0
\\[2pt] 
      \pbrak{r^\en}{K} \,=\, m^{\en \a} \plin_\a = \tfrac{1}{m} \plin^\en 
\end{array}
&&
  \pbrak{K}{K} \,=\, 0
&&
\begin{array}{lllll}
      \pbrak{ \lang^{ij} }{K} \,=\, -(r^i \emet^{jk} - r^j \emet^{ik})\pd_k V^\ss{1}  
\\[2pt] 
      \pbrak{ \ttfrac{1}{2}\lang^2 }{K}
      \,=\, - (\nrmtup{r}^2 \emet^{ij} - r^i r^j)\plin_i \pd_j V^\ss{1}
      \,=\, -\lang^{ij} r_i \pd_j V^\ss{1}
\end{array}
\end{align}
\end{small}
As for any Hamiltonian system, \eq{K} is an integral of motion of \eq{\sfb{X}^\ss{K}} iff \eq{\pd_t K =0 } (which, for the present case, is true iff \eq{\pd_t V =0 }). 
Note that if \eq{V^\ss{1}=0} then  \eq{ \pbrak{ \lang^{ij} }{K} =  \pbrak{ \ttfrac{1}{2}\lang^2 }{K} =0 }, verifying that angular momentum is conserved in the case that only central forces act on the system (i.e., in the case \eq{V=V^\zr(\rfun)}). 
Here, we will focus only on the above brackets for \eq{r^\en} and \eq{\plin_\en}. 
We see immediately that \eq{r^\en} is ignorable/cyclic  (\eq{\pd_\en K = 0}) such that \eq{\plin_\en} is an integral of motion (as is \eq{\plin^{\en}=\emet^{\en\en} \plin_\en \equiv \plin_\en}).\footnote{More precisely, \eq{\plin_\en(\bar{\kap}_t)= \plin_\en(\bar{\kap}_{\zr})} for any integral curve \eq{\bar{\kap}_t\in\cotsp\vsp{E}} of \eq{\sfb{X}^\ss{K}}. Equivalently, \eq{\varphi_t^* \plin_\en = \plin_\en} where \eq{\varphi_t} is the flow of \eq{\sfb{X}^\ss{K}}.   
}
As such, the ``\eq{r^\en} part'' of the solution is given simply by  \eq{r^\en_t = r^\en_0 + (\plin^\en_0/m) t}. That is,
\begin{small}
\begin{align} \label{pbraks_xpN_nom}
\begin{array}{lllll}
       \dot{\plin}_\en = \pbrak{\plin_\en}{K} = 0
        &\qquad \Rightarrow \qquad 
       \plin_\en(t) = \plin_\en(0) 
\\[2pt] 
      \dot{r}^\en =  \pbrak{r^\en}{K} \,=\, \tfrac{1}{m} \plin^\en 
    &\qquad \Rightarrow \qquad 
    r^\en(t) \,=\, r^\en(0) + \tfrac{\plin^\en(0)}{m} t
\end{array}
\end{align}
\end{small}
Thus, our system (a particle) in \eq{\vsp{E}} has the feature that it moves at constant speed \eq{\plin^\en(0)/m} along the \eq{\envec} direction (i.e., along the ``\eq{r^\en}-axis'', that is, along \eq{\vsp{N}\perp \bs{\Sigup}}). But we do not actually care about the motion in this extra dimension. We are free to assume the initial condition \eq{\plin_\en(0) = 0} such that \eq{\plin_\en(t)=0} for all time and, thus,  \eq{r^\en(t)=r^\en(0)} is also constant for all time. That is,  the particle moves in the (\eq{\en-1})-dim hyperplane, \eq{\Sig_\ii{r^\en(0)}=\bs{\Sigup}\oplus \{r^\en(0) \} \subset \vsp{E}}, that is normal to \eq{\envec}. This hyperplane is determined by the initial condition \eq{r^\en(0)} which, like \eq{\plin_\en(0)}, we are free to choose however we please (we do not care about motion in the  extra \eq{\en^\tx{th}} direction).  This is summarized in Remark \ref{rem:E3dyn_subman} below for which we recall  the hyperplane \eq{\Sig_\ii{b}\subset\vsp{E}} defined in 
Eq.\eqref{3surf_0} – \ref{3surf_3} for any \eq{b\in\mbb{R}} as follows (in the present context, \eq{b} corresponds to \eq{r^\en_0}):
 \begin{small}
    \begin{align}
          \Sig_\ii{b}  := \inv{(r^\en)}\{b\} 
    \qquad,\qquad 
          \cotsp[\cdt] \Sig_\ii{b} = \ker \envec 
        \qquad,\qquad 
         \cotsp \Sig_\ii{b} = \big\{ \bar{\kap}_{\barpt{x}}=(\barpt{x},\barbs{\kap})\in\cotsp\vsp{E} \;\big|\; 
         r^\en(\barpt{x}) = x^\en = b \;, \; \plin_\en(\bar{\kap}_{\barpt{x}}) = \envec(\barbs{\kap}) = \kap_\en =0  \big\}
    \end{align}
    \end{small}     

\begin{small}
\begin{remrm} \label{rem:E3dyn_subman}
     For any \eq{b\in\mbb{R}}, the \eq{(2\en -2)}-dim \eq{\cotsp\Sig_\ii{b}\subset \tsp^* \vsp{E}} is an \eq{\sfb{X}^\ss{K}}-invariant submanifold; integral curves that start on  \eq{\cotsp\Sig_\ii{b}} 
     remain on \eq{\cotsp\Sig_\ii{b}}.    
     More precisely, if \eq{\bar{\kap}_t=(\barpt{x}_t,\barbs{\kap}_t)} satisfies \eq{\dt{\bar{\kap}}_t = \sfb{X}^\ss{K}_{\bar{\kap}_t}} and if \eq{\bar{\kap}_{\zr}\in\cotsp\Sig_\ii{x^\en(0)}} — that is, if \eq{\plin_\en(\bar{\kap}_{\zr}) = \kap_\en(0)=0} — then \eq{\bar{\kap}_t\in\cotsp\Sig_\ii{x^\en(0)}} for all \eq{t}  (here, \eq{x^\en(0):=r^\en(\bar{\kap}_{\zr})\equiv r^\en(\barpt{x}_{\zr})}). 
        Although this holds for any value \eq{x^\en(0)}, later developments will require \eq{x^\en(0)\neq 0} and it will be most convenient to limit consideration to \eq{x^\en(0) =1}. 
\end{remrm}
\end{small}

\noindent Thus, if we let \eq{\mscr{K}} denote the restriction of \eq{K\in\fun(\cotsp\Evec)} to the hypersurface in \eq{\cotsp\vsp{E}} defined by \eq{\plin_\en = 0}, then the original system simplifies to the following 
usual case of a particle in Euclidean \eq{(\en-1)}-space:
\begin{small}
\begin{align} \label{Kprj_3dim_cart}
\begin{array}{lllll}
   &   \mscr{K} := K|_{\plin_\en=0}
    \,=\, 
      \tfrac{1}{2}m^{ij}\plin_i \plin_j \,+\, V^\zr(\rfun) + V^\ss{1}(\tup{r})
\\[2pt] 
      &\sfb{X}^\sscr{K} \,=\, \tfrac{1}{m} \plin^i \hbpart{i} \,-\, \pd_i V \hbpartup{i}
\end{array}
     && \Rightarrow &&
    \begin{array}{lllll}
         \dot{r}^i =  \tfrac{1}{m} \plin^i   &,\quad \dot{\plin}_i = -\pd_i V \,=\, -(\hat{r}_i \pd_{\rfun} V^\zr + \pd_i V^\ss{1})
    \end{array}
\end{align}
\end{small}
with \eq{r^\en} and \eq{\plin_\en} now irrelevant, having trivial solutions \eq{\plin_\en(t)=0} and \eq{r^\en(t) = r^\en(0)}. That is, integral curves of the above \eq{\sfb{X}^\sscr{K}} always lie in one of the \eq{(2\en-2)}-dim invariant submanifolds \eq{\cotsp\Sig_\ii{b}}, with \eq{b} determined by \eq{r^\en(0)}  (we can, and will, limit consideration to \eq{b =r^\en(0)=1}). Further,  \eq{\cotsp\Sig_\ii{b} \cong \cotsp\bs{\Sigup}} can easily be identified with the phase space  \eq{\cotsp\bs{\Sigup}} of the actual \eq{(\en-1)}-\sc{dof} system in which we are interested, 
as described in Eq.\eqref{Hnom_actual}.\footnote{Recall that \eq{\Sig_\ii{b}=\bs{\Sigup}+b\envec} is just a parallel copy of \eq{\bs{\Sigup}} that has been translated some distance \eq{b} along \eq{\envec}. That is, \eq{\bs{\Sigup}\perp \vsp{N}} intersects \eq{\vsp{N}} at \eq{r^\en=0} whereas \eq{\Sig_\ii{b}\perp \vsp{N}} intersects \eq{\vsp{N}} at \eq{r^\en= b}.}
The above \eq{\mscr{K}} is the same as the ``actual'' Hamiltonian in Eq.\eqref{Hnom_actual}.

\paragraph{Example:~2-Dim Kepler Problem in $\Evec^3$.} 
Let us pretend, for visual purposes, that we live in a 2-dim Euclidean universe, \eq{\bs{\Sigup}} (regarded as a vector space, for simplicity). The Kepler potential has the form \eq{V^\zr = - \sck/\rfun\in\fun(\bs{\Sigup})} where \eq{\sck\in\mbb{R}} and where  
\eq{\rfun =\sqrt{x^2 + y^2}} is the norm function and \eq{\tup{r}=(x,y)} are cartesian coordinates on \eq{\bs{\Sigup}}. The actual Hamiltonian and equations of motion for this problem are given in cartesian coordinates \eq{(\tup{r},\tup{\plin})=(x,y,\plin_x,\plin_y)} as follows (for \eq{i=1,2}): 
\begin{small}
\begin{align} \label{2dkep_in2}
    \mscr{K} = \tfrac{1}{2m}(\plin_x^2 + \plin_y^2) -\tfrac{\sck} {r}
    \qquad\quad,\qquad\quad  
    \dot{x}^i = \tfrac{1}{m} \plin^i \quad,\quad \dot{\plin}_i = -\tfrac{\sck}{r^3}r_i
\end{align}
\end{small}
(the above has already reduced the nominally \eq{4}-\sc{dof} problem to a \eq{2}-\sc{dof} problem with \eq{m} the reduced mass). 
However, we instead view our 2-dim universe, \eq{\bs{\Sigup}},  as a plane in 
\eq{\vsp{E} \equiv\Evec^3}. We then think of \eq{\vsp{E}} as \eq{\vsp{E} =\bs{\Sigup} \oplus \vsp{N}} where \eq{\bs{\Sigup}} can be seen as some ``\eq{xy}-plane'' and \eq{\vsp{N}} as the ``\eq{z}-axis'' (where \eq{z} corresponds to \eq{r^\en} in the preceding developments). The 2-dim Kepler problem is then viewed in \eq{\vsp{E}} with a new Hamiltonian and equations of motion expressed in  cartesian coordinates \eq{(\bartup{r},\bartup{\plin})=(x,y,z,\plin_x,\plin_y,\plin_z)} as follows (still for \eq{i=1,2}): 
\begin{small}
\begin{align} \label{2dkep_in3}
    K = \tfrac{1}{2m}(\plin_x^2 + \plin_y^2 + \plin_z^2) -\tfrac{\sck} {r}
    \qquad\quad,\qquad\quad  
    \dot{r}^i = \tfrac{1}{m} \plin^i \quad,\quad \dot{\plin}_i = -\tfrac{\sck}{r^3}r_i
    \quad,\quad
    \dot{z} = \tfrac{1}{m} \plin^z
    \quad,\quad \dot{\plin}_z = 0
\end{align}
\end{small}
where \eq{V^\zr =-\sck/\rfun} is \textit{still the same potential}. That is, \eq{r} is still given by \eq{\rfun =\sqrt{x^2 + y^2}} such that it is now interpreted as a function on \eq{\vsp{E}} giving the perpendicular distance from the \eq{z}-axis (i.e., the usual meaning when using cylindrical coordinates). 
As such, it is important to note that the above is \textit{not} the classic 3-dim spherically symmetric Kepler problem! 
To clarify:
\begin{small}
\begin{itemize}[nosep]
    \item The Hamiltonian \eq{\mscr{K}\in\fun(\cotsp\bs{\Sigup})} describes the usual 2-dim circularly symmetric  Kepler problem of a particle in a 2-dim Euclidean universe moving in the gravitational field of an attracting point-mass located at the origin with mass proportional to \eq{\sck} (up to a scaling by constants). The system is unchanged by rotations in \eq{\bs{\Sigup}} (circular symmetry).  
    \item The Hamiltonian \eq{K\in\fun(\cotsp\Evec)} describes a fictitious 3-dim cylindrically symmetric problem of a particle in a 3-dim Euclidean universe moving in a gravity-like force field \eq{-\tfrac{\sck}{r^2}\hsfb{r}} where \eq{\hsfb{r}\perp\hbe[z]} is directed towards the ``\eq{z}-axis'' (the \eq{z}-axis being \eq{\vsp{N}}).   The system is now cylindrically symmetric: it is unchanged by rotations in \eq{\bs{\Sigup}} (i.e., about the ``\eq{z}-axis'') as well as translations along the ``\eq{z}-axis'' (i.e., along \eq{\vsp{N}}).
\end{itemize}
\end{small}
We reiterate that, in the present example of a 2-dim Euclidean universe, \eq{\bs{\Sigup}}, the true system in which we are interested is the ``full'' 2-dim Kepler problem described by \eq{\mscr{K}} (at no point in this example are we interested in anything about the 3-dim Kepler problem).  The above 3-dim system on \eq{\vsp{E}} described by \eq{K} is a fictitious system which is equivalent to the true system in the following ways:
\begin{small}
\begin{enumerate}
  \item When the the dynamics on \eq{\vsp{E}} are projected back down to the true configuration space, \eq{\bs{\Sigup}}, we recover our original 2-dim Kepler problem. In other words, the ``\eq{\bs{\Sigup}}-part'' of any solution to Eq.\eqref{2dkep_in3} is a solution to Eq.\eqref{2dkep_in2}. This is obvious from the fact that the ODEs for \eq{(x,y,\plin_x,\plin_y)} in Eq.\eqref{2dkep_in3} are identical to those in Eq.\eqref{2dkep_in2} and are uncoupled from the ODEs for \eq{(z,\plin_z)}.  
    \item The fictitious higher dimensional system in Eq.\eqref{2dkep_in3} has the following property: if we limit consideration of initial conditions in phase space to  only those \eq{\bar{\kap}_{\zr}=(\barpt{x}_{\zr},\barbs{\kap}_{\zr})\in\cotsp\vsp{E}} for which \eq{\plin_z(\bar{\kap}_{\zr})=0} — i.e., zero initial velocity along  the ``\eq{z} direction'', \eq{\vsp{N}} — then the resulting base integral curve, \eq{\barpt{x}_t\in\vsp{E}}, is automatically confined to the ``\eq{xy}-plane'' in which it starts. That is, if \eq{\barpt{x}_t} starts with  \eq{x^z_0:=z(\barpt{x}_{\zr})} then \eq{x^z_t = x^z_0} remains fixed such that the motion in \eq{\vsp{E}} remains in the plane \eq{\Sig_\ii{x^z_\zr} = \bs{\Sigup}\oplus \{x^z_\zr\}} (where we identify \eq{\bs{\Sigup}\equiv\Sig_\zr}). Moreover, this integral curve \eq{\barpt{x}_t=\ptvec{x}_t+x^z_0 \envec \in\Sig_\ii{x^z_0}} is such that \eq{\ptvec{x}_t} is a solution to our original 2-dim Kepler problem.
\end{enumerate}
\end{small}

\section{A BF-LIKE PROJECTIVE TRANSFORMATION:~GEOMETRIC APPROACH} \label{sec:prj_geomech}

\noindent We will specify a point transformation on \eq{\vsp{E}} that is analogous to the author's previous  ``projective coordinate transformation'' summarized in section \ref{sec:prj_sum2} of this work (which itself a modified version of the BF coordinate transformation \cite{ferrandiz1987general}). 
Although we will re-use the generic name ``projective transformation'', the following transformation is not the same as previous work in that (1) it is an ``active'' transformation (not a coordinate transformation) and (2) it is a (local) diffeomorphism (rather than a submersion). After defining the point transformation and obtaining its cotangent lift (i.e., the position and momentum transformations), we then use the developments of section \ref{sec:Hmech_xform_gen} to transform the original Hamiltonian mechanical system, \eq{(\cotsp\Evec,\nbs{\omg}, K )} described in section \ref{sec:Hnom_prj}, to a new Hamiltonian system, \eq{(\cotsp\Evec,\nbs{\omg}, H )}, which will, ultimately, have more desirable properties for certain central force dynamics. In particular, the new system will be linear for central force dynamics (Kepler and Manev dynamics), but not until the evolution parameter is transformed in section \ref{sec:prj_regular} using a Sundman-like transformation.


\subsection{The Projective Point Transformation \& (Co)Tangent-Lift} \label{sec:prj_gen_active}

Everything starts by specifying a point transformation at the configuration level (analogous to the projective transformation of Burdet, Sperling, Vitins, and Bond \cite{burdet1969mouvement,Burdet+1969+71+84,vitins1978keplerian,bond1985transformation}). We  can view the transformation in two different ways:
\begin{small}
\begin{enumerate}
    \item The projective transformation as an ``active'' diffeomorphism on the configuration manifold, \eq{\vsp{E}\to\vsp{E}}. The end result, after a cotangent lift, is a new Hamiltonian mechanical system which is \textit{different} from, but diffeomorphic to,  the original system detailed in section \ref{sec:Hnom_prj}. Many properties of the original system are transformed in such a way that the new system is ``better'' in some way (linearized, in our case). This is the view emphasized in the present work.
    \item The projective transformation as a ``passive'' coordinate transformation of configuration coordinates, \eq{\rchart{R}{q}^\en \to\rchart{R}{x}^\en}. The end result, after a cotangent lift, is a new set of symplectic (i.e., canonical) coordinates such that the \textit{same} mechanical Hamiltonian system we started with can be represented in a \textit{different} coordinate chart such that the new coordinate dynamics are ``better'' (more linear) than those of the original coordinates. No properties of the system are altered, only their coordinate representation. This was essentially the formulation taken by the authors in \cite{peterson2022nonminimal,peterson2025prjCoord,peterson2023regularized}, though using a more mathematically casual presentation. 
\end{enumerate}
\end{small}
With either approach, constructing the initial configuration space point transformation is the only part which requires some creativity. If the point transformation can be defined as a diffeomorphism (perhaps local), then everything else required to transform (actively or passively) some given, original, Hamiltonian system follows in a known, prescribed, manner (as detailed in section \ref{sec:Hmech_xform_gen}).

\begin{notesq}
    For reasons which will soon become clear, we will limit consideration to the region of \eq{\vsp{E} =\bs{\Sigup}\oplus\vsp{N}} defined by:\footnote{Since the norm is defined from a \textit{positive-definite} Euclidean metric, then \eq{\nrm{\pt{x}}\neq 0} implies \eq{\ptvec{x}\neq \ptvec{0}} and vice versa. Also, note that \eq{\nrm{\pt{x}} \neq 0} and \eq{x^\en \neq 0} together imply that \eq{\nrm{\barpt{x}}\neq 0 } (but \textit{not} vice versa). }
    \begin{small}
    \begin{align} \label{E4positive}
         \bvsp{E} :=\, \bs{\Sigup}_\nozer \oplus \vsp{N}_\ii{+} 
     \,=\, \big\{ \barpt{x} = \ptvec{x} + x^\en\envec \in \Evec^{\en} \;\big|\;\,  r(\ptvec{x}) = \nrm{\pt{x}} \neq 0  \;\, , \;\,  x^\en >0  \big\}  \,\subset\Evec_\nozer 
     &&,&&
     \dim \bvsp{E} = \dim \vsp{E} = \en 
    \end{align}
    \end{small}
\end{notesq}

\paragraph{The Projective Point Transformation.} In the following, \eq{\barpt{q},\barpt{x},\barpt{s}\in\bvsp{E}\subset\vsp{E}} are arbitrary displacement vectors with no particular significance (for now). We define a point transformation \eq{\psi:\bvsp{E} \to \bvsp{E}} by the following diffeomorphism (this is our new ``projective transformation''):
\begin{small}
\begin{align} \label{prj_def_act}
    \psi \in\Dfism(\bvsp{E}) 
     \qquad
    \boxed{\begin{array}{lclll}
            \barpt{q} = \ptvec{q} + q^\en \envec 
           &\mapsto & \;\psi(\barpt{q}) \,:=\, \tfrac{1}{q^\en \nrm{\pt{q}}} \ptvec{q} +\nrm{\pt{q}} \envec 
             \,=\, \tfrac{1}{q^\en } \hpt{q} +\nrm{\pt{q}} \envec 
        \\[8pt]
            \barpt{x} = \ptvec{x} + x^\en \envec  &\mapsto &  \inv{\psi}(\barpt{x}) 
        \,=\, \tfrac{x^\en}{\nrm{\pt{x}}} \ptvec{x} + \tfrac{1}{\nrm{\pt{x}}}\envec 
        \,=\,  x^\en\hpt{x} + \tfrac{1}{\nrm{\pt{x}}}\envec 
     \end{array} }
     &&
     \begin{array}{lllll}
         \nrm{\psi(\barpt{q})}^2 = \tfrac{1}{q_\en^2} + \nrm{\pt{q}}^2
     \\[8pt]
          \nrm{\inv{\psi}(\barpt{x})}^2 =  x_\en^2 + \tfrac{1}{\nrm{\pt{x}}^2} 
    \end{array}
\end{align}
\end{small}
where \eq{ \inv{\psi}\circ\psi = \psi \circ  \inv{\psi} = \Id_\ii{\bvsp{E}}}.
For the inverse, note that if \eq{\barpt{x}=\psi(\barpt{q})} then  \eq{\ptvec{x}=\ptvec{q}/(q^\en \nrm{\pt{q}}) = \hpt{q}/q^\en} and \eq{x^\en = \nrm{\pt{q}}}. If \eq{q^\en >0}, this can be
inverted\footnote{For \eq{\barpt{x}=\psi(\barpt{q})} then \eq{x^\en = \nrm{\pt{q}}} and \eq{\ptvec{x}=\ptvec{q}/(q^\en \nrm{\pt{q}})} which leads to  \eq{\ptvec{q}=x^\en q^\en \ptvec{x} = \nrm{\pt{q}}  q^\en \ptvec{x}} such that \eq{\nrm{\pt{q}}^2=(\nrm{\pt{q}}  q^\en \nrm{\pt{x}})^2} and thus \eq{q^\en = \pm 1/\nrm{\pt{x}}}. Since the norm is given by the Euclidean metric, and we restrict consideration to \eq{\bvsp{E}}, we must have \eq{0<\nrm{\pt{x}}} \eq{0<q^{\en}} and thus \eq{q^\en = 1/\nrm{\pt{x}}>0}.   }
for \eq{q^\en= 1/\nrm{\pt{x}} } and \eq{\ptvec{q} = x^\en \ptvec{x}/ \nrm{\pt{x}} = x^\en \hpt{x}}.
That is,  if \eq{\barpt{x} = \psi(\barpt{q})} for any \eq{\barpt{q}\in\bvsp{E}} then we have the following one-to-one relation:
\begin{small}
\begin{align} \label{prj_active2}
   \psi(\barpt{q}) \,=\, \barpt{x} \quad \leftrightarrow \quad \barpt{q} \,=\, \inv{\psi}(\barpt{x})
    \qquad\qquad
    \fnsize{i.e.,} \qquad
    \ptvec{x} = \tfrac{1}{q^\en} \hpt{q} 
    \;\;,\;\; x^\en = \nrm{\pt{q}}
    \qquad \leftrightarrow \qquad 
    \ptvec{q} = x^\en \hpt{x} \;\;,\;\; q^\en = \tfrac{1}{\nrm{\pt{x}}}  
\end{align}
\end{small}
Before proceeding, we should further clarify the domain of \eq{\psi}:
\begin{small}
\begin{itemize}[nosep]
    \item  \eq{\psi\in\Dfism(\bvsp{E})} is a diffeomorphism not on all of \eq{\vsp{E}} but, rather, on the region \eq{\bvsp{E}=\bs{\Sigup}_\nozer\oplus\vsp{N}_\ii{+}\subset \vsp{E}} (Eq.\eqref{E4positive}). 
    For instance, although \eq{\psi(\barpt{q})} is defined for \eq{q^\en <0}, if this were permitted then \eq{\psi(\barpt{q}) =\psi(-\barpt{q})} and \eq{\inv{\psi}\circ\psi(\barpt{q})=-\barpt{q}} such that \eq{\psi} would not be a 
    diffeomorphism.\footnote{If we allow for \eq{q^\en<0} then we see that \eq{\psi(-\barpt{q}) = \tfrac{1}{-q^\en \nrm{\pt{q}}} (-\ptvec{q})+\nrm{\pt{q}}\envec = \tfrac{1}{q^\en \nrm{\pt{q}} \ptvec{q}}+\nrm{\pt{q}}\envec = \psi(\barpt{q})} such that \eq{\psi(-\barpt{q}) =\psi(\barpt{q})} and thus \eq{\inv{\psi}\circ\psi(\barpt{q})=-\barpt{q} }. The latter is also verified by considering \eq{\barpt{q}=\ptvec{q} - k\envec} for any \eq{k>0}, Direct substitution leads to \eq{\inv{\psi}\circ\psi(\barpt{q})=-\ptvec{q} + k\envec = -\barpt{q}}. }
    \item \eq{\psi} and \eq{\inv{\psi}} are actually defined on slightly larger domains of \eq{\vsp{E}} than just \eq{\bvsp{E}} and it is technically \eq{\psi|_\ii{\bvsp{E}}\in\Dfism(\bvsp{E})} that is the diffeomorphism. 
    In particular, we see from the above that \eq{\psi(\barpt{q})} is \textit{un}defined only if \eq{\nrm{\pt{q}}=0} or \eq{q^\en=0}, and that \eq{\inv{\psi}(\barpt{x})} is \textit{un}defined only if \eq{\nrm{\pt{x}}=0}.
    That is, \eq{\psi} and \eq{\inv{\psi}} have actual domains \eq{\man{D}_{\psi}} and \eq{\man{D}_{\inv{\psi}}} as follows:
    \begin{small}
    \begin{align}
    \man{D}_{\psi} = \bs{\Sigup}_\nozer\oplus\vsp{N}_{\nozer}
        \quad,\quad 
        \man{D}_{\inv{\psi}}=\bs{\Sigup}_\nozer\oplus\vsp{N}
        \quad,\quad 
        \psi(\man{D}_{\psi}) = \inv{\psi}(\man{D}_{\inv{\psi}}) =  \bvsp{E}
         \qquad,\quad 
        \bvsp{E} \subset \man{D}_{\psi}\subset \man{D}_{\inv{\psi}}\subset \vsp{E}
    \end{align}
    \end{small}
    \item  Despite the above, we will write \eq{\psi\in\Dfism(\bvsp{E})} understanding that \eq{\psi} and \eq{\inv{\psi}} are taken to mean \eq{\psi\equiv \psi|_\ii{\bvsp{E}}} and \eq{\inv{\psi}\equiv \inv{\psi}|_\ii{\bvsp{E}}}. 
\end{itemize}
\end{small}

\paragraph{In Cartesian Coordinates.}
Let \eq{\bartup{r}=(\tup{r},r^\en):\vsp{E} \to\mbb{R}^{\en}  }  be global cartesian coordinates on \eq{\vsp{E}} for some inertial \eq{\bsfb{\emet}_\ii{\!\Evec}}-orthonormal basis, \eq{\hbe_\a \in\vsp{E}}, such that  \eq{\emet_{\a\b}:=\bsfb{\emet}_\ii{\!\Evec}(\hbe_\a,\hbe_\b)=\kd_{\a\b}}  
as described in Eq.\eqref{emet_cart}-Eq.\eqref{rhat_cart}. 
With \eq{\sfb{r}=r^i\hbe_i} and \eq{r = \nrm{\sfb{r}} \equiv \nrm{\tup{r}}\in\fun(\Evec)} the  displacement ``vector field'' and norm on \eq{\bs{\Sigup}\subset\vsp{E}}, 
we note that \eq{\psi,\inv{\psi}\in\Dfism(\bvsp{E})} may then be expressed 
as vector-valued maps:\footnote{\eq{\psi} and \eq{\inv{\psi}} are smooth maps from \eq{\bvsp{E}} to \eq{\bvsp{E}}. They are \textit{not} vector fields in the usual geometric sense; the fact that they may be expressed as in Eq.\eqref{proj_cartesian} is simply a convenience of vector spaces.}
\begin{small}
\begin{align} \label{proj_cartesian}
       \psi \,=\, \tfrac{1}{r^\en r}r^i \hbe_i  \,+\, r \envec 
       \;=\; 
       \tfrac{1}{r^\en} \hsfb{r} \,+\, r \envec 
&&,&& 
        \inv{\psi} \,=\, \tfrac{r^\en}{r}r^i \hbe_i \,+\, \tfrac{1}{r} \envec 
        \;=\; 
        r^\en \hsfb{r} \,+\,  \tfrac{1}{r} \envec 
\end{align}
\end{small}
where, as per Eq.\eqref{E4_rhat_vec}, 
\eq{\hsfb{r}=\tfrac{1}{r}\sfb{r}\equiv\be_{\rfun}\in\vect(\vsp{E})} is the radial coordinate (unit) vector field on \eq{\bs{\Sigup}\subset \vsp{E}} which, above, is instead regarded as a map \eq{\hsfb{r}:\vsp{E}\to\vsp{E}} such that \eq{\hsfb{r}_{\barpt{s}} \equiv \hsfb{r}_{\ptvec{s}} = \hpt{s}} for any \eq{\barpt{0}\neq\barpt{s}\in\vsp{E}}. 
From the above, we note the following:\footnote{As usual, we define \eq{\hat{r}^i:=r^i/\rfun}.}
\begin{small}
\begin{align} \label{proj_cartesian_i}
     \begin{array}{llllll}
        r^i \circ \psi 
        = \tfrac{1}{r^\en}\hat{r}^i
        \\[4pt]
          r^\en \circ \psi = r  
      \\[4pt]
          r\circ\psi = \tfrac{1}{r^\en}
    \end{array}
&&,&&
     \begin{array}{llllll}
         r^i \circ \inv{\psi} 
         = r^\en \hat{r}^i 
     \\[4pt] 
         r^\en \circ \inv{\psi} = \tfrac{1}{r}
     \\[4pt] 
         r\circ \inv{\psi} =  r^\en
     \end{array}
&&,&&  
     \begin{array}{rlll}
          &\hsfb{r}\circ \psi \,=\, \hsfb{r}\circ \inv{\psi}  \,=\,  \hsfb{r}
    \\[4pt]
        &\hat{r}^i \circ \psi  \,=\,  \hat{r}^i\circ\inv{\psi}  \,=\,  \hat{r}^i
     \\[4pt]
        \fnsize{i.e.,} \!\! &  \barpt{x}=\psi(\barpt{q}) \;\; \Rightarrow \;\; \hpt{x} = \hpt{q} 
    \end{array}
\end{align}
\end{small}
Now, for any given point transformation (diffeomorphism at the level of configuration manifolds), we know that objects associated with a mechanical Hamiltonian system on \eq{(\cotsp\bvsp{E},\nbs{\omg})} will transform as detailed in section \ref{sec:Hmech_xform_gen}. 
Most of these transformations will need the differentials of \eq{\psi} and \eq{\inv{\psi}},  expressed in the \eq{r^\a} frame fields, \eq{\hbe_\a \equiv \be[r^\a]}, by
\begin{small}
\begin{align} \label{dprj_gen}
    \dif \psi = \pderiv{(r^\a\circ\psi)}{r^\b}(\hbe_\a)_\ii{\psi} \otms \hbep^\b  \,\equiv\,  \pderiv{\psi^\a}{r^\b}\hbe_\a \otms \hbep^\b
    &&,&&
    \dif \inv{\psi} = \pderiv{(r^\a \circ \inv{\psi})}{r^\b}(\hbe_\a)_\ii{\inv{\psi}} \otms \hbep^\b  \,\equiv\, \pderiv{(\inv{\psi})^\a}{r^\b}\hbe_\a \otms \hbep^\b
\end{align}
\end{small}
where \eq{\hbe_\a \equiv \be[r^\a]} are homogeneous such that we may ignore compositions like \eq{(\hbe_\a)_\ii{\psi}:=\hbe_\a \circ\psi}
(see footnote\footnote{The last equalities in Eq.\eqref{dprj_gen} use the fact that \eq{\hbe_\a} are homogeneous and all (co)tangent spaces are the ``same'' (e.g., \eq{\tsp[\barpt{x}]\bvsp{E}\cong \tsp[\barpt{q}]\bvsp{E}\cong \vsp{E}}), such that \eq{\dif \psi} may be viewed as a sort of \eq{(1,1)}-tensor field on \eq{\bvsp{E}}. In general, the differential of a smooth map is most definitely \textit{not} a \eq{(1,1)}-tensor field. Our ability to treat \eq{\dif \psi} as such is a privilege of vector spaces.  }).
We note that the matrix representations of \eq{\dif \psi} and \eq{\dif \inv{\psi}} in the \eq{r^\a} coordinates,  \eq{\crd{\dif \psi}{\bartup{r}}=[ \pderiv{(r^\a\circ\psi)}{r^\b}]} and \eq{\crd{\dif \inv{\psi}}{\bartup{r}}=[ \pderiv{(r^\a\circ\inv{\psi})}{r^\b}]}, are 
obtained from Eq.\eqref{proj_cartesian_i} as follows:\footnote{For cartesian coordinates \eq{\bartup{r}=(\tup{r},r^\en)=(r^1,\dots,r^{\en-\ii{1}},r^\en):\Evec^4\to\mbb{R}^4}, we use the following notation, as described in Eq.\eqref{cord_norm}:
\begin{align} \nonumber 
    r_\en^2 \equiv (r^\en)^2
  &&,&&
    \hat{r}^i := r^i/\nrmtup{r} = r^i/ r
      &&,&&
    r_i := \emet_{ij} r^j = \kd_{ij} r^j
      &&,&&
    \htup{r}:=\tup{r}/\nrmtup{r}
    &&,&&
    \tup{r}^\flt =  [r_i] \cong \trn{\tup{r}}
    &&,&&
    \tup{r}\otms \tup{r}^\flt = [r^i r_j] \cong \tup{r}\trn{\tup{r}}
\end{align} }
\begin{small}
\begin{align} \label{dproj_cartesian}
 \crd{\dif \psi}{\bartup{r}} 
     \,=\, \begin{pmatrix}
         \tfrac{1}{r^\en \nrmtup{r}}( \kd^i_j - \hat{r}^i \hat{r}_j) & -\tfrac{1}{r_\en^2}\hat{r}^i 
         \\
        \hat{r}_i  & 0 
     \end{pmatrix} 
     \equiv  \pderiv{\psi}{\bartup{r}}  
&&,&&
\begin{array}{rllllll}
  \crd{\dif \inv{\psi}}{\bartup{r}} 
  &\!\!\!= 
      \begin{pmatrix}
         \tfrac{r^\en }{\nrmtup{r}}(\kd^i_j - \hat{r}^i \hat{r}_j) & \hat{r}^i 
         \\
        -\tfrac{1}{\nrmtup{r}^2} \hat{r}_i   & 0 
     \end{pmatrix} 
      \equiv \pderiv{\inv{\psi}}{\bartup{r}}
  \\[16pt]
  \crd{\dif \inv{\psi}_\ii{\psi}}{\bartup{r}} 
  &\!\!\!= 
      \begin{pmatrix}
         r^\en \nrmtup{r}( \kd^i_j - \hat{r}^i \hat{r}_j) & \hat{r}^i
         \\[3pt]
        -r_\en^2 \hat{r}_i   & 0 
     \end{pmatrix} \equiv \inv{\big(\pderiv{\psi}{\bartup{r}}\big)}  
\end{array}
\end{align}
\end{small}
where \eq{\dif \inv{\psi}_\ii{\psi}:= (\dif \inv{\psi})\circ\psi} satisfies
\eq{\dif \inv{\psi}_\ii{\psi(\barpt{q})} =\inv{(\dif \psi_\ss{\barpt{q}})}} and
\eq{\dif \inv{\psi}_\ii{\psi} \cdot  \dif \psi = \dif(\inv{\psi}\circ \psi) = \bar{\iden}_\ii{\!\vecE}}.  
Not that  \eq{  \crd{\dif \inv{\psi}_\ii{\psi}}{\bartup{r}} \cdot  \crd{\dif \psi}{\bartup{r}} = \imat_{\en}} but that  \eq{ \crd{\dif \inv{\psi}}{\bartup{r}} \cdot  \crd{\dif \psi}{\bartup{r}} \neq \imat_{\en}}.\footnote{For any diffeomorphism of smooth manifolds, \eq{\varphi\in\Dfism(\man{N};\man{M})}, then \eq{\dif \inv{\varphi} \cdot \dif \varphi  \neq \iden_\ii{\man{N}}}  and \eq{\dif \varphi\cdot\dif \inv{\varphi} \neq \iden_\ii{\man{M}}} but, rather: 
\begin{align} \nonumber 
    \dif \inv{\varphi}_\ii{\varphi} \cdot  \dif \varphi = \dif(\inv{\varphi}\circ \varphi) = \dif \Id_\ii{\man{N}}  =  \iden_\ii{\man{N}} 
    \qquad,\qquad 
    \dif \varphi_\ii{\inv{\varphi}}\cdot\dif \inv{\varphi} = \dif(\varphi\circ\inv{\varphi})  = \dif \Id_\ii{\man{M}} = \iden_\ii{\man{M}}
\end{align} }

\paragraph{(Co)Tangent-Lifted Projective Transformation.}
Our eventual goal is to obtain how mechanical Hamiltonian systems on \eq{\cotsp\bvsp{E}}  are transformed by \eq{\psi\in\Dfism(\bvsp{E})}. As per section \ref{sec:Hmech_xform_gen}, this will require the cotangent lift, \eq{\colift\psi\in\Spism(\cotsp\bvsp{E},\nbs{\omg})}. Although the tangent lift, \eq{\tlift\psi\in\Dfism(\tsp\bvsp{E})}, is not required at the moment, we will include it here for comparison and future reference. 

Now, recall that the (co)tangent lifts of any \eq{\psi\in\Dfism(\bvsp{E})} are defined as follows for any \eq{(\barpt{q},\bsfb{u})\in\tsp\bvsp{E}} and  \eq{(\barpt{q},\barbs{\mu})\in\cotsp\bvsp{E}}:
\begin{small}
\begin{align}
\begin{array}{llllll}
     \tlift\psi \,=\, (\psi\circ\tpr, \dif \psi(\slot,\sblt) ) \in\Dfism(\tsp\bvsp{E}) 
     &,\qquad 
     \tlift\psi (\barpt{q},\bsfb{u}) \,=\, ( \psi(\barpt{q}), \dif \psi_\ss{\barpt{q}}\cdot\bsfb{u}) \,=\, ( \psi(\barpt{q}), \psi_*\bsfb{u})
     &\!\!=\, (\barpt{x},\bsfb{v})
\\[5pt]
      \colift\psi \,=\, (\psi\circ\copr, \inv{\dif \psi_\ii{\psi}} (\sblt,\slot)) \in \Spism(\cotsp\bvsp{E},\nbs{\omg})
      &,\qquad 
      \colift\psi (\barpt{q},\barbs{\mu}) 
      \,=\, 
       ( \psi(\barpt{q}), \barbs{\mu}\cdot \inv{ (\dif \psi_\ss{\barpt{q}})} )
      \,=\, ( \psi(\barpt{q}), \psi_*\barbs{\mu})
       &\!\!=\, (\barpt{x},\barbs{\kap})
\end{array}
\end{align}
\end{small}
where, as always, \eq{\dif \inv{\psi}_\ii{\psi(\barpt{q})} = \inv{(\dif \psi_\ss{\barpt{q}})}}. 
Above, we have defined \eq{(\barpt{x},\bsfb{v})} as the \eq{\tlift\psi}-image of some \eq{(\barpt{q},\bsfb{u})\in\tsp\bvsp{E}}, and  \eq{(\barpt{x},\barbs{\kap})} as the \eq{\colift\psi}-image of some \eq{(\barpt{q},\barbs{\mu})\in\cotsp\bvsp{E}}. 
In actuality, \eq{(\barpt{x},\barbs{\kap})} will correspond to a phase space point of some original system and the above implicitly defines \eq{(\barpt{q},\barbs{\mu})} as the corresponding phase space point of a new system transformed by \eq{\colift\psi} (as detailed in sections \ref{sec:Hmech_xform_gen} and \ref{sec:prj_Xform_gen}). For now, these points are arbitrary.
Regardless, we will want explicit expressions for the above (co)vector transformations.
Using \eq{\dif \psi} from Eq.\eqref{dprj_gen} – \ref{dproj_cartesian} we find, for \eq{\eq{(\barpt{x},\bsfb{v})} \leftrightarrow (\barpt{q},\bsfb{u}) } and  \eq{(\barpt{x},\barbs{\kap})\leftrightarrow (\barpt{q},\barbs{\mu})} related as above, the (co)vectors transform as:
\begin{small}
\begin{align} \label{lift_prj_act1}
   \bsfb{v}_\ii{\barpt{x}} \leftrightarrow\bsfb{u}_\ss{\barpt{q}} \;\; 
  & \left\{\;\;
\begin{array}{rlllll}
         \bsfb{v}_\ii{\barpt{x}} \,= \;   \psi_* (\bsfb{u}_\ss{\barpt{q}}) \,=\,  \dif \psi_\ss{\barpt{q}}\cdot\bsfb{u} 
      &\!\!\! =\, 
    \tfrac{1}{q^\en \nrm{\pt{q}}}( \iden_\ii{\!\Sigup} - \hsfb{q}\otms \hsfb{q}^\flt)\cdot\sfb{u} \,-\, \tfrac{1}{q_\en^2} \enform (\bsfb{u})  \hsfb{q}
    \,+\, \hsfb{q}^\flt (\sfb{u})\envec  
    &\in\tsp[\psi(\barpt{q})]\bvsp{E}
    \\[6pt]
         \bsfb{u}_\ss{\barpt{q}} \,= \;   \psi^* (\bsfb{v}_\ii{\barpt{x}}) \,=\,  \inv{(\dif \psi_\ss{\barpt{q}})}\cdot\bsfb{v} 
      &\!\!\! =\, 
     \tfrac{x^\en}{ \nrm{\pt{x}}}( \iden_\ii{\!\Sigup} - \hsfb{x}\otms \hsfb{x}^\flt)\cdot\sfb{v} \,+\,  \enform (\bsfb{v})  \hsfb{x} \,-\, \tfrac{1}{\nrm{\pt{x}}^2}  \hsfb{x}^\flt (\sfb{v}) \envec 
      &\in\tsp[\barpt{q}]\bvsp{E}
\end{array} \right.
\\[5pt] \label{lift_prj_act3}
    \barbs{\kap}_\ii{\barpt{x}} \leftrightarrow \barbs{\mu}_\ss{\barpt{q}} 
     \;\; 
  & \left\{\;\; 
\begin{array}{llllll}
        \barbs{\kap}_\ii{\barpt{x}} \,=\;  \psi_* (\barbs{\mu}_\ss{\barpt{q}}) 
     \,=\, \barbs{\mu}  \cdot \inv{(\dif \psi_\ss{\barpt{q}})} 
      &\!\!\! =\, 
        \bs{\mu} \cdot  q^\en \nrm{\pt{q}} ( \iden_\ii{\!\Sigup} - \hsfb{q}\otms \hsfb{q}^\flt) \,-\, q_\en^2 \barbs{\mu}(\envec ) \hsfb{q}^{\flt} \,+\, \bs{\mu}(\hsfb{q}) \enform 
       &\in\cotsp[\psi(\barpt{q})]\bvsp{E}
  \\[6pt]
         \barbs{\mu}_\ss{\barpt{q}}\,=\;  \psi^* (\barbs{\kap}_\ii{\barpt{x}}) 
    \,=\, \barbs{\kap}  \cdot \dif \psi_\ss{\barpt{q}}
      &\!\!\! =\, 
        \bs{\kap} \cdot \tfrac{\nrm{\pt{x}}}{x^\en} ( \iden_\ii{\!\Sigup} - \hsfb{x}\otms \hsfb{x}^\flt) \,+\, \barbs{\kap}(\envec )  \hsfb{x}^\flt
        \,-\, \nrm{\pt{x}}^2 \bs{\kap}(\hsfb{x}) \enform     
       &\in\cotsp[\barpt{q}]\bvsp{E} 
\end{array} \right.
\end{align}
\end{small}
where \eq{\iden_\ii{\!\Sigup}\equiv \bs{\Pi}\in\tens^\ss{1}_\ss{1}(\vsp{E})},
where \eq{\psi(\barpt{q})=\barpt{x}} such that \eq{\hsfb{r}_{\!\pt{q}} = \hsfb{q} =\hsfb{x} = \hsfb{r}_{\!\pt{x}} }.  
The tangent lift \eq{\tlift\psi\in\Dfism(\tsp\bvsp{E})} is then given explicitly by:
\begin{small} 
\begin{align} \label{lift_prj_act6}
\begin{array}{rclll}
      (\barpt{x},\bsfb{v}) \,=\, \colift \psi (\barpt{q},\bsfb{u}) \,=\, ( \psi(\barpt{q}), \psi_* \bsfb{u} ) 
    &\qquad 
    &\qquad
    (\barpt{q},\bsfb{u}) \,=\, \colift \inv{\psi} (\barpt{x},\bsfb{v}) \,=\, ( \inv{\psi}(\barpt{x}), \psi^* \bsfb{v} )
\\[5pt]
       \barpt{x} 
       \,=\,  \tfrac{1}{q^\en}\hpt{q} + \nrm{\pt{q}}\envec 
       &\qquad \leftrightarrow &\qquad 
       \barpt{q} 
       \,=\, x^\en \hpt{x} + \tfrac{1}{\nrm{\pt{x}}} \envec 
\\[5pt]
     \bsfb{v} \,=\,  \tfrac{1}{q^\en \nrm{\pt{q}}}\big(  \sfb{u} - \inner{\hsfb{q}}{\sfb{u}}\hsfb{q} \big)
    \,-\, \tfrac{u^\en}{q_\en^2}  \hsfb{q}
    \,+\, \inner{\hsfb{q}}{\sfb{u}}\envec 
 &\qquad 
 &\qquad
     \bsfb{u}  \,=\, \tfrac{x^\en}{ \nrm{\pt{x}}} \big( \sfb{v} - \inner{\hsfb{x}}{\sfb{v}} \hsfb{x} \big) \,+\, \v^\en \hsfb{x} \,-\, \tfrac{1}{\nrm{\pt{x}}^2} \inner{\hsfb{x}}{\sfb{v}} \envec 
\end{array}
\end{align}
\end{small}
Similarly, the cotangent lift \eq{\colift\psi\in\Spism(\cotsp\bvsp{E},\nbs{\omg})} is given explicitly by:
\begin{small}
\begin{align} \label{colift_prj_def}
 \begin{array}{rclll}
      (\barpt{x},\barbs{\kap}) \,=\, \colift \psi (\barpt{q},\barbs{\mu}) \,=\, ( \psi(\barpt{q}), \psi_* \barbs{\mu}) 
    &\qquad 
    &\qquad
    (\barpt{q},\barbs{\mu}) \,=\, \colift \inv{\psi} (\barpt{x},\barbs{\kap}) \,=\, ( \inv{\psi}(\barpt{x}), \psi^* \barbs{\kap} )
\\[5pt]
       \barpt{x} 
       \,=\,  \tfrac{1}{q^\en}\hpt{q} + \nrm{\pt{q}}\envec 
       &\qquad \leftrightarrow &\qquad
       \barpt{q} 
       \,=\, x^\en \hpt{x} + \tfrac{1}{\nrm{\pt{x}}} \envec 
\\[5pt]
     \barbs{\kap} \,=\,  q^\en \nrm{\pt{q}} \big( \bs{\mu} - (\bs{\mu}\cdot\hsfb{q}) \hsfb{q}^\flt \big) \,-\, q_\en^2 \mu_\en \hsfb{q}^{\flt} \,+\, (\bs{\mu}\cdot \hsfb{q})\enform 
 &\qquad 
 &\qquad
     \barbs{\mu}  \,=\,  \tfrac{\nrm{\pt{x}}}{x^\en} \big( \bs{\kap} - (\bs{\kap}\cdot\hsfb{x}) \hsfb{x}^\flt \big) \,+\, \kap_\en \hsfb{x}^\flt \,-\, \nrm{\pt{x}}^2 (\bs{\kap}\cdot\hsfb{x})\enform  
\end{array} 
\end{align}
\end{small}
It may be helpful to express the above as follows, with the \eq{\en^\mrm{th}} components (normal components) separated: 
\begin{small}
\begin{align} \label{colift_prj_def1}
\boxed{ \begin{array}{rlcllll}
       \ptvec{x} =   \tfrac{1}{q^\en}\hpt{q} 
       &,\quad x^\en =  \nrm{\pt{q}}
\\[5pt]
     \bs{\kap} 
     \,=\,  q^\en \nrm{\pt{q}} \big( \trn{\iden}_\ii{\!\Sig}-   \hsfb{q}^\flt \otms \hsfb{q} \big) \cdot \bs{\mu} \,-\, q_\en^2 \mu_\en \hsfb{q}^{\flt} 
     &,\quad \kap_\en = \bs{\mu}\cdot \hsfb{q} 
\end{array} 
\qquad \leftrightarrow \qquad 
\begin{array}{lllll}
     \ptvec{q} =  x^\en \hpt{x} \quad,\quad q^\en = \tfrac{1}{\nrm{\pt{x}}} 
\\[5pt]
     \bs{\mu}  
      \,=\,  \tfrac{\nrm{\pt{x}}}{x^\en} \big( \trn{\iden}_\ii{\!\Sig}-   \hsfb{x}^\flt \otms \hsfb{x} \big) \cdot \bs{\kap} \,+\, \kap_\en \hsfb{x}^\flt
     \quad,\quad \mu_\en = 
     -\nrm{\pt{x}}^2  \bs{\kap}\cdot\hsfb{x}  
\end{array}  }
\end{align}
\end{small}
We note that any such \eq{\colift\psi}-related points then satisfy the following:\footnote{where  \eq{\nrm{\sfb{v}}^2 := \inner{\sfb{v}}{\sfb{v}} = \bsfb{\emet}_\ii{\!\Evec}(\sfb{v},\sfb{v}) = \sfb{\emet}_\ii{\!\Sigup} (\sfb{v},\sfb{v})} for any \eq{\bsfb{v}=\sfb{v} + \v^\en \envec\in\tsp[\cdt]\vsp{E}}
(or any \eq{\barpt{s}=\ptvec{s}+s^\en\envec \in\vsp{E}}) and where \eq{\nrm{\bs{\eta}}^2 := \inner{\bs{\eta}}{\bs{\eta}} = \inv{\bsfb{\emet}_\ii{\!\Evec}}(\bs{\eta},\bs{\eta}) = \inv{\sfb{\emet}_\ii{\!\Sigup}} (\bs{\eta},\bs{\eta})} for any \eq{\barbs{\eta}=\bs{\eta}+\eta_\en \enform\in\cotsp[\cdt]\vsp{E}}. }
\begin{small}
\begin{align} \label{some_Prels_prj}
\bar{\kap}_{\barpt{x}} = \colift \psi(\bar{\mu}_{\barpt{q}}) \;\; \left\{ \quad \begin{array}{lllll}
     \nrm{\bs{\kap}}^2 \,=\, \inv{\sfb{\emet}_\ii{\!\Sigup}}(\bs{\kap},\bs{\kap}) \,=\,  q_\en^2 \big( \nrm{\pt{q}}^2 \nrm{\bs{\mu}}^2 - (\ptvec{q}\cdot\bs{\mu})^2 \,+\, q_\en^2 \mu_\en^2 \big)
     &\;,\;\;\quad 
     \ptvec{x}\cdot\bs{\kap} \,=\, -q^\en \mu_\en
     &\;,\;\;\quad 
     \envec\cdot\barbs{\kap} = \hsfb{q}\cdot\bs{\mu}
\\[5pt]
     \nrm{\bs{\mu}}^2 \,=\, \inv{\sfb{\emet}_\ii{\!\Sigup}}(\bs{\mu},\bs{\mu}) \,=\,  \tfrac{1}{x_\en^2} \big( \nrm{\pt{x}}^2 \nrm{\bs{\kap}}^2 - (\ptvec{x}\cdot\bs{\kap})^2 \,+\, x_\en^2 \kap_\en^2 \big)
      &\;,\;\;\quad 
     \ptvec{q}\cdot\bs{\mu} \,=\,  x^\en \kap_\en 
      &\;,\;\;\quad 
     \envec\cdot\barbs{\mu} =  -\nrm{\pt{x}}^2  \bs{\kap}\cdot\hsfb{x}
\end{array} \right.
\end{align}
\end{small}
The above, along with \eq{\nrm{\pt{x}}^2=1/q_\en^2} and \eq{\nrm{\pt{q}}^2=x_\en^2}, leads to the following relation for the angular momentum: 
\begin{small}
\begin{align} \label{angMom_prj0}
\bar{\kap}_{\barpt{x}} = \colift \psi(\bar{\mu}_{\barpt{q}}) \;\; \left\{ \quad \begin{array}{lllll}
     \ptvec{x} \wedge \bs{\kap}^\shrp \,=\, \ptvec{q} \wedge \bs{\mu}^\shrp 
 \\[4pt]
      \ptvec{x}\,^\flt \wedge \bs{\kap} \,=\, \ptvec{q}\,^\flt \wedge \bs{\mu}
  \end{array} \right. 
\qquad,\qquad 
    \lang^2(\kap_\pt{x}) \,=\,
    \nrm{\pt{x}}^2 \nrm{\bs{\kap}}^2 - (\ptvec{x}\cdot\bs{\kap})^2 \,=\,  \nrm{\pt{q}}^2 \nrm{\bs{\mu}}^2 - (\ptvec{q}\cdot\bs{\mu})^2 \,=\, \lang^2(\mu_\ss{\pt{q}})
\end{align}
\end{small}
where, as was detailed in Eq.\eqref{angMoment_prj_def}, \eq{\lang^2 = \tfrac{1}{2} \emet_{ik}\emet_{js}\lang^{ij}\lang^{ks} = \tfrac{1}{2} \lang^{ij} \lang_{ij} = \lang^{ij} r_i \plin_j} with \eq{\lang^{ij} =  r^i \plin^j - \plin^i r^j\in\fun(\cotsp\vsp{E})} the angular momentum functions on \eq{\cotsp\bs{\Sigup}\subset\cotsp\vsp{E}} (where \eq{\plin^i:=\emet^{ij} \plin_j}).

\begin{small}
\begin{notesq}
    Eq.\eqref{some_Prels_prj}-Eq.\eqref{angMom_prj0} hold for arbitrary  \eq{\colift\psi}-related points and thus imply the following relations for cartesian coordinates \eq{(\tup{r},r^\en,\tup{\plin},\plin_\en)}:
    \begin{small}
    \begin{align} \label{angMom_prj}
    \boxed{\begin{array}{llllll}
        \colift\psi^* \nrmtup{\plin}^2 = r_\en^2 ( \lang^2 + r_\en^2 \plin_\en^2 )
         &\;\;,\;\;\quad  
         \colift\psi^*(r^i \plin_i) = - r^\en \plin_\en
        &\;\;,\;\;\quad  
         \colift\psi^* \plin_\en =  \hat{r}^i \plin_i
         &\;\;,\;\;\quad   
        \colift\psi^* \lang^{ij} = \lang^{ij} =  r^i \plin^j - \plin^i r^j
    \\[5pt]
         \colift\psi_* \nrmtup{\plin}^2 = \tfrac{1}{r_\en^2} ( \lang^2 + r_\en^2 \plin_\en^2 ) 
           &\;\;,\;\;\quad  
         \colift\psi_*(r^i \plin_i) = r^\en \plin_\en
         &\;\;,\;\;\quad  
           \colift\psi_* \plin_\en =  -r^2 \hat{r}^i \plin_i
         &\;\;,\;\;\quad  
         \colift\psi^* \lang^2 = \lang^2 = \nrmtup{r}^2 \nrmtup{\plin}^2 - (r^i \plin_i)^2
    \end{array}}
    \end{align}
    \end{small}
\end{notesq}
\end{small}

\subsection{Transformed Riemannian \& Symplectic Structures} \label{sec:prj_metric}

 As detailed in section \ref{sec:Hmech_xform_gen}, we are interested in using \eq{\psi} and \eq{\colift\psi} to transform  the original mechanical Hamiltonian system on \eq{\cotsp\bvsp{E}}, where \eq{\psi} is used to pull back the original metric on \eq{\bvsp{E}} and \eq{\colift\psi} is used to pullback the original symplectic form on \eq{\cotsp\bvsp{E}} (which, in this case, is the canonical symplectic form). We will start with the latter since it is trivial:

\paragraph{$\colift\psi$-Induced Symplectic Form on $\cotsp\bvsp{E}$}
It automatically holds that the cotangent lift of any diffeomorphism is a symplectomorphism (with respect to the canonical symplectic form on the cotangent bundle) with the additional property that it also preserves the canonical 1-form. Therefore, for the projective transformation in Eq.\eqref{prj_def_act} or Eq.\eqref{proj_cartesian}, we know \textit{a priori} that \eq{\colift\psi\in\Spism(\cotsp\bvsp{E},\nbs{\omg})} satisfies:
\begin{small}
\begin{align}
    \colift{\psi}\in \Spism(\cotsp\bvsp{E},\nbs{\omg}) 
\qquad ,\qquad 
    \colift{\psi}^*\bs{\theta} =\bs{\theta}
\qquad ,\qquad 
      \colift{\psi}^*\nbs{\omg} =\nbs{\omg}
\qquad ,\qquad 
      \colift{\psi}^*\inv{\nbs{\omg}} = \inv{\nbs{\omg}}
\end{align}
\end{small}
That is, in contrast to the \eq{\psi}-induced Riemannian structure on \eq{\bvsp{E}} (given below), there is no \eq{\colift\psi}-induced symplectic structure on \eq{\cotsp\bvsp{E}};  it is just the original, canonical, symplectic structure on the cotangent bundle. This holds for the cotangent lift (but not tangent lift) of any diffeomorphism on any finite-dimensional smooth manifold.

\paragraph{$\psi$-Induced Metric on $\bvsp{E}$.} Recall that the inner product, \eq{\bsfb{\emet}\equiv \inner{\cdot}{\cdot}\in \botimes^0_2\vsp{E}}, is a symmetric non-degenerate (in this case, positive-definite)  twice-covariant tensor on \eq{\vsp{E}}. Using the identification \eq{\tsp[\cdt]\vsp{E} \cong \vsp{E}}, this same \eq{\bsfb{\emet}} may be also viewed as a tensor \textit{field}, \eq{\bsfb{\emet}\in \tens^0_2(\vsp{E})}, which is homogeneous (the same everywhere) such that \eq{(\vsp{E},\bsfb{\emet}=\inner{\cdot}{\cdot})} is simultaneously an inner product space and a Riemannian manifold. We will now obtain the \eq{\psi}-induced metric \eq{\sfb{g}:=\psi^*\bsfb{\emet}\in \tens^0_2(\bvsp{E})} which, we will see, is truly a tensor \textit{field} (not just a tensor) such that \eq{(\bvsp{E},\sfb{g})} is a Riemannian manifold but \textit{not} an inner product
space.\footnote{However, for some given \eq{\barpt{q}\in\bvsp{E}}, one \textit{could} choose to use \eq{\sfb{g}_{\barpt{q}}\in\botimes^0_2\tsp[\cdt]\bvsp{E}\cong \vsp{E}} as an inner product on \eq{\vsp{E}}. That is, \eq{(\vsp{E}, \sfb{g}_{\barpt{q}})} is a valid inner product space, but we will have no reason to consider it as such. In contrast, we \textit{will} have reason to consider the Riemannian manifold \eq{(\bvsp{E},\sfb{g})}.}

\begin{small}
\begin{notesq}
    We note that \eq{\psi\in\Dfism(\bvsp{E})} is not an isometry in the sense of inner product spaces (that is, \eq{\inner{\barpt{q}}{\barpt{q}}\neq \inner{\psi(\barpt{q})}{\psi(\barpt{q})}}), nor is it an isometry in the sense of Riemannian manifolds (that is, \eq{\psi^*\bsfb{\emet}_\ii{\!\Evec}\neq\bsfb{\emet}_\ii{\!\Evec}}).
\end{notesq}
\end{small}

Now, recall from Eq.\eqref{metric_pback_gen} that   
the pullback of \eq{\bsfb{\emet}\in\tens^0_2(\vsp{E})} is  \eq{\psi^* \bsfb{\emet} = \trn{\dif \psi} \cdot \bsfb{\emet}_\ii{\psi} \cdot\dif \psi } 
(where the composition \eq{\bsfb{\emet}_\ii{\psi}=\bsfb{\emet}\circ\psi} is superfluous since \eq{\bsfb{\emet}} is homogeneous\footnote{It should be noted, however, that in an arbitrary coordinate basis, then \eq{\psi^*\bsfb{\emet}= (\emet_{\a\b} \circ\psi)\pd_\gamma \psi^\a \pd_\sig \psi^\b  \btau^\gamma \otms \btau^\sig } where the composition \eq{\emet_{\a\b} \circ\psi} is \textit{not} superfluous unless \eq{\emet_{\a\b}} happen to be constant (as is indeed the case in a  linear coordinate basis).}). 
For cartesian coordinates \eq{r^\a} with frame fields 
\eq{\hbe_\a \equiv \be[r^\a]}, then \eq{\emet_{\a\b}} are constant such that \eq{\emet_{\a\b}\circ\psi \equiv \emet_{\a\b}} and the \eq{\psi}-induced metric may be found from:
\begin{small}
\begin{align} \label{g_prj}
       \sfb{g} := \psi^*\bsfb{\emet}_\ii{\!\Evec} 
       \,=\, \emet_{\a\b} \pderiv{\psi^\a}{r^\gamma} \pderiv{\psi^\b}{r^\sig}    \hbep^\gamma \otms  \hbep^\sig  \, \in\tens^0_2(\bvsp{E})
       \qquad,\qquad 
       \emet_{\a\b} :=\bsfb{\emet}_\ii{\!\Evec}(\hbe_\a,\hbe_\b) = \kd^\a_\b
         \qquad,\qquad 
       \txi{g}_{\gam \sig} :=\sfb{g}(\hbe_\gam,\hbe_\sig) = \emet_{\a\b} \pderiv{\psi^\a}{r^\gamma} \pderiv{\psi^\b}{r^\sig}
\end{align}
\end{small}
where \eq{\psi^\a:=r^\a\circ\psi} are given in Eq.\eqref{proj_cartesian_i} and \eq{\pderiv{\psi^\a}{r^\gamma}} are given in Eq.\eqref{dproj_cartesian}. This leads to \eq{\sfb{g}} expressed in the cartesian coordinate basis (which is \eq{\bsfb{\emet}_\ii{\!\Evec}}-orthonormal but not \eq{\sfb{g}}-orthonormal) 
as:\footnote{Note that if we regard the \eq{(\en-1)}-space metric as a degenerate bilinear form on \eq{\affE}, \eq{\sfb{\emet}_\ii{\!\Sigup} = \bsfb{\emet}_\ii{\!\Evec} - \enform\otms\enform}, then we may also find \eq{\psi^*\sfb{\emet}_\ii{\!\Sigup}} but the result is  degenerate: 
\begin{align} \nonumber 
    \psi^* \sfb{\emet}_\ii{\!\Sigup} \,=\, \tfrac{1}{r_\en^2\, r^2} (\sfb{\emet}_\ii{\!\Sigup}-\hsfb{r}^\flt\otms \hsfb{r}^\flt)  \,+\, \tfrac{1}{r_\en^4 }\enform \otms \enform 
    \;=\;  \tfrac{1}{r_\en^2\, r} \nab \hsfb{r}^\flt  \,+\, \tfrac{1}{r_\en^4 }\enform \otms \enform
\end{align}
}
\begin{small}
\begin{align}\label{g_proj_active}
\begin{array}{rllll}
  \sfb{g} \,:=\, \psi^*\bsfb{\emet}_\ii{\!\Evec}  &\!\!\! =\,   \tfrac{1}{r_\en^2 \nrmtup{r}^2}( \emet_{ij} - \hat{r}_i\hat{r}_j) \hbep^i \otms \hbep^j \,+\, \hat{r}_i\hat{r}_j \hbep^i \otms \hbep^j \,+\, \tfrac{1}{r_\en^4 }\enform  \otms \enform
 \\[6pt]
    &\!\!\! =\, \tfrac{1}{r_\en^2\, r^2} (\sfb{\emet}_\ii{\!\Sigup}-\hsfb{r}^\flt\otms \hsfb{r}^\flt) \,+\, \hsfb{r}^\flt \otms \hsfb{r}^\flt \,+\, \tfrac{1}{r_\en^4 }\enform \otms \enform
\\[6pt]
    &\!\!\! =\, \tfrac{1}{r_\en^2\, r} \nab \hsfb{r}^\flt \,+\, \hsfb{r}^\flt \otms \hsfb{r}^\flt \,+\, \tfrac{1}{r_\en^4 }\enform \otms \enform
\end{array}
&&
\begin{array}{ccc}
     \cord{\txi{g}}{\bartup{r}} = \fnsz{ \begin{pmatrix}
    \tfrac{1}{r_\en^2 \nrmtup{r}^2}( \emet_{ij} - \hat{r}_i\hat{r}_j) + \hat{r}_i\hat{r}_j & \tup{0} \\
    \trn{\tup{0}} & \tfrac{1}{r_\en^4}
\end{pmatrix}}
\\[12pt]
     \det \cord{\txi{g}}{\bartup{r}} =  \tfrac{1}{r_\en^8 \nrmtup{r}^4}
\end{array}
\end{align}
\end{small}
where \eq{\hsfb{r}^\flt = \dif r \in\forms(\bvsp{E})} and where \eq{\nab} the Euclidean LC connection for \eq{\bsfb{\emet}_\ii{\!\Evec}} such that \eq{  \nab \hsfb{r}^\flt \,=\, \tfrac{1}{r}(\sfb{\emet}_\ii{\!\Sigup}-\hsfb{r}^\flt\otms \hsfb{r}^\flt)}. 
The inverse \eq{\psi}-induced metric is  \eq{\inv{\sfb{g}} = \inv{\psi}_* \inv{\bsfb{\emet}} = \dif \inv{\psi}_\ii{\psi} \cdot \inv{\bsfb{\emet}}_\ii{\psi}\cdot \dif \invtrn{\psi}_\ii{\psi} }. 
Using \eq{\crd{\dif \inv{\psi}_\ii{\psi}}{\bartup{r}}\;} (not \eq{\crd{\dif \inv{\psi}}{\bartup{r}}})  given as in Eq.\eqref{dproj_cartesian}, we obtain
\begin{small}
\begin{align} \label{gin_proj_active}
\begin{array}{rlllll}
    \inv{\sfb{g}} = \inv{\psi}_* \inv{\bsfb{\emet}_\ii{\!\Evec}} &\!\!\!
    =\,    r_\en^2 \nrmtup{r}^2 ( \emet^{ij} - \hat{r}^i\hat{r}^j) \hbe_i \otms \hbe_j \,+\, \hat{r}^i\hat{r}^j \hbe_i \otms \hbe_j \,+\, r_\en^4 \envec \otms \envec 
    \\[5pt]
     &\!\!\! =\,  r_\en^2 r^2 ( \inv{\sfb{\emet}_\ii{\!\Sigup}} - \hsfb{r}\otms \hsfb{r}) \,+\, \hsfb{r}\otms\hsfb{r} \,+\, r_\en^4 \envec \otms \envec 
\end{array}
&&
\begin{array}{ccc}
      \cord{\inv{\txi{g}}}{\bartup{r}} = \fnsz{ \begin{pmatrix}
     r_\en^2 \nrmtup{r}^2 ( \emet^{ij} - \hat{r}^i\hat{r}^j) + \hat{r}^i \hat{r}^j & \tup{0}
     \\[3pt]
    \trn{\tup{0}}   & r_\en^4
 \end{pmatrix} }
\\[12pt]
     \det \cord{\inv{\txi{g}}}{r} =  r_\en^8 \nrmtup{r}^4
\end{array}
\end{align}
\end{small}
To clarify, 
at any \eq{\barpt{q}=\ptvec{q}+q^\en \envec \in\bvsp{E}}, then  \eq{\sfb{g}_\ss{\barpt{q}}\in \botimes^0_2\tsp[\barpt{q}]\bvsp{E}} and \eq{\inv{\sfb{g}}_\ss{\barpt{q}}\in\botimes^2_0 \tsp[\barpt{q}]\bvsp{E}} are given as follows (where \eq{\hsfb{r}_\ss{\barpt{q}} \equiv \hsfb{r}_\ss{\pt{q}} = \hsfb{q}}):
\begin{small}
\begin{align}
\begin{array}{rllllll}
    \sfb{g}_\ss{\barpt{q}}  &\!\!\!\! =\, \tfrac{1}{q_\en^2\, \nrm{\pt{q}}^2} (\sfb{\emet}_\ii{\!\Sigup}-\hsfb{q}^\flt\otms \hsfb{q}^\flt) \,+\, \hsfb{q}^\flt \otms \hsfb{q}^\flt \,+\, \tfrac{1}{q_\en^4 } \enform \otms \enform
\\[10pt]
 \sfb{g}_\ss{\barpt{q}}(\barsfb{u}) 
      &\!\!\!\! =\, \tfrac{1}{q_\en^2\, \nrm{\pt{q}}^2} (\sfb{u}^\flt - \inner{\hsfb{q}}{\sfb{u}} \hsfb{q}^\flt )  + \inner{\hsfb{q}}{\sfb{u}} \hsfb{q}^\flt \,+\,  \tfrac{1}{q_\en^4 } u^\en \enform 
\\[3pt]
     &\!\!\!\! =\, - \tfrac{1}{q_\en^2\, \nrm{\pt{q}}^4} (\ptvec{q}^\flt \wedge \sfb{u}^\flt)\cdot \ptvec{q} + \inner{\hsfb{q}}{\sfb{u}} \hsfb{q}^\flt \,+\,  \tfrac{1}{q_\en^4 } u^\en \enform 
\end{array}
&&,&&
\begin{array}{rllllll}
     \inv{\sfb{g}}_\ss{\barpt{q}}  &\!\!\!\! =\, q_\en^2 \nrm{\pt{q}}^2 ( \inv{\sfb{\emet}_\ii{\!\Sigup}} - \hsfb{q}\otms \hsfb{q}) \,+\, \hsfb{q}\otms \hsfb{q} \,+\, q_\en^4 \envec \otms \envec
\\[10pt]
     \inv{\sfb{g}}_\ss{\barpt{q}}(\barbs{\mu}) 
      &\!\!\!\! =\,   q_\en^2 \nrm{\pt{q}}^2 \big(\bs{\mu}^\shrp - \hsfb{q}(\bs{\mu}) \hsfb{q}  \big)  \,+\, \hsfb{q}(\bs{\mu})\hsfb{q}  \,+\, q_\en^4 \mu_\en  \envec 
 \\[3pt]
     &\!\!\!\! =\,   -q_\en^2 (\ptvec{q}\wedge \bs{\mu}^\shrp)\cdot\ptvec{q}^\flt  \,+\, \hsfb{q}(\bs{\mu})\hsfb{q}  \,+\, q_\en^4 \mu_\en  \envec 
\end{array}
\end{align}
\end{small}
(we do not bother indicating the evaluation of \eq{\sfb{\emet}_\ii{\!\Sigup}} or \eq{\envec} at a point since they are homogeneous). We have also indicated above the isomorphism \eq{\tsp[\barpt{q}]\bvsp{E} \leftrightarrow \cotsp[\barpt{q}]\bvsp{E}} given by \eq{\sfb{g}_\ss{\barpt{q}}} (this also extends to an isomorphism  \eq{\vect(\bvsp{E})\leftrightarrow \forms(\bvsp{E})} in the usual way).   
Interestingly, if we blur the distinction between \eq{\tsp[\cdt]\bvsp{E}\cong \vsp{E}} and \eq{\vsp{E}} such that \eq{\sfb{g}_\ss{\barpt{q}}} is ``allowed'' to be fed  \eq{\barpt{q}=\ptvec{q}+q^\en \envec\in\bvsp{E}} itself, then:
\begin{small}
\begin{align} \label{g_prj_interesting}
    \sfb{g}_\ss{\barpt{q}}(\ptvec{q}) 
    = \sfb{\emet}_\ii{\!\Sigup} (\ptvec{q}) = \ptvec{q}\,^\flt  
&&,&&
    \sfb{g}_\ss{\barpt{q}}(\ptvec{q},\ptvec{q}) = \sfb{\emet}_\ii{\!\Sigup}(\ptvec{q},\ptvec{q}) = \inner{\ptvec{q}}{\ptvec{q}} 
&&,&&
    \sfb{g}_\ss{\barpt{q}}(\ptvec{q},\sfb{u}) 
    = \sfb{\emet}_\ii{\!\Sigup} (\ptvec{q},\sfb{u})
    = \inner{\ptvec{q}}{\sfb{u}}
&&,&&
      \sfb{g}_\ss{\barpt{q}}(\barpt{q},\barpt{q}) = \nrm{{q}}^2 + \tfrac{1}{q_\en^2} = \nrm{\psi(\barpt{q})}^2 
\end{align}
\end{small}
Note, however, that for some other \eq{\barpt{x}\in\bvsp{E}}, then \eq{\sfb{g}_{\barpt{x}}(\ptvec{q})\neq \sfb{g}_{\barpt{q}}(\ptvec{q})} and \eq{\sfb{g}_{\barpt{q}}(\ptvec{x})\neq \sfb{g}_{\barpt{x}}(\ptvec{x})} would \textit{not} give the same relations as above.  This illustrates the previously-mentioned detail that, unlike \eq{\bsfb{\emet}_\ii{\!\Evec}\equiv \inner{\cdot}{\cdot}}, the metric \eq{\sfb{g}} is \textit{not} an inner product on \eq{\bvsp{E}}. Rather, \eq{\sfb{g}\in\tens^0_2(\bvsp{E})} is well and truly a tensor \textit{field}.
The following example for the classical case \eq{\en -1 = 3} may clarify the nature of \eq{\sfb{g}}:

\begin{small}
\begin{itemize}[nosep]
    \item \sbemph{Relation to Spherical Metric on $\man{S}^{\en-\two}$.}  Consider the (\eq{\en-1})-dim Euclidean hyperplane \eq{\bs{\Sigup}\subset\vsp{E}} and let \eq{\sfb{s}} denote the metric on the unit (\eq{\en-2})-sphere \eq{\man{S}^{\en -\two}\subset \bs{\Sigup}}. Note we have a collection of Riemannian (sub)manifolds, \eq{(\man{S}^{\en-\two},\sfb{s})\subset (\bs{\Sigup}^{\en-\one},\sfb{\emet}_\ii{\!\Sigup})\subset (\vsp{E}^\en,\bsfb{\emet}_\ii{\!\Evec})}, where we will continue to take the extrinsic view of everything in \eq{\vsp{E}\equiv\vsp{E}^\en} such that the metrics can be expressed in terms of one another as: 
    \begin{small}
    \begin{align} \label{2sphere_met}
    \begin{array}{cccccc}
         (\man{S}^{\en-\two},\sfb{s}) 
        & \subset &
        (\bs{\Sigup},\sfb{\emet}_\ii{\!\Sigup})
        & \subset &
        (\vsp{E},\bsfb{\emet}_\ii{\!\Evec})
    \\[5pt]
      \begin{array}{llllll}
          \sfb{s} 
           \,=  \tfrac{1}{r^2}( \sfb{\emet}_\ii{\!\Sigup} - \hsfb{r}^\flt \otms \hsfb{r}^\flt  ) = \tfrac{1}{r} \nab \hsfb{r}^\flt
      \\[5pt]
         \inv{\sfb{s}} 
         = r^2( \inv{\sfb{\emet}_\ii{\!\Sigup}} - \hsfb{r} \otms \hsfb{r} )
      \end{array}
     &,&
      \begin{array}{llllll}
           \sfb{\emet}_\ii{\!\Sigup} = \hsfb{r}^\flt \otms \hsfb{r}^\flt  + r^2 \sfb{s}
           \,=\, \bsfb{\emet}_\ii{\!\Evec} - \enform\otms\enform
      \\[5pt]
          \inv{\sfb{\emet}_\ii{\!\Sigup}} =  \hsfb{r} \otms \hsfb{r} + \tfrac{1}{r^2}\inv{\sfb{s}}
           \,=\, \inv{\bsfb{\emet}_\ii{\!\Evec}} - \envec\otms\envec
      \end{array}
      &,&
       \begin{array}{llllll}
           \bsfb{\emet}_\ii{\!\Evec} 
        =  \hsfb{r}^\flt \otms \hsfb{r}^\flt + r^2 \sfb{s} + \enform \otms\enform
     \\[5pt]
         \inv{\bsfb{\emet}_\ii{\!\Evec}} 
         = \hsfb{r} \otms \hsfb{r} + \tfrac{1}{r^2} \inv{\sfb{s}} + \envec \otms\envec
      \end{array}
    \end{array}
    \end{align}
    \end{small}
    where \eq{\sfb{s}=\tfrac{1}{r} \nab \hsfb{r}^\flt} gives the metric inherited from \eq{\sfb{\emet}_\ii{\!\Sigup}} on a codimension-1 sphere embedded in \eq{\bs{\Sigup}}. 
    Note, however, that this is the \textit{extrinsic} view of \eq{\sfb{s}} as a \textit{degenerate} tensor field on \eq{\bs{\Sigup}} (actually, on \eq{\vsp{E}} since we regard \eq{\bs{\Sigup}} itself as embedded in \eq{\vsp{E}}) 
    and the notation \eq{\inv{\sfb{s}}} should then be interpreted with care.\footnote{Note that \eq{\inv{\sfb{s}}\cdot\sfb{s}= \iden_\ii{\!\Sigup} - \be_{r}\otimes\bep^{r}} is not the identity on \eq{\bs{\Sigup}} but is, rather, the identity on \eq{\man{S}^{\en - 2}} viewed extrinsically in \eq{\bs{\Sigup}}. Further, since \eq{\bs{\Sigup}\subset\vsp{E}} is itself viewed extrinsically in \eq{\bvsp{E}}, we can write this as:\\
    \eq{\qquad }  \eq{\inv{\sfb{s}}\cdot\sfb{s}= \iden_\ii{\!\Sigup} - \be_{r}\otms\bep^{r} \,=\, \bar{\iden}_\ii{\!\Evec} - \be_{r}\otms\bep^{r} - \envec \otms \enform }  .}
      Using the above, we then find that the  \eq{\psi}-induced metric, \eq{\sfb{g}:=\psi^*\bsfb{\emet}_\ii{\!\Evec} \in\tens^0_2(\bvsp{E})}, from Eq.\eqref{g_proj_active} may now be written as:
      \begin{small}
    \begin{align} \label{g_prj_sphere}
    \begin{array}{lllll}
              \sfb{g} 
             \,=\, \hsfb{r}^\flt \otms \hsfb{r}^\flt \,+\, \tfrac{1}{r_\en^2} \sfb{s} \,+\, \tfrac{1}{r_\en^4 }\enform  \otms \enform 
       \qquad,\qquad 
             \inv{\sfb{g}} 
              \,=\,  \hsfb{r} \otms \hsfb{r} \,+\, r_\en^2 \inv{\sfb{s}} \,+\, r_\en^4 \envec \otms \envec 
    \end{array}
    \end{align}
    \end{small}
        \item \textit{Example for \eq{\man{S}^2} (\eq{\en =4}).} 
        To make clear that \eq{\sfb{s}} is the metric on a sphere, consider the case \eq{\en=4} such that \eq{\man{S}^2\subset\bs{\Sigup}^3\subset \vsp{E}^4} is the standard unit 2-sphere embedded in Euclidean 3-space (which is then embedded in 4-space but this has no impact on \eq{\man{S}^2}). Take local coordinates  \eq{(r,\theta,\phi,r^\en):\chart{E}{}^4\to\mbb{R}^4}  where  \eq{(r,\theta,\phi)} are the usual spherical coordinates on the 3-dim hyperplane \eq{\bs{\Sigup}\subset \vsp{E}}, with \eq{\rfun =\nrm{\sfb{r}}}. Writing \eq{\sfb{\emet}_\ii{\!\Sigup}} in the spherical coordinate basis \eq{(\be[r],\be[\theta],\be[\phi])}, it is quick to verify that the equations for \eq{\sfb{s}} and \eq{\inv{\sfb{s}}} in  Eq.\eqref{2sphere_met} lead to the familiar expressions:
        \begin{small}
        \begin{align}
             \sfb{s} 
           \,=\,  \tfrac{1}{r^2}( \sfb{\emet}_\ii{\!\Sigup} - \bep[r] \otms \bep[r]  ) 
            \,=\,  \bep[\theta]\otms \bep[\theta] + \sin^2 \theta \bep[\phi]\otms\bep[\phi] 
        &&,&&
             \inv{\sfb{s}} 
            \,=\, r^2( \inv{\sfb{\emet}_\ii{\!\Sigup}} - \be[r] \otms \be[r] )
            \,=\,   \be[\theta]\otms \be[\theta] \,+\, \tfrac{1}{\sin^2\theta} \be[\phi]\otms \be[\phi]
        \end{align}
        \end{small}
        where \eq{\bep[r]=\dif r \equiv \hsfb{r}^\flt} and \eq{\be[r]\equiv\hsfb{r}}. The above is the usual local expression for the metric on \eq{\man{S}^2} and whether it is the extrinsic or intrinsic view depends on which view is applied to the \eq{(\theta,\phi)} frame fields (we have implied the extrinsic view).  
\end{itemize}
\end{small}

\subsection{Transformation of the Original Hamilton System} \label{sec:prj_Xform_gen}

We now use the projective transformation, \eq{\psi\in\Dfism(\bvsp{E})} and \eq{\colift\psi\in\Spism(\cotsp\bvsp{E},\nbs{\omg})}, to transform the original Hamiltonian system that was detailed in section \ref{sec:Hnom_prj}. The general method and important details for doing so using \textit{any} configuration manifold diffeomorphism was developed in section \ref{sec:Hmech_xform_gen}. Now, all we must do is apply that procedure for the case at hand.

\paragraph{Review:~the Original System.}  
Recall that the original system is a mechanical system of a particle of mass \eq{m} moving in \eq{(\vsp{E},\bsfb{\emet}_\ii{\!\Evec}} subject to conservative forces corresponding to a 
potential
\eq{V=V^\zr(\rfun)+V^\ss{1}(\tup{r})}
(non-conservative forces are added later). 
Thus, our original Hamiltonian  system is \eq{(\cotsp\vsp{E},\nbs{\omg}, K )} where \eq{K\in\fun(\cotsp\vsp{E})} is a mechanical Hamiltonian for the mass-scaled Euclidean metric, \eq{m\bsfb{\emet}_\ii{\!\Evec}\in\tens^0_2(\vsp{E})}, and
potential
\eq{V\in\fun(\vsp{E})}\footnote{As usual, for formulations on \eq{\cotsp\vsp{E}}, the potential is treated as \eq{V\equiv\copr^*V\in\fun(\cotsp\vsp{E})}. }
given as follows for any \eq{\bar{\kap}_{\barpt{x}}=(\barpt{x},\barbs{\kap})\in\cotsp\vsp{E}}:
\begin{small}
\begin{align} \label{Kprj_4dim_2} 
    K(\barpt{x},\barbs{\kap}) = \tfrac{1}{2}\inv{\sfb{m}}(\barbs{\kap},\barbs{\kap}) + V(\ptvec{x})
    && &&
    \fnsize{i.e., } \;\;  K = \tfrac{1}{2} m^{\a\b} \plin_\a \plin_\b + V
    = \tfrac{1}{2m} (\nrmtup{\plin}^2 + \plin_\en^2)  + V
    &&,&&
    \begin{array}{lll}
         \sfb{m} := m \bsfb{\emet}_\ii{\!\Evec} = m \inner{\cdot}{\cdot}
     \\[4pt]
         V = V^\zr(\rfun) + V^\ss{1}(\tup{r})
    \end{array}
\end{align}
\end{small}
where \eq{(\bartup{r},\bartup{\plin}) =(\tup{r},r^\en,\tup{\plin},\plin_\en) = \colift \bartup{r}:\cotsp\vsp{E} \to \mbb{R}^{\ii{2}\en}} are cotangent-lifted inertial cartesian coordinates such that \eq{m_{\a\b}=m \emet_{\a\b} = m \kd^\a_\b} and \eq{\pd_\a m_{\b\gam} = 0 = \pd_\a m^{\b\gam}}. Recall also that the potential is such that  \eq{\pd_\en V = 0} and  \eq{V=V^\zr+V^\ss{1}} where \eq{V^\zr =V^\zr(\rfun)} depends only on \eq{\rfun=\nrm{\sfb{r}}}, and \eq{V^\ss{1}=V^\ss{1}(\tup{r})} accounts for all other conservative perturbations (see Remark \ref{rem:U4}, Eq.\eqref{U4_r}).
The original dynamics, \eq{ \sfb{X}^\ss{K} = \inv{\nbs{\omg}}(\dif K,\cdot)\in\vechm(\cotsp\vsp{E},\nbs{\omg})},  are then given as follows (where \eq{\plin^\a:=\emet^{\a\b} \plin_\b}):
\begin{small}
\begin{align}       
    \sfb{X}^\ss{K} 
    \,=\,  \upd^\a K \hbpart{\a} \,-\,  \pd_\a K\hbpartup{\a} 
    \;=\; m^{\a\b} \plin_\b \hbpart{\a} \,-\, \pd_i V \hbpartup{i} 
    && \Rightarrow &&
    \begin{array}{lllll}
         \dot{r}^i = \tfrac{1}{m} \plin^i   &,\quad \dot{\plin}_i = -\pd_i V \,=\, -(\hat{r}_i \pd_{\rfun} V^\zr + \pd_i V^\ss{1})
     \\[4pt]
         \dot{r}^\en = \tfrac{1}{m} \plin^\en  &,\quad \dot{\plin}_\en = 0
    \end{array}
\end{align}
\end{small}
If \eq{\bar{\kap}_t=(\barpt{x}_t,\barbs{\kap}_t)\in\cotsp\vsp{E}} is an integral curve (i.e., \eq{\dt{\bar{\kap}}_t=\sfb{X}^\ss{K}_{\bar{\kap}_t}}) then \eq{\barbs{\kap}_t=\sfb{m}(\dt{\bsfb{x}}_t)\in\cotsp[\barpt{x}_t]\vsp{E}} is the (Euclidean) kinematic momentum covector along the base curve \eq{\barpt{x}\in\vsp{E}}. 
Also, as previously discussed in Remark \ref{rem:E3dyn_subman}, \eq{r^\en} is cyclic/ignorable such that \eq{\plin_\en} is an integral of motion and each \eq{\cotsp\Sig_\ii{b}\subset\cotsp\vsp{E}} is a  \eq{(2\en -2)}-dim  \eq{\sfb{X}^\ss{K}}-invariant submanifold (where each \eq{\Sig_\ii{b}:= \inv{(r^\en)}\{b\} = \bs{\Sigup}\oplus\{b\}} is a Euclidean hyperplane in \eq{\vsp{E}}, normal to \eq{\vsp{N}}).

\begin{small}
\begin{notesq}
    We are going to transform the above system using the projective transformation,  \eq{\psi\in\Dfism(\bvsp{E})}, which is only a diffeomorphism on the region \eq{\bvsp{E}=\bs{\Sigup}_\nozer\oplus\vsp{N}_\ii{+}\subset\vsp{E}}. Thus, to expedite subsequent developments, we will from now on restrict the original system described above to the phase space \eq{(\cotsp\bvsp{E},\nbs{\omg}\equiv\nbs{\omg}|_\ii{\cotsp\bvsp{E}})},
    which differs from \eq{(\cotsp\vsp{E},\nbs{\omg})} only in that \eq{\bvsp{E}=\bs{\Sigup}_\nozer\oplus \vsp{N}_\ii{+}} excludes any \eq{\barpt{x}=\ptvec{x}+x^\en\envec\in\vsp{E}} for which \eq{\nrm{\pt{x}}=0} or \eq{x^\en\leq 0}.\footnote{At the end of the day, we anyway do not care about what happens to the original system in the extra dimension, \eq{\vsp{N}}, so the restriction to \eq{x^\en > 0} does not exclude any physically meaningful configurations.}
\end{notesq}
\end{small}

\paragraph{The Transformed System.}
To transform the above system by the projective transformation, we simply follow the procedure given in section \ref{sec:Hmech_xform_gen} using the map \eq{\psi\in\Dfism(\bvsp{E})} specified in Eq.\eqref{prj_def_act}, along with  its cotangent lift, \eq{\colift\psi\in\Spism(\cotsp\bvsp{E},\nbs{\omg})}, 
given by Eq.\eqref{colift_prj_def} – \ref{colift_prj_def1} :
\begin{small}
\begin{align} \label{prj_active2_again}
\begin{array}{ccc}
      (\barpt{x},\barbs{\kap}) \,=\, \colift \psi (\barpt{q},\barbs{\mu}) \,=\, ( \psi(\barpt{q}), \psi_*(\barbs{\mu})) 
    \\[5pt]
    \begin{pmatrix}
          \barpt{x}  \,=\,  \tfrac{1}{q_\en}\hpt{q} \,+\, \nrm{\pt{q}}\envec 
    \\
          \barbs{\kap} \,=\,  q^\en \nrm{\pt{q}} \big( \bs{\mu} - (\bs{\mu}\cdot\hsfb{q}) \hsfb{q}^\flt \big) \,-\, q_\en^2 \mu_\en \hsfb{q}^{\flt} \,+\, (\bs{\mu}\cdot \hsfb{q})\enform 
    \end{pmatrix}
\end{array}    
  && \leftrightarrow &&
\begin{array}{ccc}
    (\barpt{q},\barbs{\mu}) \,=\, \colift \inv{\psi} (\barpt{x},\barbs{\kap}) \,=\, ( \inv{\psi}(\barpt{x}), \psi^*(\barbs{\kap}))
\\[5pt]
\begin{pmatrix}
     \barpt{q}   \,=\, x^\en \hpt{x} \,+\, \tfrac{1}{\nrm{\pt{x}}} \envec 
\\
    \barbs{\mu}  \,=\,  \tfrac{\nrm{\pt{x}}}{x_\en} \big( \bs{\kap} - (\bs{\kap}\cdot\hsfb{x}) \hsfb{x}^\flt \big) \,+\, \kap_\en \hsfb{x}^\flt \,-\, \nrm{\pt{x}}^2 (\bs{\kap}\cdot\hsfb{x})\enform 
\end{pmatrix}
\end{array}
\end{align}
\end{small}
We know from Eq.\eqref{Hsystem_gen} that our 
\eq{\colift\psi}-related Hamiltonian system is \eq{(\cotsp\bvsp{E},\nbs{\omg}, H )} with \eq{H:=\colift\psi^* K \in\fun(\cotsp\bvsp{E})} a mechanical Hamiltonian for \textit{non-Euclidean} metric \eq{\sfg:=\psi^*\sfb{m}\in\tens^0_2(\bvsp{E})} and 
potential \eq{U:=\psi^* V\in\fun(\bvsp{E})}.\footnote{Again, \eq{U\in\fun(\bvsp{E})} is treated as a basic function \eq{U\equiv\copr^* U\in\fun(\cotsp\bvsp{E})}.}
For any \eq{\bar{\mu}_{\barpt{q}}=(\barpt{q},\barbs{\mu})\in\cotsp\bvsp{E}}:
\begin{small}
\begin{align} \label{Hfun_prj}
 \left. \begin{array}{lllll}
    H:= \colift\psi^* K 
\\[3pt]
      \sfg := \psi^* \sfb{m} 
\\[3pt]
      U:= \psi^* V 
 \end{array}
 \right\} \quad \Rightarrow 
 \qquad
 \begin{array}{cccc}
         H(\barpt{q},\barbs{\mu}) \,=\, \tfrac{1}{2}\inv{\sfg}_\ss{\barpt{q}}(\barbs{\mu},\barbs{\mu}) \,+\, U(\barpt{q})
 \\[6pt]
       \sfb{X}^\ss{H} \,=\, \inv{\nbs{\omg}}(\dif H,\cdot) \,=\, \colift\psi^* \sfb{X}^\ss{K}  
 \end{array}
 \end{align}
 \end{small}
In cotangent-lifted cartesian coordinates, \eq{(\bartup{r},\bartup{\plin})}, the new Hamiltonian and its associated Hamiltonian vector field are then:
 \begin{small}
 \begin{align}\label{Hdyn_prj}
 \begin{array}{cc}
      H = \tfrac{1}{2}g^{\a\b} \plin_\a \plin_\b \,+\, U
 \\[5pt]
      \sfb{X}^\ss{H} 
       \,=\, \upd^\a H \hbpart{\a} \,-\,  \pd_\a H\hbpartup{\a} 
    \;=\; 
    g^{\a\b} \plin_\b \hbpart{\a} \,+\, (  g^{\gam \sig} \Gamma^\b_{\a\gam} \plin_\b \plin_\sig \,-\, \pd_\a U) \hbpartup{\a} 
 \end{array}
 &&\Rightarrow &&
 \begin{array}{llll}
      \dot{r}^\a \,=\,  \upd^\a H \,=\, g^{\a\b} \plin_\b
    \\[5pt] 
     \dot{\plin}_\a \,=\, - \pd_\a H \,=\,  g^{\gam \sig}\Gamma^\b_{\a \gam} \plin_\b \plin_\sig \,-\, \pd_\a U 
 \end{array}
 \end{align}
 \end{small}
 Where the ODEs on the right determine the \eq{(\bartup{r},\bartup{\plin})}-representation of the integral curves of \eq{\sfb{X}^\ss{H}}, where \eq{\Gamma^{\a}_{\b\gam}=\Gamma^{\a}_{\gam\b}} are the \eq{r^\a} basis Levi-Civita connection coefficients for \eq{\sfg}, and we note that  \eq{\sfg = m \sfb{g}} and \eq{ \inv{\sfg}=\tfrac{1}{m}\inv{\sfb{g}}} are just a scaling of what was given in Eq.\eqref{gin_proj_active}. Importantly, as will soon be shown, the new potential \eq{U=\psi^* V} has the form \eq{U=U^\zr(r^\en)+U^\ss{1}(\bartup{r})}.  
The coordinate expressions for the above math the be written as 
follows:\footnote{The second terms in the \eq{\dot{r}^i} and \eq{\dot{\plin}_i} equations are a result of the \eq{(\htup{r}\cdot\tup{\plin})^2} term in \eq{H=\colift\psi^*K}. This term is, in turn, a direct consequence of including the ``extra'' \eq{\plin_\en^2} term in the original Hamiltonian, \eq{K}.  }
\begin{small}
\begin{gather} \nonumber
  H :=\colift\psi^* K \,=\,  \tfrac{1}{2}g^{\a\b} \plin_\a \plin_\b \,+\, U \,=\, \tfrac{r_\en^2}{2m} ( \lang^2 + r_\en^2 \plin_\en^2 ) \,+\, \tfrac{1}{2m} (\htup{r} \cdot\tup{\plin})^2 \,+\,  U^\zr(r^\en) + U^\ss{1}(\bartup{r}) 
 \\[3pt]  \label{Hprj_full_alt} 
\begin{array}{lllllll}
     \dot{r}^i =  \upd^i H 
     \,=\, -\tfrac{r_\en^2}{m} \lang^{ij}r_j \,+\,  \tfrac{1}{m}(\htup{r}\cdot\tup{\plin})\hat{r}^i
\\[4pt]
   \dot{r}^\en =  \upd^\en H 
   \,=\, \tfrac{r_\en^4}{m}  \plin^\en
\end{array}
\qquad,\qquad 
 \begin{array}{lllllll}  
      \dot{\plin}_i  = - \pd_i H 
     \,=\, -\tfrac{r_\en^2}{m} \lang_{ij} \plin^j 
        \,+\, \tfrac{(\htup{r}\cdot\tup{\plin})}{m \nrmtup{r}^3} \lang_{ij}r^j
        \,-\, \pd_i U^\ss{1} 
\\[4pt]
      \dot{\plin}_\en = - \pd_\en H 
      \,=\, - \tfrac{r_\en}{m}  ( \lang^2 + 2 r_\en^2 \plin_\en^2 ) \,-\, \pd_\en (U^\zr + U^\ss{1})
\end{array}
\end{gather}
\end{small}
where \eq{\lang^{ij}=r^i \plin^j- \plin^i r^j} and  \eq{\lang^2 =\nrmtup{r}^2 \nrmtup{\plin}^2 - (\tup{r}\cdot\tup{\plin})^2 } satisfy Eq.\eqref{angmoment_rels_prj}. 
Explicit expressions in the footnote\footnote{Expressing everything explicitly in terms of \eq{(\bartup{r},\bartup{\plin})}, then \eq{ H = \tfrac{r_\en^2}{2m} \big( \nrmtup{r}^2 \nrmtup{\plin}^2 - (\tup{r}\cdot\tup{\plin})^2 + r_\en^2 \plin_\en^2 \big) + \tfrac{1}{2m} (\htup{r} \cdot\tup{\plin})^2 + U^\zr(r^\en) + U^\ss{1}(\bartup{r}) }, and the dynamics are:
\begin{align} \label{Hprj_full}
  \begin{array}{lllllll}
     \dot{r}^i 
     \,=\,  \tfrac{r_\en^2}{m} \big( \nrmtup{r}^2 \plin^i  - (\tup{r}\cdot \tup{\plin} ) r^i \big) \,+\,  \tfrac{1}{m}(\htup{r}\cdot\tup{\plin})\hat{r}^i
\\[4pt]
   \dot{r}^\en 
   \,=\, \tfrac{r_\en^4}{m}  \plin^\en
\end{array}
\qquad,\qquad 
 \begin{array}{lllllll}  
      \dot{\plin}_i 
      \,=\,  - \tfrac{r_\en^2}{m}   \big( \nrmtup{\plin}^2 r_i - (\tup{r}\cdot\tup{\plin}) \plin_i \big) 
      \,-\, \tfrac{(\htup{r}\cdot\tup{\plin})}{m \nrmtup{r}^3}  \big( \nrmtup{r}^2 \plin_i - (\tup{r}\cdot\tup{\plin}) r_i \big)
      \,-\, \pd_i U^\ss{1}
\\[4pt]
      \dot{\plin}_\en 
      \,=\, 
     - \tfrac{r_\en}{m}  \big( \nrmtup{r}^2 \nrmtup{\plin}^2 - (\tup{r}\cdot\tup{\plin})^2 + 2 r_\en^2 \plin_\en^2 \big) \,-\, \pd_\en (U^\zr + U^\ss{1})
\end{array}  
\end{align}
}.


\begin{small}
\begin{itemize}[nosep]
    \item[{}]  \textit{Looking ahead:}
    Other than the fact that \eq{U^\zr} does not appear in the equations for \eq{\dot{\plin}_i}, the above dynamics for \eq{H} are rather hideous and it is not at all clear how this system is ``better'' than the original system for \eq{K}. Although the above will soon be simplified (section \ref{sec:IOMs_prj_act}), the advantages will still not be clear until later sections when the transformation \eq{\colift\psi} is combined with a transformation of the evolution parameter. Then, 
    the above dynamics for \eq{(r^i,\plin_i)} will end up being fully linear in the case  of arbitrary central force dynamics (i.e., for the case \eq{V=V^\zr(\rfun)}, which transforms to \eq{U=\psi^* V^\zr =U^\zr(r^\en)}).  The dynamics for \eq{(r^\en,\plin_\en)} will be also be linear but only for central forces corresponding to a \textit{original} potential of the particular form \eq{V^\zr = -\sck_1/\rfun - \sck_2/\rfun^2 } for any \eq{\sck_1,\sck_2\in\mbb{R}} (which transforms to \eq{U^\zer = -\sck_1 r_\en - \sck_2 r_\en^2}). 
    The Kepler-Coulomb problem corresponds to the case \eq{V^\zr = -\sck/\rfun \mapsto U^\zr = -\sck r_\en}. 
\end{itemize}
\end{small}

\vspace{1ex} \noindent Now, the above dynamics are not particularly illuminating nor will we have much use for them in their present form. It will, however, be useful to summarize how various objects (metric, forces, momentum covectors, etc.) transform between the original system for \eq{K} and the new system for \eq{H=\colift\psi^* K }:
 \begin{small}
\begin{itemize}[nosep]
  \item  \sbemph{Kinetic Energy Metric.}  The induced kinetic energy metric seen above is defined by \eq{\sfg:= \psi^*\sfb{m} = m \psi^*\bsfb{\emet}_\ii{\!\Evec} = m \sfb{g} \in\tens^0_2(\bvsp{E})} and \eq{\inv{\sfg}=\psi^*\inv{\sfb{m}} = \tfrac{1}{m} \psi^* \inv{\bsfb{\emet}_\ii{\!\Evec}} = \tfrac{1}{m}\inv{\sfb{g}} \in\tens^2_0(\bvsp{E})}, where \eq{\sfb{g}:=\psi^*\bsfb{\emet}_\ii{\!\Evec}} and \eq{\inv{\sfb{g}}=\psi^*\inv{\bsfb{\emet}_\ii{\!\Evec}}} were  given in Eq.\eqref{g_proj_active} – \ref{gin_proj_active}. 
  We simply scale by mass:\footnote{Recall that \eq{\tfrac{1}{r}(\sfb{\emet}_\ii{\!\Sigup}-\hsfb{r}^\flt\otms \hsfb{r}^\flt)= \nab\hsfb{r}^\flt } such that \eq{\sfg} can also be written as: \\
    \eq{\qquad\qquad \qquad} \eq{\sfg =m\big( \tfrac{1}{r_\en^2\, r} \nab\hsfb{r}^\flt + \hsfb{r}^\flt \otms \hsfb{r}^\flt + \tfrac{1}{r_\en^4 }\enform  \otms \enform  \big)}. }
     \begin{small}
     \begin{align} \label{prj_metric_again}
    \begin{array}{rlllll}
          \sfg \,:=\, \psi^*\sfb{m}  &\!\!\! =\, m\Big( \tfrac{1}{r_\en^2\, r^2} (\sfb{\emet}_\ii{\!\Sigup}-\hsfb{r}^\flt\otms \hsfb{r}^\flt) \,+\, \hsfb{r}^\flt \otms \hsfb{r}^\flt \,+\, \tfrac{1}{r_\en^4 }\enform  \otms \enform  \Big)
    \\[10pt]
         \inv{\sfg} = \inv{\psi}_* \inv{\sfb{m}}  \! &\!\!\! =\, \tfrac{1}{m} \Big( r_\en^2 r^2 ( \inv{\sfb{\emet}_\ii{\!\Sigup}} - \hsfb{r}\otms \hsfb{r}) \,+\, \hsfb{r}\otms\hsfb{r} \,+\, r_\en^4 \envec \otms \envec \Big)
    \end{array}
    &&
    \begin{array}{rllll}
           \cord{g}{\bartup{r}} &\!\!\!\! =  m\, \fnsz{ \begin{pmatrix}
        \tfrac{1}{r_\en^2 \nrmtup{r}^2}( \emet_{ij} - \hat{r}_i\hat{r}_j) + \hat{r}_i\hat{r}_j &  \tup{0} \\
        \trn{\tup{0}} & \tfrac{1}{r_\en^4}
        \end{pmatrix}}
      \\[12pt] 
        \cord{\inv{g}}{\bartup{r}} &\!\!\!\! = \tfrac{1}{m}\fnsz{ \begin{pmatrix}
         r_\en^2 \nrmtup{r}^2 ( \emet^{ij} - \hat{r}^i\hat{r}^j) + \hat{r}^i \hat{r}^j & \tup{0}
         \\[3pt]
        \trn{\tup{0}}   & r_\en^4
        \end{pmatrix} }
    \end{array}
     \end{align}
     \end{small}
\item  \sbemph{Flows \& Integral Curves.} Since \eq{\sfb{X}^\ss{K}=\colift\psi_*\sfb{X}^\ss{H}} and \eq{\sfb{X}^\ss{H} = \colift\psi^* \sfb{X}^\ss{K}} are \eq{\colift\psi}-related, we know from Remark \ref{rem:HK_related} that: 
    \begin{small}
    \begin{itemize}[nosep]
        \item If \eq{\varphi_t \in \Spism(\cotsp\bvsp{E},\nbs{\omg})} is the flow of \eq{\sfb{X}^\ss{K}}, then the flow of \eq{\sfb{X}^\ss{H}} is given by \eq{\phi_t =  \inv{\colift\psi} \circ \varphi_t \circ \colift\psi\in\Spism(\cotsp\bvsp{E},\nbs{\omg}) }. 
        \item If \eq{\bar{\kap}_t=(\barpt{x}_t,\barbs{\kap}_t)\in\cotsp\bvsp{E}} is an integral curve of \eq{\sfb{X}^\ss{K}}, then \eq{\bar{\mu}_t=(\barpt{q}_t,\barbs{\mu}_t) = \inv{\colift\psi}(\bar{\kap}_t)\in\cotsp\bvsp{E}} is an integral curve of \eq{\sfb{X}^\ss{H}}.
        Additionally, \eq{ \barbs{\kap}_t=\sfb{m}_{\barpt{x}}(\dt{\bsfb{x}}_t)} and \eq{\barbs{\mu}_t  =\sfg_\ss{\!\barpt{q}}(\dt{\bsfb{q}}_t)} are are the kinematic momentum covectors along their respective base curves,  \eq{\barpt{x}_t} and \eq{\barpt{q}_t}, but with \eq{\barbs{\mu}_t} given by the non-Euclidean kinetic energy metric, \eq{\sfg=\psi^*\sfb{m}} (more details in section \ref{sec:prj_momentum}).
        The explicit expressions for the transformation \eq{\bar{\kap}_t\leftrightarrow \bar{\mu}_t} is given in Eq.\eqref{prj_active2} but the key point is:
        \begin{small}
        \begin{align} \label{intcurv_xform_prj}
           \fnsize{if:}\quad \dt{\bar{\kap}}_t = \sfb{X}^\ss{K}_{\bar{\kap}_t} 
            \quad \fnsize{\&} \quad 
            \dt{\bar{\mu}}_t = \sfb{X}^\ss{H}_{\bar{\mu}_t}
            && \Rightarrow  &&
            \begin{array}{rlclll}
                  &\bar{\kap}_t = \colift\psi(\bar{\mu}_t) & \leftrightarrow & \bar{\mu}_t = \inv{\colift\psi}(\bar{\kap}_t)
              \\[5pt]
                \fnsize{with:} &\barbs{\kap}_t=\sfb{m}(\dt{\bsfb{x}}_t)
                 & \leftrightarrow &
                 \barbs{\mu}_t  =\sfg_\ss{\!\barpt{q}}(\dt{\bsfb{q}}_t)
            \end{array}
        \end{align}
        \end{small}
        \item As shown in Eq.\eqref{Hp_rels_gen}, the above implies the following hold along any such \eq{\bar{\kap}_t} and \eq{\bar{\mu}_t} (the potential functions are discussed next):
        \begin{small}
        \begin{align} \label{Hp_rels_prj}
             H(\barpt{q}_t,\barbs{\mu}_t) \,=\,  K(\barpt{x}_t,\barbs{\kap}_t) 
          \quad\;\; , \;\;\quad 
             \inv{\sfg}_\ss{\!\barpt{q}}(\barbs{\mu},\barbs{\mu}) = \sfg_\ss{\!\barpt{q}}(\dt{\bsfb{q}}, \dt{\bsfb{q}}) \,=\, \sfb{m}_{\barpt{x}}(\dt{\bsfb{x}}, \dt{\bsfb{x}}) = \inv{\sfb{m}}_{\barpt{x}}(\barbs{\kap},\barbs{\kap}) 
        \quad\;\; , \;\;\quad 
        \begin{array}{rll}
            &U(\barpt{q}) = V(\ptvec{x}) 
            \\[3pt]
            \fnsize{with:}\!\!\! & U^\zr(q^\en) = V^\zr(\nrm{\pt{x}}) 
        \end{array}
        \end{align}
        \end{small}
  \end{itemize}
  \end{small}
 \item  \sbemph{Potential \& Conservative Forces.} Unlike the original potential, \eq{V}, the transformed potential, \eq{U:=\psi^*V}, is no longer independent of \eq{r^\en}. Whereas \eq{V} was of the form \eq{V=V^\zr(\rfun)+V^\ss{1}(\tup{r})}, the new \eq{U} is of the form \eq{U=U^\zr(r^\en)+ U^\ss{1}(\bartup{r})} where the radial potential, \eq{V^\zr}, is transformed to a potential \eq{U^\zr = \psi^* V^\zr} that depends \textit{only} on \eq{r^\en} (and \eq{U^\ss{1}= \psi^* V^\ss{1}} accounts for all other conservative forces):
 \begin{small}
 \begin{align} \label{Uprj}
 \begin{array}{rllll}
     \fnsize{original:} &V \,=\, V^\zr(\rfun) \,+\, V^\ss{1}(\tup{r})
      &\quad,\qquad 
      \pd_\en V = 0
      &\quad,\qquad 
      \dif V \,=\, \pd_i V \hbep^i \,=\, \pd_{\rfun} V^\zr \hsfb{r}^\flt \,+\, \pd_i V^\ss{1} \hbep^i
 \\[5pt]
        \fnsize{transformed:} &U :=\psi^* V = U^\zr(r^\en) + U^\ss{1}(\bartup{r})
     &\quad,\qquad 
     \pd_i U^\zr = 0  
      &\quad,\qquad  
     \dif U \,=\, \pd_\a U \hbep^\a \,=\,  \pd_\en U^\zr \enform \,+\, \pd_\a U^\ss{1} \hbep^\a
 \end{array}
 \end{align}
 \end{small}
 In particular, we will see that \eq{\pd_\en V =\dif V \cdot \envec =0} leads to \eq{\dif U \cdot \hsfb{r}=0}:
\begin{small}
\begin{align}
    \lderiv{\envec}V = \dif V \cdot \envec = \pd_\en V  =0 \qquad \Leftrightarrow \qquad  \lderiv{\hsfb{r}} U =  \dif U \cdot \hsfb{r} = \hat{r}^i \pd_i U = 0
    &&
    (\fnsize{and }  \; r^i \pd_i U = 0)
\end{align}
\end{small}
This is verified by expressing \eq{\dif U} in terms of  \eq{\dif V} via  \eq{\dif U = \psi^* \dif V  = \trn{\dif \psi}\cdot \dif V_\ii{\psi}}, 
leading to (using \eq{\envec\cdot\dif V=0}):\footnote{For \eq{\barpt{x}=\psi(\barpt{q})}, the actual expression for \eq{\dif U = \psi^* \dif V = \dif V_\ii{\psi} \cdot \dif \psi} is given as follows, using the same pullback expression as in Eq.\eqref{lift_prj_act3}:
 \begin{align}
     \dif U_\ss{\barpt{q}} \,=\, \pd_i U_\ss{\barpt{q}} \hbep^i + \pd_\en U_\ss{\barpt{q}}\,  \enform 
       \;=\;  \trn{\dif \psi}_\ss{\barpt{q}}   \cdot  \dif V_{\barpt{x}} 
         \,=\,  \tfrac{1}{q^\en \nrm{\pt{q}}} \big( \trn{\iden}  - \hpt{q}^\flt \otms \hpt{q} \big) \cdot \dif V_{\pt{x}} \,-\, \tfrac{1}{q_\en^2} (\dif V_{\pt{x}} \cdot\hpt{q})\enform   
         \;=\;  \tfrac{1}{q^\en \nrm{\pt{q}}} \big( \pd_i V_{\pt{x}}  - \hat{q}_i\hat{q}^j \pd_j V_{\pt{x}} \big) \hbep^i 
         \,-\,\tfrac{1}{q_\en^2} \hat{q}^i\pd_i V_{\pt{x}}\, \enform
 \end{align} 
 Equating coefficients then leads to Eq.\eqref{dU_xform_prj}.  }
\begin{small}
\begin{align}
\begin{array}{rllll}
         & \dif U  \,=\, \trn{\dif \psi}\cdot \dif V_\ii{\psi}
      \,=\, \tfrac{1}{r_\en r} ( \trn{\iden}_{\ii{\!\Sig}} - \hsfb{r}^\flt \otms \hsfb{r}) \cdot \dif V_\ii{\psi} \,-\, \tfrac{1}{r_\en^2} (\hsfb{r}\cdot \dif V_\ii{\psi}) \enform
    \\[5pt]
         &  \dif U_\ss{\barpt{q}} \,=\, \trn{\dif \psi}_\ss{\barpt{q}} \cdot \dif V_{\pt{x}}
      \,=\,   \tfrac{1}{q^\en \nrm{\pt{q}}} ( \dif V_{\pt{x}}   -  \pd_{\rfun} V_{\pt{x}} \hsfb{q}^\flt ) 
         \,-\,\tfrac{1}{q_\en^2} \pd_{\rfun} V_{\pt{x}}\, \enform 
         \;=\;  \tfrac{\nrm{{x}}}{x^\en }  ( \dif V_{\pt{x}}  -  \pd_{\rfun} V_{\pt{x}} \hsfb{x}^\flt ) 
         \,-\, \nrm{{x}}^2 \pd_{\rfun} V_{\pt{x}}\, \enform 
         \; \in\cotsp[\cdt]\bvsp{E}
\end{array}
\end{align}
\end{small}
where \eq{\barpt{x}=\psi(\barpt{q})} and thus  \eq{\hsfb{q} \equiv \hsfb{x}}. 
 The inverse relation, \eq{\dif V = \psi_* \dif U = \invtrn{\dif \psi}\cdot\dif U_\ii{\inv{\psi}}}, leads to the following 
  (using \eq{\hsfb{r}\cdot\dif U =0}):\footnote{Using the expressions in Eq.\eqref{lift_prj_act3} for the pushforward of a 1-form by \eq{\psi}, along with \eq{\hsfb{r}\cdot\dif U =0} and \eq{\barpt{x}=\psi(\barpt{q})}, we obtain: \\
  \eq{ \dif V \,=\, \psi_* \dif U \,=\, \invtrn{ \dif \psi} \cdot \dif U_\ii{\inv{\psi}} 
      \,=\, \tfrac{r^\en}{r}( \trn{ \iden_\ii{\!\Sigup} } - \hsfb{r}^\flt\otms \hsfb{r}) \cdot \dif U_\ii{\inv{\psi}} \,-\, \tfrac{1}{r^2}( \envec \cdot \dif U_\ii{\inv{\psi}} ) \hsfb{r}^\flt  \,+\, (\hsfb{r}\cdot \dif U_\ii{\inv{\psi}}) \enform
        \;=\;   \tfrac{r^\en}{r} \trn{ \iden_\ii{\!\Sigup} } \cdot \dif U_\ii{\inv{\psi}}  \,-\, \tfrac{1}{r^2}( \envec \cdot \dif U_\ii{\inv{\psi}} ) \hsfb{r}^\flt  } \\
    \eq{ \dif V_{\pt{x}} \,=\, \psi_* (\dif U_\ss{\barpt{q}}) 
        \,=\,  \invtrn{ (\dif \psi_\ss{\barpt{q}})} \cdot \dif U_\ss{\barpt{q}} \,=\, q^\en \nrm{\pt{q}} ( \trn{ \iden_\ii{\!\Sigup} } - \hsfb{q}^\flt\otms \hsfb{q}) \cdot \dif U_\ss{\barpt{q}} 
       \,-\, q_\en^2 (\envec \cdot \dif U_\ss{\barpt{q}}) \hsfb{q}^\flt  \,+\, (\hsfb{q} \cdot \dif U_\ss{\barpt{q}})  \enform
       \;=\;  q^\en \nrm{\pt{q}} \trn{ \iden_\ii{\!\Sigup} } \cdot \dif U_\ss{\barpt{q}} \,-\, q_\en^2 \pd_\en U_\ss{\barpt{q}} \hsfb{q}^\flt
       \;=\; 
        \tfrac{x^\en}{\nrm{\pt{x}}} \trn{ \iden_\ii{\!\Sigup} } \cdot \dif U_\ss{\barpt{q}} \,-\, \tfrac{1}{\nrm{\pt{x}}^2} \pd_\en U_\ss{\barpt{q}} \hsfb{x}^\flt  }.
    }
    \begin{small}
    \begin{align}
    \begin{array}{llll}
       \dif V \,=\, 
        \invtrn{ \dif \psi} \cdot \dif U_\ii{\inv{\psi}} 
       \,=\, \tfrac{r^\en}{r} \trn{ \iden_\ii{\!\Sigup} } \cdot \dif U_\ii{\inv{\psi}}  \,-\, \tfrac{1}{r^2}( \envec \cdot \dif U_\ii{\inv{\psi}} ) \hsfb{r}^\flt  
    \\[5pt]
       \dif V_{\pt{x}} \,=\, 
            \invtrn{ (\dif \psi_\ss{\barpt{q}})} \cdot \dif U_\ss{\barpt{q}} 
       \,=\, 
        \tfrac{x^\en}{\nrm{\pt{x}}} \trn{ \iden_\ii{\!\Sigup} } \cdot \dif U_\ss{\barpt{q}} \,-\, \tfrac{1}{\nrm{\pt{x}}^2} \pd_\en U_\ss{\barpt{q}} \hsfb{x}^\flt
       \;=\; q^\en \nrm{\pt{q}} \trn{ \iden_\ii{\!\Sigup} } \cdot \dif U_\ss{\barpt{q}} \,-\, q_\en^2 \pd_\en U_\ss{\barpt{q}} \hsfb{q}^\flt
        \; \in \cotsp[\cdt]\bs{\Sigup} \subset \cotsp[\cdt]\bvsp{E}
    \end{array}
    \end{align}
    \end{small}
    where \eq{\trn{\iden_\ii{\!\Sigup} } \cdot \dif U_\ss{\barpt{q}} =  \dif U_\ss{\barpt{q}} - (\envec\cdot  \dif U_\ss{\barpt{q}})\enform =  \pd_i U_\ss{\barpt{q}} \hbep^i \in\cotsp[\cdt]\bs{\Sigup}\subset \cotsp[\cdt]\bvsp{E}}.
     The  components of \eq{\dif V_{\pt{x}}} and \eq{\dif U_\ss{\barpt{q}}} in the \eq{r^\a} basis, \eq{\hbep^\a},  transform as:
    \begin{small}
    \begin{align} \label{dU_xform_prj}
    \begin{array}{llllll}
          \pd_i V_{\pt{x}} \,=\, \tfrac{x^\en}{\nrm{\pt{x}}} \pd_i U_\ss{\barpt{q}} \,-\, \tfrac{1}{\nrm{\pt{x}}^2}\hat{x}_i \pd_\en U_\ss{\barpt{q}}
        \,=\,  q^\en \nrm{\pt{q}}  \pd_i U_\ss{\barpt{q}} \,-\, q_\en^2 \hat{q}_i \pd_\en U_\ss{\barpt{q}}
        &,\qquad 
        \pd_\en V_{\pt{x}} \,=\, 0  
     \\[5pt]
           \pd_i U_\ss{\barpt{q}} \,=\,  \tfrac{1}{q^\en \nrm{\pt{q}}} \big( \kd^j_i   - \hat{q}_i\hat{q}^j  \big) \pd_j V_{\pt{x}}
         \,=\,  \tfrac{\nrm{\pt{x}}}{x^\en } \big( \pd_i V_{\pt{x}}  - \hat{x}_i \pd_{\rfun} V_{\pt{x}} \big)
       &,\qquad 
          \pd_\en U_\ss{\barpt{q}}  \,=\, -\tfrac{1}{q_\en^2} \hat{q}^i\pd_i V_{\pt{x}}
          \,=\, - \nrm{\pt{x}}^2 \pd_{\rfun} V_{\pt{x}}
    \end{array}
    \end{align}
    \end{small}
    where \eq{ \pd_\a U_\ss{\barpt{q}}:=\pderiv{U}{r^\a}|_\ss{\barpt{q}} } and  \eq{\pd_\a V_{\pt{x}} := \pderiv{V}{r^\a}|_{\pt{x}}\,} and where \eq{\barpt{x}=\psi(\barpt{q})} such that:
    \begin{small}
    \begin{align}
        x^\en = \nrm{\pt{q}} 
        &&,&&
        q^\en = 1/\nrm{\pt{x}}
        &&,&&
        \hsfb{x} = \hsfb{q}
        &&,&&
        \pd_{\rfun} V_{\pt{x}} \,=\,  \pderiv{V}{r}|_{\pt{x}} \,=\, \hsfb{x} \cdot \dif V_{\pt{x}}  = \hsfb{q} \cdot \dif V_{\pt{x}}   \,=\, \hat{x}^i \pd_i V_{\pt{x}} \,=\, \hat{q}^i \pd_i V_{\pt{x}}
    \end{align}
    \end{small}
\end{itemize}
\end{small}

\subsection{Transformed Momentum, Velocity, \& Newton-Riemann Dynamics} \label{sec:prj_momentum}

\begin{small}
\begin{notation} For the following developments:
\begin{small}
\begin{itemize}[nosep]
    \item[–] \eq{\nab = \nab^\ii{\bsfb{\emet}_\ii{\!\Evec}} = \nab^\ss{\sfb{m}} } denotes the LC connection for the Euclidean kinetic energy metric, \eq{\sfb{m}= m \bsfb{\emet}_\ii{\!\Evec} \in\tens^0_2(\vsp{E}) }, and \eq{\nab^\ss{\sfb{g}}=\nab^\ss{\sfg}} denotes the LC connection for the \eq{\psi}-induced metric, \eq{\sfg:=\psi^*\sfb{m}\in\tens^0_2(\bvsp{E})}.\footnote{For scalar constant \eq{c\in\mbb{R}}, then scaling a metric by \eq{c} does not affect the LC connection of that metric. }  
    \item[–] Much of the following is for various objects evaluated along curves but, for brevity, we often omit the argument \eq{t} from expressions. 
    \item[–] for any curve \eq{\barpt{s}_t= \ptvec{s}_t + s^\en_t\envec\in \vsp{E}=\bs{\Sigup}\oplus \vsp{N}}, we denote by \eq{\dot{s}(t)} what one might think of as \eq{\diff{}{t}\nrm{\ptvec{s}_t}},  and which
    ``really means'' \eq{\dot{s}(t) := \dtsfb{s}_t \cdot \dif r_{\pt{s}_t}}:\footnote{\eq{\lderiv{\dtsfb{s}} r} is an abuse of notation; the Lie derivative is defined only for vector \textit{fields}. Yet, the lie derivative of a function is simply \eq{\lderiv{\sfb{X}}f = \dif f \cdot \sfb{X}} which does not depend on any derivative of \eq{\sfb{X}}. Thus, for any \eq{\sfb{u}\in\tsp[\pt{r}]\man{X}} we define the abuse of notation \eq{\lderiv{\sfb{u}}f := \dif f_\pt{r}\cdot\sfb{u}} (sometimes seen as \eq{\sfb{u}(f)} or \eq{\sfb{u}[f]}). }
\begin{small}
\begin{align} \label{magdot_def}
\forall \; \barpt{s}_t = \ptvec{s}_t + s^\en_t\envec\in \vsp{E}:
&&
\begin{array}{llllll}
        \dot{s} :=\, \lderiv{\dtsfb{s}} r \,=\, \dtsfb{s} \cdot \dif r_{\pt{s}} \,=\, \dtsfb{s} \cdot \hsfb{r}^\flt_{\pt{s}} \,=\, \dtsfb{s} \cdot \hsfb{s}^\flt  \,=\, \bsfb{\emet}_\ii{\!\Evec}(\dtsfb{s},\hsfb{s})  
     \,=\, \inner{\dtsfb{s}}{\hsfb{s}}
     \,\equiv\, \diff{}{t} \nrm{\ptvec{s}_t}
 \\[5pt]
        \sfg_{\barpt{s}}(\dtsfb{s},\hsfb{s}) \,=\, \sfb{m}_{\barpt{s}} (\dtsfb{s},\hsfb{s}) \,=\, m \inner{\dtsfb{s}}{\hsfb{s}}   \,=\,  m \dot{s} 
\end{array}
\end{align}
\end{small}
where the relation \eq{\sfg_{\barpt{s}}(\dtsfb{s},\hsfb{s}) = \sfb{m}_{\barpt{s}} (\dtsfb{s},\hsfb{s})} may not be immediately obvious but is quickly verified from Eq.\eqref{momentum_prj_act0}. 
\end{itemize}
\end{small}
\end{notation}
\end{small}

\paragraph{Translational Momentum \& Velocity.} 
Let \eq{\bar{\kap}_t= (\barpt{x}_t,\barbs{\kap}_t)} and \eq{\bar{\mu}_t=(\barpt{q}_t,\barbs{\mu}_t)} be integral curves satisfying  \eq{\dt{\bar{\kap}}_t=\sfb{X}^\ss{K}_{\bar{\kap}_t}} and  \eq{\dt{\bar{\mu}}_t=\sfb{X}^\ss{H}_{\bar{\mu}_t}}, respectively. They are \eq{\colift\psi}-related with \eq{(\barpt{x}_t,\barbs{\kap}_t)=\colift\psi(\barpt{q},\barbs{\mu}_t)\leftrightarrow (\barpt{q}_t,\barbs{\mu}_t)= \inv{\colift\psi}(\barpt{x}_t,\barbs{\kap}_t) }
given as in Eq.\eqref{prj_active2}.  That is, the base points transform as:
\begin{small}
\begin{gather} \label{prj_active3}
\barpt{x}_t=\psi(\barpt{q}_t)  = \left\{ \begin{array}{lll}
       \ptvec{x} =   \tfrac{1}{q^\en}\hpt{q} 
    \\[4pt]
   x^\en =  \nrm{\pt{q}}
\end{array} \right.
\qquad \leftrightarrow \qquad
  \left. \begin{array}{lll}
       \ptvec{q} =  x^\en \hpt{x} 
\\[4pt]
   q^\en = \tfrac{1}{\nrm{\pt{x}}} 
\end{array} \right\} =  \barpt{q}_t = \inv{\psi}( \barpt{x}_t)
\end{gather}
\end{small}
and the covectors transform 
    as:\footnote{Note that the ``\eq{\cotsp[\cdt]\bs{\Sigup}}-part''  of the covector transformation in Eq.\eqref{prj_active33} may also be written in terms of \eq{\lang_{ij}=r_i \plin_j - \plin_i r_j} (where \eq{r_i:=\emet_{ij}r^j}) as:
    \begin{align} \nonumber
        \bs{\kap} = \tfrac{q^\en}{\nrm{\pt{q}}} \lang_{ij}(\mu_\pt{q})q^i \hbep^j - q_\en^2 \mu_\en \hat{q}_j \hbep^j 
        \qquad,\qquad 
        \bs{\mu} = \tfrac{1}{x^\en \nrm{\pt{x}}} \lang_{ij} x^i \hbep^j + \kap_\en \hat{x}_j \hbep^j 
    \end{align} }
\begin{small}
\begin{align} \label{prj_active33}
    \barbs{\kap}_t=\psi_* (\barbs{\mu}_t)  = \left\{ \begin{array}{lll}
       \bs{\kap} \,=\,  q^\en \nrm{\pt{q}} \big( \bs{\mu} - (\bs{\mu}\cdot\hsfb{q}) \hsfb{q}^\flt \big) - q_\en^2 \mu_\en \hsfb{q}^{\flt} 
    \\[4pt]
    \kap_\en = \bs{\mu}\cdot \hsfb{q} 
\end{array} \right.
&&  \leftrightarrow &&
  \left. \begin{array}{lll}
       \bs{\mu}  \,=\,  \tfrac{\nrm{\pt{x}}}{x^\en} \big( \bs{\kap} - (\bs{\kap}\cdot\hsfb{x}) \hsfb{x}^\flt \big) + \kap_\en \hsfb{x}^\flt
\\[4pt]
   \mu_\en = 
     -\nrm{\pt{x}}^2  \bs{\kap}\cdot\hsfb{x}
\end{array} \right\} =  \barbs{\mu}_t =\psi^* ( \barbs{\kap}_t)
\end{align}
\end{small}
Additionally, as mentioned previously in Eq.\eqref{intcurv_xform_prj}, it also holds that
 \eq{\barbs{\kap}_t=\sfb{m}(\dt{\bsfb{x}}_t)\in \cotsp[\barpt{x}_t]\bvsp{E}} and \eq{\barbs{\mu}_t  =\sfg_\ss{\!\barpt{q}}(\dt{\bsfb{q}}_t) \in \cotsp[\barpt{q}_t]\bvsp{E} } are the kinematic
 (as well as conjugate\footnote{For a mechanical Hamiltonian or Lagrangian function, the kinematic momentum (given by the metric) and conjugate momentum (given by the fiber derivative of the Lagrangian) are the same.}) 
 momentum covectors along their respective base curves, \eq{\barpt{x}_t} and \eq{\barpt{q}_t}, with velocity tangent vectors 
 \eq{\dt{\bsfb{x}}_t\in\tsp[\barpt{x}_t]\bvsp{E}} and \eq{\dt{\bsfb{q}}_t\in\tsp[\barpt{q}_t]\bvsp{E}}. These velocity vectors are also \eq{\psi}-related, with \eq{\dt{\bsfb{x}}_t=\psi_* (\dt{\bsfb{q}}_t) \leftrightarrow \dt{\bsfb{q}}_t =\psi^* (\dt{\bsfb{x}}_t) } obtained from Eq.\eqref{lift_prj_act6}: 
\begin{small}
\begin{align}
\dt{\bsfb{x}}_t=\psi_* (\dt{\bsfb{q}}_t)  = \left\{ \begin{array}{lll}
      \dtsfb{x} = \tfrac{1}{q^\en \nrm{\pt{q}}}( \dtsfb{q} - \inner{\hsfb{q}}{\dtsfb{q}} \hsfb{q} ) - \tfrac{1}{q_\en^2} \dot{q}^\en \hsfb{q}
    \\[4pt]
     \dot{x}^\en =  \inner{\hsfb{q}}{\dtsfb{q}} = \dot{q}
\end{array} \right.
&&  \leftrightarrow &&
  \left. \begin{array}{lll}
      \dtsfb{q} = \tfrac{x^\en}{\nrm{\pt{x}}}( \dtsfb{x} - \inner{\hsfb{x}}{\dtsfb{x}} \hsfb{x}) + \dot{x}^\en \hsfb{x} 
\\[4pt]
    \dot{q}^\en = -\tfrac{1}{\nrm{\pt{x}}^2} \inner{\hsfb{x}}{\dtsfb{x}} = -\tfrac{1}{\nrm{\pt{x}}^2} \dot{x}
\end{array} \right\} = \dt{\bsfb{q}}_t =\psi^* (\dt{\bsfb{x}}_t)
\end{align}
\end{small}
\sloppy with \eq{\dot{q} \equiv \diff{}{t} \nrm{\ptvec{q}_t}} and \eq{\dot{x} \equiv \diff{}{t} \nrm{\ptvec{x}_t}} as in Eq.\eqref{magdot_def}. 
To clarify, 
    the \eq{\colift\psi}-related Hamiltonian
    systems \eq{(\cotsp\bvsp{E},\nbs{\omg}, K )} and \eq{(\cotsp\bvsp{E},\nbs{\omg} =\colift\psi^*\nbs{\omg}, H =\colift\psi^* K )} are equivalent, respectively,  to the \eq{\psi}-related Newton-Riemann mechanical systems \eq{(\bvsp{E},\sfb{m},V)} and \eq{(\bvsp{E},\sfg=\psi^*\sfb{m},U=\psi^* V)} such that integral curves  \eq{\bar{\kap}_t= (\barpt{x}_t,\barbs{\kap}_t)} and \eq{\bar{\mu}_t=(\barpt{q}_t,\barbs{\mu}_t)} related as in Eq.\eqref{prj_active3}-Eq.\eqref{prj_active33} satisfy:
    \begin{small}
    \begin{align}
   \begin{array}{rccclccc}
        &\dt{\bar{\kap}}_t  \,=\, -\inv{\nbs{\omg}}_{\bar{\kap}_t}(\dif K)
        &\quad  \xleftrightarrow{\;\;\;\colift\psi\;\;\;}  
        &\quad  \dt{\bar{\mu}}_t  \,=\,  -\inv{\nbs{\omg}}_{\bar{\mu}_t}(\dif H)
  \\[8pt] 
        &\!\!\! \copr {\Bigg\downarrow}  &\quad    
         &\quad {\Bigg\downarrow} \copr  
    \\[12pt] 
       \del{\dt{\bsfb{x}}}^\ss{\sfb{m}} \dt{\bsfb{x}}_t \,=\, -\inv{\sfb{m}}_{\barpt{x}_t}(\dif V)
        \quad \xleftrightarrow[\;\; \sfb{m} \;\; ]{}
        & 
              \del{\dt{\bsfb{x}}}^\ss{\sfb{m}} \barbs{\kap}_t \,=\, -\dif V_{\barpt{x}_t}
        &\quad  \xleftrightarrow[\quad\psi\quad]{}  
        &\quad 
               \del{\dt{\bsfb{q}}}^\ss{\sfg} \barbs{\mu}_t \,=\, -\dif U_{\barpt{q}_t}
        & \xleftrightarrow[\;\; \sfg \;\; ]{} \quad 
        \del{\dt{\bsfb{q}}}^\ii{\sfg}  \dt{\bsfb{q}}_t \,=\, -\inv{\sfg}_{\!\barpt{q}_t}(\dif U)
   \end{array}
    \end{align}
    \end{small}
(the above is not a commutative diagram, it is merely a concise depiction of the dynamics in different settings). 
While \eq{\barbs{\kap}_t=\sfb{m}(\dt{\bsfb{x}}_t) = m \bsfb{\emet}_\ii{\!\Evec}(\dt{\bsfb{x}}_t)} is just the usual translational momentum along a curve, \eq{\barpt{x}_t}, in Euclidean space:
\begin{small}
\begin{align}
    \barbs{\kap}_t = \sfb{m}(\dt{\bsfb{x}}_t) = m \dt{\bsfb{x}}^\flt_t = m \emet_{\a\b}\dot{x}^\b \hbep^\a = m \dot{x}_\a \hbep^\a
   \qquad,\qquad 
   \dt{\bsfb{x}}_t = \inv{\sfb{m}}(\barbs{\kap}_t) = \tfrac{1}{m}  \barbs{\kap}^\shrp_t  = \tfrac{1}{m} \emet^{\a\b} \kap_\b \hbe_\a = \tfrac{1}{m}  \kap^\a \hbe_\a 
\end{align}
\end{small}
we note that \eq{\barbs{\mu}_t=\sfg_\ss{\!\barpt{q}}(\dt{\bsfb{q}}_t)} is instead given by the non-Euclidean   \eq{\psi}-induced  metric \eq{\sfg=\psi^*\sfb{m}} (given in  Eq.~\ref{prj_metric_again}):\footnote{Recall that \eq{\nab \hsfb{r}^\flt = \tfrac{1}{r}( \sfb{\emet}_\ii{\!\Sigup} - \hsfb{r}^\flt \otms \hsfb{r}^\flt)} and \eq{\nab \hsfb{r} = \tfrac{1}{r}( \iden_\ii{\!\Sigup} - \hsfb{r} \otms \hsfb{r}^\flt)}. }
\begin{small}
\begin{align}\label{momentum_prj_act0}
\begin{array}{rllllll}  
      \barbs{\mu}_t \,=\,  \sfg_\ss{\!\barpt{q}}(\dt{\bsfb{q}}_t) 
    &\!\!\! =\, m\big( \tfrac{1}{q_\en^2\, \nrm{\pt{q}}} \nab \hsfb{r}^\flt_\ss{\!\pt{q}} \,+\, \hsfb{r}^\flt_\ss{\!\pt{q}} \otms \hsfb{r}^\flt_\ss{\!\pt{q}} \,+\, \tfrac{1}{q_\en^4 }\enform  \otms \enform  \big)\cdot \dt{\bsfb{q}}_t
     &=\,  m\big( \tfrac{1}{q_\en^2\, \nrm{\pt{q}}} \del{\dtsfb{q}} \hsfb{r}^\flt_\ss{\!\pt{q}} \,+\, \hsfb{r}^\flt_\ss{\!\pt{q}} (\dtsfb{q}) \hsfb{r}^\flt_\ss{\!\pt{q}} \,+\, \tfrac{1}{q_\en^4 } \enform (\dt{\bsfb{q}}) \enform  \big)
  \\[8pt]
      &\!\!\!=\, \sfb{m} \cdot \big( \tfrac{1}{q_\en^2\, \nrm{\pt{q}}} \del{\dtsfb{q}} \hsfb{q}  \,+\, \hsfb{q}^\flt  (\dtsfb{q}) \hsfb{q} \,+\, \tfrac{1}{q_\en^4 } \enform  (\dt{\bsfb{q}}) \envec  \big)
      &=\, \sfb{m} \cdot \big(  \tfrac{1}{q_\en^2 \nrm{\pt{q}}^2} ( \dtsfb{q} - \inner{\hsfb{q}}{\dtsfb{q}} \hsfb{q} ) \,+\,  \inner{\hsfb{q}}{\dtsfb{q}} \hsfb{q}
    \,+\,  \tfrac{1}{q_\en^4} \inner{\envec}{\dt{\bsfb{q}}} \envec \big)
\end{array}
\end{align}
\end{small}
where \eq{\hsfb{r}_\ss{\!\pt{q}} =\hsfb{q}}. 
Separating out the \eq{\en^\tx{th}} component, \eq{\mu_\en:= \plin_\en(\bar{\mu}_t)=\envec (\barbs{\mu}_t)}, we may write \eq{ \barbs{\mu} = \bs{\mu} + \mu_\en \enform = \sfg_\ss{\!\barpt{q}}(\dt{\bsfb{q}}_t) } as:
\begin{small}
\begin{align}\label{momentum_prj_act}
         \bs{\mu} 
          \,=\, m \big( \tfrac{1}{q_\en^2\nrm{\pt{q}}}  \del{\dtsfb{q}} \hsfb{r}^\flt + \inner{\hsfb{q}}{\dtsfb{q}} \hsfb{q}^\flt \big)
         \,=\, \sfb{m} \cdot \big(  \tfrac{1}{q_\en^2 \nrm{\pt{q}}^2} ( \dtsfb{q} - \dot{q} \hsfb{q} ) +  \dot{q} \hsfb{q} \big)
         \,=\, \sfg_\ss{\!\barpt{q}}(\dtsfb{q}) 
    &&,&& 
         \mu_\en \,=\,  \tfrac{m}{q_\en^4 } \enform (\dt{\bsfb{q}}) \,=\,  \tfrac{m}{q_\en^4} \dot{q}^\en 
\end{align}
\end{small} 
Now, the momentum covectors are also \eq{\psi}-related with \eq{\barbs{\kap}_t=\psi_*(\barbs{\mu}_t)} and \eq{\barbs{\mu}_t=\psi^*(\barbs{\kap}_t)} given as in  Eq.\eqref{prj_active33}. When combined with Eq.\eqref{momentum_prj_act}, this leads to:
\begin{small}
\begin{align}
 \begin{array}{llll}
         \bs{\kap} 
         \,=\, \sfb{m}\cdot \tfrac{1}{q^\en} ( \del{\dtsfb{q}} \hsfb{r} - \tfrac{\dot{q}^\en}{q^\en} \hsfb{q} )
     \\[3pt]
         \kap_\en \,=\, \bs{\mu}\cdot \hsfb{q}  \,=\, \sfb{m}({\dtsfb{q}},{\hsfb{q}})
         =  m\dot{q}
    \end{array}
 &&,&&  
     \begin{array}{llll}
         \bs{\mu} 
          \,=\,
          \sfb{m}\cdot \big( \tfrac{\nrm{\pt{x}}^2 }{x^\en}  \del{\dtsfb{x}} \hsfb{r} \,+\, \dot{x}^\en \hsfb{x}  \big) 
     \\[3pt]
         \mu_\en \,=\,   -\nrm{\pt{x}}^2 \bs{\kap}\cdot\hsfb{x} 
         \,=\, -m \nrm{\pt{x}}^2 \dot{x}
    \end{array}
\end{align}
\end{small}
and we note that all of the following are equal (in value) when evaluated along \eq{\colift\psi}-related integral curves of \eq{\sfb{X}^\ss{K}} and \eq{\sfb{X}^\ss{H}}:\footnote{The explicit expressions for \eq{ \sfb{m}_{\barpt{x}}(\dt{\bsfb{x}}, \dt{\bsfb{x}}) \,=\, \inv{\sfb{m}}_{\barpt{x}}(\barbs{\kap},\barbs{\kap})  \,=\, \inv{\sfg}_\ss{\!\barpt{q}}(\barbs{\mu},\barbs{\mu}) \,=\, \sfg_\ss{\!\barpt{q}}(\dt{\bsfb{q}}, \dt{\bsfb{q}})} are: 
\begin{align}
\begin{array}{rlcrllllll}
          \sfb{m}_{\barpt{x}}(\dt{\bsfb{x}}, \dt{\bsfb{x}}) 
      &\!\!\!\!=\, m( \nrm{\dtsfb{x}}^2 + \dot{x}_\en^2)
       &=&  
          \sfg_\ss{\!\barpt{q}}(\dt{\bsfb{q}}, \dt{\bsfb{q}})  &\!\!\!\! =\, 
           \tfrac{m}{q_\en^2} \big( \tfrac{1}{\nrm{\pt{q}}^2} ( \nrm{\dtsfb{q}}^2  -  \inner{\hsfb{q}}{\dtsfb{q}}^2)  + \tfrac{1}{q_\en^2}\dot{q}_\en^2 \big) \,+\, m\inner{\hsfb{q}}{\dtsfb{q}}^2
    \\[6pt]
       =\; \inv{\sfb{m}}_{\barpt{x}}(\barbs{\kap},\barbs{\kap})  
       &\!\!\!\!=\, \tfrac{1}{m}(\nrm{\bs{\kap}}^2 + \kap_\en^2)
        &=&  
          \inv{\sfg}_\ss{\!\barpt{q}}(\barbs{\mu},\barbs{\mu})  &\!\!\!\!=\, \tfrac{q_\en^2}{m} \big( \nrm{\pt{q}}^2\nrm{\bs{\mu}}^2 - (\ptvec{q}\cdot\bs{\mu})^2 \,+\, q_\en^2 \mu_\en^2 \big) \,+\, \tfrac{1}{m}( \hsfb{q}\cdot\bs{\mu})^2
\end{array}
\end{align} }
\begin{small}
\begin{align}
\begin{array}{cccc}
     H(\barpt{q}_t,\barbs{\mu}_t) \,=\,  K(\barpt{x}_t,\barbs{\kap}_t) 
     \,=\, E(\barpt{x}_t,\dt{\bsfb{x}}_t)
\quad\;\;,\;\;\quad 
     \inv{\sfg}_\ss{\!\barpt{q}}(\barbs{\mu},\barbs{\mu}) \,=\, \sfg_\ss{\!\barpt{q}}(\dt{\bsfb{q}}, \dt{\bsfb{q}}) \,=\,  \sfb{m}_{\barpt{x}}(\dt{\bsfb{x}}, \dt{\bsfb{x}}) \,=\, \inv{\sfb{m}}_{\barpt{x}}(\barbs{\kap},\barbs{\kap})  
\quad\;\;,\;\;\quad 
      U(\barpt{q}) = V(\barpt{x}) \equiv V(\ptvec{x}) 
\end{array}
\end{align}
\end{small}
That is, the value of  \eq{H(\barpt{q}_t,\barbs{\mu}_t)} is, in fact, equal to that of total mechanical
``Euclidean energy\footnote{\rmsb{Euclidean Energy.} For any particle with mass \eq{m} moving in \eq{(\vsp{E},\bsfb{\emet}_\ii{\!\Evec})}, subject to forces corresponding to a potential function \eq{W\in\fun(\vsp{E})}, the ``Euclidean energy'' evaluated along the particle's velocity curve, \eq{(\barpt{s}_r,\dt{\bsfb{s}}_t)\in\tsp\vsp{E}}, is simply the total mechanical energy with the kinetic term given by the metric \eq{\sfb{m}:=m\bsfb{\emet}_\ii{\!\Evec}}: \\  \eq{\qquad\qquad\qquad\qquad} \eq{E(\barpt{s}_r,\dt{\bsfb{s}}_t)=  \tfrac{1}{2} \sfb{m}_{\barpt{s}}(\dt{\bsfb{s}}, \dt{\bsfb{s}}) + V(\barpt{s}_t) = \tfrac{m}{2} \inner{\dt{\bsfb{s}}_t}{\dt{\bsfb{s}}_t} + V(\barpt{s}_t) \in \mbb{R}}. }''
of the \textit{original} system, \eq{E(\barpt{x}_t,\dt{\bsfb{x}}_t) = K(\barpt{x}_t,\barbs{\kap}_t) }, along the curve \eq{(\barpt{x}_t,\barbs{\kap}_t)=\colift\psi(\barpt{q}_t,\barbs{\mu}_t)}.

\paragraph{Angular Momentum.}
Recall the angular momentum functions, \eq{\lang^{ij},\lang^2\in\fun(\cotsp\vsp{E})}, defined earlier in Eq.\eqref{angMoment_prj_def}. In cartesian coordinates (where \eq{\plin^i:=\emet^{ij} \plin_j}):  
\begin{small}
\begin{align} \label{angMom_def_again}
    \lang^{ij} :=\, r^i \plin^j - \plin^i r^j 
    \qquad,\qquad 
      \lang^2 \,:=\, \tfrac{1}{2} \emet_{ik} \emet_{jl} \lang^{ij}\lang^{kl} \,=\, \tfrac{1}{2} \lang^{ij}\lang_{ij}  \nrmtup{r}^2 \nrmtup{\plin}^2 \,-\, (r^i \plin_i)^2
\end{align}
\end{small}
The above are functions on \eq{\cotsp\bs{\Sigup}\subset \cotsp\vsp{E}} and, as such, we will write \eq{\lang^{ij}(\kap_{\pt{x}})} rather than  \eq{\lang^{ij}(\bar{\kap}_{\barpt{x}})}.\footnote{As mentioned, the functions in Eq.\eqref{angMom_def_again} do not depend on \eq{r^\en} or \eq{\plin_\en}. That is, \eq{\lang^{ij}} are functions on \eq{\cotsp\bs{\Sigup}}, but regarded on \eq{\cotsp\vsp{E}}. Therefore, for any \eq{\bar{\kap}_{\barpt{x}}=(\barpt{x},\barbs{\kap})\in\cotsp\vsp{E}} we write \eq{\lang^{ij}(\kap_{\pt{x}}) \equiv \lang^{ij}(\bar{\kap}_{\barpt{x}})} where \eq{\kap_{\pt{x}}=(\ptvec{x},\bs{\kap})\in\cotsp\bs{\Sigup}}. } 
Now, recall that the projective transformation conveniently satisfies \eq{\colift\psi^* \lang^{ij} = \lang^{ij}} and, likewise, \eq{\colift\psi^* \lang^2 = \lang^2} (this was shown in Eq.\eqref{angMom_prj}). Thus, for integral curves  \eq{\bar{\kap}_t = (\barpt{x}_t,\barbs{\kap}_t)} and \eq{\bar{\mu}_t=(\barpt{q}_t,\barbs{\mu}_t)} of \eq{\sfb{X}^\ss{K}} and \eq{\sfb{X}^\ss{H}}, where \eq{\bar{\kap}_t= \colift \psi(\bar{\mu}_t)}, we then have:
\begin{small}
\begin{align} \label{angMom_prj_redundant}
\begin{array}{llllll}
    \colift\psi^* \lang^{ij} = \lang^{ij}
\\[5pt]
   \colift\psi^* \lang^2 = \lang^2 
\end{array}
\;\left\{\;\begin{array}{llllll}
       \ptvec{q}_t \wedge \bs{\mu}^\shrp_t \,=\, \ptvec{x}_t \wedge \bs{\kap}^\shrp_t 
       &=\, \ptvec{x}_t\wedge m \dtsfb{x}_t 
\\[5pt]
    \lang^{ij}(\mu_t)  \,=\,  q^i \mu^j - \mu^i q^j \,=\, \lang^{ij}(\kap_t) \,=\,  x^i \kap^j - \kap^i x^j 
    &=\, m(  x^i \dot{x}^j - \dot{x}^i x^j)
\\[5pt]
   \lang^2(\mu_t) \,=\,  \nrm{\pt{q}}^2 \nrm{\bs{\mu}}^2 - ( \ptvec{q}\cdot\bs{\mu})^2
  =  \lang^2(\kap_t) \,=\,  \nrm{\pt{x}}^2 \nrm{\bs{\kap}}^2 - (\ptvec{x}\cdot\bs{\kap})^2 
   &=\, m^2( \nrm{\pt{x}}^2 \nrm{\dtsfb{x}}^2 - \inner{\ptvec{x}}{\dtsfb{x}}^2 ) 
\end{array}\;\right. 
\end{align}
\end{small}
where \eq{(\cdot)^\shrp:=\inv{\bsfb{\emet}_\ii{\!\Evec}}(\cdot)} such that \eq{\bs{\kap}^\shrp=\inv{\bsfb{\emet}_\ii{\!\Evec}}\cdot\sfb{m}(\dtsfb{x}) = m \dtsfb{x}}.
However, \eq{\bs{\mu}^\shrp=\inv{\bsfb{\emet}_\ii{\!\Evec}}\cdot\sfg(\dtsfb{q}) \neq m \dtsfb{q}\,}(!) such that the above is \textit{not} equivalent under the exchange \eq{(\ptvec{x},\dtsfb{x})\leftrightarrow (\ptvec{q},\dtsfb{q})}. Instead, one finds using Eq.\eqref{momentum_prj_act} that the above is equivalent to:
\begin{small}
\begin{align}
  \ptvec{q}_t \wedge \bs{\mu}^\shrp_t 
  \,=\,  \tfrac{m}{q_\en^2 \nrm{\pt{q}}^2}(\ptvec{q}_t\wedge \dtsfb{q}_t)  
\qquad,\qquad 
  \lang^2(\mu_t) \,=\,  \nrm{\pt{q}}^2 \nrm{\bs{\mu}}^2 - ( \ptvec{q}\cdot\bs{\mu})^2
    \,=\,  \tfrac{m^2}{q_\en^4 \nrm{\pt{q}}^4} \big( \nrm{\pt{q}}^2 \nrm{\dtsfb{q}}^2 - \inner{\ptvec{q}}{\dtsfb{q}}^2 \big) 
\end{align}
\end{small}

\begin{small}
\begin{notesq}
    In the case that \eq{\en-1=3}, then \eq{\lang} is the usual angular momentum magnitude function on \eq{\cotsp\bs{\Sigup}^3 \subset \cotsp\Evec^4}, with:
 \begin{small}
 \begin{align} \label{angMom_prj_3D}
       \fnsize{for } \, \en-1=3: \qquad \lang^2(\kap_{\pt{x}})  \,=\, \nrm{\ptvec{x}\times \bs{\kap}^{\shrp} }^2 \,=\, \nrm{\ptvec{q}\times \bs{\mu}^{\shrp} }^2 \,=\, \lang^2(\mu_\ss{\pt{q}})
 \end{align}
 \end{small}
\end{notesq}
\end{small}

\subsection{Integrals of Motion \& Invariant Submanifolds} \label{sec:IOMs_prj_act}

 Recall from section \ref{sec:Hnom_prj} that each \eq{\cotsp\Sig_\ii{b}\subset \cotsp\bvsp{E}} is an \eq{\sfb{X}^\ss{K}}-variant submanifold of the original system. We will now show that each  \eq{\cotsp\man{Q}_\ii{b}\subset \cotsp\bvsp{E}} is an \eq{\sfb{X}^\ss{H}}-variant submanifold of the transformed system and, furthermore, that \eq{\cotsp\Sig_\ii{b}=\colift\psi (\cotsp\man{Q}_\ii{b})} and \eq{\cotsp\man{Q}_\ii{b} = \inv{\colift\psi}(\cotsp\Sig_\ii{b})}. Before proceeding, we first review these submanifolds (defined previously in Eq.\eqref{3surf_0} – \ref{3surf_3}).

\paragraph{Some Important Submanifolds (\textit{redux}).}
Recall the \eq{(\en-1)}-dim hypersurfaces  \eq{\man{Q}_\ii{b}\subset \bvsp{E}} and \eq{\Sig_\ii{b}\subset \bvsp{E}} defined in Eq.\eqref{3surf_0} – \ref{3surf_3} as follows for any positive \eq{ b\in\mbb{R}_\ii{+}}: 
\begin{small}
\begin{align} 
 \begin{array}{rllllll}
         \man{Q}_\ii{b}  \,=\,   \inv{\rfun}\{b\} &\!\!\!\! = \big\{  \barpt{q} = \ptvec{q} + q^\en\envec \in \bvsp{E} \;\big|\;  r(\ptvec{q}) = \nrm{\pt{q}} = b   \big\}  
         &\!\!\! \cong\,\man{S}^{\en-\ii{2}}_\ii{b} \times \vsp{N}_\ii{+} 
         &\quad,\qquad \man{Q}_\ii{b} \,=\, \inv{\psi}(\Sig_\ii{b}) 
     \\[5pt]
     \Sig_\ii{b}  \,= \inv{(r^\en)}\{b\} &\!\!\!\! =  \big\{ \barpt{x}=\ptvec{x}+x^\en \envec \in\bvsp{E} \;\big| \;\, r^\en(\barpt{x}) = x^\en = b \big\}   
     &\!\!\! \cong\, \bs{\Sigup}_\nozer \oplus \{b\} 
     &\quad,\qquad   \Sig_\ii{b}  \,=\,  \psi(\man{Q}_\ii{b}) 
    \end{array}
\end{align}
\end{small}
 It is important for the present developments that these are defined as subsets of \eq{\bvsp{E}=\bs{\Sigup}_\nozer\oplus \vsp{N}_\ii{+}} 
     (rather than \eq{\vsp{E}=\bs{\Sigup}\oplus \vsp{N}}).\footnote{In order for Eq.\eqref{prj_subman_rels0} to hold,   \eq{\Sig_\ii{b}} and \eq{\man{Q}_\ii{b}} must be taken as hypersurfaces in \eq{\bvsp{E}=\bs{\Sigup}_\nozer\oplus \vsp{N}_\ii{+}}; they are not ``just'' the level sets \eq{r^\en} and \eq{r} but, more specifically:
    \begin{align} \label{prj_subman_actualDef_again}
        \man{Q}_\ii{b} := \big(\inv{\rfun}\{b\} \cap \bvsp{E} \big)\subset \bvsp{E} \subset \vsp{E}
        \qquad,\qquad 
        \Sig_\ii{b} := \big(\inv{(r^\en)}\{b\} \cap  \bvsp{E} \big) \subset \bvsp{E} \subset \vsp{E}
    \end{align}}
Note  \eq{\Sig_\ii{b}\perp \vsp{N}} is just the Euclidean hyperplane in \eq{\bvsp{E}} that orthogonally intersects \eq{\vsp{N}} (the \eq{r^\en}-axis) at \eq{r^\en= b\in\mbb{R}_\ii{+}}, and  \eq{\man{Q}_\ii{b}} can be identified with the product manifold \eq{\man{S}^{\en-\ii{2}}_\ii{b}\times \vsp{N}_\ii{+}}, where  \eq{\man{S}^{\en-\ii{2}}_\ii{b}\subset \bs{\Sigup}} is the \eq{(\en-2)}-sphere or radius \eq{b},  viewed as hypersurface in \eq{\bs{\Sigup}} 
    (with \eq{\bs{\Sigup}}, in turn, a hyperplane in \eq{\bvsp{E}}).\footnote{For instance, let \eq{\en =3} and take \eq{\vsp{N}} to be the ``\eq{z} axis'' in \eq{\vsp{E}^3=\bs{\Sigup}\oplus\vsp{N}} with \eq{\bs{\Sigup}} the ``\eq{xy}-plane''. Then \eq{\man{Q}_\ii{b} = \man{S}^1_\ii{b}\times \vsp{N}_\ii{+}} would be a radius-\eq{b} cylinder centered around the \eq{z}-axis. Actually, since we are defining \eq{\man{Q}_\ii{b}} as a submanifold of \eq{\bvsp{E}=\bs{\Sigup}_\nozer\oplus\vsp{N}_\ii{+}}, it would be the upper half of the cylinder with \eq{r^\en = z >0}. } 
Their (co)tangent bundles are \eq{(2\en -2)}-dim  subbundles of \eq{\cotsp\bvsp{E}} characterized by:
    \begin{small}
    \begin{align} \label{prj_subs_again}
    \begin{array}{rcllll}
         \tsp\man{Q}_\ii{b} \,=&\!\!\!\! \big\{ (\barpt{q},\bsfb{u})\in\tsp\bvsp{E} \;\big|\; r(\ptvec{q}) = b \;,\; \hsfb{r}^\flt_\ss{\!\pt{q}}(\sfb{u}) 
         = 0 \big\}
    \\[5pt]
          \cotsp\man{Q}_\ii{b} \,=&\!\!\!\! \big\{ (\barpt{q},\barbs{\mu})\in\cotsp\bvsp{E} \;\big|\; r(\ptvec{q}) = b \;,\; \hsfb{r}_\ss{\!\pt{q}}(\bs{\mu}) 
          = 0 \big\}
      \\[5pt]
          &\!\!\! \fnsize{(i.e, \eq{\; \nrm{\pt{q}}= b\;,\;\; \inner{\hsfb{q}}{\sfb{u}} = 0 \;,\;\; \hsfb{q}(\bs{\mu})=0} )}
    \end{array}
    &&,&&
    \begin{array}{rcllll}
         \tsp\Sig_\ii{b} \,=&\!\!\!\! \big\{ (\barpt{x},\bsfb{v})\in\tsp\bvsp{E} \;\big|\;  r^\en(\barpt{x}) = b \;,\;  \enform(\bsfb{v}) = 0 \big\}
    \\[5pt]
          \cotsp\Sig_\ii{b} \,= &\!\!\!\! \big\{ (\barpt{x},\barbs{\kap})\in\cotsp\bvsp{E} \;\big|\; r^\en(\barpt{x}) = b \;,\;  \envec (\barbs{\kap}) = 0 \big\}
    \\[5pt]
         & \fnsize{(i.e, \eq{\; x^\en= b\;,\;\; \v^\en = 0 \;,\;\; \kap_\en =0} )}
    \end{array}
    \end{align}
    \end{small}
    where \eq{\hsfb{r}^\flt = \dif r} and \eq{\hsfb{r}_{\!\pt{q}} = \hsfb{q}}, and where \eq{\enform = \dif r^\en} is homogeneous such that \eq{\plin_\en(\barpt{x},\barbs{\kap}) = \envec(\barbs{\kap})=\kap_\en} does not  depend on the base  point \eq{\barpt{x}}. Now, a very convenient property of the projective transformation
    is that \eq{\psi(\man{Q}_\ii{b}) = \Sig_\ii{b}} and \eq{\inv{\psi}(\Sig_\ii{b}) = \man{Q}_\ii{b}}, which 
    can be verified by inspection of \eq{\psi\in\Dfism(\bvsp{E})} in Eq.\eqref{prj_def_act}, or from the expressions \eq{\psi=\tfrac{1}{r^\en}\hsfb{r} + r \envec} and \eq{\inv{\psi}=r^\en \hsfb{r} + \tfrac{1}{r}\envec}.\footnote{Again, we stress that the relations \eq{\psi(\man{Q}_\ii{b}) = \Sig_\ii{b}} and \eq{\inv{\psi}(\Sig_\ii{b}) = \man{Q}_\ii{b}} require that \eq{\Sig_\ii{b}} and \eq{\man{Q}_\ii{b}} be defined as hypersurfaces in \eq{\bvsp{E}=\bs{\Sigup}_\nozer \oplus \vsp{N}_\ii{+}\subset \vsp{E}_\nozer \subset \vsp{E}}.}
    This relation extends to the (co)tangent bundles;  when \eq{\psi}, \eq{\tlift\psi}, or \eq{\colift\psi} is restricted to \eq{\man{Q}_\ii{b}}, \eq{\tsp\man{Q}_\ii{b}}, or \eq{\cotsp\man{Q}_\ii{b}}, we then have:
    \begin{small}
    \begin{align} \label{prj_subman_rels0}
         \psi|_\ii{\man{Q}_\ii{b}} \in\Dfism(\man{Q}_\ii{b};\Sig_\ii{b})
    &&,&&
        \tlift\psi|_\ii{\tsp\man{Q}_\ii{b}} \in\Dfism(\tsp\man{Q}_\ii{b} ; \tsp \Sig_\ii{b} )
    &&,&&
          \colift\psi|_\ii{\cotsp\man{Q}_\ii{b}} \in\Dfism(\cotsp\man{Q}_\ii{b} ;\cotsp\Sig_\ii{b})
    \end{align}
    \end{small}
    For instance, if \eq{\barpt{u}_{\barpt{q}}= \inv{\tlift\psi}(\barpt{\v}_{\barpt{x}})} for some \eq{\barpt{\v}_{\barpt{x}}\in\tsp\Sig_\ii{b}}, then \eq{\barpt{u}_{\barpt{q}} \in\tsp\man{Q}_\ii{b}}, and vice versa.
    Similarly, if \eq{\bar{\mu}_{\barpt{q}}= \inv{\colift\psi}(\bar{\kap}_{\barpt{x}})} for some \eq{\bar{\kap}_{\barpt{x}}\in\cotsp\Sig_\ii{b}}, then \eq{\bar{\mu}_{\barpt{q}} \in\cotsp\man{Q}_\ii{b}}, and vice versa. These transformations for the case \eq{b =1} are given below in Eq.\eqref{tlift_prj_E3}-Eq.\eqref{colift_prj_E3}.

\paragraph{Some Important Poisson Brackets.} For the original dynamics \eq{\sfb{X}^\ss{K}\in\vechm(\cotsp\bvsp{E})}, recall from Eq.\eqref{pbraks_nom} the following Poisson brackets with the original Hamiltonian function, \eq{K}, given in cartesian coordinates as \eq{K=\tfrac{1}{2}m^{\a\b} \plin_\a \plin_\b + V^\zr(\rfun) + V^\ss{1}(\tup{r})}: 
\begin{small}
\begin{align} \label{iom_nom_again}
\begin{array}{rlllll}
     1.& \pbrak{\plin_\en}{K} = 0
\\[4pt]
      2.& \pbrak{r^\en}{K} 
      = \tfrac{1}{m} \plin^\en 
\end{array}
&&
 3.\quad \pbrak{K}{K} \,=\, 0
  &&
\begin{array}{rllll}
      \mrm{4a.}& \pbrak{ \lang^{ij} }{K} \,=\, -(r^i \emet^{jk} - r^j \emet^{ik})\pd_k V^\ss{1}  
\\[4pt]
      \mrm{4b.}& \pbrak{ \ttfrac{1}{2}\lang^2 }{K}
       \,=\,  -( r^2 \emet^{ij} - r^i r^j) \plin_i \pd_j V^\ss{1}
        \,=\, -\lang^{ij} r_i \pd_j V^\ss{1}
\end{array}
\end{align}
\end{small}
The above were discussed in section \ref{sec:Hnom_prj};  the first two brackets imply that each \eq{\cotsp\Sig_\ii{b}} is \eq{\sfb{X}^\ss{K}}-invariant; the third is the usual property that \eq{K} is an integral of motion of \eq{\sfb{X}^\ss{K}} (iff \eq{\pd_t K =0 } which, for the present case, is true iff \eq{\pd_t V =0\,}\footnote{Nothing so far has required \eq{\pd_t V =0}, although it has been implied by our notation.});
the fourth verifies the expected result that, in the case only central forces do work (i.e., \eq{V^\ss{1}=0}), then the angular momentum functions \eq{\lang^{ij}=r^i \plin^j - \plin^i r^j\in\fun(\cotsp\vsp{E})} are integrals of motion and, thus, so too is the norm \eq{\lang^2=\ttfrac{1}{2}\lang^{ij}\lang_{ij}=\nrmtup{r}^2\nrmtup{\plin}^2-(r^i \plin_i)^2}.

Now consider the \eq{\colift\psi}-related Hamiltonian dynamics, \eq{\sfb{X}^\ss{H} = \colift\psi^*\sfb{X}^\ss{K}\in\vechm(\cotsp\bvsp{E})}, for \eq{ H := \colift\psi^* K} given in Eq.\eqref{Hprj_full} 
(and repeated here\footnote{The Hamiltonian \eq{H := \colift\psi^* K} was shown to be expressed in cartesian coordinates as:\\ 
\eq{\qquad \quad  H \,=\,  \tfrac{1}{2}g^{\a\b} \plin_\a \plin_\b + U \;=\; \tfrac{r_\en^2}{2m}  \big( \lang^2 + r_\en^2 \plin_\en^2 \big) \,+\, \tfrac{1}{2m} (\htup{r} \cdot\tup{\plin})^2 \,+\, U \;=\; \tfrac{r_\en^2}{2m}  \big( \nrmtup{r}^2 \nrmtup{\plin}^2 - (\tup{r}\cdot\tup{\plin})^2 + r_\en^2 \plin_\en^2 \big) \,+\, \tfrac{1}{2m} (\htup{r} \cdot\tup{\plin})^2 \,+\, U^\zr(r_\en) + U^\ss{1}(\bartup{r}) }\;. }).  
It will be shown that the brackets with \eq{K} in Eq.\eqref{iom_nom_again} lead to the following brackets with \eq{ H}: 
\begin{small}
\begin{align}  \label{iom_Hprj_0}
\begin{array}{rllll}
    1.&  \pbrak{ \hat{r}^i \plin_i }{H} = 0
\\[4pt]
    2.&  \pbrak{r}{H} 
    \,=\, \tfrac{1}{m}\hat{r}^i \plin_i
\end{array}
&&
  3.\quad \pbrak{H}{H} \,=\, 0
  &&
\begin{array}{rlllll}
     \mrm{4a.}& \pbrak{ \lang^{ij} }{H} \,=\, -(r^i \emet^{jk} - r^j \emet^{ik})\pd_k U^\ss{1}  
\\[4pt]
     \mrm{4b.}&  \pbrak{ \ttfrac{1}{2}\lang^2 }{H}
      \,=\,  -r^2 \plin^i \pd_i U^\ss{1} 
\end{array}
\end{align}
\end{small}
We will show that the first two brackets above imply that each \eq{\cotsp\man{Q}_\ii{b}} is \eq{\sfb{X}^\ss{H}}-invariant. 
The third bracket is the standard property that \eq{H} is an integral of motion of \eq{\sfb{X}^\ss{H}} (iff \eq{\pd_t H =0} which, for the present case, is true if \eq{\pd_t K =0} since \eq{\colift\psi} is not time-dependent). 
The fourth shows that that angular momentum is again conserved in the case \eq{U^\ss{1}=0}  (corresponding to an original \eq{V^\ss{1}=0}).
In addition to the above, 
it is useful to note brackets of the following functions
(which are not all independent of the above\footnote{The functions \eq{\rfun =\nrm{\sfb{r}}, \nrmtup{\plin}, r^i \plin_i}, and \eq{\lang} must satisfy the relation \eq{\lang^2 = \nrmtup{r}^2\nrmtup{\plin}^2 - (r^i \plin_i)^2}. Clearly, \eq{r^i \plin_i = r \hat{r}^i \plin_i} must also hold. }):
\begin{small}
\begin{align} \label{iom_Hprj_more}
    \pbrak{r^i \plin_i}{H} \,=\, \tfrac{1}{m}(\hat{r}^i \plin_i)^2
    \qquad,\qquad 
     \pbrak{ \ttfrac{1}{2}r^2}{H} \,=\, \tfrac{1}{m}r^i \plin_i
      \qquad,\qquad 
     \pbrak{ \ttfrac{1}{2}\nrmtup{\plin}^2}{H} \,=\, - \plin^i \pd_i U^\ss{1} -\tfrac{(\htup{r}\cdot\tup{\plin})}{m r^3} \lang^2
\end{align}
\end{small}

\begin{small}
\begin{itemize}[noitemsep]
    \item[{}] \textit{Derivation of Eq.\eqref{iom_Hprj_0}.} We could derive the brackets in Eq.\eqref{iom_Hprj_0} and Eq.\eqref{iom_Hprj_more} by direct coordinate calculation 
     (e.g.,  \eq{\pbrak{r}{H}} and  \eq{\pbrak{ \ttfrac{1}{2}\nrmtup{\plin}^2}{H}} as derived in the footnote\footnote{Using \eq{H} given in cartesian coordinates \eq{(\bartup{r},\bartup{\plin})} by Eq.\eqref{Hprj_full}, we find the Poisson brackets for \eq{r^2=\nrmtup{r}^2=\emet_{ij}r^i r^j} and \eq{\nrmtup{\plin}^2=\emet^{ij} \plin_i \plin_j} as follows (using \eq{\pd_i \nrmtup{r}^2/2 = r_i} and \eq{\upd^i \nrmtup{\plin}^2/2 = \plin^i}): 
    \begin{align} \label{rmag_pbrak_act}
    \begin{array}{lllll}
          \pbrak{\ttfrac{1}{2} r^2}{H} = \pbrak{\tfrac{1}{2}\nrmtup{r}^2}{H} = r_i  \upd^i H  = r_i \dot{r}^i = -\tfrac{r_\en^2}{m} \cancel{\lang^{ij}r_i r_j} + \tfrac{1}{m}(\htup{r}\cdot\tup{\plin})\hat{r}^i r_i 
          =  \tfrac{1}{m}(\htup{r}\cdot\tup{\plin})\nrmtup{r}
          =  \tfrac{1}{m}(\tup{r}\cdot\tup{\plin})
    \\[2pt]
          \pbrak{  \ttfrac{1}{2}\nrmtup{\plin}^2}{H} = -  \plin^j \pd_j H = \plin^j \dot{\plin}_j = -\tfrac{r_\en^2}{m} \cancel{\lang_{jk} \plin^j \plin^k} - \tfrac{(\htup{r}\cdot\tup{\plin})}{m \nrmtup{r}}  \big( \plin_j - (\htup{r}\cdot\tup{\plin}) \hat{r}_j \big) \plin^j - \plin^j \pd_j U^\ss{1}
         =  -\tfrac{(\htup{r}\cdot\tup{\plin})}{m \nrmtup{r}^3} \lang^2  - \plin^j \pd_j U^\ss{1} 
    \end{array}
    \end{align} 
    From which we also obtain \eq{\pbrak{r}{H} = \tfrac{1}{r}\pbrak{\ttfrac{1}{2} r^2}{H} =\tfrac{1}{m}(\htup{r}\cdot\tup{\plin})} and \eq{\pbrak{\nrmtup{\plin}}{H}  =  \tfrac{1}{\nrmtup{\plin}} \pbrak{\tfrac{1}{2} \nrmtup{\plin}^2}{H}}.  }).
    However, some of these are more easily realized as a transformation of the original system's brackets in Eq.\eqref{iom_nom_again}. 
    We know that if \eq{f\in\fun(\cotsp\bvsp{E})} is an integral of motion of \eq{\sfb{X}^\ss{K}}, then \eq{\colift\psi^*f = f \circ \colift \psi} is an integral of motion of \eq{\sfb{X}^\ss{H}}, and  vice versa 
     (this was shown in Remark \ref{rem:HK_related}, Eq.\eqref{iom_xform_gen}, as repeated in the footnote\footnote{The proof from Remark \ref{rem:HK_related} is repeated here: it is a known property of Lie derivatives that \eq{\lderiv{\colift\psi^* \sfb{X}^\ss{K}}(\colift\psi^*f) = \colift\psi^* \lderiv{\sfb{X}^\ss{K}} f }. Thus, for  \eq{\sfb{X}^\ss{H}= \colift\psi^* \sfb{X}^\ss{K}} it follows that \eq{\lderiv{\sfb{X}^\ss{H}}(\colift\psi^*f) = \lderiv{\colift\psi^* \sfb{X}^\ss{K}}(\colift\psi^*f) = \colift\psi^* \lderiv{\sfb{X}^\ss{K}} f }. Therefore, if \eq{\lderiv{\sfb{X}^\ss{K}} f=0} then \eq{\lderiv{\sfb{X}^\ss{H}}(\colift\psi^*f) = \colift\psi^*\lderiv{\sfb{X}^\ss{K}} f=0} such that \eq{\colift\psi^*f} is an integral of motion of  \eq{\sfb{X}^\ss{H}}. Equivalently, since \eq{\colift\psi} is automatically a symplectomorphism, it follows that  \eq{\pbrak{\colift\psi^*f}{\colift\psi^*g} = \colift\psi^* \pbrak{f}{g}} for any \eq{f,g\in\fun(\cotsp\bvsp{E})}. Thus, if \eq{\lderiv{\sfb{X}^\ss{K}} f \equiv \pbrak{f}{K}=0}, then \eq{\pbrak{\colift\psi^*f}{H} = \pbrak{\colift\psi^*f}{\colift\psi^* K } = \colift\psi^* \pbrak{f}{K} =0 }. }). 
     More generally, since \eq{\colift\psi\in\Spism(\cotsp\bvsp{E},\nbs{\omg})}, we know that \eq{\pbrak{\colift\psi^*f}{\colift\psi^*g} = \colift\psi^*\pbrak{f}{g}} for any \eq{f,g\in\fun(\cotsp\bvsp{E})}. 
    Using the fact that, by definition,  \eq{\colift\psi^* K  =: H}, this gives the following relation for the canonical symplectic Poisson bracket on \eq{\cotsp\bvsp{E}}:
    \begin{small}
    \begin{align}\label{pbrak_HK_prj}
         \pbrak{\colift\psi^*f}{\colift\psi^*g} = \colift\psi^*\pbrak{f}{g}
         \qquad \Rightarrow \qquad 
          \pbrak{\colift\psi^*f}{H} \,=\,  \colift\psi^* \pbrak{f}{K}
    \end{align}
    \end{small}
    Using the above, the brackets in Eq.\eqref{iom_Hprj_0} then follow from those in Eq.\eqref{iom_nom_again}, by way of the following:
    \begin{align} \label{pbraks_pullbacks}
    \begin{array}{rllll}
        1.& \colift\psi^* \plin_\en = \tfrac{1}{r}  r^i \plin_i = \hat{r}^i \plin_i   
    \\[4pt]
          2.&  \colift\psi^* r^\en \equiv \psi^* r^\en = r = \nrmtup{r}
    \end{array}
    &&
      3.\quad  \colift\psi^* K =: H 
      &&
    \begin{array}{rllll}
          \mrm{4a.}& \colift\psi^* \lang^{ij} = \lang^{ij}
    \\[4pt]
        \mrm{4b.}&  \colift\psi^* \lang^2 = \lang^2
    \end{array}
    \end{align}
    where 3 is just the definition of \eq{H}, where 2 was shown at Eq.\eqref{proj_cartesian}, and where 1, 4a, and 4b were shown in Eq.\eqref{angMom_prj0} – \ref{angMom_prj}
    (an alternative derivation of \eq{1} is given in the   footnote\footnote{From the  definition of \eq{\colift\psi} and \eq{\plin_\en =\pfun^{\envec} \in \fun(\cotsp\Evec)}, it follows that, for an arbitrary point \eq{(\barpt{q},\barbs{\mu})\in\cotsp\bvsp{E}}, then: \\
    \eq{\qquad\qquad\qquad  ( \colift \psi^* \plin_\en)(\barpt{q},\barbs{\mu}) = \plin_\en\circ \colift \psi(\barpt{q},\barbs{\mu}) = \pfun^{\envec}\big( \colift \psi(\barpt{q},\barbs{\mu}) \big) = {\hbe_{\en}}_\ii{|\psi(\barpt{q})} \cdot \psi_*(\barbs{\mu}) \equiv \envec \cdot \psi_*(\barbs{\mu})} . \\
    Using the expression for \eq{\psi_*(\barbs{\mu})} from Eq.\eqref{lift_prj_act3} – \ref{colift_prj_def1}, we  have \eq{\envec \cdot\psi_*(\barbs{\mu}) = \bs{\mu}\cdot\hsfb{q}}. Since the point \eq{(\barpt{q},\barbs{\mu})} was arbitrary, this is expressed generally in cartesian coordinates as \eq{\colift\psi^* \plin_\en = \plin_\en \circ\colift\psi = \plin_i \hat{r}^i}. }).
   The above then verifies the brackets in Eq.\eqref{iom_Hprj_0} \textit{except} 
   for the right-hand-side of the expressions for \eq{\pbrak{\lang^{ij}}{H}} and \eq{\pbrak{\ttfrac{1}{2}\lang^2}{H}}; these are given in section \ref{app:prj_geo}. 
   \item[]  For the brackets in Eq.\eqref{iom_Hprj_more},  the bracket of \eq{\nrmtup{\plin}^2/2} was just derived in Eq.\eqref{rmag_pbrak_act} of the footnote, the bracket of \eq{r^2/2} follows trivially from that of \eq{r}, and the bracket of \eq{r^i \plin_i = r \hat{r}^i \plin_i} 
   follows\footnote{That is, since \eq{\pbrak{\hat{r}^i \plin_i}{H}=0}, we find \eq{\pbrak{r^i \plin_i}{H} = \pbrak{r \hat{r}^i \plin_i}{ H} = \hat{r}^i \plin_i \pbrak{r }{ H} = 
    \tfrac{1}{m} (\htup{r}\cdot\tup{\plin})^2}. } 
   from the brackets of \eq{r} and \eq{\hat{r}^i \plin_i} seen in Eq.\eqref{iom_Hprj_0}. 
\end{itemize}
\end{small}

\noindent Note that the integral of motion \eq{\hat{r}^i \plin_i=\colift\psi^* \plin_\en} of the transformed Hamiltonian, \eq{H:=\colift\psi^* K}, corresponds to the Lagrange multiplier in first part of this work (which we denoted \eq{\lambda} there). In the present context, we may observe that \eq{\coVlift{\envec}=\sfb{X}^{\plin_\en} = \hbpart{\en}} was a Noether symmetry of the original system (corresponding to the ignorable coordinate \eq{r^\en} and integral of motion \eq{\plin_\en}), and that it is transformed by the lifted projective transformation, \eq{\colift\psi}, to the integral of motion \eq{\hat{r}^i \plin_i}, where the Noether symmetry \eq{\sfb{X}^{(\hat{r}^i \plin_i)}} is the complete cotangent lift of \eq{\hsfb{r}\equiv \be_{\rfun}}: 
\begin{small}
\begin{align}
      \hat{r}^i \plin_i = \pfun^{\hsfb{r}} \qquad,\qquad   \sfb{X}^{(\hat{r}^i \plin_i)} \,=\, \sfb{X}^{\pfun^{\hsfb{r}}} \,=\, \coVlift{\hsfb{r}} \,=\, \hat{r}^i \hbpart{i} \,-\, \tfrac{1}{\rfun}(\kd^j_i - \hat{r}^j\hat{r}_i) \plin_j \hbpartup{i} 
      \;=\; \hat{r}^i \hbpart{i} \,+\, \tfrac{1}{\rfun^3}\lang_{ij}r^j \hbpartup{i}
\end{align}
\end{small}

\paragraph{Invariant Submanifolds \& Simplified Dynamics on $\cotsp\man{Q}_\ii{1}$.}
The relations \eq{ \pbrak{K}{K} = 0 = \pbrak{H}{H}} are standard for any Hamiltonian system and we will, for now, set them aside and focus on the first two brackets in Eq.\eqref{iom_nom_again} and Eq.\eqref{iom_Hprj_0}. For the original system, the brackets  \eq{\pbrak{\plin_\en}{K} =0} and \eq{ \pbrak{r^\en}{K} = \tfrac{1}{m} \plin^\en} in Eq.\eqref{iom_nom_again} show that \eq{\plin_\en} is an integral of motion of \eq{\sfb{X}^\ss{K}} and each \eq{\cotsp\Sig_\ii{b}\subset \cotsp\bvsp{E}} is an \eq{\sfb{X}^\ss{K}}-invariant submanifold (this was already detailed in \ref{rem:E3dyn_subman}). In a similar manner,  
the brackets of \eq{\hat{r}^i \plin_i} and \eq{\rfun =\nrm{\sfb{r}}=\nrm{\tup{r}}} with \eq{H} in Eq.\eqref{iom_Hprj_0} show that \eq{\hat{r}^i \plin_i} is an integral of motion of \eq{\sfb{X}^\ss{H}=\colift\psi^*\sfb{X}^\ss{K}} and each \eq{\cotsp\man{Q}_\ii{b}=\inv{\colift\psi}(\cotsp\Sig_\ii{b})} is a \eq{\sfb{X}^\ss{H}}-invariant submanifold. 
To summarize, Eq.\eqref{prj_subman_rels0} along with 
the first two brackets in Eq.\eqref{iom_nom_again} and \ref{iom_Hprj_0} leads to:

\begin{small}
\begin{remrm} \label{rem:invSub_prj}
   For any \eq{ b\in\mbb{R}_\ii{+}}, it was shown previously that each \eq{\cotsp\Sig_\ii{b}  \subset \cotsp\bvsp{E}} is a \eq{(2\en -2)}-dim \eq{\sfb{X}^\ss{K}}-invariant submanifold. We see now that and each \eq{\cotsp\man{Q}_\ii{b}\subset \cotsp\bvsp{E}} is a \eq{(2\en -2)}-dim \eq{\sfb{X}^\ss{H}}-invariant submanifold. Furthermore, if
   \eq{\bar{\mu}_t\in\cotsp\man{Q}_\ii{b}} is an integral curve of \eq{\sfb{X}^\ss{H}}, then \eq{\bar{\kap}_t= \colift\psi(\bar{\mu}_t) \in \cotsp\Sig_\ii{b}} is an integral curve of the original dynamics \eq{\sfb{X}^\ss{K}=\colift\psi_*\sfb{X}^\ss{H}}, and vice versa.\footnote{because \eq{ \colift\psi|_\ii{\cotsp\man{Q}_\ii{b}} \in\Dfism(\cotsp\man{Q}_\ii{b} ;\cotsp\Sig_\ii{b})}.}
   We will only consider the case \eq{b =1} such that these curves transform as
    in Eq.\eqref{colift_prj_E3} below.
\end{remrm}
\end{small}

\noindent The above will be clarified further but, first, let us note an important implication of Remark \ref{rem:invSub_prj}. 
In cartesian coordinates \eq{(\bartup{r},\bartup{\plin})=(\tup{r},r^\en,\tup{\plin},\plin_\en)}, the coordinate representation of general integral curves of \eq{\sfb{X}^\ss{H}} are found by solving the rather hideous system of ODEs in Eq.\eqref{Hprj_full}. However, if we limit consideration to integral curves on the  \eq{\sfb{X}^\ss{H}}-invariant submanifold \eq{\cotsp\man{Q}_\ii{1}}, then any such curve has cartesian coordinate representation satisfying \eq{\nrm{\tup{r}_{\zr}}=\nrm{\tup{r}_t}=1} and \eq{\htup{r}_{\zr}\cdot\tup{\plin}_{\zr}=\htup{r}_t\cdot\tup{\plin}_t=0} (and thus \eq{\tup{r}_{\zr}\cdot\tup{\plin}_{\zr}=\tup{r}_t\cdot\tup{\plin}_t=0}). Therefore, the cartesian coordinate representation of such an integral curve on \eq{\cotsp\man{Q}_\ii{1}} can be found by solving the following, greatly simplified,  system of ODEs (which follow from Eq.\eqref{Hprj_full} with \eq{\rfun =1} and \eq{\hat{r}^i \plin_i=0}):  
\begin{small}
\begin{align} \label{prj_xpdot_simp}
\begin{array}{cc}
     \fnsize{for integral curves} \,   \\
     \fnsize{of }\, \sfb{X}^\ss{H}  \fnsize{ on }\, \cotsp\man{Q}_\ii{1}
\end{array} \;\;
\left\{ \;\;
\boxed{ \begin{array}{lllllll}
     \dot{r}^i \,=\, -\tfrac{r_\en^2}{m} \lang^{ij}r_j  &\!\!=\,  \tfrac{r_\en^2}{m}  \plin^i
     &,\qquad  
      \dot{r}^\en  \,=\, \tfrac{r_\en^4}{m}  \plin^\en
\\[4pt]
     \dot{\plin}_i \,=\, -\tfrac{r_\en^2}{m} \lang_{ij}\plin^j  - \pd_i U^\ss{1} 
     &\!\!=  - \tfrac{r_\en^2}{m}  \nrmtup{\plin}^2 r_i  
      - \pd_i U^\ss{1}
   &,\qquad   
   \dot{\plin}_\en  \,=\, - \tfrac{r_\en}{m} (  \nrmtup{\plin}^2 + 2 r_\en^2 \plin_\en^2 ) \,-\, \pd_\en (U^\zr + U^\ss{1})
\end{array} } \right.
\end{align}
\end{small}
Importantly, if \eq{\bar{\mu}_t\in\cotsp\man{Q}_\ii{1}} is such an integral curve of \eq{\sfb{X}^\ss{H}} — that is, \eq{(\bartup{r},\bartup{\plin})\circ \bar{\mu}_t\in\mbb{R}^{\ii{2}\en}} is a solution to the above —  then \eq{\bar{\kap}_t=\colift\psi(\bar{\mu}_t)\in \cotsp\Sig_\ii{1}} is an integral curve of the original dynamics, \eq{\sfb{X}^\ss{K}}, on the \eq{\sfb{X}^\ss{K}}-invariant submanifold \eq{\cotsp\Sig_\ii{1}} (the converse is true as well). That is,   \eq{(\bartup{r},\bartup{\plin})\circ \bar{\kap}_t\in\mbb{R}^{\ii{2}\en}} satisfies \eq{r^\en=1} and \eq{\plin_\en =0} and is a solution to \eq{\dot{r}^i= \plin_i/m} and \eq{\dot{\plin}_i = -\pd_i(V^\zr+ V^\ss{1})}.  Recall that, in reality, for the \textit{original} system,  we are only interested in the ``\eq{(r^i,\plin_i)}-part'' of the system  (i.e., the dynamics on \eq{\cotsp\bs{\Sigup}\cong\cotsp\Sig_\ii{1}}), with the  ``\eq{(r^\en,\plin_\en)}-part''  (i.e., the dynamics on \eq{\cotsp\vsp{N}}) being physically irrelevant. As such, we are free to limit consideration of the transformed system to the above ODEs (i.e., to the submanifold \eq{\cotsp\man{Q}_\ii{1}}) as this still allows us to fully recover the the original system solutions on \eq{\cotsp\Sig_\ii{1}} (the ``\eq{(r^i,\plin_i)}-part'' of the original system).

\begin{small}
\begin{itemize}[nosep]
    \item[{}] \rmsb{Clarifying Remark \ref{rem:invSub_prj}.} Let us attempt to mitigate any confusion regarding Remark \ref{rem:invSub_prj}. Let \eq{\bar{\mu}_t=(\barpt{q}_t,\barbs{\mu}_t)\in\cotsp\bvsp{E}}  be some integral curve of \eq{\sfb{X}^\ss{H}}, and \eq{\bar{\kap}_t=(\barpt{x}_t,\barbs{\kap}_t)=\colift\psi(\bar{\mu}_t)\in\cotsp\bvsp{E}} the corresponding integral curve of \eq{\sfb{X}^\ss{K}}.
    It follows from Eq.\eqref{iom_nom_again} and \ref{iom_Hprj_0} that \eq{\kap_\en=\bs{\mu}\cdot\hsfb{q}} is constant and that \eq{x^\en=\nrm{\pt{q}}} evolves linearly in time:\footnote{Note that \eq{ \pbrak{r^\en}{K}|_\ii{\cotsp\Sig_\ii{b}} =0 } and \eq{ \pbrak{r}{H}|_\ii{\cotsp\man{Q}_\ii{b}} =0}. }
    \begin{small}
    \begin{align}
    \begin{array}{rlllllllll}
        \fnsize{original:} &
        \left.\begin{array}{llll}
               \pbrak{\plin_\en}{K}_{\bar{\kap}_t} \,=\,  0 
            \\[4pt]
            \pbrak{r^\en}{K}_{\bar{\kap}_t} \,=\,  \tfrac{1}{m} \kap^\en(t) 
        \end{array}\right.
         &\qquad \Rightarrow \qquad 
         \begin{array}{lllll}
            \plin_\en(\bar{\kap}_t) =  \envec \cdot\barbs{\kap}_t = \kap_\en(t) = \kap_\en(0) = \fnsize{const.} =: \kap^\iio_\en 
         \\[4pt]
            r^\en(\barpt{x}_t) 
            = x^\en(t) = x^\en(0) + \tfrac{\kap_\iio^\en}{m} t
         \end{array}
    \\ \;\;
    \\
         \fnsize{transformed:} & 
        \left.\begin{array}{llll}
              \pbrak{\plin_i\hat{r}^i}{H}_{\bar{\mu}_t} \,=\,  0
         \\[4pt]
            \pbrak{r}{H}_{\bar{\mu}_t} \,=\,  \tfrac{1}{m} \bs{\mu}_t\cdot \hsfb{q}_t 
        \end{array}\right.
          &\qquad \Rightarrow \qquad 
       \begin{array}{llll}
             \plin_i(\bar{\mu}_t)\hat{r}^i(\barpt{q}_t) =  \bs{\mu}_t\cdot \hsfb{q}_t =   \bs{\mu}_{\zr}\cdot \hsfb{q}_{\zr} = \kap^\iio_\en 
         \\[4pt]
           r(\barpt{q}_t) = \nrm{\ptvec{q}_t} =  \nrm{\ptvec{q}_{\zr}} + \tfrac{\kap_\iio^\en }{m}  t
        \end{array}
    \end{array}
    \end{align}
    \end{small}
    \item[] \textit{Original:}
    The above brackets with \eq{K} were already discussed in section \ref{sec:Hnom_prj}, Eq.\eqref{pbraks_xpN_nom}:  \eq{\plin_\en(\bar{\kap}_t) = \kap_\en(t)= \kap_\en(0)=:\kap^\iio_\en} is a constant determined by the initial conditions for translational momentum of the \textit{original} system along the ``extra'' dimension \eq{\vsp{N}\perp\bs{\Sigup}} (i.e., the \eq{\envec} direction).  Since we do not care about the \textit{original} dynamics in this fictitious dimension, we are free to limit consideration to  \eq{\kap^\iio_\en= 0} and it then follows that \eq{r^\en(\barpt{x}_t) = x^\en(t)=x^\en(0)=:x_\iio^\en} is also a constant and thus
    \eq{\bar{\kap}_t\in\cotsp\Sig_{x_\iio^\en}}. That is, if  \eq{\bar{\kap}_{\zr}\in\cotsp\Sig_{x_\iio^\en}} — which simply means \eq{\kap_\en(0)=\kap^\iio_\en = 0} —  then  \eq{\bar{\kap}_t\in\cotsp\Sig_{x_\iio^\en}}. In other words, each \eq{\cotsp\Sig_\ii{b}\subset\cotsp\bvsp{E}} is \eq{\sfb{X}^\ss{K}}-invariant (with \eq{b} corresponding to \eq{x_\iio^\en}). 
     \item[] \textit{Transformed:}
    Now consider the corresponding integral curve \eq{\bar{\mu}_t=(\barpt{q}_t,\barbs{\mu}_t) =\inv{\colift\psi}(\bar{\kap}_t)} of the transformed dynamics, \eq{\sfb{X}^\ss{H}}. From \eq{\bar{\kap}_t=\colift\psi(\bar{\mu}_t)}, it follows that \eq{\kap_\en = \hsfb{q}\cdot\bs{\mu} } 
     — that is, \eq{\plin_\en(\bar{\kap}_t) =  (\hat{r}^i \plin_i)(\bar{\mu}_t)} — such that the constant \eq{\kap^\iio_\en} is the same as
    \eq{\kap^\iio_\en  = \kap_\en = \bs{\mu}\cdot\hsfb{q}} (for all \eq{t}). From the above, we see that considering the case \eq{\kap^\iio_\en=0} means that \eq{r(\ptvec{q}_t)=\nrm{\ptvec{q}_t}=\nrm{\ptvec{q}_{\zr}}=:\nrm{\pt{q}_\iio}} is also constant. In other words, if \eq{\bar{\mu}_{\zr} \in \cotsp\man{Q}_\ii{\nrm{\pt{q}_\iio}}} — which simply means \eq{\hsfb{q}_{\zr}\cdot\bs{\mu}_{\zr}=0} — then \eq{\bar{\mu}_t \in  \cotsp\man{Q}_\ii{\nrm{\pt{q}_\iio}}}. This is equivalent to saying that each \eq{\cotsp\man{Q}_\ii{b}} is \eq{\sfb{X}^\ss{H}}-invariant (where \eq{b} corresponds to \eq{\nrm{\pt{q}_\iio}}). But, since \eq{(\barpt{x}_t,\barbs{\kap}_t)=\colift\psi(\barpt{q}_t,\barbs{\mu}_t)}, we also have that \eq{\nrm{\pt{q}_\iio} = x_\iio^\en}.
    Therefore,  \eq{\cotsp\Sig_\ii{b}\ni \bar{\kap}_t \Leftrightarrow \bar{\mu}_t \in \cotsp\man{Q}_\ii{b}} with \eq{b =x_\iio^\en = \nrm{\pt{q}_\iio}}. 
    Lastly, recalling again that we do not care about the motion or configuration of the \textit{original} system along ``extra'' \eq{\envec} direction, we are free limit consideration to the case that \eq{x^\en(0)=x_\iio^\en=1}, which maps to \eq{\nrm{\ptvec{q}_{\zr}}=\nrm{\pt{q}_\iio}=1}. The value of \eq{1} is chosen since it is the simplest case compatible with the projective transformation, \eq{\psi} (\eq{0} is not compatible with \eq{\psi}). 
\end{itemize}
\end{small}

\begin{small}
\begin{notesq}
 We re-iterate that, in reality, we are only concerned with the evolution of, and the initial conditions of,  the \eq{(2\en -2)} coordinates \eq{(\tup{r},\tup{\plin})} for the \textit{original} system. The initial conditions for \eq{(r^\en,\plin_\en)} of the \textit{original} system are arbitrary; we may choose them however we  please
 (almost\footnote{In order to use the projective transformation, we require \eq{r^\en >0}}). 
 These correspond to \eq{\psi^*r^\en = \nrmtup{r}} and \eq{\colift\psi^* \plin_\en = \hat{r}^i \plin_i} for the transformed system. Thus, we may choose initial conditions of  \eq{\nrmtup{r}} and \eq{\hat{r}^i \plin_i= \tfrac{1}{r} r^i \plin_i} for the \textit{transformed} system however we please (so long as \eq{r > 0}). 
\end{notesq}
\end{small}

\subsection*{Simplified Relations on the Submanifolds} 

\paragraph{Transformations \& Kinematics on the Submanifolds.} We now examine the relations between integral curves of \eq{\sfb{X}^\ss{K}} and \eq{\sfb{X}^\ss{H}} on the invariant submanifolds \eq{\cotsp\Sig_\ii{b}} and \eq{\cotsp\man{Q}_\ii{b}}.\footnote{The general case for  integral curves on \eq{\cotsp\bvsp{E}} was detailed in section \ref{sec:prj_Xform_gen}, \ref{sec:prj_momentum}. }
First, recall the following properties of the projective transformation:  
\begin{small}
\begin{align} \label{prj_subman_rels_again}
     \psi|_\ii{\man{Q}_\ii{b}} \in\Dfism(\man{Q}_\ii{b};\Sig_\ii{b})
&&,&&
    \tlift\psi|_\ii{\tsp\man{Q}_\ii{b}} \in\Dfism(\tsp\man{Q}_\ii{b} ; \tsp \Sig_\ii{b} )
&&,&&
      \colift\psi|_\ii{\cotsp\man{Q}_\ii{b}} \in\Dfism(\cotsp\man{Q}_\ii{b} ;\cotsp\Sig_\ii{b})
\end{align}
\end{small}
For instance, if 
\eq{(\barpt{q},\bsfb{u}) = \inv{\tlift\psi}(\barpt{x},\bsfb{v})} for some \eq{(\barpt{x},\bsfb{v})\in\tsp\Sig_\ii{b}}, then \eq{(\barpt{q},\bsfb{u})\in\tsp\man{Q}_\ii{b}}, and vice versa. The explicit transformation is given by  \eq{\tlift\psi} in Eq.\eqref{lift_prj_act6} which, for the present case, can be simplified using \eq{ x^\en = \nrm{\pt{q}} = b} and \eq{ \v^\en = \inner{\sfb{u}}{\hsfb{q}} = 0 }.
In particular, for \eq{b =1}, then \eq{\tlift\psi}-related points in \eq{\tsp\Sig_\ii{1}} and \eq{\tsp\man{Q}_\ii{1}} transform by:
\begin{small}
\begin{align} \label{tlift_prj_E3}
\begin{array}{rclll}
   \tsp\Sig_{1} \ni \;  (\barpt{x},\bsfb{v}) \,=\, \tlift\psi(\barpt{q},\bsfb{u}) 
    &\quad 
    &\quad
  (\barpt{q},\bsfb{u}) \,=\, \inv{\tlift\psi}(\barpt{x},\bsfb{v})  
  \; \in \tsp\man{Q}_{1}
\\[5pt]
       \barpt{x} 
   \,=\,  \tfrac{1}{q^\en}\ptvec{q} +  \envec 
       &\quad \leftrightarrow &\quad 
       \barpt{q} 
   \,=\,  \tfrac{1}{\nrm{\pt{x}}} (\ptvec{x} +  \envec ) 
\\[5pt]
    \bsfb{v} = \sfb{v} 
    \,=\, 
   \tfrac{1}{q^\en} (\sfb{u} 
    - \tfrac{u^\en}{q^\en} \hsfb{q} )
 &\quad 
 &\quad
     \bsfb{u} 
    \,=\, 
  \tfrac{1}{ \nrm{\pt{x}}} \big( \sfb{v} - \inner{\hsfb{x}}{\sfb{v}} \hsfb{x} \big)  \,-\, \tfrac{1}{\nrm{\pt{x}}^2} \inner{\hsfb{x}}{\sfb{v}} \envec 
\end{array}
\qquad 
\left| \qquad \begin{array}{llll}
     x^\en = \nrm{\pt{q}} = 1
 \\[4pt]
     \v^\en = \inner{\sfb{u}}{\hsfb{q}} = 0 
\\[4pt]
         \nrm{\sfb{v}}^2 = \tfrac{1}{q_\en^2}( \nrm{\sfb{u}}^2  + u_\en^2/q_\en^2 )
  \\[5pt]
     \nrm{\sfb{u}}^2 = \tfrac{1}{\nrm{\pt{x}}^2}( \nrm{\sfb{v}}^2 - \inner{\hsfb{x}}{\sfb{v}}^2)
\end{array}\right.
\end{align}
\end{small}
Similarly, for the cotangent lift, if \eq{(\barpt{q},\barbs{\mu})= \inv{\colift\psi}(\barpt{x},\barbs{\kap})} for some \eq{(\barpt{x},\barbs{\kap})\in\cotsp\Sig_\ii{b}}, then \eq{(\barpt{q},\barbs{\mu})\in\cotsp\man{Q}_\ii{b}}, and vice versa. In particular, for \eq{b =1} then Eq.~\ref{colift_prj_def} – \ref{colift_prj_def1} for \eq{\colift\psi} simplifies to the following for any \eq{\colift\psi}-related points in \eq{\cotsp\Sig_\ii{1}} and \eq{\cotsp\man{Q}_\ii{1}}:
\begin{small}
\begin{align} \label{colift_prj_E3}
\boxed{ \begin{array}{rclll}
     \cotsp\Sig_{1} \ni \;   (\barpt{x},\barbs{\kap}) \,=\, \colift \psi (\barpt{q},\barbs{\mu}) 
    &\quad 
    &\quad
    (\barpt{q},\barbs{\mu}) \,=\, \colift \inv{\psi} (\barpt{x},\barbs{\kap}) 
     \; \in\cotsp\man{Q}_{1}
\\[5pt]
       \barpt{x} 
   \,=\,  \tfrac{1}{q^\en}\ptvec{q} +  \envec 
       &\quad \leftrightarrow &\quad 
       \barpt{q} 
    \,=\, \tfrac{1}{\nrm{\pt{x}}} (\ptvec{x} +  \envec )
\\[5pt]
    \barbs{\kap}  =  \bs{\kap} 
    \,=\, 
    q^\en (\bs{\mu} - q^\en \mu_\en \hsfb{q}^{\flt} )
 &\quad 
 &\quad
    \barbs{\mu} 
    \,=\, 
  \nrm{\pt{x}} \big( \bs{\kap} - (\bs{\kap}\cdot\hpt{x}) \hsfb{x}^\flt \big) - \nrm{\pt{x}}^2 (\bs{\kap}\cdot\hsfb{x})\enform  
\end{array} }
\qquad 
\left| \qquad \begin{array}{llll}
     x^\en = \nrm{\pt{q}} = 1
 \\[4pt]
     \kap_\en = \bs{\mu}\cdot \hsfb{q} = 0  
\\[4pt]
        \nrm{\bs{\kap}}^2 = q_\en^2( \nrm{\bs{\mu}}^2 + q_\en^2 \mu_\en^2) 
 \\[4pt]
    \nrm{\bs{\mu}}^2 = \nrm{\pt{x}}^2 \big( \nrm{\bs{\kap}}^2 - (\hsfb{x}\cdot\bs{\kap})^2 \big) 
\end{array}\right.
\end{align}
\end{small} 
    (see footnote for arbitrary \eq{b} \footnote{\eq{\forall\,  b\in\mbb{R}_\ii{+}}, the transformation for \eq{\colift\psi}-related points, \eq{ (\barpt{x},\barbs{\kap}) = \colift \psi (\barpt{q},\barbs{\mu}) \in \cotsp \Sig_\ii{b}} and \eq{(\barpt{q},\barbs{\mu})= \inv{\colift \psi} (\barpt{x},\barbs{\kap}) \in\cotsp \man{Q}_\ii{b}} is as follows (with \eq{\hpt{q}=\ptvec{q}/b} and \eq{x^\en = b}):
    \begin{align} 
      \begin{array}{rclll}
           \barpt{x}  \,=\,  \tfrac{1}{q^\en b }\ptvec{q} +  b \envec 
           &\leftrightarrow & 
           \barpt{q} 
       \,=\,  b\hpt{x} +  \tfrac{1}{\nrm{\pt{x}}}\envec 
    \\[3pt]
        \barbs{\kap}  =  \bs{\kap} 
        \,=\, 
        b q^\en \bs{\mu} \,-\, q_\en^2 \mu_\en \hsfb{q}^{\flt} 
     &\leftrightarrow &
        \barbs{\mu}  \,=\, 
      \tfrac{\nrm{\pt{x}}}{b} \big( \bs{\kap} - (\bs{\kap}\cdot\hsfb{x}) \hsfb{x}^\flt \big) \,-\, \nrm{\pt{x}}^2 (\bs{\kap}\cdot\hsfb{x})\enform  
    \end{array} 
    \end{align} }).
Note the angular momentum magnitude, \eq{\lang^2=\nrmtup{r}^2 \nrmtup{\plin}^2 - (r^i \plin_i)^2\in\fun(\cotsp\vsp{E})}, satisfies: 
\begin{small}
\begin{align} \label{ell_submans}
    \begin{array}{ccc}
           & &  \fnsize{for $\en-1=3$:} \\
      \lang^2|_\ii{\cotsp\man{Q}_{1}} \,=\, \nrmtup{\plin}^2
      \qquad,\qquad 
       \nrm{\bs{\mu}}^2 \,=\, \lang^2(\mu_\ss{\pt{q}}) \,=\,  \nrm{\pt{q}}^2 \nrm{\bs{\mu}}^2 - (\ptvec{q}\cdot\bs{\mu})^2  \,=\, \nrm{\pt{x}}^2 \nrm{\bs{\kap}}^2 - (\ptvec{x}\cdot\bs{\kap})^2 \,=\, \lang^2(\kap_{\pt{x}}) \qquad 
       &  &
         =\, \nrm{\ptvec{q}\times \bs{\mu}^\shrp} \,=\,
        \nrm{\ptvec{x}\times \bs{\kap}^\shrp}
    \end{array}
\end{align}
\end{small}
The above relations in Eq.\eqref{colift_prj_E3} and E.\ref{ell_submans} are valid for any  \eq{\colift\psi}-related points in \eq{\cotsp \Sig_\ii{1}} and  \eq{\cotsp \man{Q}_\ii{1}}; they are not limited to integral curves of \eq{\sfb{X}^\ss{K}} and \eq{\sfb{X}^\ss{H}} (though that is our present interest).
We will now assume that \eq{(\barpt{q}_t,\barbs{\mu}_t)\in\cotsp\man{Q}_\ii{1}} and \eq{(\barpt{x}_t,\barbs{\kap}_t)\in\cotsp\Sig_\ii{1}}
are indeed integral curves of \eq{\sfb{X}^\ss{H}} and \eq{\sfb{X}^\ss{K}}, respectively.  It then follows that, in addition to the above, \eq{\barbs{\kap}_t = \sfb{m}_{\barpt{x}}(\dt{\bsfb{x}}_t)} and  \eq{\barbs{\mu}_t = \sfg_\ss{\barpt{q}}(\dt{\bsfb{q}}_t)} are also kinematic momentum covectors  satisfying the various relations detailed in section \ref{sec:prj_momentum}
which,  for the preset case,  may be simplified 
 using \eq{ x^\en = \nrm{\pt{q}} = 1} and  \eq{\kap_\en = m\dot{x}_\en = \bs{\mu}\cdot \hsfb{q} = m \dot{q}= 0 }.
For instance, the expressions for \eq{\barbs{\mu}_t = \sfg_\ss{\barpt{q}}(\dt{\bsfb{q}}_t)} and \eq{\barbs{\kap}_t = \sfb{m}_{\barpt{x}}(\dt{\bsfb{x}}_t)} from Eq.\eqref{momentum_prj_act0} simplify to the following
(suppressing \eq{t}:\footnote{The relations \eq{ x^\en = \nrm{\pt{q}} = 1} and  \eq{\kap_\en = \bs{\mu}\cdot \hsfb{q} = 0 }  (which define \eq{\cotsp\Sig_\ii{1}} and \eq{\cotsp\man{Q}_\ii{1}})  also give: 
\begin{align}
   \kap_\en = m\dot{x}^\en \,=\,   \bs{\mu}\cdot \hsfb{q} = \sfg_\ss{\!\barpt{q}}(\dtsfb{q},\hsfb{q}) = m\inner{\dtsfb{q}}{\hsfb{q}} = m\dot{q} = 0
    &&,&&
    \hsfb{r}^\flt_\ss{\!\pt{q}} \cdot \dtsfb{q} = \hsfb{q}^\flt \cdot \dtsfb{q} = \inner{\dtsfb{q}}{\hsfb{q}}  = \dot{q} = 0
   &&,&&
    \del{\dtsfb{q}} \hsfb{r}_\ss{\!\pt{q}} \,=\, \dtsfb{q}
    &&,&&
    \hsfb{q} \equiv \ptvec{q}
\end{align} }).
\begin{small}
\begin{align} \label{momentum_prj_subman}
\begin{array}{rlllllll}
      \barbs{\kap} \,=\, \sfb{m}(\dt{\bsfb{x}}) \,=\, \sfb{m}(\dtsfb{x}) \,=\,  \bs{\kap} \in\cotsp[\barpt{x}]\Sig_\ii{1} \,=\, \cotsp[\cdt]\bs{\Sigup} 
      \!\!\!& \left\{ \begin{array}{llll}
            \bs{\kap} = \sfb{m}(\dtsfb{x}) = m \dtsfb{x}^\flt = m\dot{x}_i\hbep^i \in \cotsp[\cdt]\bs{\Sigup}
       \\[3pt]
            \kap_\en = m \dot{x}_\en = 0
      \end{array} \right.
\\[16pt]
     \barbs{\mu} = \sfg_\ss{\barpt{q}}(\dt{\bsfb{q}})
       \,=\,  \tfrac{1}{q_\en^2} \sfb{m}(\dtsfb{q})  +  \tfrac{1}{q_\en^4} \sfb{m}(\envec,\dt{\bsfb{q}}) \enform 
      \in\cotsp[\barpt{q}]\man{Q}_\ii{1}
       \!\!\! & \left\{ \begin{array}{llll}
           \bs{\mu} = \tfrac{m}{q_\en^2} \del{\dtsfb{q}} \hsfb{r}^\flt  \,=\, \tfrac{1}{q_\en^2} \sfb{m}(\dtsfb{q}) = \tfrac{m}{q_\en^2}\dtsfb{q}^\flt = \tfrac{m}{q_\en^2}\dot{q}_i\hbep^i    \in \cotsp[\pt{q}]\man{S}^{\en-\ii{2}}_\ii{1}
       \\[3pt]
           \mu_\en = m \dot{q}^\en / q_\en^4
      \end{array} \right.
\end{array}
\end{align} 
\end{small}
where \eq{m_{ij}= m \emet_{ij} = m \kd^i_j}, \eq{\dot{x}_i=\emet_{ij}\dot{x}^j}, and \eq{\dot{q}_i=\emet_{ij}\dot{q}^j}. 
Combining Eq.\eqref{momentum_prj_subman} with the relations in Eq.\eqref{colift_prj_E3} also leads to:
\begin{small}
\begin{align}
\begin{array}{llll}
      \bs{\kap} 
      \,=\,  \sfb{m} \cdot \tfrac{1}{q^\en}(\dtsfb{q} - \tfrac{1}{q^\en} \dot{q}^\en \hsfb{q} ) 
\\[4pt]
     \kap_\en \,=\, \bs{\mu}\cdot \hsfb{q}  \,=\,  m\dot{q} \,=\, 0 
\end{array}
\qquad,\qquad 
\begin{array}{llll}
     \bs{\mu} \,=\, \sfb{m}\cdot \nrm{\pt{x}}^2  \del{\dtsfb{x}} \hsfb{r} 
     \,=\, \sfb{m}\cdot \nrm{\pt{x}}( \dtsfb{x} -\dot{x}\hsfb{x})
\\[4pt]
      \mu_\en \,=\,   -\nrm{\pt{x}}^2 \bs{\kap}\cdot\hsfb{x} 
         \,=\, -m \nrm{\pt{x}}^2 \dot{x}
\end{array}
\end{align}
\end{small}
We also note that the velocity tangent vectors,\footnote{Recall that, for our purposes, any arbitrary \eq{\tsp[\cdt]\Sig_\ii{b}} is equivalent to  \eq{\tsp[\cdt]\bs{\Sigup}}. Neither the base point nor the value of \eq{b} matters (in practice).}
\begin{small}
\begin{align}
  \dt{\bsfb{x}} = \dtsfb{x} = \dot{x}^i \hbe_i \in \tsp[\cdt]\Sig_\ii{1} = \tsp[\cdt]\bs{\Sigup}
  &&,&&
  \dt{\bsfb{q}} = \dot{q}^\a \hbe_\a = \dtsfb{q} + \dot{q}^\en \envec \in \tsp[\barpt{q}]\man{Q}_\ii{1}
  \qquad
  \fnsize{with} \quad \dtsfb{q} = \dot{q}^i \hbe_i =  \del{\dtsfb{q}} \hsfb{r}_\ss{\!\pt{q}} \in\tsp[\pt{q}]\man{S}^{\en-\ii{2}}_\ii{1}
\end{align}
\end{small}
are related as in Eq.\eqref{tlift_prj_E3} such that \eq{\dt{\bsfb{x}}_t=\dtsfb{x}_t=\psi_*(\dt{\bsfb{q}}_t)} and \eq{\dt{\bsfb{q}}_t = \psi^*(\dtsfb{x}_t)}:  
\begin{small}
\begin{align}
\begin{array}{lllll}
     \dtsfb{x} \,=\, \tfrac{1}{q^\en} (\dtsfb{q} - \tfrac{1}{q^\en}\dot{q}^\en \hsfb{q} )
\\[4pt]
     \dot{x}^\en \,=\, \inner{\dtsfb{q}}{\hsfb{q}}  \,=\, \dot{q}   \,=\, 0 
\end{array}
\qquad,\qquad 
\begin{array}{lllll}
    \dtsfb{q} \,=\, \tfrac{1}{\nrm{\pt{x}}}( \dtsfb{x} - \dot{x} \hsfb{x})
 \\[4pt]
    \dot{q}^\en \,=\, - \dot{x} / \nrm{\pt{x}}^2 
\end{array}
\end{align}
\end{small}



\paragraph{Reduced Dynamics \& Hamiltonian.}
Let \eq{\mscr{K}} denote the restriction of \eq{K\in\fun(\cotsp\bvsp{E})} to the hypersurface in \eq{\cotsp\bvsp{E}} defined by \eq{\plin_\en = 0} such that both \eq{\plin_\en} and \eq{r^\en} are integrals of motion of \eq{\sfb{X}^\sscr{K}} (as was detailed in section \ref{sec:Hnom_prj}):
\begin{small}
\begin{align}
       \mscr{K} := K|_{\plin_\en=0}
    \,=\, 
      \tfrac{1}{2}m^{ij} \plin_i \plin_j + V \,=\,   \tfrac{1}{2m} \nrmtup{\plin}^2 + V^\zr(\rfun)  + V^\ss{1}(\tup{r})
      &&,&&
    \begin{array}{lllllll}
           \pbrak{\mscr{K}}{\mscr{K}} \,=\,  \pbrak{r^\en}{\mscr{K}} \,=\,  \pbrak{\plin_\en}{\mscr{K}} \,=\, 0
    \\[4pt]
            \pbrak{ \lang^{ij} }{\mscr{K}} \,=\, -(r^i \emet^{jk} - r^j \emet^{ik})\pd_k V^\ss{1} 
    \\[4pt]
           \pbrak{ \ttfrac{1}{2}\lang^2 }{\mscr{K}}
          \,=\, - \nrmtup{r}^2( \emet^{ij} - \hat{r}^i \hat{r}^j) \plin_i \pd_j V^\ss{1}
           \,=\, -\lang^{ij} r_i \pd_j V^\ss{1}
\end{array}
\end{align}
\end{small}
We then  define \eq{\mscr{H}:= \colift\psi^*\mscr{K}} which is equivalent to \eq{\eq{\mscr{H}\cotsp\psi^*( K -\tfrac{1}{2m} \plin_\en^2)}}. Recalling that \eq{\cotsp\psi^* \plin_\en = \hat{r}^i \plin_i}, we have \eq{\mscr{H} = H - \tfrac{1}{2m}(\htup{r}\cdot\tup{\plin})^2} 
leads to the following cartesian coordinate ODEs and noteworthy Poisson brackets (note \eq{\mscr{H}\neq \tfrac{1}{2}g^{ij} \plin_i \plin_j +U}):
\begin{small}
\begin{gather}\label{Hams_reduce_act}
\begin{array}{llllll}
     \boxed{ \mscr{H} :=\,  \colift\psi^*\mscr{K} \,=\,  \tfrac{r_\en^2}{2m} ( \lang^2 + r_\en^2 \plin_\en^2 )  +  U  }
     \;=\;   \tfrac{r_\en^2}{2m}  \big( \nrmtup{r}^2 \nrmtup{\plin}^2 - (\tup{r}\cdot\tup{\plin})^2 \,+\, r_\en^2 \plin_\en^2 \big)  +  U^\zr(r^\en)  + U^\ss{1}(\bartup{r})  
\end{array}
\\ \nonumber 
\begin{array}{lllllll}
     \dot{r}^i =  \upd^i \mscr{H} \,=\, - \tfrac{r_\en^2}{m} \lang^{ij} r_j
\\[4pt]
   \dot{r}^\en =  \upd^\en \mscr{H} \,=\, \tfrac{r_\en^4}{m}  \plin^\en
\\[4pt]
      \dot{\plin}_i = - \pd_i \mscr{H} \,=\, - \tfrac{r_\en^2}{m} \lang_{ij} \plin^j -\pd_i U^\ss{1}
\\[4pt]
      \dot{\plin}_\en = - \pd_\en \mscr{H} \,=\, - \tfrac{r_\en}{m} ( \lang^2 + 2 r_\en^2 \plin_\en^2 ) - \pd_\en (U^\zr + U^\ss{1})
\end{array}
\qquad
\left| \qquad\begin{array}{lllllll}
    \pbrak{\mscr{H}}{\mscr{H}} \,=\,  \pbrak{r}{\mscr{H}} \,=\, \pbrak{\hat{r}^i \plin_i }{\mscr{H}}  \,=\, \pbrak{r^i \plin_i}{\mscr{H}} 
    \,=\, 0
 \\[4pt]
        \pbrak{ \lang^{ij} }{\mscr{H}} \,=\, -(r^i \emet^{jk} - r^j \emet^{ik})\pd_k U^\ss{1}  
\\[4pt]
      \pbrak{ \ttfrac{1}{2}\lang^2 }{\mscr{H}} \,=\, 
   - \nrmtup{r}^2 \plin^j \pd_j U^\ss{1} 
\\[4pt]
   \pbrak{  \ttfrac{1}{2}\nrmtup{\plin}^2 }{\mscr{H}} 
    \,=\,  - \plin^j \pd_j U^\ss{1} 
\end{array} \right.
\end{gather}
\end{small}
where \eq{-\lang^{ij} r_j = \nrmtup{r}^2 \plin^i  - (r^j \plin_j ) r^i} and \eq{\lang_{ij} \plin^j =  \nrmtup{\plin}^2 r_i - (r^j \plin_j) \plin_i}.  
In contrast to the brackets with \eq{H} in Eq.\eqref{iom_Hprj_0}, we see from the above that both \eq{\htup{r}\cdot\tup{\plin}=\colift\psi^* \plin_\en} and \eq{r = \nrmtup{r} =  \psi^* r^\en} are integrals of motion of \eq{\sfb{X}^\sscr{H}} and, thus, so too is \eq{\tup{r}\cdot\tup{\plin}= r \htup{r}\cdot\tup{\plin}}.
We have also included the Poisson bracket for \eq{\nrmtup{\plin}^2:=\emet^{ij} \plin_i \plin_j}
(derivation\footnote{The Poisson bracket \eq{\tfrac{1}{2}\pbrak{\nrmtup{\plin}^2}{H}} was derived in Eq.\eqref{rmag_pbrak_act}. For the simplified  Hamiltonian \eq{\mscr{H}=H -\tfrac{1}{2m}(\htup{r}\cdot\tup{\plin})^2}, the bracket \eq{\tfrac{1}{2}\pbrak{\nrmtup{\plin}^2}{\mscr{H}}} is then easily obtained by dropping the \eq{\tfrac{1}{2m}(\htup{r}\cdot\tup{\plin})^2} term such that Eq.\eqref{rmag_pbrak_act} leads to  \eq{\tfrac{1}{2}\pbrak{\nrmtup{\plin}^2}{\mscr{H}} = -  \emet^{ij} \plin_i \pd_j U^\ss{1} = - \plin^j \pd_j U^\ss{1}}. This is equivalent to  \eq{\pbrak{\nrmtup{\plin}}{\mscr{H}} = -\hat{\plin}^j \pd_j U^\ss{1}}. })
which verifies \eq{ \pbrak{ \ttfrac{1}{2}\lang^2 }{\mscr{H}}|_\ii{\cotsp\man{Q}_{1}} = \pbrak{  \ttfrac{1}{2}\nrmtup{\plin}^2 }{\mscr{H}}} (as expected from \eq{\lang^2|_\ii{\cotsp\man{Q}_{1}} = \nrmtup{\plin}^2}).  

\subsection{Including Nonconservative Forces} \label{sec:prj_noncon}

We now consider the more general case that, in addition to conservative forces (which are built into the Hamiltonian function), arbitrary nonconservative forces also act on the system. The general geometric framework for Hamiltonian systems with nonconservative forces is detailed in appendix \ref{sec:noncon}, and the relations for transforming such systems were detailed in section \ref{sec:Hxform_noncon}.

\paragraph{Original System.}  Suppose the original system is a nonconservative  Hamiltonian system, \eq{(\cotsp\bvsp{E},\nbs{\omg},K,\bs{f})}, where \eq{(\cotsp\bvsp{E},\nbs{\omg})} is still the same Hamiltonian system considered thus far and where \eq{\bs{f}=f_\a \hbdel^\a\in\formsh(\cotsp\bvsp{E})} is a horizontal and non-closed 1-form accounting for all nonconservative forces (see sections \ref{sec:noncon} and \ref{sec:Hxform_noncon}). As previously discussed, it could be that \eq{\bs{f}} is not a true 1-form on \eq{\cotsp\vsp{E}} as it may have some domain other than \eq{\cotsp\vsp{E}} (e.g., \eq{\bs{f}} could be nonautonomous or some non-feedback control force). Yet, what remains true — and what is important for the below developments — is that \eq{\bs{f}} \textit{does} take values in the horizontal cotangent subspaces, \eq{\cotsph[\cdt](\cotsp\bvsp{E})\subset \cotsp[\cdt](\cotsp\bvsp{E})}, such that \eq{\inv{\nbs{\omg}}(\bs{f})=f_\a \hbpartup{\a}} takes values in the vertical tangent subspaces, \eq{\tspv[\cdt](\cotsp\bvsp{E})}.
With this caveat in mind, we will continue to write \eq{\bs{f}\in\formsh(\cotsp\bvsp{E})} and \eq{\inv{\nbs{\omg}}(\bs{f})\in\vectv(\cotsp\bvsp{E})} for simplicity. 
The total dynamics for the original system, including the nonconservative forces, are then given by a non-Hamiltonian vector field, \eq{\sfb{X}^\ss{K,\bs{f}}=  \sfb{X}^\ss{K} + \inv{\nbs{\omg}}(\bs{f}) \in\vect(\cotsp\bvsp{E})}, such that the derivative of any \eq{h\in\fun(\cotsp\bvsp{E})} along the original flow, \eq{\lderiv{\sfb{X}^{K,\bs{f}}} h},  is no longer simply the Poisson bracket with \eq{K} but, rather, as seen below: 
\begin{small}
\begin{align}  \nonumber
       \sfb{X}^\ss{K,\bs{f}} :=\, \inv{\nbs{\omg}}( -\dif K + \bs{f}) \,=\,    
    \sfb{X}^\ss{K} + \inv{\nbs{\omg}}(\bs{f})   \,=\, \upd^\a K \hbpart{\a} \,-\, (\pd_\a K - f_\a) \hbpartup{\a}
    \quad\;,\quad\;  \inv{\nbs{\omg}}(\bs{f}) = f_\a \hbpartup{\a} \in\vectv(\cotsp\bvsp{E})
\\ \label{dfdt_noncon} 
    \dt{h} \,\equiv\,  \lderiv{\sfb{X}^{K,\bs{f}}} h \,=\, \dif h \cdot \sfb{X}^\ss{K,\bs{f}}  \,=\, -\inv{\nbs{\omg}}(\dif h,\dif K) + \inv{\nbs{\omg}}(\dif h,\bs{f})   \,=\, \pbrak{h}{K} \,+\, f_\a \upd^\a h 
\end{align}
\end{small}
where the additional term is \eq{\lderiv{\inv{\bs{\omg}}(\bs{f})}h = f_\a \upd^\a h }. 
The above holds for any nonconservative Hamiltonian system on the cotangent bundle of \textit{any} (real) smooth manifold, and the local expressions hold for \textit{any} cotangent-lifted coordinate basis. 
Here, we are concerned specifically with the original mechanical Hamiltonian system on Euclidean \eq{\en}-space detailed in section \ref{sec:Hnom_prj}; \eq{K} is expressed in  cotangent-lifted cartesian coordinates \eq{(\bartup{r},\bartup{\plin})} as \eq{ K =\tfrac{1}{2}m^{\a\b} \plin_\a \plin_\b + V} with \eq{m^{\a\b}=\tfrac{1}{m}\emet^{\a\b}=\tfrac{1}{m} \kd^{\a\b}}.
Recall that the original potential is of the form  \eq{V=V^\zr(\rfun)+V^\ss{1}(\tup{r})} and, importantly, we impose the property \eq{\pd_\en V=  0} (i.e., \eq{V=V(\tup{r})} does not depend on \eq{r^\en}). Or, as conservative forces correspond to the exact horizontal 1-form \eq{-\dif V \equiv -\copr^*\dif V \in \formsex\cap\formsh(\cotsp\bvsp{E})}, the property  \eq{\pd_\en V=  0} is equivalent 
to  \eq{\dif V \cdot \hbpart{\en} = 0}.\footnote{We have often expressed this as \eq{\dif V \cdot \envec = 0}. Recalling that \eq{\envec=\be[r^\en]\in\vect(\bvsp{E})} and \eq{\hbpart{\en} = \Vlift{\envec}\in\vect(\cotsp\bvsp{E})},  we have \eq{\pd_\en V = \dif V \cdot \envec = \copr^* \dif V \cdot \Vlift{\envec} \equiv \dif V \cdot \hbpart{\en}}. Here, as usual, we re-use notation \eq{V\equiv \copr^* V\in\fun(\cotsp\bvsp{E})} and \eq{\dif V \equiv \copr^* \dif V\in \formsex\cap\formsh(\cotsp\bvsp{E})}.  }
We now impose the analogous condition on the nonconservative forces: 

\begin{small}
\begin{remrm}
    For the original system, \eq{(\cotsp\bvsp{E},\nbs{\omg}, K ,\bs{f})}, we limit consideration of any nonconservative forces, \eq{\bs{f}\in\formsh(\cotsp\bvsp{E})}, to only those satisfying \eq{\bs{f}\cdot\hbpart{\en}=0} (that is, \eq{\bs{f}=f_i\hbdel^i}) such that the  total dynamics \eq{\sfb{X}^\ss{K,\bs{f}}}, and corresponding cartesian coordinate ODEs, are: 
    \begin{small}
    \begin{align} \label{f_noncon_nom}
    \begin{array}{lllll}
         f_\en := \bs{f}\cdot\hbpart{\en}=0
      \qquad  \Rightarrow
     \\[4pt]
         \phantom{X}
    \end{array}
    &&
     \begin{array}{rlllll}
          \sfb{X}^\ss{K,\bs{f}}  
        &\!\!\! =\, \upd^\a K \hbpart{\a} - (\pd_\a K - f_\a) \hbpartup{\a}
        \\[4pt]
         &\!\!\! =\, m^{\a\b} \plin_\b \hbpart{\a} - (\pd_i V - f_i ) \hbpartup{i} 
     \end{array}
        && \Rightarrow && 
    \begin{array}{llll}
        \dot{r}^i = \tfrac{1}{m} \plin^i
        \;, &\; 
        \dot{\plin}_i = -\pd_i V + f_i
        \\[3pt]
         \dot{r}^\en = \tfrac{1}{m} \plin^\en 
         \;, &\;
         \dot{\plin}_\en = 0 
    \end{array}
    \end{align}
    \end{small}
   Note \eq{\plin_\en} is still an integral of motion of \eq{ \sfb{X}^\ss{K,\bs{f}}} and \eq{\cotsp\Sig_\ii{b}} is still \eq{\sfb{X}^\ss{K,\bs{f}}}-invariant, as verified from  \eq{ \lderiv{\sfb{X}^{K,\bs{f}}} (\plin_\en)  = 0} 
    and \eq{ \lderiv{\sfb{X}^{K,\bs{f}}} (r^\en ) = \tfrac{1}{m} \plin^\en  }.\footnote{\textit{Derivation.}  The derivative of any function \eq{h} along \eq{\sfb{X}^\ss{K,\bs{f}}} is given by \eq{\lderiv{\sfb{X}^{K,\bs{f}}} h = \pbrak{h}{K} + f_\a \upd^\a h}. Then property \eq{f_\en=0} then leads to  \eq{\lderiv{\sfb{X}^{K,\bs{f}}} \plin_\en =  \pbrak{\plin_\en}{K} + f_\a\upd^\a \plin_\en = f_\en =0}. For the basic function \eq{r^\en\equiv \copr^* r^\en} we have simply \eq{\lderiv{\sfb{X}^{K,\bs{f}}} r^\en = \lderiv{\sfb{X}^{K}} r^\en =  \pbrak{r^\en}{K} = \tfrac{1}{m} \plin^\en}.}
\end{remrm}
\end{small}

Above, \eq{f_\a:=\bs{f}\cdot\hbpart{\a}} are the components in the (lifted) \eq{r^\a \equiv \copr^* r^\a} basis on \eq{\cotsp\bvsp{E}}. Since \eq{r^\a} are linear coordinates, these \eq{f_\a} are simply the force components one would see in the standard equation \eq{\ddot{r}^\a = m^{\a\b}f_\b} (or, when \eq{r^\a} are further cartesian one chooses to ignore distinction between co/contra-variant indices, then simply \eq{\ddot{r}^\a = \tfrac{1}{m}f_\a}).
 The requirement \eq{f_\en =0} means these forces do no work along the ``extra'' dimension \eq{\vsp{N}} (that is, the \eq{\envec} direction). This is the same property we imposed on the conservative 
 forces.\footnote{For conservative forces, this property is conveniently distilled to the relation \eq{\pderiv{V}{r^\en}=0}. Yet, if we define a the associated force 1-form \eq{\bs{f}^{\mrm{con.}}:= - \copr^* \dif V \in\formsex\cap\formsh(\cotsp\bvsp{E})}, then we find that \eq{\pderiv{V}{r^\en}=0} is equivalent to \eq{\bs{f}^{\mrm{con.}}\cdot\hbpart{\en}=0}.} 
 It follows that \eq{\cotsp\Sig_\ii{b}} is still an invariant submanifold of the original dynamics \eq{\sfb{X}^\ss{K,\bs{f}}}, as stated above. 
This fact also follows immediately from the coordinate ODEs in Eq.\eqref{f_noncon_nom} where \eq{\dot{r}^\en} and \eq{\dot{\plin}_\en} ``actually'' mean \eq{\lderiv{\sfb{X}^{K,\bs{f}}} (\plin_\en)} and \eq{\lderiv{\sfb{X}^{K,\bs{f}}} (r^\en )}, respectively.  
Recall that the derivative of any function \eq{h} along the flow of \eq{\sfb{X}^\ss{K,\bs{f}}} is \eq{\lderiv{\sfb{X}^{K,\bs{f}}} h}, given by Eq.\eqref{dfdt_noncon} with \eq{f_\en = 0}. For instance, along with \eq{r^\en} and \eq{\plin_\en}, we find the derivatives of \eq{K,\,\lang^{ij}}, and \eq{\lang^2} as follows:\footnote{We have used \eq{ \upd^i \tfrac{1}{2} \lang^2 = \nrmtup{r}^2 \plin^i  - (\tup{r}\cdot\tup{\plin})r^i = -\lang^{ij}r_j}. }
\begin{small}
\begin{align} \label{iom_nom_noncon} 
     &\dt{\square} \equiv \lderiv{\sfb{X}^{K,\bs{f}}} \square = \pbrak{\square}{K} + f_i \upd^i \square 
 \\  \nonumber 
 & \; \Rightarrow \;\; 
 \left\{\quad  \begin{array}{llllll}
  \lderiv{\sfb{X}^{K,\bs{f}}} (\plin_\en) =  0 
\\[4pt]
   \lderiv{\sfb{X}^{K,\bs{f}}} (r^\en) =  \tfrac{1}{m} \plin^\en
\end{array} \right.
  \quad, \quad  \lderiv{\sfb{X}^{K,\bs{f}}} (K) 
     = \tfrac{1}{m} \plin^i f_i
  \quad, \quad  
    \begin{array}{lllll}
     \lderiv{\sfb{X}^{K,\bs{f}}} (\lang^{ij}) \,=\,   ( r^i \emet^{jk} - r^j \emet^{ik}) (-\pd_k V^\ss{1} + f_k) 
\\[4pt]
     \lderiv{\sfb{X}^\ii{K,\,\bs{f}}} (\ttfrac{1}{2}\lang^2) 
      \,=\, 
     r^2 (\emet^{ij} - \hat{r}^i\hat{r}^j) \plin_i (-\pd_j V^\ss{1} + f_j)   \,=  \lang^{ij} r_i(-\pd_j V^\ss{1} + f_j) 
\end{array}
\end{align}
\end{small}
where the various expressions for \eq{\pbrak{\square}{K}} have  been given several times previously (for instance, Eq.\eqref{pbraks_nom} or \ref{iom_nom_again}). 
Note the level sets of \eq{K} (the original Euclidean mechanical energy) are generally not invariant submanifolds of \eq{\sfb{X}^\ss{K,\bs{f}}}. 

\paragraph{Transformed System.} Now, as detailed in section \ref{sec:Hmech_xform_gen}, we use \eq{\psi} and \eq{\colift\psi\in\Spism(\cotsp\bvsp{E})} to transform the original Hamiltonian system, \eq{(\cotsp\bvsp{E},\nbs{\omg},K,\bs{f})}, to a new system, \eq{(\cotsp\bvsp{E},\nbs{\omg},H,\bs{\alpha})},  with \eq{H:=\colift\psi^* K =\tfrac{1}{2}g^{\a\b} \plin_\a \plin_\b + U(\bartup{r})} the same as before and where the original nonconservative forces
must\footnote{\eq{\bs{\alpha}:=\colift\psi^*\bs{f}} means that  \eq{\sfb{X}^\ss{H,\bs{\alpha}} = \colift\psi^* \sfb{X}^\ss{K,\bs{f}}} which ensures that if \eq{\bar{\mu}_t\in\cotsp\bvsp{E}} is an integral curve of \eq{\sfb{X}^\ss{H,\bs{\alpha}}} then \eq{\bar{\kap}_t:=\colift\psi(\bar{\mu}_t)} is an integral curve of \eq{\sfb{X}^\ss{K,\bs{f}}}, and vice versa. That is what we mean by saying the forces ``must'' transform as \eq{\bs{\alpha}:=\colift\psi^*\bs{f}}.}
transform as \eq{\bs{\alpha}:=\colift\psi^*\bs{f} \in\formsh(\cotsp\bvsp{E})}. The new total dynamics are given by \eq{ \sfb{X}^\ss{H,\bs{\alpha}} = \colift\psi^* \sfb{X}^\ss{K,\bs{f}}}:
\begin{small}
\begin{align}
     \sfb{X}^\ss{H,\bs{\alpha}} =\, \colift\psi^* \sfb{X}^\ii{K, \barbs{f}} =\,
     \inv{\nbs{\omg}}(-\dif H + \bs{\alpha}) = \sfb{X}^\ss{H} + \inv{\nbs{\omg}}(\bs{\alpha})
     &&,&&
     \begin{array}{llll}
            \bs{\alpha}:= \colift\psi^* \bs{f} = \alpha_\a \hbdel^\a \in\formsh(\cotsp\bvsp{E})
     \\[4pt]
         \inv{\nbs{\omg}}(\bs{\alpha}) = \alpha_\a \hbpartup{\a} \in\vectv(\cotsp\bvsp{E})
     \end{array}
\end{align}
\end{small}
Leading to the following coordinate ODEs which determine the (cartesian coordinate representation of) the integral curves:
\begin{small}
\begin{align}
\!\!\!\! \boxed{ \begin{array}{llllll}
     \sfb{X}^\ss{H,\bs{\alpha}} 
    &\!\!\!\! =\, \upd^\a H \hbpart{\a} - (\pd_\a H - \alpha_\a) \hbpartup{\a}
\\[4pt]
    &\!\!\!\! =\, g^{\a\b} \plin_\b \hbpart{\a} + (  g^{\gam \sig}\Gamma^\b_{\a \gam} \plin_\b \plin_\sig - \pd_\a U + \alpha_\a) \hbpartup{\a} 
\end{array} }
&&\Rightarrow &&
 \begin{array}{llll}
        \dot{r}^i = \upd^i H
        \;, &\; 
        \dot{\plin}_i = -\pd_i H + \alpha_i
        \\[3pt]
         \dot{r}^\en = \upd^\en H 
         \;, &\;
         \dot{\plin}_\en = -\pd_\en H + \alpha_\en 
    \end{array}
\end{align}
\end{small}
where the expressions for the partials of \eq{H} are the same as given previously in Eq.\eqref{Hprj_full}-Eq.\eqref{Hprj_full_alt}. The components in a cotangent-lifted coordinate basis of the horizontal 1-forms, \eq{\bs{f},\bs{\alpha}\in\formsh(\cotsp\bvsp{E})}, transform the same as 1-forms on \eq{\bvsp{E}} (see Eq.\eqref{Fnoncon_gen}) such that \eq{\bs{\alpha}} is given in terms of \eq{\bs{f}} as follows:
\begin{small}
\begin{align} \label{Fnc_xform_prj}
\begin{array}{lllllll}
    \bs{\alpha} := \colift\psi^*\bs{f}  
    &\!\!\!\! \,=\,  \tfrac{1}{r^\en \nrmtup{r}} ( f_i  - \hat{r}_i\hat{r}^j f_j ) \hbdel^i \,-\, \tfrac{1}{r_\en^2} \hat{r}^i f_i \hbdel^\en 
    \\[5pt]
      &\!\!\!\! \,=\,  \tfrac{1}{r^\en r}( \bs{f} - f_{\rfun} \bdel^{\rfun} )
      \,-\, \tfrac{1}{r_\en^2} f_{\rfun} \hbdel^\en
\end{array}
&&,&&
\begin{array}{llll}
      \alpha_i  \,=\,  \tfrac{1}{r^\en \nrmtup{r}} ( \kd^j_i  - \hat{r}_i\hat{r}^j )f_j
\\[5pt] 
      \alpha_\en  \,=\, -\tfrac{1}{r_\en^2} \hat{r}^i f_i \,=\, -\tfrac{1}{r_\en^2} f_{\rfun}
\end{array}
\end{align}
\end{small}
where \eq{f_{\rfun}:=\bs{f}\cdot\bpart{r}=f_i\hat{r}^i} is the radial component (with ``radial'' referring to the hyperplane  \eq{\bs{\Sigup}\subset\Evec}) and \eq{\bdel^{\rfun}=\copr^* \dif r = \copr^*\hbs{r}^\flt}. Unlike the original forces, note that  \eq{\bs{\alpha}\cdot\hbpart{\en}\neq 0}. However, it is readily verified from the above that, in place of \eq{\bs{f}\cdot\hbpart{\en}=0}, we now have \eq{\bs{\alpha}\cdot\bpart{r}=0} (this was also the case for the conservative forces):
\begin{small}
\begin{align}
    \bs{f}\cdot\hbpart{\en}=0 \quad \Leftrightarrow \quad \bs{\alpha}\cdot \bpart{r} = 0
    &&
    \fnsize{i.e.,}
    \quad f_\en =0  \quad \Leftrightarrow \quad  \hat{r}^i \alpha_i = \alpha_{\rfun} = 0
\end{align}
\end{small}
and we find that each  \eq{\cotsp\man{Q}_\ii{b}} is still \eq{ \sfb{X}^\ss{H,\bs{\alpha}}}-invariant:


\begin{small}
\begin{remrm}
    It was shown that \eq{\plin_\en} is still an integral of motion of \eq{ \sfb{X}^\ss{K,\bs{f}}} and each \eq{\cotsp\Sig_\ii{b}} is still \eq{\sfb{X}^\ss{K,\bs{f}}}-invariant. Similarly, for \eq{\bs{\alpha}:=\colift\psi^*\bs{f}} as above, then \eq{\colift\psi^* \plin_\en = \hat{r}^i \plin_i} is still an integral of motion of \eq{ \sfb{X}^\ss{H,\bs{\alpha}}} and \eq{\cotsp\man{Q}_\ii{b}} is still an \eq{ \sfb{X}^\ss{H,\bs{\alpha}}}-invariant submanifold, as verified by: 
    \begin{small}
    \begin{align}
    \begin{array}{lllll}
        \lderiv{\sfb{X}^{K,\bs{f}}} (\plin_\en)  = 0
    \\[5pt]
        \lderiv{\sfb{X}^{K,\bs{f}}} (r^\en ) \,=\, \tfrac{1}{m} \plin^\en  
    \end{array}
    \qquad \Rightarrow \qquad 
    \begin{array}{lllllll}
         \lderiv{\sfb{X}^{H,\bs{\alpha}}} ( \plin_i \hat{r}^i) \,=\, \alpha_i \hat{r}^i = \bs{\alpha}\cdot \bpart{r} \,=\, 0
     \\[5pt]
          \lderiv{\sfb{X}^{H,\bs{\alpha}}}(\rfun) \,=\,   \tfrac{1}{m} \plin_i \hat{r}^i
    \end{array}
    \end{align}
    \end{small}
    \begin{footnotesize}
    \begin{itemize}[nosep]
        \item[{}] \textup{\textit{Derivation.}  
        The above relations for the original system, \eq{\sfb{X}^\ss{K,\bs{f}}} were already shown. 
        Now, for the transformed system, the derivative of any function \eq{h} along \eq{\sfb{X}^\ss{H,\bs{\alpha}}} is given by \eq{\lderiv{\sfb{X}^{H,\bs{\alpha}}} h = \pbrak{h}{H} + \alpha_\a \upd^\a h}. For \eq{\hat{r}^i \plin_i}, we already know that \eq{\pbrak{\hat{r}^i \plin_i}{H}=0}, leading to: }
        \begin{itemize}[nosep]
            \item[{}] \eq{\qquad\qquad \colift\psi^* \plin_\en = \hat{r}^i \plin_i \qquad,\qquad   \lderiv{\sfb{X}^{H,\bs{\alpha}}} (\hat{r}^i \plin_i) = \pbrak{\hat{r}^i \plin_i}{H} + \alpha_\a \upd^\a (\hat{r}^i \plin_i) \,=\, 0 + \alpha_i \hat{r}^i \,=\, \tfrac{1}{r^\en \nrmtup{r}} ( f_i   - \hat{r}_i\hat{r}^j f_j ) \hat{r}^i \,=\, 0 }  
        \end{itemize}
        \textup{And, again,  for the basic function  \eq{\rfun =\nrm{\sfb{r}} \equiv \copr^* r}, we have \eq{ \lderiv{\sfb{X}^{H,\bs{\alpha}}} r =  \lderiv{\sfb{X}^{H}} r = \pbrak{r}{H} = \tfrac{1}{m}\hat{r}^i \plin_i}. }
    \end{itemize}
    \end{footnotesize}
\end{remrm}
\end{small}

\noindent Along with \eq{\hat{r}^i \plin_i} and \eq{\rfun =\nrm{\sfb{r}}} given above, we also note the derivatives of \eq{H}, \eq{\lang^{ij}}, \eq{\lang^2} and \eq{\nrmtup{\plin}^2} along the flow 
of \eq{\sfb{X}^\ss{H,\bs{\alpha}}}:\footnote{The property \eq{\hat{r}^i \alpha_i=0} (and thus \eq{r^i \alpha_i = 0}) has been used in simplifying \eq{\lderiv{\sfb{X}^{H,\bs{\alpha}}} H = \alpha_\a \upd^\a H } to the expression seen in Eq.\eqref{iom_prj_noncon}.
We have also used \eq{ \upd^i \tfrac{1}{2} \lang^2 = \nrmtup{r}^2 \plin^i  - (\tup{r}\cdot\tup{\plin})r^i = -\lang^{ij}r_j }, along with \eq{r^i \alpha_i = 0},  to obtain \eq{ \alpha_i\upd^i \ttfrac{1}{2}\lang^2 =  -\lang^{ij}\alpha_i r_j = \lang^{ij} r_i \alpha_j = \nrmtup{r}^2 \plin^j \alpha_j}. }
\begin{small}
\begin{align} \label{iom_prj_noncon} 
     & \dt{\square} \equiv \lderiv{\sfb{X}^{H,\bs{\alpha}}} \square = \pbrak{\square}{H} + \alpha_\a \upd^\a \square
 \\  \nonumber 
 & \; \Rightarrow \;\; 
 \left\{\quad  \begin{array}{llllll}
        \lderiv{\sfb{X}^{H,\bs{\alpha}}} ( \hat{r}^i \plin_i)  = 0
    \\[4pt]
          \lderiv{\sfb{X}^{H,\bs{\alpha}}}(\rfun) =   \tfrac{1}{m} \hat{r}^i \plin_i 
    \\[4pt]
       \lderiv{\sfb{X}^{H,\bs{\alpha}}} (H) 
       = \tfrac{r_\en^2}{m} (r^2 \plin^i \alpha_i + r_\en^2 \plin^\en \alpha_\en) 
\end{array} \right.
  \quad, \quad  
    \begin{array}{lllll}
      \lderiv{\sfb{X}^{H,\bs{\alpha}}} (\lang^{ij}) \,=\,  (r^i \emet^{jk} - r^j \emet^{ik})(-\pd_k U^\ss{1} + \alpha_k)  
 \\[4pt]
      \lderiv{\sfb{X}^{H,\bs{\alpha}}} (\ttfrac{1}{2}\lang^2)    
      \,=\, r^2 \plin^j( -\pd_j U^\ss{1} + \alpha_j)
  \\[4pt]
     \lderiv{\sfb{X}^{H,\bs{\alpha}}} (\ttfrac{1}{2}\nrmtup{\plin}^2)   
       \,=\, \plin^j (-\pd_j U^\ss{1} + \alpha_j) \,-\, \tfrac{(\htup{r}\cdot\tup{\plin})}{m r^3} \lang^2
\end{array}
\end{align}
\end{small}
where the above expressions for \eq{\pbrak{\square}{H}} were given in Eq.\eqref{iom_Hprj_0} and where the property \eq{\hat{r}^i \alpha_i=0} (and thus \eq{r^i \alpha_i = 0}) has been used to simplify several of the above (for instance, \eq{\alpha_i\upd^i \ttfrac{1}{2}\lang^2 = \lang^{ij} r_i \alpha_j = \nrmtup{r}^2 \plin^j \alpha_j}).

\subsection{Projective Transformation as a Passive Coordinate Transformation} \label{sec:prj_gen_passive}

\begin{small}
\textit{Notation.} Regarding notation for the following developments:
\begin{small}
\begin{itemize}[nosep]
    \item Our indexing is again \eq{i,j,k = 1,\dots,\en-1} and \eq{\a,\beta,\gam = 1,\dots, \en}. 
    \item  As before, \eq{\bartup{r}=(\tup{r},r^\en)=(r^1,\dots,r^{\en -1},r^\en):\Evec\to\mbb{R}^{\en} } will denote global cartesian coordinates for an (autonomous and homogeneous) orthonormal basis \eq{\hbe_\a\in\Evec^{\en}} as described previously, with the corresponding cotangent-lifted global cartesian coordinates on \eq{\cotsp\Evec^{\en}} denoted \eq{(\bartup{r},\bartup{\uplin})=\colift\bartup{r}:\cotsp\Evec^{\en}\to\cotsp\mbb{R}^{\en}\cong\mbb{R}^{2\en}}. 
    \item  Previously, \eq{\barpt{q}} was used for a displacement vector in \eq{\Evec} with cartesian components \eq{q^\a=r^\a(\barpt{q})}. 
    In the following, we will recycle the same notation but with a different meaning; \eq{q^\a} will now denote new configuration coordinate \textit{functions}, \eq{\bartup{q}=(\tup{q},q^\en):\chart{E}{\bartup{q}} \to\mbb{R}^{\en}}, and  \textit{not} displacement vector components. The corresponding cotangent-lifted coordinates are denoted \eq{(\bartup{q},\bartup{\pf})=\colift\bartup{q}:\cotsp\chart{E}{\bartup{q}}^{\en}\to\mbb{R}^{2\en}}. \\
\end{itemize}
\end{small}
\end{small}

\noindent We now take a different approach and formulate a ``passive'' coordinate transformation  which is, in a certain sense, equivalent to the ``active'' projective transformation detailed in the preceding sections. The general theory was detailed in section \ref{sec:Hmech_xform_passive} and, in the following, we simply apply those developments for the specific case of the projective point transformation, \eq{\psi\in\Dfism(\bvsp{E})}, that was specified in Eq.\eqref{prj_def_act}.
It is worth reiterating an important result of section \ref{sec:Hmech_xform_passive} (see Remark \ref{rem:act_v_pass}):

\begin{small}
\begin{notesq}
    Consider some original coordinates \eq{\bartup{r}=(\tup{r},r^\en):\bvsp{E}\to\mbb{R}^{\en}} with cotangent-lifted coordinates \eq{\bartup{\xi}:=(\bartup{r},\bartup{\uplin})=\colift\bartup{r}:\cotsp\bvsp{E}\to\cotsp\mbb{R}^{\en}\cong\mbb{R}^{2\en}}. We use \eq{\psi\in\Dfism(\bvsp{E})} to define new configuration coordinates by \eq{\bartup{q}:=\psi_* \bartup{r} = \bartup{r}\circ\inv{\psi}}, with corresponding cotangent-lifted coordinates denoted \eq{\bartup{z}:=(\bartup{q},\bartup{\pf})=\colift\psi_*(\bartup{r},\bartup{\uplin}) = (\bartup{r}\circ\inv{\psi}, \bartup{\uplin}\circ \inv{\colift\psi})}. 
     The \eq{\bartup{q}}-representation of any \eq{\sfb{N}\in\tens^r_s(\bvsp{E})} then ``looks the same'' as the \eq{\bartup{r}}-representation of \eq{\psi^*\sfb{N}} and, likewise, the \eq{\bartup{z}}-representation of any \eq{\sfb{T}\in\tens^r_s(\cotsp\bvsp{E})} ``looks the same'' as the  \eq{\bartup{\xi}}-representation of 
    \eq{\,\colift \psi^*\sfb{T}}:\footnote{Eq.\eqref{prj_act_v_pass} is quick to verify:  using \eq{\bartup{q}=\bartup{r}\circ\inv{\psi}} and \eq{\bartup{z}=\bartup{\xi}\circ\inv{\colift\psi}}: \\
        \eq{\qquad\qquad \bartup{q}_*\sfb{N} = (\bartup{r} \circ \inv{\psi})_*\sfb{N} = \bartup{r}_*(\inv{\psi}_*\sfb{N}) = \bartup{r}_*( \psi^* \sfb{N})
        \qquad,\qquad 
        \bartup{z}_*\sfb{T} = (\bartup{\xi} \circ \inv{\colift\psi})_*\sfb{T} =  \bartup{\xi}_*(\inv{\colift\psi}_*\sfb{T})  =
        \bartup{\xi}_*( \colift\psi^*\sfb{T}) } .
    } 
    \begin{small}
    \begin{align}\label{prj_act_v_pass}
    \begin{array}{rllll}
        \forall\, \sfb{N}\in\tens^r_s(\bvsp{E}): 
        &\qquad 
        \bartup{q}_*\sfb{N} 
        \,=\, \bartup{r}_*( \psi^*\sfb{N}) 
        &\qquad \qquad 
        \fnsize{i.e.,} \quad \crd{\sfb{N}}{\bar{q}} \,=\, \crd{\psi^*\sfb{N}}{\bar{r}} 
    \\[5pt]
          \forall\, \sfb{T}\in\tens^r_s(\cotsp\bvsp{E}): 
        &\qquad 
        \bartup{z}_*\sfb{T} 
        \,=\, \bartup{\xi}_*( \colift\psi^*\sfb{T}) 
        &\qquad \qquad 
        \fnsize{i.e.,} \quad \crd{\sfb{T}}{\bar{z}} \,=\, \crd{\colift\psi^*\sfb{T}}{\bar{\xi}}  
    \\ 
        \;\;
    \end{array}
    \end{align}
    \end{small}
\end{notesq}
\end{small}
\noindent  The above relations are enough to deduce nearly everything that follows in section \ref{sec:prj_gen_passive}. The preceding sections consisted of pulling back various tensor fields by \eq{\psi} or \eq{\colift\psi} to transform a original Hamiltonian system to a new ``transformed Hamiltonian system''. When coordinates were used, we expressed the original and transformed objects using the \textit{same} original coordinates, \eq{\bartup{\xi}=(\bartup{r},\bartup{\uplin})}.  In the following, we instead define new cotangent-lifted coordinates, \eq{\bartup{z}=(\bartup{q},\bartup{\pf})}, as described above and then express the \textit{same} original Hamiltonian system in these new coordinates. That is, the preceding sections entailed transforming various tensors,  say  \eq{\sfb{T}\in\tens^r_s(\cotsp\bvsp{E})}, 
 to some new tensor,
 \eq{\colift\psi^*\sfb{T}}, and we always considered the \textit{same} \eq{\bartup{\xi}=(\bartup{r},\bartup{\uplin})} coordinate representations of the two \textit{different} tensors, \eq{\crd{\sfb{T}}{\bar{\xi}}} and \eq{\crd{\colift\psi^*\sfb{T}}{\bar{\xi}}} (here, \eq{\sfb{T}} is standing in for, say, a symplectic form, a Hamiltonian function, a metric, an integral of motion, etc.).
 In the following, we instead consider the same original Hamiltonian system throughout but use two \textit{different} coordinate representations. E.g., the same \eq{\sfb{T}\in\tens^r_s(\cotsp\bvsp{E})} and coordinate representations  \eq{\crd{\sfb{T}}{\bar{\xi}}} and \eq{\crd{\sfb{T}}{\bar{z}}}.
 
To summarize: before, we were concerned with transforming some \eq{\sfb{T}} to \eq{\colift\psi^*\sfb{T}} whereas, in the following, we are concerned with transforming \eq{\crd{\sfb{T}}{\bar{\xi}}} to \eq{\crd{\sfb{T}}{\bar{z}}}, where  \eq{\bartup{z}:=\bartup{\xi}\circ\inv{\colift\psi}}.
Thus, all of the work is already done;   \eq{\crd{\sfb{T}}{\bar{z}}= \crd{\colift\psi^*\sfb{T}}{\bar{\xi}}} and we already found  \eq{\colift\psi^*\sfb{T}} and \eq{\crd{\colift\psi^*\sfb{T}}{\bar{\xi}}} in the preceding sections (again, \eq{\sfb{T}} is just a place-holder for various tensors or functions we care about).

\subsubsection{Symplectic Coordinate Transformation for Cartesian and Projective Coordinates}

Following section \ref{sec:Hmech_xform_passive}, 
we use \eq{\psi\in\Dfism(\bvsp{E})} as defined in Eq.\eqref{prj_def_act} to define an associated coordinate transformation \eq{\uppsi= \ns{c}^\ss{\bar{r}}_\ss{\bar{q}}\in\Dfism(\rchart{R}{\bartup{q}}^{\en} ;\rchart{R}{\bartup{r}}^{\en})}. 
Recall that \eq{\psi,\inv{\psi}\in\Dfism(\bvsp{E})} can be expressed in the basis \eq{\hbe_\a\in\Evec^{\en}}, and the associated cartesian coordinates \eq{r^\a\in\fun(\Evec^{\en})}, as
\begin{small}
\begin{align} \label{proj_cartesian_2}
      \bsfb{r} \,=\, \sfb{r} + r^\en\envec \,=\,  r^i\hbe_i + r^\en \envec 
 &&,&& 
       \psi \,=\, \tfrac{1}{r^\en\nrmtup{r}}r^i \hbe_i \,+\, \nrmtup{r}\envec 
       \,\equiv\, 
       \tfrac{1}{r^\en} \hsfb{r} \,+\, r \envec 
&&,&& 
        \inv{\psi} \,=\, \tfrac{r^\en}{\nrmtup{r}}r^i \hbe_i \,+\, \tfrac{1}{\nrmtup{r}} \envec 
        \,\equiv\, 
        r^\en \hsfb{r} \,+\,  \tfrac{1}{r} \envec 
\end{align}
\end{small}
where  \eq{\rfun =\sqrt{\inner{\sfb{r}}{\sfb{r}}} \in\fun(\Evec)}  and \eq{\hsfb{r}=\tfrac{1}{r}\sfb{r}\cong \be_{\rfun}\in\vect(\Evec)}.  
From the original cartesian coordinates \eq{\bartup{r}=(\tup{r},r^\en)}, we now define new coordinates, \eq{\bartup{q}=(\tup{q},q^\en):\bvsp{E}\to\mbb{R}^{\en}},  by \eq{q^\a := \psi_* r^\a = r^\a \circ\inv{\psi} \in\fun(\bvsp{E})}. 
These  \eq{q^\a} are \textit{not} components of a 4-vector in \eq{\vsp{E}}; they are coordinate functions on \eq{\bvsp{E}} defined as follows for any \eq{\barpt{x}\in\bvsp{E}}:
\begin{small}
\begin{align} \label{prj_pass2}
\begin{array}{llllll}
      r^\a  = q^\a \circ \psi
      &, \qquad  r^\a (\barpt{x}) =  \hbep^\a \cdot \barpt{x} = x^\a 
       &,\qquad
    r^i(\barpt{x}) = x^i 
    &,\quad  r^\en(\barpt{x}) = x^\en 
\\[5pt]
     q^\a := r^\a \circ\inv{\psi}
         &,\qquad q^\a(\barpt{x}) = r^\a\circ \inv{\psi}(\barpt{x}) =  \hbep^\a \cdot \inv{\psi}(\barpt{x}) 
           &,\qquad
      q^i(\barpt{x}) = \tfrac{x^\en}{\nrm{\pt{x}}}x^i &,\quad  q^\en(\barpt{x}) = \tfrac{1}{\nrm{\pt{x}}}
\end{array}  
\end{align}
\end{small}
As previously shown (section \ref{sec:Hmech_xform_passive}), the transition function \eq{\ns{c}^\ss{\bar{r}}_\ss{\bar{q}}=\bartup{r}\circ\inv{\bartup{q}}:\rchart{R}{\bartup{q}}^{\en} \to \rchart{R}{\bartup{r}}^{\en} } is simply the \eq{\bartup{q}}-representation of \eq{\psi:\bvsp{E}\to\bvsp{E}}, which is also the \eq{\bartup{r}}-representation of \eq{\psi}: 
\begin{small}
\begin{align} \label{prj_trnfun}
    \uppsi:=\bartup{q}\circ\psi\circ\inv{\bartup{q}} = \bartup{r}\circ\psi\circ\inv{\bartup{r}} = \bartup{r}\circ \inv{\bartup{q}} = \ns{c}^\ss{\bar{r}}_\ss{\bar{q}} \in\Dfism(\rchart{R}{\bartup{q}}^{\en} ;\rchart{R}{\bartup{r}}^{\en})
    \qquad,\qquad 
    \begin{array}{llll}
          \bartup{r} 
          = \uppsi \circ \bartup{q}  \equiv \uppsi(\bartup{q}) 
      \\[4pt]
         \bartup{q} 
         = \inv{\uppsi}\circ\bartup{r} \equiv \inv{\uppsi}(\bartup{r})
    \end{array}
\end{align}
\end{small}

\begin{small}
\begin{notesq}
    \rmsb{Configuration Coordinate Transformation.} The transformation between cartesian coordinates, \eq{\bartup{r}=(\tup{r},r^\en}), and the ``projective coordinates\footnote{We will refer to the configuration coordinates \eq{\bartup{q}=(\tup{q},q^\en)} as ``projective coordinates''. 
    It should be noted that these coordinates are not necessarily the same as so-called ``homogeneous coordinates'' encountered in projective geometry. Yet, the way that we end up using the projective coordinates \eq{\bartup{q}=(\tup{q},q^\en)} turns out to have many similarities with said homogeneous coordinates.
    For brevity, the corresponding cotangent-lifted projective coordinates, \eq{\bartup{z}=(\bartup{q},\bartup{\pf})=\cotsp \bartup{q}}, will sometimes be referred to simply as ``projective coordinates'' again.}'',
   \eq{\bartup{q}:\chart{E}{\bartup{q}}\to\rchart{R}{\bartup{q}}^{\en}} — defined on the domain \eq{\chart{E}{\bartup{q}} = \bs{\Sigup}_\nozer\oplus \vsp{N}\subset \vsp{E}}
   (\footnote{That is, \eq{\bartup{q}=(\tup{q},q^\en)} are defined for all \eq{\barpt{x}=\ptvec{x}+x^\en \envec\in\vsp{E}=\bs{\Sigup}\oplus\vsp{N}} for which \eq{\nrm{\ptvec{x}}\neq 0} (since our metric is positive-definite, this is equivalent to \eq{\ptvec{x}\neq 0}). })
   —  is given explicitly as follows (compare to Eq.~\ref{prj_active2}):
\begin{small}
\begin{align} \label{prj_passive2}
\qquad 
&\begin{array}{rclccc}
      \bartup{r} \,=\, \bartup{q}\circ\psi \,=\,  \uppsi(\bartup{q}) 
      &\leftrightarrow&  \bartup{q} :=\, \bartup{r}\circ\inv{\psi} \,=\,  \inv{\uppsi}(\bartup{r})
\\[4pt]
      \fnsize{i.e.,} \quad
    \tup{r} = \tfrac{1}{q^\en} \htup{q} 
    \;\;,\;\; r^\en = \nrmtup{q}
   &\leftrightarrow&
    \tup{q} := r^\en \htup{r} 
    \;\;,\;\; q^\en := \tfrac{1}{\nrmtup{r}} = \tfrac{1}{r}
\end{array}  
&&
    \htup{r} = \tup{r}/\nrmtup{r} = \tup{q}/\nrmtup{q}= \htup{q}
\end{align}
\end{small}
\end{notesq}
\end{small}

\noindent For configuration coordinates \eq{\bartup{r}} and  \eq{\bartup{q}} as above,  
let \eq{(\bartup{r},\bartup{\uplin})=\colift\bartup{r}\in\fun^{2\en}(\cotsp\bvsp{E})} and \eq{(\bartup{q},\bartup{\pf})=\colift\bartup{q}\in\fun^{2\en}(\cotsp\bvsp{E})} denote the corresponding cotangent-lifted coordinates on \eq{\cotsp\bvsp{E}}. 
For \textit{any} coordinate point transformation \eq{\bartup{r}=\uppsi(\bartup{q})\leftrightarrow \bartup{q}=\inv{\uppsi}(\bartup{r})}, we know the corresponding conjugate momenta coordinates (i.e., the cotangent-lifted fiber coordinates) transform as \eq{\bartup{\uplin}= \bartup{\pf}\cdot \pderiv{\bartup{q}}{\bartup{r}} \leftrightarrow \bartup{\pf} = \bartup{\uplin}\cdot \pderiv{\bartup{r}}{\bartup{q}} }.
The Jacobian matrices \eq{ \rmd \uppsi =\pderiv{\bartup{r}}{\bartup{q}} } and \eq{ \rmd\inv{\uppsi}= \pderiv{\bartup{q}}{\bartup{r}}} are found from Eq.\eqref{prj_passive2} as:
\begin{small}
\begin{align} \label{dprj_coords}
\begin{array}{llllll}
     \rmd\uppsi = \pderiv{\bartup{r}}{\bartup{q}} 
     & =\;
     \begin{pmatrix}
         \tfrac{1}{q^\en \nrmtup{q}}( \imat_{\en\ii{-1}} - \htup{q}\otms \htup{q}^\flt) & -\tfrac{1}{q_\en^2}\htup{q} 
         \\
         \htup{q}^\flt  & 0 
     \end{pmatrix} 
      & =\;
     \begin{pmatrix}
         \tfrac{\nrmtup{r}}{r^\en}( \imat_{\en\ii{-1}} - \htup{r}\otms \htup{r}^\flt) & -\nrmtup{r}^2\htup{r} 
         \\
         \htup{r}^\flt  & 0 
     \end{pmatrix} 
\\[16pt]
     \rmd\inv{\uppsi} = \pderiv{\bartup{q}}{\bartup{r}}
      & =\;
      \begin{pmatrix}
         \tfrac{r^\en }{\nrmtup{r}}( \imat_{\en\ii{-1}} - \htup{r}\otms \htup{r}^\flt) & \htup{r} 
         \\
        -\tfrac{1}{\nrmtup{r}^2} \htup{r}^\flt   & 0 
     \end{pmatrix} 
      & =\;
     \begin{pmatrix}
         \nrmtup{q} q^\en( \imat_{\en\ii{-1}} - \htup{q}\otms \htup{q}^\flt) & \htup{q} 
         \\
        -q_\en^2 \htup{q}^\flt   & 0 
     \end{pmatrix} 
\end{array}
\end{align}
\end{small}
which are used to obtain the momenta coordinate transformation given below
    (derivation in footnote\footnote{Using Eq.\eqref{dprj_coords}, the momenta coordinate transformation is found from the standard relations as follows:  
    \begin{align}
    \begin{array}{lcllllll}
        \bartup{\uplin} = \bartup{\pf} \cdot \pderiv{\bartup{q}}{\bartup{r}} 
     &\Rightarrow &
         \plin_i \,=\, \tfrac{r^\en}{\nrmtup{r}}( 
        \kd^j_i - \hat{r}^j \hat{r}_i) \pf_j
         \,-\, \tfrac{1}{\nrmtup{r}^2} \hat{r}_i \pf_\en
         \;=\;
          q^\en\nrmtup{q}( 
        \kd^j_i - \hat{q}^j \hat{q}_i) \pf_j
         \,-\, q^2_\en \hat{q}_i \pf_\en
    &,\qquad  
         \plin_\en \,=\, \hat{r}^i \pf_i
         \;=\; \hat{q}^i \pf_i
    \\[6pt]
        \bartup{\pf} = \bartup{\uplin} \cdot \pderiv{\bartup{r}}{\bartup{q}}  
     &\Rightarrow &
         \pf_i \,=\, \tfrac{1}{q^\en\nrmtup{q}}( 
      \kd^j_i - \hat{q}^j \hat{q}_i) \plin_j \,+\,  \plin_\en \hat{q}_i
       \;=\; \tfrac{\nrmtup{r}}{r^\en} ( 
     \kd^j_i - \hat{r}^j \hat{r}_i) \plin_j  \,+\,  \plin_\en \hat{r}_i
    &,\qquad   
         \pf_\en \,=\, -\tfrac{1}{q_\en^2} \hat{q}^i \plin_i
         \;=\; -\nrmtup{r}^2 \hat{r}^i \plin_i
    \end{array}
    \end{align}
    }):

\begin{small}
\begin{notesq}
\rmsb{Symplectic Cotangent Coordinate Transformation.}
The transformation \eq{(\bartup{r},\bartup{\uplin})\leftrightarrow (\bartup{q},\bartup{\pf})} between cartesian and projective coordinates on \eq{\cotsp\bvsp{E}} is a cotangent-lifted coordinate transformation, \eq{\upcolift\uppsi\in\Spism(\cotsp\rchart{R}{\bartup{q}}^{\en} ;\cotsp\rchart{R}{\bartup{r}}^{\en})}, with \eq{\uppsi} given in Eq.\eqref{prj_passive2}. The explicit transformation is: 
\begin{small}
\begin{gather} \label{prj_full_coord}
    (\bartup{r},\bartup{\uplin})= \upcolift\uppsi(\bartup{q},\bartup{\pf}) \; \leftrightarrow \;(\bartup{q},\bartup{\pf}) = \inv{\upcolift\uppsi}(\bartup{r},\bartup{\uplin}) 
    \qquad 
    \phantom{ \lang^2 \,=\, \nrmtup{r}^2 \nrmtup{\plin}^2 - (\tup{r}\cdot\tup{\uplin})^2 \,=\,  \nrmtup{q}^2 \nrmtup{\pf}^2 - (\tup{q}\cdot\tup{\pf})^2 }
\\ \nonumber 
\boxed{ \begin{array}{llll}
    \tup{r} = \tfrac{1}{q^\en}\htup{q} 
\\[4pt]
      r^\en =    \nrmtup{q} 
\\[4pt]
     \tup{\uplin} = q^\en \nrmtup{q} \big( 
     \tup{\pf} - (\htup{q}\cdot\tup{\pf})\htup{q}^\flt \big)  - q_\en^2 \pf_\en \htup{q}^\flt 
\\[4pt]
     \plin_\en = \htup{q}\cdot \tup{\pf}
\end{array}
\quad \leftrightarrow \quad
\begin{array}{lllllll}
    \tup{q} = r^\en \htup{r} 
\\[4pt]
      q^\en = 1/ \nrmtup{r} 
\\[4pt]
     \tup{\pf} = \tfrac{\nrmtup{r}}{r^\en} \big( 
      \tup{\uplin} - (\htup{r}\cdot\tup{\uplin})\htup{r}^\flt \big) +  \plin_\en \htup{r}^\flt
 \\[4pt]
     \pf_\en = -\nrmtup{r}^2 \htup{r}\cdot \tup{\uplin}
\end{array} }
\quad
\begin{array}{llll}
          \nrmtup{\plin}^2 
         = q_\en^2  \big( 
         \nrmtup{q}^2\nrmtup{\pf}^2 -  (\tup{q}\cdot\tup{\pf})^2 + q_\en^2 \pf_\en^2 \big) 
      \\[4pt]
            \tup{r}\cdot\tup{\uplin} = -q^\en \pf_\en
      \\[4pt]
           \nrmtup{\pf}^2 
          = \tfrac{1}{r_\en^2} \big( \nrmtup{r}^2\nrmtup{\plin}^2 -  (\tup{r}\cdot\tup{\uplin})^2 + r_\en^2 \plin_\en^2 \big) 
        \\[4pt]
             \tup{q}\cdot\tup{\pf} = r^\en \plin_\en
       \\[4pt]
            \lang^{ij} = r^i \plin^j  - \plin^i r^j  =  q^i \pf^j  - \pf^i q^j
        \\[4pt]
            \lang^2 = \nrmtup{r}^2 \nrmtup{\plin}^2 - (\tup{r}\cdot\tup{\uplin})^2 =  \nrmtup{q}^2 \nrmtup{\pf}^2 - (\tup{q}\cdot\tup{\pf})^2 
\end{array}
\end{gather}
\end{small}
(this corresponds to a type-2 generating function
\eq{G(\bartup{q},\bartup{\uplin}) = \uppsi(\bartup{q}) \cdot \bartup{\uplin} = \tfrac{1}{q^\en}\hat{q}^i \plin_i + \nrmtup{q} \plin_\en}, satisfying \eq{ \pf_\a = \pderiv{G}{q^\a}} and  \eq{r^\a = \pderiv{G}{\plin_\a}}).
\end{notesq}
\end{small}

Now, all cotangent-lifted coordinates are symplectic with respect to the canonical symplectic form, \eq{\nbs{\omg}\in\formsex^2(\cotsp\vsp{E})}.
That is, the above is a symplectic coordinate transformation (canonical transformation) such that:
\begin{small}
\begin{align}
    \pbrak{r^\a}{\plin_\b} = \pbrak{q^\a}{\pf_\b} = \kd^\a_\b 
    \qquad,\qquad 
     \pbrak{r^\a}{r^\b} = \pbrak{\plin_\a}{\plin_\b} =  \pbrak{q^\a}{q^\b} = \pbrak{\pf_\a}{\pf_\b} = 0
\end{align}
\end{small}
Cartesian coordinates \eq{(\bartup{r},\bartup{\uplin})} are clearly symplectic such that \eq{\cord{\ns{\txw}}{\bartup{r},\bartup{\uplin}}=J}. The above relations for \eq{(\bartup{q},\bartup{\uplin})} are then 
equivalent to:\footnote{The full expression for the Jacobians \eq{\pderiv{(\bartup{r},\bartup{\uplin})}{(\bartup{q},\bartup{\pf})}} and \eq{\pderiv{(\bartup{q},\bartup{\pf})}{(\bartup{r},\bartup{\uplin})}} are quite lengthy and cumbersome. However, the relation \eq{\trn{\pderiv{(\bartup{r},\bartup{\uplin})} {(\bartup{q},\bartup{\pf})}} J  \pderiv{(\bartup{r},\bartup{\uplin})} {(\bartup{q},\bartup{\pf})} = J } (which we already know must be true by design) has indeed  been verified using symbolic manipulation in \sc{matlab}. Since the inverse and transpose of any symplectic matrix is also symplectic, it follows that \eq{\trn{\pderiv{(\bartup{q},\bartup{\pf})}{(\bartup{r},\bartup{\uplin})} } J  \pderiv{(\bartup{q},\bartup{\pf})}{(\bartup{r},\bartup{\uplin})} = J } also holds. }
\begin{small}
\begin{align}
    \cord{\ns{\txw}}{\bartup{r},\bartup{\uplin}}=J
&&,&&
    \cord{\ns{\txw}}{\bartup{q},\bartup{\pf}} \,=\, \upcolift\uppsi^*  \cord{\ns{\txw}}{\bartup{r},\bartup{\uplin}} 
    \,=\,
    \trn{\pderiv{(\bartup{r},\bartup{\uplin})} {(\bartup{q},\bartup{\pf})}} J  \pderiv{(\bartup{r},\bartup{\uplin})} {(\bartup{q},\bartup{\pf})} = J 
&&,&&
    \fnsize{i.e., } \quad  \pderiv{(\bartup{r},\bartup{\uplin})}{(\bartup{q},\bartup{\pf})} \, ,\,  \pderiv{(\bartup{q},\bartup{\pf})}{(\bartup{r},\bartup{\uplin})} \in\Spmat{2\en}
\end{align}
\end{small}
The above can indeed be verified and it re-affirms that both \eq{(\bartup{r},\bartup{\uplin})} and \eq{(\bartup{q},\bartup{\pf})} are symplectic coordinates. 
For instance, the canonical 1-form \eq{\bs{\theta}\in\formsh(\cotsp\vsp{E})}, symplectic form \eq{\nbs{\omg}\in\formsex^2(\cotsp\vsp{E})}, and Poisson bivector \eq{\inv{\nbs{\omg}} \in\vect^2(\cotsp\vsp{E})}, take the 
form:\footnote{These relations are  quick to verify using the usual formula for transformation of 1-form components: \eq{\theta_\ii{q^\a} = \pderiv{r^\b}{q^\a}\theta_\ii{r^\b} + \pderiv{\plin_\b}{q^\a}\theta^\ii{\plin_\b}}. With \eq{\theta_\ii{r^\b} = \plin_\b} and \eq{\theta^\ii{\plin_\b}=0} this gives \eq{\theta_\ii{q^\a} = \pderiv{r^\b}{q^\a} \plin_\b=\pf_\a} (this is simply the standard formula for cotangent-lifted fiber coordinate transformations). Therefore \eq{\bs{\theta}=\pf_\a\dif q^\a = \pf_\a \hbdel[q^\a]} and it follows immediately that  \eq{\nbs{\omg} = \hbdel[q^\a] \wedge \hbdeldn[\pf_\a]}.    }
\begin{small}
\begin{align} 
    \bs{\theta} = \plin_i \hbdel[r^i]\! = \pf_i \hbdel[q^i] 
    &&,&&
    \nbs{\omg}:=- \exd\bs{\theta} = \hbdel[r^i] \!\! \wedge\hbdeldn[\plin_i] = \hbdel[q^i] \!\! \wedge\hbdeldn[\pf_i] 
  &&,&&
    \inv{\nbs{\omg}} = -\hpdii{r^i} \wedge \hpdiiup{\plin_i} =  -\hpdii{q^i} \wedge \hpdiiup{\pf_i}
\end{align}
\end{small}
Lastly, before proceeding to Hamiltonian dynamics formulated in projective coordinates, we collect some useful relations   (all of which follow from the active view):
\begin{small}
\begin{itemize}[nosep]
     \item \sbemph{Metric Components.} Dynamics expressed in the coordinates \eq{q^\a :=\psi_* r^\a} will require the \eq{q^\a}-representations of \eq{\bsfb{\emet}} and \eq{\inv{\bsfb{\emet}}}. We know from the discussion around Eq.~\ref{prj_act_v_pass} that these have exactly the same form as the \eq{r^\a}-representation of \eq{\sfb{g}:=\psi^*\bsfb{\emet}} and \eq{\inv{\sfb{g}}=\psi^*\inv{\bsfb{\emet}}}:
     \begin{small}
    \begin{align} \label{metric_qu}
           \cord{\bar{I}}{\bartup{q}} \,:=\, \bartup{q}_* \bsfb{\emet} \,=\,   \bartup{r}_*(\psi^*\bsfb{\emet}) \,=\, \bartup{r}_* \sfb{g} \,=\, \cord{\txi{g}}{\bartup{r}}  
        \qquad,\qquad 
              \cord{\inv{\bar{I}}}{\bartup{q}} \,=\, \cord{\inv{\txi{g}}}{\bartup{r}}  
    \end{align}
    \end{small}
    Using the \eq{r^\a} basis expression for \eq{\sfb{g}}  found previously in Eq.\eqref{g_proj_active}, this implies the \eq{q^\a} basis expression for \eq{\bsfb{\emet}} is given by
    \begin{footnotesize}
    \begin{align} \label{metric_prj_pass}
    &\begin{array}{rllll}
         \bsfb{\emet} = \emet_{\hat{\a}\hat{\b}} \hbep^\a \otms \hbep^\b  &\!\!\!\! =\,   \tfrac{1}{q_\en^2 \nrmtup{q}^2}( \emet_{\hat{\imath}\hat{\jmath}} - \hat{q}_i\hat{q}_j) \btau^i \otms \btau^j \,+\, \hat{q}_i\hat{q}_j \btau^i \otms \btau^j \,+\, \tfrac{1}{q_\en^4 }\btau^\en \otms \btau^\en
     \\[5pt]
        &\!\!\!\! =\,  \tfrac{1}{q_\en^2 \nrmtup{q}^2} \emet_{\hat{\imath}\hat{\jmath}} \btau^i \otms \btau^j \,+\, \tfrac{q_\en^2 \nrmtup{q}^2 -1}{q_\en^2 \nrmtup{q}^2} \hat{q}_i\hat{q}_j \btau^i \otms\btau^j 
        \,+\, \tfrac{1}{q_\en^4 }\btau^\en \otms \btau^\en
    \end{array}
        &&,&&
         \cord{\bar{I}}{\bartup{q}} = \fnsz{\begin{pmatrix}
               \tfrac{1}{q_\en^2 \nrmtup{q}^2}( \emet_{\hat{\imath}\hat{\jmath}} - \hat{q}_i \hat{q}_j ) + \hat{q}_i \hat{q}_j & 0 \\
               \trn{0} & \tfrac{1}{q_\en^4}
           \end{pmatrix}}
    \\[6pt] \nonumber 
    &\begin{array}{rllll}
         \inv{\bsfb{\emet}} = \emet^{\hat{\a}\hat{\b}} \hbe_\a \otms \hbe_\b 
         &\!\!\!\! =\,   q_\en^2\nrmtup{q}^2( \emet^{\hat{\imath}\hat{\jmath}} - \hat{q}^i\hat{q}^j) \bt_i \otms \bt_j    \,+\,  \hat{q}^i\hat{q}^j \bt_i \otms \bt_j \,+\, q_\en^4 \bt_\en \otms \bt_\en
         \\[5pt]
        &\!\!\!\! =\,  q_\en^2\nrmtup{q}^2 \emet^{\hat{\imath}\hat{\jmath}} \bt_i \otms \bt_j    \,+\,  (1- q_\en^2\nrmtup{q}^2 )\hat{q}^i\hat{q}^j\bt_i \otms \bt_j  \,+\, q_\en^4 \bt_\en \otms \bt_\en
    \end{array}
        &&,&&
         \cord{\inv{\bar{I}}}{\bartup{q}}  \,=\, 
            \fnsz{ \begin{pmatrix}
               q_\en^2\nrmtup{q}^2( \emet^{\hat{\imath}\hat{\jmath}} - \hat{q}^i \hat{q}^j) + \hat{q}^i \hat{q}^j & 0 \\
               \trn{0} & q_\en^4
           \end{pmatrix}}
    \end{align}
    \end{footnotesize}
    where \eq{\bt_\a:=\bt[q^\a]\in\vect(\bvsp{E})} and  and \eq{\btau^\a:=\btau[q^\a]\in\forms(\bvsp{E})},  are the \eq{q^\a} frame fields (given below).  The above matrix representations indeed agree with the usual tensor component transformations 
    (see footnote\footnote{That is,   \eq{ \cord{\bar{I}}{\bartup{q}} = \uppsi^* \cord{\bar{I}}{\bartup{r}} = \trn{\pderiv{\bartup{r}}{\bartup{q}}} \cdot \cord{\bar{I}}{\bartup{r}} \cdot \pderiv{\bartup{r}}{\bartup{q}}} and \eq{ \cord{\inv{\bar{I}}}{\bartup{q}} = \uppsi^* \cord{\inv{\bar{I}}}{\bartup{r}} =  \pderiv{\bartup{q}}{\bartup{r}} \cdot  \cord{\inv{\bar{I}}}{\bartup{r}} \cdot \trn{\pderiv{\bartup{q}}{\bartup{r}}} }, with the Jacobians given in Eq.\eqref{dprj_coords} and where, for cartesian coordinates, \eq{\cord{\bar{I}}{\bartup{r}}=\cord{\inv{\bar{I}}}{\bartup{r}}=\imat_{\en}}. }).
    Interestingly, it turns out that, for the \eq{\en-1} coordinates \eq{q^i}: 
    \begin{small}
    \begin{align}
          \emet_{ij}q^j \,=\, \emet_{\hat{\imath}\hat{\jmath}} q^j = \kd_{ij}q^j 
            \qquad,\qquad
             \emet_{ij}q^iq^j \,=\, \emet_{\hat{\imath}\hat{\jmath}} q^i q^j = \nrmtup{q}^2
    \end{align}
    \end{small}
    \item \sbemph{Frame Fields on $\Evec$.} 
    For the configuration coordinate transformation \eq{r^\a \leftrightarrow q^\a}, the frame fields on \eq{\bvsp{E}} transform in the usual way:
    \begin{small}
    \begin{align} 
    \begin{array}{llll}
       \hbe_\a \equiv \be[r^\a]   \,=\, \pderiv{q^\b}{r^\a}\bt_\b \in \vect(\bvsp{E})
    \\[8pt]
        \hbep^\a \equiv \bep[r^\a] \,=\, \pderiv{r^\a}{q^\b} \btau^\b \in \forms(\bvsp{E})
    \end{array}
    \qquad \leftrightarrow \qquad 
    \begin{array}{llll}
       \bt_\a \equiv \bt[q^\a]  \,=\, \pderiv{r^\b}{q^\a} \hbe_\b \in \vect(\bvsp{E})
    \\[8pt]
       \btau^\a \equiv \btau[q^\a] \,=\, \pderiv{q^\a}{r^\b} \hbep^\b \in \forms(\bvsp{E})
    \end{array}
    \end{align}
    \end{small}
    where, as mentioned several times, we re-use the notation \eq{\hbe_\a} and \eq{\hbep^\a} for the \eq{r^\a} homogeneous frame fields on \eq{\bvsp{E}} (which are actually defined on all of \eq{\vsp{E}^{\en}}). Using the above expressions for the Jacobians, we obtain:
    \begin{small}
    \begin{align} \label{prj_basis}
    \begin{array}{llllll}
        \hbe_i 
         \,=\, q^\en \nrmtup{q} ( 
         \kd^j_i - \hat{q}^j \hat{q}_i)\bt_j 
         \,-\, q^2_\en \hat{q}_i \bt_\en
         \\[4pt] 
           \envec 
           \,=\, \hat{q}^i \bt_i
    \\[8pt]
         \hbep^i \,=\, \tfrac{1}{q^\en\nrmtup{q}}( 
         \kd^i_j - \hat{q}^i \hat{q}_j) \btau^j \,-\, \tfrac{1}{q_\en^2}\hat{q}^i \btau^\en
         \\[4pt] 
         \enform = \hat{q}_i \btau^i
    \end{array}
    \qquad \leftrightarrow \qquad 
    \begin{array}{llllll}
         \bt_i 
         \,=\, \tfrac{\nrmtup{r}}{r^\en}( 
         \kd^j_i - \hat{r}^j \hat{r}_i) \hbe_j \,+\, \hat{r}_i \envec 
          \\[4pt] 
          \bt_\en 
         \,=\, -\nrmtup{r}^2 \hat{r}^i\hbe_i
         \,=\, -r^2 \hsfb{r}
    \\[8pt]
         \btau^i \,=\,   \tfrac{r^\en}{\nrmtup{r}}( 
         \kd^i_j - \hat{r}^i \hat{r}_j) \hbep^j \,+\, \hat{r}^i \enform 
         \\[4pt] 
         \btau^\en \,=\, -\tfrac{1}{\nrmtup{r}^2}\hat{r}_i \hbep^i
         \,=\, -\tfrac{1}{r^2} \hsfb{r}^\flt
    \end{array}
    \end{align}
    \end{small}
    where \eq{\hat{r}^i=r^i/\nrmtup{r} = q^i/\nrmtup{q} = \hat{q}^i} and  \eq{x_i=\emet_{\hat{\imath}\hat{\jmath}}r^j} and \eq{q_i=\emet_{ij} q^j = \emet_{\hat{\imath}\hat{\jmath}}q^j}.
     From the above expressions for \eq{\envec} and \eq{\bt_\en} we see
    that:\footnote{From \eq{\bt_\en = -r^2 \hsfb{r} = -r\sfb{r}}, we have that \eq{\sfb{r} = -\tfrac{1}{r} \bt_\en = -q^\en \bt_\en} and \eq{ \hsfb{r} = - q^2_\en \bt_\en } (using \eq{\rfun =\nrm{\tup{r}}=1/q^\en}). Also, \eq{\envec =\envec =\hat{q}^i\bt_i} such that \eq{r^\en \envec = \nrmtup{q}\hat{q}^i\bt_i = q^i\bt_i}  (using \eq{r^\en=\nrmtup{q}}).}
     \begin{small}
    \begin{align}
        \sfb{r} \,=\, r^i\hbe_i \,=\, -q^\en \bt_\en 
     \quad,\quad 
         r^\en \envec \,=\, q^i \bt_i 
     \quad,\quad 
         \hsfb{r} \,=\, \hat{r}^i\hbe_i \,=\, - q^2_\en \bt_\en 
         \quad,\quad 
           \hsfb{r}^\flt = \dif r = \hat{r}_i \hbep^i = -\tfrac{1}{q_\en^2}\btau^\en
    \end{align}
    \end{small}
    such that \eq{\bsfb{r}\cong\Id_\ii{\Evec}} and \eq{\psi,\inv{\psi}\in\Dfism(\bvsp{E})} from Eq.\eqref{proj_cartesian_2} can also be expressed as any of the following:
     \begin{small}
    \begin{align}
    \begin{array}{rllllll}
          \bsfb{r} &\!\!\!=\, \sfb{r} + r^\en \envec &\!\! =\; r^i\hbe_i + r^\en \envec 
         &\!\! =\;  \tfrac{1}{q^\en} \hat{q}^i \hbe_i + \nrmtup{q} \envec 
        &\!\! =\; -q^\en\bt_\en + q^i\bt_i
        &,\qquad
        q^\a \bt_\a =  -\sfb{r}  +  r^\en \envec
    \\[4pt]
         \psi &\!\!\!=\, \tfrac{1}{r^\en} \hsfb{r} + r \envec 
         &\!\! =\; \tfrac{1}{r^\en}\hat{r}^i\hbe_i + \nrmtup{r} \envec 
          &\!\! =\;  \tfrac{1}{\nrmtup{q}}\hat{q}^i \hbe_i + \tfrac{1}{q^\en} \envec 
         &\!\! =\; -\tfrac{q^2_\en}{\nrmtup{q}} \bt_\en + \tfrac{1}{q^\en}\hat{q}^i \bt_i
     \\[5pt]
         \inv{\psi} \!\!&\!\!\!=\, r^\en \hsfb{r} \,+\,  \tfrac{1}{r} \envec 
         &\!\! =\; r^\en \hat{r}^i \hbe_i \,+\, \tfrac{1}{\nrmtup{r}} \envec 
           &\!\! =\; q^i \hbe_i + q^\en \envec 
         &\!\! =\; - q^2_\en \nrmtup{q} \bt_\en  + q^\en \hat{q}^i \bt_i
    \end{array}
    \end{align}
    \end{small}
    \item \sbemph{Submanifolds.}  Recall that, for any  positive \eq{b\in\mbb{R}_\ii{+}}, then \eq{\psi|_\ii{\man{Q}_\ii{b}}\in\Dfism(\man{Q}_\ii{b};\Sig_\ii{b})} where the hypersurface  \eq{\man{Q}_\ii{b}\subset\bvsp{E}} is defined by \eq{r(\barpt{x})=\nrm{\pt{x}}=b} and the hyperplane \eq{\Sig_\ii{b}\subset\bvsp{E}} is defined by \eq{r^\en(\barpt{x})= x^\en =b}. The property \eq{\nrm{\pt{x}}=b} for \eq{\man{Q}_\ii{b}} is expressed in cartesian and projective coordinates as \eq{ \nrm{\tup{r}(\ptvec{x})} = 1/q^\en(\barpt{x})= b}  (note for the case \eq{b=1} that\eq{1/q^\en(\barpt{x}) =1} is equivalent to \eq{q^\en(\barpt{x}) = 1}). Conversely, the property \eq{x^\en=b} for  \eq{\Sig_\ii{b}} is expressed as \eq{r^\en(\barpt{x}) = \nrm{\tup{q}(\barpt{x})} = b}.
    That is, 
    \begin{small}
    \begin{gather} \label{subs_prj_pass}
     \left. \begin{array}{rllll}
             \man{Q}_\ii{b}  = \inv{\psi}(\Sig_\ii{b}) \,=  &\!\!\!\! \big\{  \barpt{x}  \in \bvsp{E} \;\big|\; r(\ptvec{x})= \nrm{\pt{x}} =b   \big\}
             &\!\!\!\! =\,  \inv{\rfun}\{b\} 
         \\[5pt]
             =  &\!\!\!\!   \big\{  \barpt{x}  \in \bvsp{E} \;\big|\;  \nrm{\tup{r}(\barpt{x})} =b  \big\} 
              &\!\!\!\! =\,\inv{\nrmtup{r}}\{1\} 
        \\[5pt]
             = &\!\!\!\!   
              \big\{  \barpt{x}  \in \bvsp{E} \;\big|\;   q^\en(\barpt{x}) = 1/b \big\} 
                &\!\!\!\! =\, \inv{(q^\en)}\{1/b\} 
    \end{array} \;\; \right| \;\; 
    \begin{array}{rlllll}
        \Sig_\ii{b} = \psi(\man{Q}_\ii{b}) \,= &\!\!\!\!  \big\{ \barpt{x} \in \bvsp{E} \;\big|\;  
         \enform (\barpt{x}) = b
        \big\} 
    \\[5pt]
         \,= &\!\!\!\!  \big\{  \barpt{x}  \in \bvsp{E} \;\big|\; r^\en(\barpt{x}) = b      \big\} 
         &\!\!\!\! =\, \inv{(r^\en)}\{b\}
     \\[5pt]
          \,= &\!\!\!\!  \big\{  \barpt{x}  \in \bvsp{E} \;\big|\;  \nrm{\tup{q}(\barpt{x})} = b   \big\} 
         &\!\!\!\! =\, \inv{\nrmtup{q}}\{b\} 
    \end{array}
    \end{gather}
    \end{small}
    with (co)tangent spaces then given as follows (where \eq{\btau^\a = \dif q^\a} and \eq{\hbep^\a =\dif r^\a} and \eq{\hsfb{r}=\dif r}):
    \begin{small}
    \begin{align}
    \begin{array}{lllll}
          \tsp[\barpt{x}]\man{Q}_\ii{b} 
          \,=\, \ker \hsfb{r}^\flt_{\barpt{x}}
          \,=\, \ker \btau^\en_{\barpt{x}}
    \\[5pt]
          \cotsp[\barpt{x}]\man{Q}_\ii{b} 
          \,=\, \ker \hsfb{r}_{\barpt{x}} 
          \,=\, \ker \bt_{\en|\barpt{x}}
    \end{array}
    &&
    \begin{array}{lllll}
          \tsp[\cdt]\Sig_\ii{b} 
           \,=\, \ker \enform 
            \,\cong\, \tsp[\cdt]\bs{\Sigup}
    \\[5pt]
          \cotsp[\cdt]\Sig_\ii{b} 
           \,=\, \ker \envec  \,\cong\, \cotsp[\cdt]\bs{\Sigup}
    \end{array}
    \end{align}
    \end{small}
    where we do not bother to distinguish between  different (co)tangent spaces of some given \eq{\Sig_\ii{b}}, nor between (co)tangent spaces of different \eq{\Sig_\ii{b}} (all can be identified with the (co)tangent spaces of \eq{\bs{\Sigup}}.
   From Eq.\eqref{subs_prj_pass} we see that \eq{\bartup{q}=(\tup{q},q^\en)} are slice coordinates for each \eq{\man{Q}_\ii{b}\subset\bvsp{E}} and  that
     \eq{\bartup{r}=(\tup{r},r^\en)} are   slice coordinates for each \eq{\Sig_\ii{b} \subset\bvsp{E}}.  That is, \eq{\tup{q}=(q^1,\dots,q^{\en-1}):\man{Q}_\ii{b}\to\mbb{R}^{\en -1}} are local coordinates on \eq{\man{Q}_\ii{b}} and \eq{\tup{r}=(r^1,\dots,r^{\en-1}):\Sig_\ii{b}\to\mbb{R}^{\en-1}} are local \textit{cartesian} coordinates on \eq{\Sig_\ii{b}}. We are mostly concerned with the case \eq{b=1} for which we note the restriction of \eq{\bsfb{r}\cong\Id_\ii{\Evec}} to these submanifolds (with \eq{b=1}) may be viewed as the following inclusions:
     \begin{small}
     \begin{align}
     \begin{array}{lllll}
          \bsfb{r}|_\ii{\man{Q}_\ii{1}} \,=\, \hsfb{r} + r^\en \envec \,=\,  \hat{q}^i\hbe_i + \nrmtup{q}\envec \,=\,  q^i \bt_i - \bt_\en : \man{Q}_\ii{1} \hookrightarrow \bvsp{E}
        \\[4pt]
         \bsfb{r}|_\ii{\Sig_\ii{1}} \,=\,  \sfb{r} + \envec \,=\, r^i\hbe_i + \envec \,=\, q^i \bt_i - q^\en \bt_\en 
         : \Sig_\ii{1} \hookrightarrow \bvsp{E}
     \end{array}
     \end{align} 
     \end{small}
     Also note the expression for \eq{\nrmtup{\pf}^2} in Eq.\eqref{prj_full_coord}, when restricted to \eq{\cotsp\Sig_\ii{1}} such that \eq{r^\en=\nrmtup{q}=1} and \eq{\plin_\en=\htup{q}\cdot\tup{\pf}=\tup{q}\cdot\tup{\pf}=0}, leads to
     \begin{small}
     \begin{align}
     \begin{array}{ccccc}
            &  &\qquad   \fnsize{for $\en-1=3$:}
     \\
         \lang^2|_\ii{\cotsp\Sig_\ii{1}} =\,   \nrmtup{\pf}^2 \,=\, \nrmtup{q}^2\nrmtup{\pf}^2 -  (\tup{q}\cdot\tup{\pf})^2  \,=\,   \nrmtup{r}^2\nrmtup{\plin}^2 -  (\tup{r}\cdot\tup{\uplin})^2   
       &\quad & \qquad  =\, \nrm{\tup{q} \times \tup{\pf}}^2   =\, \nrm{\tup{r} \times \tup{\uplin}}^2 
    \end{array}
     \end{align}
     \end{small}
     Note that, for our present view of the projective transformation as a passive coordinate transformation, nothing  \textit{actually} changes and, as such, the above \eq{\man{Q}_\ii{1}} will no longer play a role; we only ever deal with the same original system with invariant submanifold \eq{\cotsp\Sig_\ii{1}}. However, the above gives us another way of conceptualizing \eq{\man{Q}_\ii{1}}, which plays an important role when taking the active view of things (as we primarily do during this part of the dissertation). 
\end{itemize}
\end{small}

\subsubsection{Hamiltonian Dynamics in Projective Coordinates}

Recall that our mechanical Hamiltonian system, \eq{(\cotsp\bvsp{E},\nbs{\omg}, K )}, is that of particle of mass \eq{m} moving in \eq{(\Evec^4,\bsfb{\emet})} subject to conservative forces corresponding to
a potential \eq{V=V^\zr(\rfun)+V^\ss{1}(\tup{r})\in\fun(\bvsp{E})}.\footnote{As detailed in Remark \ref{rem:U4}, the abuse of notation \eq{V=V^\zr(\rfun)+V^\ss{1}(\tup{r})\in\fun(\bvsp{E})} simply means that \eq{V^\zr} depends only on the 3-space radial distance, \eq{\rfun =\nrm{\sfb{r}}=\nrm{\tup{r}}\in\fun(\bvsp{E})}, and that \eq{V^\ss{1}} does not depend on \eq{r^\en}. The total potential, \eq{V}, is therefore also independent of \eq{r^\en}, that is,  \eq{V=V(\tup{r})}. More precisely: 
     \begin{align} \label{U4_pass}
        \envec \cdot \dif V =  \pderiv{V}{r^\en} = 0 
        &&,&&
         \dif V^\zr \,=\,   \pderiv{V^\zr}{r^i}\hbep^i \,=\, \pderiv{V^\zr}{r}\pderiv{r}{r^i} \hbep^i 
         \,=\,  \pderiv{V^\zr}{r} \tfrac{x_i}{r}\hbep^i 
         \,=\, \pderiv{V^\zr}{r} \dif r
         \,=\, \pderiv{V^\zr}{r} \hsfb{r}^\flt
        &&,&&
         \dif V \,=\, (\pderiv{V^\zr}{r} \hat{r}_i + \pderiv{V^\ss{1}}{r^i} )  \hbep^i  
     \end{align} }
That is, for any \eq{\bar{\kap}_{\barpt{x}}=(\barpt{x},\barbs{\kap})\in\cotsp\bvsp{E}}:
\begin{small}
\begin{align}
    K(\barpt{x},\barbs{\kap})=\tfrac{1}{2}\inv{\sfb{m}}(\barbs{\kap},\barbs{\kap}) + V(\ptvec{x})
    \qquad,\qquad 
   \inv{\sfb{m}}=\tfrac{1}{m}\inv{\bsfb{\emet}}\in\tens^2_0(\bvsp{E})
     \qquad,\qquad 
     V(\barpt{x})\equiv V(\ptvec{x}) = V^\zr(\nrm{\pt{x}}) +V^\ss{1}(\ptvec{x})
\end{align}
\end{small}
where the kinetic energy metric, \eq{\sfb{m}=m\bsfb{\emet}}, is a homogeneous tensor field. 
Now, we know the canonical symplectic form, \eq{\nbs{\omg}\in\formsex^2(\cotsp\bvsp{E})}, takes the same form in both the original cartesian coordinates, \eq{(\bartup{r},\bartup{\uplin})}, and the projective coordinates, \eq{(\bartup{q},\bartup{\pf})}.
We also know that the Hamiltonian \eq{ K \in\fun(\cotsp\bvsp{E})} and corresponding Hamiltonian vector field \eq{\sfb{X}^\ss{K}:= \inv{\nbs{\omg}}(\dif K,\cdot) \in\vechm(\cotsp\bvsp{E},\nbs{\omg})} may then be expressed in these coordinates as follows (cf.~section \ref{sec:Hmech_xform_passive}):
\begin{small}
\begin{align}
\begin{array}{rlllll}
        K &\!\!\! =\,  \tfrac{1}{2}m^{\a\b} \plin_\a \plin_\b \,+\, V
\\[5pt]
       &\!\!\! =\,  \tfrac{1}{2}g^{\a\b}\pf_\a \pf_\b \,+\, V
\end{array}
&&
\begin{array}{rlllll}
      \sfb{X}^\ss{K} &\!\!\! =\,   \pderiv{K}{\plin_\a}\hpdii{r^\a} \,-\,  \pderiv{K}{r^\a}\hpdiiup{\plin_\a} 
     \;=\; 
        m^{\a\b} \plin_\b \hpdii{r^\a}  \,-\, \pderiv{V}{r^\a} \hpdiiup{\plin_\a} 
\\[5pt]
       &\!\!\!  =\, \pderiv{K}{\pf_\a}\hpdii{q^\a} \,-\,  \pderiv{K}{q^\a} \hpdiiup{\pf_\a}
    \;=\; 
    g^{\a\b} \pf_\b \hpdii{q^\a}  \,+\, ( g^{\sig\gam} \Gamma^\b_{\a\sig} \pf_\gam \pf_\b  \,-\, \pderiv{V}{q^\a}) \hpdiiup{\pf_\a}
\end{array}
\end{align}
\end{small}
where \eq{m^{\a\b}=\inv{\sfb{m}}(\hbep^\a,\hbep^\b)} are the inverse metric components in the cartesian \eq{r^\a} basis and \eq{g^{\a\b}:= \inv{\sfb{m}}(\btau^\a,\btau^\b)}  are now the components of the \textit{same} metric in the \eq{q^\a} basis.
Note \eq{\sfb{m}}'s Levi-Civita connection coefficients (Christoffel symbols) vanish in the \eq{r^\a} basis but not in the \eq{q^\a} basis (\eq{g^{\a\b}} are not constant). 
The \eq{r^\a} and \eq{q^\a} matrix representations of \eq{\inv{\sfb{m}}} are just a scaling by \eq{1/m} of what was given in Eq.\eqref{metric_prj_pass}:
\begin{small}
\begin{align}
    \cord{\inv{m}}{\bartup{r}} = [m^{\a\b}] = \tfrac{1}{m}[\kd^{\a\b}]
    \qquad\quad,\qquad\quad 
     \cord{\inv{m}}{\bartup{q}} =: [g^{\a\b}]  
    \,=\, \tfrac{1}{m} \fnsz{\begin{pmatrix}
           q_\en^2\nrmtup{q}^2( \kd^{ij} - \htup{q}\otms\htup{q}) + \htup{q}\otms\htup{q} & 0 \\
           \trn{0} & q_\en^4
       \end{pmatrix} }
\end{align}
\end{small}
That is,  in cartesian coordinates \eq{(\bartup{r},\bartup{\uplin})} — for which \eq{\pd_\sig m^{\a\b}=0} and \eq{\pd_\en V =0} — the Hamiltonian and dynamics are expressed:
\begin{small}
\begin{gather}
    K_\ii{\bartup{r},\bartup{\uplin}} \,=\, \tfrac{1}{2}m^{\a\b} \plin_\a \plin_\b + V(\tup{r})  \;=\;  \tfrac{1}{2m}( \nrmtup{\plin}^2 + \plin_\en^2 )    + V^\zr(\rfun) + V^\ss{1}(\tup{r}) 
\\[3pt] \nonumber 
 \begin{array}{lllll}
     \dot{r}^i = m^{i\a} \plin_\a = \tfrac{1}{m} \plin_i   &,\quad 
     \dot{\plin}_i = -\pd_\ii{r^i} V \,=\, -(\tfrac{x_i}{r} \pd_{\rfun} V^\zr + \pd_\ii{r^i} V^\ss{1})
 \\[4pt]
     \dot{r}^\en = m^{\a \en} \plin_\a = \tfrac{1}{m} \plin_\en  &,\quad \dot{\plin}_\en = -\pd_\ii{r^\en} V  =  0
\end{array}
\end{gather}
\end{small}
and the \textit{same} dynamics are expressed in the projective coordinates \eq{(\bartup{q},\bartup{\pf})} as 
\begin{small}
\begin{gather} \label{Hprj_full_passive} 
  K_\ii{\bar{q},\bar{\pf}} \;=\; \tfrac{1}{2}g^{\a\b}\pf_\a \pf_\b + V(\bartup{q})
     \;=\; \tfrac{1}{2m} q_\en^2 \big( \nrmtup{q}^2 \nrmtup{\pf}^2 - (\tup{q}\cdot\tup{\pf})^2 \,+\, q_\en^2 \pf_\en^2 \big) \,+\, \tfrac{1}{2m} (\htup{q} \cdot\tup{\pf})^2 \,+\, V^\zr(q^\en) + V^\ss{1}(\bartup{q}) 
 \\[5pt] \nonumber 
\begin{array}{lllllll}
     \dot{q}^i 
     \,=\,  \tfrac{q_\en^2}{m} \big( \nrmtup{q}^2 \kd^{ij} \pf_j  \,-\, (\tup{q}\cdot \tup{\pf} ) q^i \big) \,+\,  \tfrac{1}{m}(\htup{q}\cdot\tup{\pf})\hat{q}^i
     &\quad =\, -\tfrac{q_\en^2}{m} \lang^{ij}q_j \,+\,  \tfrac{1}{m}(\htup{q}\cdot\tup{\pf})\hat{q}^i
\\[4pt]
   \dot{q}^\en 
   \,=\, \tfrac{q_\en^4}{m}  \pf_\en
\\[4pt]
      \dot{\pf}_i 
      \,=\,  - \tfrac{q_\en^2}{m}   \big( \nrmtup{\pf}^2 q_i - (\tup{q}\cdot\tup{\pf}) \pf_i \big) 
      \,-\, \tfrac{(\htup{q}\cdot\tup{\pf})}{m \nrmtup{q}}  \big( \pf_i - (\htup{q}\cdot\tup{\pf}) \hat{q}_i \big)
      \,-\, \pd_\ii{q^i} V^\ss{1} 
     &\quad =\, -\tfrac{q_\en^2}{m} \lang_{ij} \pf^j 
        \,+\, \tfrac{(\htup{q}\cdot\tup{\pf})}{m \nrmtup{q}^3} \lang_{ij}q^j
        \,-\, \pd_\ii{q^i} V^\ss{1} 
\\[4pt]
      \dot{\pf}_\en 
      \,=\,  - \tfrac{q_\en}{m}  \big( \nrmtup{q}^2 \nrmtup{\pf}^2 \,-\, (\tup{q}\cdot\tup{\pf})^2 \,+\, 2 q_\en^2 \pf_\en^2 \big) \,-\, \pd_\ii{q^\en} (V^\zr + V^\ss{1})
      &\quad =\, - \tfrac{q_\en}{m}  ( \lang^2 + 2 q_\en^2 \pf_\en^2 ) \,-\, \pd_\ii{q^\en} (V^\zr + V^\ss{1})
\end{array}
\end{gather}
\end{small}
Note that \eq{V^\zr} drops out of the ODEs for \eq{\pf_i}. Aside from that, the above equations look significantly worse than the cartesian coordinate ODEs. However, as before, the above are be simplified significantly using the fact that \eq{\plin_\en = \htup{q}\cdot\tup{\pf}} is an integral of motion and we are free to limit consideration to the case \eq{\plin_\en = \htup{q}\cdot\tup{\pf}=0} for which the above simplifies to 
\begin{small}
\begin{align}
   \begin{array}{lllllll}
     \dot{q}^i  =\, -\tfrac{q_\en^2}{m} \lang^{ij}q_j 
     &,\qquad 
     \dot{\pf}_i =\, -\tfrac{q_\en^2}{m} \lang_{ij} \pf^j 
        \,-\, \pd_\ii{q^i} V^\ss{1} 
\\[4pt]
   \dot{q}^\en 
   \,=\, \tfrac{q_\en^4}{m}  \pf_\en
   &,\qquad  
   \dot{\pf}_\en =\, - \tfrac{q_\en}{m}  ( \lang^2 + 2 q_\en^2 \pf_\en^2 ) \,-\, \pd_\ii{q^\en} (V^\zr + V^\ss{1})
\end{array} 
\end{align}
\end{small}
Further, when the projective coordinate transformation is combined with a transformation of the evolution parameter, the above ODEs will turn out to be linear in the special case of central force dynamics for \eq{V=V^\zr = \pm\sck_1/\rfun \pm \ttfrac{1}{2}\pm\sck_2/\rfun^2}.

\begin{small}
\begin{notesq}
    Note the potential \eq{V=V^\zr+V^\ss{1}\in\fun(\bvsp{E})} is independent of \eq{r^\en} but \textit{not} \eq{q^\en}. In fact, since \eq{V^\zr} depends only on the \eq{(\en-1)}-space radial distance, \eq{\rfun = \nrm{\sfb{r}}=\nrm{\tup{r}}}, it follows that, in the new coordinates, \eq{V^\zr} depends \textit{only} on \eq{\eq{q^\en= 1/\nrm{\tup{r}}=1/\rfun}}. By abuse of notation, this is expressed as
    \begin{small}
    \begin{align}
        V(\tup{r}) = V^\zr(\nrmtup{r}) +V^\ss{1}(\tup{r}) \;\;=\;\; V(\bartup{q}) =  V^\zr(q^\en) +V^\ss{1}(\bartup{q})
        &&
        \begin{array}{lllll}
             \pderiv{V}{q^i} \,=\, \pderiv{r^j}{q^i}\pderiv{V}{r^j} \,=\, \tfrac{1}{q^\en \nrmtup{q}}(\kd^j_i - \hat{q}^j \hat{q}_i)\pderiv{V}{r^j} 
         \\[5pt]
             \pderiv{V}{q^\en} \,=\, \pderiv{r^j}{q^\en}\pderiv{V}{r^j} \,=\,  -\tfrac{1}{q^2_\en}\hat{q}^j \pderiv{V}{r^j} \,=\, -\tfrac{1}{q^2_\en}\pderiv{V}{r} 
        \end{array}
    \end{align}
    \end{small}
    where the relations on the right may be used to transform the cartesian \eq{r^i} basis components, \eq{\pd_\ii{r^i}V}, of the conservative force 1-form \eq{\dif V} to the \eq{q^\a} basis components, \eq{\pd_\ii{q^\a}V} (of the same 1-form).  
    Note the following relations: 
    \begin{small}
   \begin{align}
        \pd_{\rfun}V  = \hsfb{r}\cdot \dif V \,=\, \hat{r}^j \pd_\ii{r^j}V \,=\, \hat{q}^j \pd_\ii{r^j}V 
        \qquad,\qquad 
      0 =  \pd_{r^\en} V =  \envec\cdot \dif V \,=\, \hat{q}^i \bt_i \cdot \dif V =   \hat{q}^i \pd_\ii{q^i} V
   \end{align}
    \end{small}
\end{notesq}
\end{small}

We could go to analyze the Hamiltonian dynamics, integrals of motion, and various properties in terms of the projective coordinates.  However, there is little need; the ``passive'' view follows from the ``active'' view as previously described (not to mention that the passive view is, essentially, equivalent to what was already presented by the authors in previous work \cite{peterson2022nonminimal,peterson2025prjCoord,peterson2023regularized}). Comparing the above initial stages of the coordinate-based formulation with the corresponding active formulation given prior should make this clear.

\section{REGULARIZATION OF KEPLER \& MANEV DYNAMICS} \label{sec:prj_regular}


 We now present a conformal scaling of the transformed Hamiltonian dynamics developed in section \ref{sec:prj_geomech} which will result linear equations of motion for certain central force potentials (of the \textit{original} system). For brevity and clarity, we will often present things in terms of cartesian coordinates. Much of this section will recover, by different and more clear means, the key results developed in earlier work \cite{peterson2025prjCoord,peterson2022nonminimal}. 

\begin{notation}
Regarding notation in this section:
\begin{small}
\begin{itemize}[nosep]
    \item The same notation we have been using (much of which is summarized in section \ref{sec:notation_prj}) still applies, with the same index ranges:
    \begin{small}
        \begin{align}
            i,j = 1,\dots, \en-1
            \qquad,\qquad
            \a, \beta = 1,\dots, \en
        \end{align}
    \end{small}
    \item \eq{\psi\in\Dfism(\bvsp{E})} denotes the projective transformation defined in Eq.\eqref{prj_def_act}-Eq.\eqref{proj_cartesian}, with cotangent lift \eq{\colift\psi\in\Spism(\cotsp\bvsp{E})} given in Eq.\eqref{colift_prj_def}-Eq.\eqref{colift_prj_def1}.  
     \item For a curve \eq{\gam_t \in \man{X}} on a smooth manifold \eq{\man{X}}, we will denote by \eq{\gam_s} the ``same'' curve parameterized by some other evolution parameter \eq{s}. That is, \eq{\gam_s \equiv \gam_{t(s)} \neq \gam_{t=s}} is just \eq{\gam_t} with \eq{t} expressed in 
     terms of \eq{s}.\footnote{However, notation such as \eq{\gam_{t_\iio}} or \eq{\gam_\iio} simply means \eq{\gam_{t=t_\iio}} for some fixed \eq{t_\iio\in\mbb{R}} (it does \textit{not} mean re-parameterization by some new evolution parameter \eq{t_\iio}) and likewise for \eq{\gam_{s_\iio}}.}
     This is an abuse of notation as the \eq{s}-parameterized curve is, technically,  an entirely different curve, \eq{\til{\gam}:\mbb{R}\to\man{X}}, such that \eq{\til{\gam}_s=\gam_{t(s)}} and \eq{\gam_t = \til{\gam}_{s(t)}}.  Note \eq{t(s)} and \eq{s(t)} are also abuses of notation which simply indicate one evolution parameter expressed in terms of the other.
  \item[] \eq{\;\;}
\end{itemize}
\end{small} 
\end{notation}

\paragraph{Review:~Conformally Related Vector Fields.} 
 Subsequent developments will involve the transformation of the evolution parameter of a dynamical system by means of a conformal scaling of the vector field. The general theory is detailed in section \ref{sec:confScale}, some of which is reviewed here for convenience. 
 Recall that, on some smooth manifold \eq{\man{P}}, for any \eq{\sfb{X}\in\vect(\man{P})} and any non-vanishing \eq{ 0\neq\cf \in\fun(\man{P})}, the exchange \eq{\sfb{X} \mapsto  \cf\sfb{X}} amounts to a transformation of the evolution parameter \eq{t\mapsto s} determined by \eq{\mrm{d} {t} =  \cf  \mrm{d} {s}}.\footnote{More specifically, the relation is not simply \eq{\mrm{d} {t} =  \cf  \mrm{d} {s}}, but rather \eq{\mrm{d} {t} =  \cf(\til{\gam}_s)  \mrm{d} {s}} or \eq{\mrm{d} {s} =  (1/\cf(\gam_t) )  \mrm{d} {t}} where \eq{\til{\gam}_s} is an integral curve of \eq{\cf \sfb{X}} and \eq{\gam_t} is an integral curve of \eq{\sfb{X}}; they are the ``same'' curve in the sense \eq{\til{\gam}_s = \gam_t}, that is, \eq{\til{\gam}_{s} = \gam_{t(s)}} and \eq{\gam_t = \til{\gam}_{s(t)}}. }
 In this context, \eq{\cf\sfb{X}} is said to be a \textit{conformal scaling} of \eq{\sfb{X}} and \eq{\cf} is called the \textit{conformal factor}.  
 For transforming first and second-order derivatives ``with respect to'' \eq{t} or \eq{s}, it is useful to note the following relations for any \eq{F\in\fun(\man{P})}:\footnote{The relation for \eq{ \lderiv{\cf\sfb{X}}^2 F} follows from \eq{ \lderiv{\cf\sfb{X}}^2 F \,=\, \lderiv{\cf\sfb{X}}(  \cf  \lderiv{\sfb{X}} F) \,=\,  \cf   \lderiv{\sfb{X}}(  \cf  \lderiv{\sfb{X}} F) \,=\,  \cf^2  \lderiv{\sfb{X}}^2 F \,+\,  \cf  (\lderiv{\sfb{X}} \cf  ) (\lderiv{\sfb{X}} F)}. }
\begin{small}
\begin{align}\label{lderiv_rels_prj}
\begin{array}{lllllll}
       \lderiv{\cf\sfb{X}} F \,=\,  \cf  \lderiv{\sfb{X}} F 
\\[8pt]
       \lderiv{\cf\sfb{X}}^2 F 
     \,=\,  \cf^2  \lderiv{\sfb{X}}^2 F \,+\,  \cf  \lderiv{\sfb{X}}( \cf  ) \lderiv{\sfb{X}}F
\end{array}
\qquad\quad
\xRightarrow[\lderiv{\cf\sfb{X}} \equiv \diff{}{s}]{\lderiv{\sfb{X}} \equiv \diff{}{t}}
\qquad\quad
\begin{array}{lllllll} 
      \diff{F}{s}  = \cf  \diff{F}{t}
\\[8pt]
    \ddiff{F}{s}  = \cf^2 \ddiff{F}{t}  +  \cf   \diff{\cf}{t}  \diff{F}{t} 
\end{array}
\end{align}
\end{small}
where the relations on the right — written in common, though less precise, notation —  follow from those on the left if one makes the identifications  \eq{\lderiv{\sfb{X}} \equiv \diff{}{t}}  and  \eq{\lderiv{\cf\sfb{X}} \equiv \diff{}{s}} (where \eq{\mrm{d} {t} = f \mrm{d} {s}}). 

In the following developments, we are are concerned with the  case that \eq{\sfb{X}} is a Hamiltonian vector field on a symplectic manifold, \eq{(\man{P},\nbs{\omg})}. In this context a \textit{conformally-Hamiltonian} vector field  
is any \eq{\sfb{Z}\in\vect(\man{P})} which satisfies \eq{\sfb{Z} = \cf\sfb{X}^h} for any  Hamiltonian vector field \eq{\sfb{X}^h=\inv{\nbs{\omg}}(\dif h,\slot)\in\vechm(\man{P},\nbs{\omg})} and conformal factor \eq{0 \neq  \cf  \in\fun(\man{P})}. Note \eq{\cf\sfb{X}^h} satisfies:
\begin{small}
\begin{align}
    \lderiv{\cf\sfb{X}^h}\nbs{\omg} 
    \,=\, \dif  \cf  \wedge \dif h
   \qquad,\qquad 
    \nbs{\omg}( \cf\sfb{X}^h,\slot) =  \cf  \dif h
    \qquad,\qquad 
    \div[\bs{\spvol}]( \cf\sfb{X}^{h})= \lderiv{\sfb{X}^{h}}  \cf  = \pbrak{\cf}{h}
\end{align}
\end{small}
Thus,  \eq{\cf\sfb{X}^{h}} is \textit{not} \eq{\nbs{\omg}}-Hamiltonian. It is \eq{\nbs{\omg}}-symplectic (locally Hamiltonian) only in the special case that \eq{\exd ( \cf  \dif h)=\dif  \cf  \wedge \dif h=0} (meaning \eq{\cf} is some "function of \eq{h}'', as in \eq{ \cf  =h^2},  \eq{ \cf  =\mrm{e}^h}, etc.).  
However, if  we define \eq{\til{\nbs{\omg}}:=(1/ \cf  )\nbs{\omg}} and \eq{\tsfb{X}^h:= \cf\sfb{X}^h}, then:
\begin{small}
\begin{align} \nonumber 
  \tsfb{X}^h:= \cf\sfb{X}^h 
  \quad,\quad 
  \til{\nbs{\omg}}:=\tfrac{1}{\cf}\nbs{\omg}
   \quad,\quad 
 \inv{\til{\nbs{\omg}}} =  \cf  \inv{\nbs{\omg}}
    && \Rightarrow &&
     \til{\nbs{\omg}}(\tsfb{X}^h,\slot)  = \dif  h
     \quad,\quad 
      \inv{\til{\nbs{\omg}}}(\dif h,\slot) \,=\, \tsfb{X}^h
      \quad\;\; \fnsize{but,} \;\; 
      \lderiv{\tsfb{X}^h}\til{\nbs{\omg}} 
      \neq 0
\end{align}
\end{small}
and it would almost seem that \eq{\tsfb{X}^{h}} is \eq{\til{\nbs{\omg}}}-Hamiltonian. However, there is no guarantee that \eq{\til{\nbs{\omg}}} is a valid symplectic form; it must be globally non-degenerate and closed. Global non-degeneracy is assured since \eq{\nbs{\omg}} is symplectic and \eq{ \cf  \neq 0} was already required.  
Yet, closedness further requires  \eq{\exd\til{\nbs{\omg}} = -\cf^\ss{-2}\dif  \cf  \wedge \nbs{\omg} = 0}, which does not generally hold for arbitrary \eq{\cf}. As such,  \eq{\tsfb{X}^h= \cf\sfb{X}^h} is generally not \eq{\til{\nbs{\omg}}}-symplectic (nor Hamiltonian) since \eq{\lderiv{\tsfb{X}^h}\til{\nbs{\omg}} = \tsfb{X}^h\cdot \exd \til{\nbs{\omg}}} is generally non-zero (unless \eq{\til{\nbs{\omg}}} is closed) .\footnote{From Cartan's identity, along with \eq{\tsfb{X}^h\cdot \til{\nbs{\omg}}=\dif h \in\formsex(\man{P})}, we have \eq{\lderiv{\tsfb{X}^h}\til{\nbs{\omg}} = \exd( \tsfb{X}^h\cdot \til{\nbs{\omg}}) + \tsfb{X}^h\cdot \exd \til{\nbs{\omg}} = \tsfb{X}^h\cdot \exd \til{\nbs{\omg}} = \cf\sfb{X}^h\cdot \exd(\tfrac{1}{\cf}\nbs{\omg})}.  } 
Last, note the relations in Eq.\eqref{lderiv_rels_prj} are still valid for \eq{\cf\sfb{X}^h } and, in this case, they may now also be expressed as follows (using the fact that \eq{\lderiv{\sfb{X}^h}F = -\inv{\nbs{\omg}}(\dif F,\dif h) = \pbrak{F}{h}}):
\begin{small}
\begin{align}\label{lderiv_rels_prj_HAM}
       \lderiv{\cf\sfb{X}^h} F 
       \,=\, \cf \pbrak{F}{h}
\qquad,\qquad 
       \lderiv{\cf\sfb{X}^h}^2 F 
     \,=\, \cf^2  \pbrak{ \pbrak{F}{h} }{h} + \cf \pbrak{\cf}{h} \pbrak{F}{h} 
\end{align}
\end{small}

\subsection{Linearization of (Originally) Central Force Dynamics by Conformal Scaling}

\paragraph{Review:~The $\colift\psi$-Transformed Hamiltonian System.}
Returning our attention to the specific case at hand, we begin with the Hamiltonian vector field \eq{\sfb{X}^\sscr{H}\in\vechm(\cotsp\bvsp{E})} for the simplified Hamiltonian function \eq{\mscr{H}:=\, \colift\psi^*\mscr{K}\in\fun(\cotsp\bvsp{E})} that was defined in Eq.\eqref{Hams_reduce_act}  and expressed for cotangent-lifted cartesian coordinates, \eq{(\bartup{r},\bartup{\plin})=(\tup{r},r^\en,\tup{\plin},\plin_\en):\cotsp\vsp{E}\to\mbb{R}^{\ii{2}\en}}, as follows:
\begin{small}
\begin{gather}  \label{H_reduced_again}
\begin{array}{llll}
    \mscr{H}  \,=\,
   \tfrac{r_\en^2}{2m} ( \lang^2 + r_\en^2 \plin_\en^2 )  +  U  \;=\;  \tfrac{r_\en^2 }{2m} \big( \nrmtup{r}^2 \nrmtup{\plin}^2 -(\tup{r}\cdot\tup{\plin})^2 + r_\en^2 \plin_\en^2 \big)  \,+\,  U^\zr(r_\en)  \,+\, U^\ss{1}(\bartup{r}) 
\end{array}
\\ \nonumber
 \sfb{X}^\sscr{H} =\, \inv{\nbs{\omg}}(\dif \mscr{H},\slot) \,=\, \partial^\a \mscr{H} \hbpart{\a} -   \pd_\a \mscr{H} \hbpartup{\a} 
\quad\left\{ \quad \begin{array}{lllllll}
     \dot{r}^i =  \partial^i \mscr{H} \,=\,  -\tfrac{r_\en^2}{m}\lang^{ij} r_j
     \quad=\, \tfrac{r_\en^2}{m} \big( \nrmtup{r}^2 \plin^i  - (\tup{r}\cdot \tup{\plin} ) r^i \big) 
\\[4pt]
   \dot{r}^\en =  \partial^\en \mscr{H} \,=\, \tfrac{r_\en^4}{m}  \plin^\en
\\[4pt]
      \dot{\plin}_i = - \pd_i \mscr{H} \,=\,  -\tfrac{r_\en^2}{m}\lang_{ij} \plin^j - \pd_i U^\ss{1}
      \quad=\, - \tfrac{r_\en^2}{m}   \big( \nrmtup{\plin}^2 r_i - (\tup{r}\cdot\tup{\plin})\plin_i \big) - \pd_i U^\ss{1}
\\[4pt]
      \dot{\plin}_\en = - \pd_\en \mscr{H} \,=\, - \tfrac{r_\en}{m} ( \lang^2 + 2 r_\en^2 \plin_\en^2 ) - \pd_\en (U^\zr + U^\ss{1})
\end{array} \right.
\end{gather}
\end{small}
where \eq{r_i:=\emet_{ij} r^j} and \eq{\plin^i:=\emet^{ij} \plin_j}.  
The above non-linear ODEs determine the (cartesian coordinate representation of) integral curves of \eq{\sfb{X}^\sscr{H}}. 
Recall that the above permits several useful integrals of motion, as indicated by the following:
\begin{small}
\begin{align} \label{prj_ioms_rev}
\begin{array}{lllll}
     \pbrak{\mscr{H}}{\mscr{H}} \,=\,
    \pbrak{r}{\mscr{H}} \,=\, \pbrak{\hat{r}^i \plin_i }{\mscr{H}}  \,=\, \pbrak{r^i \plin_i }{\mscr{H}} 
    \,=\, 0
\\[4pt]
      \pbrak{  \ttfrac{1}{2}\nrmtup{\plin}^2 }{\mscr{H}} 
    \,=\,  -\plin^j\pd_j U^\ss{1}  
\end{array}
\qquad\qquad\qquad
\begin{array}{lllllll}
        \pbrak{ \lang^{ij} }{\mscr{H}} \,=\, -(r^i \emet^{jk} - r^j \emet^{ik})\pd_k U^\ss{1}  
\\[4pt]
      \pbrak{ \ttfrac{1}{2}\lang^2 }{\mscr{H}} 
      \,=\, 
   - \nrmtup{r}^2 \plin^j \pd_j U^\ss{1} 
\end{array}
\end{align}
\end{small}
where  \eq{\lang^{ij}\in\fun(\cotsp\bvsp{E})} the cartesian coordinate angular momentum functions on \eq{\cotsp\bs{\Sigup}\subset\cotsp\bvsp{E}}. It will be useful for subsequent developments to recall that \eq{\lang^{ij}} and \eq{\lang^2}
satisfy the following (these were given in  Eq.\eqref{angmoment_rels_prj}): 
\begin{small}
\begin{align} \label{angmoment_rels_again}
\begin{array}{lllllll}
      \lang^{ij} = r^i \plin^j - \plin^i r^j
  \\[3pt]
      \lang_{ij} = r_i \plin_j - \plin_i r_j = \emet_{ki}\emet_{sj}\lang^{ks} 
\\[6pt]
    \lang^{is}\lang_{sj} r^j \,=\, -\lang^2 r^i 
\\[4pt]
       \lang^{is}\lang_{sj} r_i \,=\, -\lang^2 r_j
\\[4pt]
       \lang^{is}\lang_{sj} \plin^j \,=\, -\lang^2 \plin^i
\\[4pt]
       \lang^{is}\lang_{sj} \plin_i \,=\, -\lang^2 \plin_j
\end{array}
&& && 
\begin{array}{lllllll}
    \lang^2 = \ttfrac{1}{2} \lang^{ij} \lang_{ij} =  \lang^{ij} r_i \plin_j = \lang_{ij} r^i \plin^j  = \nrmtup{r}^2 \nrmtup{\plin}^2 - (r^i \plin_i)^2 
\\[4pt] 
      0 \,=\, \lang^{ij}r_i r_j = \lang^{ij}\plin_i \plin_j = \lang_{ij} \plin^i \plin^j = \lang_{ij} r^i r^j
\\[6pt]
    \lang_{ij}\plin^j \,=\, \nrmtup{\plin}^2 r_i  - (\tup{r}\cdot\tup{\plin})\plin_i 
    \,=\, (\nrmtup{\plin}^2 \emet_{ij} - \plin_i \plin_j) r^j
    &\!\!=  \pd_\ii{r^i}  \ttfrac{1}{2}\lang^2
\\[4pt]
   \lang^{ji}r_j  \,=\, \nrmtup{r}^2 \plin^i  - (\tup{r}\cdot\tup{\plin})r^i 
    \,=\, (\nrmtup{r}^2 \emet^{ij} - r^i r^j) \plin_j
    &\!\!= \pd_\ii{\plin_i}  \ttfrac{1}{2}\lang^2 
\\[4pt]
     \lang^{ij} \plin_j \,=\, \nrmtup{\plin}^2 r^i - (\tup{r}\cdot\tup{\plin}) \plin^i \,=\, (\nrmtup{\plin}^2 \kd^i_j - \plin^i \plin_j)r^j
     &\!\!=  \pd_\ii{r_i}  \ttfrac{1}{2}\lang^2
\\[4pt]
      \lang_{ji}r^j \,=\,  \nrmtup{r}^2 \plin_i - (\tup{r}\cdot\tup{\plin}) r_i \,=\, (\nrmtup{r}^2 \kd^j_i - r^j r_i) \plin_j 
       &\!\!= \pd_\ii{\plin^i} \ttfrac{1}{2}\lang^2 
\end{array}
\end{align}
\end{small}
It will often be useful to express relations in terms of the \textit{specific} angular momentum functions (i.e., per unit mass), denoted \eq{\slang^{ij}}, which obey the same relations as above (up to mass scaling) and are defined by:\footnote{Note, for instance, 
\begin{align}
     \pbrak{\slang}{\mscr{H}} = -\tfrac{\rfun^2}{m^2 \slang}  \plin^j \pd_j U^\ss{1}  
   \quad,\quad
   \pbrak{\lang}{\mscr{H}} = -\tfrac{\rfun^2}{\lang}  \plin^j \pd_j U^\ss{1}  
\end{align} }
\begin{small}
\begin{align}
   \slang^{ij} := \tfrac{1}{m} \lang^{ij} = (r^i m^{js}-r^j m^{is})\plin_s \qquad,\qquad  \slang_{ij} := \tfrac{1}{m} \lang_{ij} \qquad,\qquad  \slang^2 = \tfrac{1}{m^2} \lang^2
\end{align}
\end{small}
(we would not need to distinguish \eq{\slang} and \eq{\lang} if we had used \eq{\sfb{m}=m \bsfb{\emet}_\ii{\!\Evec}} rather than \eq{\bsfb{\emet}_\ii{\!\Evec}} for the metric musical isomorphism).

Looking at Eq.\eqref{prj_ioms_rev}, note that \eq{\lang^{ij}}, \eq{\lang^2}, and \eq{\nrmtup{\plin}^2} are also integrals of motion in the special case that \eq{U^\ss{1}=0}, which has the following interpretation. 
Recall the potential of the \textit{original} system is of the form \eq{V=V^\zr(\rfun) + V^\ss{1}(\tup{r}) } where  \eq{V^\zr} depends only on \eq{\rfun =\nrm{\sfb{r}}=\sqrt{\emet_{ij}r^i r^j}} and accounts for all purely \textit{radial} conservative forces. It was shown that the transformed potential, \eq{U:=\psi^* V}, appearing in \eq{\mscr{H}} is then of the form  \eq{U=U^\zr(r_\en)  + U^\ss{1}(\bartup{r})} where \eq{U^\zr =\psi^* V^\zr} depends only on \eq{r^\en}, corresponding to purely \textit{normal} forces (meaning normal to \eq{\bs{\Sigup}}, tangent to \eq{\vsp{N}}). The case \eq{U^\ss{1}=0}, meaning \eq{U=U^\zr(r_\en)}, then corresponds to the case \eq{V^\ss{1}=0}, meaning \eq{V=V^\zr(\rfun)}. That is, the original dynamics include only radial forces, while the transformed dynamics include only forces normal to \eq{\bs{\Sigup}} (and thus normal to all \eq{\Sig_{b}\subset\bvsp{E}}).

\begin{small}
\begin{itemize}[nosep]
    \item  \rmsb{Simplifications on the invariant submanifold $\cotsp\man{Q}_\ii{1}$.} 
    Recall from section \ref{sec:IOMs_prj_act} that the \eq{\colift\psi}-transformed Hamiltonian system reviewed above has \eq{(2\en-2)}-dim invariant submanifold \eq{\cotsp\man{Q}_\ii{1}\subset\cotsp\bvsp{E}}, where \eq{\man{Q}_\ii{1} \cong \man{S}^{\en-\two}_\ii{1} \times \vsp{N}_\ii{+}}. For any points or curves on \eq{\cotsp\man{Q}_\ii{1}}, the following relations hold
    (where \eq{\simeq} will be used as shorthand for \eq{|_\ii{\cotsp\man{Q}}}, denoting relations that hold on \eq{\cotsp\man{Q}_\ii{1}} but, generally, not on \eq{\cotsp\bvsp{E}}):
    \begin{small}
    \begin{gather} \label{T*Q_reg_rev}
        \cotsp\man{Q}_\ii{1} \,=\, \big\{  \bar{\mu}_\ss{\barpt{q}}\in\cotsp\bvsp{E} \;\big|\; r(\ptvec{q}) = \nrm{\pt{q}} = 1 \;, \; \hsfb{r}_\ss{\!\pt{q}} \cdot \bs{\mu} = \hsfb{q}\cdot \bs{\mu} = 0    \big\}
        \cong \cotsp(\man{S}^{\en-\two}_\ii{1} \times \vsp{N}_\ii{+}) \,\subset \cotsp(\bs{\Sigup}_\nozer \times \vsp{N}_\ii{+})
     \\[3pt] \nonumber   
    \begin{array}{llll}
             \rfun^2  \simeq 1 
     \qquad,\qquad 
             \hat{r}^i \plin_i \simeq r^i \plin_i \simeq 0 
     \qquad,\qquad
              \lang^2 = m^2 \slang^2 
              \simeq \nrmtup{\plin}^2
     \qquad,\qquad 
             -\lang^{ij}r_j 
             \simeq \plin^i
    \qquad,\qquad 
           \lang_{ij}\plin^j 
           \simeq \nrmtup{\plin}^2 r_i
           \simeq \lang^2 r_i
    \end{array}
    \end{gather}
    \end{small} 
    Recall also that the integral curves of \eq{\sfb{X}^\sscr{H}} in \eq{\cotsp\man{Q}_\ii{1}} are mapped by \eq{\colift\psi} to integral curves of the original \eq{\sfb{X}^\sscr{K}} in \eq{\cotsp\Sig_\ii{1} \cong \cotsp\bs{\Sigup}_\nozer},  where the latter are precisely the integral curves which correspond to the actual physical motion in which we are interested.
    Therefore, \textit{in order to recover the physically meaningful solutions on \eq{\cotsp\bs{\Sigup}_\nozer} of the original system \eq{\sfb{X}^\sscr{K}}, it is enough to limit consideration only to solutions on \eq{\cotsp\man{Q}_\ii{1}} of the transformed system \eq{\sfb{X}^\sscr{H}}}. In this case, the transformed cartesian coordinate equations of motion and solutions may be simplified using the above relations. Yet, in the following, we will be able to obtain closed form solutions for general integral curves of \eq{\sfb{X}^\sscr{H}} (and therefore also of \eq{\sfb{X}^\sscr{K}}) \textit{without} considering these simplifications on \eq{\cotsp\man{Q}_\ii{1}}. As we go, we will periodically indicate how the general relations on \eq{\cotsp\bvsp{E}} are simplified when limiting consideration to \eq{\cotsp\man{Q}_\ii{1}}. Since we are interested in the original system dynamics on \eq{\cotsp\Sig_\ii{1} \cong \cotsp\bs{\Sigup}_\nozer}, in practice, there is no reason not to limit consideration of the transformed system to \eq{\cotsp\man{Q}_\ii{1}}. 
\end{itemize}
\end{small}

\paragraph{Two Conformal Factors for Linearizing (Originally) Inverse Square \& Inverse Cubic Central Force Dynamics.}
 For \eq{\sfb{X}^\sscr{H}\in\vechm(\cotsp\bvsp{E})} given in Eq.\eqref{H_reduced_again}, 
we will consider its scaling by two different conformal factors (where \eq{\slang=\lang/m}):
\begin{small}
\begin{align} \label{dt_funs}
        \cf  := \tfrac{1}{r_\en^2} \in\fun(\cotsp\bvsp{E}) 
\qquad,\qquad 
       \cff := \tfrac{m}{\lang r_\en^2} \,=\, \tfrac{1}{\slang r_\en^2}  \in\fun(\cotsp\bvsp{E}) 
\end{align}
\end{small}
where \eq{ \cf  \equiv \copr^*  \cf  } is regarded as a basic function on \eq{\cotsp\bvsp{E}}. Each of the above defines a transformation of the evolution parameter where \eq{\sfb{X}^\sscr{H}}, \eq{\cf\sfb{X}^\sscr{H}}, and \eq{\cff\sfb{X}^\sscr{H}} have the same integral curves, up to re-parameterization, by their respective evolution parameters.  We will denote the evolution parameter defined by \eq{\cf} as \eq{s}, and that defined  by \eq{\cff} as \eq{\tau}. Recall that, for integral curve \eq{\bar{\mu}_t} of \eq{\sfb{X}^\sscr{H}}, then \eq{s} and \eq{\tau} may be defined by:
\begin{small}
\begin{align} \label{dtds_def_prj}
\boxed{ \begin{array}{llll}
        \cf  := \tfrac{1}{r_\en^2}
&,\qquad 
    \dot{s} \,=\, \diff{s}{t} = \tfrac{1}{ \cf  (\bar{\mu}_t)} 
&,\qquad 
   \pdt{t} \,=\,  \diff{t}{s} =  \cf  (\bar{\mu}_s) 
\\[5pt]    
    \cff := \tfrac{1}{\lang r_\en^2}
&,\qquad 
     \dot{\tau} \,=\, \diff{\tau}{t} \,=\, \tfrac{1}{\cff(\bar{\mu}_t)} 
&,\qquad 
    \rng{t} \,=\, \diff{t}{\tau} = \cff(\bar{\mu}_{\tau}) 
\end{array} }
\end{align}
\end{small}
where \eq{\bar{\mu}_{t}\equiv \bar{\mu}_s \equiv \bar{\mu}_\tau} are all the ``same'' curve, just re-parameterized (technically, this makes them different curves). That is, \eq{\bar{\mu}_s= \bar{\mu}_{t(s)}}, and \eq{\bar{\mu}_\tau= \bar{\mu}_{t(\tau)}}, and \eq{\bar{\mu}_\tau= \bar{\mu}_{s(\tau)}}, etc. To clarify, 
consider the function \eq{\cf} defining the evolution parameter \eq{s}. Then \eq{\bar{\mu}_s=\bar{\mu}_{t(s)}} is an integral curve of the conformally-Hamiltonian vector field \eq{\cf\sfb{X}^\sscr{H}}. That is, scaling \eq{\sfb{X}^\sscr{H}} by a conformal factor \eq{ \cf  \in\fun(\cotsp\bvsp{E})} corresponds to transforming the evolution parameter from \eq{t} (e.g., time) to a new parameter, \eq{s}, were  \eq{s} is the solution to  the ODE \eq{\dot{s} = 1/ \cf  (\bar{\mu}_t)} for an integral curve \eq{\bar{\mu}_t} of \eq{\sfb{X}^\sscr{H}}. Conversely, \eq{t} is recovered as the solution to \eq{\pdt{t}= \cf  (\bar{\mu}_s)} for the corresponding integral curve \eq{\bar{\mu}_s} of \eq{\cf\sfb{X}^\sscr{H}} (where, for our purposes, \eq{\bar{\mu}_t\equiv\bar{\mu}_s} are the ``same'' curve). The same holds for the conformal factor \eq{\cff}, evolution parameter \eq{\tau}, and conformally Hamiltonian vector field \eq{\cff\sfb{X}^\sscr{H}}. 

Generally, any non-zero function on the carrier manifold can be used as a conformal factor to implement a transformation of the evolution parameter as just described. For the case at hand, the two functions given above in Eq.\eqref{dt_funs} are of particular interest for the following reason (as will be shown):

\begin{small}
\begin{remrm}
Either \eq{\cf} or \eq{\cff} can be used as a conformal factor to \textit{linearize} the dynamics of \eq{\sfb{X}^\sscr{H}} seen in Eq.\eqref{H_reduced_again} under the following conditions:    
\begin{small}
\begin{enumerate}[topsep=1pt,itemsep=0pt]
    \item   The dynamics on \eq{\cotsp\bs{\Sigup}_\nozer} (the ``\eq{(r^i,\plin_i)}-part'') are linearized if \eq{U^\ss{1}=0}. That is, for any potential of the form \eq{U=U^\zr(r_\en)} such that \eq{\pd_i U = 0} and \eq{\dif U  = \pd_\en U^\zr \enform}. This corresponds to any \textit{original} potential \eq{V=V^\zr(\rfun)} such that \eq{\pd_\en V=0} and   \eq{\dif V =\pd_{\rfun} V^\zr \hsfb{r}^\flt}. That is, the original system is subject only to arbitrary radial forces such that the transformed system is subject only to arbitrary normal forces.
    \item  The the dynamics on \eq{\cotsp\vsp{N}_\ii{+}} (the ``\eq{(r^\en,\plin_\en)}-part'')  are fully linearized if, in addition to the above, \eq{U=U^\zr(r_\en)} is some version of the form seen below (implying that \eq{V=V^\zr(\rfun)} is the Manev potential):
    \begin{small}
    \begin{align} \label{V0U0_form}
        U= \psi^* V :
        \qquad\;\;
        V=V^\zr(\rfun) = -\sck_1/\rfun - \ttfrac{1}{2}\sck_2/ \rfun^2 
        \quad\;\; \leftrightarrow \quad\;\;
        U=U^\zr(r_\en)= -\sck_1 r_\en -\ttfrac{1}{2}\sck_2 r_\en^2
    \end{align}
    \end{small}
    where \eq{\sck_1,\sck_2\in\mbb{R}} are any scalars and where the factor of \eq{1/2} and negative signs are included for 
    convenience.\footnote{The negative signs in \eq{V^\zr} and \eq{U^\zr} seen in Eq.\eqref{V0U0_form} are
    included simply such that the central forces resulting from \eq{V^\zr} are attractive (directed towards the origin in \eq{\bs{\Sigup}} or, rather, towards the \eq{\envec} axis, \eq{\vsp{N}}) if \eq{\sck_1,\sck_2 >0}, and are repulsive (directed away from the origin in \eq{\bs{\Sigup}} or, rather, away from the \eq{\envec} axis, \eq{\vsp{N}}) if \eq{\sck_1,\sck_2<0}. This means the transformed normal forces resulting from \eq{U^\zr} are repulsive (directed away from the \eq{\bs{\Sigup}} plane, along the \eq{+\envec} direction) if \eq{\sck_1,\sck_2 >0}, and are attractive (directed towards to the \eq{\bs{\Sigup}} plane, along the \eq{-\envec} direction) if \eq{\sck_1,\sck_2<0}. }
    The above means the \textit{original} potential gives a \textit{radial} force 1-form \eq{-\dif V = -(\sck_1/\rfun^2 + \sck_2/\rfun^3)\hsfb{r}^\flt}, which transforms to a new potential that gives a \textit{normal} force 1-form \eq{-\dif U =(\sck_1 + \sck_2 r_\en)\enform}. 
    The Kepler problem (for the \textit{original} system) corresponds to \eq{\sck_2=0}. 
\end{enumerate}
\end{small} 
\end{remrm}
\end{small}

\noindent Before proceeding, let us define the abbreviated notations \eq{\dt{\square}}, \eq{\pdt{\square}}, and \eq{\rng{\square}} for derivatives of functions along \eq{\sfb{X}^\sscr{H}}, \eq{\cf\sfb{X}^\sscr{H}}, and \eq{\cff \sfb{X}^\sscr{H}} or, analogously, derivatives ``with respect to'' \eq{t}, \eq{s}, and \eq{\tau}:
\begin{small}
\begin{align} \label{dd_rels_prj}
\begin{array}{rllrlllll}
   \pbrak{\square}{\mscr{H}} = \lderiv{\sfb{X}^{\mscr{H}}} \equiv&\!\! \diff{\square}{t} =:\;
     \dt{\square} 
 &\quad, & \quad
    \lderiv{\sfb{X}^{\mscr{H}}}^2 \equiv&\!\! \ddiff{\square}{t} =:\;
     \ddt{\square}
\\[5pt]
      \cf  \pbrak{\square}{\mscr{H}} =  \lderiv{\cf\sfb{X}^\sscr{H}} \equiv &\!\! \diff{\square}{s} =:\;
      \pdt{\square}  =  \cf  \dt{\square}
  &\quad, & \quad
       \lderiv{\cf\sfb{X}^\sscr{H}}^2 \equiv&\!\! \ddiff{\square}{s} =:\;
      \pddt{\square}  
     =  \cf^2 (\ddt{\square} +  \tfrac{\dt{\cf}}{\cf} \dt{\square} )
\\[5pt]
     \cff\pbrak{\square}{\mscr{H}} =  \lderiv{\cff \sfb{X}^\sscr{H}} \equiv&\!\! \diff{\square}{\tau} =:
      \rng{\square} = \cff \dt{\square} = \tfrac{1}{\lang}  \pdt{\square}
  &\quad, & \quad
      \lderiv{\cff \sfb{X}^\sscr{H}}^2 \equiv&\!\! \ddiff{\square}{\tau} =:\;
      \rrng{\square}  = \cff^2 (\ddt{\square} + \tfrac{\dt{\cff}}{\cff} \dt{\square})
      \,=\, \tfrac{1}{\slang^2}( \pddt{\square} - \tfrac{\pdt{\slang}}{\slang} \pdt{\square} )
\end{array}
\end{align}
\end{small}
where the above derivatives ``with respect to'' \eq{t}, \eq{s}, and \eq{\tau} actually mean the Lie derivatives (of functions) along  \eq{\sfb{X}^\sscr{H}},  \eq{\cf\sfb{X}^\sscr{H}}, and  \eq{\cff\sfb{X}^\sscr{H}}, respectively. The above is simply an informal restatement of Eq.\eqref{lderiv_rels_prj}.

\begin{small}
\begin{itemize}
    \item  \rmsb{Relation to original system.} Although we are working with the \eq{\colift\psi}-transformed Hamiltonian system for \eq{\mscr{H}=\colift\psi^*\mscr{K}} throughout this section, keep in mind that it is still the original Hamiltonian system for \eq{\mscr{K}} in which we are ultimately interested. Recall that if \eq{\bar{\mu}_t=(\barpt{q}_t,\barbs{\mu}_t)} is an integral curve of \eq{\sfb{X}^\sscr{H}} then \eq{\bar{\kappa}_t=(\barpt{x}_t,\barbs{\kappa}_t)=\colift\psi(\bar{\mu}_t)} is an integral curve of \eq{\sfb{X}^\sscr{K}} (and vice versa). Any such \eq{\colift\psi}-related curves always satisfy \eq{q^\en = 1/\nrm{\pt{x}}} and  \eq{\lang(\mu_t)=\lang(\kappa_t)} (and thus \eq{\slang(\mu_t)=\slang(\kappa_t)}). 
    As such, the above differential relations defining \eq{s} and \eq{\tau} are as follows when evaluated along any such \eq{\colift\psi}-related curves:
    \begin{small}
    \begin{align} \label{dt_funs_details}
    \begin{array}{llll}
     \cf  := \tfrac{1}{r_\en^2}
    &\quad,\qquad
        \pdt{t} \,=\,  \cf  (\bar{\mu}_s) \,=\, 1/q_\en^2(s) \,=\, \nrm{\ptvec{x}_s}^2
    &\quad,\qquad 
          \dot{s} \,=\, \tfrac{1}{ \cf  (\bar{\mu}_t)}    
          \,=\, q_\en^2(t) \,=\, 1/ \nrm{\ptvec{x}_t}^2  
    \\[4pt]
     \cff := \tfrac{1}{ \slang r_\en^2 }
    &\quad,\qquad
        \rng{t} \,=\, \cff(\bar{\mu}_\tau) \,=\, \tfrac{1}{\slang(\mu_\tau) q_\en^2(\tau)} \,=\, \tfrac{\nrm{\ptvec{x}_\tau}^2}{\slang(\kappa_\tau)}
    &\quad,\qquad
        \dot{\tau} \,=\, \tfrac{1}{\cff(\bar{\mu}_t)} 
        \,=\, \slang(\mu_t) q_\en^2(t)  
        \,=\, \slang(\kappa_t)/ \nrm{\ptvec{x}_t}^2  
    \end{array}
    \end{align}
    \end{small}
    In other words, if we wish to define the evolution parameters \eq{s} and \eq{\tau} in term of the \textit{original} \eq{\mscr{K}}-system, then \eq{\cf} and \eq{\cff} are not the appropriate conformal factors; the appropriate factors are instead \eq{\til{\cf}:=\colift\psi_*  \cf } and \eq{\til{\cff}:=\colift\psi_* \cff } such that:
    \begin{small}
    \begin{align}
               \cf\sfb{X}^\sscr{H} = \colift\psi^*( \til{\cf} \sfb{X}^\sscr{K} ) 
              \quad\quad  \til{\cf} := \colift\psi_*  \cf \equiv \psi_*  \cf   =  \rfun^2
     &&,&&  
               \cff\sfb{X}^\sscr{H} = \colift\psi^*( \til{\cff} \sfb{X}^\sscr{K} ) 
                \quad\quad  \til{\cff} := \colift\psi_* \cff = \rfun^2/\slang
    \end{align}
    \end{small}
    where \eq{ \cf  (\bar{\mu}_t)=\til{\cf}(\bar{\kappa}_t)} and \eq{\cff(\bar{\mu}_t)=\til{\cff}(\bar{\kappa}_t)}, as seen in Eq.\eqref{dt_funs_details}. The above functions \eq{\til{\cf}} and \eq{\til{\cff}} are precisely the functions used in previous works taking a passive, coordinate-transformation (e.g., \cite{ferrandiz1987general,deprit1994linearization,peterson2025prjCoord}). 
    \item  \rmsb{Relation to true anomaly.}  For the case \eq{\en -1=3} and \eq{U^\zr = -\sck r_\en} (i.e., the original potential is the 3-dim Kepler potential, \eq{V^\zr = -\sck/\rfun}), then the parameter \eq{\tau} corresponds the true anomaly (up to an additive constant) of the \textit{original} system. 
\end{itemize}
\end{small}
\vspace{1ex} 
 \noindent  
 We will first go through the details of conformal scaling by \eq{\cf}, after which it is a quick process to obtain the analogous results using \eq{\cff = \cf/\lang}. Both lead to linear dynamics in the case that the potential is of the form in Eq.\ref{V0U0_form} (the Manev potential).    

\subsubsection{Linearization Using a Sundman-Like Evolution Parameter}

 We will first show the linearization via a conformal scaling of \eq{\sfb{X}^\sscr{H}} by \eq{ \cf  := 1/r_\en^2}, which defines a new evolution parameter \eq{s} by \eq{\mrm{d} {t} = \cf \mrm{d} {s}}, as described in Eq.\eqref{dtds_def_prj}.  We re-iterate that we use ``linear'' to mean ``linear in the unperturbed case \eq{U^\ss{1}=0}''. 
Starting with the Hamiltonian vector field \eq{\sfb{X}^\sscr{H}\in\vechm(\cotsp\bvsp{E})}, 
we consider the conformally-Hamiltonian vector field \eq{ \cf\sfb{X}^\sscr{H}\in\vect(\cotsp\bvsp{E})} whose \eq{s}-parameterized integral curves are represented in cartesian coordinates by solutions to the following ODEs:
\begin{small}
\begin{align} \label{d_ds_prj}
\begin{array}{cccc}
        \cf\sfb{X}^\sscr{H}
    \,=\,  \cf  \partial^\a \mscr{H} \hbpart{\a} \,-\,    \cf  \pd_\a \mscr{H} \hbpartup{\a} 
\\[5pt]
      \cf  := 1/r_\en^2  
\end{array}
\quad\left\{ \quad \begin{array}{lllllll}
     \pdt{r}^i =\,   \cf  \partial^i \mscr{H} 
     &\!\!\!\! =\,    -  \slang^{ij} r_j
     &, \quad  \pdt{\plin}_i = -  \cf  \pd_i \mscr{H} 
      &\!\!\!\! =\, - \slang_{ij} \plin^j  -  \cf  \pd_i U^\ss{1}
\\[4pt]
    \pdt{r}^\en =\,   \cf  \partial^\en \mscr{H} 
   &\!\!\!\! =\, \tfrac{r_\en^2}{m} \plin^\en
     &,\quad   \pdt{\plin}_\en = -  \cf  \pd_\en \mscr{H} 
      &\!\!\!\! =\, - \tfrac{1}{m r_\en} (\lang^2 + 2 r_\en^2 \plin_\en^2 ) -  \cf  \pd_\en (U^\zr + U^\ss{1}) 
\end{array} \right.
\end{align}
\end{small}
where \eq{\slang^{ij}=\lang^{ij}/m} and \eq{\diff{}{s}=:\pdt{\square}=  \cf  \dot{\square}} and  where \eq{\cf\sfb{X}^\sscr{H}} has the same  
integrals of motion as \eq{\sfb{X}^\sscr{H}}, as indicated in Eq.\eqref{prj_ioms_rev}:
\begin{small}
\begin{align} \nonumber 
     \pbrak{r}{\mscr{H}} = \pbrak{\hat{r}^i \plin_i }{\mscr{H}}  = \pbrak{r^i \plin_i }{\mscr{H}} = 0
 &&,&&
  \pbrak{ \slang^{ij} }{\mscr{H}} = -(r^i m^{jk} - r^j m^{ik})\pd_k U^\ss{1} 
\;\;,\;\; 
      \pbrak{ \ttfrac{1}{2}\slang^2 }{\mscr{H}} =
   - \tfrac{\rfun^2}{m^2} \plin^j \pd_j U^\ss{1} 
\;\;,\;\; 
\pbrak{\ttfrac{1}{2}\nrmtup{\plin}^2 }{\mscr{H}}  
    =  -\plin^j\pd_j U^\ss{1} 
\end{align}
\end{small}
Thus, in the unperturbed case that \eq{U^\ss{1}=0}, then \eq{\slang^{ij}} are integrals of motion and the ODEs for \eq{\pdt{r}^i} and \eq{\pdt{\plin}_i} (that is, the dynamics on \eq{\cotsp\bs{\Sigup}_\nozer}) are linear. More specifically, they describe a family of linear equations parameterized by the angular momentum values. Although less obvious from the above, we will show that the same is true of the dynamics described by \eq{\pdt{r}^\en} and \eq{\pdt{\plin}_\en} (that is, the dynamics on \eq{\cotsp\vsp{N}_\ii{+}}). Before going into further detail, we slightly adjust our chosen description of things for later convenience:
\begin{small}
\begin{enumerate}[topsep=1pt,itemsep=1pt]
   \item It will be convenient to decompose \eq{\cf\sfb{X}^\sscr{H}} into a part tangent to \eq{\cotsp\bs{\Sigup}_\nozer \subset \cotsp\bvsp{E}} (the ``\eq{(r^i,\plin_i)}-part'') and a part tangent to \eq{\cotsp\vsp{N}_\ii{+} \subset \cotsp\bvsp{E}} (the ``\eq{(r^\en,\plin_\en)}-part''). 
    In other words, we slightly modify our view of \eq{\cotsp\bvsp{E}}, regarding it as a 
    product bundle \eq{\cotsp\bs{\Sigup}_\nozer \oplus \cotsp\vsp{N}_\ii{+}}:\footnote{With \eq{\cotsp\bvsp{E}} as in Eq.\eqref{T*E_split}, the \eq{(2\en-2)}-dim submanifold \eq{\cotsp\Sig_\ii{b},\cotsp\man{Q}_\ii{b}\subset\cotsp\bvsp{E}} are then likewise viewed as 
        \begin{align}\label{T*Q_split}
            \cotsp\Sig_\ii{b} = \cotsp(\bs{\Sigup}_\nozer \times \{b\}) \cong \cotsp\bs{\Sigup}_\nozer
            \qquad,\qquad 
            \cotsp\man{Q}_\ii{b} = \cotsp(\man{S}^{\en-\two}_\ii{b} \times \vsp{N}_\ii{+}) \cong \cotsp\man{S}^{\en-\two}_\ii{b} \oplus \cotsp\vsp{N}_\ii{+}
            \qquad\quad \fnsize{with} \;\;  \cotsp\man{S}^{\en-\two}_\ii{b} \subset \cotsp\bs{\Sigup}_\nozer
        \end{align}  }
    \begin{small}
    \begin{align} \label{T*E_split}
        \cotsp\bvsp{E}  \,=\, \cotsp( \bs{\Sigup}_\nozer \oplus \vsp{N}_\ii{+} ) \,\cong\, 
        \cotsp\bs{\Sigup}_\nozer \oplus \cotsp\vsp{N}_\ii{+} 
    &&
    \begin{array}{rllllll}
           \fnsize{this:}&   \bar{\mu}_\ss{\barpt{q}} =  (\barpt{q},\barbs{\mu}) = (\ptvec{q}+\ptvec{q}^\ii{\perp}, \bs{\mu} + \bs{\mu}^\ii{\perp}) \in \cotsp\bvsp{E} 
     \\[3pt]
         \fnsize{becomes:}&   \bar{\mu}_\ss{\barpt{q}} = (\mu_\ss{\pt{q}},\mu_\ss{\pt{q}}^\ii{\perp}) =   ((\ptvec{q}, \bs{\mu}), (\ptvec{q}^\ii{\perp}, \bs{\mu}^\ii{\perp})) \in \cotsp\bs{\Sigup}_\nozer \oplus \cotsp\vsp{N}_\ii{+} 
    \end{array}
    \end{align}
    \end{small}
    (\eq{ \ptvec{q}^\ii{\perp} = q^\en \envec} and \eq{\bs{\mu}^\ii{\perp} = \mu_\en \enform}). This corresponds to a re-ordering of indices/coordinates;  \eq{(\tup{r},r^\en,\tup{\plin},\plin_\en)} becomes \eq{(\tup{r},\tup{\plin},r^\en,\plin_\en)}:
    \begin{small}
    \begin{align}
    \begin{array}{rlllll}
            \fnsize{this:}&  \bartup{z}:=(\bartup{r},\bartup{\plin}) =(\tup{r},r^\en,\tup{\plin},\plin_\en) 
            \;\;,\;\; 
             \cord{\ns{\txw}}{\bartup{z}} = J_{\two \en}
     \qquad\quad\qquad 
              \fnsize{becomes:}& 
              \bartup{z}:=(\tup{z},\tup{z}^\ii{\perp})= (\tup{r},\tup{\plin},r^\en,\plin_\en) 
               \;\;,\;\; 
             \cord{\ns{\txw}}{\bartup{z}} = 
             \fnsz{\begin{pmatrix}
              J_{\two(\en-\one)} & 0 \\
              0 & J_{\two}
             \end{pmatrix}}
    \end{array}
    \end{align}
    \end{small}
    \item Additionally, the dynamics of \eq{\cf\sfb{X}^\sscr{H}} on \eq{\cotsp\vsp{N}_\ii{+} \subset \cotsp\bvsp{E}} are more conveniently expressed — specifically, the linear nature is more obvious — if we consider a new cofiber coordinate (i.e., momenta-level coordinate), \eq{ \tilplin_\en \in \fun(\cotsp\bvsp{E})}, defined by 
    \begin{small}
    \begin{gather} \label{newpn_def0}
    \phantom{xxxxxxxxxxxx}
        \boxed{\; \tilplin_\en :=  r_\en^2 \plin_\en = \plin_\en / \cf 
        \quad \leftrightarrow \quad 
        \plin_\en = \tilplin_\en / r_\en^2 = \cf \plin_\en
       \qquad,\qquad
        \begin{array}{cc}
            (\tup{z},\tiltup{z}^\ii{\perp}):=(\tup{r},\tup{\plin},r^\en,\tilplin_\en)
        \\[3pt]
             (\tup{z},\tup{z}^\ii{\perp})=(\tup{r},\tup{\plin},r^\en,\plin_\en)
        \end{array} }
    \end{gather}
    \end{small}
   It follows from the above 
   — rather, the associated frame field transformation\footnote{The coordinate transformation \eq{(\tup{z},\tup{z}^\ii{\perp}) \,\leftrightarrow \, (\tup{z},\tiltup{z}^\ii{\perp})}, which is fully described by \eq{\plin_\en = \tilplin_\en/r_\en^2 \,\leftrightarrow \, \tilplin_\en = r_\en^2 \plin_\en}, gives a frame field transformation: 
    \begin{align} \nonumber 
        \begin{array}{llllll}
             \hbpart{r^\en} = \bpart{r^\en} + 2 r_\en \plin_\en \bpartup{\tilplin_\en} 
        \\[2pt]
              \hbpartup{\plin_\en} = r_\en^2 \bpartup{\tilplin_\en}
        \end{array}
        \;\;\; \leftrightarrow \;\;\;
        \begin{array}{llllll}
             \bpart{r^\en} = \hbpart{r^\en}   -2\tfrac{\tilplin_\en}{r_\en^3} \hbpartup{\plin_\en} 
        \\[2pt]
              \bpartup{\tilplin_\en} = \tfrac{1}{r_\en^2} \hbpartup{\plin_\en}
        \end{array}
        &&,&&
        \begin{array}{llllll}
             \hbdel^{r^\en} = \bdel^{r^\en} = \dif r^\en 
        \\[2pt]
              \hbdeldn_{\plin_\en} = -2\tfrac{\tilplin_\en}{r_\en^3} \bdel^{r^\en} + \tfrac{1}{r_\en^2} \bdeldn_{\tilplin_\en}
        \end{array}
        \;\;\; \leftrightarrow \;\;\;
         \begin{array}{llllll}
             \bdel^{r^\en} = \hbdel^{r^\en}  
        \\[2pt]
              \bdeldn_{\tilplin_\en} = 2r_\en \plin_\en \hbdel^{r^\en} + r_\en^2 \hbdeldn_{\plin_\en}
        \end{array}
    \end{align}
    (the frame fields for \eq{r^i} and \eq{\plin_i} are unchanged). Above, \eq{\hat{\square}} denotes frame fields for the cotangent-lifted symplectic coordinates \eq{(\tup{z},\tup{z}^\ii{\perp})=(\tup{r},\tup{\plin},r^\en,\plin_\en)}. } —
    that the canonical symplectic form and bivector are expressed in the non-symplectic frame fields for  \eq{(\tup{z},\tiltup{z}^\ii{\perp}):=(\tup{r},\tup{\plin},r^\en,\tilplin_\en)}
    as follows (where \eq{ \nbs{\omg}_\ii{\!\Sig} = \hbdel^{r^i} \wedge \hbdeldn_{\plin_i}} and \eq{ \inv{\nbs{\omg}_\ii{\!\Sig}} = - \hbpart{r^i} \wedge \hbpartup{\plin_i}}):
    \begin{small}
    \begin{align} \label{sp_newpn}
    \begin{array}{lllllll}
         \nbs{\omg} = \nbs{\omg}_\ii{\!\Sig} + \hbdel^{r^\en} \wedge \hbdeldn_{\plin_\en} 
         &\!\!\!\! =\, \nbs{\omg}_\ii{\!\Sig} +  r_\en^\ss{-2} \bdel^{r^\en} \wedge \bdeldn_{\tilplin_\en} 
    \\[3pt]
       \inv{\nbs{\omg}} = \inv{\nbs{\omg}_\ii{\!\Sig}}  - \hbpart{r^\en} \wedge \hbpartup{\plin_\en}  
        &\!\!\!\! =\, \inv{\nbs{\omg}_\ii{\!\Sig}}  -  r_\en^2 \bpart{r^\en} \wedge \bpartup{\tilplin_\en} 
    \end{array}
    &&
    \fnsize{i.e.,}
    \quad 
        \til{\ns{\txw}} =  \fnsz{\begin{pmatrix}
              J_{\two(\en-\one)} & 0 \\
              0 & r_\en^{-2} J_{\two}
             \end{pmatrix}}
             \;\;,\;\;
              \inv{\til{\ns{\txw}}} =  \fnsz{\begin{pmatrix}
              -J_{\two(\en-\one)} & 0 \\
              0 & - r_\en^2 J_{\two}
             \end{pmatrix}}
    \end{align}
    \end{small}
    \item[]  (Nothing in Eq.\eqref{T*E_split}-Eq.\eqref{sp_newpn} above is \emph{necessary} for any of the following developments; it is merely convenient.) \\
\end{enumerate}
\end{small}
Now,  \eq{\tilplin_\en} is not the cofiber coordinate conjugate to  \eq{r^\en}, that is, \eq{(\tup{z},\tiltup{z}^\ii{\perp})=(\tup{r},\tup{\plin},r^\en,\tilplin_\en)} are \textit{not} \eq{\nbs{\omg}}-symplectic coordinates (not canonical coordinates) and thus do not obey the classic, \textit{coordinate} form, of Hamilton's equations of motion. Fortunately, the function \eq{\mscr{H}\in\fun(\cotsp\bvsp{E})} and  the vector fields \eq{\sfb{X}^\sscr{H}\in\vechm(\cotsp\bvsp{E},\nbs{\omg})} and \eq{\cf\sfb{X}^\sscr{H}\in\vect(\cotsp\bvsp{E})}  do not care about our coordinates (nor do any other function or tensor field). We can ``do'' Hamiltonian mechanics using any conceivable coordinates; they need not be symplectic. 
To clarify, consider \eq{\inv{\nbs{\omg}}} expressed in the non-symplectic frame fields for  \eq{(\tup{z},\tiltup{z}^\ii{\perp}):=(\tup{r},\tup{\plin},r^\en,\tilplin_\en)} as given in Eq.\eqref{sp_newpn}. Then,  for arbitrary \eq{h\in\fun(\cotsp\bvsp{E})}, the Hamiltonian vector field \eq{\sfb{X}^h} and the conformally-related \eq{\cf\sfb{X}^h = r_\en^\ss{-2} \sfb{X}^h} are expressed as 
\begin{small}
\begin{align} 
\begin{array}{rlllll}
     \sfb{X}^h \!:= \inv{\nbs{\omg}}(\dif h,\slot) 
      &\!\!=\; \pderiv{h}{\plin_i} \hbpart{r^i} -   \pderiv{h}{r^i}  \hbpartup{\plin_i} +   \pderiv{h}{\plin_\en} \hbpart{r^\en} -   \pderiv{h}{r^\en} \hbpartup{\plin_\en}
      &\!\! =\; \pderiv{h}{\plin_i} \hbpart{r^i} -   \pderiv{h}{r^i}  \hbpartup{\plin_i} \,+\,  r_\en^2\big(  \pderiv{h}{\tilplin_\en} \bpart{r^\en} -   \pderiv{h}{r^\en} \bpartup{\tilplin_\en} \big)
\\[6pt]
    \cf\sfb{X}^h =  r_\en^\ss{-2} \sfb{X}^h
      &\!\! =\; r_\en^\ss{-2} \big( \pderiv{h}{\plin_i} \hbpart{r^i} -   \pderiv{h}{r^i}  \hbpartup{\plin_i} +   \pderiv{h}{\plin_\en} \hbpart{r^\en} -   \pderiv{h}{r^\en} \hbpartup{\plin_\en} \big)
  &\!\! =\;   r_\en^\ss{-2} \big( \pderiv{h}{\plin_i} \hbpart{r^i} -   \pderiv{h}{r^i}  \hbpartup{\plin_i} \big) +   \pderiv{h}{\tilplin_\en} \bpart{r^\en}  -   \pderiv{h}{r^\en} \bpartup{\tilplin_\en} 
\end{array}
\end{align}
\end{small}
Thus, instead of using symplectic coordinates \eq{(\tup{z},\tup{z}^\ii{\perp})} as in Eq.\eqref{d_ds_prj},  an alternative, equally valid,  coordinate representation of the dynamics of \eq{\cf \sfb{X}^\sscr{H} = \cf \inv{\nbs{\omg}}(\dif \mscr{H},\slot) } is given using the non-symplectic coordinates  \eq{(\tup{z},\tiltup{z}^\ii{\perp})=(\tup{r},\tup{\plin},r^\en,\tilplin_\en)} as 
follows:\footnote{Similarly, the  dynamics of \eq{\sfb{X}^\sscr{H}} itself are represented in the non-symplectic coordinates  \eq{(\tup{z},\tiltup{z}^\ii{\perp})=(\tup{r},\tup{\plin},r^\en,\tilplin_\en)} as:
\begin{align} \nonumber 
    \sfb{X}^\sscr{H} \!:= \inv{\nbs{\omg}}(\dif \mscr{H},\slot) 
    \;\; \left\{ \;\; \begin{array}{llllllll}
     \dot{r}^i =\, \partial^i \mscr{H} 
     \,=\,    -  r_\en^2 \slang^{ij} r_j
     &,\quad 
     \dot{\plin}_i =  - \pd_i \mscr{H} 
      \,=\,  - r_\en^2  \slang_{ij} \plin^j  -  \pd_i U^\ss{1}
    \\[2pt] 
        \dot{r}^\en =\,  r_\en^2  \partial^{\tilplin_\en}  \mscr{H} 
       \,=\, \tfrac{r_\en^2}{m} \tilplin^\en
         &,\quad   \dot{\tilplin}_\en = - r_\en^2 \pd_\en \mscr{H} 
          \,=\, - m\slang^2 r_\en^3   -  r_\en^2  \pd_\en (U^\zr + U^\ss{1}) 
    \end{array} \right.
\end{align} }
\begin{small}
\begin{align} \label{d_ds_prj_altCords}
 \begin{array}{rcllllllllll}
        &  \mscr{H} = \tfrac{r_\en^2}{2m} ( \lang^2 + r_\en^2 \plin_\en^2) + U  \;=\;  \tfrac{1}{2m} (r_\en^2 \lang^2 +  \tilplin_\en^2) + U
\\[6pt]
        \cf\sfb{X}^\sscr{H} = r_\en^\ss{-2} \sfb{X}^\sscr{H}
 &\!\! \left\{ \;\; \begin{array}{llllllll}
         \pdt{r}^i =\,   r_\en^\ss{-2} \partial^i \mscr{H} 
         \,=\,    -  \slang^{ij} r_j
         &,\quad 
         \pdt{\plin}_i =  - r_\en^\ss{-2}  \pd_i \mscr{H} 
          \,=\, - \slang_{ij} \plin^j  -  r_\en^\ss{-2} \pd_i U^\ss{1}
     \\[3pt]
        \pdt{r}^\en =\,    \partial^{\tilplin_\en} \mscr{H} 
       \,=\, \tfrac{1}{m} \tilplin^\en
         &,\quad   \pdt{\tilplin}_\en = -   \pd_\en \mscr{H} 
          \,=\, - m \slang^2 r_\en  -   \pd_\en (U^\zr + U^\ss{1}) 
    \end{array}  \right.
\end{array}
\end{align}
\end{small}
Note that using \eq{\tilplin_\en= r_\en^2 \plin_\en} rather than \eq{\plin_\en}  has no affect on the coordinate representation of \eq{\cf\sfb{X}^\sscr{H}}  on \eq{\cotsp\bs{\Sigup}_\nozer} (the ``\eq{(r^i,\plin_i})-part''), 
but, on \eq{\cotsp\vsp{N}_\ii{+}}, it happens to cancel out the conformal factor in such a way that 
the  coordinate representation on \eq{\cotsp\vsp{N}_\ii{+}} (the ``\eq{(r^\en,\tilplin_\en)}-part'')  looks like it obeys Hamilton's canonical equations, even though \eq{\cf\sfb{X}^\sscr{H}} is not Hamiltonian. More to the point, using \eq{\tilplin_\en} more clearly reveals the linear nature of the \eq{\cotsp\vsp{N}_\ii{+}} part of  \eq{\cf\sfb{X}^\sscr{H}}, \textit{but only for certain forms of \eq{U^\zr=U^\zr(r_\en)}}. 
Specifically, as mentioned in Eq.\eqref{V0U0_form}, consider the case that \eq{U^\zr(r_\en)=\psi^*V^\zr(\rfun)} is any linear  combination of \eq{r_\en} and \eq{r_\en^2} — this corresponds to the \textit{original} \eq{V^\zr} being any linear  combination of \eq{1/\rfun} and \eq{1/\rfun^2}.  The above then leads to the following (only \eq{\pdt{\tilplin}_\en} is affected by \eq{U^\zr}):
\begin{small}
\begin{align} \label{d_ds_prj_altCords_U0} 
   \fnsz{\left(\begin{array}{cc}
          \fnsize{for} \;\;   U^\zr = - \sck_1 r_\en - \ttfrac{1}{2}\sck_2 r_\en^2 
          \\[2pt]
        \fnsize{i.e.,} \;\;  V^\zr = -\sck_1 / \rfun - \ttfrac{1}{2} \sck_2/ \rfun^2 
    \end{array}\right)}
\;\;\, \xRightarrow{\text{Eq.\eqref{d_ds_prj_altCords}}} \;\;\,
 \boxed{\begin{array}{llllllll}
         \pdt{r}^i =  -  \slang^{ij} r_j
         &,\quad 
         \pdt{\plin}_i = - \slang_{ij} \plin^j  -  r_\en^\ss{-2} \pd_i U^\ss{1}
     \\[3pt]
        \pdt{r}^\en = \tfrac{1}{m} \tilplin^\en
         &,\quad  
         \pdt{\tilplin}_\en = -m \beta^2 r_\en  + \sck_1  -  \pd_\en U^\ss{1}
    \end{array}}
 &&
 \begin{array}{llllll}
      \slang=\lang/m   \\[2pt]
      \beta^2 := \slang^2 - \txkbar_2 \\[2pt]
      \txkbar_{1,2} := \sck_{1,2}/m 
 \end{array}
\end{align}
\end{small}
where \eq{\sck_1,\sck_2\in\mbb{R}} are any scalars. 
When \eq{U^\ss{1}=0} (i.e., \eq{V^\ss{1}=0} such that the original system is subject to purely radial forces for the above potential), it holds that \eq{\slang^{ij}} and \eq{\slang} are integrals of motion such that the above describes a family of linear ODEs, parameterized by the angular momentum value (determined by initial conditions). Note the constant ``driving force'' term \eq{m\txkbar_1=\sck_1} in the equation for \eq{\pdt{\tilplin}_\en}.  
\begin{small}
\begin{notesq}
    \rmsb{simplifications on $\cotsp\man{Q}_\ii{1}$.} For integral curves of \eq{\cf\sfb{X}^\sscr{H}} on the \eq{(2\en-2)}-dim invariant submanifold $\cotsp\man{Q}_\ii{1}\subset\cotsp\bvsp{E}$, the above may be simplified using the relations given in Eq.\eqref{T*Q_reg_rev} (for instance,  \eq{\lang^2 = m^2\slang^2 \simeq \nrmtup{\plin}^2}). In particular, the dynamics on \eq{\cotsp\bs{\Sigup}_\nozer} (the ``\eq{(r^i,\plin_i)}-part'') 
    evolve on the cotangent bundle of a sphere, \eq{\cotsp\man{S}^{\en-\two}\subset \cotsp\bs{\Sigup}_\nozer}, with the above ODEs for \eq{(\pdt{r}^i,\pdt{\plin}_i)} simplifying to
    \begin{small}
    \begin{align} \label{d_ds_T*S}
          \phantom{l=p}  &&
         \pdt{r}^i \simeq \tfrac{1}{m} \plin^i
         \qquad,\qquad 
         \pdt{\plin}_i \simeq -\tfrac{1}{m} \nrmtup{\plin}^2 r_i - r_\en^\ss{-2}\pd_i U^\ss{1} \,\simeq\, - m\slang^2 r_i - r_\en^\ss{-2}\pd_i U^\ss{1}
         &&
         \begin{array}{lll}
              \slang \simeq \nrm{\pdt{\tup{r}}} \\
              \lang \simeq \nrm{\tup{\plin}}
         \end{array}
    \end{align}
    \end{small}
\end{notesq}
\end{small}

\paragraph{Second-Order ODEs ``Equivalent'' to $\cf\sfb{X}^\sscr{H}$.}  We note that the dynamics in Eq.\eqref{d_ds_prj_altCords} also lead to the following second-order ODE — which are derived further down — for \eq{r^i} and \eq{r^\en}, with \eq{s} as the evolution parameter: 
  \begin{small}
  \begin{align} \label{dd_ds_prj} 
    \qquad\quad &&
        \fnsize{Eq.\eqref{d_ds_prj_altCords}}
           \qquad \Rightarrow  \qquad
           \begin{array}{llllll}
                 \pddt{r}\,^i 
                 \,+\, \slang^2 r^i \,=\,   -\tfrac{r^2}{r_\en^2} m^{ij}\pd_j U^\ss{1}
          \\[3pt]
                \pddt{r}\,^\en 
                 \,+\, \slang^2 r^\en \,=\,   - m^{\en\en}\pd_\en( U^\zr + U^\ss{1})
           \end{array} 
       &&,&&
        \pdt{\slang} \,=\, -\tfrac{r^2}{r_\en^2 m^2 \slang } \plin^i \pd_i U^\ss{1} 
    \end{align}
    \end{small}
    where \eq{\pddt{r}\,^\a =\lderiv{\cf\sfb{X}^\sscr{H}}^2 r^\a \equiv \ddiff{}{s} r^\a} . In the unperturbed case (\eq{U^\ss{1}=0}), then \eq{\slang} is an integral of motion and the above equations for \eq{\pddt{r}\,^i} are that of a (family of) \eq{(\en-1)}-dim linear harmonic oscillator with natural frequency \eq{\slang}, whose constant value is determined from initial conditions. This is true for any arbitrary \eq{U^\zr(r_\en)=\psi^*V^\zr(\rfun)}. While the above equation for \eq{\pddt{r}\,^\en} is, generally, nonlinear for arbitrary \eq{U^\zr(r_\en)}, it is indeed linear in the case that \eq{U^\zr} is given as in Eq.\eqref{V0U0_form}, which leads to: 
    \begin{small}
    \begin{align} \label{ddu_ds_prj} 
        \left(\begin{array}{cc}
              \fnsize{for} \;\;  U^\zr = - \sck_1 r_\en - \ttfrac{1}{2}\sck_2 r_\en^2 
              \\[2pt]
            \fnsize{i.e.,} \;\;  V^\zr = -\sck_1 / \rfun - \ttfrac{1}{2} \sck_2/ \rfun^2 
        \end{array}\right)
       \qquad 
       \Rightarrow \qquad
        \boxed{ \begin{array}{llllll}
                 \pddt{r}\,^i 
                 \,+\, \slang^2 r^i \,=\,   -\tfrac{r^2}{r_\en^2} m^{ij}\pd_j U^\ss{1}
          \\[3pt]
                \pddt{r}\,^\en  \,+\, \beta^2 r^\en \,-\, \txkbar_1 \,=\,  -  m^{\en\en} \pd_\en U^\ss{1} 
           \end{array} }
         &&,&&
           \begin{array}{llllll}
              \beta^2 := \slang^2 - \txkbar_2 \\
              \txkbar_{1,2} := \sck_{1,2}/m 
         \end{array}
    \end{align}
    \end{small}
    Note for the Kepler problem (\eq{\sck_2=0}) that \eq{\beta^2=\slang^2} and the above all have the same  naturally frequency, \eq{\slang}.  For integral curves on the \eq{(2\en-2)}-dim invariant submanifold $\cotsp\man{Q}_\ii{1}\subset\cotsp\bvsp{E}$, we further have \eq{\rfun=\nrm{\tup{r}}\simeq 1}, \eq{\lang\simeq \nrmtup{\plin}}, and \eq{\slang \simeq \nrm{\pdt{\tup{r}}}}. 
\begin{footnotesize}
\begin{itemize}[nosep]
    \item  \textit{Derivation of Eq.\eqref{dd_ds_prj}.}   Since Eq.\eqref{ddu_ds_prj}  follows immediately from Eq.\eqref{dd_ds_prj} so we need only derive the latter. It follows simply by substitution of the appropriate derivatives from  Eq.\eqref{d_ds_prj} or Eq.\eqref{d_ds_prj_altCords} into the following:
    \begin{align} \nonumber 
    \begin{array}{rllllll}
       \pddt{r}\,^i =
       \lderiv{\cf\sfb{X}^\sscr{H}}^2 r^i  \,=&\!\!\!
        \diff{}{s}(  \slang^{ij} r_j ) 
        \,=\,
        -\slang^{ij} \pdt{r}_j -  \pdt{\slang}^{ij} r_j 
        &\!\!\! =\, -\slang^2 r^i  
        - \tfrac{r^2}{m r_\en^2}(\emet^{ik} -  \hat{r}^i \hat{r}^k)\pd_k U^\ss{1} 
        \,=\, -\slang^2 r^i  
       - \tfrac{r^2}{ r_\en^2}m^{ik} \pd_k U^\ss{1} 
    \\[4pt]
      \pddt{r}\,^\en =
       \lderiv{\cf\sfb{X}^\sscr{H}}^2 r^\en  \,=&\!\!\!
       \diff{}{s}( \tfrac{1}{m} r_\en^2 \plin^\en )
       \;=\; \tfrac{1}{m} \pdt{\tilplin}^\en
      &\!\!\! =\;  -\slang^2 r^\en  -  m^{\en\en}\pd_\en ( U^\zr +  U^\ss{1})
    \end{array}
    \end{align}
     where \eq{\pdt{\slang}^{ij} = \tfrac{\cf}{m} \pbrak{\lang^{ij}}{\mscr{H}}} and \eq{\pdt{\slang} = \tfrac{\cf}{m} \pbrak{\lang}{\mscr{H}}} are obtained from Eq.\eqref{prj_ioms_rev}.
    Note the \eq{\pddt{r}\,^i} equation has been simplified further using the property \eq{\hat{r}^k \pd_k U^\ss{1} = \hsfb{r} \cdot \dif U^\ss{1}=0} (cf., section \ref{sec:prj_Xform_gen}). The intermediate steps for the above relations are given explicitly in 
   the footnote\footnote{From \eq{\pdt{r}^i = -\ttfrac{1}{m}\lang^{ij}r_j} we get \eq{\pddt{r}\,^i = -\ttfrac{1}{m}(\lang^{ij}\pdt{r}_j + \pdt{\lang}^{ij}r_j)}. Substituting \eq{\pdt{r}_j = -\ttfrac{1}{m}\lang_{js} r^s } and \eq{\pdt{\lang}^{ij} =  \cf  \pbrak{\lang^{ij}}{\mscr{H}} = - \cf  (r^i \emet^{jk} - r^j \emet^{ik})\pd_k U^\ss{1} } leads to:\\
    \eq{\quad \pddt{r}\,^i =   -\tfrac{1}{m} \lang^{ij} \pdt{r}_j- \tfrac{1}{m} \pdt{\lang}^{ij} r_j \,=\, \tfrac{1}{m^2} \lang^{ij}\lang_{jk} r^k  + \tfrac{\cf}{m}( r^i\emet^{jk} - r^j \emet^{ik})r_j\pd_k U^\ss{1} \,=\, -\tfrac{\lang^2}{m^2} r^i   - \tfrac{\cf r^2}{m}(\emet^{ik} -  \hat{r}^i \hat{r}^k)\pd_k U^\ss{1} \,=\, -\tfrac{\lang^2}{m^2} r^i   - \tfrac{r^2}{m r_\en^2}\emet^{ik}\pd_k U^\ss{1} }.\\
    where \eq{\lang^{ij}\lang_{jk} r^k = -\lang^2 r^i} and \eq{\hat{r}^k \pd_k U^\ss{1} =  \hsfb{r} \cdot \dif U^\ss{1}=0}.
    Equivalently, from \eq{\pdt{r}^i =  - \tfrac{1}{m} \lang^{ij} r_j = \tfrac{1}{m} \big( \nrmtup{r}^2 \plin^i  - (\tup{r}\cdot \tup{\plin} ) r^i \big) }  then \eq{\pddt{r}\,^i = \tfrac{1}{m} \big( \nrmtup{r}^2 \emet^{ij} \pdt{\plin}_j  - (\tup{r}\cdot \tup{\plin} ) \pdt{r}^i \big)}.
    Substituting back in the expressions from Eq.\eqref{d_ds_prj}, with \eq{ \pdt{\plin}_i=  - \tfrac{1}{m} \lang_{ij} \plin^j -  \cf  \pd_i U^\ss{1} = - \tfrac{1}{m} \big( \nrmtup{\plin}^2 r_i - (\tup{r} \cdot \tup{\plin})\plin_i \big) -  \cf  \pd_i U^\ss{1}}  leads to:\\
    \eq{\pddt{r}\,^i =   -\tfrac{\nrmtup{r}^2}{m^2} \big( \nrmtup{\plin}^2 r^i -  (\tup{r} \cdot \tup{\plin}) \emet^{ij} \plin_j \big) -  \tfrac{\nrmtup{r}^2}{m}  \cf  \pd_i U^\ss{1}  -  \tfrac{(\tup{r} \cdot \tup{\plin})}{m^2} \big(\nrmtup{r}^2 \emet^{ij} \plin_j - (\tup{r} \cdot \tup{\plin}) r^i \big) = - \tfrac{1}{m^2} \big( \nrmtup{r}^2 \nrmtup{\plin}^2 -  (\tup{r} \cdot \tup{\plin})^2 \big) r^i  -  \tfrac{\nrmtup{r}^2}{m}  \cf  \pd_i U^\ss{1}
     =  - \tfrac{\lang^2}{m^2} r^i  -  \tfrac{\nrmtup{r}^2}{m}  \cf  \pd_i U^\ss{1}. }\\
     Likewise, the equation for \eq{\pddt{r}\,^\en} is obtained from Eq.\eqref{d_ds_prj} as \eq{\pddt{r}^\en = \tfrac{r_\en^2}{m}  \pdt{\plin}_\en + 2 \tfrac{r^\en}{m} \plin_\en \pdt{r}^\en }. Substituting back in \eq{\pdt{r}^\en} and \eq{\pdt{\plin}_\en} leads to:\\
    \eq{ \pddt{r}\,^\en = \tfrac{r_\en^2}{m}  \pdt{\plin}_\en + 2 \tfrac{r^\en}{m} \plin_\en \pdt{r}^\en = \tfrac{r_\en^2}{m}  \pdt{\plin}_\en + 2 \tfrac{r_\en^3}{m^2} \plin_\en^2 = -\tfrac{r_\en^2}{m^2}( \tfrac{1}{r^\en}\lang^2 + 2 r_\en \plin_\en^2) - \tfrac{r_\en^2}{m}  \cf  \pd_\en(U^\zr+U^\ss{1}) + 2 \tfrac{r_\en^3}{m^2} \plin_\en^2 = -\tfrac{\lang^2}{m^2}r^\en - \tfrac{r_\en^2}{m}  \cf  \pd_\en(U^\zr+U^\ss{1}).}
}.
\item[] $\;$
\end{itemize}
\end{footnotesize}
\noindent   It should be noted that such second-order equations yield the (cartesian coordinate representation of) the \textit{base} integral curves of \eq{\cf\sfb{X}^\sscr{H}\in\vect(\cotsp\bvsp{E})} (and of \eq{\sfb{X}^\sscr{H}}) on \eq{\bvsp{E}=\bs{\Sigup}_\nozer\oplus\vsp{N}_\ii{+}}. However, they should not be considered a true representation of \eq{\cf\sfb{X}^\sscr{H}} itself. Being second-order, they are more properly seen as a representation of a vector field on the \textit{tangent} bundle (velocity phase space), \eq{\tsp\bvsp{E}}. Specifically, a \eq{g^\shrp}-related Hamiltonian vector field on the tangent bundle (velocity phase space), \eq{\bar{\cf}\bs{\Gamma}^\ss{E}= g^{\shrp}_*(\cf\sfb{X}^\sscr{H}) \in\vect(\tsp\bvsp{E})}, where \eq{\bs{\Gamma}^\ss{E} = g^{\shrp}_*\sfb{X}^\sscr{H} \in\vechm(\tsp\bvsp{E},\nbs{\varpi}^\sscr{L})}.
The Lagrangian/tangent bundle description is omitted from this already-lengthy work. 


Before examining the case \eq{U^\ss{1}=0} in more detail, we first consider another conformal factor which linearizes \eq{\sfb{X}^\sscr{H}}.

\subsubsection{Linearization Using a True-Anomaly-Like Evolution Parameter}

The above developments for the conformally Hamiltonian vector field \eq{\cf\sfb{X}^\sscr{H}} can be carried out in a similar manner, with similar results, for a new vector field, \eq{\cff \sfb{X}^\sscr{H}}, with  conformal factor \eq{\cff =  \cf  /\slang = 1/(\slang r_\en^2)\in\fun(\cotsp\bvsp{E})}. As per Eq.\eqref{dtds_def_prj}, this \eq{\cff}  defines an evolution parameter, \eq{\tau}, that is the
the solution to \eq{ \dot{\tau} = \diff{\tau}{t} = 1/\cff(\bar{\mu}_t)}
for integral curve \eq{\bar{\mu}_t} of \eq{\sfb{X}^\sscr{H}} (recall that, in the case the \textit{original} system has potential \eq{V^\zr=-\sck/\rfun}, then \eq{\tau} corresponds to the true anomaly along a \eq{\colift\psi}-related integral curve of the \textit{original} dynamics, \eq{\sfb{X}^\sscr{K}}).
 As before, we consider the conformally-Hamiltonian vector field, \eq{\cff \sfb{X}^\sscr{H}}, whose \eq{\tau}-parameterized integral curves are represented in cartesian coordinates \eq{(\tup{z},\tup{z}^\ii{\perp})=(\tup{r},\tup{\plin},r^\en,\plin_\en)} by solutions to the following ODEs
 (with \eq{\diff{\square}{\tau}=:\rng{\square}= \tfrac{1}{\slang} \pdt{\square} = \cff \dot{\square}}):
\begin{small}
\begin{align} \label{d_dta_prj}
\begin{array}{cccc}
      \cff \sfb{X}^\sscr{H}
\\[5pt]
    \cff := 1/(\slang r_\en^2) = m/(\lang r_\en^2)
\end{array}
\quad\left\{ \quad \begin{array}{lllllll}
     \rng{r}^i =  \cff\partial^i \mscr{H} 
    \,=\, -\tfrac{1}{\slang} \slang^{ij}r_j
     &,\qquad 
      \rng{\plin}_i = - \cff\pd_i \mscr{H} 
      \,=\, - \tfrac{1}{\slang} \slang_{ij} \plin^j  - \cff \pd_i U^\ss{1}
\\[4pt]
    \rng{r}^\en =  \cff\partial^\en \mscr{H} 
   \,=\, \tfrac{r_\en^2}{\lang}  \plin^\en
      &,\qquad  \rng{\plin}_\en = - \cff\pd_\en \mscr{H} 
      \,=\, - \tfrac{1}{\lang r_\en} ( \lang^2 + 2 r_\en^2 \plin_\en^2 ) - \cff \pd_\en (U^\zr + U^\ss{1})
\end{array} \right.
\end{align}
\end{small}
 where \eq{\tfrac{1}{\lang} \lang^{ij} = \tfrac{1}{\slang}\slang^{ij}}. The above dynamics of \eq{\cff\sfb{X}^\sscr{H}} may also be represented using the modified coordinates \eq{(\tup{z},\tiltup{z}^\ii{\perp})=(\tup{r},\tup{\plin},r^\en,\tilplin_\en)}, with  \eq{\tilplin_\en :=r_\en^2 \plin_\en} from Eq.\eqref{newpn_def0}.  The same reasoning that lead to Eq.\eqref{d_ds_prj_altCords} now leads to:
\begin{small}
\begin{flalign} \label{d_dta_prj_altCords}
\qquad\qquad\qquad
 \begin{array}{llllllll}
         \rng{r}^i 
         \,=\,    -\tfrac{1}{\lang} \lang^{ij}r_j
         &,\qquad 
         \rng{\plin}_i 
          \,=\,  - \tfrac{1}{\lang} \lang_{ij} \plin^j  -  \cff \pd_i U^\ss{1}
 \\[3pt]
        \rng{r}^\en 
       \,=\, \tfrac{1}{\lang} \tilplin^\en
         &,\qquad   \rng{\tilplin}_\en 
          \,=\, - \lang r_\en  -   \tfrac{1}{\slang} \pd_\en (U^\zr + U^\ss{1}) 
\end{array} 
&&
 \left|\qquad \begin{array}{rllll}
      \fnsize{for} & U^\zr = -\sck_1 r_\en - \ttfrac{1}{2}\sck_2 r_\en^2
 \\[3pt]
    \Rightarrow  & \rng{\tilplin}_\en
      \,=\, - \tfrac{m}{\slang} \beta^2 r_\en  +   \tfrac{\sck_1}{\slang} -  \tfrac{1}{\slang} \pd_\en U^\ss{1} 
\end{array} \right. \quad
\end{flalign}
\end{small}
where \eq{U^\zr} appears only in the equation for \eq{\rng{\tilplin}_\en}, which is given above for arbitrary  \eq{U^\zr} and for the particular form of \eq{U^\zr} from Eq.\eqref{V0U0_form}. For the latter, \eq{\beta^2 := \slang^2 - \txkbar_2} with \eq{\txkbar_{1,2}:=\sck_{1,2}/m} (for the Kepler potential, \eq{\sck_2=0} and \eq{\beta^2 = \slang^2}). 
\begin{small}
\begin{notesq}
\rmsb{simplifications on $\cotsp\man{Q}_\ii{1}$.} For integral curves on $\cotsp\man{Q}_\ii{1}\subset\cotsp\bvsp{E}$, the same reasoning as for Eq.\eqref{d_ds_T*S} still applies, leading to:
    \begin{small}
    \begin{align} \label{d_dta_T*S}
         \lang^2 = m^2\slang^2 \simeq \nrmtup{\plin}^2 
         \qquad,\qquad 
         \rng{r}^i \simeq \tfrac{1}{\lang} \plin^i 
         \simeq \hat{\plin}^i
         \qquad,\qquad 
         \rng{\plin}_i \simeq -\tfrac{1}{\lang} \nrmtup{\plin}^2 r_i - \cff \pd_i U^\ss{1} \,\simeq\, - \nrmtup{\plin} r_i - \cff \pd_i U^\ss{1}
    \end{align}
    \end{small}
\end{notesq}
\end{small}

\paragraph{Second-Order ODEs ``Equivalent'' to $\cff\sfb{X}^\sscr{H}$.}
All of the previous remarks about linearity of \eq{\cf \sfb{X}^\sscr{H}} apply similarly to \eq{\cff \sfb{X}^\sscr{H}}. However, the equivalent second order coordinate ODEs for the above \eq{\cff\sfb{X}^\sscr{H}} (which are not a direct representation of \eq{\cff \sfb{X}^\sscr{H}} itself) are slightly, but notably, different from those for \eq{\cf\sfb{X}^\sscr{H}} (given previously in Eq.\eqref{dd_ds_prj} and Eq.\eqref{ddu_ds_prj}). 
The second-order ODEs for \eq{r^\a}, with \eq{\tau} as the evolution parameter may be obtained from Eq.\eqref{d_dta_prj_altCords} as follows:
\begin{small}
\begin{align} \label{dd_dta_prj}
 & \begin{array}{llllll}
             \rrng{r}\,^i   +  r^i \,=\,   - \tfrac{m \rfun^2}{\lang^2 r_\en^2}  \big( \emet^{ik}  + \tfrac{1}{\lang^2} \lang^{ij}r_j  \plin^k \big) \pd_k U^\ss{1} 
             \;=\;
             - \tfrac{m \rfun^2}{\lang^2 r_\en^2 } \big( \emet^{ik} - \tfrac{\rfun^2}{\lang^2} \plin^i \plin^k + \tfrac{\tup{r} \cdot \tup{\plin}}{\lang^2}  r^i \plin^k \big) \pd_k U^\ss{1} 
      \\[6pt]
            \rrng{r}\,^\en 
            + r^\en  + \tfrac{m}{\lang^2} \emet^{\en\en}\pd_\en U^\zr 
            \,=\, - \tfrac{m}{\lang^2}\big(  \emet^{\en\en}\pd_\en U^\ss{1} - \tfrac{\rfun^2}{r_\en^2 \lang^2} \tilplin^\en  \plin^j \pd_j U^\ss{1} \big)
            \,=\, -  \tfrac{m}{\lang^2}\big(  \emet^{\en\en}\pd_\en U^\ss{1} - \tfrac{\rfun^2}{\lang^2} \plin^\en  \plin^j \pd_j U^\ss{1} \big)
   \end{array} 
&&,&&
    \rng{\lang} \,=\, -\tfrac{m r^2}{\lang^2  r_\en^2} \plin^i \pd_i U^\ss{1}
\end{align}
\end{small}
In the case that \eq{U^\ss{1}=0}, the above equation for \eq{r^i} is again that of a linear \eq{(\en-1)}-dim harmonic oscillator but, unlike before, it has \textit{unit} natural frequency (rather than \eq{\slang}). This holds for any arbitrary \eq{U^\zr(r_\en)}. The equation for \eq{r^\en} again involves \eq{U^\zr} and,
for the case that \eq{U^\zr} is  given as in Eq.\eqref{V0U0_form}, the above leads to:
\begin{small}
\begin{align} \label{ddu_dta_prj}
  \left(\begin{array}{cc}
          \fnsize{for} \;\;  U^\zr = - \sck_1 r_\en - \ttfrac{1}{2}\sck_2 r_\en^2 
          \\[2pt]
        \fnsize{i.e.,} \;\;  V^\zr = -\sck_1 / \rfun - \ttfrac{1}{2} \sck_2/ \rfun^2 
    \end{array}\right)
   \quad \Rightarrow  \quad
  \boxed{ \begin{array}{llllll}  
      \rrng{r}\,^i   \,+\,  r^i \,=\,   - \tfrac{m\rfun^2}{\lang^2 r_\en^2}  \big( \emet^{ik}  + \tfrac{1}{\lang^2} \lang^{ij}r_j  \plin^k \big) \pd_k U^\ss{1} 
   \\[4pt]
        \rrng{r}\,^\en  + \tfrac{\beta^2}{\slang^2} r^\en - \tfrac{\txkbar_1}{\slang^2} \,=\, - \tfrac{m}{\lang^2}  \big(\emet^{\en\en}  \pd_\en U^\ss{1} - \tfrac{r^2}{\lang^2} \plin^\en \plin^j \pd_j U^\ss{1} \big) 
  \end{array}  }  
&&,&&
   \begin{array}{llllll}
      \beta^2/\slang^2 = 1 - \txkbar_2/\slang^2 \\
      \txkbar_{1,2} := \sck_{1,2}/m 
 \end{array}
\end{align}
\end{small}
Note for the Kepler problem (\eq{\sck_2=0}) that \eq{\beta^2/\slang^2=1} and the above all have the same unit frequency.
\begin{footnotesize}
\begin{itemize}[nosep]
    \item  \textit{Derivation of Eq.\eqref{dd_dta_prj}.} 
    The equation for \eq{\rrng{r}\,^i} is found by direct differentiation of Eq.\eqref{d_dta_prj_altCords}, leading to: 
    \begin{align} \nonumber
    \begin{array}{lllll}
         \rrng{r}\,^i = \lderiv{\cff \sfb{X}^\sscr{H}}^2 r^i \,=\,  \diff{}{\tau}(-\tfrac{1}{\lang}\lang^{ij}r_j) 
         &\!\!\!\! =\,
         -\tfrac{1}{\lang}\lang^{ij}\rng{r}_j -  \tfrac{1}{\lang}(\rng{\lang}^{ij} - \tfrac{1}{\lang}\rng{\lang}\lang^{ij})r_j
         \,=\,
         \tfrac{1}{\lang^2}\lang^{ij}\lang_{jk}r^k
         - \tfrac{\cff}{\lang} \big( -( r^i\emet^{jk}-r^j\emet^{ik})\pd_k U^\ss{1}
         + \tfrac{\rfun^2}{\lang^2}  \plin^k \pd_k U^\ss{1} \lang^{ij}\big) r_j
     \\[3pt]
         &\!\!\!\! =\, - r^i \,-\, \tfrac{\cff}{\lang} \big( \rfun^2 \emet^{ik} -  r^ir^k + \tfrac{\rfun^2}{\lang^2} \lang^{ij}r_j  \plin^k \big) \pd_k U^\ss{1} 
        \;=\;  - r^i \,-\, \tfrac{\cff \rfun^2}{\lang} \big( \emet^{ik}  + \tfrac{1}{\lang^2} \lang^{ij}r_j  \plin^k \big) \pd_k U^\ss{1} 
      \\[3pt]
         &\!\!\!\!  =\,  
          - r^i \,-\,  \tfrac{\cff \rfun^2}{\lang} \big( \emet^{ik} - \tfrac{\rfun^2}{\lang^2} \plin^i \plin^k + \tfrac{(\tup{r} \cdot \tup{\plin})}{\lang^2}  r^i \plin^k \big) \pd_k U^\ss{1} 
    \end{array}
    \end{align}
    where \eq{\cff=1/(\slang r_\en^2)=m/(\lang r_\en^2)} and  where \eq{\rng{\lang}^{ij} = \cff \pbrak{\lang^{ij}}{\mscr{H}}} and \eq{\rng{\lang} = \cff \pbrak{\lang}{\mscr{H}}} are obtained from Eq.\eqref{prj_ioms_rev}. Note the above has used \eq{r^k \pd_k U^\ss{1}=0} (cf., section \ref{sec:prj_Xform_gen}) along with \eq{\lang^{ik}r_k = -(\nrmtup{r}^2 \plin^i -(\tup{r} \cdot \tup{\plin})r^i)}. 
    Next, \eq{\rrng{r}\,^\en} is similarly obtained from Eq.\eqref{d_dta_prj_altCords} as:
    \begin{align} \nonumber 
    \begin{array}{lllll}
         \rrng{r}\,^\en  = \lderiv{\cff \sfb{X}^\sscr{H}}^2 r^\en \,=\, 
         \diff{}{\tau}(\tfrac{1}{\lang}\tilplin^\en) 
          &\!\!\!\! =\, \tfrac{1}{\lang}\rng{\tilplin}^\en - \tfrac{1}{\lang^2}\rng{\lang} \tilplin^\en 
         \,=\;
          - r^\en  -   \tfrac{m}{\lang^2} \emet^{\en\en} \pd_\en (U^\zr + U^\ss{1})
          + \tfrac{\cff \rfun^2}{\lang^3} \tilplin^\en  \plin^j \pd_j U^\ss{1} 
          \,=\,
           - r^\en  -  \tfrac{m}{\lang^2} \big(  \emet^{\en\en} \pd_\en (U^\zr + U^\ss{1}) - \tfrac{\rfun^2}{r_\en^2 \lang^2} \tilplin^\en  \plin^j \pd_j U^\ss{1} \big)
    \end{array}
    \end{align}
    substitution of  \eq{\tilplin^\en=r_\en^2 \plin^\en} then leads to \eq{\rrng{r}\,^\en} as seen in Eq.\eqref{dd_dta_prj}. 
\end{itemize}
\end{footnotesize}

\begin{small}
\begin{notesq}
\rmsb{simplifications on $\cotsp\man{Q}_\ii{1}$.} For integral curves on $\cotsp\man{Q}_\ii{1}\subset\cotsp\bvsp{E}$, we may simplify using Eq.\eqref{T*Q_reg_rev}; with \eq{\tfrac{\beta^2}{\slang^2} \simeq 1- \tfrac{m\sck_2}{\nrmtup{\plin}^2}}, and \eq{\tfrac{\txkbar_1}{\slang^2} \simeq \tfrac{m\sck_1}{\nrmtup{\plin}^2}}:
    \begin{small}
    \begin{align}
    \begin{array}{llllll}
         \lang  \simeq \nrmtup{\plin}
         &,\;\;   \rfun \simeq 1  \\
         \plin^i/\lang \simeq \hat{\plin}^i
        &,\;\;   \tup{r} \cdot \tup{\plin} \simeq 0 
    \end{array}
         \qquad \Rightarrow \qquad 
     \begin{array}{llllll}
         \rrng{r}\,^i   +  r^i \,\simeq\,   - \tfrac{m}{\nrmtup{\plin}^2 r_\en^2}  ( \emet^{ik}  -\hat{\plin}^i \hat{\plin}^k ) \pd_k U^\ss{1} 
         &\!\!\! \simeq \,  - \tfrac{m}{\nrmtup{\plin}^2 r_\en^2}  ( \emet^{ik}  -\rng{r}^i \rng{r}^k ) \pd_k U^\ss{1} 
    \\[3pt]
        \rrng{r}\,^\en  + \tfrac{\beta^2}{\slang^2} r^\en - \tfrac{\txkbar_1}{\slang^2} \,\simeq\, - \tfrac{m}{\nrmtup{\plin}^2}  \big(\emet^{\en\en}  \pd_\en U^\ss{1} - \tfrac{1}{\nrmtup{\plin}} \plin^\en \hat{\plin}^j \pd_j U^\ss{1} \big) 
        &\!\!\! \simeq \,  - \tfrac{m}{\nrmtup{\plin}^2 r_\en^2}  ( r_\en^2 \emet^{\en\en}  \pd_\en U^\ss{1} - \rng{r}^\en \rng{r}^j \pd_j U^\ss{1} ) 
    \end{array}
    \end{align}
    \end{small}
\end{notesq}
\end{small}

\subsection{Closed-Form Solutions} \label{sec:prj_sols}

We now consider more closely the case that \eq{U^\ss{1}=0} (corresponding to \eq{V^\ss{1}=0} for the \textit{original} system). That is, the original dynamics, \eq{\sfb{X}^\sscr{K}}, include central/radial forces and, therefore, the \eq{\colift\psi}-transformed dynamics, \eq{\sfb{X}^\sscr{H}=\colift\psi^*\sfb{X}^\sscr{K}}, include only normal forces (where ``normal'' means normal to \eq{\bs{\Sigup}_\nozer\subset \bvsp{E}}).
When the particular form of these forces matter, we will consider the case that \eq{U^\zr} has the form seen in Eq.\eqref{V0U0_form}, meaning  the original potential, \eq{V}, and transformed potential \eq{U=\psi^* V} are as follows:
\begin{small}
\begin{align} \label{V0U0_yetagain}
\fnsize{for} \; U = \psi^* V:
\qquad\quad 
\begin{array}{lllll}
    V =  V^\zr = -\tfrac{\sck_1}{\rfun} -  \tfrac{\sck_2}{2\rfun^2}
 \\[3pt]
    - \dif V^\zr =  -(\tfrac{\sck_1}{\rfun^2} +  \tfrac{\sck_2}{\rfun^3}) \dif \rfun
\end{array}
\qquad \Leftrightarrow \qquad
\begin{array}{lllll}
   U =   U^\zr = -\sck_1 r_\en - \ttfrac{1}{2}\sck_2 r_\en^2 
 \\[3pt]
     -\dif U^\zr = (\sck_1 +  \sck_2 r_\en) \dif r^\en  
\end{array}
&& \phantom{X}
\end{align}
\end{small}
where \eq{\dif r \equiv \hsfb{r}^\flt } takes values in \eq{\bs{\Sigup}} and \eq{\dif r^\en \equiv \enform \perp \bs{\Sigup}}.
The Kepler potential is a special case of the above corresponding to \eq{\sck_2=0} and \eq{\sck_1>0} (the Coulomb potential further permits \eq{\sck_1<0}); this is summarized in section \ref{sec:prj_KEPgeomech}.


\subsubsection{Summary of Unperturbed Dynamics \& Solutions}
 We summarize the preceding develops, and  and give closed-form solutions in terms of \eq{s} and \eq{\tau}, for the special case \eq{U^\ss{1}=0} and \eq{U^\zr} is given by Eq.\eqref{V0U0_yetagain}. 
 Here, we will simply sate, without derivation, the key equations of motion, solutions, and transformations need to use the projective regularization scheme. The derivations and details are given in section \ref{sec:prj_sols_derivation}.

 With \eq{U^\ss{1}=0}, the transformed Hamiltonian \eq{\mscr{H}=\colift\psi^*\mscr{K}\in\fun(\cotsp\bvsp{E})} is then given as follows, along with some Poisson brackets 
(which follow from Eq.\eqref{prj_ioms_rev}):\footnote{Also, as usual, \eq{\pbrak{\mscr{H}}{\mscr{H}}=0} such that \eq{\mscr{H}} is also an integral of motion iff \eq{\pd_t \mscr{H}=0} (which is true for the present case since \eq{\pd_t U^\zr =0}).}
\begin{small}
\begin{gather} \label{Hprj_U0}
\begin{array}{cccc}
      \mscr{H}  \,=\,
   \tfrac{r_\en^2}{2m} ( \lang^2 + r_\en^2 \plin_\en^2 )  +  U^\zr
   \;=\;  \tfrac{1}{2m} ( r_\en^2 \lang^2 + \tilplin_\en^2 )  +  U^\zr 
\\[5pt]
    \pbrak{\rfun^2 }{\mscr{H}} = \pbrak{ \nrmtup{\plin}^2 }{\mscr{H}} = \pbrak{\hat{r}^i \plin_i }{\mscr{H}}  = \pbrak{r^i \plin_i }{\mscr{H}}  
     =  \pbrak{ \lang^{ij} }{\mscr{H}} 
     =  \pbrak{ \lang^2 }{\mscr{H}}
    = 0
\end{array}
\end{gather}
\end{small}
where \eq{ \lang^2 = \lang^{ij}r_i \plin_j = \nrmtup{r}^2 \nrmtup{\plin}^2 - (\tup{r} \cdot \tup{\plin})^2 } and where 
\eq{\tilplin_\en:=r_\en^2 \plin_\en} is the quasi-momenta coordinate defined in Eq.\eqref{newpn_def0}; the pair   \eq{(r^\en,\tilplin_\en)} is not symplectic/canonical, yet, the dynamics are ``nicer'' when expressed using \eq{\tiltup{z}^\ii{\perp}=(r^\en,\tilplin_\en)} rather than \eq{\tup{z}^\ii{\perp}=(r^\en,\plin_\en)}. 
The ODEs corresponding to the conformally-Hamiltonian vector fields \eq{\cf\sfb{X}^\sscr{H}} and \eq{\cff\sfb{X}^\sscr{H}} are then expressed in terms of  \eq{(\tup{z},\tiltup{z}^\ii{\perp})=(\tup{r},\tup{\plin},r^\en,\tilplin_\en)} as follows 
(note \eq{U^\zr} does not affect the dynamics on \eq{\cotsp\bs{\Sigup}_\nozer \subset \cotsp\bvsp{E}}, which are now purely kinematic):
\begin{small}
\begin{align} \label{d_ds_unpert}
  \begin{array}{cccc}
        \cf\sfb{X}^\sscr{H}
\\[2pt]
      \cf  := 1/r_\en^2  
\end{array}\;
&\left\{ \;\; \begin{array}{lllllll}
     \pdt{r}^i \,=\,   \cf  \partial^i \mscr{H} 
     &\!\!\!\! =\,    -  \slang^{ij} r_j
\\[4pt]
       \pdt{\plin}_i = -  \cf  \pd_i \mscr{H} 
      &\!\!\!\! =\, - \slang_{ij} \plin^j  
\end{array}\right. 
\quad\;, \quad 
\begin{array}{lllll}
    \pdt{r}^\en  \,=\;     \partial^{\til{\en}} \mscr{H} 
   &\!\!\!\! =\, \tfrac{1}{m} \tilplin^\en
\\[4pt]
       \pdt{\tilplin}_\en = -  \pd_\en \mscr{H} 
      &\!\!\!\! =\,  - m\slang^2 r_\en  -   \pd_\en U^\zr  
      &\!\!\!
      =\, -m \beta^2 r_\en  + \sck_1
\end{array} \quad
\\[6pt] \label{d_dta_unpert}
\begin{array}{cccc}
        \cff\sfb{X}^\sscr{H}
\\[2pt]
      \cff  := 1/(\slang r_\en^2) 
\end{array}\; 
&\left\{ \;\; \begin{array}{lllllll}
     \rng{r}^i \,=\,  \cff\partial^i \mscr{H} 
     &\!\!\!\! =\, -\tfrac{1}{\slang} \slang^{ij}r_j 
\\[4pt]
       \rng{\plin}_i = - \cff\pd_i \mscr{H} 
      &\!\!\!\! =\, - \tfrac{1}{\slang} \slang_{ij} \plin^j 
\end{array}\right. 
\quad, \quad 
\begin{array}{lllll}
    \rng{r}^\en \,=\;  \tfrac{1}{\slang} \partial^{\til{\en}} \mscr{H} 
   &\!\!\!\! =\, \tfrac{1}{\lang}  \tilplin^\en 
\\[4pt]
       \rng{\tilplin}_\en = - \tfrac{1}{\slang}  \pd_\en \mscr{H} 
      &\!\!\!\! =\, -  \lang r_\en  -   \tfrac{1}{\slang}\pd_\en U^\zr  
      &\!\!\!
      =\, -\tfrac{m}{\slang} \beta^2 r_\en  + \tfrac{1}{\slang}\sck_1
\end{array} 
\end{align}
\end{small}
where \eq{\slang=\lang/m}, where \eq{\pd_\en U^\zr = -\sck_1 - \sck_2 r_\en}, and \eq{\beta^2:=\slang^2-\txkbar_2} (we will assume \eq{\slang^2\geq \txkbar_2} such that \eq{\beta\in\mbb{R}}).
The usual, conformally scaled, canonical equations of motion for the symplectic coordinate set \eq{(\tup{z},\tup{z}^\ii{\perp})=(\tup{r},\tup{\plin},r^\en,\plin_\en)} differ only in the equations for \eq{\tup{z}^\ii{\perp}=(r^\en,\plin_\en)}, given previously and again in the footnote\footnote{With \eq{\pd_\en U^\zr = -\sck_1 - \sck_2 r_\en}, Eq.\eqref{d_ds_prj} and Eq.\eqref{d_dta_prj} lead to the following (with \eq{\beta^2:=\slang^2-\txkbar_2}): 
\begin{align}
\begin{array}{rlllll}
   \cf\sfb{X}^\sscr{H}:  &
   \pdt{r}^\en =\,   \cf  \partial^\en \mscr{H} 
   \,=\, \tfrac{r_\en^2}{m} \plin^\en
 &,\qquad 
       \pdt{\plin}_\en = -  \cf  \pd_\en \mscr{H} 
      =  - \tfrac{1}{m r_\en} (\lang^2 + 2 r_\en^2 \plin_\en^2 ) -  \cf  \pd_\en U^\zr 
      &\!\!\!\! =\,  -m\beta^2 r_\en^\ii{-1} + \sck_1 r_\en^\ss{-2} -   \tfrac{2}{m}\plin_\en^2 r_\en
\\[4pt]
   \cff\sfb{X}^\sscr{H}: & 
    \rng{r}^\en =  \cff\partial^\en \mscr{H} 
   \,=\, \tfrac{r_\en^2}{\lang}  \plin^\en
 &,\qquad 
       \rng{\plin}_\en = - \cff \pd_\en \mscr{H} 
      \,=\, - \tfrac{1}{\lang r_\en} (\lang^2 + 2 r_\en^2 \plin_\en^2 ) - \cff \pd_\en U^\zr 
      &\!\!\!\! =\,  -\tfrac{m}{\slang}\beta^2 r_\en^\ii{-1} + \tfrac{1}{\slang}\sck_1 r_\en^\ss{-2} -   \tfrac{2}{\lang}\plin_\en^2 r_\en
\end{array}
\end{align}
}. 
As a non-Hamiltonian alternative to the above, the \eq{(\tup{r},r^\en)}-representation of the base integral curves on \eq{\bvsp{E}=\bs{\Sigup}_\nozer\oplus\vsp{N}_\ii{+}} satisfy the following second order ODEs:
\begin{small}
\begin{align}
\phantom{XXXXXX}
\begin{array}{rllllll}
     \cf\sfb{X}^\sscr{H} :\;\; 
     &\quad\pddt{r}\,^i 
         \,+\, \slang^2 r^i = 0
  &\;\;,\qquad 
        \pddt{r}\,^\en  \,+\, \beta^2 r^\en  =  \txkbar_1
\\[5pt]
      \cff\sfb{X}^\sscr{H} :\;\;   
       &\quad \rrng{r}\,^i   \,+\,  r^i \,=\,  0
  &\;\;,\qquad 
        \rrng{r}\,^\en  +  \tfrac{1}{\slang^2}\beta^2 r^\en \,=\, \tfrac{1}{\slang^2} \txkbar_1
\end{array} 
&&
 \fnsz{\begin{array}{llllll}
      \beta^2 := \slang^2 - \txkbar_2 \\[2pt]
      \txkbar_{1,2} := \sck_{1,2}/m 
 \end{array}}
\end{align}
\end{small}
Now, the dynamics in Eq.\eqref{d_ds_unpert} and Eq.\eqref{d_dta_unpert}  have the same integrals of motion as \eq{\sfb{X}^\sscr{H}} seen in Eq.\eqref{Hprj_U0}.\footnote{\eq{\sfb{X}^\sscr{H}}, \eq{\cf\sfb{X}^\sscr{H}}, and \eq{\cff\sfb{X}^\sscr{H}} have the same integrals of motion.} 
In particular, since \eq{\slang=\lang/m} is an integral of motion, then the evolution parameters \eq{s} and \eq{\tau} are related simply by
\begin{small}
\begin{align}
    \mrm{d} \tau =  \slang \mrm{d} s 
    \qquad
   \xRightarrow{ U^\ss{1}=0}  \qquad 
    \tau = \slang s \;\; \leftrightarrow \;\; s = \tau/ \slang 
\end{align}
\end{small}
where the above \eq{\slang} is really the constant value, \eq{\slang_\iio\in\mbb{R}}, along some integral curve in question.\footnote{That is, that the relation is more specifically \eq{\tau = \slang(\bar{\mu}_s) s} where \eq{\bar{\mu}_s} is an integral curve of \eq{\cf\sfb{X}^\sscr{H}}. Equivalently, \eq{s=\tau/\slang(\bar{\mu}_\tau)} where \eq{\bar{\mu}_\tau} is an integral curve of \eq{\cff\sfb{X}^\sscr{H}} (\eq{\bar{\mu}_\tau} and \eq{\bar{\mu}_s} are the same curve, just re-parameterized). We will assume this to be implicitly understood.}
Thus, when \eq{U^\ss{1}=0}, any integral curve (solution) parameterized by \eq{s} is easily re-parameterized in terms \eq{\tau}, and vice versa, using the above. 
These integral curves are represented in coordinates \eq{(\tup{z},\tiltup{z}^\ii{\perp})=(\tup{r},\tup{\plin},r^\en,\tilplin_\en)} by solutions to the ODEs in Eq.\eqref{d_ds_unpert} and Eq.\eqref{d_dta_unpert} (they have the same solutions, up to parameterization by \eq{s} or \eq{\tau}). 
It will will be shown that, for some fixed value \eq{\lang^{ij}_\iio\in\mbb{R}}, these coordinate solutions are given as follows for any initial conditions \eq{(\tup{z}_\zr,\tiltup{z}^\ii{\perp}_\zr)} satisfying \eq{\lang^{ij}(\tup{z}_\zr)=\lang^{ij}_\iio}:
\begin{small}
\begin{align} \label{prj_sols_gen}
\boxed{\begin{array}{rlllllllll}
    \tup{r}_s &\!\!\!\!\equiv\,  \tup{r}_\tau =\,
     \tup{r}_\zr \cos \tau - \tfrac{1}{\lang_\iio} L_\iio \tup{r}_\zr \sin  \tau 
     &,
\\[5pt]
     \tup{\plin}_s &\!\!\!\!\equiv\, \tup{\plin}_\tau =\, \tup{\plin}_\zr \cos \tau - \tfrac{1}{\lang_\iio} \til{\Lang}_\iio \tup{\plin}_\zr \sin \tau 
      &,
\end{array}
\quad
\begin{array}{rlllllllll}
      r^\en_s \,\equiv\, r^\en_\tau  
      \,= &\!\!\!\!  r^\en_\zr \cos{\varep} + \tfrac{1}{m\beta_\iio} \tilplin^\en_\zr \sin{\varep}  + \tfrac{\txkbar_1}{\beta_\iio^2}(1-\cos{\varep})   
\\[5pt]
     \tilplin_{\en_s} \equiv\, \tilplin_{\en_\tau}
      = &\!\!\!\!\!  - m \beta_\iio r_{\en_\zr}  \sin{\varep} +  \tilplin_{\en_\zr} \cos{\varep} + \tfrac{\sck_1}{\beta_\iio} \sin{\varep}
\end{array} }
&&
\begin{array}{llll}
     \tau = \slang_\iio s  \\
     \varep :=  \beta_\iio s = \tfrac{\beta_\iio}{\slang_\iio} \tau
     \\[3pt]
     L:= [\lang^{ik}\emet_{kj}] \\
     \til{\Lang}:= [\lang_{ik}\emet^{kj}]
\end{array} 
\end{align}
\end{small}
 where \eq{\varep:=  \beta_\iio s} is defined merely for brevity. For the Kepler problem (\eq{\sck_2=0}), we note \eq{\beta=\slang} such that  \eq{\varep = \tau}: 
\begin{small}
\begin{align}
\fnsize{Kepler:}
\qquad \sck_2 = 0
\qquad \Rightarrow \qquad  
\beta = \slang
 \;\;\;, \;\;\; 
 \varep = \tau = \slang_\iio s
\end{align}
\end{small}
where the form of \eq{U^\zr} only affects the dynamics on \eq{\cotsp\vsp{N}_\ii{+}} (that is, the ``\eq{(r^\en,\plin_\en)}-part'') for which we recall that the  
the solutions for the symplectic pair \eq{(r^\en,\plin_\en)} are easily recovered from any solutions for \eq{(r^\en,\tilplin_\en)} simply by  substitution of: 
\begin{small}
\begin{align}
    \plin_\en = \tilplin_\en/r_\en^2 \quad \leftrightarrow \quad  \tilplin_\en := r_\en^2 \plin_\en
\end{align}
\end{small}

\begin{small}
\begin{notesq}
    \rmsb{simplifications on $\cotsp\man{Q}_\ii{1}$.} For integral curves of \eq{\sfb{X}^\sscr{H}} (or any conformal scaling \eq{\sfb{X}^\sscr{H}}) that lie in the invariant submanifold $\cotsp\man{Q}_\ii{1}\subset\cotsp\bvsp{E}$, the dynamics may be simplified using  Eq.\eqref{T*Q_reg_rev}. In this case, the coordinate ODEs for \eq{(r^i,\plin_i)} may be simplified as in Eq.\eqref{d_ds_T*S} or Eq.\eqref{d_dta_T*S} and the solutions (for \eq{U^\ss{1}=0}) are still as in  Eq.\eqref{prj_sols_gen} above, which now simplify to:  
    \begin{small}
    \begin{align} \label{prj_sols_gen_T*Q}
    \begin{array}{lll}
         \fnsize{with } s:&  \pdt{r}^i \simeq \tfrac{1}{m} \plin^i
         &,\quad 
         \pdt{\plin}_i \simeq -\tfrac{1}{m} \nrmtup{\plin}^2 r_i \simeq - m\slang^2 r_i 
     \\[3pt]
     \fnsize{with } \tau:& 
      \rng{r}^i \simeq \tfrac{1}{\lang} \plin^i
         &,\quad 
         \rng{\plin}_i \simeq -\tfrac{1}{\lang} \nrmtup{\plin}^2 r_i \simeq - \lang r_i 
    \end{array}
    && \Rightarrow && 
    \begin{array}{rlllllllll}
        \tup{r}_s &\!\!\!\!\equiv\,  \tup{r}_\tau \simeq\,
         \tup{r}_\zr \cos \tau \,+\, \tfrac{1}{\lang_\iio}\tup{\plin}_\zr^{\shrp} \sin  \tau 
         &\!\!\!\!
         \simeq\; \tup{r}_\zr \cos \tau \,+\, \tfrac{1}{\nrmtup{\plin}_\zr}\tup{\plin}_\zr^{\shrp} \sin  \tau
    \\[3pt]
         \tup{\plin}_s &\!\!\!\!\equiv\, \tup{\plin}_\tau \simeq\, \tup{\plin}_\zr \cos \tau - \tfrac{\nrmtup{\plin}^2_\zr}{\lang_\iio}  \tup{r}_\zr^{\flt} \sin \tau 
         &\!\!\!\!
         \simeq\; \tup{\plin}_\zr \cos \tau - \nrmtup{\plin}_\zr \tup{r}_\zr^{\flt} \sin \tau 
    \end{array}
    \end{align}
    \end{small}
\end{notesq}
\end{small}

\paragraph{Recovering Solutions for the Original System.}  Although this was addressed previously in section \ref{sec:prj_geomech}, it is  reproduced here for ease of reference. 
Recall that the integrals curves (solutions) for the original dynamics,  \eq{\sfb{X}^\sscr{K}},  are recovered from those of the transformed dynamics, \eq{\sfb{X}^\sscr{H}}, using the cotangent lift of the projective transformation, \eq{\colift\psi\in\Spism(\cotsp\bvsp{E},\nbs{\omg})}:
\begin{small}
\begin{align} \label{intcurv_xform_prj_reg}
   \fnsize{for:}\quad \dt{\bar{\kappa}}_{\barpt{x}} = \sfb{X}^\sscr{K}_{\bar{\kappa}_{\barpt{x}}} 
    \quad \fnsize{\&} \quad 
    \dt{\bar{\mu}}_{\barpt{q}} = \sfb{X}^\sscr{H}_{\bar{\mu}_{\barpt{q}}}
    \;: 
    &&
    \begin{array}{rccclll}
          &\bar{\kappa}_{\barpt{x}} = \colift\psi(\bar{\mu}_{\barpt{q}}) 
         =  ( \psi(\barpt{q}), \psi_* \barbs{\mu}) 
          & \leftrightarrow &
          \bar{\mu}_{\barpt{q}} = \inv{\colift\psi}(\bar{\kappa}_{\barpt{x}})
         = ( \inv{\psi}(\barpt{x}), \psi^* \barbs{\kappa} )
      \\[4pt]
        \fnsize{with:} &\barbs{\kappa}=\sfb{m}(\dt{\bsfb{x}}) = m \dt{\bsfb{x}}^{\flt}
         & \leftrightarrow &
         \barbs{\mu}  =\sfg_\ss{\!\barpt{q}}(\dt{\bsfb{q}})
    \end{array}
\end{align}
\end{small}
where \eq{\bar{\mu}_{\barpt{q}_t}\in\cotsp\bvsp{E}} is an integral curve of \eq{\sfb{X}^\sscr{H}} 
and \eq{\bar{\kappa}_{\barpt{x}_t}:=\colift\psi(\bar{\mu}_{\barpt{q}}) \in\cotsp\bvsp{E} } is an integral curve of \eq{\sfb{X}^\sscr{K}}. 
As previously shown, the above transformations for \eq{\barpt{x}=\psi(\barpt{q})\leftrightarrow \barpt{q}=\inv{\psi}(\barpt{x})} and \eq{\barbs{\kappa}=\psi_* \barbs{\mu} \leftrightarrow \barbs{\mu}= \psi^* \barbs{\kappa} }
are given explicitly as follows (separating the \eq{\en^{\tx{th}}} components): 
\begin{small}
\begin{align} \label{colift_prj_reg}
 \begin{array}{rlcllll}
       \ptvec{x} =   \tfrac{1}{q^\en}\hpt{q} 
       &,\quad x^\en =  \nrm{\pt{q}}
\\[5pt]
     \bs{\kappa} 
    =  q^\en \nrm{\pt{q}} \big( \trn{\iden}_\ii{\!\Sig}-   \hsfb{q}^\flt \otms \hsfb{q} \big) \cdot \bs{\mu} - q_\en^2 \mu_\en \hsfb{q}^{\flt} 
     &,\quad \kappa_\en = \bs{\mu}\cdot \hsfb{q} 
\end{array} 
\qquad \leftrightarrow \qquad 
\begin{array}{lllll}
     \ptvec{q} =  x^\en \hpt{x} \quad,\quad q^\en = \tfrac{1}{\nrm{\pt{x}}} 
\\[5pt]
     \bs{\mu}  
      =  \tfrac{\nrm{\pt{x}}}{x^\en} \big( \trn{\iden}_\ii{\!\Sig} -   \hsfb{x}^\flt \otms \hsfb{x} \big) \cdot \bs{\kappa} + \kappa_\en \hsfb{x}^\flt
     \quad,\quad \mu_\en = 
     -\nrm{\pt{x}}^2  \bs{\kappa}\cdot \hsfb{x}  
\end{array}  
\end{align}
\end{small}
Note \eq{\barbs{\kappa} \leftrightarrow \barbs{\mu}} may be expressed using \eq{\til{\mu}_\en := \tilplin_\en(\bar{\mu}_{\barpt{q}})=q_\en^2 \mu_\en = \mu_\en/ \nrm{\pt{x}}^2}, rather than the cartesian component \eq{\mu_\en := \plin_\en(\bar{\mu}_{\barpt{q}})}, as:
\begin{small}
\begin{align} \label{colift_prj_reg_pnew}
\begin{array}{llllll}
     \bs{\kappa} 
     =  q^\en \nrm{\pt{q}} \big( \trn{\iden}_\ii{\!\Sig}-   \hsfb{q}^\flt \otms \hsfb{q} \big) \cdot \bs{\mu} -  \til{\mu}_\en \hsfb{q}^{\flt} 
     &,\quad \kappa_\en = \bs{\mu}\cdot \hsfb{q} 
\end{array} 
\qquad \leftrightarrow \qquad 
\begin{array}{lllll}
     \bs{\mu}  
      =  \tfrac{\nrm{\pt{x}}}{x^\en} \big( \trn{\iden}_\ii{\!\Sig} -   \hsfb{x}^\flt \otms \hsfb{x} \big) \cdot \bs{\kappa} + \kappa_\en \hsfb{x}^\flt
     \quad,\quad \til{\mu}_\en = 
     - \bs{\kappa}\cdot \hsfb{x}  
\end{array} 
\end{align}
\end{small}
Now, the \eq{s}- or \eq{\tau}-parameterized integral curves of \eq{\sfb{X}^\sscr{K}} (which are technically integral curves of a conformal scaling of \eq{\sfb{X}^\sscr{K}}) are recovered from the \eq{s}- or \eq{\tau}-parameterized integral curves of \eq{\sfb{X}^\sscr{H}} (which are really integral curves of  \eq{\cf\sfb{X}^\sscr{H}} and \eq{\cff\sfb{X}^\sscr{H}}) using \eq{\colift\psi} just as given above. 
And we now know the closed-form solutions for \eq{s}- or \eq{\tau}-parameterized integral curves of \eq{\sfb{X}^\sscr{H}}; they are precisely the integral curves of \eq{\cf\sfb{X}^\sscr{H}} and \eq{\cff\sfb{X}^\sscr{H}}, represented in the coordinates \eq{(\tup{z},\tiltup{z}^\ii{\perp})=(\tup{r},\tup{\plin},r^\en,\tilplin_\en)} as given in Eq.\eqref{prj_sols_gen}. 

\begin{notesq}
   \rmsb{simplifications on $\cotsp\man{Q}_\ii{1}$ and \eq{\cotsp\Sig_\ii{1}}.}
   In the above, consider the case that \eq{\barpt{\mu}_{\barpt{q}}\in\cotsp\man{Q}_\ii{1}} such that \eq{\nrm{\pt{q}}\simeq 1} and \eq{\bs{\mu}\cdot \hsfb{q}\simeq 0}. It follows that \eq{\barpt{\kappa}_{\barpt{x}}=\colift\psi(\bar{\mu}_{\barpt{q}})\in\cotsp\Sig_\ii{1}\cong \cotsp\bs{\Sigup}_\nozer} such that \eq{x^\en \simeq 1} and \eq{\kappa_\en=\envec\cdot \bs{\kappa}\simeq 0}. The above transformation simplifies to: 
    \begin{small}
    \begin{gather} \nonumber 
       \cotsp\Sig_\ii{1} \ni  \bar{\kappa}_{\barpt{x}}
        \qquad \leftrightarrow \qquad  \bar{\mu}_{\barpt{q}}\in\cotsp\man{Q}_\ii{1}
        \phantom{XXXXXX}
    \\ \label{colift_prj_reg_T*Q}
     \boxed{ \begin{array}{rlcllll}
           \ptvec{x}  
           \simeq \tfrac{1}{q^\en}\ptvec{q}
           &,\quad x^\en 
           \simeq 1
    \\[4pt]
         \bs{\kappa}  \simeq 
         q^\en \bs{\mu} - \til{\mu}_\en \sfb{q}^{\flt} 
         &,\quad \kappa_\en 
         \simeq 0 
    \end{array} 
    \qquad \leftrightarrow \qquad 
    \begin{array}{lllll}
         \ptvec{q} 
         \simeq  \hpt{x} 
         \quad,\quad q^\en = \tfrac{1}{\nrm{\pt{x}}} 
    \\[4pt]
         \bs{\mu}  
          \simeq  \nrm{\pt{x}} \big( \trn{\iden}_\ii{\!\Sig} -   \hsfb{x}^\flt \otms \hsfb{x} \big) \cdot \bs{\kappa} 
         \quad,\quad 
           \til{\mu}_\en = -\bs{\kappa}\cdot \hsfb{x}  
    \end{array} } 
    \end{gather}
    \end{small}
    with \eq{\eq{\til{\mu}_\en = \tilplin_\en(\bar{\mu}_{\barpt{q}})=q_\en^2 \mu_\en}}. 
     The \eq{s}- or \eq{\tau}-parameterized integral curves of \eq{\sfb{X}^\sscr{H}} (i.e., integral curves of \eq{\cf\sfb{X}^\sscr{H}} and \eq{\cff\sfb{X}^\sscr{H}})  to be used in the above are represented in the coordinates \eq{(\tup{z},\tiltup{z}^\ii{\perp})=(\tup{r},\tup{\plin},r^\en,\tilplin_\en)} as given in Eq.\eqref{prj_sols_gen}; on \eq{\cotsp\man{Q}_\ii{1}}, the ``\eq{(r^i,\plin_i)}-part'' further simplifying as in Eq.\eqref{prj_sols_gen_T*Q}. 
\end{notesq}

\subsubsection{Derivations and Details}
\label{sec:prj_sols_derivation}

We now go through the derivations and details of the solutions to the unperturbed problem (\eq{V^\ss{1}=U^\ss{1}=0}) that were stated above. We address the dynamics on \eq{\cotsp\bs{\Sigup}_\nozer} and \eq{\cotsp\vsp{N}_\ii{+}} separately.

\paragraph{Derivation and Details:~Motion on  $\cotsp\bs{\Sigup}_\nozer\subset \cotsp\bvsp{E}$.}
With \eq{U^\ss{1}=0}, 
the projection of \eq{\cf\sfb{X}^\sscr{H}= r_\en^\ss{-2}\sfb{X}^\sscr{H}\in\vect(\cotsp\bvsp{E})} on  \eq{\cotsp\bs{\Sigup}_\nozer \subset \cotsp\bvsp{E}} is described in terms of the \eq{2\en-2} (cartesian) coordinates, \eq{\tup{z}=(\tup{r},\tup{\plin}):\cotsp\bs{\Sigup}\to\mbb{R}^{\two(\en-\one)} }, by the following ODEs:
\begin{small}
\begin{align} \label{drp_ds_E3}
\tup{z}=(\tup{r},\tup{\plin}) \;
\left\{\; \begin{array}{lllllll}
     \pdt{r}^i  = -\slang^{ij} r_j  
     \,=\,
     -{\slang^i}_j r^j
\\[4pt]
       \pdt{\plin}_i  = -\slang_{ij}\plin^j 
       \,=\, 
       -{\slang_i}^j \plin_j
\end{array} \right.
&& \fnsize{i.e.,} \qquad
 \begin{array}{lllllll}
     \pdt{\tup{r}}_s =  -\Slang_\ss{\tup{z}_s} \tup{r}_s 
\\[4pt]
      \pdt{\tup{\plin}}_s =  -\til{\Slang}_\ss{\tup{z}_s} \tup{\plin}_s
\end{array}
\qquad,\qquad 
 \begin{array}{lllllll}
     \Slang := [{\slang^i}_j]  
\\[4pt]
     \til{\Slang} :=  [{\slang_i}^j] 
\end{array}
\end{align}
\end{small}
where \eq{{\slang^i}_j:=\slang^{ik}\emet_{kj} = \emet^{ik}\slang_{kj}} and \eq{{\slang_i}^j:=\slang_{ik}\emet^{kj} = \emet_{ik}\slang^{kj}} are collected into the matrices \eq{\Slang} and \eq{\til{\Slang}}, which warrent a brief comment:
\begin{small}
\begin{itemize}[nosep]
\item  With \eq{{\slang^i}_j} and \eq{{\slang_i}^j} so defined, it holds numerically that \eq{{\slang_i}^j=-{\slang^i}_j} and, thus,  \eq{\til{\Slang}=-\trn{\Slang}} and \eq{ \Slang=-\trn{\til{\Slang}}} 
    (see footnote for details\footnote{We use \eq{\sfb{\emet}_\ii{\Sig}} to raise/lower indices (or, since we view things extrinsically in \eq{\bvsp{E}}, we technically use \eq{\bsfb{\emet}_\ii{\!\Evec}}, but the result is the same here). We then find:\\
    \eq{\qquad \quad  {\lang_i}^j:= \lang_{ik}\emet^{kj} = \emet^{kj}\lang_{ik} = -\emet^{jk}\lang_{ki} = - {\lang^j}_i \;\;,   }
    \eq{\qquad}
    or:
    \eq{\quad {\lang_i}^j:= \lang_{ik}\emet^{kj} = \emet_{ia}\emet_{kb}\lang^{ab} \emet^{kj}  = \emet_{ia} \lang^{ab} \kd^j_b =   \emet_{ia} \lang^{aj} = -  \emet_{ia} \lang^{ja} =  - \lang^{ja} \emet_{ai} = - {\lang^j}_i}  .\\ 
    (\eq{{\slang_i}^j =- {\slang^j}_i } follows the same way). Thus, letting \eq{\trn{\Slang}=\trn{[{\slang^i}_j]} = [{\slang^j}_i]}  and  \eq{\trn{\til{\Slang}} = \trn{[{\slang_i}^j]} = [{\slang_j}^i]} denote simply the \textit{matrix} transpose, we have that \eq{\til{\Slang}=-\trn{\Slang}} and  \eq{\Slang=-\trn{\til{\Slang}}}. Eq.\eqref{drp_ds_E3} may then be expressed equivalently by any of the following:
    \begin{align}
    \til{\Slang}=-\trn{\Slang} \;\;,\;\; \Slang=-\trn{\til{\Slang}}
    \qquad \Rightarrow \qquad 
     \begin{array}{llll}
         \pdt{r}^i  = -\slang^{ij} r_j \,= -{\slang^i}_j r^j 
         \;\;\,=  {\slang_j}^i r^j 
         &,\qquad 
         \pdt{\tup{r}} \,=\, -\Slang \cdot \tup{r}  \,=\, \tup{r} \cdot \til{\Slang} \,=\, \trn{\til{\Slang}}\cdot \tup{r}
    \\
         \pdt{\plin}_i  = -\slang_{ij}\plin^j  = -{\slang_i}^j \plin_j 
           \;\; = {\slang^j}_i \plin_j  
            &,\qquad 
            \pdt{\tup{\plin}} \,=\, -\til{\Slang}\cdot \tup{\plin}  \,=\, \tup{\plin}\cdot \Slang \,=\, \trn{\Slang}\cdot \tup{\plin}
    \end{array}   
    \end{align}
    and we could, if we wished, choose to express \eq{(\pdt{r}^i,\pdt{\plin}_i)} solely in terms of \eq{{\slang^i}_j} and \eq{\Slang}, or solely in terms of \eq{{\slang_i}^j} and \eq{\til{\Slang}}. Finally, note that if we ``forget'' that these represent geometric objects and we fully immerse ourselves in the cartesian world of matrices (where co/contra-variant indices have no meaning), then we further have \eq{\til{\Slang}=-\trn{\Slang}\equiv \Slang} and  \eq{\Slang=-\trn{\til{\Slang}}\equiv \til{\Slang}} such that there is no distinction to be made at all.  
    }). 
As such, we could express the above solely in terms of one or the other; e.g., the above is equivalent  to \eq{\pdt{r}^i=  -{\slang^i}_j r^j } and \eq{\pdt{\plin}_i={\slang^j}_i \plin_j}, and also  to 
\eq{\pdt{r}^i = {\slang_j}^i r^j } and \eq{\pdt{\plin}_i = -{\slang_i}^j \plin_j }. In matrix form:
\item[]  \eq{\qquad\qquad \qquad \qquad \qquad} 
    \eq{\pdt{\tup{r}} \,=\, -\Slang \cdot \tup{r}  \,=\, \tup{r} \cdot \til{\Slang} \,=\, \trn{\til{\Slang}}\cdot \tup{r} \,\equiv\,  - \til{\Slang}\cdot \tup{r}
    \qquad,\qquad  
    \pdt{\tup{\plin}} \,=\, -\til{\Slang}\cdot  \tup{\plin}  \,=\, \tup{\plin}\cdot  \Slang \,=\, \trn{\Slang}\cdot \tup{\plin} \,\equiv\,  -\Slang\cdot \tup{\plin} }
\item[] (the last equalities hold in the cartesian world of matrices where co/contra-variant indices have no meaning). However, \eq{{\slang^i}_j} and  \eq{{\slang_i}^j} should be seen as representing different \eq{(1,1)}-tensors taking values in \eq{\tsp\!\pmb{.}\bs{\Sigup} \otimes \cotsp\!\pmb{.}\bs{\Sigup} }  and \eq{\cotsp\!\pmb{.}\bs{\Sigup} \otimes \tsp\!\pmb{.}\bs{\Sigup} }, respectively (which are \textit{not} the same spaces). As such, we will continue to write \eq{\Slang = [{\slang^i}_j]} and \eq{ \til{\Slang} =  [{\slang_i}^j]} even though all expressions could be written solely in terms of one or the other. 
\end{itemize}
\end{small}
Now, for linear coordinates, \eq{\emet_{ij},\emet^{ij}\in\mbb{R}} are simply scalar constants (for \textit{cartesian} coordinates, then \eq{\emet_{ij}=\emet^{ij}=\kd^i_j} and we could, if  we wished, ignore any distinction between co/contra-variant indices). As such, \eq{{\slang^{i}}_j} and \eq{{\slang_i}^j} are still integrals of motion and we note that \eq{\Slang} and \eq{\til{\Slang}} are \eq{\somat{\en-\one}}-\textit{valued} matrices 
whose entries are integrals of motion; they are constant along any given integral curve of \eq{\cf\sfb{X}^\sscr{H}} (but they are not true scalar constants). 
That is, for some fixed values \eq{\slang^{ij}_\iio\in\mbb{R}}, the above ODEs are linear, with those of \eq{\tup{r}} uncoupled from those of \eq{\tup{\plin}}. For some initial value \eq{\tup{z}_\zr=(\tup{r}_\zr,\tup{\plin}_\zr)} such that \eq{\slang^{ij}(\tup{z}_\zr)=\slang^{ij}_\iio}, then the matrix exponentials of \eq{\Slang_\iio, \til{\Slang}_\iio \in \somat{\en-\one}}, which are elements of \eq{\Somat{\en-\one}},  define a coordinate curve \eq{\tup{z}_s=(\tup{r}_s,\tup{\plin}_s)} that is a solution to the above (i.e., is the coordinate representation of the \eq{\cotsp\bs{\Sigup}_\nozer}-part of an integral curve of \eq{\cf\sfb{X}^\sscr{H}}) and therefore it holds that \eq{\Slang_\ss{\tup{z}_s}=\Slang_\ss{\tup{z}_\zr} = \Slang_\iio}.  More specifically:
\begin{small}
\begin{align} \label{E3sol0_s_prj}
\begin{array}{lllllll}
     \pdt{\tup{r}}_s =   -\Slang_\ss{\tup{z}_s} \tup{r}_s = -\Slang_\iio \tup{r}_s  
\\[4pt]
      \pdt{\tup{\plin}}_s =  -\til{\Slang}_\ss{\tup{z}_s} \tup{\plin}_s    =  -\til{\Slang}_\iio \tup{\plin}_s  
\end{array}
\quad \Rightarrow \quad
\begin{array}{lllllll}
     \tup{r}_s =  \mrm{e}^{-\Slang_\iio s} \tup{r}_\zr  
\\[4pt]
      \tup{\plin}_s =  \mrm{e}^{-\til{\Slang}_\iio s} \tup{\plin}_\zr 
\end{array}
\qquad\quad ,\qquad\quad
 \begin{array}{lllllll}
     \Slang_\iio  
, 
     \til{\Slang}_\iio   \in \somat{\en-\one}
\end{array}
\quad,\quad 
\begin{array}{llllllll}
     \mrm{e}^{-\Slang_\iio s}  
, 
     \mrm{e}^{-\til{\Slang}_\iio s} \in \Somat{\en-\one}
\end{array}
\end{align}
\end{small}
A closed form solution may be obtained using the Rodrigues rotation formula (note \eq{\tfrac{1}{\slang}\Slang = \tfrac{1}{\lang}L}): 
\begin{small}
\begin{align} \label{E3sol_s_prj}
\begin{array}{rlllllll}
      \mrm{e}^{-\Slang s} &\!\!\!\! =\imat_{\en-\one} - \tfrac{\sin{\slang s}}{\slang} \Slang + \tfrac{1-\cos{\slang s}}{\slang^2} \Slang \Slang 
\\[2pt]
    &\!\!\!\! =\imat_{\en-\one} - \tfrac{\sin{\slang s}}{\lang} {\Lang} + \tfrac{1-\cos{\slang s}}{\lang^2} {\Lang} {\Lang}
\end{array}
&& \Rightarrow &&
\begin{array}{lllllll}
     \tup{r}_s =  \mrm{e}^{-\Slang_\iio s} \tup{r}_\zr  
      &\!\!\!\! =\,  \tup{r}_\zr \cos{\slang_\iio s} - \tfrac{1}{\slang_\iio} \Slang_\iio \tup{r}_\zr \sin {\slang_\iio s} 
\\[4pt]
      \tup{\plin}_s =  \mrm{e}^{-\til{\Slang}_\iio s} \tup{\plin}_\zr  
      &\!\!\!\! =\,   \tup{\plin}_\zr \cos{\slang_\iio s} - \tfrac{1}{\slang_\iio} \til{\Slang}_\iio \tup{\plin}_\zr \sin {\slang_\iio s} 
\end{array} 
\quad
\left\{ \;\begin{array}{lllll}
     \tup{r}_s\cdot \tup{\plin}_s = \tup{r}_\zr\cdot \tup{\plin}_\zr 
\\[3pt]
    \tup{r}_s\cdot \tup{r}_s = \tup{r}_\zr\cdot \tup{r}_\zr 
\\[3pt]
     \tup{\plin}_s\cdot \tup{\plin}_s = \tup{\plin}_\zr\cdot \tup{\plin}_\zr 
\end{array}\right. 
\end{align}
\end{small}
with intermediate steps given in the footnote\footnote{Note \eq{\mrm{e}^{-\til{\Slang} s}} is given the same as  \eq{\mrm{e}^{-\Slang s}} (just with \eq{\til{\Slang}} rathe than \eq{\Slang}). We have then used \eq{\tfrac{1}{\slang}\Slang = \tfrac{1}{\lang}L}, along with \eq{{\Lang}{\Lang} \tup{r} = -\lang^2 \tup{r}} and \eq{\til{\Lang}\til{\Lang} \tup{\plin} = -\lang^2 \tup{\plin}}. Likewise, \eq{\Slang\Slang \tup{r} = -\slang^2 \tup{r}} and \eq{\til{\Slang}\til{\Slang} \tup{\plin} = -\slang^2 \tup{\plin}}; the solutions may be expressed using either \eq{\tfrac{1}{\lang}{\Lang}} or \eq{\tfrac{1}{\slang}\Slang}.}. 
The above may also be combined and expressed in terms of \eq{\tup{z}=(\tup{r},\tup{\plin})}, illustrating the analogies between antisymmetric/orthogonal linear systems and Hamiltonian/symplectic linear systems:
\begin{small}
\begin{align} \label{dz_ds_E3}
    \pdt{\tup{z}}_s \,=\, \Lambda_\ss{\tup{z}_s} \tup{z}_s =  \Lambda_\iio \tup{z}_s 
\quad \Rightarrow \quad
    \tup{z}_s = \mrm{e}^{\Lambda_\iio s} \tup{z}_\zr
&&,&&
      \Lambda_\iio := \begin{pmatrix}
        -\Slang_\iio  & 0 \\  0 & -\til{\Slang}_\iio
    \end{pmatrix}   \in \spmat{\two(\en-\one)}
\;\;\;,\;\;\;
    \mrm{e}^{\Lambda_\iio s} = \begin{pmatrix}
          \mrm{e}^{-\Slang_\iio s} & 0 \\
          0 &   \mrm{e}^{- \til{\Slang}_\iio s}
    \end{pmatrix}
    \in \Spmat{\two(\en-\one)}
\end{align}
\end{small}
\begin{small}
\begin{itemize}[nosep]
    \item \rmsb{with $\tau$ as the evolution parameter.}  The \eq{\tau}-parameterized integral curves of \eq{\cff\sfb{X}^\sscr{H}} on \eq{\cotsp\bs{\Sigup}_\nozer} are the same as the \eq{s}-parameterized integral curves of \eq{\cf\sfb{X}^\sscr{H}} seen above, just with \eq{\slang_\iio s} replaced by \eq{\tau} (recall \eq{\slang_\iio s = \tau} when \eq{U^\ss{1}=0}). The coordinate representation of the dynamics of \eq{\cff\sfb{X}^\sscr{H}} on \eq{\cotsp\bs{\Sigup}_\nozer} is similar  those of \eq{\cf\sfb{X}^\sscr{H}} seen above, just with \eq{\pdt{\square}:=\diff{}{s}} replaced by \eq{\rng{\square}:=\diff{}{\tau}} and with \eq{\slang^{ij}} replaced by \eq{\tfrac{1}{\slang}\slang^{ij}=\tfrac{1}{\lang}\lang^{ij}} (and likewise for anything built from \eq{\slang^{ij}}). That is, for some fixed  values \eq{\slang^{ij}_\iio\in\mbb{R}}: 
    \begin{small}
    \begin{align} \label{E3sol_ta_prj}
    \begin{array}{rlllllll}
        \rng{\tup{r}}_\tau &\!\!\!\! =   -\tfrac{1}{\slang(\tup{z}_\tau)}\Slang_\ss{\tup{z}_\tau} \tup{r}_\tau  \,=\, -\tfrac{1}{\slang_\iio}\Slang_\iio \tup{r}_\tau  
    \\[3pt]
          \rng{\tup{\plin}}_\tau  &\!\!\!\! =  -\tfrac{1}{\slang(\tup{z}_\tau)}\til{\Slang}_\ss{\tup{z}_\tau} \tup{\plin}_\tau \,=\,  -\tfrac{1}{\slang_\iio}\til{\Slang}_\iio \tup{\plin}_\tau 
    \end{array}
    \qquad \Rightarrow \qquad 
   \begin{array}{lllllll}
         \tup{r}_\tau =  \mrm{e}^{-\Slang_\iio \tau/\slang_\iio} \tup{r}_\zr  
         &\!\!\!\! =\,
         \tup{r}_\zr \cos \tau \,-\, \tfrac{1}{\slang_\iio} \Slang_\iio \tup{r}_\zr \sin  \tau 
    \\[3pt]
          \tup{\plin}_\tau =  \mrm{e}^{-\til{\Slang}_\iio \tau/\slang_\iio} \tup{\plin}_\zr 
          &\!\!\!\! =\,   \tup{\plin}_\zr \cos \tau \,-\, \tfrac{1}{\slang_\iio} \til{\Slang}_\iio \tup{\plin}_\zr \sin \tau 
    \end{array} 
    \;\;
    \begin{array}{lllllll}
        \equiv\, \tup{r}_s   
    \\[3pt]
         \equiv\,  \tup{\plin}_s 
    \end{array}
    \end{align}
    \end{small}
    where \eq{\tau = \slang_\iio s} and
    where \eq{\mrm{e}^{-\Slang \tau/\slang} =\imat_{\en-\one} - \tfrac{1}{\slang}(\sin \tau) \Slang + \tfrac{1}{\slang^2}(1-\cos \tau) \Slang \Slang = \mrm{e}^{-\Slang s} } is \eq{ \Somat{\two(\en-\one)} }-valued  (likewise for \eq{\mrm{e}^{-\til{\Slang} \tau/\slang} = \mrm{e}^{-\til{\Slang} s}}). 
    The above can also be combined into a single equation for \eq{\tup{z}_\tau=(\tup{r}_\tau,\tup{\plin}_\tau)}, which is equivalent to  the above Eq.\eqref{dz_ds_E3}:
    \begin{small}
    \begin{align}
         \rng{\tup{z}}_\tau = \tfrac{1}{\slang(\tup{z}_\tau)} \Lambda_\ss{\tup{z}_\tau} \tup{z}_\tau = \tfrac{1}{\slang_\iio} \Lambda_\iio \tup{z}_\tau 
    \qquad &\Rightarrow \qquad 
        \tup{z}_\tau = \mrm{e}^{\Lambda_\iio \tau / \slang_\iio} \tup{z}_\zr
        \quad\equiv\,\tup{z}_s
    \end{align}
    \end{small}
    \item \rmsb{alternative approach.} 
    Recall that the ODEs for \eq{\tup{z}=(\tup{r},\tup{\plin})} in Eq.\eqref{drp_ds_E3} or Eq.\eqref{dz_ds_E3} can also be expressed as
    \begin{small}
    \begin{align}
        \tup{z}=(\tup{r},\tup{\plin})
    \; \left\{ \;\begin{array}{lllllll}
         \pdt{r}^i  = - \slang^{ij} r_j   =  \tfrac{1}{m} \big( \nrmtup{r}^2 \emet^{ij}\plin_j  -(\tup{r} \cdot \tup{\plin} ) r^i \big)
    \\[3pt]
           \pdt{\plin}_i   = - \slang_{ij}\plin^j  = - \tfrac{1}{m}   \big( \nrmtup{\plin}^2 \emet_{ij}r^j - (\tup{r} \cdot \tup{\plin})\plin_i \big) 
    \end{array} \right.
    && \Rightarrow &&
        \pdt{\tup{z}}_s = A_\ss{\tup{z}_s} \tup{z}_s 
        \quad,\quad 
        A:= 
         \tfrac{1}{m} 
        \fnsz{\begin{pmatrix}
            -(\tup{r} \cdot \tup{\plin} ) \kd^i_j & \nrmtup{r}^2 \emet^{ij} \\
                -\nrmtup{\plin}^2 \emet_{ij}  &  (\tup{r} \cdot \tup{\plin})  \kd^j_i
        \end{pmatrix}}
    \end{align}
    \end{small}
    where, again, \eq{A} is a \eq{\spmat{\two(\en-\one)}}-\textit{valued} matrix of integrals of motion (when \eq{U^\ss{1}=0}) such that, for some fixed values, \eq{A_\iio \in\spmat{\two(\en-\one)}} is a Hamiltonian matrix whose exponential is a symplectic matrix giving solutions, \eq{\tup{z}_s}, to the above (where \eq{A_\iio = A_\ss{\tup{z}_\zr}=A_\ss{\tup{z}_s}}). That is:
    \begin{small}
    \begin{align}
    \begin{array}{rllll}
         \fnsize{for} \; \cf\sfb{X}^\sscr{H}:& \pdt{\tup{z}}_s =\, A_\ss{\tup{z}_s} \tup{z}_s \,=\, A_\iio \tup{z}_s
     \\[3pt]
         \fnsize{for} \; \cff\sfb{X}^\sscr{H}:&   \rng{\tup{z}}_\tau = \tfrac{1}{\slang(\tup{z}_\tau)} A_\ss{\tup{z}_\tau} \tup{z}_\tau  = \tfrac{1}{\slang_\iio} A_\iio \tup{z}_\tau  
    \end{array}
        \;\; \Rightarrow \;\;
    \begin{array}{llll}
         \tup{z}_s = \mrm{e}^{A_\iio s} \tup{z}_\zr 
     \\[3pt]
         \tup{z}_\tau = \mrm{e}^{A_\iio \tau/\slang_\iio} \tup{z}_\zr 
    \end{array}
     &&,&&
         A_\iio, \,\tfrac{1}{\slang_\iio} A_\iio  \in \spmat{\two(\en-\one)}
      \;\;\;,\;\;\;
          \mrm{e}^{A_\iio s} = \mrm{e}^{A_\iio \tau/\slang_\iio} 
           \in \Spmat{\two(\en-\one)}
    \end{align}
    \end{small}
     where, as before, the ODEs and \eq{\tau}-parameterized solutions for \eq{\cff\sfb{X}^\sscr{H}} are related to those of \eq{\cf\sfb{X}^\sscr{H}} as above. 
    The exponential is:
    \begin{small}
    \begin{align}
         \mrm{e}^{A s}  =  \mrm{e}^{A \tau/\slang} = 
      \fnsz{\begin{pmatrix}
        \big(\cos \tau - \tfrac{(\tup{r} \cdot \tup{\plin})}{\lang} \sin \tau \big) \kd^i_j 
        & \big(\tfrac{\nrmtup{r}^2}{\lang}\sin \tau \big) \emet^{ij} \\
        -\big( \tfrac{\nrmtup{\plin}^2}{\lang} \sin \tau \big) \emet_{ij} & 
          \big( \cos \tau +  \tfrac{(\tup{r} \cdot \tup{\plin})}{\lang} \sin \tau \big) \kd^j_i
     \end{pmatrix}}
     \qquad,\qquad 
     \tau = \slang s
    \end{align}
    \end{small}
        \begin{small}
        \begin{itemize}[nosep]
            \item \textit{simplifications on $\cotsp\man{Q}_\ii{1}$.} Note that, for integral curves on \eq{\cotsp\man{Q}_\ii{1}}, the matrix \eq{A} and its exponential simplify to:
            \begin{small}
            \begin{align}
                 A \simeq \fnsz{\tfrac{1}{m} \begin{pmatrix}
                    0 & \nrmtup{r}^2 \emet^{ij} \\
                        -\nrmtup{\plin}^2 \emet_{ij}  &  0
                \end{pmatrix}}
                \simeq  \fnsz{\tfrac{1}{m} \begin{pmatrix}
                    0 &  \emet^{ij} \\
                        -\lang^2 \emet_{ij}  &  0
                \end{pmatrix}}
                \qquad,\qquad 
                  \mrm{e}^{A s}  =  \mrm{e}^{A \tau/\slang} 
                  \simeq \fnsz{\begin{pmatrix}
                 \cos \tau \, \kd^i_j 
                & \tfrac{1}{\lang}\sin \tau \,\emet^{ij} \\
                -\lang \sin \tau \, \emet_{ij} & 
                   \cos \tau \, \kd^j_i
             \end{pmatrix}}
            \end{align}
            \end{small}
        \end{itemize}
        \end{small}
\end{itemize}
\end{small}
It can be verified that the solution \eq{\tup{z}_s = \mrm{e}^{A_\iio s}\tup{z}_\zr} agrees with the previous solution \eq{\tup{z}_s = \mrm{e}^{\Lambda_\iio s}\tup{z}_\zr}, and likewise with \eq{\tau}.  For \eq{A}, \eq{\Lambda}, \eq{s}, and \eq{\tau} discussed above, the following hold anywhere along a coordinate solution \eq{\tup{z}_s\equiv \tup{z}_\tau}:
\begin{small}
\begin{align}
\begin{array}{llll}
    \Lambda \,\neq\,  \tfrac{1}{\slang}\Lambda  \,\neq\,  A \neq \tfrac{1}{\slang} A
\\[4pt]
    \mrm{e}^{\Lambda s} 
    \,=\, \mrm{e}^{\Lambda \tau /\slang} 
     \,\neq\, \mrm{e}^{A s } 
    \,=\, \mrm{e}^{A \tau/\slang} 
\end{array}
&&
 \fnsize{but,}  \quad 
\begin{array}{llll}
    \Lambda \tup{z} \,=\, A \tup{z} \,\neq\, \tfrac{1}{\slang}\Lambda \tup{z} \,=\, \tfrac{1}{\slang} A \tup{z}
\\[4pt]
     \mrm{e}^{\Lambda s} \tup{z}
        \,=\, \mrm{e}^{\Lambda \tau /\slang} \tup{z}
         \,=\, \mrm{e}^{A s } \tup{z}
        \,=\, \mrm{e}^{A \tau/\slang} \tup{z}
\end{array}
\end{align}
\end{small}
In particular, since any solution \eq{\tup{z}_s\equiv \tup{z}_\tau} must always satisfy \eq{\slang^{ij}_\iio = \slang^{ij}(\tup{z}_\zr)=\slang^{ij}(\tup{z}_s)=\slang^{ij}(\tup{z}_\tau)} (and therefore also \eq{\Lambda_\iio=\Lambda_\ss{\tup{z}_\zr}=\Lambda_\ss{\tup{z}_s}=\Lambda_\ss{\tup{z}_\tau}} and \eq{A_\iio= A_\ss{\tup{z}_\zr}= A_\ss{\tup{z}_s}= A_\ss{\tup{z}_\tau}}), we have
    that:\footnote{We also note the eigenvalues of the matrices (given here as if \eq{\en-1=3}):
    \begin{align}\nonumber 
    \begin{array}{lllll}
        \fnsize{eig}(\Lambda) = (0,0, \mrm{i}\slang, \mrm{i}\slang, -\mrm{i}\slang, -\mrm{i}\slang)
    \\[2pt]
        \fnsize{eig}(\tfrac{1}{\slang}\Lambda) = (0,0,\mrm{i}, \mrm{i}, -\mrm{i}, -\mrm{i})
    \end{array}
        &&,&&
    \begin{array}{lllll}
       \fnsize{eig}(A) = (\mrm{i}\slang,\mrm{i}\slang,\mrm{i}\slang,-\mrm{i}\slang,-\mrm{i}\slang,-\mrm{i}\slang)
    \\[2pt]
        \fnsize{eig}(\tfrac{1}{\slang}A) = (\mrm{i},\mrm{i},\mrm{i},-\mrm{i},-\mrm{i},-\mrm{i})
    \end{array}
    &&,&& \slang \simeq \nrmtup{\plin}
    \end{align} }
\begin{small}
\begin{align}
    \pdt{\tup{z}}_s =  \Lambda_\iio \tup{z}= A_\iio \tup{z}_s \;\,\neq\,\; \rng{\tup{z}}_\tau = \tfrac{1}{\slang_\iio}\Lambda_\iio \tup{z}_\tau = \tfrac{1}{\slang_\iio}A_\iio \tup{z}_\tau 
    \qquad\quad,\qquad\quad 
    \tup{z}_s =  \mrm{e}^{\Lambda_\iio s} \tup{z}_\zr = \mrm{e}^{A_\iio s } \tup{z}_\zr \;\,=\,\; 
    \tup{z}_\tau = \mrm{e}^{\Lambda_\iio \tau /\slang_\iio} \tup{z}_\zr  = \mrm{e}^{A_\iio \tau/\slang_\iio} \tup{z}_\zr
\end{align}
\end{small}


\paragraph{Derivation and Details:~Motion on $\cotsp\vsp{N}_\ii{+}\subset \cotsp\bvsp{E}$.} 
For \eq{U=U^\zr=-\sck_1 r_\en -\ttfrac{1}{2}\sck_2 r_\en^2}, the projection of \eq{\cf\sfb{X}^\sscr{H}= r_\en^\ss{-2}\sfb{X}^\sscr{H}\in\vect(\cotsp\bvsp{E})} on  \eq{\cotsp\vsp{N}_\ii{+} \subset \cotsp\bvsp{E}} is described in cartesian coordinates, \eq{\tup{z}^\ii{\perp}=(r^\en,\plin_\en):\cotsp\vsp{N}\to\mbb{R}^{\two}}, or the non-symplectic pair \eq{\tiltup{z}^\ii{\perp}=(r^\en,\tilplin_\en):\cotsp\vsp{N}\to\mbb{R}^{\two}}, 
by the ODEs: 
\begin{small}
\begin{align} 
\tup{z}^\ii{\perp} = (r^\en,\plin_\en) 
\left\{\begin{array}{llll}
       \pdt{r}^\en =  \tfrac{r_\en^2}{m} \plin^\en 
\\[4pt]
       \pdt{\plin}_\en 
    = -m\beta^2 r_\en^\ii{-1} + \sck_1 r_\en^\ss{-2} -   \tfrac{2}{m}\plin_\en^2 r_\en
\end{array} \right. 
\qquad,\qquad 
\tiltup{z}^\ii{\perp} = (r^\en,\tilplin_\en) 
\left\{ \begin{array}{llll}
        \pdt{r}^\en = \tfrac{1}{m} \tilplin^\en 
    \\[3pt]
         \pdt{\tilplin}_\en = -m \beta^2 r_\en  + \sck_1 
\end{array} \right. 
\qquad\;\;\;
\beta^2 := \slang^2-\txkbar_2
\end{align}
\end{small}
 where \eq{\tilplin_\en:=r_\en^2 \plin_\en} was defined in Eq.\eqref{newpn_def0} and \eq{\txkbar_{1,2}:=\sck_{1,2}/m} (we assume \eq{\slang^2\geq \txkbar_2} such that \eq{\beta\in\mbb{R}}). 
 We will use the pair \eq{\tiltup{z}^\ii{\perp}=(r^\en,\tilplin_\en)} for which the above dynamics may be expressed 
 as:\footnote{In this case, note that \eq{N} is invertible with $\inv{N} = \mscale[0.8]{\begin{pmatrix}0 & -\beta^\ss{-2}m^{\en\en} \\  m_{\en\en} & 0 \end{pmatrix} = m \begin{pmatrix} 0 & - (m\beta)^\ss{-2} \\  1 & 0\end{pmatrix}} $ also a \eq{\spmat{\two}}-valued matrix.  }
\begin{small}
\begin{align} \label{rnpn_ds}
 \pdt{\tiltup{z}}^\ii{\perp}_s \,=\, N_\ss{\tiltup{z}^{\perp}_s} \tiltup{z}^\ii{\perp}_s
 +  \tup{b}
 \quad,\qquad 
 N :=  \fnsz{\begin{pmatrix}
       0 & m^{\en\en} \\  -\beta^2 m_{\en\en} & 0
    \end{pmatrix}} 
\quad,\quad 
     \tup{b} := \fnsz{\begin{pmatrix}
     0 \\ \sck_1
 \end{pmatrix}} \in \mbb{R}^\two
\quad,\quad 
\mrm{e}^{N s} = 
\fnsz{\begin{pmatrix}
    \cos(\beta s)\kd^\en_\en & \tfrac{1}{\beta} \sin(\beta s) m^{\en\en} \\
    -\beta \sin(\beta s) m_{\en\en} &  \cos(\beta s)\kd^\en_\en
\end{pmatrix}}
\end{align}
\end{small}
where \eq{m_{\en\en} = m \emet_{\en\en} = m} and \eq{m^{\en\en} = (1/m) \emet^{\en\en} = 1/m}. 
Again, \eq{N} is a \eq{\spmat{\two}}-\textit{valued} matrix — and thus \eq{\mrm{e}^{N s}} is \eq{\Spmat{\two}}-valued —  depending on the value of \eq{\slang} (through \eq{\beta^2 = \slang^2-\txkbar_2}) which is an integral of motion such that \eq{N_\ss{\tiltup{z}^\ii{\perp}_s} = N_\ss{\tiltup{z}^\ii{\perp}_\zr} = N_\iio} for any coordinate solution \eq{\tiltup{z}^\ii{\perp}_s}. As before, \eq{N_\ss{\tiltup{z}^\ii{\perp}_s}} may be replaced by the constant \eq{N_\iio}; the above reduces to a (nonhomogeneous) linear ODE: 
\begin{small}
\begin{align} \label{rnpn_ds_2}
     \pdt{\tiltup{z}}^\ii{\perp}_s 
     \,=\, N_\iio \tiltup{z}^\ii{\perp}_s
      + \tup{b}
 \quad \Rightarrow \quad 
    \tiltup{z}^\ii{\perp}_s \,=\, \mrm{e}^{N_\iio s} \tiltup{z}^\ii{\perp}_\zr \,+\, \tup{y}_s
\qquad,\qquad 
\begin{array}{llll}
   N_\iio \in \spmat{\two}
 \\[3pt]
    \mrm{e}^{N_\iio s} \in \Spmat{\two} 
\end{array}
\quad,\quad 
\tup{y}_s :=  
     \int_{\zr}^{s}( \mrm{e}^{-N_\iio s } \tup{b} ) \mrm{d} s
     = \fnsz{\begin{pmatrix}
        \tfrac{\txkbar_1}{\beta_\iio^2}(1-\cos{\beta_\iio s}) \\
         \tfrac{\sck_1}{\beta_\iio} \sin{\beta_\iio s}
    \end{pmatrix}}
\end{align}
\end{small} 
(with \eq{\txkbar_1=\sck_1/m}). 
With \eq{\mrm{e}^{N_\iio s}} given in Eq.\eqref{rnpn_ds}, the above leads to \eq{r^\en(s)} and \eq{\tilplin_\en(s)} given as follows (writing \eq{m_{\en\en}r^\en = m r_\en\,(\equiv m r^\en)} and \eq{m^{\en\en}\tilplin_\en = \tilplin^\en/m \,(\equiv \tilplin_\en /m)}):
\begin{small}
\begin{align}
\begin{array}{llll}
     r^\en_s \,=\, r^\en_\zr \cos{\beta_\iio s} \,+\, \tfrac{1}{m\beta_\iio} \tilplin^\en_\zr \sin{\beta_\iio s}  \,+\, \tfrac{\txkbar_1}{\beta_\iio^2}(1-\cos{\beta_\iio s})   
     &=\,  (r^\en_\zr-\tfrac{\txkbar_1}{\beta_\iio^2}) \cos{\beta_\iio s} \,+\, \tfrac{1}{m\beta_\iio} \tilplin^\en_\zr \sin{\beta_\iio s}  \,+\, \tfrac{\txkbar_1}{\beta_\iio^2}
 \\[3pt]
     \tilplin_{\en_s} \,=\,  - m \beta_\iio r_{\en_\zr}  \sin{\beta_\iio s} \,+\,  \tilplin_{\en_\zr} \cos{\beta_\iio s} \,+\, \tfrac{\sck_1}{\beta_\iio} \sin{\beta_\iio s}
      &=\, - m \beta_\iio( r_{\en_\zr} - \tfrac{\txkbar_1}{\beta_\iio^2}) \sin{\beta_\iio s}  \,+\,  \tilplin_{\en_\zr} \cos{\beta_\iio s} 
\end{array}
\end{align}
\end{small}
The solution for \eq{\plin_\en} is then easily recovered using \eq{\plin_\en = \tilplin_\en/r_\en^2 \leftrightarrow \tilplin_\en = r_\en^2 \plin_\en } (and the same for \eq{\plin^\en\equiv \plin_\en} and \eq{\tilplin^\en \equiv \tilplin_\en}). 
\begin{small}
\begin{itemize}[nosep]
    \item  \rmsb{with \eq{\tau} as the evolution parameter.}  As before, the \eq{\tau}-parameterized solutions for \eq{\cff\sfb{X}^\sscr{H}} on \eq{\cotsp\vsp{N}_\ii{+}} have \eq{(r^\en, \tilplin_\en)}-representations the same as those given above for \eq{\cf\sfb{X}^\sscr{H}}, 
    just with \eq{s} replaced by \eq{\tau/\slang_\iio} (we denoted this in Eq.\eqref{prj_sols_gen} using \eq{\varep:=\beta_\iio s = \beta_\iio \tau/\slang_\iio}).  The  \eq{\tau}-parameterized ODEs themselves are given as in Eq.\eqref{rnpn_ds}-Eq.\eqref{rnpn_ds_2} above, just with \eq{N} and \eq{\tup{b}} replaced by \eq{\tfrac{1}{\slang}N} and  \eq{\tfrac{1}{\slang}\tup{b}}. As before, we then have:
    \begin{small}
    \begin{align}
         \rng{\tiltup{z}}^\ii{\perp}_\tau = \tfrac{1}{\slang_\iio}N_\iio \tiltup{z}^\ii{\perp}_\tau + \tfrac{1}{\slang_\iio}\tup{b}
         \qquad \Rightarrow \qquad 
         \tiltup{z}^\ii{\perp}_\tau =  \mrm{e}^{N_\iio \tau/\slang_\iio} \tiltup{z}^\ii{\perp}_\zr + \tup{y}_\tau
         \;\equiv\;
         \tiltup{z}^\ii{\perp}_s =  \mrm{e}^{N_\iio s} \tiltup{z}^\ii{\perp}_\zr + \tup{y}_s
    \end{align}
    \end{small}
\end{itemize}
\end{small}

\paragraph{Solutions Using Second-Order Equations.}
The following is merely another way of obtaining the solutions given previously; it is included simply for the sake of completeness.
As shown several times, with \eq{U^\ss{1}=0}, the \eq{(r^i,r_\en)} coordinate representation of the base integral curves on \eq{\bvsp{E}} of \eq{\cf\sfb{X}^\sscr{H}} and  \eq{\cff\sfb{X}^\sscr{H}} satisfy the following second order coordinate ODEs: 
\begin{small}
\begin{align} \label{dd_ds_prj_unpert}
\begin{array}{rlllllll}
\fnsize{for any $\,U^\zr(r_\en)$}: &\quad
        \pddt{r}^{\,i}_s 
      \,+\, \slang_\iio^2 r^i_s \,=\,  0
      &\quad,\qquad 
        \rrng{r}^{\,i}_\tau 
      \,+\,  r^i_\tau \,=\,  0
\\[4pt]
\fnsize{for $\,U^\zr=-\sck_1 r_\en -\ttfrac{1}{2}\sck_2 r_\en^2$}: &\quad
     \pddt{r}^{\,\en}_s \,+\, \beta_\iio^2 r^\en_s \,=\, \txkbar_1
      &\quad,\qquad 
       \rrng{r}^{\,\en}_\tau  \,+\, \tfrac{\beta_\iio^2}{\slang_\iio^2} r^\en_\tau \,=\, \tfrac{\txkbar_1}{\slang_\iio^2} 
\end{array}
&& 
 \fnsz{\left| \quad \begin{array}{llll}
  \slang^2 := \lang^2 / m^2
\\[2pt]
     \beta^2 := \slang^2 - \txkbar_2 
\\[2pt]
    \txkbar_{1,2} := \sck_{1,2} / m  
\end{array}\right.}
\end{align}
\end{small}
where, as before, for some fixed  value \eq{\slang_\iio\in\mbb{R}}, the above ODEs for are those of a linear harmonic oscillator — with a constant driving term present in the \eq{r^\en} equation — where the natural frequency either depends on \eq{\slang_\iio}, or, is a unit natural frequency. 
Again, the \eq{s}-parameterized solutions and \eq{\tau}-parameterized solutions are the "same", using \eq{\slang_\iio s =\tau}. 
The solution to the above (using \eq{s} or \eq{\tau}) is readily found from the following well-known formula for arbitrary \eq{\tup{R}_t, \tup{b}\in\mbb{R}^m} and \eq{\omega\in\mbb{R}_\ii{+}}:
\begin{small}
\begin{align}
    \ddot{\tup{R}}_t  + \omega^2 \tup{R}_t - \tup{b} = 0 
    \qquad \Rightarrow \qquad 
\begin{array}{lllll}
     \tup{R}_t \,=\, \tup{R}_\zr \cos{\omega t} \,+\, \tfrac{1}{\omega} \dottup{R}_\zr \sin{\omega t} \,+\, \tfrac{1}{\omega^2} \tup{b} (1-\cos\omega t)
\\[4pt]
     \dot{\tup{R}}_t \,=\,  - \omega \tup{R}_\zr \sin{\omega t} \,+\, \dottup{R}_\zr \cos{\omega t} \,+\,  \tfrac{1}{\omega} \tup{b} \sin{\omega t}
\end{array}
\end{align}
\end{small}
That is, the solutions to Eq.\eqref{dd_ds_prj_unpert}
for  \eq{(\tup{r}_s,r^\en_s) \equiv (\tup{r}_\tau,r^\en_\tau)} and \eq{(\pdt{\tup{r}}_s,\pdt{r}^\en_s) = \slang_\iio(\rng{\tup{r}}_\tau,\rng{r}^\en_\tau)}  are given as follows:
\begin{small}
\begin{align}
\begin{array}{lllllll}
     \tup{r}_s = \tup{r}_\zr \cos{\slang_\iio s} + \tfrac{1}{\slang_\iio} \pdt{\tup{r}}_\zr \sin{\slang_\iio s} 
     &\equiv&
     \tup{r}_\tau = \tup{r}_\zr \cos{\tau} + \rng{\tup{r}}_\zr \sin{\tau} 
\\[4pt]
     \pdt{\tup{r}}_s =  - \slang_\iio \tup{r}_\zr \sin{\slang_\iio s} + \pdt{\tup{r}}_\zr \cos{\slang_\iio s} 
     &\nequiv &
     \rng{\tup{r}}_\tau =  -\tup{r}_\zr \sin{\tau} + \rng{\tup{r}}_\zr \cos{\tau} 
 \\[3pt]
     r^\en_s = r^\en_\zr \cos{\beta_\iio s} + \tfrac{1}{\beta_\iio} \pdt{r}^\en_\zr \sin{\beta_\iio s}+ \tfrac{\txkbar_1}{\beta_\iio^2}(1 - \cos{\beta_\iio s})
      &\equiv&
      r^\en_\tau = r^\en_\zr \cos{\varep} + \tfrac{\slang_\iio}{\beta_\iio}\rng{r}^\en_\zr \sin{\varep} 
     +  \tfrac{\txkbar_1}{\beta_\iio^2}(1 - \cos{\varep})
\\[4pt]
     \pdt{r}^\en_s =  - \beta_\iio r^\en_\zr \sin{\beta_\iio s} + \pdt{r}^\en_\zr \cos{\beta_\iio s} +  \tfrac{\txkbar_1}{\beta_\iio} \sin{\beta_\iio s}
     &\nequiv&
     \rng{r}^\en_\tau =  - \tfrac{\beta_\iio}{\slang_\iio} r^\en_\zr \sin{\varep} + \rng{r}^\en_\zr \cos{\varep} 
     +  \tfrac{\txkbar_1}{\slang_\iio \beta_\iio} \sin{\varep}
\end{array}
\end{align}
\end{small}
where \eq{ \tau = \slang_\iio s } and  \eq{ \varep :=  \beta_\iio s = \tfrac{\beta_\iio}{\slang_\iio} \tau}. 

\subsection{Compendium of Relations for the Two-Body Problem Problem}
\label{sec:prj_KEPgeomech}

We collect here, for ease of reference, how the previous developments simplify for the case that the \textit{original} potential, \eq{V=V^\zr(\rfun) +V^\ss{1}(\tup{r})}, has only a Kepler-type central force term, \eq{V^\zr=-\sck_1/\rfun}.  
As before, ``\eq{\simeq}'' will be used to indicate simplified relations that hold for integral curves of \eq{\sfb{X}^\sscr{H}} (or any conformal scaling of \eq{\sfb{X}^\sscr{H}}) on  $\cotsp\man{Q}_\ii{1}\subset\cotsp\bvsp{E}$ but generally not on \eq{\cotsp\bvsp{E}} (see Eq.\eqref{T*Q_reg_rev}).

\paragraph{Transformed Two-Body Dynamics.}
The original Kepler potential, \eq{V^\zr=-\sck_1/\rfun}, transforms to a normal force potential \eq{U^\zr = \psi^* V^\zr = -\sck_1 r_\en}. The transformed Hamiltonian system for the general two-body problem is obtained simply by setting \eq{\sck_2=0} in the previous developments (and thus \eq{\beta = \slang}). The \eq{s}-parameterized dynamics for the \textit{transformed} two-body problem, \eq{\cf\sfb{X}^\sscr{H}}, are then expressed in the coordinates \eq{(\tup{z},\tiltup{z}^\ii{\perp})=(\tup{r},\tup{\plin},r^\en,\tilplin_\en)}, where \eq{\tilplin_\en:= r_\en^2 \plin_\en}, as follows:
\begin{small}
\begin{align} \label{d_ds_KEPprj_altCords}
    \cf\sfb{X}^\sscr{H} = r_\en^\ss{-2} \sfb{X}^\sscr{H}
\left\{ \; \begin{array}{llllllll}
     \pdt{r}^i =  -  \slang^{ij} r_j
     \phantom{ - \cf \pd_i U^\ss{1}}
     \qquad  \simeq  \tfrac{1}{m}\plin^i  
\\[5pt]  
     \pdt{\plin}_i = - \slang_{ij} \plin^j  - \cf \pd_i U^\ss{1}
     \quad\;  \simeq  -m\slang^2 r_i  - \cf \pd_i U^\ss{1}
 \\[5pt]
    \pdt{r}^\en = \tfrac{1}{m} \tilplin^\en
 \\[5pt]  
     \pdt{\tilplin}_\en = -m \slang^2 r_\en  + \sck_1  -  \pd_\en U^\ss{1}
\end{array}  \right.
&& \; &&
\left|\quad \begin{array}{llllll}
         \pddt{r}\,^i  + \slang^2 r^i =   -\tfrac{r^2}{r_\en^2} m^{ij}\pd_j U^\ss{1}
  \\[6pt]
        \pddt{r}\,^\en  + \slang^2 r^\en = \txkbar_1  \,-\,  m^{\en\en} \pd_\en U^\ss{1} 
    \\[3pt]
     \;\;\; ( \rfun^2 \simeq 1 \;\;,\;\;
     \slang^2\simeq \nrmtup{\plin}^2/m^2) 
\end{array} \right.
\end{align}
\end{small}
where \eq{\slang=\lang/m} and \eq{\txkbar_1:=\sck_1/m}, and where we have also indicated the second-order ODEs that produce the same configuration coordinate dynamics for \eq{(\tup{r},r^\en)}. 
The equivalent \eq{\tau}-parameterized dynamics — where \eq{\tau} is the true anomaly, up to an additive constant, of the \textit{original} system — are expressed in these same coordinates as:
\begin{small}
\begin{align} \label{d_dta_KEPprj_altCords}
    \cff\sfb{X}^\sscr{H} = \tfrac{1}{\slang r_\en^2} \sfb{X}^\sscr{H}
\left\{ \; \begin{array}{llllllll}
      \rng{r}^i 
         =    -\tfrac{1}{\lang} \lang^{ij}r_j
         &\!\!\!\!  \simeq  \tfrac{1}{\lang}\plin^i \simeq \hat{\plin}^i  
\\[5pt]  
         \rng{\plin}_i  =  - \tfrac{1}{\lang} \lang_{ij} \plin^j - \cff \pd_i U^\ss{1}
          &\!\!\!\! \simeq  -\lang r_i  - \cff \pd_i U^\ss{1}
 \\[5pt]
        \rng{r}^\en 
       = \tfrac{1}{\lang} \tilplin^\en
  \\[5pt]  
      \rng{\tilplin}_\en = - \lang r_\en  +   \tfrac{\sck_1}{\slang} -  \tfrac{1}{\slang} \pd_\en U^\ss{1}  
\end{array}  \right.
&&\qquad\quad
\left|\;\;\; \begin{array}{rllllll}
      \rrng{r}\,^i   +  r^i \!\!\!\! &=    - \tfrac{m\rfun^2}{\lang^2 r_\en^2}  \big( \emet^{ik}  + \tfrac{1}{\lang^2} \lang^{ij}r_j  \plin^k \big) \pd_k U^\ss{1} 
      \\[2pt]
      &\simeq - \tfrac{m}{\nrmtup{\plin}^2 r_\en^2} ( \emet^{ik}  - \hat{\plin}\hat{\plin}^k  ) \pd_k U^\ss{1} 
\\[6pt]
        \rrng{r}\,^\en  +  r^\en   
        \!\!\!\! &= \tfrac{\txkbar_1}{\slang^2} - \tfrac{m}{\lang^2}  \big(\emet^{\en\en}  \pd_\en U^\ss{1} - \tfrac{r^2}{\lang^2} \plin^\en \plin^j \pd_j U^\ss{1} \big) 
        \\[2pt]
         &\simeq \tfrac{\txkbar_1}{m^2\nrmtup{\plin}^2} - \tfrac{m}{\nrmtup{\plin}^2}  \big(\emet^{\en\en}  \pd_\en U^\ss{1} - \tfrac{1}{\nrmtup{\plin}} \plin^\en \hat{\plin}^j \pd_j U^\ss{1} \big) 
\end{array}\right.
\end{align}
\end{small}
where \eq{\sfb{X}^\sscr{H}}, \eq{\cf\sfb{X}^\sscr{H}}, and \eq{\cff\sfb{X}^\sscr{H}} have the ``same'' integral curves, up to parameterization by \eq{t}, \eq{s}, or \eq{\tau} (and likewise for coordinate solutions for the ODEs in Eq.\eqref{d_ds_KEPprj_altCords} or Eq.\eqref{d_dta_KEPprj_altCords}).

\paragraph{Transformed Kepler Solutions.}
For the case \eq{U^\ss{1}=0}, then \eq{\lang^{ij}} are integrals of motion and, for some fixed values \eq{\lang^{ij}_\iio\in\mbb{R}}, the above \eq{s}-parameterized coordinate ODEs for \eq{(\tup{z},\tiltup{z}^\ii{\perp})=(\tup{r},\tup{\plin},r^\en,\tilplin_\en)} can be expressed in matrix form as:
\begin{small}
\begin{align} \label{something_s}
 \cf\sfb{X}^\sscr{H} 
\left\{ \;
\begin{array}{llllllll}
     \pdt{\tup{z}}_s = \Lambda_\iio \tup{z}_s  \;=\; A_\iio \tup{z}_s
     &,\qquad 
     \Lambda_\iio \in \spmat{\two(\en-\one)}
     &,\quad 
     \Lambda = \fnsz{\begin{pmatrix}
        -\Slang  & 0 \\  0 & -\til{\Slang}
    \end{pmatrix} }  
 \\[8pt]
     \pdt{\tiltup{z}}^\ii{\perp}_s = N_\iio \tiltup{z}^\ii{\perp}_s \,+\, \tup{b}
     &,\qquad 
     N_\iio \in \spmat{\two}
      &,\quad
     N =  
    \fnsz{\tfrac{1}{m} \begin{pmatrix}
       0 & \emet^{\en\en} \\  -\lang^2 \emet_{\en\en} & 0
    \end{pmatrix}}
    \;\;,\;\;
    \tup{b} = \fnsz{\begin{pmatrix}
       0 \\ \sck_1
    \end{pmatrix}}
\end{array} \right.
&&
 \left|\quad
 \begin{array}{llllll}
         \pddt{r}\,^i  + \slang_\iio^2 r^i = 0
  \\[8pt]
        \pddt{r}\,^\en  + \slang_\iio^2 r^\en = \txkbar_1 
\end{array} \right.
\end{align}
\end{small}
where \eq{\Slang:= [\slang^{ik}\emet_{kj}]} and \eq{\til{\Slang}:= [\slang_{ik}\emet^{kj}] \equiv -\trn{\Slang}}. 
Note \eq{\Lambda}, \eq{A}, and \eq{N} are matrices of integrals of motion
(\eq{A} is given again in the footnote\footnote{As shown previously, \eq{\pdt{\tup{z}}_s} and \eq{\rng{\tup{z}}_\tau} may be expressed using either \eq{\Lambda} or \eq{A} where \eq{\Lambda} is given as above an \eq{A}  is given by (where \eq{\tau=\slang s}):  
\begin{align}\nonumber
     & A_\iio, \tfrac{1}{\slang_\iio} A_\iio \in \spmat{\two(\en-\one)} \;,
  &&
       \mrm{e}^{A_\iio s} = \mrm{e}^{A_\iio \tau/\slang_\iio} \in \Spmat{\two(\en-\one)} 
\\ \nonumber 
    & A =  \tfrac{1}{m} \mscale[0.9]{\begin{pmatrix}
        -(\tup{r} \cdot \tup{\plin} ) \kd^i_j & \nrmtup{r}^2 \emet^{ij} \\
            -\nrmtup{\plin}^2 \emet_{ij}  &  (\tup{r} \cdot \tup{\plin})  \kd^j_i
    \end{pmatrix}}
     \simeq 
        \tfrac{1}{m} \mscale[0.9]{\begin{pmatrix}
            0 &  \emet^{ij} \\
                -\lang^2 \emet_{ij}  &  0
        \end{pmatrix}} \;,   
 &&
     \mrm{e}^{A s}  =  \mrm{e}^{A \tau/\slang} = 
      \mscale[0.9]{\begin{pmatrix}
       (\cos \tau - \tfrac{(\tup{r} \cdot \tup{\plin})}{\lang} \sin \tau) \kd^i_j 
        &(\tfrac{\nrmtup{r}^2}{\lang}\sin \tau) \emet^{ij} \\
        -\big( \tfrac{\nrmtup{\plin}^2}{\lang} \sin \tau) \emet_{ij} & 
         ( \cos \tau +  \tfrac{(\tup{r} \cdot \tup{\plin})}{\lang} \sin \tau) \kd^j_i
     \end{pmatrix}}
     \simeq
     \mscale[0.9]{\begin{pmatrix}
         \cos \tau \, \kd^i_j 
        & \tfrac{1}{\lang}\sin \tau \,\emet^{ij} \\
        -\lang \sin \tau \, \emet_{ij} & 
           \cos \tau \, \kd^j_i
     \end{pmatrix}}
\end{align}
where \eq{\simeq} is used for relations that hold on \eq{\cotsp\man{Q}_\ii{1}}. 
}).
The ODEs for \eq{\tup{z}=(\tup{r},\tup{\plin})} may be expressed using either \eq{\Lambda} or \eq{A};
we will use \eq{\Lambda}, though the result is the same either way. 
The analogous \eq{\tau}-parameterized coordinate ODEs are also given in matrix form as:
\begin{small}
\begin{align} \label{something_ta}
 \cff\sfb{X}^\sscr{H} 
\left\{ \;
\begin{array}{llllllll}
     \rng{\tup{z}}_\tau = \tfrac{1}{\slang_\iio}\Lambda_\iio \tup{z}_\tau  \;=\; \tfrac{1}{\slang_\iio}A_\iio \tup{z}_\tau
     &,\qquad 
     \tfrac{1}{\slang_\iio}\Lambda_\iio \in \spmat{\two(\en-\one)}
     &,\quad
     \mrm{e}^{\Lambda_\iio \tau/\slang_\iio} =\mrm{e}^{\Lambda_\iio s} 
     \in \Spmat{\two(\en-\one)}
 \\[8pt]
     \rng{\tiltup{z}}^\ii{\perp}_\tau = \tfrac{1}{\slang_\iio}N_\iio \tiltup{z}^\ii{\perp}_\tau \,+\, \tfrac{1}{\slang_\iio}\tup{b}
     &,\qquad 
     \tfrac{1}{\slang_\iio}N_\iio \in \spmat{\two}
     &,\quad  
     \mrm{e}^{N_\iio \tau/\slang_\iio}  = \mrm{e}^{N_\iio s}  \in \Spmat{\two}
\end{array} \right.
&&
 \left|\quad \begin{array}{llllll}
      \rrng{r}\,^i   +  r^i =    0
\\[8pt]
        \rrng{r}\,^\en  +  r^\en   
        = \tfrac{\txkbar_1}{\slang^2} 
\end{array} \right.
\end{align}
\end{small} 
Either of the above ODE systems may be solved for closed-form coordinate solutions:
\begin{small}
\begin{gather}
     \tup{z}_s = \mrm{e}^{\Lambda_\iio s} \tup{z}_\zr
     \;\equiv\; \tup{z}_\tau = \mrm{e}^{\Lambda_\iio \tau/\slang_\iio} \tup{z}_\zr
\qquad,\qquad
      \tiltup{z}^\ii{\perp}_s =  \mrm{e}^{N_\iio s} \tiltup{z}^\ii{\perp}_\zr + \tup{y}_s
     \;\equiv\; \tiltup{z}^\ii{\perp}_\tau =  \mrm{e}^{N_\iio \tau/\slang_\iio} \tiltup{z}^\ii{\perp}_\zr + \tup{y}_\tau
\\ \nonumber 
      \mrm{e}^{\Lambda_\iio s} 
     =( \fnsize{ Eq.\eqref{E3sol_s_prj}-Eq.\eqref{dz_ds_E3} } )
\quad,\quad 
     \mrm{e}^{N_\iio s} 
     = 
    \fnsz{\begin{pmatrix}
        \cos{\tau}\, \kd^\en_\en & \tfrac{1}{\lang_\iio} \sin{\tau}\, \emet^{\en\en} \\
        -\lang_\iio \sin{\tau} \,\emet_{\en\en} &  \cos{\tau} \,\kd^\en_\en
    \end{pmatrix}} 
\quad,\quad 
    \tup{y}_\tau = \tup{y}_s  = 
      \int_{\zr}^{s}( \mrm{e}^{-N_\iio s } \tup{b} ) \mrm{d} s  
     \,=\, \fnsz{\begin{pmatrix}
        \tfrac{\txkbar_1}{\slang_\iio^2}(1-\cos{\tau}) \\
         \tfrac{\sck_1}{\slang_\iio} \sin{\tau}
    \end{pmatrix} } 
\end{gather}
\end{small}
The above is written explicitly as:  
\begin{small}
\begin{align} \label{prj_KEPsols_apx}
\begin{array}{rrlllllllll}
    \tup{r}_s \equiv&\!\!\!  \tup{r}_\tau &\!\!\!\!=\,
     \tup{r}_\zr \cos \tau - \tfrac{1}{\lang_\iio} L_\iio \tup{r}_\zr \sin  \tau 
     & 
        \simeq \tup{r}_\zr \cos \tau \,+\, \tfrac{1}{\lang_\iio}\tup{\plin}_\zr^{\shrp} \sin  \tau
\\[4pt]
     \tup{\plin}_s \equiv&\!\!\! \tup{\plin}_\tau &\!\!\!\!=\, \tup{\plin}_\zr \cos \tau - \tfrac{1}{\lang_\iio} \til{\Lang}_\iio \tup{\plin}_\zr \sin \tau 
     &  
       \simeq \tup{\plin}_\zr \cos \tau - \lang_\iio \tup{r}_\zr^{\flt} \sin \tau 
\\[4pt] 
      r^\en_s \equiv&\!\!\! r^\en_\tau  
      &\!\!\!\! =\,  r^\en_\zr \cos{\tau} + \tfrac{1}{\lang_\iio} \tilplin^\en_\zr \sin{\tau}  + \tfrac{\txkbar_1}{\slang_\iio^2}(1-\cos{\tau})   
      &=  (r^\en_\zr-\tfrac{\txkbar_1}{\slang_\iio^2}) \cos{\tau} + \tfrac{1}{\lang_\iio} \tilplin^\en_\zr \sin{\tau}  + \tfrac{\txkbar_1}{\slang_\iio^2}
\\[4pt]
     \tilplin_{\en_s} \equiv&\!\!\! \tilplin_{\en_\tau}
      &\!\!\!\! =\,  -\lang_\iio r_{\en_\zr}  \sin{\tau} +  \tilplin_{\en_\zr} \cos{\tau} + \tfrac{\sck_1}{\slang_\iio} \sin{\tau}
      &=\, - \lang_\iio( r_{\en_\zr} - \tfrac{\txkbar_1}{\slang_\iio^2}) \sin{\tau}  +  \tilplin_{\en_\zr} \cos{\tau} 
      \;\;=\, (r_\en^2 \plin_\en)_\tau
\end{array} 
&&
 \left(\!\begin{array}{llll}
      \tau = \slang_\iio s  \\
      \lang_\iio \simeq \nrm{\tup{\plin}_\zr}
 \end{array}\!\right) 
\end{align}
\end{small}
where the \eq{s}-parameterized solutions are easily expressed using \eq{\tau = \slang_\iio s} (\eq{\slang} is an integral of motion when \eq{U^\ss{1}}). 
The solutions for the symplectic pair \eq{(r^\en,\plin_\en)} are recovered from the \eq{(r^\en,\tilplin_\en)} solutions using  
\eq{\plin_\en = \tilplin_\en/r_\en^2 \leftrightarrow \tilplin_\en = r_\en^2 \plin_\en }.
Lastly, we also note that the second-order ODEs in Eq\ref{something_s} and Eq.\eqref{something_ta} lead to the following (where \eq{\pdt{r}^\a_s = \slang_\iio \rng{r}^\a_\tau}):
\begin{small}
\begin{align}
\begin{array}{lllllll}
     \tup{r}_s = \tup{r}_\zr \cos{\slang_\iio s} + \tfrac{1}{\slang_\iio} \pdt{\tup{r}}_\zr \sin{\slang_\iio s} 
     &\equiv&
     \tup{r}_\tau = \tup{r}_\zr \cos{\tau} + \rng{\tup{r}}_\zr \sin{\tau} 
\\[4pt]
     \pdt{\tup{r}}_s =  - \slang_\iio \tup{r}_\zr \sin{\slang_\iio s} + \pdt{\tup{r}}_\zr \cos{\slang_\iio s} 
     &\nequiv &
     \rng{\tup{r}}_\tau =  -\tup{r}_\zr \sin{\tau} + \rng{\tup{r}}_\zr \cos{\tau} 
 \\[3pt]
     r^\en_s = r^\en_\zr \cos{\slang_\iio s} + \tfrac{1}{\slang_\iio} \pdt{r}^\en_\zr \sin{\slang_\iio s}+ \tfrac{\txkbar_1}{\slang_\iio^2}(1 - \cos{\slang_\iio s})
      &\equiv&
      r^\en_\tau = r^\en_\zr \cos{\tau} + \rng{r}^\en_\zr \sin{\tau} 
     +  \tfrac{\txkbar_1}{\slang_\iio^2}(1 - \cos{\tau})
\\[4pt]
     \pdt{r}^\en_s =  - \slang_\iio r^\en_\zr \sin{\slang_\iio s} + \pdt{r}^\en_\zr \cos{\slang_\iio s} +  \tfrac{\txkbar_1}{\slang_\iio} \sin{\slang_\iio s}
     &\nequiv&
     \rng{r}^\en_\tau =  - r^\en_\zr \sin{\tau} + \rng{r}^\en_\zr \cos{\tau} 
     +  \tfrac{\txkbar_1}{\slang^2_\iio} \sin{\tau}
\end{array}
\end{align}
\end{small}

\begin{small}
\begin{itemize}[nosep]
    \item  \rmsb{clarification.} To clarify, let us consider the \eq{\tau}-parametrized dynamics given by \eq{\cff\sfb{X}^\sscr{H}} (all that follows applies analogously for the \eq{s}-parametrized dynamics of \eq{\cf\sfb{X}^\sscr{H}}). 
Consider any phase space point 
\eq{\bar{\mu}_\iio \in\cotsp\bvsp{E} \cong \cotsp\bs{\Sigup}_\nozer \oplus \cotsp\vsp{N}_\ii{+}}, decomposed as: 
\begin{small}
\begin{align}
    \bar{\mu}_\iio \equiv \bar{\mu}_{\barpt{q}_\iio}
    =(\mu_\iio,\mu^\ii{\perp}_\iio) \,=\, ((\ptvec{q}_\iio,\bs{\mu}_\iio), (\ptvec{q}^\ii{\perp}_\iio, \bs{\mu}^\ii{\perp}_\iio ))
     &\!\!\! \in \cotsp\bs{\Sigup}_\nozer \oplus \cotsp\vsp{N}_\ii{+}
     \qquad,\qquad 
     \ptvec{q}^\ii{\perp} = q^\en \envec \;\;,\;\;  \bs{\mu}^\ii{\perp} = \mu_\en \enform = (q_\en^\ss{-2} \til{\mu}_\en)\enform
\end{align}
\end{small}
We then note the following:
\begin{small}
\begin{align}
   \begin{array}{lllll}
          \langb_\iio := \langb_{\mu_\iio} 
           = \ptvec{q}_\iio \otms \bs{\mu}^{\shrp}_\iio - \bs{\mu}^{\shrp}_\iio \otms \ptvec{q}_\iio
          \,= \ptvec{q}_\iio \wedge \bs{\mu}^{\shrp}_\iio
          &\!\!\! \in \bwedge{2}\tsp[\pt{q}_\iio]\bs{\Sigup}_\nozer
    \\[2pt]
         \Langb_\iio := \Langb_{\mu_\iio} =   \ptvec{q}_\iio \otms \bs{\mu}_\iio - \bs{\mu}^{\shrp}_\iio \otms \ptvec{q}^\flt_\iio
         \,= \langb_\iio \cdot \sfb{\emet}_\ii{\Sig}
        &\!\!\!\in \tsp[\pt{q}_\iio]\bs{\Sigup}_\nozer \otimes \cotsp[\pt{q}_\iio]\bs{\Sigup}_\nozer
        &,\qquad 
        \Langb_\iio \cdot \ptvec{q}_\iio = -\trn{\til{\Langb}}_\iio  \cdot \ptvec{q}_\iio = -\ptvec{q}_\iio \cdot  \til{\Langb}_\iio
    \\[2pt]
         \til{\Langb}_\iio := \til{\Langb}_{\mu_\iio}  =  \ptvec{q}^\flt_\iio \otms \bs{\mu}^{\shrp}_\iio - \bs{\mu}_\iio \otms \ptvec{q}_\iio
         \,= -\trn{\Langb}_\iio
         &\!\!\! \in \cotsp[\pt{q}_\iio]\bs{\Sigup}_\nozer \otimes \tsp[\pt{q}_\iio]\bs{\Sigup}_\nozer
         &,\qquad 
         \til{\Langb}_\iio \cdot \bs{\mu}_\iio = -\trn{\Langb}_\iio  \cdot \bs{\mu}_\iio = -\bs{\mu}_\iio \cdot  \Langb_\iio
    \end{array} 
\end{align}
\end{small}
If \eq{\phi_\tau:\cotsp\bvsp{E} \to \cotsp\bvsp{E}} denotes the flow of \eq{\cff\sfb{X}^\sscr{H}},
then any point generates an integral curve  of \eq{\cff\sfb{X}^\sscr{H}} given by \eq{\bar{\mu}_\tau := \phi_\tau(\bar{\mu}_\iio)} for which the following holds:  
\begin{small}
\begin{align}
    \bar{\mu}_\tau := \phi_\tau(\bar{\mu}_\iio)
    \quad \Rightarrow \quad 
    \rng{\bar{\mu}}_\tau = \cff\sfb{X}^\sscr{H}_{\bar{\mu}_\tau}
    \qquad \fnsize{with: } \;\; \bar{\mu}_\zr = \bar{\mu}_\iio
   \quad,\quad 
   \langb_{\mu_\tau} =   \langb_{\mu_\zr} =  \langb_\iio
\end{align}
\end{small}
where all other objects built from \eq{\langb_{\mu_\tau}} are also conserved along the curve.  From Eq.\eqref{d_dta_KEPprj_altCords}, we know that the relation \eq{ \rng{\bar{\mu}}_\tau = \cff\sfb{X}^\sscr{H}_{\bar{\mu}_\tau}} implies that the various "parts" of \eq{\bar{\mu}_\tau =(\mu_\tau,\mu^\ii{\perp}_\tau) = ((\ptvec{q}_\tau,\bs{\mu}_\tau), (\ptvec{q}^\ii{\perp}_\tau, \bs{\mu}^\ii{\perp}_\tau ))} obey the following relations
\begin{small}
\begin{align}
    \rng{\bar{\mu}}_\tau = \cff\sfb{X}^\sscr{H}_{\bar{\mu}_\tau}
\;\left\{ \; \begin{array}{llllllll}
  \rng{\ptvec{q}}_\tau 
     =    -\tfrac{1}{\lang_\iio} \Langb_\iio \cdot \ptvec{q}_\tau
     &\!\!\!\!  \simeq  \tfrac{1}{\lang_\iio} \bs{\mu}^{\shrp}_\tau 
\\[4pt]  
     \rng{\bs{\mu}}_\tau  =  - \tfrac{1}{\lang_\iio} \til{\Langb}_\iio \cdot \bs{\mu}_\tau  
      &\!\!\!\! \simeq  -\lang_\iio \ptvec{q}^\flt  
\end{array} \right.
\qquad,\qquad 
\begin{array}{llllll}
    \rng{q}^\en_\tau 
   = \tfrac{1}{\lang_\iio} \til{\mu}^\en_\tau
\\[4pt]  
  \rng{\til{\mu}}_{\en_\tau} = - \lang_\iio q_{\en_\tau}  +   \tfrac{1}{\slang_\iio} \sck_1
\end{array} 
&& (\lang_\iio = m\slang_\iio  \simeq \nrm{\bs{\mu}_\zr})
\end{align}
\end{small}
where the last is expressed using \eq{\til{\mu}_\en :=\tilplin_\en(\bar{\mu}) = q_\en^2 \mu_\en} rather than the cartesian momentum component \eq{\mu_\en := \plin_\en(\bar{\mu})}. The relations ``\eq{\simeq}'' hold if \eq{\bar{\mu}_\iio\in\cotsp\man{Q}_\ii{1}\subset\cotsp\bvsp{E}}, in which case \eq{\bar{\mu}_\tau \in \cotsp\man{Q}_\ii{1}} for all \eq{\tau}. 
From the coordinate solutions in Eq.\eqref{prj_KEPsols_apx}, we know that the curve \eq{\bar{\mu}_\tau} is then given in terms of \eq{\bar{\mu}_\iio = \bar{\mu}_\zr} as follows:
\begin{small}
\begin{align} \label{prj_actualKEPsols_apx}
\boxed{ \begin{array}{rrlllllllll}
    &\ptvec{q}_\tau &\!\!\!\!=\,
     \ptvec{q}_\zr \cos \tau - \tfrac{1}{\lang_\iio} (\Langb_\iio \cdot \ptvec{q}_\zr) \sin  \tau 
     & 
        \simeq \ptvec{q}_\zr \cos \tau \,+\, \tfrac{1}{\lang_\iio}\bs{\mu}_\zr^{\shrp} \sin  \tau
\\[4pt]
    &\bs{\mu}_\tau &\!\!\!\!=\, \bs{\mu}_\zr \cos \tau - \tfrac{1}{\lang_\iio} (\til{\Langb}_\iio \cdot \bs{\mu}_\zr) \sin \tau 
     &  
       \simeq \bs{\mu}_\zr \cos \tau \,-\, \lang_\iio \ptvec{q}_\zr^{\flt} \sin \tau 
\\[4pt] 
    \ptvec{q}^\ii{\perp}_\tau = q^\en_\tau \envec \,:&
      q^\en_\tau  
      &\!\!\!\! =\,  q^\en_\zr \cos{\tau} + \tfrac{1}{\lang_\iio} \til{\mu}^\en_\zr \sin{\tau}  + \tfrac{\txkbar_1}{\slang_\iio^2}(1-\cos{\tau})   
      &=  (q^\en_\zr-\tfrac{\txkbar_1}{\slang_\iio^2}) \cos{\tau} + \tfrac{1}{\lang_\iio} \til{\mu}^\en_\zr \sin{\tau}  + \tfrac{\txkbar_1}{\slang_\iio^2}
\\[4pt]
    \bs{\mu}^\ii{\perp}_\tau =  ( q_\en^\ss{-2}\til{\mu}_\en)_\tau \enform \,:&
    \til{\mu}_{\en_\tau}
      &\!\!\!\! =\,  -\lang_\iio q_{\en_\zr}  \sin{\tau} +  \til{\mu}_{\en_\zr} \cos{\tau} + \tfrac{\sck_1}{\slang_\iio} \sin{\tau}
      &=\, - \lang_\iio( q_{\en_\zr} - \tfrac{\txkbar_1}{\slang_\iio^2}) \sin{\tau}  +  \til{\mu}_{\en_\zr} \cos{\tau}  
      \;\;\;= ( q_\en^2 \mu_\en)_\tau
\end{array} } 
\end{align}
\end{small}
And the \eq{s}-parameterization of this same curve (denoted simply  \eq{\bar{\mu}_s \equiv \bar{\mu}_\tau}) is given exactly as above using \eq{\tau=\slang_\iio s}; it is an integral curve of \eq{\cf\sfb{X}^\sscr{H}} such that \eq{\pdt{\bar{\mu}}_s = \cf\sfb{X}^\sscr{H}_{\bar{\mu}_s}}. From Eq.\eqref{d_ds_KEPprj_altCords}, this implies the following: 
\begin{small}
\begin{align}
    \pdt{\bar{\mu}}_s = \cf\sfb{X}^\sscr{H}_{\bar{\mu}_s}
\;\left\{ \; \begin{array}{llllllll}
  \pdt{\ptvec{q}}_s 
     =    -\Slangb_\iio \cdot \ptvec{q}_s
     &\!\!\!\!  \simeq  \tfrac{1}{m} \bs{\mu}^{\shrp}_s 
\\[4pt]  
     \pdt{\bs{\mu}}_s  =  -\til{\Slangb}_\iio \cdot \bs{\mu}_s  
      &\!\!\!\! \simeq  -m \slang^2_\iio \ptvec{q}^\flt  
\end{array} \right.
\qquad,\qquad 
\begin{array}{llllll}
    \pdt{q}^\en_s 
   = \tfrac{1}{m} \til{\mu}^\en_s
\\[4pt]  
  \pdt{\til{\mu}}_{\en_s} = - m\slang^2_\iio q_{\en_s}  +   \sck_1
\end{array} 
&& \phantom{ (\lang_\iio = m\slang_\iio \simeq \nrm{\bs{\mu}_\zr}) }
\end{align}
\end{small}
\end{itemize}
\end{small}

\paragraph{Recovering Solutions for the Original Kepler System.} 
The integral curves of the original Kepler system, \eq{\sfb{X}^\sscr{K}} are related to those of \eq{\sfb{X}^\sscr{H}} given above by \eq{\colift\psi}, where \eq{\bar{\kappa}_{\barpt{x}} = \colift\psi(\bar{\mu}_{\barpt{q}})} is given as follows (cf.~Eq.\eqref{colift_prj_reg}-Eq.\eqref{colift_prj_reg_T*Q}):
\begin{small}
\begin{align} \label{colift_prj_KEP}
 \begin{array}{llllllll}
       \ptvec{x} \,=\,   \tfrac{1}{q^\en \nrm{\pt{q}}}\ptvec{q}\, \simeq\, \tfrac{1}{q^\en}\ptvec{q} 
       &,\quad x^\en =  \nrm{\pt{q}} \simeq 1
\\[5pt]
     \bs{\kappa} 
     =  q^\en \nrm{\pt{q}} \big( \trn{\iden}_\ii{\!\Sig}-   \hsfb{q}^\flt \otms \hsfb{q} \big) \cdot \bs{\mu} -  \til{\mu}_\en \hsfb{q}^{\flt} 
     \,\simeq\, q^\en \bs{\mu} - \til{\mu}_\en \sfb{q}^\flt
     &,\quad \kappa_\en = \bs{\mu}\cdot \hsfb{q} \simeq 0
\end{array} 
\end{align}
\end{small}
For an integral curve \eq{\bar{\mu}_\tau\in \cotsp\bvsp{E}} of \eq{\cff\sfb{X}^\sscr{H}}, then \eq{\bar{\kappa}_\tau=\colift\psi(\bar{\mu}_\tau)\in \cotsp\bvsp{E}} is an integral curve of \eq{\til{\cff}\sfb{X}^\sscr{K}}, where \eq{\til{\cff}:= \colift\psi_* \cff = \rfun^2/\slang} (that is, \eq{\bar{\mu}_\tau} and \eq{\bar{\kappa}_\tau} are just the \eq{\tau}-parameterizations of integral curves \eq{\bar{\mu}_t} and \eq{\bar{\kappa}_t} of \eq{\sfb{X}^\sscr{H}} and \eq{\sfb{X}^\sscr{K}}). 
In particular, if \eq{\bar{\mu}_\tau\in\cotsp\man{Q}_\ii{1}\subset\cotsp\bvsp{E}}, then \eq{\bar{\kappa}_\tau\in\cotsp\Sig_\ii{1}\subset\cotsp\bvsp{E}} and the closed-form solutions for \eq{\bar{\kappa}_\tau} are obtained from those of \eq{\bar{\mu}_\tau} given in Eq.\eqref{prj_actualKEPsols_apx}:
\begin{small}
\begin{align}
\boxed{\begin{array}{rllll}
     \ptvec{x}_\tau \,\simeq\, \tfrac{1}{q^\en_\tau}\ptvec{q}_\tau
     &=\, 
     \dfrac{ \ptvec{q}_\zr \cos \tau \,+\, \tfrac{1}{\lang_\iio}\bs{\mu}_\zr^{\shrp} \sin  \tau  }{ (q^\en_\zr- {\txkbar_1}/{\slang_\iio^2}) \cos{\tau} + ({\til{\mu}^\en_\zr}/{\lang_\iio})  \sin{\tau}  + {\txkbar_1}/{\slang_\iio^2}   }
\\[18pt]
     \bs{\kappa}_\tau \,\simeq\,  q^\en_\tau \bs{\mu}_\tau - \til{\mu}_{\en_\tau} \sfb{q}^\flt_\tau
     & =\, 
     \bs{\kappa}_\zr 
     \,+\, \tfrac{\txkbar_1}{\slang_\iio^2}\big( \bs{\mu}_\zr (\cos{\tau}-1) - \lang_\iio \ptvec{q}^\flt_\zr \sin{\tau} \big)
\\[2pt]
     &=\,   \bs{\kappa}_\zr \,+\, \tfrac{\txkbar_1}{\slang_\iio^2}( \bs{\mu}_\tau - \bs{\mu}_\zr) 
\end{array} }
&&
\begin{array}{llll}
      x^\en_\tau = \nrm{\ptvec{q}_\tau} = \nrm{\ptvec{q}_\zr}  \simeq 1
\\[3pt]
      \kappa_{\en_\tau} = \bs{\mu}_\tau \cdot \hsfb{q}_\tau = \bs{\mu}_\zr \cdot \hsfb{q}_\zr   \simeq 0
\\[3pt]
    \bs{\kappa}_\zr \simeq q^\en_\zr \bs{\mu}_\zr - \til{\mu}_{\en_\zr} \sfb{q}^\flt_\zr
\\[3pt]
    \lang_\iio \simeq \nrm{\bs{\mu}_\zr} 
\end{array}
\end{align}
\end{small}
where \eq{\bar{\kappa}_\tau=(\barpt{x}_\tau,\barbs{\kappa}_\tau)\in\cotsp\Sig_\ii{1}} is an integral curve of the Kepler problem. 
As usual, everything  may easily be parameterized in terms of \eq{s} using \eq{\tau=\slang_\iio s}. Everything can also be parameterized in terms of \eq{t}, but not as a closed-form expression; one must integrate \eq{t-t_\zr = \int_\zr^{s} (q^\en_s)^\ss{-2} \mrm{d} s =  \tfrac{1}{\slang_\iio}\int_\zr^{\tau} (q^\en_\tau)^\ss{-2} \mrm{d} \tau}, where \eq{q^\en_\tau = 1/\nrm{\ptvec{x}_\tau} = 1/\rfun(\ptvec{x}_\tau)}.

\phantomsection
\addcontentsline{toc}{section}{CONCLUSION}
\section*{CONCLUSION} 

This work has presented a projective transformation, lifted to a phase space symplectomorphism, which fully linearizes central force dynamics for both the Kepler-type dynamics and Manev-type dynamics.  By incorporating the latter, we demonstrate that this projective regularization method extends beyond the classic Kepler problem to include relativity-motivated corrections, while still maintaining full linearization.

We frame the initial projective point transformation as a diffeomorphism, rather than a submersion, which allows us to sidestep the usual issues of redundant dimensions and over-parameterization encountered in previous work. 
By formulating Hamiltonian dynamics in a coordinate-agnostic, tensorial framework, we establish a more general, geometric approach to the problem which not only clarifies the deeper structural properties of the transformation but also connects our results to broader themes in symplectic and Riemannian geometry, offering a more universal formulation of the regularization process. Although, the symmetries, reduction, and integrability of the transformed system warrant a more detailed treatment.
Beyond intrinsic mathematical interest, the developments presented here have applications in celestial mechanics, astrodynamics, and space mission design where algorithms benefit from underlying linear and symplectic structures in the dynamics.



\phantomsection
\addcontentsline{toc}{section}{REFERENCES}
\begin{small}

\end{small}

\begin{appendices}
\renewcommand{\appendixname}{Appx.}


\section{BASIC GEOMETRY OF PHASE SPACE, \txi{T*Q}} \label{sec:phase space}

\noindent Let \eq{\man{Q}} be some smooth \eq{n}-dim configuration manifold for some \eq{n}-\sc{dof} mechanical system. The cotangent bundle is naturally a non-compact, \eq{2n}-dim, \textit{exact} symplectic manifold,  \eq{(\cotsp \man{Q},\nbs{\omg}=-\exd\bs{\spform})}. 


\begin{notation}
    Regarding notation in this section: 
\begin{itemize}[topsep=1pt, itemsep=1pt]
    \item We use the following indexing ranges:
    \begin{small}
    \begin{align} \nonumber
          i,j,k = 1,2,\dots , n
        \qquad,\qquad 
         \fnsz{I,J,K} \,=\, 1,2,\dots, 2n
    \end{align}
    \end{small}
    \item \sloppy For any configuration coordinates, \eq{\tp{q}=(q^1,\dots,q^n):\chart{Q}{q}\to\mbb{R}^n}, the corresponding cotangent-lifted coordinates will typically be labeled \eq{\tp{z}=(\tp{q},\tp{\pf}):=\colift\tp{q}:\tsp\chart{Q}{q}\to\cotsp \mbb{R}^n\cong\mbb{R}^{2n}}, where 
     we use the same notation, \eq{q^i}, for both the original  configuration coordinates \eq{q^i \in \fun(\man{Q})} as well as the same \eq{q^i} lifted to 
     phase space, \eq{q^i\equiv \copr^* q^i \in\fun(\cotsp \man{Q})}. 
    When the \eq{q^i}-coordinate frame fields on \eq{\man{Q}} are needed, they will be denoted as \eq{\bt[q^i]\equiv \bt_i\in\vect(\man{Q})} and \eq{\btau[q^i]\equiv \btau^i\in\forms(\man{Q})}, or similar. The corresponding cotangent-lifted frame fields for \eq{(\tp{q},\tp{\pf})=\colift\tp{q}} will be denoted by any of the following: 
     \begin{small}
    \begin{align} \label{qcord_difs_T*Q}
     \begin{array}{lllll}
      \bt_i\equiv \bt[q^i] \in \vect(\man{Q})
      \\[4pt]
           \btau^i\equiv \btau[q^i] \in \forms(\man{Q}) 
     \end{array}
     &&,&&
     \begin{array}{lllll}
        \hbpart{i} \equiv \hpdii{q^i} = \coVlift{(\bt[q^i])}\,\in \vect(\cotsp \man{Q})
      \\[4pt]
         \hbdel^i\equiv \hbdel[q^i] \equiv  \dif q^i = \copr^* \btau[q^i] \,\in \formsh(\cotsp \man{Q})
     \end{array}
      &&,&&
     \begin{array}{lllll}
        \hbpartup{i} \equiv \hpdiiup{\pf_i} = \vlift{(\btau[q^i])} \,\in \vectv(\cotsp \man{Q})
      \\[4pt]
         \hbdeldn_i \equiv \hbdeldn[\pf_i] \equiv \dif \pf_i  \,\in \forms(\cotsp \man{Q})
     \end{array} 
 \end{align}
 \end{small}
 We have preemptively given the cotangent-lifted frame fields a ``hat'' (circumflex), \eq{\,\hat{\sblt}\,}, which we use for a symplectic basis (i.e., Darboux basis); we will see that all cotangent-lifted coordinates are, in fact, symplectic coordinates.  
\end{itemize}
\end{notation}

\subsection{Canonical Symplectic Structure on \txi{T*Q}}

The cotangent bundle, \eq{\cotsp \man{Q}} (phase space), of any smooth manifold, \eq{\man{Q}} (configuration space), is canonically an exact symplectic manifold. It comes naturally equipped with five important tensor fields, expressed in any local cotangent-lifted frame fields:
\begin{small}
\begin{align} \label{T*Q_forms_sum} 
\begin{array}{rllll}
     \fnsize{1-form (canonical)} &\quad \bs{\spform} \,=\, \pf_i \hbdel^i
      &\in \formsh(\cotsp \man{Q})
\\[4pt] 
      \fnsize{symplectic form (canonical)}   &\quad 
     \nbs{\omg}  :=\, -\exd \bs{\spform} \,=\, \hbdel^i \wedge \hbdeldn_i \,=\, J_\ssc{IJ}\hbdel^\ssc{I}\otimes \hbdel^\ssc{J} 
    &\in \formsex^2(\cotsp \man{Q})
\\[4pt] 
     \fnsize{Poisson bivector (canonical)} &\quad 
     \inv{\nbs{\omg}} \,=\, -\hbpart{i}\wedge\hbpartup{i} \,=\,   J^\ssc{IJ}\hbpart{\ssc{I}}\otimes \hbpart{\ssc{J}}  
    &\in \vect^2(\cotsp \man{Q})
\\[4pt] 
       \fnsize{Liouville vol.~form (canonical)}  &\quad 
     \bs{\spvol} :=\, \tfrac{n^\ii{\pm}}{n!}\nbs{\omg}^{\wedge n} \,=\,  \hbdel^1 \wedge \cdots \wedge \hbdel^{2n} 
    &\in \forms^{2n}(\cotsp \man{Q})
\\[4pt]
    \fnsize{Euler vec.~field (canonical)} &\quad 
 \bscr{P} \,:=\, \inv{\nbs{\omg}}\cdt\bs{\spform} \,=\,  \pf_i\hbpartup{i} 
     &\in \vectv(\cotsp \man{Q})
\end{array}
\end{align}
\end{small}
where \eq{\tp{z}:=(\tp{q},\tp{\pf}):= \colift\tp{q}} are any arbitrary cotangent-lifted coordinates with frame fields denoted as in Eq.\eqref{qcord_difs_T*Q}. However, the above local expressions for \eq{\nbs{\omega},\;\inv{\nbs{\omega}}}, and \eq{\bs{\spvol}} hold for any arbitrary symplectic coordinates (not necessarily cotangent-lifted). 
Note that the above are not al independent; the 1-form \eq{\bs{\spform}} defines the other four. 
We will next review the above in a bit more more detail.

\paragraph{Canonical Tensor Fields on Cotangent Bundles (Momentum Phase Space).}
Let \eq{(\chart{Q}{q},\tp{q})} be any local configuration coordinates for a smooth \textit{n}-dim configuration manifold and let  \eq{\tp{z}:=(\tp{q},\tp{\pf}):= \colift\tp{q}} be the cotangent-lifted coordinates with frame fields denoted as in Eq.\eqref{qcord_difs_T*Q}. 
The  \sbemph{canonical 1-form}\footnote{Also called the tautological 1-form, the Liouville 1-form, the Poincaré 1-form, or the symplectic potential.}, \eq{\bs{\spform}\in\formsh(\cotsp \man{Q})}, is characterized as follows: 
\begin{small}
\begin{align}
\!\!\! \begin{array}{cc}
     \fnsize{canonical}  \\[-1pt]
     \fnsize{1-form}
\end{array} \!\!\! :
 \quad\bs{\spform} \,=\, \pf_i\hbdel^i  \in\formsh(\cotsp \man{Q}) 
 &&
\left| \quad \begin{array}{rllll}
   \fnsize{$\forall \, \mu_\pt{r}\in \cotsp \man{Q}$:}&
      \bs{\spform}_{\mu_\pt{r}} = \copr^* \bs{\mu} 
      =  \mu_i \hbdel^i_{\pt{r}}   \, \in \tsp[\mu_\pt{r}]^*(\cotsp \man{Q})
\\[5pt]
     \fnsize{$\forall \, \sfb{w}\in\tsp[\mu_\pt{r}](\cotsp \man{Q})$:}& 
   \bs{\spform}_{\mu_\pt{r}} \cdt \sfb{w} = (\copr^*\bs{\mu})\cdt \sfb{w} = \bs{\mu}\cdt (\copr_* \sfb{w}) = \bs{\mu}\cdt \dif \copr_{\mu_\pt{r}} \cdt \sfb{w} = \mu_i w^i
\end{array}  \right.
\end{align}
\end{small}
The  coordinate-agnostic relation \eq{\bs{\spform}_{\mu_\pt{r}} = \copr^* \bs{\mu}} is a more appropriate \textit{definition} of \eq{\bs{\spform}}. Another common definition is that \eq{s_{\bs{\alpha}}^* \bs{\spform}= \bs{\alpha}} for any \eq{\bs{\alpha}\in\forms(\man{Q})} 
(details in footnote\footnote{\textbf{``Pullback by Sections'' Definition of $\bs{\spform}$.}  For any \eq{\bs{\alpha}\in\forms(\man{Q})}, we define the associated ``section of \eq{\cotsp \man{Q}}'' as \eq{s_{\bs{\alpha}}:\man{Q}\to\cotsp \man{Q}} such that \eq{s_{\bs{\alpha}}(\pt{r})=(\pt{r},\bs{\alpha}_{\pt{r}})} and \eq{\copr\circ s_{\bs{\alpha}}=\Id_{\man{Q}}} (i.e., \eq{s_{\bs{\alpha}}(\man{Q})\subset\cotsp \man{Q}} is the graph of \eq{\bs{\alpha}}). Then, the canonical 1-form, \eq{\bs{\spform}\in\forms(\cotsp \man{Q})}, is the \textit{unique} 1-form on \eq{\cotsp \man{Q}} with the  property \eq{ s_{\bs{\alpha}}^* \bs{\spform} = \bs{\alpha} } for \emph{all} \eq{\bs{\alpha}\in \forms(\man{Q})}. This may be taken as the definition of \eq{\bs{\spform}}.  For \eq{\nbs{\omg} :=-\exd\bs{\spform}}, it leads to
\begin{align} \label{section_1form}
 \forall\, \bs{\alpha}\in \forms(\man{Q}) \;:
  \qquad\qquad  s_{\bs{\alpha}}^* \bs{\spform}  \,=\, \bs{\alpha}  \qquad,\qquad 
       s_{\bs{\alpha}}^* \nbs{\omg}  \,=\, -\exd \bs{\alpha} 
\end{align} }).
Yet, the local expression \eq{\bs{\spform}=\pf_i\hbdel^i} is indeed valid (but only for \textit{cotangent-lifted} coordinates); if \eq{(\tp{q},\tp{\pf}):=\colift\tp{q}} and \eq{(\chart{Q}{\til{q}},\tiltp{q})} are any two sets of cotangent-lifted coordinates (with overlapping domain), then it is quick to 
verify\footnote{The relation \eq{\bs{\spform} = \pf_i \hbdel[q^i] = \til{\pf}_i\hbdel[\til{q}^i]} is easily verified by substitution of the usual transformations \eq{\pf_i = \pderiv{\til{q}^j}{q^i}\til{\pf}_j} and \eq{ \hbdel[q^i] =\pderiv{q^i}{\til{q}^j}\hbdel[\til{q}^j]}, which are true for any cotangent-lifted coordinates and frame fields). }
that  \eq{\bs{\spform} = \pf_i \hbdel[q^i] = \til{\pf}_i\hbdel[\til{q}^i]}.
The canonical 1-form is sometimes called the \textit{symplectic potential} for the following reason.  
Taking its
(negative\footnote{This is a matter of convention. Some sources instead define \eq{\nbs{\omg}=\exd\bs{\spform}=\dif \pf_i\wedge\dif q^i}. When reading other work, one must pay attention to which sign convention is being used as it will affect the sign on many other relations integral to geometry and dynamics on \eq{\cotsp \man{Q}}. The convention here agrees with \cite{abraham2008foundations,marsden2013introduction,holm2009geometric}. }) 
exterior  derivative,  \eq{-\exd \bs{\spform}}, and using the fact that  \eq{\exd \hbdel^i = \exd( \dif q^i)  = 0},  gives a closed and non-degenerate 2-form (i.e., a symplectic form) which we will denote by \eq{\nbs{\omg}}. This is the  \sbemph{canonical symplectic form} (or canonical 2-form) on \eq{\cotsp \man{Q}}, and its inverse, \eq{\inv{\nbs{\omg}}}, is the \sbemph{canonical Poisson bivector}: 
\begin{small}
\begin{align} \label{w_canon0}
\begin{array}{rr}
    \fnsize{symplectic form:}
\\[5pt]
\fnsize{Poisson bivector:}
\end{array}
\;
\begin{array}{lll}
    \nbs{\omg}  \,:=\, -\exd  \bs{\spform}   \,=\, \hbdel^i \!\wedge \hbdeldn_i \,=\, \tfrac{1}{2} J_{\ssc{IJ}}\,\hbdel^\ssc{I} \!\wedge \hbdel^\ssc{J}  
    &\in\formsex^2(\cotsp \man{Q})
\\[5pt]
  \inv{\nbs{\omg}} \,=\, - \hbpart{i} \wedge \hbpartup{i} \,=\, \tfrac{1}{2}J^\ssc{IJ} \hbpart{\ssc{I}}\wedge\hbpart{\ssc{J}}
  &\in\vect^2(\cotsp \man{Q})
\end{array} 
 &&,&&
 \begin{array}{lll}
   \cord{\ns{\txw}}{z} = J = \invtrn{J} = -\trn{J} = -\inv{J} =  -\cord{\inv{\ns{\txw}}}{z}
\end{array}
\end{align}
\end{small}
The above 2-form is symplectic (closed and non-degenerate) such that \eq{(\cotsp \man{Q},\nbs{\omg})} is a symplectic manifold. It is further an \textit{exact} symplectic manifold\footnote{Any symplectic manifold, \eq{(\man{P},\nbs{\omg})} is said to be an \textit{exact} symplectic manifold if the symplectic form is not only closed but is further exact such that \eq{\nbs{\omg}=\pm \exd \bs{\spform}} for some globally defined 1-form \eq{\bs{\spform}\in\forms(\man{P})}.  }
since \eq{\nbs{\omg}=-\exd \bs{\spform}} holds globally. 
Moreover, all cotangent-lifted coordinates,  \eq{(\tp{q},\tp{\pf})=\colift\tp{q}}, are clearly symplectic with respect to \eq{\nbs{\omg}}; it was already shown that, for  arbitrary \eq{(\tp{q},\tp{\pf})=\colift\tp{q}} and \eq{(\tiltp{q},\tiltp{\pf})=\colift\tiltp{q}}, then  \eq{\bs{\spform} = \pf_i \hbdel[q^i] = \til{\pf}_i\hbdel[\til{q}^i] } and it immediately follows that \eq{\nbs{\omg}  = -\exd \bs{\spform}= \hbdel[q^i] \!\wedge \hbdeldn[\pf_i] = \hbdel[\til{q}^i]\!\wedge \hbdeldn[\til{\pf}_i] }. It is usefull to note the following:

\begin{small}
\begin{itemize}
      \item The canonical 1-form \eq{\bs{\spform}\in\forms(\cotsp \man{Q})} is the unique 1-from satisfying \eq{s_{\bs{\alpha}}^*\bs{\spform}=\bs{\alpha}}.  However, \eq{\bs{\spform}} is \textit{not} unique as a ``symplectic potential'' in the sense that \eq{\nbs{\omg}=-\exd \bs{\spform} = -\exd(\bs{\spform}+\bs{\eta})} for any \textit{closed} 1-form \eq{\bs{\eta}\in\formscl(\cotsp \man{Q})}. That is, \eq{\bs{\spform}+\bs{\eta}} is an equivalent symplectic potential iff \eq{\exd\bs{\eta}=0}.
    \item \textit{All cotangent-lifted coordinates are \eq{\nbs{\omg}}-symplectic.} If 
\eq{\tp{q}:\chart{Q}{q}\to\mbb{R}^n} are any local configuration coordinates, then the cotangent lift, \eq{\colift\tp{q}:\cotsp \chart{Q}{q} \to \cotsp \mbb{R}^n\cong\mbb{R}^{2n} }, 
   is a (local) symplectomorphism such that \eq{\nbs{\omg}=(\colift\tp{q})^* J_{2n}}. Thus, for all \eq{(\chart{Q}{q},\tp{q})}, there is an induced \emph{symplectic} coordinate chart \eq{(\cotsp \chart{Q}{q},\colift\tp{q})}, where  \eq{\cotsp \chart{Q}{q} = \inv{\copr}(\chart{Q}{q})}. We often re-name the lifted coordinate functions \eq{\tp{z}:=(\tp{q},\tp{\pf}) := \colift\tp{q}}.
   Cotangent-lifted coordinates are actually a subset of all symplectic coordinates on \eq{(\cotsp \man{Q},\nbs{\omg})} with the additional feature that the canonical 1-form is expressed in any such coordinate basis as \eq{\bs{\spform}=\pf_i\dif q^i}. That is,  
   \begin{small}
   \begin{align} \label{eq:colift}
         (\tp{q},\tp{\pf})=\colift\tp{q} 
         \qquad \Leftrightarrow  \qquad 
         \bs{\spform}= \pf_i\dif q^i  
         \qquad 
         \Rnlarrow  \qquad \nbs{\omg}=\dif q^i\wedge\dif \pf_i
          \qquad \Leftrightarrow  \qquad 
           (\tp{q},\tp{\pf}) \;\; \fnsize{are symplectic}
   \end{align}
   \end{small}
    \item \textit{Not all \eq{\nbs{\omg}}-symplectic coordinates are cotangent-lifted.} 
   Some \eq{(\tp{s},\tp{\pi})\in\fun^{2n}(\cotsp \man{Q})} are local symplectic coordinates iff \eq{\nbs{\omg}=\dif s^i\wedge\dif \pi_i}. But this does \textit{not} imply that \eq{(\tp{s},\tp{\pi})=\colift\tp{s}} for some configuration coordinates \eq{(\chart{Q}{s},\tp{s})}.  Nor does it imply that \eq{\bs{\spform}=\pi_i\dif s^i}. In fact,   
   if  \eq{(\tp{s},\tp{\pi})} are such that the canonical 1-form can be expressed locally as  \eq{\bs{\spform} =\pi_i \dif s^i  + \dif f} for any (perhaps local) function \eq{f\in\fun(\cotsp \man{Q})}, then these coordinates are symplectic (but \textit{not} necessarily cotangent-lifted):\footnote{Eq.\eqref{dtheta_exact_T*Q} actually applies more generally for the case \eq{ \bs{\spform} = \pi_i\dif s^i  + \bs{\eta}} for any closed \eq{\bs{\eta}\in\formscl(\cotsp \man{Q})} which, by the Poincaré lemma, can always be written locally as \eq{\bs{\eta}=\dif f}.} 
   \begin{small}
   \begin{align} \label{dtheta_exact_T*Q}
         \bs{\spform} = \pi_i\dif s^i  + \dif f
         \qquad 
        \Leftrightarrow    \qquad \nbs{\omg}=\dif s^i\wedge\dif \pi_i
          \qquad \Leftrightarrow   \qquad 
           (\tp{s},\tp{\pi}) \;\; \fnsize{are symplectic}
   \end{align}
   \end{small}
  Cotangent-lifted coordinates are a subset of \eq{\nbs{\omg}}-symplectic coordinates, satisfying the above for the case
   \eq{f=0}.  
\end{itemize}
\end{small}

\vspace{1ex} 
\noindent Next, as on any symplectic manifold, the symplectic form lets us define the corresponding \sbemph{symplectic/Liouville volume form},
\eq{\bs{\spvol}\in\forms^{2n}(\cotsp \man{Q})}, in the usual way: 
\begin{small}
\begin{align}
    \bs{\spvol} :=\, \tfrac{n^\ii{\pm}}{n!} \nbs{\omg}^{\wedge n} 
    \,=\,  \mag{\cord{\txw}{\zeta}}^{\sfrac{1}{2}} \bdel^1 \wedge \cdots \wedge \bdel^{2n} \;=\; 
     \hbdel^1 \wedge \cdots \wedge \hbdel^{2n}
    &&
    n^\ii{\pm} := (-1)^{\frac{n}{2}(n-1)}
\end{align}
\end{small}
That is, in an arbitrary coordinate basis, \eq{\bs{\spvol}} has components \eq{\spvol_{\ssc{I}_1 \dots \ssc{I}_{2n}} = \mag{\cord{\txw}{\zeta}}^{\sfrac{1}{2}} \lc_{\ssc{I}_1 \dots \ssc{I}_{2n}}} and, in any symplectic  basis, it has components \eq{\hat{\spvol}_{\ssc{I}_1 \dots \ssc{I}_{2n}}=\lc_{\ssc{I}_1 \dots \ssc{I}_{2n}}} (where \eq{\lc_{\ssc{I}_1 \dots \ssc{I}_{2n}}} is the \eq{2n}-dim antisymmetric Levi-Civita \textit{symbol}). The volume form is central to the notion of divergence and Liouville's theorem in Hamiltonian mechanics. 



Lastly, on any exact symplectic manifold, \eq{(\man{P},\nbs{\omg}=-\exd\bs{\spform})},    we may define the \sbemph{Euler field} as  \eq{\bscr{P}:=\inv{\nbs{\omg}}\cdt \bs{\spform}\in\vect(\man{P})} \cite[p.171]{marsden2013introduction}. For the case \eq{\man{P}=\cotsp \man{Q}} with any cotangent-lifted coordinates \eq{(\tp{q},\tp{\pf})}, we then have the following for any \eq{\mu_\pt{r}=(\pt{r},\bs{\mu})\in\cotsp \man{Q}}:
\begin{small}
\begin{align}
     \bscr{P} \,:=\,   \inv{\nbs{\omg}}\cdt \bs{\spform} 
     \,=\, \pf_i\hbpartup{i} \in \vectv(\cotsp \man{Q})
&&,&&
      \lderiv{\bscr{P}} \bs{\spform} \,=\,  \bs{\spform} \,=\, \nbs{\omg}\cdt\bscr{P}
&&,&&
       \lderiv{\bscr{P}} \nbs{\omg} \,=\,  \nbs{\omg}
 &&,&&
    \bscr{P}_{\!\mu} =  \vlift{\bs{\mu}} = \mu_i\hbpartup{i}
\end{align}
\end{small}
where the coordinate basis  expression holds for any cotangent-lifted coordinates (for arbitrary symplectic coordinates \eq{(\tp{s},\tp{\kappa})}, generally \eq{\bs{\spform}\neq \kappa_i\dif s^i}). From the above, we see that integral curves of \eq{\bscr{P}} are determined, in any cotangent-lifted coordinates,  by 
\eq{\dot{q}^i=0} and \eq{\dot{\pf}_i = \pf_i}. That is, if \eq{\mu_t=(\pt{r}_t,\bs{\mu}_t)} with \eq{\bs{\mu}=\mu_i\btau[q^i]} is an integral integral curve of \eq{\bscr{P}} then,
\begin{small}
\begin{align}
      \dt{\mu}_t \,=\, \bscr{P}_{\!\mu_t} 
      &&  \Rightarrow &&
      \dot{\mu}_i \,=\, \mu_i
       &&  \Rightarrow &&
       \mu_i(t) \,=\, \mu_i(0)\tx{e}^{t} 
      &&  \Rightarrow &&
      \mu_t = \varep_t(\mu_\zr)= (\pt{r}_\zr,  \mrm{e}^t \bs{\mu}_\zr  )
\end{align}
\end{small}
where \eq{\varep_t} is the flow of \eq{\bscr{P}}. 
Further, for any \eq{\varphi\in\Spism(\cotsp \man{Q},\nbs{\omg})} that satisfies \eq{\bs{\spform}=\varphi^*\bs{\spform}} (i.e., \eq{\varphi=\colift\sigma} for some \eq{\sigma\in\Dfism(\man{Q})}, as discussed at Eq.~\eqref{eq:lift_taut}),  this field has the property \eq{\varphi_*\bscr{P}=\bscr{P}} such that \eq{\varphi_*(\bscr{P}_{\!\mu}) := \dif \varphi_{\mu}\cdt \bscr{P}_{\!\mu} = \bscr{P}_{\!\varphi(\mu)} }.  That is,
\begin{small}
\begin{align}
    \fnsize{if } \; \bs{\spform}=\varphi^*\bs{\spform}
    \qquad\quad \Rightarrow \qquad\quad
    \varphi_*\bscr{P}=\bscr{P}
    \qquad\quad \Rightarrow \qquad\quad
    \dif \varphi\cdt \bscr{P} = \bscr{P}_\ss{\!\varphi}
    \quad\;\;,\quad\;\;
    \varphi\circ\varep_t = \varep_t \circ\varphi
\end{align}
\end{small}

\subsection{(Co)Tangent Lifts} \label{sec:lifts_short}

\begin{notation}
    In the following, \eq{\tpr:\tsp\man{Q}\to\man{Q}} and \eq{\copr:\cotsp\man{Q}\to\man{Q}} denote the natural (co)tangent bundle projections for some \eq{n}-dim smooth manifold, \eq{\man{Q}}. 
    For some configuration coordinates \eq{\tp{q}=(q^1,\dots,q^n):\chart{Q}{q}\to \mbb{R}^n}, we denote the tangent-lifted coordinates on \eq{\tsp\man{Q}} by  \eq{\tp{\xi}=(\tp{q},\tp{v}):=\tlift\tp{q}:\tsp\chart{Q}{q}\to \tsp\mbb{R}^n}, and the cotangent-lifted coordinates on \eq{\cotsp \man{Q}}
     by \eq{ \tp{z} = (\tp{q},\tp{\pf}):= \colift\tp{q}:\cotsp \chart{Q}{q}\to \cotsp \mbb{R}^n}.\footnote{As usual, we abuse notation slightly by re-using \eq{q^i} to mean \eq{q^i\equiv q^i\circ \tpr\in\fun(\tsp\man{Q})} and also \eq{q^i\equiv q^i\circ \copr\in\fun(\cotsp \man{Q})}. That is, the same symbol, \eq{q^i}, is used for three subtly different functions: the original generalized coordinates \eq{q^i\in\fun(\man{Q})}, as well as \eq{q^i\circ \tpr\in\fun(\tsp\man{Q})} and also \eq{q^i\circ \copr\in\fun(\cotsp \man{Q})}. However, for any \eq{\pt{r}\in\man{Q}}, any \eq{(\pt{r},\sfb{u})\in\tsp\man{Q}}, and any \eq{(\pt{r},\bs{\mu})\in\cotsp \man{Q}}, it is always true that   \eq{q^i(\pt{r}) = q^i \circ \tpr(\pt{r},\sfb{u}) =  q^i\circ \copr (\pt{r},\bs{\mu})}. Thus, we will simply use \eq{q^i} for all three interpretations.}
    It should be clear from context whether \eq{q^i} is being regarded as a function on \eq{\man{Q}}, \eq{\tsp\man{Q}}, or \eq{\cotsp \man{Q}}.  
    What is less clear, and warrants greater notational clarity, is  whether \eq{\pdii{q^i}} and \eq{\bdel[q^i]=\dif q^i} are to be regarded as  the frame fields on \eq{\man{Q}}, \eq{\tsp\man{Q}}, or \eq{\cotsp \man{Q}}. To avoid ambiguity, we adopt the following notation 
    (for \eq{i=1,\dots, n}):
    The \eq{q^i} frame fields on the base are denoted \eq{\bt_i\equiv\bt[q^i]\in\vect(\man{Q})} and \eq{\btau^i\equiv\btau[q^i]\in\forms(\man{Q})}; the \eq{(q^i,v^i)} frame fields the tangent bundle are denoted \eq{\pdii{q^i},\pdii{v^i}\in \vect(\tsp\man{Q})} and \eq{\bdel[q^i]\equiv \dif q^i, \bdel[v^i]\equiv \dif v^i \in\forms(\tsp\man{Q})}; the \eq{(q^i,\pf_i)} frame fields the cotangent bundle are denoted \eq{\hbpart{i}\equiv\hpdii{q^i},\hbpartup{i}\equiv\hpdiiup{\pf_i}\in \vect(\cotsp\man{Q})} and  \eq{\hbdel^i\equiv \hbdel[q^i] \equiv \dif q^i, \hbdeldn_i\equiv \hbdeldn[\pf_i] \equiv \dif \pf_i \in\forms(\cotsp\man{Q})}.     
\end{notation}

 \noindent There are a number of different ways that maps and tensors on \eq{\man{Q}} can be ``lifted''  to a map or tensor on \eq{\tsp\man{Q}} and/or \eq{\cotsp\man{Q}}.
A simple example: a  1-form \eq{\bs{\alpha}} on \eq{\man{Q}} can be lifted to a function (0-form) \eq{\vfun^{\bs{\alpha}}} on \eq{\tsp\man{Q}} and, similarly, a vector field \eq{\sfb{w}} on \eq{\man{Q}} can be  lifted to a function \eq{\pfun^{\sfb{w}}}  on \eq{\cotsp\man{Q}}, defined as follows: 
\begin{small}
\begin{align} \label{PVfuns_def0} 
\begin{array}{lllllll}
   \bs{\alpha}\mapsto \vfun^{\bs{\alpha}}  \in \fun(\tsp\man{Q}) 
    &, \qquad 
    \vfun^{\bs{\alpha}}(\pt{u}_{\pt{r}}) :=\, \bs{\alpha}_\pt{r}\cdot\sfb{u} \,=\, \alpha_i(\pt{r})u^i
    &\qquad  \fnsize{i.e.,} \;\;  \vfun^{\bs{\alpha}} = v^i \alpha_i 
\\[4pt]
   \sfb{w} \mapsto \pfun^{\sfb{w}} \in \fun(\cotsp\man{Q})
     &, \qquad 
    \pfun^{\sfb{w}}(\mu_{\pt{r}}) :=\, \sfb{w}_\pt{r}\cdot\bs{\mu} \,=\, w^i(\pt{r})\mu_i
     &\qquad \fnsize{i.e.,} \;\; \pfun^{\sfb{w}} = \pf_i w^i 
\end{array}
\end{align}
\end{small}
In the following, we detail several other, more interesting, "lifts" that may be defined on any (co)tangent bundle.

\paragraph{(Co)Tangent Lifts of Diffeomorphisms.}
Recall that any diffeomorphism, \eq{\varphi\in\Dfism(\man{Q};\tman{Q})}, can be lifted to a a diffeomorphism of (co)tangent bundles — called the tangent lift, \eq{\tlift{\varphi}\in\Dfism(\tsp\man{Q};\tsp\tman{Q})}, and cotangent lift, \eq{\colift{\varphi}\in\Dfism(\cotsp\man{Q};\cotsp\tman{Q})} — given as follows for any \eq{\pt{u}_\pt{r}=(\pt{r},\sfb{u})\in\tsp\man{Q}} and \eq{\mu_\pt{r}=(\pt{r},\bs{\mu})\in\cotsp\man{Q}}:
\begin{small}
\begin{align} \label{Tlift_review1}
\begin{array}{rlllll}
        \tlift \varphi =   
     ( \varphi \circ \tpr ,\, \dif \varphi(\cdot,\sblt ) ) \!\!\!\!\!\!\! &:\tsp \man{Q} \to \tsp \tman{Q}
        &,\qquad 
         (\pt{r},\sfb{u}) \;\mapsto\; \tlift \varphi(\pt{r},\sfb{u})   \,=\, ( \varphi(\pt{r}), \varphi_*\sfb{u})  \,=\, ( \varphi(\pt{r}),  \dif \varphi_{\pt{r}}\cdot\sfb{u}  ) 
\\[4pt]
         \colift \varphi  =   
     ( \varphi \circ \copr ,\, \dif \inv{\varphi}_\ii{\varphi}(\sblt,\cdot ) ) \!\!\!\!\!\!\! & :\cotsp\man{Q} \to \cotsp\tman{Q}
           &,\qquad 
            (\pt{r},\bs{\mu}) \;\mapsto\; \colift \varphi(\pt{r},\bs{\mu}) 
             \,=\,
         (\varphi(\pt{r}), \varphi_* \bs{\mu}) 
         \,=\, ( \varphi(\pt{r}),\bs{\mu}\cdot \inv{(\dif \varphi_\pt{r})}) 
\end{array}
\end{align}
\end{small} 
where we have used \eq{\dif \inv{\varphi}_\ii{\varphi(\pt{r})} = \inv{(\dif \varphi_\pt{r})}}. The following properties can be verified: 
\begin{small}
\begin{align} \label{Tlift_review2}
\begin{array}{lllllll}
    \tpr\circ \tlift \varphi = \varphi \circ \tpr
    &,&
    \inv{(\tlift \varphi)} = \tlift (\inv{\varphi})
     &,&  \tlift ( \psi \circ\varphi   ) \,=\, \tlift  \psi \circ  \tlift \varphi 
\\[4pt]
 \copr\circ \colift \varphi = \varphi \circ \copr
     &,&
     \inv{(\colift \varphi)} = \colift (\inv{\varphi})
     &,&
    \colift  (\psi \circ \varphi) \,=\,   \colift \psi \circ \colift \varphi
\end{array}
\end{align}
\end{small} 
\begin{small}
\begin{itemize}[topsep=1pt]
    \item[{}]  \textit{Coordinate representations.} Let \eq{(\chart{Q}{q},\tp{q})} and \eq{(\chart{\til{Q}}{\til{q}},\tiltp{q})} be local charts such that \eq{\varphi:\man{Q}\to\tman{Q}} has coordinate representation in these charts denoted \eq{\ubar{\upvarphi} := \tiltp{q}\circ\varphi\circ\inv{\tp{q}}:\rchart{R}{q}^n \to \rchart{R}{\til{q}}^n}. Consider the  (co)tangent-lifted coordinates, \eq{\tp{\xi}:=(\tp{q},\tp{v})=\tlift\tp{q}} and \eq{\tp{z}:=(\tp{q},\tp{\pf})=\colift\tiltp{q}} (and \eq{\tiltp{\xi}} and \eq{\tiltp{z}} defined likewise from \eq{\tiltp{q}}). 
     The coordinate representations of the (co)tangent-lifted diffeomorphism, \eq{\tlift\varphi} and  \eq{\colift\varphi}, are given
    in these (co)tangent-lifted coordinates simply by \eq{\tlift \ubar{\upvarphi}} and  \eq{\colift \ubar{\upvarphi}} 
    (see footnote\footnote{The coordinate representation of $\tlift\varphi:\tsp\man{Q}\to\tsp\tman{Q}$, in tangent-lifted coordinates, is given by $\ubar{\tlift \varphi}=\tlift \tiltp{q} \circ \tlift \varphi \circ \inv{(\tlift\tp{q})}$. From the relations $\inv{(\tlift \varphi)}= \tlift \inv{(\varphi)}$ and $\tlift ( \psi \circ \varphi   ) = \tlift  \psi \circ  \tlift \varphi$,   it follows that $\ubar{\tlift\varphi} = \tlift \tp{q}\circ \tlift \varphi \circ \inv{(\tlift \tp{q})} = \tlift( \tiltp{q}\circ\varphi\circ\inv{\tp{q}}) = \tlift \ubar{\upvarphi}$. The same reasoning leads to $\ubar{\colift\varphi}  =  \colift \ubar{\upvarphi}$. }).
    That is, if \eq{\mrm{q}_\pt{r}:=\tp{q}(\pt{r})}, \eq{\cord{\tp{\mrm{u}}}{q}:=\crd{\sfb{u}}{q}}, and \eq{\cord{\muup}{q}:=\crd{\bs{\mu}}{q} } then
    \begin{small}
    \begin{align}
        \begin{array}{lll}
           \ubar{\tlift\varphi} \,=\, \tiltp{\xi}\circ \tlift\varphi\circ\inv{\tp{\xi}} =   \tlift \tiltp{q} \circ \tlift \varphi \circ \inv{(\tlift \tp{q})} \,=\, \tlift \ubar{\upvarphi}  : \tsp\rchart{R}{q}^n \to  \tsp\rchart{R}{\til{q}}^n 
              &, \qquad  
            \tlift\ubar{\upvarphi} (\mrm{q}_\pt{r}, \cord{\tp{\mrm{u}}}{q})  \,=\, ( \ubar{\upvarphi}(\mrm{q}_\pt{r}), \dif \ubar{\upvarphi}_{\mrm{q}}\cdot  \cord{\tp{\mrm{u}}}{q} ) 
      \\[4pt]
            \ubar{\colift\varphi} \,=\, \tiltp{z}\circ \colift\varphi\circ\inv{\tp{z}} =   \colift \tiltp{q} \circ \colift \varphi \circ \inv{(\colift \tp{q})} \,=\, \colift \ubar{\upvarphi} : \cotsp\rchart{R}{q}^n \to  \cotsp\rchart{R}{\til{q}}^n 
               &, \qquad  
            \colift\ubar{\upvarphi} (\mrm{q}_\pt{r}, \cord{\tp{\muup}}{q})  \,=\, ( \ubar{\upvarphi}(\mrm{q}_\pt{r}),   \cord{\tp{\muup}}{q} \cdot \inv{ (\dif \ubar{\upvarphi}_{\mrm{q}}) }  ) 
        \end{array}
    \end{align}
    \end{small}
\end{itemize}
\end{small}

\paragraph{Complete (Co)Tangent Lifts of Vector Fields.} For some \eq{\sfb{w}=w^i\bt[q^i]\in\vect(\man{Q})}, its  \sbemph{complete (co)tangent lift} (or simply \textit{complete lift}) to \eq{\vect(\tsp\man{Q})} or \eq{\vect(\cotsp\man{Q})}, is given in a local (co)tangent-lifted basis as 
\begin{small}
\begin{align} \label{complete_lift_def}
\begin{array}{llllllll}
   \Vlift{(\cdot)} : \vect(\man{Q}) \to \vect(\tsp\man{Q})
    &,\qquad 
      \sfb{w}=w^i \bt[q^i]  & \mapsto & 
      \Vlift{\sfb{w}} \,:=\, w^i\pdii{q^i}  \,+\, v^j\pderiv{w^i}{q^j} \pdii{v^i} 
      \,\in\vect(\tsp\man{Q})
      &, \qquad 
      \Vlift{(\bt[q^i])} = \pdii{q^i}
\\[5pt]
 \coVlift{(\cdot)}  : \vect(\man{Q}) \to \vect(\cotsp\man{Q})
    &,\qquad 
    \sfb{w}=w^i \bt[q^i]  & \mapsto & 
   \coVlift{\sfb{w}} \!:=\, w^i\hpdii{q^i}  \,-\, \pf_j \pderiv{w^j}{q^i}\hpdiiup{\pf_i}  \,\in\vecsp(\cotsp\man{Q})
     &, \qquad 
      \coVlift{(\bt[q^i])} = \hpdii{q^i}
\end{array} 
\qquad\quad
\end{align}
\end{small}
(where we should technically write \eq{w^i\circ\tpr} and \eq{w^i\circ\copr}) such that at any points \eq{\pt{u}_\pt{r}=(\pt{r},\sfb{u})\in\tsp\man{Q}} and \eq{\mu_\pt{r}=(\pt{r},\bs{\mu})\in\cotsp\man{Q}}:  
\begin{small}
\begin{align}
\begin{array}{llllll}
      \Vlift{\sfb{w}}_{\pt{u}_\pt{r}}  \,=\, w^i(\pt{r}) \pdii{q^i} + u^j\pderiv{w^i}{q^j}\big|_{\pt{r}}  \pdii{v^j}
       &,\quad 
       \Vlift{\sfb{w}}_r := (\Vlift{\sfb{w}})_{\pt{w}_\pt{r}} \,=\, w^i(\pt{r}) \pdii{q^i} + w^j(\pt{r}) \pderiv{w^i}{q^j}\big|_{\pt{r}} \pdii{v^j}
      &,\quad  \coVlift{\sfb{w}}_{\mu_\pt{r}}  \,=\, w^i(\pt{r})\hpdii{q^i} - \mu_j \pderiv{w^j}{q^i}\big|_{\pt{r}} \hpdiiup{\pf_i}
\end{array}
\end{align}
\end{small}
These lifts are Lie algebra homomorphisms and the images \eq{\Vlift{\vect(\man{Q})} \subset \vect(\tsp\man{Q})} and  \eq{\coVlift{\vect(\man{Q})} \subset \vect(\cotsp\man{Q})} are Lie subalgebras: 
\begin{small}
\begin{align}
    \lbrak{\Vlift{\sfb{v}}}{\Vlift{\sfb{w}}} = \Vlift{ \lbrak{\sfb{v}}{\sfb{w}} }
    \qquad,\qquad 
      \lbrak{\coVlift{\sfb{v}}}{\coVlift{\sfb{w}}} = \coVlift{ \lbrak{\sfb{v}}{\sfb{w}} }
\end{align}
\end{small}
In terms of the canonical 1-form and symplectic form, the complete cotangent lift is given by:
\begin{small}
\begin{align}
     \coVlift{\sfb{w}} \,=\, \sfb{W}^{\pfun^{\sfb{w}}} \,=\, \inv{\nbs{\omg}}(\dif \pfun^{\sfb{w}},\cdot) 
     \in\vechm(\cotsp\man{Q},\nbs{\omg})
&&
\begin{array}{lll}
     \lderiv{\coVlift{\sfb{w}}} \bs{\theta} = 0
 \\[3pt]
     \bs{\theta}\cdot \coVlift{\sfb{w}} =   \pfun^{\sfb{w}} 
\end{array}
&&
\begin{array}{lll}
    \lderiv{\coVlift{\sfb{w}}} \pfun^{\sfb{w}} =  \coVlift{\sfb{w}} \cdot \dif \pfun^{\sfb{w}} = 0
  \\[3pt]
     \lderiv{\coVlift{\sfb{v}}} \pfun^{\sfb{w}} =  \coVlift{\sfb{v}} \cdot \dif \pfun^{\sfb{w}} = \pfun^{\lbrak{\sfb{v}}{\sfb{w}}}
\end{array}
\end{align}
\end{small} 
That is, \eq{\coVlift{\sfb{w}}} is the Hamiltonian vector field for \eq{\sfb{w}}'s ``momentum function''   \eq{\pfun^{\sfb{w}}=\pf_i w^i \in\fun(\cotsp\man{Q})}, as defined in Eq.\ref{PVfuns_def0}. The above relations are derived later in  section \ref{sec:colift_TQsp} where it is shown that cotangent-lifted vector fields are Hamiltonian, and cotangent-lifted diffeomorphisms are \eq{\nbs{\omg}}-symplectomorphisms. 
\begin{small}
\begin{itemize}[topsep=2pt,noitemsep]
    \item \sbemph{(Co)Tangent-Lifted Flows.}    If \eq{\sfb{w}\in\vect(\man{Q})} has flow \eq{\sigma_t\in\Dfism(\man{Q})}, then \eq{\Vlift{\sfb{w}}\in\vect(\tsp\man{Q})} has flow \eq{\tlift\sigma_t \in \Dfism(\tsp\man{Q})} and \eq{\coVlift{\sfb{w}}\in\vect(\cotsp\man{Q})} has flow \eq{\colift\sigma_t \in \Dfism(\cotsp\man{Q})} \cite[p.493]{fecko2006differential}. That is,  the flow of the (co)tangent lift of \eq{\sfb{w}} is the (co)tangent lift of the flow of \eq{\sfb{w}}. If \eq{\pt{r}_t=\sigma_t(\pt{r}_\zr)} is some integral curve of \eq{\sfb{w}\in\vect(\man{Q})} and \eq{\bs{\Sigma}_t:=\dif \sigma_t|_{\pt{r}_\zr}}, then recall that the (co)tangent lifts \eq{\tlift{\sigma_t}} and \eq{\colift{\sigma_t}} are given by:
    \begin{small}
    \begin{align}
    \begin{array}{llllll}
         &\tlift \sigma_t = (\sigma_t\circ \tpr , \dif \sigma_t(\slot,\sblt)) 
         &,\qquad
         \tlift \sigma_t(\pt{r}_\zr,\sfb{u}_\zr) \,=\, (\sigma_t(\pt{r}_\zr), \dif \sigma_t|_{\pt{r}_\zr}\cdot\sfb{u}_\zr) \,=\, (\pt{r}_t, \bs{\Sigma}_t\cdot\sfb{u}_\zr)
    \\[5pt]
        &\colift \sigma_t = (\sigma_t\circ \copr, \dif \inv{\sigma}_t(\sblt,\slot)) 
        &,\qquad
        \colift \sigma_t(\pt{r}_\zr,\bs{\mu}_\zr) 
        \,=\, (\sigma_t(\pt{r}_\zr), \bs{\mu}_\zr \cdot \dif \inv{\sigma}_t|_{\pt{r}_t} )
        \,=\, (\pt{r}_t, \bs{\mu}_\zr \cdot \inv{\bs{\Sigma}_t}  ) 
    \end{array}
    \end{align}
    \end{small}
\end{itemize}
\end{small}

\subsection{Symplectomorphisms \& Hamiltonian Vector Fields on \txi{T*Q}} \label{sec:T*Q_CT}


On any symplectic manifold, \eq{(\man{P},\nbs{\omg})}, a symplectomorphism, \eq{\varphi\in\Spism(\man{P},\nbs{\omg})}, is any diffeomorphism which satisfies \eq{\nbs{\omg}=\varphi^*\nbs{\omg}}, and a symplectic vector field, \eq{\sfb{X}\in\vecsp(\man{P},\nbs{\omg})}, is any vector field which satisfies  \eq{\lderiv{\sfb{X}}\nbs{\omg}=0} or, equivalently,  \eq{\nbs{\omg}(\sfb{X},\slot)\in\formscl(\man{P})} is closed. 
Such a symplectic vector field is further Hamiltonian iff \eq{\nbs{\omg}(\sfb{X},\slot)\in\formsex(\man{P})} is exact in which case we often write is as \eq{\sfb{X}^h := \inv{\nbs{\omg}}(\dif h,\slot)} for some  \eq{h\in\fun(\man{P})}. 
The canonically symplectic cotangent bundle, \eq{(\cotsp \man{Q},\nbs{\omg})}, is further an \textit{exact} symplectic manifold with \eq{\nbs{\omg}:=-\exd\bs{\spform}}. In such cases, many properties involving the 2-form \eq{\nbs{\omg}} can be re-phrased in terms of the 1-form, \eq{\bs{\spform}}. 
For instance, \eq{\Spism}, \eq{\vecsp\,}, and \eq{\vechm\,} may be defined in in terms of ether \eq{\nbs{\omg}} or \eq{\bs{\spform}} 
    as:\footnote{The definition of \eq{\Spism(\cotsp\man{Q},\nbs{\omg})} in Eq.\eqref{spT*Q_exact} is more specifically for \textit{auto}symplectomorphisms from \eq{\cotsp \man{Q}} to itself.  More generally, for any diffeomorphism \eq{\varphi\in\Dfism(\cotsp \man{Q};\cotsp \tman{Q})} between two cotangent bundles, then \eq{\varphi} is a symplectomorphism (with respect to the canonical symplectic forms)  iff \eq{\varphi^* \tbs{\spform}=\bs{\spform} + \bs{\eta}} for any \eq{\bs{\eta}\in\formscl(\cotsp \man{Q})}: 
    \begin{align} \label{colift_sp_1form_T*Q}
         \begin{array}{ccc}
       (\,  \varphi^* \tbs{\spform} = \bs{\spform} + \bs{\eta} \;\; \fnsize{and} \;\; \exd \bs{\eta} = 0 \,)
      \end{array}
          \qquad  \Leftrightarrow
           \qquad  \varphi^* \til{\nbs{\omg}} =\nbs{\omg} 
           \qquad \Leftrightarrow\qquad \varphi\in\Spism(\cotsp \man{Q};\cotsp \tman{Q})
    \end{align} }
\begin{small}
\begin{align} \label{spT*Q_exact}
\begin{array}{rlllllll}
     \Spism(\cotsp \man{Q},\nbs{\omg})  \,:=
      &\!\!\! \{ \varphi\in\Dfism(\cotsp \man{Q}) \;|\; \varphi^* \nbs{\omg} =\nbs{\omg}  \} 
      &\!\!\!=\; \{ \varphi\in\Dfism(\cotsp \man{Q}) \;|\; \varphi^* \bs{\spform} - \bs{\spform} \in \formscl(\cotsp \man{Q}) \}  
\\[4pt]
     \vecsp(\cotsp \man{Q},\nbs{\omg})
     \,:= &\!\!\! \{ \sfb{X} \in \vect(\cotsp \man{Q}) \;|\;    \nbs{\omg}(\sfb{X},\slot) \in \formscl(\cotsp \man{Q}) \} 
     &\!\!\!=\; \{ \sfb{X} \in \vect(\cotsp \man{Q}) \;|\;   \lderiv{\sfb{X}} \bs{\spform} \in \formscl(\cotsp \man{Q}) \} 
 \\[4pt]
      \vechm(\cotsp \man{Q},\nbs{\omg})
      \,:= &\!\!\! \{ \sfb{X} \in \vect(\cotsp \man{Q}) \;|\;    \nbs{\omg}(\sfb{X},\slot) \in \formsex(\cotsp \man{Q}) \} 
     &\!\!\!=\; \{ \sfb{X} \in \vect(\cotsp \man{Q}) \;|\;   \lderiv{\sfb{X}} \bs{\spform} \in \formsex(\cotsp \man{Q}) \} \,\subseteq  \vecsp
\end{array}
\end{align}
\end{small}
\begin{small}
\begin{itemize}
    \item \textit{Derivation.} We quickly verify that the usual definitions in terms of \eq{\nbs{\omg}} imply those given above in terms of \eq{\bs{\spform}}. First, using the fact that the exterior derivative and pullback of forms commute, as do the Lie and exterior derivative of forms, it is easy to show 
     (see footnote\footnote{The exterior derivative of forms commutes with the pullback and so, if \eq{\bs{\spform} = \varphi^* \bs{\spform}}, then \eq{\varphi^* \nbs{\omg} = \varphi^*(-\exd \bs{\spform}) = -\exd (\varphi^* \bs{\spform}) = -\exd\bs{\spform} = \nbs{\omg}}. Next, the exterior and Lie derivative of forms also commute and so we have \eq{\lderiv{\sfb{X}}\nbs{\omg} = - \lderiv{\sfb{X}} \exd \bs{\spform} = -\exd \lderiv{\sfb{X}} \bs{\spform}} such that if  \eq{\lderiv{\sfb{X}} \bs{\spform}=0} then \eq{\lderiv{\sfb{X}}\nbs{\omg}=0}. })
    that if \eq{\varphi^* \bs{\spform} = \bs{\spform} } then \eq{\varphi^* \nbs{\omg} = \nbs{\omg} } such that \eq{\varphi} is a symplectomorphism and, similarly, if \eq{ \lderiv{\sfb{X}} \bs{\spform} = 0 } then \eq{\lderiv{\sfb{X}} \nbs{\omg} = 0 } such that \eq{\sfb{X}} is a symplectic vector field (it is actually further Hamiltonian). The converse of these relations are not necessarily true:
    \begin{small}
    \begin{align}
    \begin{array}{llllllll}
         \varphi^* \bs{\spform} = \bs{\spform} 
         & \rnlarrow &
         \varphi^*\nbs{\omg}=\nbs{\omg}
         & \Leftrightarrow &
         \varphi \in \Spism
     \\[4pt] 
         \lderiv{\sfb{X}} \bs{\spform} = 0 
         & \rnlarrow &
         \lderiv{\sfb{X}} \nbs{\omg} = 0 
         &\Leftrightarrow &
         \sfb{X} \in \vecsp
    \end{array}
    \end{align}
    \end{small}
    That is, \eq{\varphi^* \bs{\spform} = \bs{\spform} } and \eq{\lderiv{\sfb{X}} \bs{\spform} = 0} are sufficient, but not necessary, conditions for \eq{\varphi\in\Spism} and \eq{\sfb{X}\in\vecsp} (the latter is actually further sufficient for \eq{\sfb{X}\in\vechm}). Let us first consider \eq{\Spism}. 
    As previously mentioned, \eq{\bs{\spform}} gives the same symplectic form as \eq{\bs{\spform} + \bs{\eta}} so long as \eq{\exd \bs{\eta}=0}.  As such,  \eq{\varphi \in \Spism} iff \eq{\varphi^* \bs{\spform} = \bs{\spform} + \bs{\eta}}  for any \textit{closed} \eq{\bs{\eta}\in\formscl} (including the  special case \eq{\bs{\eta}=0}):
    \begin{small}
    \begin{align} \label{socks}
       (\,  \varphi^* \bs{\spform} = \bs{\spform} + \bs{\eta}   \;\; \fnsize{and} \;\; \exd \bs{\eta} = 0 \,)
     \qquad \Leftrightarrow \qquad 
    \varphi^* \bs{\spform} - \bs{\spform} \in\formscl
         \qquad \Leftrightarrow \qquad 
         \varphi^*\nbs{\omg}=\nbs{\omg}
         \qquad  \Leftrightarrow \qquad 
         \varphi \in \Spism 
    \end{align}
    \end{small}
    The above is a necessary and sufficient condition for \eq{\varphi} to be a symplectomorphism. 
    Now, what are the conditions on \eq{\lderiv{\sfb{X}}\bs{\spform}} such that a vector field \eq{\sfb{X}} is symplectic or, further, Hamiltonian? 
    For arbitrary \eq{\sfb{X}\in\vect}, Cartan's identity leads to 
    \begin{small}
    \begin{align} \label{Xhm_1form}
        \lderiv{\sfb{X}}\bs{\spform} \,=\, \exd (\sfb{X}\cdot\bs{\spform}) \,+\, \sfb{X}\cdot\exd\bs{\spform} \,=\,  \dif (\sfb{X}\cdot\bs{\spform}) - \nbs{\omg}(\sfb{X},\cdot)
        &&
         \nbs{\omg}(\sfb{X},\cdot) \,=\, \dif (\sfb{X}\cdot\bs{\spform}) \,-\,  \lderiv{\sfb{X}}\bs{\spform}
    \end{align}
    \end{small}
    Now, \eq{\sfb{X}\in\vecsp} iff \eq{\nbs{\omg}(\sfb{X},\cdot)\in\formscl} and \eq{\sfb{X}\in\vechm} iff \eq{\nbs{\omg}(\sfb{X},\cdot)=\dif h\in\formsex} for some function \eq{h}. Thus, from the above, the necessary and sufficient conditions for some \eq{\sfb{X}} to be symplectic or, further, Hamiltonian are that  \eq{\lderiv{\sfb{X}} \bs{\spform}} be closed or, further, exact: 
    \begin{small}
    \begin{align} \label{Xhm_1form_exact}
    \begin{array}{cccccc}
         \lderiv{\sfb{X}} \bs{\spform} \in \formscl
         & \Leftrightarrow  &
         \nbs{\omg}(\sfb{X},\cdot) \in\formscl
         &\Leftrightarrow &
         \sfb{X} \in \vecsp
    \\[2pt]
         \Uparrow  \nDownarrow 
    \\[2pt]
         \lderiv{\sfb{X}} \bs{\spform}\in \formsex
         & \Leftrightarrow  &
          \nbs{\omg}(\sfb{X},\cdot) \in\formsex
         &\Leftrightarrow &
         \sfb{X} \in \vechm
    \end{array}
    \end{align}
    \end{small}  
    The condition \eq{ \lderiv{\sfb{X}} \bs{\spform}\in \formsex} 
    implies  that \eq{\lderiv{\sfb{X}} \bs{\spform} = \dif f} for some function \eq{f}. We then see from  Eq.\eqref{Xhm_1form} that this implies \eq{\sfb{X}=\sfb{X}^h} is Hamiltonian for the function \eq{h:= \sfb{X}\cdt\bs{\spform} - f}. This includes the special case \eq{\lderiv{\sfb{X}} \bs{\spform} = 0 }  (i.e., \eq{f=0}):
    \begin{small}
     \begin{align} \label{ldiv_sp_exact_T^*Q}
            \lderiv{\sfb{X}} \bs{\spform} = \dif f \in\formsex 
          \quad \Leftrightarrow \quad
           \left.\begin{array}{ccccc}
               \sfb{X}=\sfb{X}^h := \inv{\nbs{\omg}}(\dif h,\slot) \in \vechm 
            \\[3pt]
                \fnsize{for } \;  h:= \sfb{X}\cdt\bs{\spform} - f
           \end{array} \qquad \right|\qquad
           \lderiv{\sfb{X}} \bs{\spform} = 0 
           \quad \Leftrightarrow \quad
           \begin{array}{ccccc}
               \sfb{X}=\sfb{X}^h 
            \\[3pt]
                \fnsize{for } \; h:= \sfb{X}\cdt\bs{\spform}  
           \end{array}
          \quad, \quad \fnsize{and } \;  \lderiv{\sfb{X}} \bscr{P} = 0    
    \end{align}
    \end{small}
\end{itemize}
\end{small}

\subsubsection{Cotangent Lifts are Symplectic} \label{sec:colift_TQsp}

 Eq.~\eqref{spT*Q_exact}-Eq.\eqref{colift_sp_1form_T*Q} above are actually valid on \textit{any} exact symplectic manifold (including cotangent bundles). 
Since we are here concerned specifically with \eq{(\cotsp \man{Q},\nbs{\omg}=-\exd\bs{\spform})}, let us then revisit the cotangent lifts of  diffeomorphisms and vector fields on \eq{\man{Q}} which we recall are defined as follows:
\begin{small}
\begin{itemize}
    \item \sloppy The cotangent lift of any \eq{\sig\in\Dfism(\man{Q})} is given by \eq{\colift \sig  =  (\sig\circ\copr,\dif \inv{\sig}_\ii{\sig}(\sblt,\slot))\in\Dfism(\cotsp \man{Q})}  such that, for any  \eq{(\pt{r},\bs{\mu})\in\cotsp \man{Q}}, then
     \eq{\colift\sig (\pt{r},\bs{\mu}) = (\sig(\pt{r}),\invtrn{(\dif\sig_\pt{r})}\cdt\bs{\mu})}. Or, more compactly, \eq{\colift \sig = (\copr^*\sig, \sig_*)} and  \eq{\colift\sig (\pt{r},\bs{\mu}) = (\sig(\pt{r}),\sig_*\bs{\mu})}. 
    \item The (complete) cotangent lift of any \eq{\sfb{u}=u^i\bt[q^i]\in\vect(\man{Q})} is a vector field \eq{\eq{\coVlift{\sfb{u}}\in\vect(\cotsp \man{Q})}} defined such that if \eq{\sig_t\in\Dfism(\man{Q})} is the flow of \eq{\sfb{u}} then \eq{\colift\sig_t} is the flow of \eq{\coVlift{\sfb{u}}}. It is expressed in any cotangent-lifted coordinate basis as \eq{\coVlift{\sfb{u}}=u^i\hpdii{q^i} - \pf_j\pderiv{u^j}{q^i}\hpdiiup{\pf_i}}.
\end{itemize}
\end{small}
Now,  for any  \eq{\sig\in\Dfism(\man{Q})} and any \eq{\sfb{u}\in\vect(\man{Q})}, it can be shown that \eq{\colift\sig} is a symplectomorphism and that \eq{\coVlift{\sfb{u}}} is not just symplectic but is further Hamiltonian for the function \eq{\pfun^{\sfb{u}}=\pf_i u^i\in\fun(\cotsp\man{Q})}:  
\begin{small}
\begin{align} \label{spT*Q_lifts}
\begin{array}{rllllllll}
      \forall \, \sig\in\Dfism(\man{Q}):
    &\quad  
        \colift \sig^* \bs{\spform} = \bs{\spform}
     &\quad  ,   &\quad 
          \colift \sig^* \nbs{\omg} = \nbs{\omg}
    &\quad  ,   &\quad 
        \colift \sig \in \Spism (\cotsp \man{Q},\nbs{\omg})
    &\quad  ,   &\quad 
        \colift\sig_* \bscr{P} =\bscr{P}
\\[4pt]
        \forall \, \sfb{u}\in\vect(\man{Q}) :
   &\quad 
   \lderiv{\coVlift{\sfb{u}}}\bs{\spform} = 0
    &\quad  ,  &\quad
    \nbs{\omg}(\coVlift{\sfb{u}},\slot) = \dif \pfun^{\sfb{u}}
     &\quad  ,  &\quad 
     \coVlift{\sfb{u}} 
     \in\vechm(\cotsp \man{Q},\nbs{\omg})
\end{array} 
\end{align}
\end{small}
where the relation \eq{\lderiv{\coVlift{\sfb{u}}}\bs{\spform} = 0 \,\Rightarrow \, \nbs{\omg}(\coVlift{\sfb{u}},\slot) = \dif \pfun^{\sfb{u}}} follows from  Eq.\eqref{Xhm_1form} along with the fact that \eq{\bs{\spform}\cdt \coVlift{\sfb{u}} =\pfun^{\sfb{u}} = \pf_i u^i\in \fun(\cotsp \man{Q})}. 
In fact, the converse is true as well; if  \eq{\Sigma\in\Spism(\cotsp \man{Q},\nbs{\omg})} satisfies \eq{\Sigma^*\bs{\spform}=\bs{\spform}}, then \eq{\Sigma=\colift\sig} for some \eq{\sig\in\Dfism(\man{Q})} \cite[p.166]{marsden2013introduction}. 
Similarly, if \eq{\sfb{U}\in\vecsp(\cotsp \man{Q},\nbs{\omg})} satisfies  \eq{\lderiv{\sfb{U}}\bs{\spform} = 0}, then \eq{\sfb{U}=\coVlift{\sfb{u}}=\inv{\nbs{\omg}}(\dif \pfun^{\sfb{u}},\slot)} for some \eq{\sfb{u}\in\vect(\man{Q})}. That is,
\begin{small}
\begin{align} \label{spT*Q_lifts1}
\begin{array}{ccccccccccc}
        \Sigma = \colift \sig  
    &\quad  \Leftrightarrow   &\quad 
        \Sigma^* \bs{\spform} = \bs{\spform}
     &\quad  \Rnlarrow   &\quad 
     \Sigma \in \Spism(\cotsp \man{Q},\nbs{\omg})
      &\quad  \Leftrightarrow   &\quad 
        \Sigma ^* \nbs{\omg} = \nbs{\omg}
\\[4pt]
      \sfb{U} = \coVlift{\sfb{u}} 
    &\quad  \Leftrightarrow   &\quad 
   \lderiv{\sfb{U}}\bs{\spform} = 0
    &\quad  \Rnlarrow   &\quad 
    \sfb{U} \in\vechm(\cotsp \man{Q},\nbs{\omg})
     &\quad  \Rnlarrow   &\quad
    \lderiv{\sfb{U}} \nbs{\omg} = 0
\end{array} 
\end{align}
\end{small}
\begin{footnotesize}
\begin{itemize}[nosep]
    \item[{}] \textit{Derivation of $\lderiv{\sfb{U}}\bs{\spform} = 0  \,\Rightarrow \, \sfb{U} = \coVlift{\sfb{u}} $.}  
    We already showed that \eq{\lderiv{\coVlift{\sfb{u}}}\bs{\spform} = 0} always holds. We now show the converse; that \eq{\lderiv{\sfb{U}}\bs{\spform} = 0} implies \eq{\sfb{U} = \coVlift{\sfb{u}}}. 
     For any \eq{\sfb{U}=U^\ssc{I}\hpdii{z^\ssc{I}} = U^\ss{q^i}\hpdii{q^i} + U_\ss{\pf_i}\hpdiiup{\pf_i}\in\vect(\cotsp \man{Q})}, recall that \eq{\lderiv{\sfb{U}}z^\ssc{I}=U^\ssc{I}} and \eq{\lderiv{\sfb{U}}\hbdel[z^\ssc{I}] = \dif U^\ssc{I}}. 
    Using these relations, along with Leibniz's rule, we find \eq{\lderiv{\sfb{U}}\bs{\spform}=\lderiv{\sfb{U}}(\pf_i\hbdel[q^i])} is given as follows:
     \begin{itemize}[nosep]
       \item[{}]  \eq{\lderiv{\sfb{U}}\bs{\spform} \,=\, \lderiv{\sfb{U}}(\pf_i\hbdel[q^i])  \,=\, ( \lderiv{\sfb{U}} \pf_i) \hbdel[q^i] \,+\, \pf_j 
     \lderiv{\sfb{U}} \hbdel[q^j]  
        \,=\, U_\ss{\pf_i} \hbdel[q^i] \,+\, \pf_j \dif U^\ss{q^j}
        \,=\,
        ( U_\ss{\pf_i} + \pf_j\pderiv{U^\ss{q^j}}{q^i}) \hbdel[q^i] \,+\, \pf_j\pderiv{U^\ss{q^j}}{\pf_i}\hbdeldn[\pf_i] }  
    \end{itemize}
    Such that, if \eq{ \lderiv{\sfb{U}}\bs{\spform}=0}, then  this implies \eq{ U_\ss{\pf_i} = - \pf_j\pderiv{}{q^i}U^\ss{q^j}} and \eq{ \pf_j\pderiv{}{\pf_i}U^\ss{q^j}=0}. The second relation means  \eq{ \mu_j(\pderiv{}{\pf_i}U^\ss{q^j})_{\mu_\pt{r}}=0}  must hold for for all arbitrary \eq{\mu_\pt{r}\in\cotsp \man{Q}} such that it further implies \eq{\pderiv{}{\pf_i} U^\ss{q^j}=0}. In other words \eq{U^\ss{q^j}} is a \textit{basic} functions such that \eq{U^\ss{q^j} = \copr^* u^j\in\fun(\cotsp \man{Q})} for some \eq{u^j\in\fun(\man{Q})}. 
    That is, in any cotangent-lifted coordinate basis, \eq{\sfb{U}} is of the form \eq{\sfb{U} = u^i\hpdii{q^i} - \pf_j\pderiv{u^j}{q^i}\hpdiiup{\pf_i}} which is precisely the local expression for \eq{\coVlift{\sfb{u}}}:  
    \begin{itemize}[nosep]
        \item[{}] \eq{ \lderiv{\sfb{U}}\bs{\spform}=0 \qquad \Rightarrow \qquad 
        U^\ss{q^j} = \copr^* u^j \;\; \fnsize{ and } \;\; U_\ss{\pf_i} = - \pf_j\pderiv{u^j}{q^i}
        \qquad \fnsize{i.e., } \; \sfb{U} = \coVlift{\sfb{u}} }
    \end{itemize}
    \sloppy An alternative derivation: using Eq.\eqref{Xhm_1form}, we see that \eq{\lderiv{\sfb{U}}\bs{\spform}=0} leads to \eq{\sfb{U} = -\inv{\nbs{\omg}}\cdt \dif (\bs{\spform}\cdt\sfb{U})} where, in any cotangent-lifted coordinate basis,  \eq{\dif (\bs{\spform}\cdt\sfb{U}) = \dif (\pf_j U^\ss{q^j} ) = \pf_j \pderiv{U^\ss{q^j}}{q^i} \hbdel[q^i] + ( U^\ss{q^i} + \pf_j \pderiv{U^\ss{q^j}}{\pf_i})\hbdeldn[\pf_i]}. This leads to
     \begin{itemize}[nosep]
     \item[{}] \eq{\lderiv{\sfb{U}}\bs{\spform}=0 
         \qquad \Rightarrow  \qquad
         \sfb{U} = -\inv{\nbs{\omg}}\cdt \dif (\bs{\spform}\cdt\sfb{U})
          \qquad \Rightarrow  \qquad
          \sfb{U} = U^\ss{q^i}\hpdii{q^i} + U_\ss{\pf_i}\hpdiiup{\pf_i}
          \,=\, 
          ( U^\ss{q^i} + \pf_j \pderiv{U^\ss{q^j}}{\pf_i}) \hpdii{q^i} \,-\,  \pf_j \pderiv{U^\ss{q^j}}{q^i} \hpdiiup{\pf^i} }
    \end{itemize}
    Equating components verifies again that \eq{\lderiv{\sfb{U}}\bs{\spform}=0 } implies \eq{U^\ss{q^j} = \copr^* u^j} is basic and   \eq{U_\ss{\pf_i} = - \pf_j\pderiv{u^j}{q^i}} such that \eq{\sfb{U}=\coVlift{\sfb{u}}} for some \eq{\sfb{u}\in\vect(\man{Q})}.  
    \item[{}] $\;\;$
\end{itemize}
\end{footnotesize}

\begin{small}
\begin{notesq}
The above relations for \eq{\colift{\sigma}} hold more generally with \eq{\sig:\man{Q}\to\tman{Q}}  a diffeomorphism between any two manifolds, with \eq{\cotsp \man{Q}} and \eq{\cotsp \tman{Q}} equipped with the canonical 1-forms and symplectic forms.  
That is, \eq{\colift \sig \in \Spism(\cotsp \man{Q}; \cotsp \tman{Q})} for all \eq{\sig\in\Dfism(\man{Q};\tman{Q})}.  It further holds that \eq{\bs{\spform}=\Sigma^*\tbs{\spform}} iff \eq{\Sigma=\colift\sig} for some \eq{\sig\in\Dfism(\man{Q};\tman{Q})} \cite[p.166]{marsden2013introduction}:
    \begin{small}
    \begin{align} \label{eq:lift_taut}
    \begin{array}{ccccccccccc}
             \Sigma = \colift \sig  
        &\quad  \Leftrightarrow   &\quad 
            \Sigma^* \tbs{\spform} = \bs{\spform}
         &\quad  \Rnlarrow   &\quad 
         \Sigma ^* \til{\nbs{\omg}} = \nbs{\omg}
           &\quad  \Leftrightarrow   &\quad 
         \Sigma \in \Spism(\cotsp \man{Q}; \cotsp \tman{Q})
    \end{array} 
    \end{align}
    \end{small} 
\end{notesq}
\end{small}

\subsubsection{More on Canonical Transformations (Symplectic Coordinate Transformations)}

\paragraph{Cotangent-Lifted Coordinate Transformations are Canonical Transformations.}
Above, we showed that the cotangent lift of a diffeomorphism is always a symplectomorphism. The ``passive'' analog of this is that all cotangent-lifted coordinates are symplectic. Or, equivalently, any transformation of cotangent-lifted coordinates is a canonical transformation.
Recall that a canonical transformation is simply a transition function between any symplectic coordinates (i.e., a symplectic coordinate transformation). 
For any point transformation,  \eq{\tp{q}\leftrightarrow \tiltp{q}}, of configuration coordinates on \eq{\man{Q}}, the cotangent lift is a canonical transformation of cotangent-lifted coordinates on \eq{\cotsp \man{Q}} (this follows immediately from Eq.~\eqref{eq:colift} and Eq.~\eqref{eq:lift_taut}).  In more detail:  let \eq{\tp{z}=(\tp{q},\tp{\pf}) =\colift\tp{q}} be any cotangent-lifted coordinates  and consider a configuration coordinate transformation given by a transition function, \eq{\ns{c}_\ss{\til{q}q}:=\tiltp{q}\circ \inv{\tp{q}}: \rchart{R}{q}^n \to \rchart{R}{\til{q}}^n},
which we treat as \eq{\tiltp{q}=\ns{c}_\ss{\til{q}q}(\tp{q})}.\footnote{The transition function is a map between open subsets of coordinate space, \eq{\ns{c}_\ss{\til{q}q}:=\tiltp{q}\circ \inv{\tp{q}}: \rchart{R}{q}^n \to \rchart{R}{\til{q}}^n}. But, when we write \eq{\tiltp{q}=\ns{c}_\ss{\til{q}q}(\tp{q})}, we are treating \eq{\ns{c}_\ss{\til{q}q}} as map between coordinate functions, \eq{\ns{c}_\ss{\til{q}q}: \fun^{n}(\chart{Q}{q}) \to \fun^n(\chart{Q}{\til{q}})}. The notation \eq{\tiltp{q}=\ns{c}_\ss{\til{q}q}(\tp{q})} ``really means'' \eq{\tiltp{q}=\ns{c}_\ss{\til{q}q}\circ \tp{q}}, which is indeed true.}
It then follows from Eq.\eqref{eq:lift_taut} that the cotangent lift, \eq{\colift \ns{c}_\ss{\til{q}q}: \cotsp \rchart{R}{q}^n \to  \cotsp  \rchart{R}{\til{q}}^n}, is a local symplectomorphism between open subsets of \eq{(\cotsp \mbb{R}^n\cong\mbb{R}^{2n},J)}. Thus, if we define new coordinates by \eq{\tiltp{z}:=\colift \ns{c}_\ss{\til{q}q}(\tp{z}) } then these \eq{\tiltp{z}} are also symplectic and are, in fact, the cotangent-lifted coordinates \eq{\tiltp{z}= \colift\tiltp{q}} 
(derivation in footnote\footnote{Taking the cotangent lift of \eq{\ns{c}_\ss{\til{q}q}:=\tiltp{q}\circ \inv{\tp{q}}} gives \eq{\colift \ns{c}_\ss{\til{q}q}= \colift(\tiltp{q}\circ \inv{\tp{q}})= \colift\tiltp{q}\circ \colift(\inv{\tp{q}}) =\colift\tiltp{q}\circ \inv{(\colift{\tp{q}})} }. But \eq{\colift{\tp{q}}=:\tp{z}} and \eq{\colift{\tiltp{q}}=:\tiltp{z}} are, by definition, local cotangent-lifted coordinates on \eq{\cotsp \man{Q}} and so   \eq{\colift \ns{c}_\ss{\til{q}q} =\colift\tiltp{q}\circ \inv{(\colift{\tp{q}})} = \tiltp{z}\circ \inv{\tp{z}} = \ns{c}_\ss{\til{z} z}} is precisely the transition function for these cotangent lifted coordinates. Further, since cotangent lifted coordinates are always symplectic, the cotangent lift/transition function \eq{\colift \ns{c}_\ss{\til{q}q}  = \ns{c}_\ss{\til{z} z}}  is a canonical transformation. 
}).
In other words, the cotangent lift of a  transition function, \eq{\ns{c}_\ss{\til{q}q}\in\Dfism(\rchart{R}{q}^{n};\rchart{R}{\til{q}}^{n})}, is precisely the transition function for the corresponding cotangent-lifted coordinates (which are always symplectic with respect to the canonical symplectic form) such that \eq{\colift \ns{c}_\ss{\til{q}q} = \ns{c}_\ss{\til{z}z}\in\Spism(\cotsp \rchart{R}{q}^{n};\cotsp \rchart{R}{\til{q}}^{n})}. That is, 
\begin{small}
\begin{align} \label{CT_lifted}
\begin{array}{rllll}
     \fnsize{given:} &   \ns{c}_\ss{\til{q}q}:=\tiltp{q}\circ\inv{\tp{q}} \in \Dfism(\rchart{R}{q}^{n};\rchart{R}{\til{q}}^{n})
\\[5pt]
     \fnsize{then:} &  \ns{c}_\ss{\til{z}z}  = \colift \ns{c}_\ss{\til{q}q} 
     \in \Spism(\cotsp \rchart{R}{q}^{n};\cotsp \rchart{R}{\til{q}}^{n})
\end{array}
    &&
\begin{array}{lllll}
        \tiltp{q} = \ns{c}_\ss{\til{q}q}(\tp{q}) 
\\[4pt]  
 \tiltp{\pf} = \tp{\pf} \cdt \dif \inv{\ns{c}_\ss{\til{q}q}} \equiv \tp{\pf} \cdt \pderiv{\tp{q}}{\tiltp{q}}
\\[4pt]
       \tiltp{z} = \colift \ns{c}_\ss{\til{q}q}(\tp{z}) = \colift\tiltp{q}
 \end{array}
 \qquad \Leftrightarrow \qquad 
  \begin{array}{llllll}
        \tp{q} = \inv{\ns{c}_\ss{\til{q}q}}(\tiltp{q})
 \\[4pt]  
      \tp{\pf} = \ttp{\pf} \cdt \dif \ns{c}_\ss{\til{q}q} \equiv  \tiltp{\pf} \cdt \pderiv{\tiltp{q}}{\tp{q}}
\\[4pt]
        \tp{z} =  \colift\inv{\ns{c}_\ss{\til{q}q}}(\tiltp{z}) =\colift\tp{q}
 \end{array}  
\end{align}
\end{small}
where \eq{\tp{z}=(\tp{q},\tp{\pf})} and \eq{\tiltp{z}=(\tiltp{q},\tiltp{\pf})}. 
We will often refer to such transformations of cotangent-lifted coordinates as a \textit{canonical point transformation} or a  \textit{cotangent-lifted point transformation} (which is shorter than ``canonical transformation induced by a point a transformation''). These are the subset of canonical transformations for which \eq{\pf_i\dif q^i = \til{\pf}_i \dif \til{q}^i}. That is, for any canonical point transformation as above, then: 
\begin{small}
\begin{align}
\begin{array}{cccc}
     \bs{\spform}= \pf_i\dif q^i=\til{\pf}_i\dif \til{q}^i
&\Rnlarrow  &\;\;
    \nbs{\omg}=\dif q^i\wedge \dif \pf_i=\dif \til{q}^i\wedge\dif \til{\pf}_i
\\
    \fnsize{(canonical \textit{point} transforamtion)}
    & &
    \fnsize{(general canonical transforamtion)}
\end{array} 
\end{align}
\end{small}
\begin{small}
\begin{itemize}[nosep]
    \item A (type-2) \textit{generating function} for the canonical point transformation in Eq.\eqref{CT_lifted} is given by  \eq{F(\tp{q},\tiltp{\pf})= \ns{c}_\ss{\til{q}q}(\tp{q}) \cdt \tiltp{\pf}\equiv \tiltp{q}(\tp{q})\cdt \tiltp{\pf}}. 
\end{itemize}
\end{small}


\paragraph{Non-Cotangent-Lifted Symplectomorphisms.}
Not every \eq{\varphi\in\Spism(\cotsp \man{Q};\cotsp \tman{Q})} need be the cotangent lift of a diffeomorphism (only those which satisfy \eq{\varphi^*\tbs{\spform}=\bs{\spform}}). More generally, \eq{\varphi} need only satisfy the following:   
\begin{small}
 \begin{align} \label{colift_T*Q_1form}
     \begin{array}{ccc}
   (\,  \varphi^* \tbs{\spform} = \bs{\spform} + \bs{\alpha} \;\; \fnsize{and} \;\; \exd \bs{\alpha} = 0 \,)
  \end{array}
      \qquad  \Leftrightarrow
       \qquad  \varphi^* \til{\nbs{\omg}} =\nbs{\omg} 
       \qquad \Leftrightarrow\qquad \varphi\in\Spism(\cotsp \man{Q},\cotsp \tman{Q})
 \end{align}
 \end{small}
 for some \eq{\bs{\alpha}\in\formscl(\cotsp \man{Q})}. Any such \eq{\varphi} is not necessarily a cotangent lift (unless \eq{\bs{\alpha}=0}). 
The following is an example of a non-cotangent-lifted symplectomorphism: 
\begin{small}
\begin{itemize}[itemsep=1pt,topsep=1pt]
\item \sbemph{Fiber Translations.}
Consider some \eq{\bs{\alpha}\in \forms(\man{Q})} and the  (auto)diffeomorphism \eq{ t_{\bs{\alpha}}\in\Dfism(\cotsp \man{Q})}, called a \textit{fiber translation}, defined by  \eq{t_{\bs{\alpha}}:(\pt{r},\bs{\mu}) \mapsto (\pt{r}, \bs{\mu} + \bs{\alpha}_{\pt{r}})}.
Then, it can be shown that \eq{ t_{\bs{\alpha}}^* \bs{\spform} = \bs{\spform} + \copr^* \bs{\alpha}} and \eq{t_{\bs{\alpha}}^* \nbs{\omg}= \nbs{\omg} - \copr^* \exd \bs{\alpha}}. Thus,  \eq{t_{\bs{\alpha}}} is a symplectomorphism iff \eq{\bs{\alpha}} is \emph{closed} (\eq{\exd\bs{\alpha}=0}). 
\begin{small}
\begin{align}
       \fnsize{for } \;\; t_{\bs{\alpha}}:(\pt{r},\bs{\mu}) \mapsto (\pt{r}, \bs{\mu} + \bs{\alpha}_{\pt{r}}) \;,  
&&
 \fnsize{if }\; \exd \bs{\alpha} = 0 \qquad \Leftrightarrow \qquad 
  \nbs{\omg} \,=\, t_{\bs{\alpha}}^* \nbs{\omg} 
 \qquad \Leftrightarrow \qquad
t_{\bs{\alpha}}\in\Spism(\cotsp \man{Q},\nbs{\omg})
\end{align}
\end{small}
In particular,  \eq{t_{\dif f} \in\Spism(\cotsp \man{Q},\nbs{\omg}) } for any \eq{f\in \fun(\man{Q})}. However, \eq{t_{\bs{\alpha}}} is \textit{not} a cotangent-lift and \eq{\bs{\spform}\neq t_{\bs{\alpha}}^*\bs{\spform} = \bs{\spform} + \copr^* \bs{\alpha} }.
    \item \sbemph{Momenta Coordinate Translations.}
    This is the  passive view of the above. Let \eq{\tp{z}=(\tp{q},\tp{\pf}) =\colift\tp{q}} be cotangent-lifted coordinates and consider a coordinate transformation \eq{\psiup:(\tp{q},\tp{\pf})\mapsto (\tp{q},\tp{\pf}+\tp{\alpha})=:(\tp{q},\tp{\kappa})} for some \eq{n} functions \eq{\alpha_i\in\fun(\man{Q})} (treated as \eq{\alpha_i\circ\copr\in\fun(\cotsp \man{Q})}). That is, \eq{\tp{q}} is unchanged and the momenta coordinates transform as \eq{\tp{\kappa}=\tp{\pf}+\tp{\alpha} \leftrightarrow \tp{\pf} =\tp{\kappa}-\tp{\alpha} }. The canonical 1-form and symplectic form are given in the new coordinates by
     \begin{small}
    \begin{align}
        \kappa_i := \pf_i + \alpha_i(\tp{q})
        &&
         \bs{\spform} \,=\, \pf_i\dif q^i \,=\, (\kappa_i - \alpha_i)\dif q^i 
         &&
         \nbs{\omg} \,=\, \dif q^i\wedge \dif \pf_i \,=\,  \dif q^i\wedge \dif \kappa_i \,-\,  \pderiv{\alpha_i}{q^j}\dif q^i \wedge \dif q^j 
    \end{align}
    \end{small}
    Thus, \eq{(\tp{q},\tp{\kappa}):=(\tp{q},\tp{\pf}+\tp{\alpha})} is a canonical transformation iff   \eq{\pderiv{\alpha_j}{q_i}-\pderiv{\alpha_i}{q_j}=0} (and \eq{\pderiv{\alpha_i}{\pf_j}=0}, which was assumed). This always holds if \eq{\alpha_i=\pderiv{f}{q^i}} for any \eq{f\in\fun(\man{Q})}. That is,  \eq{(\tp{q},\tp{\pf})\mapsto (\tp{q},\tp{\pf}+\pderiv{f}{\tp{q}})} is a canonical (but not point) transformation for any \eq{f\in\fun(\man{Q})}. 
        \item The above view are related, with the ``passive'' view stated equivalently as follows. For any closed \eq{\bs{\alpha}=\alpha_i\btau[q^i]\in\formscl(\man{Q})} and any cotangent-lifted coordinates, \eq{(\tp{q},\tp{\pf}):\cotsp \chart{Q}{q}\to\mbb{R}^{2n}}, then \eq{(\tp{q},\tp{\kappa}):=(\tp{q},\tp{\pf})\circ t_{\bs{\alpha}}=(\tp{q},\tp{\pf}+\cord{\tp{\alpha}}{q})} are symplectic, but not cotangent-lifted, coordinates (where \eq{\cord{\tp{\alpha}}{q}=(\alpha_1,\dots,\alpha_n)} are the \eq{\btau[q^i]}-components of \eq{\bs{\alpha}}). And closedness ensures that, at least locally, \eq{\bs{\alpha}=\dif f} for some \eq{f\in\fun(\man{Q})}.
\end{itemize}
\end{small}


\paragraph{Canonical Transformations \& Generating Functions.}
  Recall from Eq.\eqref{dtheta_exact_T*Q} that local coordinates \eq{(\tp{s},\tp{\pi})\in\fun^{2n}(\cotsp \man{Q})} are \eq{\nbs{\omg}}-symplectic (but not necessarily cotangent-lifted) iff \eq{\nbs{\omg}=\dif s^i\wedge\dif \pi_i} or, equivalently, \eq{\bs{\spform}=\pi_i \dif s^i + \dif f} for some perhaps-local function \eq{f\in\fun(\cotsp \man{Q})} (and \eq{(\tp{s},\tp{\pi})} are cotangent-lifted iff \eq{f=0}). Technically, \eq{(\tp{s},\tp{\pi})} are also symplectic if \eq{\bs{\spform}} can be expressed as \eq{\bs{\spform}=-s^i\dif \pi_i + \dif f} but we ignore this for now,  assuming \eq{\bs{\spform}=\pi_i \dif s^i + \dif f}. 
  Recall that a canonical transformation is simply a transition function between any local symplectic coordinates (i.e., a symplectic coordinate transformation). Thus, if \eq{(\tp{s},\tp{\pi})} and  \eq{(\tiltp{s},\tiltp{\pi})} are any symplectic coordinates (with overlapping domains) then
\begin{small}
\begin{align}
    \bs{\spform} \,=\,  \pi_i\dif s^i + \dif f \,=\, \til{\pi}_i\dif \til{s}^i \,+\, \dif \til{f} 
    \qquad,\qquad 
    \nbs{\omg} = - \exd\bs{\spform}  \,=\, \dif s^i\wedge\dif \pi \,=\,  \dif \til{s}^i\wedge\dif \til{\pi}_i 
\end{align}
\end{small}
for some perhaps-local functions \eq{f,\til{f}\in\fun(\cotsp\man{Q})}. From the above expression for \eq{\bs{\spform}}, it follows that any canonical transformation \eq{(\tp{s},\tp{\pi})\leftrightarrow (\tiltp{s},\tiltp{\pi})} between arbitrary symplectic coordinates must satisfy the following for some perhaps-local function \eq{F}:
\begin{small}
\begin{align} \label{dtheta_exact_T*Q_2}
      \exd (\pi_i\dif s^i \,-\, \til{\pi}_i\dif \til{s}^i  ) \,=\, 0
      \qquad \Rightarrow \qquad
     \pi_i\dif s^i \,-\, \til{\pi}_i\dif \til{q}^i \,=\, \dif F  
\end{align}
\end{small}
Examples:
\begin{small}
\begin{itemize}
   \item \sloppy\textit{Exchange transformation.}
   For any symplectic coordinates \eq{(\tp{s},\tp{\pi})},  the exchange transformation \eq{(\tp{s},\tp{\pi})\mapsto (\tiltp{s},\tiltp{\pi}):=(\tp{\pi},-\tp{s}) } is canonical, leading to \eq{\bs{\spform}=\pi_i\dif s^i + \dif f = -\til{s}^i \dif \til{\pi}_i+ \dif \til{f}}). The is generated by a (type-1) generating function  \eq{F(\tp{s},\tiltp{s}) = \tp{s}\cdt\tiltp{s}}. 
     \item \textit{Cotangent-lifted point transformations.} As already  discussed, transformations of cotangent-lifted coordinates are of the form \eq{(\tp{q},\tp{\pf})\mapsto (\tiltp{q},\tiltp{\pf}) := (\ns{c}_\ss{\til{q}q}(\tp{q}),\tp{\pf}\cdt \pderiv{\tiltp{q}}{\tp{q}})} for some given \eq{\tiltp{q}=\ns{c}_\ss{\til{q}q}(\tp{q})}. These are a special subset of general canonical transformations, having the property \eq{\bs{\spform}=\pf_i\dif q^i = \til{\pf}_i\dif \til{q}^i}. Such transformations are generated by a (type-2) generating function  \eq{F(\tp{q},\tiltp{\pf})= \ns{c}_\ss{\til{q}q}(\tp{q}) \cdt \tiltp{\pf} \equiv \tiltp{q}(\tp{q})\cdt\tiltp{\pf}}.  
\end{itemize}
\end{small}

\section{HAMILTONIAN DYNAMICS WITH NONCONSERVATIVE FORCES} \label{sec:noncon}



\begin{small}
\begin{notation}
    For all the following, we consider a system with \eq{n}-dim (pseudo)Riemannian configuration manifold, \eq{(\man{Q},\sfg)}, and consider a Hamiltonian system on the cotangent bundle, \eq{(\cotsp\man{Q},\nbs{\omg},\mscr{H})} with \eq{\nbs{\omg}=-\exd\bs{\spform}\in\forms^2(\cotsp \man{Q})} the canonical symplectic form, as well as the equivalent Hamiltonian system on the tangent bundle, \eq{(\tsp\man{Q},\nbs{\varpi}^\sscr{L},E^\sscr{L})}, where \eq{E^\sscr{L}=\bscr{V}\cdt\dif\mscr{L}-\mscr{L}} is the `energy' of some specified Lagrangian \eq{\mscr{L}\in\fun(\tsp\man{Q})}, and where \eq{\nbs{\varpi}^\sscr{L}= -\exd\bs{\lagform}=\Upsilon^*\nbs{\omg}\in\forms^2(\tsp\man{Q})} is the Lagrange 2-form form, with \eq{\bs{\lagform}^\sscr{L} = \vlift{\iden}(\dif\mscr{L},\slot)} the Lagrange 1-form. We will often drop the superscript \eq{(\slot)^\sscr{L}} when there is no danger of confusion. 
    We assume \eq{\mscr{L}} is regular (\eq{\det \ppderiv{\mscr{L}}{\tp{v}}{\tp{v}} \neq 0}) such that \eq{\nbs{\varpi}} is symplectic.  The upper/lower indices in this section are \textbf{\textit{not}} indicative of a musical isomorphism from a metric or a symplectic form.
\end{notation}
\end{small}

\noindent Consider a mechanical system, subject to external forces, whose configuration evolves on a smooth manifold, \eq{\man{Q}}.
There are three ``standard'', and \textit{distinct}, settings in which we may formulate the dynamics:
\begin{small}
\begin{enumerate}
\item on the (pseudo)Riemannian configuration manifold \eq{(\man{Q},\sfg)},
        \begin{small}
        \begin{itemize}[nosep]
            \item Conservative forces are modeled as exact 1-forms  on \eq{\man{Q}}. \eq{\sfg^\shrp} gives the corresponding vector field.  
             \item Nonconservative forces are generally not viewed as tensor fields on \eq{\man{Q}} (though they can be viewed as \eq{\cotsp[\cdt]\man{Q}}-valued tensors). 
        \end{itemize}
        \end{small}
\item  on the canonically-symplectic cotangent bundle (phase space), \eq{(\cotsp \man{Q},\nbs{\omg})}, 
    \begin{small}
        \begin{itemize}[nosep]
            \item Conservative forces are modeled as \textit{basic} horizontal exact 1-forms. \eq{\nbs{\omg}^\shrp} gives a vertical Hamiltonian vector field.
             \item Nonconservative forces are modeled as horizontal, non-exact, 1-forms.  \eq{\nbs{\omg}^\shrp} gives a vertical, non-Hamiltonian, vector field.  
        \end{itemize}
        \end{small}
\item on the \eq{\mscr{L}}-adapted symplectic tangent bundle (velocity phase space),  \eq{(\tsp\man{Q},\nbs{\varpi})}.
        \begin{small}
        \begin{itemize}[nosep]
            \item Conservative forces are modeled as \textit{basic} horizontal exact 1-forms. \eq{\nbs{\varpi}^\shrp} gives a vertical Hamiltonian vector field.
             \item Nonconservative forces are modeled as horizontal, non-exact, 1-forms.  \eq{\nbs{\varpi}^\shrp} gives a vertical, non-Hamiltonian, vector field.  
        \end{itemize}
        \end{small}
\end{enumerate}
\end{small}
\noindent Note the classifications of, exact, closed, horizontal, and vertical are important for any discussion of nonconservative forces. Therefore, we will first quickly review some relations regarding horizontal 1-forms (i.e., semi-basic 1-forms) and vertical vector fields. We assume the reader is familiar with exact and closed forms.

\subsection{Horizontal Forms \& Vertical Vector Fields}

\paragraph{Horizontal 1-Forms \& Vertical Vector Fields on $\tsp\man{Q}$.} First, recall that a vertical vector field, \eq{\sfb{F}\in\vectv(\tsp\man{Q})}, is one for which \eq{\vlift{\iden}\cdt\sfb{F}=0} where \eq{\vlift{\iden}=\pdii{v^i}\otimes \bdel[q^i]\in\tens^1_1(\tsp\man{Q})} is the vertical endomorphism (this also means \eq{\dif \tpr \cdt\sfb{F}=0}). Equivalently, for any tangent-lifted coordinates \eq{(\tp{q},\tp{v})=\tlift\tp{q}}, \eq{\sfb{F}} may be written as \eq{F=F^\ss{v^i}\pdii{v^i}}. A horizontal 1-form (i.e., semi-basic 1-form), \eq{\bs{\lambda}\in\formsh(\tsp\man{Q})}, is one for which \eq{\bs{\lambda}\cdt\sfb{F}=0} for any vertical vector field \eq{\sfb{F}}. Equivalently, one for which \eq{\bs{\lambda}\cdt\vlift{\iden}=0}. Equivalently, \eq{\bs{\lambda}} may be expressed in any tangent-lifted basis as \eq{\bs{\lambda}=\lambda_i \bdel[q^i]}. That is, 
\begin{small}
\begin{align} \label{VHvect_TQ_review}
\begin{array}{llllll}
    \vectv(\tsp\man{Q}) \,:=\, \big\{ \sfb{F} \in \vect(\tsp\man{Q})\;\big|\;  
 \vlift{\iden}\cdt\sfb{F} = 0 \big\}   \,=\, \ker \dif \copr
 \\[4pt]
    \formsh(\tsp\man{Q}) \,:=\,  \big\{  \bs{\lambda} \in\forms(\tsp\man{Q}) \;\big|\;   \bs{\lambda} \cdt \vlift{\iden} = 0  \big\}
\end{array}
\qquad 
  \scrsize{$\begin{array}{cc}
     \text{tangent-lifted}  \\[-2pt]
     \Longrightarrow  \\[-2pt]
     \text{coordinates} 
    \end{array}$ }
\qquad
\begin{array}{llllll}
    \sfb{F} \,=\, F^\ss{v^i}\pdii{v^i}
\\[4pt]
    \bs{\lambda} = \lambda_i \bdel[q^i]
\end{array}
\end{align}
\end{small}
Recall also that a \textit{basic} \eq{k}-form is any horizontal  \eq{\bar{\bs{\eta}}\in\formsh^k(\tsp\man{Q})} with the additional property that \eq{\bar{\bs{\eta}}= \tpr^* \bs{\eta} } for some \eq{\bs{\eta}\in\forms^k(\man{Q})}. Likewise, a basic function is any \eq{\bar{f}\in\fun(\tsp\man{Q})} such that \eq{\bar{f}=\tpr^*f} for some \eq{f\in\fun(\man{Q})} (i.e, \eq{\pderiv{\bar{f}}{v^i}=0}). The components of a basic \eq{k}-form are basic functions.

Now,  the \textit{symplectic} musical isomorphism on  $(\tsp\man{Q},\nbs{\varpi})$, when restricted to the above subsets of vectors and 1-forms, is further an isomorphism between vertical vector fields and horizontal 1-forms; 
\textit{any} vertical vector field, \eq{\sfb{F}=F^\ss{v^i}\pdii{v^i}\in\vectv(\tsp\man{Q})}, gives rise to a horizontal 1-form, \eq{\nbs{\varpi}\cdt \sfb{F} = L_{ij}F^\ss{v^j} \bdel[q^i]}. Likewise, \textit{any} horizontal 1-form, \eq{\bs{\lambda}=\lambda_i\bdel[q^i]}, gives rise to a vertical vector field, \eq{\inv{\nbs{\varpi}}\cdt\bs{\lambda}=L^{ij}\lambda_j \pdii{v^i}}. That is, 
\begin{small}
\begin{align} \label{VH_iso_tan}
\begin{array}{llllll}
     &\; \nbs{\varpi}(\slot,\sblt) :  \vectv(\tsp\man{Q}) \to \formsh(\tsp\man{Q})
    &\qquad 
     \sfb{F}=F^\ss{v^i}\pdii{v^i}  \; \mapsto\;   \nbs{\varpi}\cdt\sfb{F} = L_{ij} F^\ss{v^j} \bdel[q^i]  
 \\[3pt] 
     &\inv{\nbs{\varpi}}(\slot,\sblt): \formsh(\tsp\man{Q}) \to   \vectv(\tsp\man{Q})
     &\qquad
      \bs{\lambda} = \lambda_i \bdel[q^i] \; \mapsto \; \inv{\nbs{\varpi}} \cdt \bs{\lambda} = L^{ij} \lambda_j \pdii{v^i}   
      &,\qquad L_{ij}:=\ppderiv{\mscr{L}}{v^j}{v^i}
\end{array}
\end{align}
\end{small}
Note we have assumed \eq{\mscr{L}} is regular such that \eq{L^{ij}} exists. 
Furthermore, from \eq{\bs{g}=g_{ij}\btau[q^i]\otimes\btau[q^j]\in\tens^0_2(\man{Q})}, we define the Sasaki metric, \eq{\barsfb{g}:= \sfg\oplus\sfg = g_{ij}( \bdel[q^i]\!\otimes\bdel[q^j] + \bDel[v^i]\!\otimes\bDel[v^j] )  \in \tens^0_2(\tsp\man{Q})}, where any direct sum, \eq{a\oplus b}, is always respect to the horizontal-vertical decomposition. The \textit{metric} musical isomorphism on \eq{(\tsp\man{Q},\barsfb{g})} is as follows when restricted to  vertical/horizontal vector fields and 1-forms
\begin{small}
\begin{align}
\begin{array}{llll}
     \barsfb{g}:\vectv(\tsp\man{Q}) \to  \formsv(\tsp\man{Q}) 
  &\qquad \sfb{F}=F^\ss{v^i}\pdii{v^i} \; \mapsto\; \barsfb{g}\cdt\sfb{F} \,=\, g_{ij}F^\ss{v^j} \bDel[v^i] 
\\[4pt]
 \barsfb{g}:\vecth(\tsp\man{Q}) \to  \formsh(\tsp\man{Q}) 
   &\qquad   \sfb{H}=H^i\sfbsc{d}_\ss{\!q^i} \;\mapsto \; \barsfb{g} \cdt\sfb{H} \,=\, g_{ij}H^j \bdel[q^i] 
\\[8pt]
     \inv{\barsfb{g}}:\formsv(\tsp\man{Q}) \to \vectv(\tsp\man{Q}) 
  &\qquad \bs{\chi}=\chi_i \bDel[v^i] \; \mapsto\;  \inv{\barsfb{g}}\cdt\bs{\chi} \,=\, g^{ij}\chi_j \pdii{v^i} 
\\[4pt]
 \inv{\barsfb{g}}: \formsh(\tsp\man{Q}) \to \vecth(\tsp\man{Q}) 
   &\qquad  \bs{\lambda}=\lambda_i \bdel[q^i]  \;\mapsto \; \inv{\barsfb{g}} \cdt \bs{\lambda} \,=\, g^{ij}\lambda_j \sfbsc{d}_\ss{\!q^i}
\end{array}
\end{align}
\end{small}

\paragraph{Horizontal Forms \& Vertical Vector Fields on $\cotsp\man{Q}$.}
 Vertical vector fields and horizontal 1-forms on \eq{\cotsp\man{Q}} are defined by
 the following relations with the canonical 1-form and 2-form: 
\begin{small}
\begin{align} \label{VHvect_T*Q_review}
    \begin{array}{llllll}
    \vectv(\cotsp\man{Q}) \,:=\, \big\{ \sfb{F} \in \vect(\cotsp\man{Q})\;\big|\;  \bs{\spform}\cdt \sfb{F}=0 \big\} \,=\, \ker \dif \copr
 \\[4pt]
    \formsh(\cotsp\man{Q}) \,:=\, \big\{  \bs{\lambda} \in\forms(\cotsp\man{Q}) \;\big|\; \inv{\nbs{\omg}}(\bs{\lambda},\bs{\spform}) =0     \big\}  
\end{array}
\qquad
  \scrsize{$\begin{array}{cc}
     \text{cotangent-lifted}  \\[-2pt]
     \Longrightarrow  \\[-2pt]
     \text{coordinates} 
    \end{array}$ }
\qquad
\begin{array}{llllll}
    \sfb{F} \,=\,  F_i \hbpartup{i} 
\\[4pt]
    \bs{\lambda} = \lambda_i \hbdel^{i} 
\end{array}
\end{align}
\end{small}
where \eq{F_i} means \eq{F^\ss{\pf_i}} and \eq{\hbpartup{i}} means \eq{\hpdii{\pf_i}}. \textit{Basic} forms on \eq{\cotsp \man{Q}} are defined the same as they were are on \eq{\tsp\man{Q}} following Eq.\eqref{VHvect_TQ_review}.  
The \textit{symplectic} musical isomorphism on \eq{(\cotsp\man{Q},\nbs{\omg})}, when restricted to the above subsets of vectors and 1-forms, gives an isomorphism between vertical vector fields and horizontal 1-forms: 
\begin{small}
\begin{align} \label{VH_iso_cotan}
\begin{array}{llllll}
     &\; \nbs{\omg}(\slot,\sblt) :  \vectv(\cotsp\man{Q}) \to \formsh(\cotsp\man{Q})
    &\qquad 
     \sfb{F} = F_i \hbpartup{i}  \; \mapsto\;   \nbs{\omg}\cdt\sfb{F}  \,=\, F_i \hbdel^{i}   
 \\[3pt] 
     & \inv{\nbs{\omg}}(\slot,\sblt): \formsh(\cotsp\man{Q}) \to   \vectv(\cotsp\man{Q})
     &\qquad
      \bs{\lambda} = \lambda_i \hbdel^{i} \; \mapsto \; \inv{\nbs{\omg}} \cdt \bs{\lambda} = \lambda_i \hbpartup{i}    
\end{array}
\end{align}
\end{small}
Further special cases of the above will be important later on.  For any \eq{\bs{\lambda}\in\forms(\cotsp \man{Q})}, let us denote the \eq{\inv{\nbs{\omg}}}-induced vector field by \eq{\sfb{X}^\ss{\bs{\lambda}}:=\inv{\nbs{\omg}} \cdt\bs{\lambda}\in\vect(\cotsp \man{Q})}. We wish to know the properties of \eq{\sfb{X}^\ss{\bs{\lambda}}} for the special cases that \eq{\bs{\lambda}} is exact,  horizontal (as above), basic, or basic \textit{and} exact. This is summarized as follows for any \eq{h\in\fun(\cotsp \man{Q})}, any  \eq{\bs{\alpha}\in\forms(\man{Q})}, and any \eq{f\in\fun(\man{Q})}:
\begin{small}
\begin{align}
\sfb{X}^\ss{\bs{\lambda}}:=\inv{\nbs{\omg}} \cdt\bs{\lambda}
\quad \left\{ \quad 
\begin{array}{rrllllll}
 \fnsize{if exact }& \bs{\lambda} = -\dif h &\!\!\! \in \formsex
    &\quad\Rightarrow&\quad
   \sfb{X}^\ss{\bs{\lambda}} \,=\, \sfb{X}^h 
     & \in \vechm
\\[4pt]
 \fnsize{if horizontal }& \bs{\lambda} &\!\!\! \in \formsh
    &\quad\Rightarrow&\quad
   \sfb{X}^\ss{\bs{\lambda}} 
    &\in \vectv
\\[4pt]
       \fnsize{if also basic }& \bs{\lambda} = \copr^* \bs{\alpha}  &\!\!\! \in \formsbh 
     &\quad\Rightarrow&\quad
   \sfb{X}^\ss{\bs{\lambda}}  = \vlift{\bs{\alpha}}  
    &\in \vectbv 
\\[4pt]
     \fnsize{if basic and exact }& \bs{\lambda} = -\copr^* \dif f   &\!\!\! \in  \forms_\ss{\mrm{bh}} \cap \formsex 
     &\quad\Rightarrow&\quad
   \sfb{X}^\ss{\bs{\lambda}} = -\vlift{\dif f} = \sfb{X}^{\hat{f}} 
     &\in \vectbv  \cap \vechm
\end{array} \right.
\end{align}
\end{small}
where, in the last line, \eq{\hat{f}:=\copr^*f}. Note that any horizontal and exact 1-form is necessarily also basic  such that, in the last line we could have said for any horizontal exact 1-form. The negative signs in the above are simply to maintain compatibility with our notation for Hamiltonian vector fields, \eq{\sfb{X}^h=\dif h\cdt \inv{\nbs{\omg}} =-\inv{\nbs{\omg}} \cdt \dif h }.

Furthermore, the \textit{metric} musical isomorphism on \eq{(\cotsp \man{Q},\hsfb{g})}, using the dual Sasaki metric, which is given in a \eq{\nab}-adapted basis as \eq{\hsfb{g} = g_{ij} \hbdel[q^i]\!\otimes\hbdel[q^j] + g^{ij}\bs{\Delta}_i\!\otimes\bs{\Delta}_i  \in \tens^0_2(\cotsp \man{Q})}, is as follows when restricted to  vertical/horizontal vector fields and 1-forms
\begin{small}
\begin{align}
\begin{array}{llll}
     \hsfb{g}:\vectv(\cotsp \man{Q}) \to  \formsv(\cotsp \man{Q}) 
  &\qquad \sfb{F}=F_i\hbpartup{i} \; \mapsto\; \hsfb{g}\cdt\sfb{F} \,=\, g^{ij}F_j \bs{\Delta}_i 
\\[4pt]
 \hsfb{g}:\vecth(\cotsp \man{Q}) \to  \formsh(\cotsp \man{Q}) 
   &\qquad   \sfb{H}=H^i\sfbsc{d}_\ss{\!q^i} \;\mapsto \; \hsfb{g} \cdt\sfb{H} \,=\, g_{ij}H^j \hbdel^i 
\\[8pt]
     \inv{\hsfb{g}}:\formsv(\cotsp \man{Q}) \to \vectv(\cotsp \man{Q})  
  &\qquad   
\\[4pt]
 \inv{\hsfb{g}}: \formsh(\cotsp \man{Q}) \to \vecth(\cotsp \man{Q}) 
   &\qquad    
\end{array}
\end{align}
\end{small}

\subsection{Conservative Forces} \label{sec:force_con}

Before discussing nonconservative forces, we should first clarify what is meant by a ``conservative'' force. 

\paragraph{Conservative Forces on $(\man{Q},\sfg)$.}
 For a system with \eq{n}-dim configuration manifold, \eq{\man{Q}}, \textit{conservative forces} (also called \textit{potential} forces or, sometimes, \textit{monogenic} forces) are naturally viewed as an \textit{exact} 1-form on \eq{\man{Q}}. That is, they are forces which are obtained from some potential function, \eq{U\in\fun(\man{Q})}, as \eq{\bs{\alpha}^\ss{U} :=-\dif U \in\formsex(\man{Q})}.   
\begin{small}
\begin{itemize}
    \item \rmsb{relation to the classical view.}  A conservative force 1-form on \eq{\man{Q}} is exact, \eq{\bs{\alpha}^\ss{U}=-\dif U}, and therefore closed, \eq{\exd\bs{\alpha}^\ss{U}=0 }.
    Recall from undergraduate mechanics that, when we consider motion on \eq{\Evec^3} — and do not discriminate between 1-forms and vectors (due to exclusive use of orthonormal bases) —  then a conservative force is one that does no work around a closed path and thus, using some integral identities, we conclude it has zero curl. This is often written as \eq{\dif \times \bs{\alpha} = 0} where \eq{\dif} is treated as a sort of 1-form  (one usually sees \eq{\nab} rather than \eq{\dif}, but we reserve the former for the covariant derivative), which is simply the 3-dim version of \eq{\dif \wedge \bs{\alpha} \equiv \exd \bs{\alpha} = 0}. 
    (In \eq{\Evec^3}, the relation between \eq{\bs{\alpha}\times\bs{\beta}} and \eq{\bs{\alpha}\wedge\bs{\beta}} is simply \eq{\bs{\alpha}\times\bs{\beta} =\hdge{(\bs{\alpha}\wedge\bs{\beta})}}, where \eq{\star} is the Hodge operator defined from the Euclidean metric volume form.)
\end{itemize}
\end{small}
\noindent \sloppy If \eq{\man{Q}} has (pseudo)Riemannian metric \eq{\sfg\in\tens^1_1(\man{Q})} — often the kinetic energy metric — then the conservative force 1-form, \eq{\bs{\alpha}^\ss{U}=-\dif U \in\formsex(\man{Q})}, gives rise to the conservative force vector field using the metric musical isomorphism, \eq{\sfb{f}^\ss{U} :=\inv{\sfg} ( \bs{\alpha}^\ss{U}) = -\inv{\sfg}(\dif U)\in\vect(\man{Q})}.  ``Newton's law'' then takes the form \eq{\del{\dt{\sfb{r}}} \dt{\sfb{r}} = \sfb{f}_{\!\pt{r}}^\ss{U} \,}  or, equivalently, \eq{\del{\dt{\sfb{r}}}\bs{\mu}= \bs{\alpha}^\ss{U}_{\!\pt{r}} } with \eq{\bs{\mu}=\sfg(\dt{\sfb{r}})\in\tsp[\pt{r}]^*\man{Q}} the kinematic momentum along the curve \eq{\pt{r}_t\in\man{Q}}. Here, \eq{\nab} is the Levi-Civita connection (covariant derivative) which allows us to view the acceleration along \eq{\pt{r}_t} as a tangent vector, \eq{\del{\dt{\sfb{r}}} \dt{\sfb{r}}\in\tsp[\pt{r}]\man{Q}}, and formulate most things in terms of \eq{\botimes^r_s \tsp[\bdt]\man{Q}}-valued tensor fields.
However, this is not the approach taken for the symplectic formulation of dynamics on \eq{\cotsp\man{Q}} or \eq{\tsp\man{Q}}, which we now describe.

\paragraph{Conservative Forces on $(\cotsp\man{Q},\nbs{\omg})$.}
For the dynamics on  \eq{(\cotsp\man{Q},\nbs{\omg})}, the potential \eq{U\in\fun(\man{Q})} is regarded as a momentum-independent function on \eq{\cotsp\man{Q}} and is technically a new function, \eq{\hat{U}:=\copr^*U = U\circ\copr \in\fun(\cotsp\man{Q})}. The differential of \eq{\hat{U}} is thus a horizontal 1-form which is further \textit{basic} and \textit{exact}, \eq{\dif \hat{U}  = \dif (\copr^*U) = \copr^* \dif U = \pderiv{\hat{U}}{q^i}\hbdel[q^i]  \in \formsbh(\cotsp\man{Q})}. In analogy to the Riemannian case, we then define the  conservative force 1-form as \eq{\bs{\alpha}^\ss{\hat{U}} := -\dif\hat{U} = \copr^*\bs{\alpha}^\ss{U}\in\formsbh(\cotsp \man{Q})} (the sign is a matter of convention). 
Note this 1-form is \textit{horizontal} and, for the conservative case, it is also \textit{basic} and \textit{exact}.  
Then, 
as per Eq.\eqref{VH_iso_cotan}, the symplectic musical isomorphism  gives a \textit{vertical} vector field, \eq{\sfb{X}^\ss{\hat{U}}:= \inv{\nbs{\omg}}\cdt \bs{\alpha}^\ss{\hat{U}} = -\inv{\nbs{\omg}}\cdt \dif \hat{U}\in\vectv\cap \vechm(\cotsp\man{Q},\nbs{\omg})} which, for the conservative case, is simply the Hamiltonian field for \eq{\hat{U}} as well as  the vertical vector lift of the original 1-form \eq{\bs{\alpha}^\ss{U}=-\dif U\in\forms(\man{Q})}. To summarize:
\begin{small}
\begin{align}
  \hat{U}:=\copr^* U \in \fun(\cotsp\man{Q}) 
&\quad  \Rightarrow \quad 
      \bs{\alpha}^\ss{\hat{U}} :=- \dif \hat{U} \in \formsbh(\cotsp\man{Q})
\\ \nonumber 
&\quad  \Rightarrow \quad 
    \sfb{X}^\ss{\hat{U}} = \inv{\nbs{\omg}}( \bs{\alpha}^\ss{\hat{U}} ) =  -\inv{\nbs{\omg}}( \dif \hat{U})  =\vlift{(\bs{\alpha}^\ss{U})} =  -\pd_i \hat{U} \hbpartup{i}
    \in \vectbv \cap \vechm(\cotsp\man{Q},\nbs{\omg})
\end{align}
\end{small}
 where the last expression holds in any cotangent-lifted basis (\eq{\hbpartup{i}=\hpdiiup{\pf_i}}).
Thus, for some nominal Hamiltonian system \eq{(\cotsp\man{Q},\nbs{\omg},\mscr{H}_\iio)}, we may include the effects of some conservative force \eq{\sfb{X}^\ss{\hat{U}}} simply by using the new Hamiltonian \eq{\mscr{H} = \mscr{H}_\iio + \hat{U}}: 
\begin{small}
\begin{align}
    \mscr{H} = \mscr{H}_\iio + \hat{U} 
    &&
   \sfb{X}^\sscr{H} \,=\,  -\inv{\nbs{\omg}} \cdt \dif (\mscr{H}_\iio+\hat{U})
   \,=\, \sfb{X}^\ss{\mscr{H}_\iio +\hat{U}} \,=\,    \sfb{X}^\ss{\mscr{H}_\iio} + \sfb{X}^\ss{\hat{U}} \, \in\vechm(\cotsp \man{Q},\nbs{\omg})
\end{align}
\end{small}

\paragraph{Conservative Forces on $(\tsp\man{Q},\nbs{\varpi})$.}
For dynamics on \eq{(\tsp\man{Q},\nbs{\varpi})}, everything proceeds the same as above starting from \eq{\bar{U}:=\tpr^* U = U\circ\tpr\in\fun(\tsp\man{Q})} as the velocity-independent potential function on \eq{\tsp\man{Q}}. We then have the exact, basic, horizontal and 1-form \eq{\bs{\alpha}^\ss{\bar{U}}:=-\dif\bar{U}\in\formsbh(\tsp\man{Q})} leading to a basic vertical Hamiltonian vector field \eq{\sfb{X}^\ss{\bar{U}}=\inv{\nbs{\varpi}}(\bs{\alpha}^\ss{\bar{U}}) = -\inv{\nbs{\varpi}}(\dif \bar{U})}. 
\begin{small}
\begin{align}
    \bar{U}:= \tpr^* U \in\fun(\tsp\man{Q}) 
&\quad  \Rightarrow \quad 
       \bs{\alpha}^\ss{\bar{U}} =- \dif \bar{U} \in \formsbh(\tsp\man{Q})
\\ \nonumber 
&\quad  \Rightarrow \quad 
    \sfb{X}^\ss{\bar{U}} 
    = \inv{\nbs{\varpi}}(\bs{\alpha}^\ss{\bar{U}}) = -\inv{\nbs{\varpi}}(\dif \bar{U}) \,=\,   -L^{ij}\pd_j \bar{U} \pdii{v^i}\in \vectbv\cap\vechm(\tsp\man{Q},\nbs{\varpi})
\end{align}
\end{small}
where the last expression holds in any tangent-lifted coordinate basis, with \eq{L^{ij}} the inverse of  \eq{L_{ij}=\ppderiv{\mscr{L}}{v^j}{v^i}} (this is often \eq{L_{ij}=g_{ij}} and \eq{L^{ij}=g^{ij}}).  As before, the effects of the conservative force are incorporated into some nominal Hamiltonian system \eq{(\tsp\man{Q},\nbs{\varpi},E_\iio)} simply by using \eq{\mscr{L} = \mscr{L}_\iio - \bar{U}} which leads to \eq{E=E_\iio + \bar{U}} such that the total Hamiltonian field is
\begin{small}
\begin{align}
    \mscr{L} = \mscr{L}_\iio - \bar{U}
    \quad \Rightarrow \quad 
    E=E_\iio + \bar{U}
    &&
   \bs{\Gamma}^\ss{E} \,=\,  -\inv{\nbs{\varpi}} \cdt \dif (E_\iio+\bar{U})  \,=\, \bs{\Gamma}^\ss{E_\iio +\bar{U}}
   \,=\, \bs{\Gamma}^\ss{E_\iio} + \sfb{X}^\ss{\bar{U}} \, \in\vechm(\tsp\man{Q},\nbs{\varpi})
\end{align}
\end{small}
 Since \eq{\sfb{X}^\ss{\bar{U}}} is vertical, the above \eq{\bs{\Gamma}^\ss{E_\iio} + \sfb{X}^\ss{\bar{U}}} is still second-order and thus has integral curves of the form \eq{\pt{\v}_t=(\pt{r}_t,\dt{\sfb{r}}_t)}. 


\paragraph{Conservative Mechanical Systems (Snapshot).}
Conservative forces are modeled as \textit{exact} 1-forms on \eq{\man{Q}}. 
For a \textit{mechanical} Lagrangian, \eq{\mscr{L}= \ttfrac{1}{2} g_{ij}v^iv^j - \bar{U}}, with corresponding energy function  \eq{E =\ttfrac{1}{2} g_{ij}v^iv^j + \bar{U}}, and 
Hamiltonian \eq{\mscr{H} =\ttfrac{1}{2} g^{ij}\pf_i \pf_j + \hat{U}}, 
 the dynamics formulated on \eq{(\man{Q},\sfg)}, \eq{(\cotsp\man{Q},\nbs{\omg})}, and \eq{(\tsp\man{Q},\nbs{\varpi})} are:
\begin{small}
\begin{align} \label{EOMs_cons}
\!\!\!&\quad \fnsize{EOMs}&& \fnsize{conservative force 1-form} &&  \fnsize{conservative force vector}
\\ \nonumber 
\fnsize{on $\man{Q}$} \;\;
  &\!\!\!\left\{\begin{array}{lllll}
         \del{\dt{\sfb{r}}} \dt{\sfb{r}}_t = \sfb{f}_{\pt{r}_t}^\ss{U} \,=\, -\inv{\sfg}_{\pt{r}}(\dif U)
    \\[5pt]
         \del{\dt{\sfb{r}}}\bs{\mu}_t = \bs{\alpha}_{\pt{r}_t}^\ss{U} \,=\, -\dif U_{\pt{r}}
    \end{array} \right.
    &&
    \bs{\alpha}^\ss{U} :=-\dif U \in\formsex(\man{Q})
    &&
    \sfb{f}^\ss{U} := \inv{\sfg}(\bs{\alpha}^\ss{U}) 
    \in\vect(\man{Q})
\\[10pt] \nonumber 
\fnsize{on $\cotsp\man{Q}$} \;\;
    &\!\!\! 
    \left\{ \begin{array}{llll}
         \dt{\mu}_t = \sfb{X}^\ss{\mscr{H} }_{\mu_t} = -\inv{\nbs{\omg}}_{\mu_t}(\dif \mscr{H}) 
     \\[5pt]
         \lderiv{\sfb{X}^\sscr{H}} \bs{\theta} = \dif E^\sscr{H}
    \end{array} \right.
    &&
    \bs{\alpha}^\ss{\hat{U}} = -\dif \hat{U} =  \copr^* \bs{\alpha}^\ss{U} \in\formsbh\cap\formsex(\cotsp\man{Q})
    &&
     \sfb{X}^\ss{\hat{U}} := \inv{\nbs{\omg}}(\bs{\alpha}^\ss{\hat{U}}) = \vlift{(\bs{\alpha}^\ss{U})} \in\vectbv\cap\vechm(\cotsp\man{Q})
\\[10pt]  \nonumber 
\fnsize{on $\tsp\man{Q}$} \;\;
    &\!\!\! 
    \left\{ \begin{array}{llll}
        \dt{\pt{\v}}_t = \bs{\Gamma}^\ss{E}_{\!\!\pt{\v}_t}  = -\inv{\nbs{\varpi}}_{\pt{\v}_t}(\dif E)
     \\[5pt]
         \lderiv{\bs{\Gamma}^E} \bs{\varth} = \dif \mscr{L}
    \end{array} \right.
    &&
     \bs{\alpha}^\ss{\bar{U}}  = -\dif \bar{U} = \tpr^* \bs{\alpha}^\ss{U} \in\formsbh\cap\formsex(\tsp\man{Q})
    &&
     \sfb{X}^\ss{\bar{U}} := \inv{\nbs{\varpi}} (\bs{\alpha}^\ss{\bar{U}}) = \vlift{(\sfb{f}^\ss{U})} \in\vectbv\cap\vechm(\tsp\man{Q})
\end{align}
\end{small}
where \eq{\pt{\v}_t=(\pt{r}_t,\dt{\sfb{r}}_t)\in\tsp\man{Q}} and \eq{\mu_t=(\pt{r}_t,\bs{\mu}_t)=\Upsilon(\pt{\v}_t)=g^\flt(\pt{\v}_t)\in\cotsp\man{Q}} with \eq{\bs{\mu}=\fibd\!\mscr{L}_\pt{r}(\dt{\sfb{r}})=\sfg_{\pt{r}}(\dtsfb{r}) } the `kinematic/conjugate momentum' covector along \eq{\pt{r}_t} (they are equivalent for a mechanical Lagrangian).  Note that \eq{E^\sscr{H}:=\bs{\theta}\cdt\sfb{X}^\sscr{H}-\mscr{H}} is \textit{not} the same as \eq{E\equiv E^\sscr{L}:= \bs{\varth}\cdt\bs{\Gamma}^\ss{E}-\mscr{L} = \mscr{H}\circ \Upsilon}. Note the following:
\begin{small}
\begin{itemize}[nosep] 
    \item  On the (co)tangent bundle, the conservative force is simply included in the energy or Hamiltonian function. That is, \eq{\sfb{X}^\sscr{H} = \sfb{X}^\ss{T+\hat{U}} = \sfb{X}^\ss{T} + \sfb{X}^\ss{\hat{U}}} and \eq{\bs{\Gamma}^\ss{E} =\bs{\Gamma}^\ss{T+\bar{U}} = \bs{\Gamma}^\ss{T} + \sfb{X}^\ss{\bar{U}}}.
    \item  The conservative force 1-form on the (co)tangent bundle is both exact and horizontal. It is also \textit{basic} as it is the natural horizontal lift of an exact 1-form on \eq{\man{Q}}. 
    \item  The conservative force vector on the (co)tangent bundle is both Hamiltonian and vertical. It is also \textit{basic} as it is the vertical vector lift of a 1-form/vector on \eq{\man{Q}}. 
\end{itemize}
\end{small}

   \noindent Often, we will not bother to make any notational distinction between \eq{U\in\fun(\man{Q})}, \eq{\hat{U}=U\circ\copr\in\fun(\cotsp\man{Q})}, and \eq{\bar{U}=U\circ\tpr\in\fun(\tsp\man{Q})}. We will simply write \eq{U} and \eq{\dif U} with the appropriate domain implied by context. 


\subsection{Nonconservative Forces} \label{sec:force_noncon}

For a mechanical system subject to non-conservative forces, the developments are analogous to the above case for conservative forces, with the caveat that the  horizontal force 1-form, \eq{\bs{\alpha}} (in whichever setting we are concerned with), is \textit{not} exact and, further, \textit{not} closed. As such, its influence on the dynamics cannot be encapsulated by adding a basic function (a potential energy) to the Lagrangian or Hamiltonian. The details are given below.

\paragraph{Nonconservative Mechanical Systems (Snapshot).}
Nonconservative forces are modeled as \textit{horizontal} 1-forms on either \eq{\cotsp \man{Q}} or \eq{\tsp\man{Q}}. 
 For a \textit{mechanical} Lagrangian,  \eq{\mscr{L}= \ttfrac{1}{2} g_{ij}v^iv^j - \bar{U}}, with corresponding energy function  \eq{E =\ttfrac{1}{2} g_{ij}v^iv^j + \bar{U}}, and 
Hamiltonian \eq{\mscr{H} =\ttfrac{1}{2} g^{ij}\pf_i \pf_j + \hat{U}}, 
 the dynamics — including nonconservative forces —  formulated on \eq{(\man{Q},\sfg)}, \eq{(\cotsp\man{Q},\nbs{\omg})}, and \eq{(\tsp\man{Q},\nbs{\varpi})}  are:
\begin{small}
\begin{align} 
& \quad\fnsize{EOMs}&& \fnsize{NC force 1-form} &&  \fnsize{NC force vector}
\\ \nonumber 
\fnsize{on $\man{Q}$} \quad 
  &\!\!\!\left\{\begin{array}{lllll}
         \del{\dt{\sfb{r}}} \dt{\sfb{r}}_t  
         = \sfb{f}_{\pt{r}_t}^\ss{U} + \sfb{f}^\ss{\bs{\alpha}}_{\pt{\v}_t}
         = \inv{\sfg}(-\dif U_{\pt{r}_t} + \bs{\alpha}_{\pt{\v}_t} )  
    \\[5pt]
         \del{\dt{\sfb{r}}}\bs{\mu}_t = -\dif U_{\pt{r}_t} + \bs{\alpha}_{\mu_t} 
    \end{array}\right.
    &&
    \quad\text{—} 
    &&
    \quad \text{—} 
\\[10pt] \nonumber 
\fnsize{on $\cotsp\man{Q}$} \quad 
    &\!\!\! 
    \left\{\begin{array}{llll}
        \dt{\mu}_t  = \sfb{X}^\ss{\mscr{H} }_{\mu_t} + \sfb{F}^{\hbs{\alpha}}_{\mu_t}
        = \inv{\nbs{\omg}}_{\mu_t}(-\dif \mscr{H} + \hbs{\alpha} ) 
         =: \sfb{X}_{\mu_t}
     \\[5pt]
         \lderiv{\sfb{X}} \bs{\theta} = \dif E^\sscr{H} + \hat{\bs{\alpha}} 
    \end{array} \right.
    &&
    \hbs{\alpha}  \in\formsh(\cotsp\man{Q})
    &&
     \sfb{F}^{\hbs{\alpha}} := \inv{\nbs{\omg}} (\hbs{\alpha}) 
     \in\vectv(\cotsp\man{Q})
\\[8pt]  \nonumber 
\fnsize{on $\tsp\man{Q}$} \quad 
     &\!\!\! \left\{ \begin{array}{llll}
       \dt{\pt{\v}}_t =  \bs{\Gamma}^\ss{E}_{\!\!\pt{\v}_t} +  \sfb{F}^{\barbs{\alpha}}_{\!\pt{\v}_t}  = \inv{\nbs{\varpi}}_{\pt{\v}_t}(-\dif E + \barbs{\alpha} ) 
       =: \bs{\Gamma}_{\!\pt{\v}_t}
     \\[5pt]
         \lderiv{\sfb{X}} \bs{\varth} =  \dif\mscr{L} + \barbs{\alpha} \; 
    \end{array} \right.
    &&
    \barbs{\alpha}  \in\formsh(\tsp\man{Q})
    &&
     \sfb{F}^{\barbs{\alpha}} := \inv{\nbs{\varpi}} (\barbs{\alpha}) 
     \in\vectv(\tsp\man{Q})
\end{align}
\end{small}
where \eq{\sfb{X}} or \eq{\bs{\Gamma}} denotes, respectively, the total vector field on \eq{\cotsp \man{Q}} or \eq{\tsp\man{Q}} (the conservative forces  are included in \eq{\sfb{X}^\sscr{H}} and \eq{\bs{\Gamma}^\ss{E}}), and  where \eq{\pt{\v}_t=(\pt{r}_t,\dt{\sfb{r}}_t)\in\tsp\man{Q}} and \eq{\mu_t=(\pt{r}_t,\bs{\mu}_t)=\Upsilon(\pt{\v}_t)\in\cotsp\man{Q}} with \eq{\bs{\mu}=\fibd\!\mscr{L}_\pt{r}(\dt{\sfb{r}}) } the `conjugate momentum' covector along \eq{\pt{r}_t} which, for a mechanical Lagrangian, coincides with the  
`kinematic momentum' covector, \eq{\bs{\mu}=\sfg_\pt{r}\cdt\dt{\sfb{r}}}, appearing in the dynamics on \eq{\man{Q}}.
In contrast to conservative forces given in Eq.\ref{EOMs_cons},  the nonconservative force 1-form/vectors have the following properties:
\begin{small}
\begin{itemize}[nosep] 
    \item  The nonconservative force 1-form on the (co)tangent bundle is \textit{not} exact (or even closed), but it is still horizontal. It is \textit{not} the lift of a 1-form on \eq{\man{Q}}. 
    \item  The nonconservative force vector on the (co)tangent bundle is \textit{not} Hamiltonian (or even symplectic), but it is still vertical. It is \textit{not} the lift of a 1-form/vector field on \eq{\man{Q}}.
    \item  On \eq{\tsp\man{Q}}, the total vector field \eq{\bs{\Gamma}=\bs{\Gamma}^\ss{E}+\sfb{F}^{\barbs{\alpha}}}  is still second-order and thus has integral curves of the form \eq{\pt{\v}_t=(\pt{r}_t,\dt{\sfb{r}}_t)}. 
    \item  The nonconservative forces on \eq{\man{Q}} can be seen as (co)vector-valued objects but they are \textit{not} tensor fields on \eq{\man{Q}}; they are generally velocity/momentum dependent.  The \eq{\bs{\alpha}}  appearing in the dynamics on \eq{\man{Q}} is obtained from the 1-form on the (co)tangent bundle using the natural isomorphisms \eq{\cotsp[\bdt]\man{Q}\cong \cotsph[\bdt](\tsp\man{Q})}  and  \eq{\cotsp[\bdt]\man{Q}\cong \cotsph[\bdt](\cotsp \man{Q})}. That is, when evaluated at a point, then \eq{\bs{\alpha}\cdt\dif\copr = \hbs{\alpha}} and  \eq{\bs{\alpha}\cdt\dif\tpr = \barbs{\alpha}}. 
\end{itemize}
\end{small}

\paragraph{Nonconservative Forces on $\cotsp\man{Q}$.} 
Following \cite{deLeon2022HJforce}, 
for dynamics on \eq{\cotsp\man{Q}}, an arbitrary external force is geometrically interpreted as a horizontal (i.e., semi-basic) 1-form, \eq{\bs{\alpha}\in\formsh(\cotsp\man{Q})}, such that it is given in any cotangent-lifted basis as \eq{\bs{\alpha}=\alpha_i\hbdel^i} with \eq{\alpha_i\in\fun(\cotsp\man{Q})}.  The corresponding vertical vector field is \eq{\sfb{F}^\ss{\bs{\alpha}}:= \inv{\nbs{\omg}}\cdt\bs{\alpha}\in\vectv(\cotsp\man{Q})}. Thus, for arbitrary coordinates \eq{\tp{\zeta}}, any general symplectic coordinates \eq{(\tp{s},\tp{\kappa})}, and any symplectic \textit{cotangent-lifted} coordinates  \eq{(\tp{q},\tp{\pf})=\colift\tp{q}}, we have 
\begin{small}
\begin{align} \label{Xpert_cotan}
\begin{array}{rccccc}
     \bs{\alpha}   & \!\!=\; 
       \alpha_\ssc{I} \bdel[\zeta^\ssc{I}]   
    & \!\!=\;
      \alpha_i \hbdel[s^i]  \,+\,  \alpha^i \hbdeldn[\kappa_i]
    & \!\!=\;
    \alpha_i \hbdel[q^i] &\in \formsh(\cotsp\man{Q})
\\[3pt]
      &\;\fnsize{arbitrary coord.} 
     &    \fnsize{symplectic coord.}
     &\quad \fnsize{cotangent-lifted coord.}
\\[3pt]
    \sfb{F}^\ss{\bs{\alpha}} := \inv{\nbs{\omg}}\cdt \bs{\alpha} & \!\!=\; 
     \omega^\ssc{IJ} \alpha_{\ssc{J}} \pdii{\zeta^\ssc{I}} 
      & \!\!=\;
      -\alpha^i\hpdii{s^i} \,+\, \alpha_i \hpdiiup{\kappa_i}
       & \!\!=\;
    \alpha_i \hpdiiup{\pf_i}   &\in \vectv(\cotsp\man{Q})
\end{array}
\end{align}
\end{small}
where \eq{\alpha_\ssc{I},\,\alpha_i,\,\alpha^i} in each of the above are obviously not the same but are the components of \eq{\bs{\alpha}} in the different  indicated bases. Note that if we were to instead use \eq{\sfb{F}^\ss{\bs{\alpha}}=\inv{\nbs{\omg}}(\bs{\alpha},\slot)} then all terms from here forward with \eq{\bs{\alpha}} would change sign. 
For some Hamiltonian system \eq{(\cotsp\man{Q},\nbs{\omg},\mscr{H})}, the external forces are included  by the addition of \eq{\sfb{F}^\ss{\bs{\alpha}}}  such that the total dynamics are given by the (generally non-symplectic) vector field:
\begin{small}
\begin{align} \label{Xham_noncon} 
\boxed{ \begin{array}{rccccc}
     \sfb{X} \,=\,  \sfb{X}^\sscr{H} + 
       \sfb{F}^\ss{\bs{\alpha}}   
       \,=\, \inv{\nbs{\omg}}\cdt(-\dif \mscr{H} + \bs{\alpha})   & \!\!=\; 
      \omega^\ssc{JI} ( \pderiv{\mscr{H}}{\zeta^\ssc{J}} \,-\, \alpha_\ssc{J} )\pdii{\zeta^\ssc{I}} 
    & \!\!=\;
      (\pderiv{\mscr{H}}{\kappa_i} - \alpha^i )\hpdii{s^i} - ( \pderiv{\mscr{H}}{s^i} - \alpha_i )\hpdiiup{\kappa_i}
    & \!\!=\;
   \pderiv{\mscr{H}}{\pf_i}\hpdii{q^i} - ( \pderiv{\mscr{H}}{q^i} - \alpha_i )\hpdiiup{\pf_i}
\\[3pt]
      &\quad \fnsize{arbitrary coord.}
     & \quad   \fnsize{symplectic coord.}
     &\quad \fnsize{cotangent-lifted coord.}
\end{array}  }  
\end{align}
\end{small}
Note the following useful relations: 
\begin{small}
\begin{align} \label{Xpert_rels}
    \sfb{X}^\sscr{H}\cdt\dif\mscr{H} = 0 
    &&
    \sfb{F}^\ss{\bs{\alpha}}\cdt\bs{\alpha} = 0
    &&
     \sfb{X}\cdt\dif\mscr{H}  = \sfb{F}^\ss{\bs{\alpha}} \cdt\dif\mscr{H} \,=\, \sfb{X}^\sscr{H}\cdt \bs{\alpha} = \sfb{X}\cdt\bs{\alpha}
     &&
      \nbs{\omg}\cdt\sfb{X} = \dif \mscr{H} - \bs{\alpha}
\end{align}
\end{small}
 If \eq{\rho_t} is an integral curve such that \eq{\dt{\rho}_t= \sfb{X}_{\rho_t}=\sfb{X}^\sscr{H}_{\rho_t} + \sfb{F}_{\rho_t}} then the derivative of some \eq{h\in\fun(\cotsp\man{Q})} along \eq{\rho_t} is given by 
\begin{small}
\begin{align} \label{Pbrakpert_cotan}
 & \dt{h} \,=\, \lderiv{\sfb{X}} h \,=\, (\sfb{X}^\sscr{H} + \sfb{F}^\bs{\alpha})\cdt \dif h \,=\, \pbrak{h}{\mscr{H}} + \inv{\nbs{\omg}}(\dif h,\bs{\alpha})
    \;=\; \pbrak{h}{\mscr{H}} + \omega^\ssc{IJ}\alpha_\ssc{J} \pd_\ssc{I} h
\end{align}
\end{small}
where the last expression holds for any local coordinates.
For any symplectic coordinates, \eq{(\tp{s},\tp{\kap})}, and any cotangent-lifted symplectic coordinates, \eq{(\tp{q},\tp{\pf})=\colift\tp{q}}, it simplifies to 
\begin{small}
\begin{align}
    \boxed{ \dt{h} \,=\, \lderiv{\sfb{X}} h \,=\, \pbrak{h}{\mscr{H}} \,+\, \lderiv{\sfb{F}^\ss{\bs{\alpha}}} h \,=\, \pbrak{h}{\mscr{H}} +  \alpha_i \pderiv{h}{\kap_i}  - \alpha^i \pderiv{h}{s^i} 
    \,=\, \pbrak{h}{\mscr{H}} +  \alpha_i \pderiv{h}{\pf_i} } 
\end{align}
\end{small}
In particular, the Hamiltonian function itself, \eq{\mscr{H}}, is no longer an integral of motion; from the above, along with Eq.\eqref{Xpert_rels}, its time-evolution is given by 
\begin{small}
\begin{align} \label{Hdot_pert}
      & \diff{}{t}\mscr{H}(\rho_t) \,=\, \lderiv{\sfb{X}}\mscr{H}_{\rho_t} \,=\, 
      \sfb{F}^\ss{\bs{\alpha}}_{\rho_t} \cdt \dif\mscr{H} \,=\, \inv{\nbs{\omg}}_{\rho_t} (\dif \mscr{H},\bs{\alpha})  \,=\,  \sfb{X}^\sscr{H}_{\rho_t} \cdt\bs{\alpha} 
      \,=\, \sfb{X}_{\rho_t}\cdt \bs{\alpha} \,=\,  \dt{\rho}_t\cdt\bs{\alpha}
      \;=\; \omega^{\ssc{IJ}}\alpha_{\ssc{J}}\pd_{\ssc{I}}\mscr{H} \,=\, \alpha_{\ssc{J}} \dot{\zeta}^{\ssc{J}} (\rho_t)
\end{align}
\end{small}

\paragraph{Nonconservative Forces on $\tsp\man{Q}$.}
The story on \eq{\tsp\man{Q}} is analogous to eq.~\eqref{Xpert_cotan}-Eq.\eqref{Hdot_pert} but with \eq{\sfb{X}^\sscr{H}} and \eq{\nbs{\omg}} replaced by \eq{\bs{\Gamma}^\ss{E}} and \eq{\nbs{\varpi}}, respectively,  where \eq{E= \bscr{V}\cdt\dif\mscr{L}-\mscr{L}} is the `energy of \eq{\mscr{L}}' (i.e., \eq{\mscr{H}} but as a function on \eq{\tsp\man{Q}}), and  where the force 1-form is some horizontal \eq{\bs{\alpha}\in\formsh(\tsp\man{Q})} (we are using the same symbol, but this is a different 1-form than the one on the cotangent bundle). The corresponding vertical vector field is \eq{\sfb{F}^\ss{\bs{\alpha}}:= \inv{\nbs{\varpi}}\cdt\bs{\alpha},\in\vectv(\tsp\man{Q})}. Thus, for arbitrary coordinates \eq{\tp{\xi}},  and any  \textit{tangent-lifted} coordinates  \eq{(\tp{q},\tp{v})=\tlift\tp{q}}, we have 
\begin{small}
\begin{align} \label{Xpert_tan}
\begin{array}{rccccc}
     \bs{\alpha}   & \!\!=\; 
       \alpha_\ssc{I} \bdel[\xi^\ssc{I}]   
    & \!\!=\;
    \alpha_i \bdel[q^i] &\in \formsh(\tsp\man{Q})
\\[3pt]
      &\;\fnsize{arbitrary coord.} 
     &\quad \fnsize{tangent-lifted coord.}
\\[3pt]
    \sfb{F}^\ss{\bs{\alpha}} := \inv{\nbs{\varpi}}\cdt \bs{\alpha} & \!\!=\; 
     \varpi^\ssc{IJ} \alpha_{\ssc{J}} \pdii{\xi^\ssc{I}} 
       & \!\!=\;
    L^{ij}\alpha_j \pdii{v^i}   &\in \vectv(\tsp\man{Q})
\end{array}
\end{align}
\end{small}
 The total dynamics are then given by \eq{\bs{\Gamma} =  \bs{\Gamma}^\ss{E} + \sfb{F}^\ss{\bs{\alpha}}},  which is still \textit{second-order} (since \eq{\sfb{F}^\ss{\bs{\alpha}}} is vertical):
\begin{small}
\begin{align}
    \bs{\Gamma} \,=\,  \bs{\Gamma}^\ss{E} + \sfb{F}^\ss{\bs{\alpha}}
    \,=\,   \bs{\Gamma}^\ss{E} + \inv{\nbs{\varpi}}\cdt\bs{\alpha}
    \,=\,   v^i\pdii{q^i} \,-\, L^{ij}\big( \ppderiv{\mscr{L}}{q^k}{v^j}v^k \,-\, \pderiv{\mscr{L}}{q^j} \,-\, \alpha_i \big) \pdii{v^i}
&&
\begin{array}{lllll}
     \vlift{\iden}\cdt \bs{\Gamma} \,=\,  \vlift{\iden}\cdt \bs{\Gamma}^\ss{E}  \,=\, \bscr{V}
 \\[2pt]
    \vlift{\iden}\cdt \sfb{F} = 0
\end{array}
\end{align}
\end{small}

\paragraph{Extrinsic View.} 
Consider a system with \eq{n}-dim (embedded) configuration (sub)manifold, \eq{\man{Q}\subseteq\Evec^\ss{N}}, with inclusion \eq{\imath \equiv \ptvec{r}=r^\a\sfb{i}_\a:\man{Q}\hookrightarrow\Evec^\ss{N}}.  
Let  \eq{\bs{f}= \bs{f}^\ss{(1)}\oplus \dots \oplus \bs{f}^\ss{(P)} = f_\a \bs{\iota}^\a \in \Evec^{N*} } denote the nonconservative forces, \textit{viewed as covectors/1-forms},  acting on the \textit{P}-particle system. Note that \eq{\sfb{i}_\a \equiv \bpart{r^\a}} and \eq{\bs{\iota}^\a \equiv \dif r^\a}. 
The generalized nonconservative forces, \eq{\bs{\alpha}}, are then given in a local coordinate basis on \eq{\man{Q}} by 
\begin{small}
\begin{align}
    \bs{\alpha} := \imath^* \bs{f}  = \bs{f} \cdt \dif \imath \,\equiv\, \bs{f} \cdt \dif \ptvec{r}   =  f_\a \pd_i r^\a \bdel^i =: \alpha_i \bdel^i   \in \forms(\man{Q})
    &&,&&
    \alpha_i = \bs{\alpha}\cdt\bpart{i}  =   \bs{f}\cdt\be_{i}  = \bs{f}\cdt \pd_i r^\a\bs{\iota}_\a = 
    f_\a \pd_i r^\a
\end{align}
\end{small}
(where \eq{i,j=1,\dots, n} and \eq{\alpha,\beta=1,\dots, N:=3P}).
That is, if \eq{f_\a} are the components of \eq{\bs{f}} in the fixed inertial \eq{\sfb{i}_\a} basis, then \eq{\alpha_i=f_\a \pderiv{r^\a}{q^i}} are simply the components of \eq{\bs{f}} in the coordinate basis \eq{\be[q^i]=\imath_* \pdii{q^i}}. That is, \eq{\bs{\alpha}} is nothing more than the intrinsic view of \eq{\bs{f}}:
\begin{small}
\begin{align}
       \bs{f} = f_\a \bs{\iota}^\a = \alpha_i \bs{\ep}^i \;\cong\; \bs{\alpha}=\alpha_i \bdel^{i} \,=\, \imath^* \bs{f}  
\end{align}
\end{small}
Then, for any symplectic coordinates \eq{(\tp{s},\tp{\kappa})} and any symplectic cotangent-lifted coordinates \eq{(\tp{q},\tp{\pf})}:
\begin{small}
\begin{align}
      \sfb{F} :=\, \inv{\nbs{\omg}}(\bs{\alpha},\slot) 
      \;=\; f_\a \inv{\nbs{\omg}}(\bdel^{r^\a},\slot) \,=\, f_\a\sfb{X}^{r^\a}
      \quad =\;   \omega^\ssc{JI}f_\a \pd_\ssc{J} r^\a \bpart{\ssc{I}} 
      \,=\,  
      f_\a \pderiv{r^a}{\kappa_i}\hpdii{s^i} \,-\,   f_\a \pderiv{r^a}{s^i}\hpdiiup{\kappa_i}
      \;=\;
      - f_\a \pderiv{r^a}{q^i}\hpdiiup{\pf_i}  
\end{align}
\end{small}
and \eq{\sfb{X}=\sfb{X}^\sscr{H} + \sfb{F}} agrees with Eq.\eqref{Xham_noncon}.

\subsection{Coordinate View:~Hamilton's Equations with Non-Conservative Forces} \label{app:non_conserv}

We outline the process — within the classic analytical (non-geometric) dynamics framework — for including non-conservative forces in Hamilton's canonical equations of motion, and for transforming such forces between different coordinate sets. Though the resulting equations of motion follow immediately from the previous geometric treatment, the following developments makes no use of geometry. We  will revert back to a more casual coordinated-based view.

\paragraph{Generalized Forces for Arbitrary Canonical Coordinates, $(\tup{Q},\tup{P})$.} 
 Suppose some non-conservative forces, \eq{\sfb{a}^\ii{\mrm{nc}}}, acts on the particle, or system of particles, of interest. In this work, \eq{\sfb{a}^\ii{\mrm{nc}}} is usually the force per unit reduced mass.
For some arbitrary canonical/symplectic coordinates, \eq{(\tup{Q},\tup{P})}, Hamilton's canonical equations of motion including \eq{\sfb{a}^\ii{\mrm{nc}}}
 are given by
\begin{small}
\begin{align} \label{dQP_nc}
\begin{array}{lll}
     & \dot{\tup{Q}} \,=\, \pderiv{\mscr{K}}{\tup{P}} \;-\; \tup{A}_\ss{Q}  &\quad,
     \\[9pt]
     & \dot{\tup{P}} \,=\, -\pderiv{\mscr{K}}{\tup{Q}} \;+\; \tup{A}_\ss{P}
     &\quad, 
\end{array} 
\qquad\qquad
\begin{array}{ll}
     &  \tup{A}_\ss{Q} \;:=\; \trn{\pderiv{\tup{r}}{\tup{P}}} \tup{a}^\ii{\mrm{nc}} 
     \\[8pt]
     & \tup{A}_\ss{P} \;:=\;  \trn{\pderiv{\tup{r}}{\tup{Q}}} \tup{a}^\ii{\mrm{nc}}
\end{array}
\end{align}
\end{small}
where \eq{\tup{r}} and \eq{\tup{a}^\ii{\mrm{nc}}} are the cartesian components of \eq{\sfb{r}} and \eq{\sfb{a}^\ii{\mrm{nc}}} in an inertial basis. 
Note that it is often the case that \eq{\tup{Q}} are
true configuration-level coordinates (e.g., cartesian position coordinates, spherical coordinates, Euler angles, etc.) and \eq{\tup{P}} are velocity-level coordinates such that \eq{\sfb{r}=\sfb{r}(\tup{Q},t)} is not a function of \eq{\tup{P}}. When this is the case, the above simplifies to  
\begin{small}
\begin{align} \label{dQP_nc_2}
\begin{array}{lll}
     & \dot{\tup{Q}} \,=\, \pderiv{\mscr{K}}{\tup{P}}  &\quad
     \\[9pt]
     & \dot{\tup{P}} \,=\, -\pderiv{\mscr{K}}{\tup{Q}} \;+\; \tup{A}
     &\quad,
\end{array} 
\qquad\qquad
\begin{array}{ll}
     &  \;\;
     \\[8pt]
     & \tup{A} \;:=\;  \trn{\pderiv{\tup{r}}{\tup{Q}}} \tup{a}^\ii{\mrm{nc}}
\end{array}
\end{align}
\end{small}
Where we have dropped the subscript on the generalized forces, which now appear only in the dynamics for the momenta. 
Although various derivations of Eq.\eqref{dQP_nc} and Eq.\eqref{dQP_nc_2} can be found in many texts on analytical mechanics \cite{lanczos2020variational,goldstein2002classical,schaub2003analytical}, we briefly summarize the process below.

\begin{small}
\begin{itemize}
    \item[] \textit{Derivation of Eq.\eqref{dQP_nc}.}  
We may account for the effects of some non-conservative force on the equations of motion by adding the work due to this force, denoted \eq{W^\ii{\mrm{nc}}},  to the action such that Hamilton's principle leads to
\begin{small}
\begin{align} \label{dIQP}
\begin{array}{rlllll}
   0\,=\,\delta I 
    \,=\, \delta\int_{t_\iio}^{t_f} (\tup{P}\cdot\dot{\tup{Q}} - \mscr{K} + W^\ii{\mrm{nc}})\,\mrm{d}t  
   &\,=\, \int_{t_\iio}^{t_f} ( \dot{\tup{Q}}\cdot\delta\tup{P} \,+\, \tup{P}\cdot\delta\dot{\tup{Q}}
    \,-\, \delta\mscr{K} 
    \,+\, \delta W^\ii{\mrm{nc}})\,\mrm{d} t 
\\[5pt] 
  & \,=\,
    \int_{t_\iio}^{t_f} ( \dot{\tup{Q}}\cdot\delta\tup{P} \,-\, \dot{\tup{P}}\cdot\delta\tup{Q}
    \,-\, \delta\mscr{K} 
    \,+\, \delta W^\ii{\mrm{nc}})\,\mrm{d} t 
\\[5pt] 
    &\,=\,
    \int_{t_\iio}^{t_f} \big( (\dot{\tup{Q}}-\pderiv{\mscr{K}}{\tup{P}})\cdot\delta\tup{P} \,+\,  (-\dot{\tup{P}}-\pderiv{\mscr{K}}{\tup{Q}})\cdot\delta\tup{Q} \,+\, \sfb{a}^\ii{\mrm{nc}}\cdot\delta\sfb{r} \big)\mrm{d} t \,=\, 0 
 \end{array}
\end{align}
\end{small}
where we have used \eq{\delta W^\ii{\mrm{nc}}=\sfb{a}^\ii{\mrm{nc}}\cdot\delta \sfb{r}} as well as fact that the variations vanish at the boundaries:  \eq{\eval{\delta\tup{Q}}{t_\iio}\!\!=\eval{\delta\tup{P}}{t_\iio}\!\!=\tup{0}}, \eq{\delta t_\iio=0}, and likewise at \eq{t_f}. 
Integration by parts was used to replace \eq{\tup{P}\cdot\delta\dot{\tup{Q}}} with \eq{-\dot{\tup{P}}\cdot\delta\tup{Q}} inside the integral.
Next, it is often assumed that \eq{\tup{Q}} are true configuration/position-level coordinates and that the momenta, \eq{\tup{P}}, are velocity-level coordinates such that \eq{\sfb{r}=\sfb{r}(\tup{Q},t)} is a function only of the configuration coordinates and, possibly, time.
Yet, this is not the most general case as a canonical transformation can render this assumption inaccurate.\footnote{Take the Delaunay variables for example. These are canonical coordinates for orbital motion for which \eq{\sfb{r}} is a function of both the configuration and momentum coordinates.}  
Thus, taking the most general case that \eq{\sfb{r}=\sfb{r}(\tup{Q},\tup{P},t)}, we find that \eq{\delta W^\ii{\mrm{nc}}} is given by 
\begin{small}
\begin{align} \label{dW_QP}
    \delta W^\ii{\mrm{nc}} \,=\, \sfb{a}^\ii{\mrm{nc}}\cdot\delta\sfb{r} \,=\,
    \tup{a}^\ii{\mrm{nc}}\cdot(\pderiv{\tup{r}}{\tup{Q}}\delta \tup{Q} \,+\, \pderiv{\tup{r}}{\tup{P}}\delta\tup{P})
    \,=\,
    (\trn{\pderiv{\tup{r}}{\tup{Q}}} \tup{a}^\ii{\mrm{nc}})\cdot\delta\tup{Q}
    \,+\, 
    ( \trn{\pderiv{\tup{r}}{\tup{P}}} \tup{a}^\ii{\mrm{nc}})\cdot\delta\tup{P}
\end{align}
\end{small}
such that Eq.\eqref{dIQP} becomes
\begin{small}
\begin{align} \label{dIQP2}
\begin{array}{llll}
     \delta I \,=\, 
    \int_{t_\iio}^{t_f} \Big( (\dot{\tup{Q}} - \pderiv{\mathscr{K}}{\tup{P}}+\trn{\pderiv{\tup{r}}{\tup{P}}} \tup{a}^\ii{\mrm{nc}} )\cdot\delta\tup{P} \;-\;
    (\dot{\tup{P}} + \pderiv{\mathscr{K}}{\tup{Q}} - \trn{\pderiv{\tup{r}}{\tup{Q}}} \tup{a}^\ii{\mrm{nc}} ) \cdot\delta\tup{Q} 
    \Big)\mrm{d} t = 0
\end{array}
\end{align}
\end{small}
for the above to hold for all arbitrary \eq{\delta\tup{Q}} and \eq{\delta\tup{P}} requires that their coefficients in the above be zero. This leads to the canonical equations of motion given in Eq.\eqref{dQP_nc}. 
\end{itemize}
\end{small}


\paragraph{Transformation of the Generalized Forces Under a Canonical Transformation.} 
Now suppose that we perform some general canonical transformation from \eq{(\tup{Q},\tup{P})} to new canonical coordinates, \eq{(\tup{q},\tup{\pf})}.\footnote{The developments of this section apply the same weather \eq{\tup{q}} and \eq{\tup{\pf}} are  minimal or non-minimal canonical coordinates.} 
Since \eq{\tup{q}} and \eq{\tup{\pf}} are also canonical coordinates, the derivation of the equations of motion including non-conservative forces exactly parallels Eq.\eqref{dQP_nc}-Eq.\eqref{dIQP2}. In general, we assume \eq{\sfb{r}=\sfb{r}(\tup{q},\tup{\pf},t)} such that \eq{\delta W^\ii{\mrm{nc}}} is now given by Eq.\eqref{dW_QP} with \eq{\tup{q}} and \eq{\tup{\pf}} taking the place of \eq{\tup{Q}} and \eq{\tup{P}}, respectively. 
 The requirement that \eq{\delta I =0} then leads to canonical equations of motion which are precisely the same \textit{form} as Eq.\eqref{dQP_nc}:
\begin{small}
\begin{align} \label{dqp_nc}
    \begin{array}{lll}
     & \dot{\tup{q}} \,=\, \pderiv{\mscr{H}}{\tup{q}} \;-\; \tup{\alpha}_{q}  &\quad,
     \\[9pt]
     & \dot{\tup{\pf}} \,=\, -\pderiv{\mscr{H}}{\tup{q}} \;+\; \tup{\alpha}_{\pf}
     &\quad,
\end{array} 
\qquad\qquad
\begin{array}{ll}
     &  \tup{\alpha}_{q} \;:=\; \trn{\pderiv{\tup{r}}{\tup{\pf}}} \tup{a}^\ii{\mrm{nc}} 
     \\[8pt]
     & \tup{\alpha}_{\pf} \;:=\;  \trn{\pderiv{\tup{r}}{\tup{q}}} \tup{a}^\ii{\mrm{nc}}
\end{array}
\end{align}
\end{small}
where \eq{\mscr{H}} is the Hamiltonian for the \eq{(\tup{q},\tup{\pf})} phase space. 
What we would like to know is, given some canonical transformation between \eq{(\tup{Q},\tup{P})} and  \eq{(\tup{q},\tup{\pf})}, what then is the relation between the above  \eq{\tup{\alpha}} and the \eq{\tup{A}}  defined in Eq.\eqref{dQP_nc}? We will present the most general case as well a more specific, yet common, case. 
\begin{itemize}
\item \rmsb{Case 1.} Consider the most general case:
    \begin{small}
    \begin{itemize}[nosep]
        \item \eq{(\tup{Q},\tup{P})} are arbitrary canonical/symplectic coordinates such that, in general, \eq{\sfb{r}=\sfb{r}(\tup{Q},\tup{P},t)}. 
        \item  \eq{(\tup{Q},\tup{P})} and  \eq{(\tup{q},\tup{\pf})} are related by a general canonical transformation such that, in general,  \eq{\tup{Q}=\tup{Q}(\tup{q},\tup{\pf},t)} and \eq{\tup{P}=\tup{P}(\tup{q},\tup{\pf},t)}.
    \end{itemize}
    \end{small}
Then the relation between \eq{\tup{A}} and \eq{\tup{\alpha}} can be found from Eq.\eqref{dqp_nc} as
\begin{small}
\begin{align}
  \begin{array}{lll}
       \tup{\alpha}_{q}\,:=\, \trn{\pderiv{\tup{r}}{\tup{\pf}}} \tup{a}^\ii{\mrm{nc}}
     \,=\,  \trn{\big( \pderiv{\tup{r}}{\tup{Q}} \pderiv{\tup{Q}}{\tup{\pf}} + \pderiv{\tup{r}}{\tup{P}} \pderiv{\tup{P}}{\tup{\pf}} \big)} \tup{a}^\ii{\mrm{nc}}
      &\,=\,
      \big( \trn{\pderiv{\tup{Q}}{\tup{\pf}}} \trn{\pderiv{\tup{r}}{\tup{Q}}} \tup{a}^\ii{\mrm{nc}}  + \trn{\pderiv{\tup{P}}{\tup{\pf}}}  \trn{\pderiv{\tup{r}}{\tup{P}}} \tup{a}^\ii{\mrm{nc}}  \big)
  \\[10pt]
    \tup{\alpha}_{\pf}\,:=\, \trn{\pderiv{\tup{r}}{\tup{q}}} \tup{a}^\ii{\mrm{nc}}
      \,=\, \trn{\big( \pderiv{\tup{r}}{\tup{Q}} \pderiv{\tup{Q}}{\tup{q}} + \pderiv{\tup{r}}{\tup{P}} \pderiv{\tup{P}}{\tup{q}} \big)} \tup{a}^\ii{\mrm{nc}}
      &\,=\,
       \big( \trn{\pderiv{\tup{Q}}{\tup{q}}} \trn{\pderiv{\tup{r}}{\tup{Q}}} \tup{a}^\ii{\mrm{nc}}  + \trn{\pderiv{\tup{P}}{\tup{q}}}  \trn{\pderiv{\tup{r}}{\tup{P}}} \tup{a}^\ii{\mrm{nc}}  \big)
\end{array}
\end{align}
\end{small}
From the above, along with the definitions of \eq{\tup{A}_\ss{Q}} and  \eq{\tup{A}_\ss{P}} in Eq.\eqref{dQP_nc}, Eq.\eqref{dqp_nc} then gives the equations of motion and generalized forces  as
\begin{small}
\begin{align} \label{genForce_CT}
  \begin{array}{lll}
     & \dot{\tup{q}} \,=\, \pderiv{\mscr{H}}{\tup{q}} \;-\; \tup{\alpha}_{q}  &\quad,
     \\[9pt]
     & \dot{\tup{\pf}} \,=\, -\pderiv{\mscr{H}}{\tup{q}} \;+\; \tup{\alpha}_{\pf}
     &\quad,
\end{array} 
&&
\begin{array}{ll}
          \tup{\alpha}_{q}\,=\, \trn{\pderiv{\tup{r}}{\tup{\pf}}} \tup{a}^\ii{\mrm{nc}} \,=\,     \trn{\pderiv{\tup{Q}}{\tup{\pf}}} \tup{A}_\ss{P}  + \trn{\pderiv{\tup{P}}{\tup{\pf}}}  \tup{A}_\ss{Q}  
 \\[8pt]
         \tup{\alpha}_{\pf}\,=\, \trn{\pderiv{\tup{r}}{\tup{q}}} \tup{a}^\ii{\mrm{nc}} \,=\, \trn{\pderiv{\tup{Q}}{\tup{q}}} \tup{A}_\ss{P}  + \trn{\pderiv{\tup{P}}{\tup{q}}}  \tup{A}_\ss{Q} 
    \end{array} 
\end{align}
\end{small}
Thus, once one knows the relation between the phase space coordinates, one may then use the above equations on the right to calculate the relation between the generalized forces. 
\item \rmsb{Case 2.}  Now consider a more specific, common, case:
    \begin{small}
    \begin{itemize}[nosep]
        \item \eq{(\tup{Q},\tup{P})} are configuration-level and velocity-level coordinates, respectively, such that \eq{\sfb{r}=\sfb{r}(\tup{Q},t)} is \textit{not} a function of \eq{\tup{P}}
        \item  \eq{(\tup{Q},\tup{P})} and  \eq{(\tup{q},\tup{\pf})} are related by a time-dependent \textit{point} transformation such that \eq{\tup{Q}=\tup{Q}(\tup{q},t)} and \eq{\tup{P}=\tup{P}(\tup{q},\tup{\pf},t)}.
    \end{itemize}
    \end{small}
The first of the above leads to \eq{\tup{A}_Q=\tup{0}} such that the equations of motion for \eq{\tup{Q}} and \eq{\tup{P}} are given by Eq.\eqref{dQP_nc_2}. 
The second condition — that the two phase spaces are related by a canonical \textit{point} transformation —  means that \eq{\pderiv{\tup{Q}}{\tup{\pf}}=O}. 
Eq.\eqref{genForce_CT} then simplifies to
\begin{small}
\begin{align} \label{dqp_nc_pt}
 \begin{array}{lll}
     & \dot{\tup{q}} \,=\, \pderiv{\mscr{H}}{\tup{q}}   &\quad
     \\[9pt]
     & \dot{\tup{\pf}} \,=\, -\pderiv{\mscr{H}}{\tup{q}} \;+\; \tup{\alpha}
     &\quad,
\end{array} 
&&
\begin{array}{ll}
     & \;\;  
     \\[8pt]
     & \tup{\alpha} \,\equiv\, \tup{\alpha}_{\pf}  \,=\,  \trn{\pderiv{\tup{r}}{\tup{q}}} \tup{a}^\ii{\mrm{nc}} \,=\, \trn{\pderiv{\tup{Q}}{\tup{q}}} \tup{A}
\end{array} 
\end{align}
\end{small}
Note that if \eq{\tup{Q}} and \eq{\tup{P}} are simply the position and velocity components in an inertial basis (if  \eq{\tup{Q}=\tup{r}}), then \eq{\tup{A}=\tup{a}^\ii{\mrm{nc}}} is simply the components of the force from non-conservative forces in that same basis. 
\end{itemize}


\section{CONFORMAL SCALING OF VECTOR FIELDS} \label{sec:confScale}


A \textit{conformal scaling} of a vector field \eq{\sfb{X}} is simply the multiplication of \eq{\sfb{X}} by any nowhere-zero function \eq{\cff} (it is sometimes further required that \eq{\cff>0}). This simple scaling is more interesting than it may first seem. 
For instance, it can be interpreted as a transformation of the evolution parameter, \eq{t}, to a new parameter, \eq{s}, determined by \eq{\mrm{d} t = \cff \mrm{d} s} (where \eq{t} parametrizes the integral curves of \eq{\sfb{X}} and \eq{s} parametrizes those of \eq{\cff \sfb{X}}). This, and more, is detailed below.

\subsection{Conformal Scaling = Transformation of the Evolution Parameter} \label{sec:t_xform}

\paragraph{Reparameterizing an ODE.}
First consider some ODE on \eq{\mbb{R}^m}, \eq{\dot{\tp{\xi}} = {X}(\tp{\xi})}, where  \eq{\dot{(\slot)}:=\diff{(\slot)}{t}} and \eq{t} is the evolution parameter (i.e., independent variable). Let \eq{\tp{\xi}_t} be some solution (integral curve). Suppose we wish to change to a new evolution parameter, \eq{s}, defined implicitly through the relation \eq{\tx{d} t = {\cff}(\tp{\xi}_t) \tx{d}s}   for some function \eq{{\cff}\in\fun(\mbb{R}^m)}. Or, since \eq{\tp{\xi}_t} is parameterized by \eq{t}, it would be more appropriate to write this as \eq{\tx{d} s = \tfrac{1}{{\cff}(\tp{\xi}_t)} \tx{d}t}. 
Then, from \eq{\diff{}{t} = \tfrac{1}{{\cff}} \diff{}{s} }, the system \eq{\dot{\tp{\xi}} = {X}(\tp{\xi})} is transformed into \eq{\rng{\tp{\xi}} = {\cff} {X}(\tp{\xi})}, where \eq{\rng{(\slot)}:=\diff{(\slot)}{s}}. If \eq{\tiltp{\xi}_s} is a solution to \eq{\rng{\tp{\xi}} = {\cff} {X}(\tp{\xi})}, 
then \eq{\tiltp{\xi}_{s(t)}=\tp{\xi}_t} is a solution to the original ODE \eq{\dot{\tp{\xi}} = {X}(\tp{\xi})}. Conversely, if \eq{\tp{\xi}_t} is a solution to the original ODE, then \eq{\tp{\xi}_{t(s)}=\tiltp{\xi}_s} is a solution to  \eq{\rng{\tp{\xi}} = {\cff} {X}(\tp{\xi})}. Note that we are  using \eq{s(t)} and \eq{t(s)} simply to denote one evolution parameter expressed in terms of the other, as determined by solving the relation  \eq{\tx{d} s = \tfrac{1}{{\cff}(\tp{\xi}_t)} \tx{d}t} or, equivalently, \eq{\mrm{d} t = {\cff}(\tiltp{\xi}_s)\mrm{d} s}.  
These same ideas carry over to vector fields on smooth manifolds, as detailed in the following.

\paragraph{Reparameterizing a Curve.} If \eq{\pt{x}_t\in\man{X}} is a curve then it is a map \eq{\pt{x}:\rchart{R}{t}\to\man{X}}, parameterized by \eq{t} (we call \eq{t} the \textit{evolution parameter}). If we have some bijective transformation \eq{t=t(s)\leftrightarrow s=s(t)} for a new parameter \eq{s} then one might interpret the notation \eq{\pt{x}_{s}} as the same curve just parameterized using \eq{s}. That is, one might think that \eq{\pt{x}_{s}} just means \eq{\pt{x}_t} with \eq{t} expressed in terms of \eq{s}. Unfortunately, the notation \eq{\pt{x}_{s}}, strictly interpreted, means the  map \eq{\pt{x}:\rchart{R}{t}\to\man{X}}   \textit{evaluated} at \eq{s}. The re-parameterized curve we wish to express is actually an entirely new curve (that is, a new map), \eq{\tilpt{x}:\rchart{R}{s}\to\man{X}}, such that \eq{\tilpt{x}_{s(t)}=\pt{x}_t} and \eq{\pt{x}_{t(s)}=\tilpt{x}_{s}} where, by abuse of notation, we are  using \eq{s(t)} and \eq{t(s)} to denote one evolution parameter expressed in terms of the other.  More specifically, if we think of \eq{t,s\in\fun(\mbb{R})} are coordinates on \eq{\mbb{R}} related by some \eq{t=c_{ts}(s)} then \eq{\tilpt{x} = \pt{x}\circ c_{ts} } such that \eq{\tilpt{x}_s=\pt{x}\circ c_{ts}(s)=\pt{x}_t}.
\begin{small}
\begin{itemize}[topsep=1pt]
    \item \textit{Example.}  Consider a curve in \eq{\man{X}=\mbb{R}} given by \eq{t\mapsto x_t := \cos \omega t} for some constant \eq{\omega\in\mbb{R}}. Consider a transformation of the evolution parameter \eq{t=s/\omega \leftrightarrow s = \omega t}. Then, technically, \eq{x_s} does \textit{not} mean \eq{\cos s} but means \eq{x_s = \cos \omega s} which is equivalent to \eq{\cos \omega^2 t}. The reparameterized curve that we want, \eq{s\mapsto \cos s}, is a technically a \textit{new} curve, \eq{\til{x}_s :=  \cos s}, such that \eq{\til{x}_s :=  \cos s = \cos \omega t = x_t} (by design).  But \eq{\til{x}:\rchart{R}{s}\to\man{X}} and \eq{x:\rchart{R}{t}\to\man{X}} are \textit{different} curves (whose images give the \textit{same} overall path). 
\end{itemize}
\end{small}

 \begin{small}
 \begin{notation}
     Despite the above careful distinction between \eq{\pt{x}_t} and \eq{\tilpt{x}_s}, we will sometimes disregard this distinction and, by abuse of notation, write \eq{\pt{x}_s} to indicate the curve \eq{\pt{x}_t} reparameterized using some new evolution parameter \eq{s}.  That is, instead of defining the new curve \eq{\tilpt{x}} such that \eq{\tilpt{x}_{s}=\pt{x}_{t(s)}}, we will simply write \eq{\pt{x}_s}, which is taken to mean \eq{\pt{x}_{t(s)}} (the curve \eq{\pt{x}_t} with \eq{t} expressed in terms of \eq{s}). Thus, when (ab)using this notation, \eq{\pt{x}_t} and \eq{\pt{x}_s} mean the ``same'' curve up to a  parameterization (which are technically two different curves, but both give the same path/trajectory). However, for the time being, we will continue to distinguish between \eq{\pt{x}_t} and \eq{\tilpt{x}_s}. 
 \end{notation}
 \end{small}

\paragraph{Reparameterizing Dynamics:~Conformally Related Vector Fields.} Two vector fields, \eq{\sfb{X},\tsfb{X}\in\vect(\man{X})} are said to be \textit{conformally related} if \eq{\tsfb{X}= {\cff}\sfb{X}} for some nowhere-vanishing function \eq{{\cff}\in\fun(\man{X})}, called the \textit{conformal factor}. We will see shortly that the dynamics of \eq{\tsfb{X}= {\cff}\sfb{X}} are the ``same'' as those of \eq{\sfb{X}} under a transformation of the evolution parameter \eq{\mrm{d} t = {\cff} \mrm{d} s}. First, note that \eq{\sfb{X}} and \eq{\tsfb{X}} have the same 
 integrals of motion;  if  \eq{\lderiv{\sfb{X}}h=0} then \eq{\lderiv{\tsfb{X}}h= {\cff}\lderiv{\sfb{X}}h= 0}. That is, \eq{\lderiv{\sfb{X}}h=0} implies \eq{\lderiv{\tsfb{X}}h=0} and vice versa.
Note this only applies to integrals of motion which are functions (which is what the term is usually assumed to mean); for some tensor field  \eq{\sfb{T}\in\tens^r_s(\man{X})} of order greater than 0, then  \eq{\lderiv{\sfb{X}}\sfb{T}=0} does \textit{not} imply \eq{\lderiv{\tsfb{X}}\sfb{T}=0}. That is, \eq{\sfb{X}} and \eq{\tsfb{X}} do not ``preserve'' the same tensor fields. In particular, for any \eq{\sfb{u}\in\vect(\man{X})}, then \eq{\lderiv{\sfb{X}}\sfb{u}=-\lderiv{\sfb{u}}{\sfb{X}}=0} does \textit{not} imply \eq{\lderiv{\sfb{u}}{\tsfb{X}}=0}. Thus,
\eq{\sfb{X}} and \eq{\tsfb{X}} have the same integrals of motion but \textit{not} the same Lie symmetries: 
\begin{small}
\begin{align} \label{fX_iom}
    \tsfb{X}:= {\cff} \sfb{X} 
    \quad 
    \left\{ \;\; \begin{array}{llllllll}
         \lderiv{\sfb{X}} h =0 & \Rightarrow &  \lderiv{\tsfb{X}} h = {\cff} \lderiv{\sfb{X}} h = 0 
     \\[4pt]
         \lderiv{\sfb{u}}{\sfb{X}} =0 & \Rightarrow &  \lderiv{\sfb{u}} \tsfb{X} =  (\lderiv{\sfb{u}} {\cff})\sfb{X} 
    \end{array} \right.
\end{align}
\end{small}
Now, as alluded to previously, conformally related vector fields also have the same integral curves, up to a reparameterization.
Consider \eq{\sfb{X}\in\vect(\man{X})} with some integral curve \eq{\pt{x}_t} such that \eq{\dtpt{x}_t=\sfb{X}_{\pt{x}_t}}. If we define a conformally related vector field \eq{\tsfb{X}:= {\cff}\sfb{X}\in\vect(\man{X})} — for some non-vanishing (usually positive) function \eq{{\cff}\in\fun(\man{Q})} — then this amounts to a transformation of the evolution parameter from \eq{t} to a new parameter \eq{s} related by 
\begin{small}
\begin{align} \label{dtdtau0}
    \mrm{d} t = {\cff}(\tilpt{x}_s)  \mrm{d} s
\qquad,\qquad 
   \mrm{d} s = \tfrac{1}{{\cff}(\pt{x}_t)} \mrm{d} t
 \qquad,\qquad
   \dt{(\slot)} := \diff{}{t} = \tfrac{1}{{\cff}}\rng{(\slot)}
\qquad,\qquad
   \rng{(\slot)} := \diff{}{s} = {\cff} \dt{(\slot)}
\end{align}
\end{small}
where \eq{\tilpt{x}_s = \pt{x}_{t(s)}} is the reparameterized curve. Importantly, \eq{\tilpt{x}_s} is an integral curve of  \eq{\tsfb{X}} such that \eq{\rng{\tilpt{x}}_s =  \tsfb{X}_{\tilpt{x}_s}}.
Thus, \eq{\sfb{X}} and \eq{\tsfb{X}= {\cff}\sfb{X}} have the ``same'' integral curves (up to a reparameterization). As mentioned in Eq.\eqref{fX_iom}, \eq{\sfb{X}} and \eq{\tsfb{X}} also have the same integrals of motion; \eq{\lderiv{\tsfb{X}} h = {\cff} \lderiv{\sfb{X}} h} for any \eq{h\in\fun(\man{X})}, that is:
\begin{small}
\begin{align}
        \lderiv{\sfb{X}} \cong \diff{}{t} = \dt{(\slot)}
        \qquad,\qquad 
        \lderiv{\tsfb{X}}  = {\cff} \lderiv{\sfb{X}} \cong {\cff} \diff{}{t} \,=\, \diff{}{s} = \rng{(\slot)}
         \qquad,\qquad 
        \dt{h}=0 \,\Leftrightarrow\, \rng{h}=0
\end{align}
\end{small} 
(\eq{\lderiv{\tsfb{X}}  = {\cff} \lderiv{\sfb{X}}} is only valid when  acting on functions!)
where the above relations hold along integral curves. 
Now, to further clarify the relation between \eq{t} and \eq{s}. Let \eq{\phi_t} and \eq{\til{\phi}_s} be the flows of \eq{\sfb{X}} and \eq{\tsfb{X}}, respectively, such that \eq{\pt{x}_t=\phi_t(\pt{x}_\zr)} and  \eq{\tilpt{x}_s=\til{\phi}_s(\pt{x}_\zr)} are integral curves. When we write \eq{t(s)} or \eq{s(t)} we mean the solutions to
\begin{small}
\begin{align} \label{st_int}
\begin{array}{lll}
      s(t): \quad  s-s_\zr \,=\, \int_{t_\zr}^t \tfrac{1}{{\cff}(\phi_t(\pt{x}_\zr)) } \mrm{d} t 
      \qquad\qquad,\qquad\qquad 
      t(s): \quad
      t-t_\zr \,=\, \int_{s_\zr}^s {\cff}(\til{\phi}_s(\pt{x}_\zr)) \mrm{d} s
\end{array}
\end{align}
\end{small}
such that, if \eq{\pt{x}_t} is known then \eq{\tilpt{x}_s} is given by \eq{\tilpt{x}_s = \pt{x}_{t(s)}} or, alternatively, if \eq{\tilpt{x}_s} is known then \eq{\pt{x}_t} is given by \eq{\pt{x}_t=\tilpt{x}_{s(t)}}. Note from that above that these should really be written \eq{t(s,\pt{x}_\zr)} and  \eq{s(t,\pt{x}_\zr)}. To summarize:
\begin{small}
\begin{align}
     \fnsize{for } \; \tsfb{X}:= {\cff} \sfb{X} : &&
      \dtpt{x}_t = \sfb{X}_{\pt{x}_t} \;\;\Leftrightarrow \;\;  \rng{\tilpt{x}}_s = \tsfb{X}_{\tilpt{x}_s} \qquad\quad 
    \fnsize{with: } \quad 
    \tilpt{x}_s = \pt{x}_{t(s,\pt{x}_\zr)}  \;\; \equiv\;\; \pt{x}_t = \tilpt{x}_{s(t,\pt{x}_\zr)} 
\end{align}
\end{small}

\begin{small}
\begin{notesq}
    For the case that \eq{{\cff}} is an integral of motion of \eq{\sfb{X}} (and thus also an integral of motion of \eq{\tsfb{X}}) then \eq{\phi_t^* {\cff}= {\cff}} and \eq{\til{\phi}_s^* {\cff} = {\cff}} is constant along integral curves and Eq.\eqref{st_int} leads to the simple relations: 
    \begin{small}
    \begin{align}
        \fnsize{if} \;\; \lderiv{\sfb{X}} {\cff}=0
         \qquad\quad \Rightarrow \qquad\quad
           s-s_\zr = \tfrac{1}{f_\zr}(t -t_\zr) \qquad,\qquad t-t_\zr = f_\zr(s-s_\zr)
    \end{align}
    \end{small}
\end{notesq}
\end{small}

\noindent Aside from transforming the evolution parameter,  other situations where conformally related vector fields arise:
\begin{small}
\begin{itemize}[topsep=1pt,itemsep=1pt]
    \item If \eq{\sfb{X}\in\vect(\man{X})} is not a complete vector field, there may exist some \eq{{\cff}\in\fun(\man{X})} such that \eq{\tsfb{X}= {\cff}\sfb{X}} is complete \cite{carinena2015geoFromDyn}. This is always true when \eq{\man{X}} is para-compact (with \eq{{\cff}>0} strictly positive).
    \item If \eq{\div[\bs{\sigma}]\sfb{X}=0} for some volume form \eq{\bs{\sigma}} then, generally, \eq{\div[\bs{\sigma}]( {\cff}\sfb{X})\neq0} (or vice versa). 
    \item If \eq{\sfb{k}\in\veckl(\man{Q},\sfg)} is a Killing vector field (\eq{\lderiv{\sfb{k}}\sfg =0}), then \eq{\tsfb{k}= {\cff}\sfb{k}} may not be (or vice versa).
    \item If \eq{\sfb{X}\in\vecsp(\man{P},\Bfi{\txw})} is a symplectic vector field (\eq{\lderiv{\sfb{X}}\Bfi{\txw} =0}), then \eq{\tsfb{X}= {\cff}\sfb{X}} may not be 
    (or vice versa).\footnote{In particular, for Hamiltonian vector field \eq{\sfb{X}^h=\inv{\Bfi{\txw}}(\dif h,\slot)}, then \eq{\Bfi{\txw}(\tsfb{X}^h,\slot) = {\cff} \Bfi{\txw}(\sfb{X}^h,\slot) = {\cff} \dif h} such that \eq{\tsfb{X}^h} is generally not Hamiltonian (or even symplectic).
    However, if  \eq{\exd ( {\cff}\dif h) = \dif {\cff} \wedge \dif h =0} then \eq{\tsfb{X}^h} is at least \textit{locally} Hamiltonian (i.e., symplectic) with respect to \eq{\Bfi{\txw}}. Further, note that if we define \eq{\til{\Bfi{\txw}}:=(1/ {\cff})\Bfi{\txw}} then \eq{\tsfb{X}^h} is \eq{\til{\Bfi{\txw}}}-Hamiltonian since \eq{\til{\Bfi{\txw}}(\tsfb{X}^h,\slot)=\dif h}. Yet, \eq{\til{\Bfi{\txw}}} is symplectic iff \eq{{\cff}} is such that \eq{\exd \til{\Bfi{\txw}} = \exd ((1/ {\cff}) \Bfi{\txw})= \dif(1/ {\cff})\wedge\Bfi{\txw} =0}.}  
     \item If \eq{\bs{\Gamma}\in\vect(\tsp\man{Q})} is  a second order vector field (\eq{\lft{\iden}\cdt\bs{\Gamma}= \bscr{V}}), then \eq{\tbs{\Gamma}:= {\cff} \bs{\Gamma}} may not be second order 
     (or vice versa).\footnote{It is easy to see that  \eq{\lft{\iden}\cdt\bs{\Gamma}= \bscr{V} \;\Rightarrow \; \lft{\iden}\cdt\tbs{\Gamma} = {\cff} \bscr{V}} such that \eq{\tbs{\Gamma}= {\cff} \bs{\Gamma}} is not second-order. But, if \eq{{\cff}\equiv\tpr^* {\cff}} is a basic function, we may define a \emph{\eq{\tbs{\Gamma}}-adapted tangent structure}, \eq{\lft{\til{\iden}}:=(1/ {\cff})\lft{\iden}}, for which \eq{\tbs{\Gamma}} is indeed ``second order'' in the sense \eq{\lft{\til{\iden}}\cdt\tbs{\Gamma} = \bscr{V}}.} 
\end{itemize}
\end{small}

\subsection{Conformally Hamiltonian Vector Fields} \label{sec:confHam}

On a symplectic manifold, \eq{(\man{P},\nbs{\omg})}, we define a 
\textit{conformally-Hamiltonian}\footnote{Recall that two vector fields, \eq{\sfb{X}} and \eq{\tsfb{X}}, on any smooth manifold are said to be conformally-related if \eq{\tsfb{X}={\cff} \sfb{X}} for some nowhere-zero function, \eq{{\cff}}.}
vector field any vector field that can be written as \eq{{\cff}\sfb{X}^\ss{h}} for some Hamiltonian vector field \eq{\sfb{X}^\ss{h}=\inv{\nbs{\omg}}(\dif h,\slot)\in\vechm(\man{P},\nbs{\omg})} and some conformal factor (nowhere-vanishing function) \eq{0 \neq {\cff}\in\fun(\man{P})}. Here, we will use the notation \eq{\tsfb{X}^\ss{h}:={\cff}\sfb{X}^\ss{h}}. 
 Recall that the exchange \eq{\sfb{X}\leftrightarrow {\cff}\sfb{X}} amounts to a transformation of the evolution parameter \eq{t\leftrightarrow s} determined by \eq{\mrm{d} t = {\cff} \mrm{d} s}.   
This \eq{\tsfb{X}^\ss{h}} is generally \textit{not} \eq{\nbs{\omg}}-Hamiltonian or, even \eq{\nbs{\omg}}-symplectic (locally Hamiltonian), as is evident by the following
    (see footnote\footnote{The relation 
    \eq{\lderiv{\tsfb{X}^{h}}\nbs{\omg} = \dif {\cff} \wedge \dif h}
    is verified by \eq{\lderiv{{\cff}\sfb{X}^\ss{h}}\nbs{\omg} = {\cff}\lderiv{\sfb{X}^\ss{h}}\nbs{\omg} \,+\, \dif {\cff} \wedge \nbs{\omg}(\sfb{X}^\ss{h},\slot) 
    = \dif {\cff} \wedge \dif h}. The derivation of  \eq{\div[\bs{\spvol}]({\cff}\sfb{X}^\ss{h})= \pbrak{{\cff}}{h}} is less obvious: 
     For any vector field \eq{\sfb{X}}, function \eq{{\cff}}, and volume form \eq{\bs{\spvol}}, it can be shown that:\\
    \eq{\qquad\qquad   \div[\bs{\spvol}]({\cff}\sfb{X}) \bs{\spvol} \,=\,  
    \lderiv{{\cff}\sfb{X}}\bs{\spvol} \,=\, \lderiv{\sfb{X}}({\cff}\bs{\spvol}) \,=\, {\cff} \lderiv{\sfb{X}}\bs{\spvol} \,+\, (\lderiv{\sfb{X}} {\cff}) \bs{\spvol}  \;=\;
    ( {\cff} \div[\bs{\sig}]\sfb{X} \,+\, \lderiv{\sfb{X}} {\cff})  \bs{\spvol} \qquad \Rightarrow \qquad  \div[\bs{\spvol}]({\cff}\sfb{X}) = {\cff} \div[\bs{\spvol}] \sfb{X} \,+\, \lderiv{\sfb{X}} {\cff}    } .\\
     for  \eq{\sfb{X}^\ss{h}\in\vechm} and \eq{\bs{\spvol}} the symplectic Liouville volume form, we then have \eq{\div[\bs{\sig}]\sfb{X}^\ss{h} =0} and the above simplifies to  \eq{\div[\bs{\spvol}]({\cff}\sfb{X}^\ss{h}) =\lderiv{\sfb{X}^\ss{h}} {\cff} = \pbrak{{\cff}}{h} }.  }):
\begin{small}
\begin{align}
    \tsfb{X}^\ss{h}:={\cff}\sfb{X}^\ss{h} :
    \qquad\qquad 
    \lderiv{\tsfb{X}^{h}}\nbs{\omg} 
    \,=\, \dif {\cff} \wedge \dif h
   \qquad,\qquad 
    \nbs{\omg}(\tsfb{X}^\ss{h},\slot) = {\cff} \dif h
    \qquad,\qquad 
    \div[\bs{\spvol}](\tsfb{X}^\ss{h})= \lderiv{\sfb{X}^\ss{h}} {\cff} = \pbrak{{\cff}}{h}
\end{align}
\end{small}
Thus, \eq{\tsfb{X}^\ss{h}} is \eq{\nbs{\omg}}-symplectic only in the special case that \eq{\exd ({\cff}\dif h)=\dif {\cff} \wedge \dif h=0} (meaning \eq{{\cff}} is some "function of \eq{h}'' as in \eq{{\cff}=h^2} or \eq{{\cff}=\mrm{e}^h} etc.). 
However, if  we define \eq{\til{\nbs{\omg}}:=(1/{\cff})\nbs{\omg}} then:
\begin{small}
\begin{align}
  \tsfb{X}^\ss{h}:={\cff}\sfb{X}^\ss{h} 
  \quad,\quad 
  \til{\nbs{\omg}}:=\tfrac{1}{{\cff}}\nbs{\omg}
   \quad,\quad 
 \inv{\til{\nbs{\omg}}} = {\cff} \inv{\nbs{\omg}}
    && \Rightarrow &&
     \til{\nbs{\omg}}(\tsfb{X}^\ss{h},\slot)  = \dif  h
     \quad,\quad 
      \inv{\til{\nbs{\omg}}}(\dif h,\slot) \,=\, \tsfb{X}^\ss{h}
\end{align}
\end{small}
such that \eq{\tsfb{X}^\ss{h}} appears to be \eq{\til{\nbs{\omg}}}-Hamiltonian. Indeed it would be, \textit{if} \eq{\til{\nbs{\omg}}} is a symplectic form which is not generally true; \eq{\til{\nbs{\omg}}=(1/{\cff})\nbs{\omg}} is symplectic iff it is globally non-degenerate and closed. Global non-degeneracy is assured since \eq{\nbs{\omg}} is symplectic and \eq{{\cff}\neq 0} was already required.  
Yet, closedness still requires  \eq{\exd\til{\nbs{\omg}} = -{\cff}^\ss{-2}\dif {\cff} \wedge \nbs{\omg} = 0}, which does not generally hold for arbitrary \eq{{\cff}}.
As such,  \eq{\tsfb{X}^\ss{h}} is generally not \eq{\til{\nbs{\omg}}}-symplectic (nor Hamiltonian) since \eq{\lderiv{\tsfb{X}^{h}}\til{\nbs{\omg}} = \tsfb{X}^\ss{h}\cdt \exd \til{\nbs{\omg}}} is generally non-zero (unless \eq{\til{\nbs{\omg}}} is closed).
With some useful identities, we find \eq{\lderiv{\tsfb{X}^{h}}\til{\nbs{\omg}}} is generally given as follows 
(derived in footnote\footnote{From Cartan's identity, along with \eq{\tsfb{X}^\ss{h}\cdt \til{\nbs{\omg}}=\dif h \in\formsex(\man{P})}, we have:\\ 
   \eq{\qquad\quad} \eq{\lderiv{\tsfb{X}^{h}}\til{\nbs{\omg}} = \exd( \tsfb{X}^\ss{h}\cdt \til{\nbs{\omg}}) + \tsfb{X}^\ss{h}\cdt \exd \til{\nbs{\omg}} = \tsfb{X}^\ss{h}\cdt \exd \til{\nbs{\omg}} = {\cff}\sfb{X}^\ss{h}\cdt \exd(\tfrac{1}{{\cff}}\nbs{\omg}) = {\cff}\sfb{X}^\ss{h}\cdt (-{\cff}^\ss{-2}\dif {\cff} \wedge \nbs{\omg}) = -\tfrac{1}{{\cff}}\sfb{X}^\ss{h}\cdt (\dif {\cff} \wedge \nbs{\omg})  }.\\
Then, we make use of the following (really only the first) for any \eq{\sfb{u}\in\vect(\man{P})} and \eq{\bs{\lambda}\in\forms^l(\man{P})} and \eq{\bs{\kap}\in\forms^k(\man{P})}:\\
    \eq{\qquad\quad}
    \eq{\sfb{u}\cdt ( \bs{\lambda}\wedge\bs{\kap}) = (\sfb{u}\cdt  \bs{\lambda})\wedge\bs{\kap} + (-1)^l \bs{\lambda}\wedge (\sfb{u}\cdt \bs{\kap})
       \qquad,\qquad
     \bs{\lambda}\wedge\bs{\kap} = (-1)^{lk}\bs{\kap}\wedge\bs{\lambda}
      \qquad,\qquad \sfb{u}\cdt  \bs{\kap} = (-1)^{k+1}\bs{\kap}\cdt  \sfb{u} = (-1)^{k+1}\bs{\kap}(\sfb{u})  }\\
Which leads to \eq{\;\; \lderiv{\tsfb{X}^{h}}\til{\nbs{\omg}} = -\tfrac{1}{{\cff}}\sfb{X}^\ss{h}\cdt (\dif {\cff} \wedge \nbs{\omg}) = -\tfrac{1}{{\cff}} \big( (\sfb{X}^\ss{h}\cdt  \dif {\cff})  \nbs{\omg} - \dif {\cff} \wedge (\sfb{X}^\ss{h}\cdt  \nbs{\omg}) \big)  = -\tfrac{1}{{\cff}} \big(  \pbrak{{\cff}}{h}\nbs{\omg} - \dif {\cff} \wedge \dif h \big)  }.  }):
\begin{small}
\begin{align}
    \lderiv{\tsfb{X}^{h}}\til{\nbs{\omg}} \,=\,   \tsfb{X}^\ss{h}\cdt \exd \til{\nbs{\omg}} 
    \,=\, -\tfrac{1}{{\cff}} \big(  \pbrak{{\cff}}{h}\nbs{\omg} - \dif {\cff} \wedge \dif h \big)
\end{align}
\end{small}

\section{SOME MISCELLANEOUS DETAILS}  \label{app:prj_geo}

Here we collect of some further details regarding the projective point transformation, \eq{\psi\in\Dfism(\bvsp{E})}, that was detailed in section \ref{sec:prj_geomech}.

\paragraph{Differentials of the Projective Point Transformation.}
The cartesian coordinate matrix representations of \eq{\dif \psi} and \eq{\inv{\dif \psi}} were given in section \ref{sec:prj_gen_active}. 
Here, we give a few more details. For any \eq{\psi\in\Dfism(\bvsp{E})},  the differentials (regarded as bilinear pairings) are given in coordinate frame fields \eq{\be[r^\a]\in\vect(\Evec)}   and \eq{\bep[r^\a]\in\forms(\Evec)} by
\begin{small}
\begin{align} \label{dprj_gen_apx}
\begin{array}{lllll}
      \dif \psi = \pderiv{(r^\a\circ\psi)}{r^\b}(\be[r^\a])_\ii{\psi} \otms \bep[r^\b] 
\qquad,\qquad 
    \dif \inv{\psi} = \pderiv{(r^\a \circ \inv{\psi})}{r^\b}(\be[r^\a])_\ii{\inv{\psi}} \otms \bep[r^\b] 
\end{array}
\end{align}
\end{small}
Specifically, for \eq{\psi} the projective transformation, then we have following relations for cartesian coordinates \eq{r^\a\in\fun(\Evec)} defined from an orthonormal basis \eq{\hbe_\a\in\Evec}:
\begin{small}
\begin{align} \label{prj_cart_apx}
\begin{array}{llll}
     \psi \,=\, \tfrac{1}{r^\en}\hsfb{r} + r \envec  \,=\,  \tfrac{1}{r^\en}\hat{r}^i \hbe_i + r \envec 
\\[4pt]
     \inv{\psi} \,=\,  r^\en \hsfb{r} + \tfrac{1}{r}\envec   \,=\,  r^\en \hat{r}^i \hbe_i + \tfrac{1}{r}\envec  
\end{array}
&&
  \begin{array}{llllll}
        r^i \circ \psi = \tfrac{1}{r^\en\nrmtup{r}}r^i 
          &,\quad r^\en \circ \psi = r 
          &,\qquad  r\circ\psi = \tfrac{1}{r^\en}
      \\[4pt]
         r^i \circ \inv{\psi} =\tfrac{r^\en}{\nrm{r}}r^i
          &,\quad r^\en \circ \inv{\psi} = \tfrac{1}{r}
          &,\qquad r\circ \inv{\psi} =  r^\en
     \end{array}
\end{align}
\end{small}
(where \eq{\hat{r}^i:=r^i/\rfun} and \eq{\rfun =\nrm{\sfb{r}}}).  
Substituting the above into Eq.\eqref{dprj_gen_apx}, and collecting the coefficients into the matrix representations of \eq{\dif \psi} and \eq{\dif \inv{\psi}} in the \eq{r^\a} coordinates,  \eq{\crd{\dif \psi}{\bar{r}}=[ \pderiv{(r^\a\circ\psi)}{r^\b}]} and \eq{\crd{\dif \inv{\psi}}{\bar{r}}=[ \pderiv{(r^\a\circ\inv{\psi})}{r^\b}]}, we have the same expressions as
given earlier:\footnote{For cartesian coordinates \eq{\bartup{r}=(\tup{r},r^\en)=(r^1,\dots,r^{\en-1},r^\en):\vsp{E}\to\mbb{R}^{\en}}, we use the following notation, as described in Eq.\eqref{cord_norm}:
\begin{align}
    r_\en^2 \equiv (r^\en)^2
  &&,&&
    \hat{r}^i := r^i/\nrmtup{r} = r^i/ r
      &&,&&
    r_i := \emet_{ij}r^j = \kd_{ij} r^j =r^i
      &&,&&
    \htup{r}:=\tup{r}/\nrmtup{r}
    &&,&&
    \tup{r}^\flt = [r_i] \cong \trn{\tup{r}}
    &&,&&
    \tup{r}\otms \tup{r}^\flt = [r^i r_j] \cong \tup{r}\trn{\tup{r}}
\end{align} }
\begin{footnotesize}
\begin{align} \nonumber 
     \crd{\dif \psi}{\bar{r}} 
    =
     \begin{pmatrix}
         \tfrac{1}{r^\en \nrm{r}}( \imat_{\en\ii{-1}} - \htup{r}\otms \htup{r}^{\flt}) & -\tfrac{1}{r_\en^2}\htup{r} 
         \\
         \htup{r}^{\flt}  & 0 
     \end{pmatrix} 
     \equiv  \pderiv{\psi}{\bartup{r}}  
&&,&& 
  \crd{\dif \inv{\psi}}{\bar{r}} 
  =  \begin{pmatrix}
         \tfrac{r^\en }{\nrm{r}}( \imat_{\en\ii{-1}} - \htup{r}\otms \htup{r}^{\flt}) & \htup{r} 
         \\
        -\tfrac{1}{\nrm{r}^2} \htup{r}^{\flt}   & 0 
     \end{pmatrix} 
      \equiv \pderiv{\inv{\psi}}{\bartup{r}}
&&,&& 
  \crd{\dif \inv{\psi}_\ii{\psi}}{\bar{r}} 
   =  \begin{pmatrix}
         r^\en \nrm{r}( \imat_{\en\ii{-1}} - \htup{r}\otms \htup{r}^{\flt}) & \htup{r} 
         \\
        -r_\en^2 \htup{r}^{\flt}   & 0 
     \end{pmatrix} \equiv \inv{\big(\pderiv{\psi}{\bartup{r}}\big)}  
\end{align}
\end{footnotesize}
where  \eq{\crd{\dif \inv{\psi}_\ii{\psi}}{\bar{r}} \cdot  \crd{\dif \psi}{\bar{r}} = \imat_{\en}} (but \eq{ \crd{\dif \inv{\psi}}{\bar{r}} \cdot  \crd{\dif \psi}{\bar{r}} \neq \imat_{\en}}). 
To clarify, for any two points related by \eq{\barpt{x}=\psi(\barpt{q})\leftrightarrow \barpt{q}=\inv{\psi}(\barpt{x})} — with cartesian coordinate vectors in the \eq{\hbe_\a} basis denoted simply \eq{\bartup{x}:=\bartup{r}(\barpt{x})\in\mbb{R}^{\en}} and \eq{\bartup{q}:=\bartup{r}(\barpt{q})\in\mbb{R}^{\en}} — the above leads to:
\begin{small}
\begin{align} \label{dproj_cartesian1_apx}
\begin{array}{rlllll}
      \crd{\dif \psi_\ss{\barpt{q}}}{\bar{r}}  \equiv \pderiv{\bartup{x}}{\bartup{q}}
    &\!\!\!=\;  \begin{pmatrix}
         \tfrac{1}{q^\en \nrm{ {q}}}( \imat_{\en\ii{-1}} - \htup{q}\otms \htup{q}^{\flt}) & -\tfrac{1}{q_\en^2}\htup{q} 
         \\
         \htup{q}^{\flt}  & 0 
     \end{pmatrix} 
     &\!\! =\;
     \begin{pmatrix}
         \tfrac{\nrm{ {x}}}{x^\en}( \imat_{\en\ii{-1}} - \htup{x}\otms \htup{x}^{\flt}) & -\nrm{ {x}}^2\htup{x} 
         \\
         \htup{x}^{\flt}  & 0 
     \end{pmatrix} 
     =\, \inv{{(\pderiv{\bartup{q}}{\bartup{x}})}}
 \\[16pt]
   \crd{\dif \inv{\psi}_{\barpt{x}}}{\bar{r}}  \equiv \pderiv{\bartup{q}}{\bartup{x}} &\!\!\!= \; 
    \begin{pmatrix}
         \tfrac{x^\en }{\nrm{ {x}}}( \imat_{\en\ii{-1}} - \htup{x}\otms \htup{x}^{\flt}) & \htup{x} 
         \\
        -\tfrac{1}{\nrm{ {x}}^2} \htup{x}^{\flt}   & 0 
     \end{pmatrix} 
      &\!\! =\;  
      \begin{pmatrix}
         q^\en\nrm{ {q}} ( \imat_{\en\ii{-1}} - \htup{q}\otms \htup{q}^{\flt}) & \htup{q} 
         \\
        -q_\en^2 \htup{q}^{\flt}   & 0 
     \end{pmatrix}  
     =\,  \crd{\inv{(\dif \psi_\ss{\barpt{q}})}}{\bar{r}} \,=\, \inv{ \crd{\dif \psi_\ss{\barpt{q}}}{\bar{r}} }  \equiv \inv{{(\pderiv{\bartup{x}}{\bartup{q}})}} 
\end{array}
\end{align}
\end{small}

\paragraph{Angular Momentum Poisson Bracket.} 

We derive the relation \eq{\pbrak{\ttfrac{1}{2}\lang^2 }{H} = -r^2 \plin^i\pd_i U^\ss{1}} that was claimed earlier. Recall  \eq{\lang^2} and \eq{H} are given in cartesian coordinates by
\begin{small}
\begin{align}
    \lang^2=\nrmtup{r}^2\nrmtup{\plin}^2 -(r^i \plin_i)^2 
    &&,&&
    H \,=\, \tfrac{r_\en^2}{2m}( \lang^2 + r_\en^2 \plin_\en^2) \,+\, \tfrac{1}{2m} (\hat{r}^i \plin_i)^2 \,+\, U^\zr \,+\, U^\ss{1} 
\end{align}
\end{small}
Let us first recall the following brackets (which were derived in section \ref{sec:IOMs_prj_act}):
\begin{small}
\begin{align}
\begin{array}{rllllll}
        &\pbrak{  \ttfrac{1}{2} \nrmtup{\plin}^2}{ H} \,=\, -  \plin^j \pd_j  H 
       \,=\,    - \plin^j \pd_j U^\ss{1}   -\tfrac{(\hat{r}^i \plin_i)}{m \nrmtup{r}^3} \lang^2
       &,\qquad 
        \pbrak{\nrmtup{\plin}}{ H} =   - \hat{\plin}^j \pd_j  -\tfrac{(\hat{r}^i \hat{\plin}_i)}{m \nrmtup{r}^3} \lang^2U^\ss{1} 
\\[5pt]
    & \pbrak{\ttfrac{1}{2}\nrmtup{r}^2}{ H} \,=\, r_i \upd^i H  \,=\, \tfrac{1}{m}\tup{r}\cdot\tup{\plin}
      &,\qquad 
        \pbrak{\nrmtup{r}}{ H} =  \pbrak{r}{ H} =  \tfrac{1}{m}\htup{r}\cdot\tup{\plin}
\\[5pt]
    &\pbrak{\hat{r}^i \plin_i}{ H} = 0
\\[5pt]
   &\pbrak{r^i \plin_i}{ H} = \pbrak{r \hat{r}^i \plin_i}{ H} \,=\,  \hat{r}^i \plin_i \pbrak{r }{ H} \,=\, 
    \tfrac{1}{m} (\htup{r}\cdot\tup{\plin})^2  \;\neq 0
    &,\qquad 
    \pbrak{  \ttfrac{1}{2}(r^i \plin_i)^2 }{ H} = r^i \plin_i  \pbrak{ r^i \plin_i }{ H} = \tfrac{r}{m} (\hat{r}^i \plin_i)^3
\end{array}
\end{align}
\end{small}
where \eq{r = \nrmtup{r}}. 
Using the above, we then obtain \eq{\pbrak{\ttfrac{1}{2}\lang^2 }{H}} as follows (where \eq{\plin^i:=\emet^{ij} \plin_j}):
\begin{small}
\begin{align}
  \pbrak{ \ttfrac{1}{2}\lang^2 }{ H} 
  \,=\,  \ttfrac{1}{2}\pbrak{ \nrmtup{r}^2\nrmtup{\plin}^2 -(\tup{r}\cdot\tup{\plin})^2 }{ H} 
   &\,=\,
    \nrmtup{\plin}^2\pbrak{  \ttfrac{1}{2}\nrmtup{r}^2 }{ H}  \,+\,   \nrmtup{r}^2\pbrak{  \ttfrac{1}{2}\nrmtup{\plin}^2 }{ H}  \,-\,   \pbrak{  \ttfrac{1}{2}(\tup{r}\cdot\tup{\plin})^2 }{ H}  
\\ \nonumber 
    &\,=\, \nrmtup{\plin}^2 \big( \tfrac{1}{m} \tup{r}\cdot\tup{\plin} \big)  \,+\, 
     \nrmtup{r}^2 \big( -\tfrac{(\tup{r}\cdot\tup{\plin})}{m \nrmtup{r}^4 } \lang^2   \,-\, \plin^j\pd_j U^\ss{1} \big)
     \,-\,  \big( \tfrac{\nrm{r}}{m} (\htup{r}\cdot\tup{\plin})^3 \big)
 \\ \nonumber 
     & \,=\, \tfrac{\tup{r}\cdot\tup{\plin}}{m}  \big(  \nrmtup{\plin}^2  - \tfrac{\lang^2}{\nrmtup{r}^2}  \big) \,-\, \nrmtup{r}^2 \plin^j\pd_j U^\ss{1} 
     \,-\,   \big(\tfrac{\nrm{r}}{m} (\htup{r}\cdot\tup{\plin})^3\big)
\\ \nonumber 
     & \,=\,
      \big( \tfrac{\nrmtup{r} }{m} (\htup{r}\cdot\tup{\plin})^3 \big) \,-\, \nrmtup{r}^2 \plin^j \pd_j U^\ss{1} 
      \,-\,   \big( \tfrac{\nrm{r}}{m} (\htup{r}\cdot\tup{\plin})^3 \big)
      \;=\; -\nrmtup{r}^2 \plin^j \pd_j U^\ss{1}
\end{align}
\end{small}

\paragraph{More on the ``Projective Metric'':~Restriction to Submanifolds.}
Let \eq{\bsfb{\emet}_\ii{\!\Evec}\equiv \inner{\cdot}{\cdot}\in\botimes^0_2\vsp{E} \cong \tens^0_2(\vsp{E})_\ii{\tx{hmg.}}} denote the homogeneous Euclidean metric. Recall the \eq{\psi}-induced metric, \eq{\sfb{g}:=\psi^*\bsfb{\emet}_\ii{\!\Evec}} (i.e., the ``projective metric''), that was given in Eq.\eqref{g_proj_active} – \ref{gin_proj_active}: 
\begin{small}
\begin{align} \label{gprj_again}
\begin{array}{rllll}
       \bsfb{\emet}_\ii{\!\Evec} &\!\!\!\! =\, \sfb{\emet}_\ii{\!\Sigup}  + \enform  \otms \enform  = \emet_{\a\b}\hbep^\a\otms\hbep^\b 
\\[6pt]
     \inv{\bsfb{\emet}_\ii{\!\Evec}} \!&\!\!\!\! =\, \inv{\sfb{\emet}}_\ii{\!\Sigup} +  \envec  \otms \envec  = \emet^{\a\b}\hbe_\a\otms\hbe_\b 
\end{array}
&&
\begin{array}{rllllll}
          \sfb{g} &\!\!\!\! =\,  \psi^*\bsfb{\emet}_\ii{\!\Evec}  &\!\!\! =\,  \tfrac{1}{r_\en^2\, r^2} (\sfb{\emet}_\ii{\!\Sigup}-\hsfb{r}^\flt\otms \hsfb{r}^\flt) \,+\, \hsfb{r}^\flt \otms \hsfb{r}^\flt \,+\, \tfrac{1}{r_\en^4 }\enform  \otms \enform   &\!\!\! \in\tens^0_2(\bvsp{E})
    \\[6pt]
         \inv{\sfb{g}} \! &\!\!\!\! =\,  \inv{\psi}_* \inv{\bsfb{\emet}_\ii{\!\Evec}}   &\!\!\!\!  =\,  r_\en^2 r^2 ( \inv{\sfb{\emet}}_\ii{\!\Sigup} - \hsfb{r}\otms \hsfb{r}) \,+\, \hsfb{r}\otms\hsfb{r} \,+\, r_\en^4 \envec \otms \envec  &\!\!\! \in\tens^2_0(\bvsp{E})
\end{array}
\end{align}
\end{small}
where \eq{\bartup{r}=(r^1,\dots,r^\en):\vsp{E} \to \mbb{R}^\en} are inertial cartesian coordinates corresponding an \eq{\bsfb{\emet}_\ii{\!\Evec}}-orthonormal basis, \eq{\hbe_\a \equiv \be[r^\a]\in\vect(\vsp{E})} such that \eq{\emet_{\a\b}=\kd_{\a\b}= \emet^{\a\b}=\kd^{\a\b}}.  
As previously shown around Eq.\eqref{2sphere_met}, the above can also be expressed in terms of the spherical metric, \eq{\sfb{s}},  on \eq{\man{S}^{\en-\two}\subset\bs{\Sigup}\subset\vsp{E}} (viewed extrinsically) as  
\begin{small}
\begin{align} \label{g_2sphere_met_apx}
\begin{array}{llllll}
   \bsfb{\emet}_\ii{\!\Evec} 
=  r^2 \sfb{s} \,+\,  \hsfb{r}^\flt \otms \hsfb{r}^\flt  \,+\, \enform \otms\enform
\\[5pt]
 \inv{\bsfb{\emet}_\ii{\!\Evec}} 
 =   \tfrac{1}{r^2} \inv{\sfb{s}} \,+\, \hsfb{r} \otms \hsfb{r} \,+\, \envec \otms\envec
\end{array}
&&,&&
\begin{array}{lllll}
          \sfb{g} 
         \,=\, \tfrac{1}{r_\en^2} \sfb{s}  \,+\, \hsfb{r}^\flt \otms \hsfb{r}^\flt \,+\, \tfrac{1}{r_\en^4 }\enform  \otms \enform 
   \\[5pt]
         \inv{\sfb{g}} 
          \,=\,  r_\en^2 \inv{\sfb{s}} \,+\,  \hsfb{r} \otms \hsfb{r} \,+\, r_\en^4 \envec \otms \envec 
\end{array}
    &&,&&
\fnsize{with}\;\; 
 \begin{array}{llllll}
      \sfb{s} 
       \,=  \tfrac{1}{r^2}( \sfb{\emet}_\ii{\!\Sigup} - \hsfb{r}^\flt \otms \hsfb{r}^\flt  ) = \tfrac{1}{r} \nab \hsfb{r}^\flt
  \\[5pt]
     \inv{\sfb{s}} 
     = r^2( \inv{\sfb{\emet}_\ii{\!\Sigup}} - \hsfb{r} \otms \hsfb{r} )
  \end{array}
\end{align}
\end{small}
where \eq{\sfb{s}=\tfrac{1}{r} \nab \hsfb{r}^\flt} gives the metric inherited from \eq{\sfb{\emet}_\ii{\!\Sigup}} on a codimension-1 unit sphere embedded in \eq{\bs{\Sigup}}. 
The case that \eq{\en-1=3} such that \eq{\man{S}^{\en-\two}=\man{S}^2} is the classic 2-sphere is given in the footnote\footnote{Consider the case \eq{\en =4} and take local coordinates  \eq{(r,\theta,\phi,r^\en):(\sman{E}\subset\vsp{E})\to\mbb{R}^4}  where  \eq{(r,\theta,\phi)} are the usual spherical coordinates on the 3-dim Euclidean hyperplane \eq{\bs{\Sigup}\subset \vsp{E}}, with \eq{\rfun =\nrm{\sfb{r}}}. Considering just \eq{(\bs{\Sigup},\sfb{\emet}_\ii{\!\Sigup})}, the metric \eq{\sfb{s}} on \eq{\man{S}^2} inherited from \eq{\sfb{\emet}_\ii{\!\Sigup}} is then given in spherical coordinate frame fields by:
\begin{align} \label{2sphere_met_apx}
\begin{array}{llllll}
       \sfb{\emet}_\ii{\!\Sigup} 
        \,=\, \bep^{\rfun}\otms \bep^{\rfun} + r^2  \bep[\theta]\otms \bep[\theta] + r^2 \sin^2\theta \bep[\phi]\otms \bep[\phi]
\\[5pt]
      \inv{\sfb{\emet}}_\ii{\!\Sigup} 
      \,=\, \be_{\rfun}\otms \be_{\rfun} + \tfrac{1}{r^2} \be[\theta]\otms \be[\theta] + \tfrac{1}{r^2 \sin^2\theta} \be[\phi]\otms \be[\phi] 
\end{array}
&& \Rightarrow &&
\begin{array}{llllll}
      \sfb{s} \,=\, \tfrac{1}{r^2}( \sfb{\emet}_\ii{\!\Sigup} - \bep^{\rfun}\otms\bep^{\rfun}  ) 
      \,=\, \bep[\theta]\otms \bep[\theta] + \sin^2 \theta \bep[\phi]\otms\bep[\phi] 
\\[5pt]
     \inv{\sfb{s}} =\, r^2( \inv{\sfb{\emet}}_\ii{\!\Sigup} - \be_{\rfun}\otms \be_{\rfun} ) 
     \,=\,   \be[\theta]\otms \be[\theta] \,+\, \tfrac{1}{\sin^2\theta} \be[\phi]\otms \be[\phi]
\end{array}
&&, &&
\begin{array}{lllll}
    \bep^{\rfun} \equiv \hsfb{r}^\flt
\\[5pt]
     \be_{\rfun} \equiv \hsfb{r} 
\end{array}
\end{align}
}. 

\paragraph{Projective Metric on $\Sig_\ii{b}\subset\bvsp{E}$.}
Let us now determine the metrics inherited from \eq{\bsfb{\emet}} and \eq{\sfb{g}} on the hypersurfaces of \eq{\bvsp{E}} given by  \eq{\Sig_\ii{b}:=\inv{(r^\en)}\{b\} = \bs{\Sigup}_\nozer \oplus \{b\}} and \eq{\man{Q}_\ii{b} := \inv{\rfun}\{b\} = \man{S}^{\en-2}_\ii{b} \times \vsp{N}_\ii{+}} (for any \eq{b \in\mbb{R}_\ii{+}}). 
Let us denote their inclusion maps by:
\begin{small}
\begin{align}
\begin{array}{lllll}
     \imath: \Sig_\ii{b} \hookrightarrow \Evec
     &,\qquad 
     \imath \,\cong\, \sfb{r} + b \envec
\\[5pt]
     \jmath:\man{Q}_\ii{b} \hookrightarrow \Evec
      &,\qquad 
     \jmath \,\cong\, b\hsfb{r} + r^\en \envec 
\end{array}
\end{align}
\end{small}
The pullbacks of \eq{\bsfb{\emet}} and \eq{\sfb{g}} to \eq{\Sig_\ii{b}} are easily obtained; it simply amounts to chopping off the \eq{\enform} component and letting \eq{r^\en =b}:
\begin{small}
\begin{align}
\begin{array}{rllll}
      \imath^* \bsfb{\emet} &\!\!\!\! =\, \sfb{\emet}_\ii{\!\Sigup} \in \tens^0_2(\Sig)
\\[5pt]
      \inv{(\imath^* \bsfb{\emet})} &\!\!\!\! =\, \inv{\sfb{\emet}}_\ii{\!\Sigup} \in \tens^2_0(\Sig) 
\end{array}
&&,&&
\begin{array}{rllll}
      \imath^* \sfb{g} &\!\!\!\! =\,  \tfrac{1}{b^2 r^2} 
     (\sfb{\emet}_\ii{\!\Sigup}-\hsfb{r}^\flt\otms \hsfb{r}^\flt) \,+\, \hsfb{r}^\flt \otms \hsfb{r}^\flt 
     &\!\!\! \in \tens^0_2(\Sig_\ii{b})
\\[5pt]
      \inv{(\imath^* \sfb{g})} &\!\!\!\! =\, {b^2 r^2} 
     (\inv{\sfb{\emet}}_\ii{\!\Sigup} - \hsfb{r} \otms \hsfb{r}) \,+\, \hsfb{r} \otms \hsfb{r}
     &\!\!\!\in \tens^2_0(\Sig_\ii{b}) 
\end{array}
\end{align}
\end{small}
where the Euclidean metric does not depend on the value of \eq{b \in \mbb{R}_\ii{+}} (i.e., is the same for any \eq{\Sig_\ii{b}})  and so we simply write \eq{\tens^0_2(\Sig)} rather than specifying \eq{\tens^0_2(\Sig_\ii{b})}.  

\paragraph{Projective Metric on $\man{Q}_\ii{b}\subset\bvsp{E}$.}
We now consider the hypersurface   \eq{\man{Q}_\ii{b} := \inv{\rfun}\{b\} = \man{S}^{\en-2}_\ii{b} \times \vsp{N}_\ii{+}} (for any \eq{b \in\mbb{R}_\ii{+}}).
To obtain expressions for \eq{\jmath^*\bsfb{\emet}} and \eq{\jmath^*\sfb{g}}, let us momentarily take \eq{\en = 4} such that \eq{\man{Q}_\ii{b} = \man{S}^2_\ii{b} \times \vsp{N}_\ii{+}}.  Consider local coordinates  \eq{(r,\theta,\phi,r^\en):(\sman{E}\subset\Evec)\to\mbb{R}^4}  where  \eq{(r,\theta,\phi)} are the usual spherical coordinates on the 3-dim Euclidean hyperplane \eq{\bs{\Sigup}}, and \eq{(\theta,\phi)} the usual coordinates on the 2-sphere.  We then have local coordinates \eq{(\theta,\phi,r^\en):(\sman{Q}\subset \man{Q}_\ii{b})\to\mbb{R}^3} and the inclusion can be viewed as:
\begin{small}
\begin{align}
     \jmath:\man{Q}_\ii{b} \hookrightarrow \Evec^4
      &,\qquad 
     \jmath \,\cong\, b\hsfb{r} + r^\en \envec   \,=\, b \big( \mrm{s}_\theta \mrm{c}_\phi \hbe_1 +   \mrm{s}_\theta \mrm{s}_\phi \hbe_2 +  \mrm{c}_\theta \hbe_3 \big)  + r^\en \envec 
\end{align}
\end{small}
We then obtain \eq{\jmath^* \bsfb{\emet}} as follows
\begin{small}
\begin{align}
\begin{array}{rllllll}
      \jmath^* \bsfb{\emet}&\!\!\!\! =\,  b^2\big(  \bep[\theta]\otms \bep[\theta] + \sin^2 \theta \bep[\phi]\otms\bep[\phi]\big) \,+\, \enform \otms \enform 
      &\!\!\!\!=\, b^2 \sfb{s}  \,+\, \enform \otms \enform 
      &\!\!\! \in \tens^0_2(\man{Q}_\ii{b} )
      &,\;\; \jmath^* \bsfb{\emet} \cong ( \bsfb{\emet} - \bep^{\rfun}\otms\bep^{\rfun})
 \\[5pt]
     \inv{(\jmath^* \bsfb{\emet})} &\!\!\!\! =\,  \tfrac{1}{b^2} \big(  \be[\theta]\otms \be[\theta] + \tfrac{1}{\sin^2 \theta} \be[\phi]\otms\be[\phi]\big) \,+\, \envec \otms \envec
    &\!\!\!\!=\, \tfrac{1}{b^2}\inv{\sfb{s}} \,+\, \envec \otms \envec
     &\!\!\!  \in \tens^2_0(\man{Q}_\ii{b} )
      &,\;\;  \inv{(\jmath^* \bsfb{\emet})} \cong ( \inv{\bsfb{\emet}} - \be_{\rfun}\otms\be_{\rfun}) 
\end{array}
\end{align}
\end{small}
where \eq{\sfb{s}} is the usual metric on the \textit{unit} 2-sphere \eq{\man{S}^2\subset \bs{\Sigup}} given in Eq.\eqref{2sphere_met_apx} 
which is being regarded in \eq{\vsp{E}^4}. The above is exactly the metric one would expect on the product manifold \eq{\man{Q}_\ii{b} = \man{S}^2_\ii{b} \times \vsp{N}}, inherited from the Euclidean metric on \eq{\Evec^4}.  
Next, we obtain \eq{\jmath^*\sfb{g}  = \jmath^*(\psi^* \bsfb{\emet}) = (\psi\circ\jmath)^*\bsfb{\emet}}: 
\begin{small}
\begin{align}
\begin{array}{rllllllll}
       \jmath^*\sfb{g} &\!\!\!\! =\, 
      \tfrac{1}{r_\en^2}\big(\bep[\theta]\otms \bep[\theta] + \sin^2 \theta \bep[\phi]\otms\bep[\phi]\big) \,+\, \tfrac{1}{r_\en^4 }\enform \otms \enform
     &\!\!\!\!=\, \tfrac{1}{r_\en^2} \sfb{s}  \,+\, \tfrac{1}{r_\en^4 }\enform \otms \enform
      &\!\!\! \in   \tens^0_2(\man{Q}_\ii{b} )
      &,\;\;
       \jmath^*\sfb{g} \cong ( \sfb{g} - \bep^{\rfun}\otms\bep^{\rfun})
\\[6pt]
     \inv{(\jmath^*\sfb{g})} &\!\!\!\! =\,   r_\en^2 \big(\be[\theta]\otms \be[\theta] + \tfrac{1}{\sin^2 \theta} \be[\phi]\otms\be[\phi]\big) \,+\, r_\en^4\envec \otms \envec
   &\!\!\!\!=\,  r_\en^2 \inv{\sfb{s}} \,+\, r_\en^4\envec \otms \envec
    &\!\!\!  \in \tens^2_0(\man{Q}_\ii{b} )
    &,\;\;
     \inv{(\jmath^* \sfb{g})} \cong ( \inv{\sfb{g}} - \be_{\rfun}\otms\be_{\rfun})
\end{array}
\end{align}
\end{small}


\paragraph{An Alternative Projective Transformation for the Burdet-Ferrándiz Formulation.}
We note that an alternative map, \eq{\til{\psi}\in\Dfism(\bvsp{E})}, corresponding to the Burdet-Ferrándiz (BF) formulation of the projective transformation, would be:
\begin{small}
\begin{align}
   \barpt{x} =  \til{\psi}(\barpt{q}) = \tfrac{1}{q^\en} \ptvec{q} +\nrm{\pt{q}} \envec 
    \qquad\leftrightarrow \qquad 
    \barpt{q} = \inv{\til{\psi}}(\barpt{x}) =  \tfrac{x^\en}{\nrm{\pt{x}}} \ptvec{x} + \tfrac{x^\en}{\nrm{\pt{x}}}\envec 
        &&,&&
        \inv{\til{\psi}}\circ \til{\psi} =\til{\psi} \circ  \inv{\til{\psi}} = \Id_\ii{\bvsp{E}}
\end{align}
\end{small}
Which we may also write in linear coordinates as  
\begin{small}
\begin{align}
    \til{\psi} \,=\, \tfrac{1}{r^\en} r^i \hbe_i + r \envec  \,=\, \tfrac{1}{r^\en} \sfb{r}  + r \envec 
    \qquad,\qquad 
    \inv{\til{\psi}} \,=\, \tfrac{r^\en}{r}( r^i\hbe_i + \envec ) \,=\, r^\en \hsfb{r}  \,+\, \tfrac{r^\en}{r}\envec  
    &&
    \begin{array}{lllll}
          r^i \circ \til{\psi} = \tfrac{r^i}{r^\en}
         &,\quad r^i \circ \inv{\til{\psi}} = r^\en \hat{r}^i
    \\[4pt]
         r^\en \circ \til{\psi} = r 
         &,\quad r^\en \circ \inv{\til{\psi}} = \tfrac{r^\en}{r}
    \\[4pt]
         r \circ \til{\psi} = \tfrac{r}{r^\en} 
          &,\quad r \circ \inv{\til{\psi}} = r^\en 
    \end{array}
\end{align}
\end{small}
Note that \eq{\hat{r}^i:=r^i/\rfun} still satisfies \eq{\hat{r}^i\circ \til{\psi} = \hat{r}^i \circ \inv{\til{\psi}} = \hat{r}^i}. 
This map induces the following metric:
\begin{small}
\begin{align}
    \sfb{g} := \til{\psi}^*\bsfb{\emet} \,=\, \tfrac{1}{r_\en^2}\sfb{\emet}_\ii{\!\Sigup} + \hsfb{r}^\flt \otms\hsfb{r}^\flt \,-\, \tfrac{1}{r_\en^3}r_i( \hbep^i\otms \enform +  \enform \otms \hbep^i) \,+\, \tfrac{r^2}{r_\en^2} \enform \otms \enform
    &&
    \cord{\txi{g}}{\bartup{r}} \,=\, 
    \begin{pmatrix}
        \tfrac{1}{r_\en^2}\kd_{ij} + \hat{r}_i\hat{r}_j & -\tfrac{1}{r_\en^3} r_i \\
        -\tfrac{1}{r_\en^3} r_i & \tfrac{r^2}{r_\en^4}
    \end{pmatrix}
\end{align}
\end{small}

\section{PREVIOUS WORK:~COORDINATE-BASED PROJECTIVE TRANSFORMATION \& REGULARIZATION} \label{sec:prj_sum2}

 Here, we present a brief summary of 
 a canonical/symplectic \textit{coordinate}  transformation for Burdet-type projective coordinates,  and its application to regularization of central force particle dynamics (in particular, the Kepler and Manev problems). This coordinate transformation is a modification of the Burdet-Ferrándiz (BF) transformation \cite{ferrandiz1987general}, and was detailed previously by the authors in \cite{peterson2025prjCoord,peterson2022nonminimal}. This section serves only for context and comparison of the present work to previous work; it may be skipped by the uninterested or impatient reader. In many ways, the present work is a reformulation, in more rigorous differential geometric terms, of what is summarized bellow.

\begin{notesq}
    The following summary is written in the coordinate-dominant manner of classical analytical mechanics, reflective of the original formulation in \cite{peterson2025prjCoord,peterson2022nonminimal}. Unlike the remainder of this work, here we avoid the mathematical formalism of differential geometry and do not shy away from abuses of notation. The only mathematical ``structure'' used are the usual structures and operations on \eq{\mbb{R}^\ii{N}}. 
    As a consequence, the contents of the present section \ref{sec:prj_sum2} should generally not be interpreted using the same mathematical framework employed in other sections. \textit{Nor should the notation employed here be assumed to carry the same meaning as in other sections.}  
\end{notesq}

\subsection{A Modified Burdet-Ferrándiz Projective Coordinate Transformation}


We begin with a modified version of Burdet's projective point transformation, given by \eq{(\tup{q},u)\mapsto\tup{r}=  \tup{q}/(\mag{\tup{q}} u)}, and then ``lift'', or, ``extend'' this point transformation to a canonical transformation, \eq{(\tup{q},u,\tup{p},p_\ss{u})\mapsto (\tup{r},\tup{\plin})}. We present the Hamiltonian formulation for particle dynamics in \eq{\Evec^3} — including arbitrary conservative and nonconservative forces —  using the redundant ``canonical projective coordinates'', \eq{(\tup{q},u,\tup{p},p_\ss{u})\in\mbb{R}^8}. The derivation and details (omitted from the following) can be found in \cite{peterson2025prjCoord,peterson2022nonminimal}. The below developments also follow from sections \ref{sec:prj_geomech} and \ref{sec:prj_regular} of the present work.


\paragraph{Original Cartesian Coordinate Formulation.}
We start with the Hamiltonian (per unit mass) and canonical equations of motion for a particle in \eq{\Evec^3}, subject to conservative forces corresponding to some potential function (per unit mass) \eq{V}, and arbitrary nonconservative forces (per unit mass), \eq{\sfb{a}^\ii{\mrm{nc}}}. The Hamiltonian dynamics are expressed in inertial \textit{cartesian} position and momentum or velocity (they are the same for the purposes of the present discussion\footnote{In this section, everything is scaled by the mass of the particle or, alternatively, you could say we use units such that the particle has unit mass. In any case, the result is that there is no substantial difference between \textit{cartesian} velocity and momentum coordinates. In geometric terms, there \textit{is} still a difference: velocity coordinates are functions on velocity phase space (a tangent bundles) whereas momentum coordinates are functions on phase space (a cotangent bundle). But, as mentioned, we are not being so precise here in section \ref{sec:prj_sum2}.})
coordinates, \eq{(\tup{r},\tup{\plin})\in\mbb{R}^6}, simply as:
\begin{align} \label{K0}
    &\mscr{K}(\tup{r},\tup{\plin},t) \,=\, \tfrac{1}{2} \plin^2 \,+\, V(\tup{r},t)
&& \begin{array}{ll}
      \dot{\tup{r}} \,=\, \pderiv{\mscr{K}}{\tup{\plin}} \,=\, \tup{\plin} 
\\[5pt] 
    \dot{\tup{\plin}} \,=\, -\pderiv{\mscr{K}}{\tup{r}} \,+\, \tup{a}^\ii{\mrm{nc}} \,=\,
    -\pderiv{V}{\tup{r}} \,+\,  \tup{a}^\ii{\mrm{nc}}
\end{array}
\end{align}
where \eq{\plin :=\mag{\tup{\plin}}},
where \eq{V(\tup{r},t)} is the cartesian coordinate representation of the (possibly time-dependent) potential, and where \eq{\tup{a}^\ii{\mrm{nc}}\in\mbb{R}^3} are the generalized \textit{non}-conservative forces which, in the case of cartesian coordinates, is nothing more than the cartesian components of the nonconservative force vector, \eq{\sfb{a}^\ii{\mrm{nc}}}.
 We consider the case that potential is comprised of a central force term, \eq{V^{0}(r)}, dependent only on the radial distance, \eq{r:=\mag{\tup{r}}}, and a second term, \eq{V^{1}(\tup{r},t)} accounting for all other conservative perturbations. The above dynamics are then equivalent to
\begin{align} \label{V0V1}
    V(\tup{r},t) \,=\, V^{0}(r) \,+\, V^{1}(\tup{r},t) 
    &&\Rightarrow &&
    \dot{\tup{r}} = \tup{\plin} 
    \quad,\quad 
    \dot{\tup{\plin}} \,=\, -\pderiv{V^{0}}{r}\htup{r} \,+\, \tup{F}
    &&,&&
     \tup{F} \,:=\, -\pderiv{V^{1}}{\tup{r}} + \tup{a}^\ii{\mrm{nc}}
\end{align}
with \eq{\htup{r}:=\tup{r}/r} the radial unit vector and where \eq{\tup{F}} is defined as the cartesian coordinate vector of the \textit{total} perturbing force (per unit mass), arising from all conservative and nonconservative perturbations (i.e., everything other than central forces from \eq{V^{0}}). 

\begin{notesq}
    When the particular form of \eq{V^{0}(r)} is relevant, we consider the  so-called \textit{Manev potential}:\footnote{The negative signs and factor of \eq{\ttfrac{1}{2}} are included simply for convenience.}
    \begin{small}
    \begin{align} \label{Vmanev_sum}
        V^{0} \,=\, -\tfrac{\txkbar_1}{r} \,-\, \tfrac{1}{2} \tfrac{\txkbar_2}{r^2}
        \qquad,\qquad 
        -\pderiv{V^{1}}{\tup{r}} \,=\,  -\tfrac{\txkbar_1}{r^2}\htup{r} \,-\,  \tfrac{\txkbar_2}{r^3}\htup{r}
    \end{align}
    \end{small}
    for scalars \eq{\txkbar_1,\txkbar_2\in\mbb{R}}. The Kepler potential corresponds to the case \eq{\txkbar_1>0} and \eq{\txkbar_2=0}.
\end{notesq}


\paragraph{Canonical Transformation for Projective Coordinates.}   For the Hamiltonian formulation of the projective decomposition, we then transform from the above \eq{(\tup{r},\tup{\plin})\in\mbb{R}^6} to
new, redundant, canonical coordinates, \eq{(\bartup{q},\bartup{p})=(\tup{q},u,\tup{p},p_\ss{u}) \in\mbb{R}^8}, where  \eq{\bartup{q}=(\tup{q},u)\in\mbb{R}^4} correspond to the classic projective coordinates.
This is done by constructing a (local) submersion, \eq{\Psiup: \mbb{R}^8 \to \mbb{R}^6}, given
as follows (where \eq{q:=\mag{\tup{q}}}, not \eq{\mag{\bartup{q}}}):\footnote{The point transformation \eq{\uppsi:\mbb{R}^4\ni(\tup{q},u)\mapsto \tup{r}\in\mbb{R}^3}  is
a \textit{submersion}; it is  surjective with \eq{\text{rnk} \rmd \uppsi \equiv \text{rnk} \pderiv{\tup{r}}{(\tup{q},u)}=3}. It is then lifted to another submersion, \eq{\Psiup:\mbb{R}^8\ni(\tup{q},u,\tup{p},p_\ss{u})\mapsto (\tup{r},\tup{\plin})\in\mbb{R}^6}, which is surjective with \eq{\text{rnk} \rmd \Psiup \equiv \rnk \pderiv{(\tup{r},\tup{\plin})}{(\tup{q},u,\tup{p},p_\ss{u})} =6} (for \eq{n\neq 0}). Note the domain and codomains of these maps are not actually all of the indicated \eq{\mbb{R}^{N}} spaces; they are subsets excluding the the case \eq{\tup{q}=0=u}.  }
\begin{small}
\begin{align} \label{PT_0}
\Psiup: \mbb{R}^8\to \mbb{R}^6 \;,\quad (\bartup{q},\bartup{p})\mapsto (\tup{r},\tup{\plin})
\quad \left\{ \;\;
\begin{array}{lllll}
      \tup{r} \,=\, \tfrac{1}{u q} \tup{q}
  \\[5pt]
     \tup{\plin} \,=\,  u q \big(  (\imat_3 - \htup{q}\otimes\htup{q})\cdot\tup{p} \,-\, \tfrac{u}{q} p_\ss{u}\htup{q} \big)
\end{array} \right. 
&&,&&
\begin{array}{llll}
     \rnk \rmd \Psiup = 6
\end{array} 
\end{align}
\end{small}
\begin{small}
\begin{notesq}
    Note we \textit{specify} the above point transformation  which then \textit{induces} the above momenta transformation. The details were given previously in \cite{peterson2025prjCoord,peterson2022nonminimal} where the authors considered a more general family of projective point transformations of the form  \eq{\tup{r} = u^n q^m \tup{q}}, for any \eq{n\neq 0,m\in\mbb{R}}. 
The authors ultimately preferred the results from choosing \eq{n=m=-1}, which is the case summarized here (and which is re-developed in a different, geometric, formulation in the bod y of this work). 
\end{notesq}
\end{small}
\noindent For the transformed dynamics, we need the Hamiltonian, \eq{\mscr{H}}, for the new coordinates, \eq{(\bartup{q},\bartup{p}}). 
Since the transformation \eq{\Psiup} is not time-dependent, the new Hamiltonian is given simply by the composition \eq{\mscr{H}=\mscr{K}\circ \Psiup}. Substituting the above into \eq{\mscr{K}} from Eq.\eqref{K0} — and assuming the potential has the form \eq{V=V^\zr+V^{1}} as in Eq.\eqref{V0V1} — leads to the following \eq{\mscr{H}} and equations of motion for the projective coordinates: 
\begin{small}
\begin{gather} \label{Ham0}
    \mscr{H} (\bartup{q},\bartup{p},t) \,=\,  \tfrac{u^2}{2} ( \slang^2 +  u^2 p_\ss{u}^2 )  +  V
    \;\;=\;\;
    \tfrac{u^2}{2}\Big( q^2 p^2 -(\tup{q}\cdot\tup{p})^2 +  u^2 p_\ss{u}^2 \Big)  \,+\, V^{0}(u) \,+\, V^{1}(\bartup{q},t)
\\
 \begin{array}{lllll}
     \dot{q}_i = \pderiv{\mscr{H}}{p_i} = - u^2 \slang_{ij}q_j
\\[10pt]
    \dot{u}  = \pderiv{\mscr{H}}{p_u} = u^4 p_\ss{u} 
 \end{array}
\quad,\quad
 \begin{array}{llll}
    \dot{p}_i = -\pderiv{\mscr{H}}{q_i} + \alpha_i 
     &\!\!\! = - u^2 \slang_{ij} p_j  \,+\, f_i
\\[10pt]
     \dot{p}_{u} = -\pderiv{\mscr{H}}{u} + \alpha_u    
      &\!\!\! = -u\big(\slang^2 +  2 u^2 p_\ss{u}^2 \big)   -  \pderiv{V^{0}}{u}  +  f_\ss{u} 
    \end{array}  
\qquad,\qquad
 \begin{array}{llll}
          \tup{\alpha} :=\,  \tup{a}^\ii{\mrm{nc}} \cdot \pderiv{\tup{r}}{\tup{q}} 
           \,=\, 
           \tfrac{1}{u q} (\imat_3 - \htup{q} \otimes \htup{q} ) \cdot \tup{a}^\ii{\mrm{nc}}
 \\[5pt]
          \alpha_\ss{u}  :=\,  \tup{a}^\ii{\mrm{nc}} \cdot \pderiv{\tup{r}}{u}  
           \,=\, -\tfrac{1}{u^2} \htup{q}\cdot \tup{a}^\ii{\mrm{nc}} 
\end{array}
\end{gather}
\end{small}
where \eq{\slang} is the magnitude of the angular momentum per unit mass (i.e., specific angular momentum), and 
where \eq{\bartup{\alpha}=(\tup{\alpha},\alpha_\ss{u})} are the new generalized nonconservative forces (per unit mass). Above, we have combined these with the perturbing conservative forces from \eq{V^{1}}, and defined the total generalized perturbing forces,  \eq{\bartup{f}=(\tup{f},f_\ss{u})}: 
\begin{small}
\begin{align} \label{Fdef_sum}
     \tup{f} :=\, -\pderiv{V^{1}}{\tup{q}} + \tup{\alpha} 
     \,=\,  \tfrac{1}{u q} (\imat_3 - \htup{q} \otimes \htup{q} ) \cdot \tup{F}
 &&,&& 
    f_\ss{u} :=\, -\pderiv{V^{1}}{u} + \alpha_\ss{u} 
    \,=\,  -\tfrac{1}{u^2} \htup{q}\cdot \tup{F}
  &&,&&   \tup{F} :=\,  -\pderiv{V^{1}}{\tup{r}}+\tup{a}^\ii{\mrm{nc}} 
\end{align}
\end{small}
where \eq{\tup{F}=-\pderiv{V^{1}}{\tup{r}}+\tup{a}^\ii{\mrm{nc}}} is the cartesian components of the total perturbing force (per unit mass).

\begin{notesq}
It is noteworthy that the (specific) angular momentum, when expressed in terms of \eq{(\bartup{q},\bartup{p})=(\tup{q},u,\tup{p},p_\ss{u})}, does not depend on \eq{u} or \eq{p_\ss{u}} and takes the same form 
in terms of \eq{(\tup{r},\tup{\plin})} or \eq{(\tup{q},\tup{p})}:\footnote{In 3-dim coordinate space, \eq{\mbb{R}^3}, we note \eq{\tup{\slang}=(\slang_1,\slang_2,\slang_3) = \tup{r}\times \tup{\plin}}, and \eq{\Slang:=[\slang_{ij}] = \tup{r}\otimes \tup{\plin} - \tup{\plin}\otimes \tup{r}}, and \eq{\slang^2}, are related by:\\
\eq{\qquad\qquad \qquad 
\slang_i = \lc_{ijk} r_j \plin_k = \tfrac{1}{2}\lc_{ijk}\slang_{jk} \;\,\leftrightarrow\;\, \slang_{ij} = \lc_{ijk} \slang_k = r_i \plin_j - \plin_i r_j 
\qquad,\qquad 
\Slang = \hdge{\tup{\slang}} \;\leftrightarrow\; \tup{\slang} = \hdge{\Slang}
\qquad,\qquad  \slang^2 = \slang_i \slang_i = \tfrac{1}{2}\slang_{ij}\slang_{ij}}.  }
\begin{small}
\begin{align}
    \slang_{ij} := r_i \plin_j - \plin_i r_j \;=\; q_i p_j - p_i q_j
    &&,&& 
      \tup{\slang} :=\, \tup{r}\times\tup{\plin} \;=\; \tup{q}\times\tup{p}  
    &&,&&
    \slang^2  =\, r^2 \plin^2 - (\tup{r}\cdot\tup{\plin})^2 \;=\; q^2 p^2 - (\tup{q}\cdot\tup{p})^2
\end{align}
\end{small}
In the case that only central forces do work, the above are integrals of motion of the original \eq{\mscr{K}(\tup{r},\tup{\plin})} and the new \eq{\mscr{H}(\bartup{q},\bartup{p})}. 
\end{notesq}

\paragraph{Inverse Transformation.}
Eq.\eqref{PT_0} is a (local) submersion, \eq{\Psiup:\mbb{R}^8 \to \mbb{R}^6}, with no unique inverse. However, we show that the new Hamiltonian \eq{\mscr{H}} for the new, redundant, phase space coordinates permits two additional ``integrals of motion'' 
— an abuse of terminology, perhaps\footnote{``Integral of motion'' is perhaps an abuse of terminology in this context; the functions in Eq.\eqref{qlam_0} are not  not true integrals of motion in the sense that they do not represent a conserved quantity for a particle moving in \eq{\Evec^3} (which is the actual system we are modeling). They are, rather,  ``kinematic constants'' which arise from artificially increasing the dimension of phase space.} — 
which allow for the inverse transformation. These are: 
\begin{small}
\begin{align} \label{qlam_0}
     \begin{array}{cc}
     \fnsize{integrals}  \\
     \fnsize{of motion} 
\end{array} 
\;\left\{ \;
\begin{array}{ll}
    k(\bartup{q}) := \mag{\tup{q}} = q 
     \\[5pt]
    \lambda(\bartup{q},\bartup{p}) :=  \tup{q}\cdot\tup{p} 
\end{array} \right.
&&
\begin{array}{ll}
    \diff{}{t} k = \pbrak{k}{\mscr{H}} + \pderiv{k}{\bartup{p}} \cdot \bartup{\alpha}  = 0 
   \\[5pt]
    \diff{}{t} \lambda  = \pbrak{\lambda}{\mscr{H}} + \pderiv{\lambda}{\bartup{p}} \cdot \bartup{\alpha}  = 0
\end{array} 
&&  \xRightarrow{\text{choose}}
&&
\begin{array}{ll}
     k = k_\zr  = 1
     \\[5pt]
       \lambda = \lambda_\zr  = 0
\end{array}
\end{align}
\end{small}
That is, the above functions \eq{k} and \eq{\lambda} are
constant along a phase space trajectory.
When specifying the initial conditions, we are free to choose/set their initial values to be any \eq{0<k_\zr\in\mbb{R}_\ii{+}}
and \eq{\lambda_\zr\in\mbb{R}} that we like and they will then remain constant at these values for all time.\footnote{Any \eq{\mag{\tup{q}}=k<0} is incompatible with a Euclidean space and also with
Eq.\eqref{PT_0}.}
We will therefore identify these \textit{functions} with their constant \textit{values}. 
These constants can be used to obtain the below inverse transformation of Eq.\eqref{PT_0}:
\begin{small}
\begin{align} \label{qp_rv_0}
\inv{\Psiup}: (\tup{r},\tup{\plin};k,\lambda) \mapsto (\bartup{q},\bartup{p})
\quad \left\{ \;\; 
\begin{array}{llll}
     \tup{q} \,=\,  k\htup{r} 
  \\[5pt]
  u \,=\, 1/r 
  %
 \\[5pt]
    \tup{p} \,=\, \tfrac{r}{k}( \imat_3 -  \htup{r}\otimes\htup{r})\cdot \tup{\plin} + k\lambda \htup{r}
 \\[5pt]
     p_\ss{u} 
     \,=\, -r^2\htup{r}\cdot\tup{\plin} \,=\, -r^2 \dot{r}
\end{array} \right.
&& \xRightarrow[\lambda=0]{k=1} &&
\begin{array}{llll}
     \tup{q} \,=\,  \htup{r} 
  \\[5pt]
 \\[5pt]
    \tup{p}  \,=\,  r\tup{\plin} -  (\htup{r}\cdot\tup{\plin})\tup{r}
    \,=\, \tup{\slang}\times\htup{r} \,=\, r^2\dot{\htup{r}}
 \\[5pt]
\end{array}
\end{align}
\end{small}
The above is not a true inverse of Eq.\eqref{PT_0}. It is only unique up to some chosen values of \eq{k} and \eq{\lambda} (constants along a phase space trajectory).  
The relations on the right follow from our 
\textit{default choices of \eq{k=k_\zr = 1} and \eq{\lambda = \lambda_\zr =0}}.\footnote{We always choose \eq{k=k_\zr = 1} and \eq{\lambda = \lambda_\zr =0}. The former ensures that \eq{\tup{q}=\htup{q}=\htup{r}}, and \eq{\lambda=0} is chosen so that \eq{\tup{p}} will have a ``nice'' interpretation.}
Assume these values are chosen for the remainder of section \ref{sec:prj_sum2}.


\subsection{Transformation of the Evolution Parameter}


We use Hamiltonian dynamics on (the coordinate view of) \textit{extended} phase space to transform the evolution parameter from the time, \eq{t}, to two new evolution parameters, \eq{s} and \eq{\tau}, defined through the following differential relations:
\begin{small}
\begin{align} \label{dtds_0}
      \mrm{d} t \,=\, r^2 \mrm{d} s \,=\, \tfrac{1}{u^2} \mrm{d} s
&&,&&
     \mrm{d} t \,=\, \tfrac{r^2}{\slang} \mrm{d} \tau 
     \,=\, \tfrac{1}{\slang u^2} \mrm{d} \tau 
  &&,&&
    \mrm{d} s \,=\, \slang \mrm{d} \tau 
&&,&&
    \slang^2 =q^2 p^2 -(\tup{q}\cdot\tup{p})^2 
\end{align}
\end{small}
When combined with the projective coordinate transformation, either of the above leads to fully linear dynamics in the unperturbed case for a certain family of cental force potentials,  \eq{V^{0}}.  
The  parameter \eq{s} is the same as used by Burdet \cite{burdet1969mouvement,Burdet+1969+71+84}, while \eq{\tau} is  is equivalent to the one used by Vitins and Ferrándiz \cite{vitins1978keplerian,ferrandiz1987general}. 
\textit{For orbital motion, \eq{\tau} is the true anomaly up to an additive constant}. 
The extended Hamiltonians with \eq{s} as the evolution parameter, \eq{\wtscr{H}}, and with \eq{\tau} as the evolution parameter, \eq{\wtscr{M}}, are then 
\begin{small}
\begin{align}
    \wtscr{H}(\bartup{q},\bartup{p},t,p_t) :=\, \diff{t}{s}(\mscr{H} + p_t ) \;=\; 
    \tfrac{1}{2}( \slang^2 + u^2 p_\ss{u}^2 )  \,+\, \wt{V}^{0}(u) \,+\, \wt{V}^{1}(\bartup{q},t) \,+\, \tfrac{1}{u^2} p_t
     &&,&&
     \wtscr{M} = \tfrac{1}{\slang}\wtscr{H}
\end{align}
\end{small}
where \eq{\diff{t}{s} =u^\ss{-2}}, where \eq{\wt{V}^{0}:= u^\ss{-2}V^{0}} (likewise for \eq{\wt{V}^{1}}), and where \eq{p_t} is the momenta conjugate to \eq{t} 
(and \eq{p_t=-\mscr{H}})\footnote{It can be shown that \eq{p_t} always has a value of \eq{p_t=-\mscr{H}} such that \eq{\wtscr{H}=0} (but \eq{\pderiv{\wtscr{H}}{q^i}\neq0\neq \pderiv{\wtscr{H}}{p_i}}) \cite{lanczos2012variational}.}. 
Using \eq{\wtscr{H}}, 
Hamilton's equations then give the following dynamics (where \eq{\pdt{\square}:=\diff{\square}{s}= u^\ss{-2}\diff{\square}{t}} and \eq{\pdt{t}=u^\ss{-2}}):\footnote{Note the equations of motion in Eq.\eqref{dqp_s_0_full} are equivalent to \eq{ \pdt{q}_\a= \pdt{t}\dot{q}_\a} and \eq{ \pdt{p}_\a= \pdt{t}\dot{p}_\a}. }
\begin{small}
\begin{align} \label{dqp_s_0_full}
  \begin{array}{lllll}
     \pdt{q}_i = \pderiv{\wtscr{H}}{p_i} = - \slang_{ij}q_j &,
\\[6pt]
    \dot{u}  = \pderiv{\wtscr{H}}{p_u} = u^2 p_\ss{u} &,
\\[6pt]
     \pdt{t}  \,=\, \pderiv{\wtscr{H}}{p_t} \,=\,  \tfrac{1}{u^2}
     &,
 \end{array}
\qquad
 \begin{array}{llll}
    \pdt{p}_i = -\pderiv{\wtscr{H}}{q_i} + \pdt{t}\alpha_i 
    &\!\!\! =
    - \slang_{ij} p_j
     \,+\, \pdt{t} f_i 
\\[6pt]
     \pdt{p}_{u} = -\pderiv{\wtscr{H}}{u} + \pdt{t}\alpha_u    
     &\!\!\! =
    -\tfrac{1}{u}\big(\slang^2 +  2 u^2 p_\ss{u}^2 \big)   -  \pdt{t}\pderiv{V^{0}}{u}  +  \pdt{t}f_\ss{u} 
\\[6pt]
     \pdt{p}_t \,=\, \pderiv{\wtscr{H}}{t} + \pdt{t}\alpha_t    &\!\!\! = 
     -\pderiv{\wt{V}^{1}}{t} - \tup{\alpha}\cdot\pdt{\tup{q}} - \alpha_\ss{u} \pdt{u}
\end{array}  
\end{align}
\end{small}
where \eq{(\tup{\alpha},\alpha_\ss{u})} and \eq{(\tup{f},f_\ss{u})} are the perturbation terms defined in Eq.\eqref{Fdef_sum}, and
where  \eq{p_t=-\mscr{H}} has been used to eliminate \eq{p_t} from the \eq{\pdt{p}_\ss{u}}
equation
and re-express it as \eq{\pdt{p}_\ss{u}=\pdt{t}\dot{p}_\ss{u}=\dot{p}_\ss{u}/u^2}.\footnote{Differentiating \eq{\wtscr{H}} directly leads to \eq{\pdt{p}_\ss{u} = -u p_\ss{u}^2   -  \pderiv{\wt{V}^{0}}{u} - \pderiv{\wt{V}^{1}}{u} +  \tfrac{2}{u^3} p_t  + \tfrac{1}{u^2}\alpha_\ss{u}}, where \eq{\wt{V}=V/u^2} and thus  \eq{\pderiv{\wt{V}}{u} = -\tfrac{2}{u^3} V + \tfrac{1}{u^2}\pderiv{V}{u}} such that  
\begin{align} \nonumber
    \pdt{p}_\ss{u} \,=\,  -u p_\ss{u}^2    +  \tfrac{1}{u^2}(-\pderiv{V^{0}}{u} -\pderiv{V^{1}}{u} + \alpha_\ss{u} ) +  \tfrac{2}{u^3}(V^{0} + V^{1} + p_t )
\end{align}
Substituting \eq{p_t= -\mscr{H}= -\tfrac{1}{2} u^2( \slang^2  +  u^2 p_\ss{u}^2 ) - V^{0} - V^{1} } into the above — and using \eq{q=1} and \eq{\tup{q}\cdot\tup{p}=0} — finally leads to the equation for \eq{\pdt{p}_\ss{u}} seen in Eq.\eqref{dqp_s_0}. }
The above dynamics, \textit{after} differentiating \eq{\wtscr{H}}, may further be simplified using the conserved quantities \eq{q=k=1} and \eq{\tup{q}\cdot\tup{p}=\lambda =0}. In fact, \eq{\lambda=0} may be used to simplify \eq{\wtscr{H}} itself, giving the effective Hamiltonian and ODEs:
\begin{small}
\begin{align}\label{dqp_s_0} 
\wtscr{H} \,=\,
     \tfrac{1}{2}\big( q^2 p^2  +  u^2 p_\ss{u}^2 \big)  \,+\, \wt{V}^{0} \,+\, \wt{V}^{1} \,+\, \tfrac{1}{u^2}p_t
&&,&&
 \begin{array}{ll}
      \pdt{q}_i 
      \,=\,  p_i  
      &,\qquad
     \pdt{p}_i 
     \,=\,  -p^2 q_i  \,+\, \pdt{t} f_i
\\[4pt]
      \pdt{u} 
      \,=\,  u^2 p_\ss{u}
      &,\qquad 
      \pdt{p}_\ss{u} 
      \,=\, -\tfrac{1}{u} (p^2 +2 u^2 p_\ss{u}^2 )   - \pdt{t} \pd_u V^{0}  \,+\, \pdt{t} f_\ss{u} 
\\[4pt]
     \pdt{t}  
     \,=\,  1/u^2
     &,\qquad
     \pdt{p}_t 
     \,=\, - \pd_t \wt{V}^{1} -\, \tup{\alpha}\cdot\pdt{\tup{q}} \,-\, \alpha_\ss{u} \pdt{u}
\end{array}
\end{align}
\end{small}
where \eq{q=k=1} has also been used to simplify the dynamics (\textit{after} differentiating \eq{\wtscr{H}}). 

\begin{notesq}
    The \eq{\tau}-parameterized dynamics
    may be obtained using extended Hamiltonian \eq{\wtscr{M} = \tfrac{1}{\slang}\wtscr{H}} with dynamics equivalent to using \eq{\diff{\square}{\tau} = \tfrac{1}{\slang}\diff{\square}{s} }.
\end{notesq}

\subsection{Linear Equations and Closed-Form Solutions}

\paragraph{Angular Motion:~$(\tup{q},\tup{p})$ Solutions.}
When  either \eq{s} or \eq{\tau} is used as the evolution parameter,  the canonical equations of motion motion for \eq{\tup{q}} and \eq{\tup{p}} in Eq.\eqref{dqp_s_0}
are equivalent to second-order equations for a perturbed  oscillator:  
\begin{small}
\begin{align} \label{dqp_ug}
\begin{array}{lllll}
      \ddiff{}{s} \tup{q}  \,+\, \slang^2 \tup{q}  
    \;=\; \tfrac{1}{u^2}\tup{f}
    &\;=\;  \tfrac{1}{u^3} (\imat_3 -  \tup{q}\otimes\tup{q} ) \cdot \tup{F} 
\\[8pt]
      \ddiff{}{\tau}\tup{q}  \,+\, \tup{q}
      \;=\; \tfrac{u^{2n}}{\slang^2}( \imat_3 - \htup{p}\otimes\htup{p} )\cdot\tup{f}
      & \;=\; \tfrac{u^{3n}}{\slang^2} \htup{\slang} \cdot \tup{F}
\end{array}
&&
\fnsize{with:}\;\;
\begin{array}{lllll}
       \diff{}{s}\slang = \tfrac{1}{u^2} \htup{p} \cdot\tup{f} = \tfrac{1}{u^3} \htup{p}\cdot\tup{F}
\\[8pt]
      \diff{}{\tau} \slang = \tfrac{1}{\slang u^2} \htup{p} \cdot\tup{f} = \tfrac{1}{\slang u^3} \htup{p}\cdot\tup{F}
\end{array}
\end{align}
\end{small}
where  \eq{\htup{\slang} =\tup{q}\times \htup{p}} and, using \eq{q=1} and \eq{\tup{q}\cdot\tup{p}=0}, then \eq{\slang=p=\mag{\pdt{\tup{q}}}}.
If only central forces are present (i.e., if \eq{\sfb{a}=a_r \hsfb{r}} such that \eq{\tup{F}=a_r \htup{r} = a_r \htup{q} }), then the right-hand-sides of the above vanish;  \eq{\slang} is an integral of motion  such that \eq{\slang=p} is constant and \eq{\tau =\slang s }.  The solutions for \eq{\tup{q}} and \eq{\tup{p}} are then given in terms of initial conditions and \eq{\tau = \slang s} by
\begin{small}
\begin{align} \label{dqp_ugug}
 \begin{array}{cc}
     \fnsize{for any central }  \\
      \fnsize{force motion} 
\end{array}  
\Big\{ \quad
     \begin{array}{llll}
         \tup{q}_\tau  \,=\,   \tup{q}_\zr\cos\tau  \,+\,  \tfrac{1}{\slang}\tup{p}_\zr\sin\tau
&,\quad 
      \tup{p}_\tau  \,=\,  -\slang\tup{q}_\zr\sin\tau
            \,+\,  \tup{p}_\zr\cos\tau   
\end{array}   
&&
    ( \tau = \slang s \;,\; \slang = \slang_\zr )
\end{align}
\end{small}
The above dynamics for \eq{(\tup{q},\tup{p})} are valid for any arbitrary form of the central force potential \eq{V^{0}} (and/or any arbitrary \eq{\tup{F}=a_r \htup{r} = a_r \htup{q} }). 

\paragraph{Radial Motion:~$(u,p_\ss{u})$ Solutions.}
Unlike the dynamics for \eq{\tup{q}=\htup{r}} and \eq{\tup{p}}, those for \eq{u=1/r} and \eq{p_\ss{u}} are linear only for certain forms of the central force potential \eq{V^{0}}. In particular, consider
the Manev potential from Eq.\eqref{Vmanev_sum}:
\begin{small}
\begin{align} \label{V_manev}
     V^{0} \,=\, -{\txkbar_1}/{r} \,-\, \tfrac{1}{2} {\txkbar_2}/{r^2} \;=\; -\txkbar_1 u \,-\, \tfrac{1}{2}\txkbar_2 u^2
\end{align}
\end{small}
where \eq{\txkbar_1,\txkbar_2\in\mbb{R}} are any scalar constants (the negative signs are for convenience). For this \eq{V^{0}}, 
the previous equations for \eq{(u,p_\ss{u})}, with either \eq{s} or \eq{\tau} as  the evolution parameter, are then equivalent to the second-order ODEs:
\begin{small}
\begin{align} \label{du_ug}
\begin{array}{lllll}
        &\ddiff{}{s}u + \omega^2 u  - \txkbar_1   \;=\, f_\ss{u} 
         &= -\tfrac{1}{u^2}\tup{F}\cdot\tup{q}
      \\[8pt]
          &\ddiff{}{\tau}u + \varpi^2 u  -  \tfrac{\txkbar_1}{\slang^2} 
         \;= \tfrac{1}{\slang^2} ( f_\ss{u} - \tfrac{p_\ss{u}}{\slang}\tup{f}\cdot\htup{p} )
        &=
         -\tfrac{1}{\slang^2 u^2}\big( \tup{q} + \tfrac{u p_\ss{u}}{\slang}\htup{p} \big) \cdot\tup{F}
    \end{array}
    &&
\fnsize{with:}\;\;
\begin{array}{lllll}
      \omega^2:= \slang^2-\txkbar_2
\\[4pt]
      \varpi^2:= {\omega^2}/{\slang^2} 
\end{array}
\end{align}
\end{small}
(the definition \eq{\omega^2:= \slang^2-\txkbar_2} assumes \eq{\slang^2-\txkbar_2 >0}). 
The natural frequencies, \eq{\omega} and \eq{\varpi}, are integrals of motion for any/all central force dynamics.  
In that case, the right-hand-side of the above ODEs vanish, and \eq{\slang} is constant such that \eq{\tau=\slang s} and \eq{\omega s=\varpi\tau}. The solution is then as follows:
\begin{small}
\begin{align} \label{du_ugug_gen}
     \begin{array}{lllllll}
      u_{\tau} \,=\, (u_\zr-\tfrac{\txkbar_1}{\omega^2})\cos\varpi\tau
    \,+\, \tfrac{1}{\omega}w_\zr\sin{\varpi\tau} \,+\, \tfrac{\txkbar_1}{\omega^2}
&,\quad 
    w_{\tau} \,=\, -\slang(u_\zr-\tfrac{\txkbar_1}{\omega^2})\sin\varpi\tau \,+\, w_\zr\cos\varpi\tau
    &,\quad p_\ss{u} = \tfrac{w}{u^2}
\end{array}  
&&
(\omega s=\varpi\tau)
\end{align}
\end{small}
The \eq{p_\ss{u}} solution is easily recovered from the above where we have, 
for convenance\footnote{Some expressions are simply more easily expressed in terms of \eq{w} rather than \eq{p_\ss{u}}. The set \eq{(\tup{q},u,\tup{p},w)} is \textit{not} canonical. }, 
defined a new momenta-level coordinate, \eq{w}, as: 
\begin{small}
\begin{align} \label{wdef}
    w  \,:=\,  u^2 p_\ss{u} \,=\, \diff{u}{s} \,=\, \slang \diff{u}{\tau} \,=\,  -\diff{r}{t}
    \qquad \leftrightarrow \qquad
    p_\ss{u} \,=\, \tfrac{1}{u^2}w \,=\, r^2 w   \,=\,  -\diff{r}{s} 
\end{align}
\end{small}
\begin{small}
\begin{notesq}
     The \rmsb{Kepler problem} is a special case of Eq.\eqref{V_manev}, along with \eq{\tup{F}=0}. Solutions are obtained from the above using: 
    \begin{small}
    \begin{align} \label{Kep_simps_0}
         V^0 = -\txkbar/r = -\txkbar u
        \qquad \Rightarrow \qquad
        \txkbar_2 = 0
        \quad,\quad \omega = \slang \quad,\quad \varpi = 1
    \end{align}
    \end{small}
\end{notesq}
\end{small}


\paragraph{Recovering Cartesian Position/Velocity Coordinate Solutions.}
To recover the cartesian position and velocity coordinate solutions from the above projective coordinate solutions, one uses the transformation \eq{\Psiup:(\bartup{q},\bartup{p})\mapsto(\tup{r},\tup{\plin})} given in Eq.\eqref{PT_0}. Although not necessarily, are free to simplify the transformation using the integrals of motion \eq{k=q=1} and \eq{\lambda=\tup{q}\cdot\tup{p}=0} such that Eq.\eqref{PT_0} simplifies to:\footnote{\textit{Caution!} the simplified transformation in Eq.\eqref{rv_qu_simp} should \textit{not} be used when forming the projective coordinate Hamiltonian \eq{\mscr{H}(\bartup{q},\bartup{p})}.} 
\begin{small}
\begin{align} \label{rv_qu_simp}
    \tup{r}(\bartup{q}) \;=\; \tfrac{1}{u}\tup{q}
    \qquad,\qquad
    \tup{\plin}(\bartup{q},\bartup{p}) \;=\; u\tup{p} \,-\, w\tup{q}
     \qquad\qquad
    (w:=u^2 p_\ss{u})
\end{align}
\end{small}
For instance, using the above, the Kepler solutions for the cartesian position and velocity coordinates, \eq{(\tup{r},\tup{\plin})}, are obtained from the solutions for \eq{(\bartup{q},\bartup{p})=(\tup{q},u,\tup{p},p_\ss{u})} in Eq.\ref{dqp_ugug} and Eq.\ref{du_ugug_gen} — and using \eq{\txkbar_1 = \txkbar},  \eq{\omega=\slang} and \eq{\varpi=1} from Eq.\eqref{Kep_simps_0} — leading to:
\begin{small}
\begin{align} \label{rv_sol_0}
      \tup{r}_\tau \;=\; \dfrac{\tup{q}_\zr\cos \tau  \,+\,  \htup{p}_\zr\sin \tau}{ (u_\zr-\tfrac{\txkbar}{\slang^2})\cos \tau \,+\, w_\zr\sin{ \tau} \,+\, \tfrac{\txkbar}{\slang^2}   }  
\qquad,\qquad 
    \tup{\plin}_\tau 
    \;=\;  \tup{\plin}_\zr + \tfrac{\txkbar}{\slang^2} (\tup{p}_\tau - \tup{p}_\zr )
&&
\begin{array}{cc}
     \fnsize{Kepler}  \\
     \fnsize{solutions} 
\end{array} 
\end{align}
\end{small}
where  \eq{\tup{\plin}_\zr=  u_\zr\tup{p}_\zr-w_\zr\tup{q}_\zr}, and where \eq{\tup{p}_\tau =  -\slang\tup{q}_\zr\sin\tau +  \tup{p}_\zr\cos\tau},  and where \eq{\slang = p_\tau = p_\zr=\slang_\zr}.

\section{NOTATION \& CONVENTIONS}

\noindent\textbf{Abbreviations \& Acronyms}
\begin{footnotesize}
 \begin{longtable}[htbp]{@{}p{0.2\textwidth} p{0.77\textwidth}@{}}
       $n$-dim  &  $n$-dimensional. 
\\[1ex]
       ODE   & ordinary differential equation.
\\[1ex]
      LC  &  Levi-Civita (one person).
\\[1ex]
      KS  &  Kustaanheimo-Stiefel (two people).
\\[1ex]
      BF  &  Burdet-Ferrándiz (two people).
\\[1ex]
    DEF & Deptrit-Elipe-Ferrer (three people).
\end{longtable}
\end{footnotesize}


\noindent\textbf{Mathematical Notation} \\
The below notation is generally used. However, some specific notation is defined/used in sections \ref{sec:prj_geomech} and \ref{sec:prj_regular}, and Appx.~\ref{sec:prj_sum2}, of this work which may not always align with the following. 

\begin{footnotesize}
\begin{longtable}[htbp]{@{}p{0.2\textwidth} p{0.77\textwidth}@{}}
    $\mbb{R}^n$   &  $n$-dim real coordinate space (regarded as a vector space).
\\[1ex]
       $\imat_n$ &  $n\times n$ identity matrix.
\\[1ex]
       $\ibase_i, \; \ibase^i $ &  standard basis for \eq{\mbb{R}^n} and \eq{\mbb{R}^{n*}\cong \mbb{R}^n} (i.e., columns/rows of \eq{\imat_n}).
\\[1ex]
     $ \kd^i_j = \kd_{ij} = \kd^{ij} $ &  Kronecker delta \textit{symbol} (i.e., elements of \eq{\imat_n}).
\\[1ex]
      $ \lc_{i_1\dots i_n} = \lc^{i_1\dots i_n}$ &  Levi-Civita permutation \textit{symbol}.
\\[1ex]
   $ \jmat_{2n} = {\scriptscriptstyle\begin{pmatrix}
     0 & \imat_n \\ -\imat_n & 0
 \end{pmatrix}} 
      $ &  standard symplectic matrix. $\inv{\jmat}=\trn{\jmat}=-\jmat\in \Spmat{2n}$. 
 \\[3ex] 
      $ \Glmat{n}, \; \glmat{n} $ &  Lie group of  \eq{n \times n} non-degenerate matrices (general linear group) and its Lie algebra.
\\[1ex]
      $ \Somat{n}, \; \somat{n} $ &  Lie group of \eq{n \times n} special orthogonal matrices and its Lie algebra.
\\[1ex]
   $ \Spmat{2n}, \; \spmat{2n} $ &  Lie group of \eq{2n \times 2n} symplectic matrices and its Lie algebra of Hamiltonian matrices.
\\[3ex]
       $\Eaf,\; \vsp{E},\; \vsp{E}^* $ & affine space \eq{\Eaf} with vector space $\vsp{E}$ and dual space $\vsp{E}^*$. (often, but not always, Euclidean.)
\\[1ex]
     $\vsp{E}_\nozer$ &  vector space, $\vsp{E}$, excluding the origin/zero vector.
\\[1ex]
       $\botimes^r_s \vsp{E}$
        &  (\textit{r,s})-tensors on $\vsp{E}$, regarded as $\botimes^r_s \vsp{E} \cong \vsp{E}^{\ss{\otimes} r} \otimes (\vsp{E}^*)^{\ss{\otimes} s}$.
\\[1ex]
       $ \bwedge{k}\vsp{E}, \; \bwedge{k}\vsp{E}^* $ 
        & \textit{k}-vectors and \eq{k}-forms on $\vsp{E}$ ($k^\mrm{th}$ exterior powers).
\\[1ex]
      $ \Glten(\vsp{E}), \; \glten(\vsp{E}) $ &  non-degenerate (1,1)-tensors (general linear group) on $\vsp{E}$ and its Lie algebra.
\\[1ex]
      $ \Soten(\vsp{E},\bemet), \; \soten(\vsp{E},\bemet) $ &  special orthogonal (1,1)-tensors on a metric space and its Lie algebra.
\\[1ex]
   $ \Spten(\vsp{P},\nbs{\omg}), \; \spten(\vsp{P},\nbs{\omg}) $ &   symplectic (1,1)-tensors on a symplectic space and its Lie algebra of Hamiltonian tensors.
\\[3ex]
    $ \man{X}, \; \sman{X}\subset\man{X}$ &  a (smooth) manifold $\man{X}$, an (open) subset  $\sman{X}$.
\\[1ex]
   $\Dfism(\man{X})$ &  Lie group of (auto)diffeomorphisms on \eq{\man{X}}, with Lie algebra \eq{\vect(\man{X})}. 
\\[1ex]
   $\vect(\man{X}),\; \vect^k  (\man{X}) $ & Lie algebra of vector fields, and \textit{k}-vector fields  on $\man{X}$. $\vect(\man{X}) \equiv \vect^1(\man{X}) \equiv \tens^1_0(\man{X})$ 
\\[1ex]
  $ \forms(\man{X}),\; \forms^k  (\man{X}) $ &  (differential) 1-form and \textit{k}-form fields  on $\man{X}$.  $\forms(\man{X})\equiv \forms^1(\man{X}) \equiv   \tens^0_1(\man{X})$. 
\\[1ex]
  $ \formsex^k(\man{X}) \subset \formscl^k(\man{X}) $ &  exact $k$-forms and closed $k$-forms.
\\[1ex]
   $\tens^r_s(\man{X}) $ & \textit{(r,s)}-tensor fields on \eq{\man{X}}, regarded as $\tens^r_s(\man{X}) \cong \vect(\man{X})^{\ss{\otimes} r} \otimes \forms(\man{X})^{\ss{\otimes} s}$. 
\\[1ex]
  $\fun(\man{X}),\;  \fun^k(\man{X}) $ &  $\mbb{R}$-valued functions, and $k$-tuples of such functions, on $\man{X}$.  $\fun(\man{X}) \equiv \tens^0_0(\man{X})\equiv C^{\infty}(\man{X})$. 
\\[1ex]
   $ (\chart{X}{\xi},\tp{\xi}) $ & local coordinate chart.  \eq{\tp{\xi}=(\xi^1,\dots,\xi^m): (\chart{X}{\xi}\subseteq \man{X})\to (\rchart{R}{\xi}^m \subseteq \mbb{R}^m)} a diffeomorphism.
\\[1ex]
   $\bpart{i} \!=\pdii{\xi^i}, \; \bdel^i \!=\dif \xi^i $ &  local coordinate basis vectors and dual 1-forms associated to \eq{(\chart{X}{\xi},\tp{\xi})}. I.e., frame fields.
\\[3ex]
    $(\man{Q},\sfg)$ & (pseudo)Riemannian manifold with metric tensor $\sfg\in \tens^0_2(\man{Q})$. 
\\[1ex]
    $\Isom(\man{Q},\sfg)$  & Lie group of (auto)isometries on $(\man{Q},\sfg)$, with Lie algebra $\veckl(\man{Q},\sfg)$.
\\[1ex]
    $\veckl(\man{Q},\sfg)$  &   Lie algebra of Killing vector fields on $(\man{Q},\sfg)$. 
\\[1ex]
    $\nab \equiv \nab^\ss{\sfg}$ & Levi-Civita Connection for metric \eq{\sfg}. 
\\[3ex]
   $ (\man{P},\nbs{\omg})$ &  symplectic manifold with symplectic form \eq{\nbs{\omg}\in\forms^2(\man{P})} and Poisson bivector \eq{\inv{\nbs{\omg}}\in\vect^2(\man{P})}.
\\[1ex]
  $ \Spism(\man{P},\nbs{\omg})$ &  Lie group of (auto)symplectomorphisms on \eq{(\man{P},\nbs{\omg})}, with Lie algebra \eq{\vecsp(\man{P}) }. 
\\[1ex]
    $\vechm(\man{P}) \subset \vecsp(\man{P})$ & Hamiltonian  and symplectic vector fields on \eq{(\man{P},\nbs{\omg})}. Lie subalgebras of \eq{\vect(\man{P})}. 
\\[1ex]
     $(\man{P},\nbs{\omg},\mscr{H})
   $ &  Hamiltonian system for some $\mscr{H}\in\fun(\man{P})$.
\\[1ex]
     $(\chart{P}{z},\tp{z}),\;  \hbpart{\ssc{I}} \,, \,\hbdel^\ssc{I}
 $ &  symplectic/canonical/Darboux coordinates, \eq{\tp{z}}, and associated basis vectors and 1-forms. 
\\[3ex]
   $ \tsp[\pt{x}]\man{Q},\;  \cotsp[\pt{x}]\man{Q}$ &  tangent and cotangent  space at \eq{\pt{x}\in\man{Q}}.
\\[1ex]
   $\tsp\man{Q}\;, \; \cotsp\man{Q}$ &  tangent and cotangent bundle of \eq{\man{Q}}, with projections $\tpr:\tsp\man{Q}\to\man{Q}$  and  $\copr:\cotsp\man{Q}\to\man{Q}$.
\\[1ex]   
    $\tp{q}$
    &  base (configuration) coordinates, $\tp{q}=(q^1,\dots, q^n):(\chart{Q}{q}\subset\man{Q})\to(\rchart{R}{q}^n \subset \mbb{R}^n) $.
\\[1ex]
    $\tp{\xi} =(\tp{q},\tp{v})=\tlift\tp{q}$ &  tangent-lifted coordinates, $\tp{\xi}=(q^1,\dots, q^n,v^1,\dots,v^n)=\tlift\tp{q}:\Tan\chart{Q}{q}\to\Tan\rchart{R}{q}^n\cong \rchart{R}{q}^n \times \mbb{R}^n $.
\\[1ex]
    $\tp{z} = (\tp{q},\tp{\pf})=\colift\tp{q}$ &  cotangent-lifted coordinates, $\tp{z}=(q^1,\dots, q^n,\pf_1,\dots,\pf_n)=\colift\tp{q}:\Tan^*\chart{Q}{q} \to\Tan^*\rchart{R}{q}^n\cong \rchart{R}{q}^n \times \mbb{R}^{n*} $.
\\[1ex]
    $\vfun^{\bs{\mu}}\in\fun(\tsp\man{Q})$ &  ``velocity function'' of $\bs{\mu}\in\forms(\man{Q})$. Locally, $\vfun^{\bs{\mu}}= v^i \mu_i$.
\\[1ex]
    $\pfun^{\sfb{u}}\in\fun(\cotsp\man{Q})$ &  ``momentum function'' of $\sfb{u}\in\vect(\man{Q})$. Locally, $\pfun^{\sfb{u}} = \pf_i u^i$.
\\[3ex]
   $ \dif,\;  \exd   $ &  differential (gradient).  $\dif (\slot) \equiv \pderiv{}{\xi^i}(\slot) \otimes \dif \xi^i$. exterior derivative. $\exd(\slot) \equiv \dif \xi^i \wedge \pderiv{}{\xi^i}(\slot)$.
\\[1ex]
    $\lderiv{\sfb{u}} $ &  Lie derivative along vector field \eq{\sfb{u}}.
\\[1ex]
    $\nab,\; \del{\sfb{u}} $ &  affine connection, covariant derivative along $\sfb{u}$ (usually the LC connection).
\\[1ex]
   $\tlift\varphi,\; \colift\varphi$ &  tangent and cotangent lift of smooth map, $\varphi$, 
\\[3ex]
   $\inner{\slot}{\slot}$ &  inner product (on an inner product space or Riemannian manifold).
\\[1ex]
      $\lbrak{\slot}{\slot}$ &  Lie bracket (on a Lie algebra).
\\[1ex]
      $\pbrak{\slot}{\slot}$ & Poisson bracket (on a symplectic or Poisson manifold).
\\[1ex]
      $\lagbrak{\slot}{\slot}$ & Lagrange bracket (on a symplectic manifold).
\\[3ex]
     $a:= b\; \fnsize{ or }\; b=:a$ & $a$ is defined as $b$. 
\\[1ex]
    $a \equiv b $ & $a$ ``means the the same thing'' as $b$ (often used informally). 
\\[1ex]
     $ a \Rightarrow b \; ,\; a \Leftrightarrow  b $ &  $a$ implies $b$. $a$ implies $b$ and vice versa. 
\\[1ex]
    $ a \;\rnlarrow \; b $ &  $a$ implies $b$ but \textit{not} vice versa.
\\[1ex]
    $ a \simeq b $ &  $a$ equals $b$ under some specified condition/assumption.
\end{longtable}
\end{footnotesize}

\paragraph{Some Mathematical Conventions.}
For ease of
reference, given below is a collection of some conventions employed throughout this work (in addition to the notation listed previously):
\begin{footnotesize}
\begin{itemize}
    \item  Tensors and tensor fields (of order \eq{> 0}) are usually in bold. Generally, but \textit{not} exclusively, we use bold Greek letters for \eq{k}-forms and bold Latin letters for all other tensor fields. 
    \item \textbf{Index Summation.} We use the (Einstein) index summation notation throughout.  For example, \eq{u^i\alpha_i:=\sum_{i=1}^n u^i \alpha_i = u^1\alpha_1  + \dots + u^n\alpha_n} (where \eq{n} is some specified dimension). This \textit{usually} adheres to the convention that summation occurs over matching upper-lower indices (as just indicated by \eq{u^i \alpha_i}), but not always. That is, \eq{u^i \alpha_i} or \eq{u_i \alpha_i} or \eq{u^i \alpha^i} all indicate summation. 
    \item \textbf{Tensor Fields vs.~Sections.}  We treat a vector field, \eq{\sfb{u}\in\vect(\man{X})},  as a smooth map \eq{\sfb{u}:\man{X}\ni \pt{x}\mapsto \sfb{u}(\pt{x})=:\sfb{u}_{\pt{x}}\in \tsp[\pt{x}]\man{X} } (though the domain of \eq{\sfb{u}} may only be some \eq{\sman{X}\subseteq\man{X}}), where \eq{\tsp[\pt{x}]\man{X}} does \textit{not} include the point \eq{\pt{x}}. The same applies for all tensor fields, that is, a \eq{(r,s)}-tensor field is regarded as \eq{\tens^r_s(\man{X})\ni\sfb{H}: \pt{x} \mapsto  \sfb{H}_{\pt{x}} \in \botimes^{r}_s (\tsp[\pt{x}]\man{X}) = \tsp[\pt{x}]\man{X}^{\ss{\otimes} r}\otimes\tsp[\pt{x}]^*\man{X}^{\ss{\otimes} s}} 
     (see footnote\footnote{Note that, more generally, the ordering of the tensor products can be ``scrambled''. We have implied that, \eq{\sfb{H}\in \tens^2_1(\man{X})} means \eq{\sfb{H}_\pt{x}\in\tsp[\pt{x}]\man{X}\otimes \tsp[\pt{x}]\man{X} \otimes \tsp[\pt{x}]^*\man{X}}. But it could also mean \eq{\sfb{H}_\pt{x}\in\tsp[\pt{x}]^*\man{X}\otimes \tsp[\pt{x}]\man{X} \otimes \tsp[\pt{x}]\man{X}} or \eq{\sfb{H}_\pt{x}\in\tsp[\pt{x}]\man{X}\otimes \tsp[\pt{x}]^*\man{X} \otimes \tsp[\pt{x}]\man{X}}. We will default to the first convention and will specify if any \eq{(r,s)}-tensor deviate from the pattern \eq{\tsp[\pt{x}]\man{X}^{\ss{\otimes} r}\otimes\tsp[\pt{x}]^*\man{X}^{\ss{\otimes} s}}. }), 
      a \eq{k}-form is regarded as \eq{\forms^k(\man{X}) \ni \bs{\kappa}:\pt{x}\mapsto \bs{\kappa}_\pt{x} \in \bwedge{k}(\tsp[\pt{x}]^*\man{X})}, and a \eq{k}-vector as  \eq{\vect^k(\man{X}) \ni \sfb{k}:\pt{x}\mapsto \sfb{k}_\pt{x} \in\bwedge{k}(\tsp[\pt{x}]\man{X})}. 
     This convention differs from some sources which instead define a vector field as a \textbf{section}  
      of \eq{\Tan\man{X}}, denoted  \eq{\Gammaup(\Tan\man{X})}, which means something slightly different; \eq{\sfb{u}\in\Gammaup(\Tan\man{X})} would imply that \eq{\sfb{u}:\pt{x}\mapsto (\pt{x},\sfb{u}_{\pt{x}})\in (\{\pt{x}\}\times\tsp[\pt{x}]\man{X}) \subset \Tan\man{X} } such that \eq{\tpr\circ\sfb{u}=\Id_\ii{\man{X}}}. We do not take this view.  If the section  associated with some \eq{\sfb{u}\in\vect(\man{X})}  (or any other tensor field) is needed, we will denote it as \eq{s_{\sfb{u}}\in\Gammaup(\Tan\man{X})}. This section can be thought of as \eq{s_{\sfb{u}}=(\Id_\ii{\man{X}},\sfb{u}):\man{X} \to \Tan\man{X}} (the codomain is a \eq{2m}-dim manifold), which is not the same as  \eq{\sfb{u}:\man{X}\to \tsp\!\pmb{.}\man{X}} (the codomain is a family of \eq{m}-dim vector spaces).   
      All of this extends to tensor fields of any order such that some \eq{\sfb{T}\in\tens^r_s(\man{X})} has an associated section \eq{s_{\sfb{T}}=(\Id_\ii{\man{X}},\sfb{T})\in\Gammaup(\botimes^r_s\tsp\man{X})} given by \eq{s_{\sfb{T}}(\pt{x})=(\pt{x},\sfb{T}_{\!\pt{x}})\in \big(\{\pt{x}\}\times \botimes^r_s \tsp[\pt{x}]\man{X} \big)}. That is, the section associated with a tensor field can be seen as the \textit{graph} of the tensor field.  For vector fields and 1-forms:
    \begin{align} \label{vec_section_def}
    \begin{array}{lllllll}
         \forall\, \sfb{u}\in\vect(\man{X}): 
         \qquad \upGamma(\tsp\man{X}) \ni  s_{\sfb{u}} = (\Id_\ii{\man{X}},\sfb{u}):\man{X}\to\tsp\man{X}
         &,\qquad s_{\sfb{u}}(\pt{x}):=(\pt{x},\sfb{u}_\pt{x})
         &,\qquad \tpr\circ s_{\sfb{u}}=\Id_\ii{\man{X}}
     \\[4pt]
         \forall\, \bs{\lambda}\in\forms(\man{X}): \qquad 
         \upGamma(\tsp^*\man{Q}) \ni s_{\bs{\lambda}}=(\Id_\ii{\man{X}},\bs{\lambda}):\man{X}\to\tsp^*\man{X}
         &,\qquad s_{\bs{\lambda}}(\pt{x}):=(\pt{x},\bs{\lambda}_\pt{x})
         &,\qquad \copr\circ s_{\bs{\lambda}}=\Id_\ii{\man{X}}
    \end{array}
    \end{align}
      \item \textbf{(Differential) \eq{k}-Forms.} Generally, but not exclusively, we use bold Greek letters for \eq{k}-forms. We treat \eq{k}-forms as totally antisymmetric (0,\textit{k})-tensor fields. Some  \eq{\bs{\kappa}\in\forms^k(\man{X})\subset\tens^0_k(\man{X})} is regarded as a map \eq{\bs{\kap}:\pt{x}\mapsto\bs{\kap}_\pt{x}\in\bwedge{k}(\cotsp[\pt{x}]\man{X})}. It may be
       expanded in any local basis   as \eq{\bs{\kappa}= {\kappa}_{i_1\dots i_k}\bep^{i_1}\otimes\dots\otimes\bep^{i_k} = \tfrac{1}{k!}\kappa_{i_1\dots i_k}\bep^{i_1}\wedge\dots\wedge\bep^{i_k} }, where the components \eq{\kappa_{i_1\dots i_k}} are totally antisymmetric. 
     For \eq{\bs{\kappa}  \in\forms^k(\man{X})}, and  \eq{\bs{\eta}\in\forms^l(\man{X})}, and \eq{\sfb{u}\in\vect(\man{X})}, note: 
    \begin{align} \nonumber
           & \bs{\kappa}\wedge \bs{\eta} = (-1)^{kl}\bs{\eta} \wedge\bs{\kappa}
       \qquad\qquad
           \sfb{u}\cdt (\bs{\kappa}\wedge\bs{\eta}) \,=\,   (\sfb{u}\cdt \bs{\kappa})\wedge\bs{\eta} \,+\, (-1)^k \bs{\kappa}\wedge (\sfb{u}\cdt\bs{\eta})
    \qquad\qquad
         \sfb{u}\cdt \bs{\kappa} \,=\, (-1)^{k+1}\bs{\kappa}\cdt\sfb{u} 
    \end{align}
   \item \textbf{Use of the ``Dot Product''.}  We use the ``dot product'', \eq{\,\cdt\,}, to denote a sort of universal notion of ``contraction''. This is somewhat unconventional so let us clarify.   
   Let \eq{\sfb{e}_i\in\vect(\man{X})} be some local basis  with dual basis \eq{\bep^i \in\forms(\man{X})} satisfying \eq{\bep^i(\sfb{e}_j) = \kd^i_j}.  For \eq{\sfb{u}=u^i\sfb{e}_i\in\vect(\man{X})} and \eq{\bs{\alpha}=\alpha_i\bep^i\in\forms(\man{X})}, we define \eq{\bs{\alpha}\cdt\sfb{u} :=  \bs{\alpha}(\sfb{u}) = \alpha_i u^i \in\fun(\man{X})}. Some refer to this as the ``natural pairing'' of vectors and 1-forms. This notation is  extended to tensors of higher order. For example,   we  adopt the following equivalent notations for the ways in which \eq{\sfb{u},\,\bs{\alpha}}, and some \eq{\sfb{H}=H^i_j \sfb{e}_i\otimes\bep^j\in \tens^1_1(\man{X})} can be contracted:  
\begin{align} \label{dot_prod}
\begin{array}{lllll}
     \bs{\alpha}\cdt\sfb{u} \,=\, \bs{\alpha}(\sfb{u})  \,=\, \sfb{u}\cdt \bs{\alpha} \,=\, \sfb{u}(\bs{\alpha})\,=\, \alpha_i u^i
       &\in \fun(\man{X}) 
\\
     \sfb{H}\cdt\sfb{u} \,=\,  \sfb{H}(\sfb{u}) \,=\,  \sfb{H}(\slot,\sfb{u}) \,=\,  \sfb{u}\cdt\trn{\sfb{H}} \,=\, \trn{\sfb{H}}(\sfb{u},\slot)
     \,=\, H^i_ju^j \sfb{e}_i 
      &\in \vect(\man{X})
  \\
     \bs{\alpha}\cdt\sfb{H}  \,=\,  \sfb{H}(\bs{\alpha},\slot) \,=\, \trn{\sfb{H}}\cdt\bs{\alpha} \,=\, \trn{\sfb{H}}(\bs{\alpha}) \,=\, \trn{\sfb{H}}(\slot,\bs{\alpha})
     \,=\, H^i_j\alpha_i \bep^j
     &\in \forms(\man{X}) 
 \\
     \bs{\alpha}\cdt\sfb{H} \cdt \sfb{u} \,=\, \sfb{H}(\bs{\alpha},\sfb{u}) \,=\, \sfb{u}\cdt\trn{\sfb{H}}\cdt\bs{\alpha}\,=\, \trn{\sfb{H}}(\sfb{u},\bs{\alpha}) 
     \,=\, H^i_j\alpha_i u^j  
    &\in \fun(\man{X}) 
\end{array}  
\end{align}
The expressions using parenthesis is the more common convention in other sources. We use any of the above interchangeably. 
Note the transpose (defined in the next point) \eq{\trn{\sfb{H}}=H^i_j \bep^j\otimes \sfb{e}_i} switches the ``slots'' of any 2-tensor.  Note that if a tensor (of any order) is fed only one argument, it is implied that it is fed into the right-most ``slot" such that, e.g., \eq{\sfb{H}(\sfb{u})=\sfb{H}\cdt\sfb{u}}.
We will also employ the dot product notation to indicate the \textbf{interior product} (often denoted with \eq{\iota} or \eq{\lrcorner\,}). For example, for some 2-form \eq{\bs{\beta}=-\trn{\bs{\beta}}=\ttfrac{1}{2}\beta_{ij}\bep^i\wedge\bep^j\in\forms^2(\man{X})}:
\begin{footnotesize}
\begin{align} \nonumber 
\begin{array}{lllll}
     \sfb{u}\cdt \bs{\beta}    := \iota_{\sfb{u}}\bs{\beta} = \bs{\beta}(\sfb{u},\slot)   
     = -\bs{\beta}\cdt\sfb{u} = -\bs{\beta}(\sfb{u})
     = \beta_{ij}u^i \bep^j 
\;\;,&
     \sfb{u}\cdt\bs{\beta} \cdt\sfb{v} :=\;   \iota_{\sfb{v}}(\iota_{\sfb{u}}\bs{\beta})   =  \bs{\beta}(\sfb{u},\sfb{v}) 
      = -\sfb{v}\cdt\bs{\beta}\cdt\sfb{u} =\ -\bs{\beta}(\sfb{v},\sfb{u})
     = \beta_{ij}u^i v^j  
\end{array}
\end{align} 
\end{footnotesize}
    \item \textbf{Transpose of \eq{2}-Tensors.} For (1,1)-tensors, unless specified otherwise, we take \eq{\sfb{H}\in\tens^1_1(\man{X})} to mean \eq{\sfb{H}_{\pt{x}}\in \tsp[\pt{x}]\man{X}\otimes \cotsp[\pt{x}]\man{X}} such that \eq{\trn{\sfb{H}}_{\pt{x}}\in \cotsp[\pt{x}]\man{X} \otimes\tsp[\pt{x}]\man{X}}. That is, in a local frame \eq{\sfb{e}_i\in\vect(\man{X})} and \eq{\bep^i \in\forms(\man{X})}, a (1,1)-tensor is assumed to be of the form \eq{\sfb{H}=H^i_j\sfb{e}_i\otimes\bep^j} such that its transpose is \eq{\trn{\sfb{H}} =H^i_j\bep^j \otimes \sfb{e}_i}.  The transpose of (0,2)-tensors and (2,0)-tensors is analogous to the usual matrix transpose; for some  \eq{\sfb{m} = m_{ij}\bep^i\otimes\bep^j\in\tens^0_2(\man{X})} and  some  \eq{ \sfb{k} = k^{ij}\sfb{e}_i\otimes\sfb{e}_j \in\tens^2_0(\man{X})}, we have \eq{\trn{\sfb{m}} = m_{ij}\bep^j\otimes\bep^i = m_{ji}\bep^i\otimes\bep^j} and \eq{ \trn{\sfb{k}} = k^{ij}\sfb{e}_j\otimes\sfb{e}_i = k^{ji}\sfb{e}_i\otimes\sfb{e}_j}.
    \item \textbf{Adjoint of $(1,1)$-Tensors.} If \eq{\man{X}} is a (pseudo)Riemannian manifold with metric tensor \eq{\sfg\in\tens^0_2(\man{X})}, then the \eq{\sfg}-adjoint of some \eq{\sfb{H}\in\tens^1_1(\man{X})} is a tensor field of the same type, denoted \eq{\sfb{H}^{\drg}\in\tens^1_1(\man{X})}, which is defined by the following for any \eq{\sfb{u},\sfb{v}\in\vect(\man{X})}: 
    \begin{align} \label{adj_def_0}
        \sfg(\sfb{H}\cdt\sfb{u},\sfb{v}) = \sfg(\sfb{u},\sfb{H}^{\drg} \cdt\sfb{v}) 
        \qquad\qquad i.e., \quad 
        \sfb{H}^{\drg} = \inv{\sfg}\cdt\trn{\sfb{H}}\cdt\sfg \in \tens^1_1(\man{X})
    \end{align}
    Note \eq{\sfb{H}^{\drg}_\pt{x}\in \tsp[\pt{x}]\man{X}\otimes \cotsp[\pt{x}]\man{X} } belongs to the same space as \eq{\sfb{H}_\pt{x}} such that \eq{\sfb{H}+\sfb{H}^{\drg}} is ``allowed'' (in contrast, \eq{\trn{\sfb{H}}_{\pt{x}}\in \cotsp[\pt{x}]\man{X} \otimes\tsp[\pt{x}]\man{X}} belongs to a different space and  \eq{\sfb{H}+\trn{\sfb{H}}} is undefined).  Similarly, for a smooth map \eq{\varphi:(\man{X},\sfg)\to(\man{Y},\sfb{h})} between (pseudo)Riemannian manifolds, the adjoint of the bilinear pairing \eq{\bs{\Phi}_\pt{x} := \dif \varphi_\pt{x}\in \tsp[\varphi(\pt{x})]\man{Y}\otimes\cotsp[\pt{x}]\man{X}} is denoted \eq{\bs{\Phi}^{\drg}_\pt{x}\in \tsp[\pt{x}]\man{X} \otimes \tsp[\varphi(\pt{x})]^*\man{Y} },  defined such that it satisfies the following for any \eq{\sfb{u}\in\vect(\man{X})} and \eq{\sfb{w}\in\vect(\man{Y})}:
    \begin{align}
        \fnsize{for } \; \bs{\Phi}=\dif\varphi \;: 
        &&
         \sfb{h}_{\varphi(\pt{x})}(\bs{\Phi}_\pt{x} \cdt \sfb{u},\sfb{w}) = \sfg_\pt{x}(\sfb{u}, \bs{\Phi}_\pt{x}^{\drg}\cdt\sfb{w})
         &&  i.e., \quad 
         \bs{\Phi}^{\drg}_\pt{x} := \inv{\sfg}_{\pt{x}} \cdt \trn{\bs{\Phi}_\pt{x}} \cdt \sfb{h}_{\varphi(\pt{x})}\in \tsp[\pt{x}]\man{X} \otimes \tsp[\varphi(\pt{x})]^*\man{Y}
    \end{align}
\end{itemize}
\end{footnotesize}

\end{appendices}







\end{document}